\documentclass[12pt,vi,oneside,makeidx]{mitthesis}
\usepackage{makeidx}
\usepackage{lgrind}
\usepackage{epsf}
\usepackage{amssymb}
\newcommand{\eref}[1]{(\ref{#1})}
\newcommand{\sref}[1]{\S\ref{#1}}
\newcommand{\tref}[1]{Table~\ref{#1}}
\newcommand{\fref}[1]{Figure~\ref{#1}}
\newcommand{\cref}[1]{Chapter~\ref{#1}}
\newcommand{\beq}{\begin{equation}}
\newcommand{\eeq}{\end{equation}}
\newcommand{\ba}{\begin{array}}
\newcommand{\ea}{\end{array}}
\newcommand{\bcenter}{\begin{center}}
\newcommand{\ecenter}{\end{center}}

\def\IB{\relax\hbox{$\inbar\kern-.3em{\rm B}$}}
\def\IC{\relax\hbox{$\inbar\kern-.3em{\rm C}$}}
\def\ID{\relax\hbox{$\inbar\kern-.3em{\rm D}$}}
\def\IE{\relax\hbox{$\inbar\kern-.3em{\rm E}$}}
\def\IF{\relax\hbox{$\inbar\kern-.3em{\rm F}$}}
\def\IG{\relax\hbox{$\inbar\kern-.3em{\rm G}$}}
\def\IGa{\relax\hbox{${\rm I}\kern-.18em\Gamma$}}
\def\IH{\relax{\rm I\kern-.18em H}}
\def\IK{\relax{\rm I\kern-.18em K}}
\def\IL{\relax{\rm I\kern-.18em L}}
\def\IP{\relax{\rm I\kern-.18em P}}
\def\IR{\relax{\rm I\kern-.18em R}}
\def\IZ{\relax\ifmmode\mathchoice
{\hbox{\cmss Z\kern-.4em Z}}{\hbox{\cmss Z\kern-.4em Z}}
{\lower.9pt\hbox{\cmsss Z\kern-.4em Z}}
{\lower1.2pt\hbox{\cmsss Z\kern-.4em Z}}\else{\cmss Z\kern-.4em Z}\fi}
\def\II{\relax{\rm I\kern-.18em I}}

\def\sCC{{\kern 0.27em\vrule height1.45ex width0.03em depth0em
          \kern-0.30em\rm C}}
\def\C{{\mathchoice
  {\sCC}
  {\sCC}
  {\kern 0.225em \vrule height1.05ex width0.025em depth0em \kern-0.25em \rm C}
  {\kern 0.180em \vrule height0.78ex width0.02em depth0em \kern-0.2em \rm C}
        }}
\def\sHH{{\rm I\kern-.16em{}H}}
\def\H{{\mathchoice
  {\sHH}
  {\sHH}
  {\rm I\kern-.13em{}H}
  {\rm I\kern-.13em{}H} }}
\def\sNN{{\rm I\kern-.16em{}N}}
\def\N{{\mathchoice
  {\sNN}
  {\sNN}
  {\rm I\kern-.12em{}N}
  {\rm I\kern-.10em{}N} }}
\def\sPP{{\rm I\kern-.16em{}P}}
\def\P{{\mathchoice
  {\sPP}
  {\sPP}
  {\rm I\kern-.12em{}P}
  {\rm I\kern-.10em{}P} }}
\def\sQQ{{\kern 0.27em \vrule height1.45ex width0.03em depth0em
          \kern-0.30em \rm Q}}
\def\Q{{\mathchoice
        {\sQQ}
        {\sQQ}
  {\kern 0.225em \vrule height1.05ex width0.025em depth0em \kern-0.25em \rm Q}
  {\kern 0.180em \vrule height0.78ex width0.020em depth0em \kern-0.20em \rm Q}
        }}
\def\sRR{{\rm I\kern-0.16em{}R}}
\def\R{{\mathchoice
  {\sRR}
  {\sRR}
  {\rm I\kern-0.12em{}R}
  {\rm I\kern-0.10em{}R} }}
\def\sZZ{{\rm Z\kern-0.32em{}Z}}
\def\Z{{\mathchoice
  {\sZZ}
  {\sZZ} 
  {\rm Z\kern-0.3em{}Z}     
  {\rm Z\kern-0.25em{}Z} }}  
\def\ZZZ{{\rm Z\kern-0.24em{}Z}}
\def\sII{{\rm I\kern-0.16em{}I}}
\def\I{{\mathchoice
  {\sII}
  {\sII}
  {\rm I\kern-0.12em{}I}
  {\rm I\kern-0.10em{}I} }}

\def\Tr{{\rm Tr}}

\def\Hom{{\rm Hom}}
\def\dim{{\rm dim}}

\def\inbar{\,\vrule height1.5ex width.4pt depth0pt}
\font\cmss=cmss10 \font\cmsss=cmss10 at 7pt

\def\odd{{\rm odd}}
\def\even{{\rm even}}
\def\rtimes{\mbox{$\times\!\rule{0.3pt}{1.1ex}\,$}}

\def\smiley{\hbox{\large$\bigcirc$\hspace{-0.80em}\raise.2ex
\hbox{$\cdot\cdot$}\kern-.61em\lower.2ex\hbox{\scriptsize$\smile$}}\ }
\def\frowny{\hbox{\large$\bigcirc$\hspace{-0.80em}\raise.2ex
\hbox{$\cdot\cdot$}\kern-.635em\lower.2ex\hbox{\scriptsize$\frown$}}\ }

\def\I{{\rlap{1} \hskip 1.6pt \hbox{1}}}
\def\sq{\framebox(10,10){~}}
\newcommand{\gen}[1]{\langle #1 \rangle}
\newcommand{\mat}[1]{\left( \matrix{#1} \right)}
\newcommand{\tmat}[1]{{\scriptsize \mat{#1}}}
\makeatletter
\let\hangafter\@hangfrom
\makeatother

\newcommand{\drawsquare}[2]{\hbox{%
\rule{#2pt}{#1pt}\hskip-#2pt
\rule{#1pt}{#2pt}\hskip-#1pt
\rule[#1pt]{#1pt}{#2pt}}\rule[#1pt]{#2pt}{#2pt}\hskip-#2pt
\rule{#2pt}{#1pt}}

\newcommand{\fund}{\raisebox{-.5pt}{\drawsquare{6.5}{0.4}}}
\newcommand{\antifund}{\overline{\fund}}

\newtheorem{definition}{\sf DEFINITION}

\newtheorem{lemma}{\sf LEMMA}

\newtheorem{theorem}{\sf THEOREM}

\newtheorem{proposition}{\sf PROPOSITION}

\newtheorem{corollary}{\sf COROLLARY}

\newtheorem{conjecture}{\sf CONJECTURE}

\newtheorem{observation}{\sf OBSERVATION}

\input psfig.sty
\makeindex

\newfont{\yinit}{yinit scaled \magstep5}        

\newcommand{\hs}[1]{\hspace{#1 mm}}
\newcommand{\ra}{\rightarrow}
\def\cK{{\cal K}}
\def\ctg{\C^2/\Gamma}

\def\mt{\widetilde{M}}
\def\GZD{$Z_k\times D_{k'}~$}

\newcommand{\bean}{\begin{eqnarray*}}
\newcommand{\eean}{\end{eqnarray*}}
\newcommand{\barrayn}{\begin{eqnarray*}}
\newcommand{\earrayn}{\end{eqnarray*}}
\newcommand{\be}{\begin{equation}}
\newcommand{\ee}{\end{equation}}
\newcommand{\bea}{\begin{eqnarray}}
\newcommand{\eea}{\end{eqnarray}}
\def\beqa{\begin{eqnarray}}
\def\eeqa{\end{eqnarray}}

\def\Asym#1#2{\vcenter{\vbox{\drawbox{#1}{#2}
              \kern-#2pt
              \drawbox{#1}{#2}}}}

\def\beq{\begin{equation}}

\def\eeq{\end{equation}}

\def\Tr{\mathop{\rm Tr}}

\def\eref#1{(\ref{#1})}

\newcommand{\bee}[1]{\left(\begin{array}{cc} i^{#1} & 0 \\ 0 & i^{#1}
	\\ \end{array} \right)}

\newcommand{\EPSFIGURE}[3][v]{\begin{figure}[#1]\psfig{file=#2}
                                        \caption{#3}\end{figure}}

\newcount\figno
\def\fig#1#2#3{
\par\begingroup\parindent=0pt\leftskip=1cm\rightskip=1cm\parindent=0pt
\baselineskip=11pt
\global\advance\figno by 1
\epsfxsize=#3
\centerline{\epsfbox{#2}}
\vskip 12pt
{\bf Figure \the\figno:} #1\par
\endgroup\par
}
\def\figlabel#1{\xdef#1{\the\figno
\mbox{ }}}
\def\encadremath#1{\vbox{\hrule\hbox{\vrule\kern8pt\vbox{\kern8pt
\hbox{$\displaystyle #1$}\kern8pt}
\kern8pt\vrule}\hrule}}
\def\drawbox#1#2{\hrule height#2pt
        \hbox{\vrule width#2pt height#1pt \kern#1pt
              \vrule width#2pt}
              \hrule height#2pt}
\begin{document}
%
%
%
%
%
%
%
\title{{On Algebraic Singularities, Finite Graphs and D-Brane Gauge
Theories: A String Theoretic Perspective}}
\author{Yang-Hui He}
\department{Department of Physics}
\degree{Doctor of Philosophy in Physics}
\degreemonth{May}
\degreeyear{2002}
\thesisdate{May 18, 2002}



\chairman{Robert Jaffe}{Director, Center for Theoretical Physics}
\chairman{Thomas J. Greytak}{Professor, Associate Department Head for
Education}





\cleardoublepage

\setcounter{savepage}{\thepage}
\begin{abstractpage}
In this writing we shall address certain beautiful inter-relations
between the construction of 4-dimensional supersymmetric gauge
theories and resolution of algebraic singularities, 
from the perspective of String Theory. We review in some
detail the requisite background in both the mathematics, such as
orbifolds, symplectic quotients and quiver representations, as well as
the physics, such as gauged linear sigma models, geometrical
engineering, Hanany-Witten setups and D-brane probes.

We investigate aspects of world-volume gauge dynamics
using D-brane resolutions of various Calabi-Yau singularities, notably
Gorenstein quotients and toric singularities. Attention will be paid
to the general methodology of constructing gauge theories for these
singular backgrounds, with and without the presence of the NS-NS
B-field, as well as the T-duals to brane setups and branes wrapping
cycles in the mirror geometry. 
Applications of such diverse and elegant mathematics as
crepant
resolution of algebraic singularities, representation of finite groups
and finite graphs, modular invariants of affine Lie algebras, 
etc.~will naturally arise.
Various viewpoints and generalisations of
McKay's Correspondence will also be considered.

The present work is a transcription of excerpts from the first three
volumes of the author's PhD thesis which was written under 
the direction of Prof.~A.~Hanany - to whom he is much indebted - 
at the Centre for Theoretical Physics of MIT, and which, at the
suggestion of friends, he posts to the ArXiv pro hac vice; it is
his sincerest wish that the ensuing pages 
might be of some small use to the beginning student.
\end{abstractpage}


\cleardoublepage
\chapter*{Pr{\ae}fatio et Agnitio}
\begin{flushright}
Forsan et haec olim meminisse iuvabit. {\it Vir. Aen. I.1.203}
\end{flushright}

{\hspace{-0.5in}{\yinit N}}{\it
ot that I merely owe this title to the font, my education,
or the clime 
wherein I was born, as being bred up either to confirm
those principles my parents instilled into my understanding, or by a 
general consent
proceed in the religion of my country; but having, in my riper years 
and confirmed judgment,
seen and examined all, I find myself
obliged, by the principles of grace, and the law of mine
own reason, to embrace no other name but this.}

So wrote Thomas Browne in {\it Religio Medici} of his conviction
to his Faith.
Thus too let me, with regard to
that title of ``Physicist,'' of which alas I am most unworthy,
with far less wit but with
equal devotion, confess my allegiance to the noble Cause of Natural Philosophy,
which I pray that in my own riper years I shall embrace none other.
Therefore prithee gentle reader, bear with this fond fool as he here leaves
his rampaging testimony to your clemency.

Some nine years have past and gone, since when the good Professor 
H.~Verlinde, of Princeton,
first re-embraced me from my straying path,
as Saul was upon the road to Damascus
- for, Heaven forbid, that in the
even greater folly of my youth I had once blindly fathomed to be my destiny 
the more pragmatic
career of an Engineer (pray mistake me not, as I hold great esteem for this
Profession, though had I pursued her my own heart and soul would have been 
greatly misplaced indeed)
- to the Straight and Narrow path leading to Theoretical Physics,
that Holy Grail of Science.

I have suffered, wept and bled sweat of labour. 
Yet the divine Bach reminds us in the Passion of Our Lord according to
Matthew,
``{\it Ja! Freilich will in uns das Fleisch und Blut zum Kreuz gezwungen
sein; Je mehr es unsrer Seele gut, Je herber geht es ein.}''
Ergo, I too have rejoiced, laughed and shed tears of
jubilation. Such is the nature of Scientific Research, and indeed the
grand {\it Principia Vit\ae}. These past half of a decade has been
constituted of thousands of nightly lucubrations, 
each a battle, each {\it une
petite mort}, each with its {\it te Deum} and {\it Non Nobis
Domine}. I carouse to these five years past, short enough to be one
day deemed a mere passing period, long enough 
to have earned some silvery strands upon my idle rank.

And thus commingled, the {\it fructus labori} of these years past,
is the humble work I shall present in the
ensuing pages. I beseech you o gentle reader, to indulge its length, I
regret to confess that what I lack in content I can only supplant with
volume, what I lack in wit I can only distract with loquacity. To
that great Gaussian principle of {\it Pauca sed Matura} let me
forever bow in silent shame.

Yet the poorest offering does still beseech painstaking preparation
and the lowliest work, a helping hand. How blessed I am, to have a
flight souls aiding me in bearing the great weight!

For what is a son, without the wings of his parent? How blessed I am,
to have my dear mother and father, my aunt DaYi and grandmother,
embrace me with four-times compounded love! Every fault, a tear, every
wrong, a guiding hand and every triumph, an exaltation.

For what is Dante, without his Virgil? How blessed I am, to have the
perspicacious guidance of the good Professor Hanany, who in these 4
years has taught me so much! His ever-lit lamp and his ever-open door
has been a beacon for home amidst the nightly storms of life and
physics.
In addition thereto, I am indebted to
Professors Zwiebach, Freedman and Jaffe, together with all
my honoured Professors and teachers, as well as the ever-supportive staff:
J.~Berggren, R.~Cohen, S.~Morley and E.~Sullivan 
at the Centre for Theoretical Physics, to have
brought me to my intellectual manhood.

For what is Damon, without his Pythias? How blessed I am, to have such
multitudes of friends! I drink to their health!
To the Ludwigs: my brother, mentor and colleague in
philosophy and mathematics, 
J.~S.~Song and his JJFS; my brother and companion in wine
and Existentialism, N.~Moeller and his Marina. To my collaborators: my
colleagues and brethren, B.~Feng, I.~Ellwood, A.~Karch, N.~Prezas and
A.~Uranga. To my brothers in
Physics and remembrances past: I.~Savonije and M.~Spradlin, may that noble
Nassau-Orange thread bind the colourless skeins of our lives.
To my
Spiritual counsellors: M.~Serna and his ever undying passion for Physics, 
D.~Matheu and his Franciscan soul, L.~Pantelidis and his worldly wisdom,
as well as the Schmidts and the Domesticity which they symbolise. 
To the fond memories of one beauteous adventuress
Ms.~M.~R.~Warden, who once wept with me at the times of sorrow and  
danced with me at the moments of delight.
And to
you all my many dear beloved friends whose names, though I could not
record here, I shall each and all engrave upon my heart.


And so composed is a fledgling, through these many years of hearty battle, 
and amidst blood, sweat and tears was formed another grain of sand
ashore the Vast Ocean of Unknown.  Therefore at this eve of my
reception of the title {\it Doctor Philosophiae}, though I myself
could never dream to deserve to
be called either ``learned'' or a ``philosopher,'' I shall fast and pray,
for henceforth I shall bear, as Atlas the weight of Earth upon his
shoulders, the name ``Physicist'' upon my soul.
And so I shall prepare for this
my initiation into a Brotherhood of Dreamers, 
as an incipient
neophyte intruding into a Fraternity of Knights, accoladed by the
sword of {\it Regina Mathematica}, who dare to uphold that Noblest calling
of {\it ``Sapere Aude''}.

Let me then embrace, not with merit but with homage, not with 
arms eager but with
knees bent, and indeed not with a mind deserving
but with a heart devout, naught else but this dear 
cherished Title of ``Physicist.''

I call upon ye all, gentle readers, my brothers and sisters, all
the Angels and Saints, and Mary, ever Virgin, to pray for me, {\it Dei
Sub Numine}, as I dedicate this humble work and my worthless self,

{\it Ad Catharinae Sanctae Alexandriae et Ad Majorem Dei Gloriam...}
%
%
\chapter*{Invocatio et Apologia}
{\vspace{-2.5in} {\bf {\it {\large De Singularitatis Algebraic\ae, 
	Graphic\ae~Finitatis, \& Theorica Mensur\ae~
 	Bran\ae~Dirichletiensis: Aspectus Theoric\ae~Chord\ae,}
	cum digressi super theorica campi chordae.
	Libellus in Quattuor Partibus, sub Auspicio CTP et LNS, MIT,
	atque DOE et NSF, sed potissimum, Sub Numine Dei.\\
\begin{flushright}
\vspace{0.5in}
	Y.-H. E. He\\
	B.~A.,~ Universitatis Princetoniensis\\
	Math.~Tripos, Universitatis Cantabrigiensis
\end{flushright}
}}}
{\yinit2 W}e live in an Age of Dualism. The Absolutism which has so long
permeated through Western Thought has been challenged in every
conceivable fashion: from philosophy to politics, from religion to
science, from sociology to aesthetics. The ideological conflicts, so
often ending in tragedy and so much a theme of
the twentieth century, had been intimately tied with the recession of
an archetypal norm of undisputed Principles.
As we enter the third millennium, the Zeitgeist is already suggestive
that we shall perhaps no longer be victims but beneficiaries, that the
uncertainties which haunted and devastated the proceeding century shall
perhaps serve to guide us instead. 

Speaking within the realms of
Natural Philosophy, beyond the wave-particle duality or the
Principle of Equivalence, is a product which originated in the 60's
and 70's, a product which by now so well exemplifies a dualistic
philosophy to its very core.

What I speak of, is the field known as String Theory, initially
invented to explain the dual-resonance behaviour of hadron
scattering. The dualism which I emphasise is more than the fact that the
major revolutions of the field, string duality and D-branes, AdS/CFT
Correspondence, etc., all involve dualities in a strict sense,
but more so the fact that the essence of the field still remains to be
defined. A chief theme of this writing shall be the dualistic nature of
String theory as a scientific endeavour: it has thus far no experimental
verification to be rendered physics and it has thus far no rigorous
formulations to be considered mathematics.
Yet String theory has by now inspired so much activity in both physics
and mathematics that, to quote C.~N.~Yang in the early days of
Yang-Mills theory, its beauty alone certainly merits our attention.

I shall indeed present you with breath-taking beauty;
in Books I and II, I shall carefully guide the readers, be them physicists
or mathematicians, to a preparatory journey to the requisite
mathematics in Liber I and to physics in Liber II. These two books
will attempt to review a tiny fraction of the many subjects
developed in the last few decades in both fields in relation to string
theory.
I quote here a saying of E.~Zaslow of which I am
particularly fond, though it applies to me far more appropriately: 
in the
Book on mathematics I shall be the physicist and the Book on physics,
I the mathematician, so as to beg the reader to forgive my inexpertise
in both.

Books III and IV shall then consist of some of my work during my very
enjoyable stay at the Centre for Theoretical Physics at MIT as a
graduate student. I regret that I shall tempt the
readers with so much elegance in the first two books and yet lead them
to so humble a work, that the journey through such a beautiful garden
would end in such a witless swamp. And I take the opportunity to
apologise again to the reader for the excruciating length, full of
sound and fury and signifying nothing. Indeed as Saramago points out
that the shortness of life is so incompatible with the verbosity of 
the world.

Let me speak no more and let our journey begin.
Come then, ye Muses nine, and with strains divine call upon mighty
Diane, that she, from her golden quiver may draw the arrow, to pierce
my trembling
heart so that it could bleed the ink with which I shall hereafter
compose this my humble work...

\pagestyle{plain}
\tableofcontents

\chapter{INTROIT}
{\vspace{-3in} {\bf {\it {\large De Singularitatis Algebraic\ae, 
	Graphic\ae~Finitatis, \& Theorica Mensur\ae~
 	Bran\ae~Dirichletiensis: Aspectus Theoric\ae~Chord\ae\\}
}}}
\vspace{2in}

{T}he two pillars of twentieth century physics, General
Relativity and
Quantum Field Theory, have brought about tremendous progress in
Physics. The former has described the macroscopic, and the latter, the
microscopic, to beautiful precision. However, the pair, in and of
themselves, stand incompatible. Standard techniques of establishing a
quantum theory of gravity have met uncancellable divergences and
unrenormalisable quantities.

As we enter the twenty-first century, a new theory, born in the
mid-1970's, has promised to be a
candidate for a Unified Theory of Everything. The theory is known
as {\bf String Theory}, whose basic tenet is that all particles are
vibrational modes of strings of Plankian length.
Such elegant
structure as the natural emergence of the graviton and embedding of
electromagnetic and large $N$ dualities, has made the theory more and
more attractive to the theoretical physics community. Moreover,
concurrent with its development in physics, string theory has prompted
enormous excitement among mathematicians. Hitherto unimagined
mathematical
phenomena such as Mirror Symmetry and orbifold cohomology have brought
about many new directions in algebraic geometry and representation
theory.

Promising to be a Unified Theory, string theory must incorporate the
Standard Model of interactions, or minimally supersymmetric extensions
thereof. The purpose of this work is to study various aspects of a wide
class of gauge theories arising from string theory in the background
of singularities, their dynamics, moduli spaces,
duality transformations etc.~as well as certain branches of associated
mathematics. We will investigate how these gauge theories, of various
supersymmetry and in various dimensions, arise as low-energy effective
theories associated with hypersurfaces in String Theory known as
D-branes.

It is well-known that the initial approach of constructing the real
world from String Theory had been the
compactification of the 10 dimensional superstring or the 10(26)
dimensional heterotic string on Calabi-Yau manifolds of complex
dimension three. These are complex manifolds described as
algebraic varieties with Ricci-flat
curvature so as to
preserve supersymmetry.
The resulting theories are ${\cal N}=1$ supersymmetric
gauge theories in 4 dimensions
that would be certain minimal extensions of the Standard Model.

This paradigm has been widely pursued since the 1980's. However, we
have a host of Calabi-Yau threefolds to choose from.
The inherent length-scale of the superstring and deformations of
the world-sheet conformal field theory,
made such violent behaviour as topology changes in space-time
natural. These changes connected vast classes of manifolds related by,
notably, mirror symmetry. For the physics, these mirror manifolds
which are markedly different mathematical objects, give rise to the
same conformal field theory.

Physics thus became equivalent with respect to various
different compactifications. Even up to this equivalence, the plethora
of Calabi-Yau threefolds (of which there is still yet no
classification) renders the precise choice of the compactification
difficult to select. A standing problem then has been this issue of
``vacuum degeneracy.''

Ever since Polchinski's introduction of
D-branes into the arena in the Second String Revolution of the
mid-90's, numerous novel techniques appeared in
the construction of gauge theories of various supersymmetries, as
low-energy effective theories of the ten dimensional superstring and
eleven dimensional M-theory (as well as twelve dimensional
F-theory). 

The natural existence of such higher dimensional surfaces from a
theory of strings proved to be crucial. The Dp-branes as well as
Neveu-Schwarz (NS) 5-branes are carriers of Ramond-Ramond and
NS-NS charges, with electromagnetic duality (in 10-dimensions) between
these charges (forms). Such a duality is well-known in supersymmetric
field theory, as exemplified by the four dimensional Montonen-Olive
Duality for ${\cal N}=4$, Seiberg-Witten for ${\cal N}=2$ and
Seiberg's Duality for ${\cal N}=1$.  These dualities are closely
associated with the underlying S-duality in the full string theory,
which maps small string coupling to the large.

Furthermore, the inherent winding modes of the string includes another
duality contributing to the dualities in the field theory, 
the so-called T-duality where small compactification radii
are mapped to large radii. By chains of applications of S and T
dualities, the Second Revolution brought about a unification of the
then five disparate models of consistent String Theories: types I,
IIA/B, Heterotic $E_8 \times E_8$ and Heterotic Spin$(32)/\IZ_2$.

Still more is the fact that these branes are actually solutions in
11-dimensional supergravity and its dimensional reduction to
10. Subsequently proposals for the enhancement for the S and T
dualities to a full so-called U-Duality were conjectured. This would
be a symmetry of a mysterious underlying M-theory of which the unified
string
theories are but perturbative limits. Recently Vafa and collaborators
have proposed even more intriguing dualities where such U-duality
structure is intimately tied with the geometric structure of blow-ups
of the complex projective 2-space, viz., the del Pezzo surfaces.

With such rich properties, branes will occupy a central theme in
this writing. We will exploit such facts as their being BPS states
which break supersymmetry, their dualisation to various pure
geometrical backgrounds and their ability to probe sub-stringy
distances. We will investigate how to construct gauge theories
empowered with them, how to realise dynamical processes in field
theory such as Seiberg duality in terms of
toric duality and brane motions, how to
study their associated open string states in bosonic string field
theory as well as many interesting mathematics that emerge.

We will follow the thread of thought of the trichotomy of methods of
fabricating low-energy effective super-Yang-Mills theories
which soon appeared in quick succession in 1996, after the D-brane
revolution.

One method was very much in the geometrical vein of
compactification: the so-named {\bf geometrical engineering} of
Katz-Klemm-Lerche-Vafa. With branes of various dimensions at their
disposal, the authors wrapped (homological) cycles in the Calabi-Yau
with branes of the corresponding
dimension. The supersymmetric cycles (i.e., cycles which preserve
supersymmetry), especially the middle dimensional
3-cycles known as Special Lagrangian submanifolds, play a crucial
r\^ole in Mirror Symmetry. 

In the context of constructing gauge
theories, the world-volume theory of the wrapped branes are described
by dimensionally reduced gauge theories inherited from the original
D-brane and supersymmetry is preserved by the special properties of
the cycles. Indeed,
at the vanishing volume limit gauge
enhancement occurs and a myriad of supersymmetric Yang-Mills theories
emerge. In this spirit, certain global issues in compactification
could be addressed in the analyses of the local behaviour of the
singularity arising from the vanishing cycles, whereby making
much of the geometry tractable.

The geometry of the homological cycles, together with the wrapped
branes, determine the precise gauge group and matter content. In the
language of sheafs, we are studying the intersection theory of
coherent sheafs associated with the cycles. We will make usage of
these techniques in the study of such interesting behaviour as ``toric
duality.'' 

The second method of engineering four dimensional gauge theories from
branes
was to study the world-volume theories of configurations of branes in
10 dimensions.
Heavy use were made
especially of the D4 brane of type IIA, placed in a specific position
with respect to various D-branes and the solitonic NS5-branes.
In the limit of low energy, the world-volume theory becomes a
supersymmetric gauge theory in 4-dimensions. 

Such configurations,
known as {\bf Hanany-Witten setups}, provided intuitive realisations of
the gauge theories. Quantities such as coupling constants and beta
functions were easily visualisable as distances and bending of the
branes in the setup. Moreover, the configurations lived directly in the
flat type II background and the intricacies involved in the
curved compactification spaces could be avoided altogether.

The open strings stretching between the branes realise as the
bi-fundamental and adjoint matter of the resulting theory while the
configurations are chosen judiciously to break down to appropriate
supersymmetry. Motions of the branes relative to each other correspond
in the field theory to moving along various Coulomb and Higgs branches
of the Moduli space. Such dynamical processes as the Hanany-Witten
Effect of 
brane creation lead to important string theoretic realisations
of Seiberg's duality.

We shall too take advantage of the insights offered by this technique
of brane setups which make quantities of the product gauge theory
easily visualisable.

The third method of engineering gauge theories
was an admixture of the above two, in the sense of
utilising both brane dynamics and singular geometry. This became known
as the {\bf brane probe} technique, initiated by Douglas and
Moore. Stacks of parallel D-branes were placed near certain local
Calabi-Yau manifolds; the world-volume theory, which would otherwise
be the uninteresting parent $U(n)$ theory in flat space, was projected
into one with product gauge groups, by the geometry of the singularity
on the open-string sector.

Depending on chosen action of the singularity, notably orbifolds, with
respect to the $SU(4)$ R-symmetry of the parent theory, various
supersymmetries can be achieved. When we choose the singularity to be
$SU(3)$ holonomy,
a myriad of gauge theories of ${\cal N}=1$
supersymmetry in 4-dimensions could be thus fabricated given local
structures of the algebraic singularities. The moduli space, as solved
by the vacuum conditions of D-flatness and F-flatness in the field
theory, is then by construction, the Calabi-Yau singularity. In this
sense space-time itself becomes a derived concept, as realised by the
moduli space of a D-brane probe theory.

As Maldacena brought about
the Third String Revolution with the AdS/CFT conjecture in 1997, new
light shone upon these probe theories. Indeed the $SU(4)$ R-symmetry
elegantly manifests as the $SO(6)$ isometry of the 5-sphere in the
$AdS_5 \times S^5$ background of the bulk string theory.
It was soon
realised by Kachru, Morrison, Silverstein et al.~that these probe
theories could be harnessed as numerous checks for the correspondence
between gauge theory and near horizon geometry.

Into various aspects of these probes theories we shall delve
throughout the writing and attention will be paid to two classes of
algebraic singularities, namely orbifolds and toric singularities,

With the wealth of dualities in String Theory
it is perhaps of no surprise that the three methods introduced above 
are equivalent
by a sequence of T-duality (mirror) transformations. Though we shall make
extensive usage of the techniques of all three throughout this writing,
focus will be on the latter two, especially the last.
We shall elucidate these three main ideas:
geometrical engineering, Hanany-Witten brane configurations and
D-branes transversely probing algebraic singularities,
respectively in Chapters 6, 7 and 8 of Book II.

The abovementioned,
of tremendous interest to the physicist, is only half the story.
In the course of this study of compactification on Ricci-flat
manifolds, beautiful and unexpected mathematics were
born. Indeed, our very understanding of classical geometry underwent
modifications and the notions of ``stringy'' or ``quantum'' geometry
emerged. 
Properties of algebro-differential geometry of the target
space-time manifested as the supersymmetric conformal field theory on
the world-sheet. Such delicate calculations as counting of holomorphic
curves and intersection of homological cycles mapped elegantly to
computations of world-sheet instantons and Yukawa couplings. 

The mirror principle, initiated by Candelas et al.~in the early 90's,
greatly simplified the aforementioned computations. Such unforeseen
behaviour as pairs of Calabi-Yau manifolds whose Hodge diamonds were
mirror reflections of each other naturally arose as spectral flow in the
associated world-sheet conformal field theory. Though we shall too
make usage of versions of mirror symmetry, viz., the {\bf local mirror},
this writing will not venture too much into the elegant
inter-relation between
the mathematics and physics of string theory through mirror geometry.

What we shall delve into, is the local model of Calabi-Yau
manifolds. These are the {\em algebraic singularities} of which we
speak. In particular we concentrate on {\bf canonical Gorenstein}
singularities that admit crepant resolutions to smooth Calabi-Yau
varieties. In particular, attention will be paid to orbifolds, i.e.,
quotients of flat space by finite groups, as well as 
toric singularities,
i.e., local behaviour of toric varieties near the singular point.

As early as the mid 80's, the string partition function 
of Dixon-Harvey-Vafa-Witten (DHVW)
proposed a resolution of orbifolds then unknown to the
mathematician and made elegant predictions on the Euler 
characteristic of orbifolds.
These gave new directions to such remarkable observations as the {\bf
McKay Correspondence} and its generalisations to beyond dimension 2 and
beyond du Val-Klein singularities. Recent work by Bridgeland, King,
and Reid on the generalised McKay from the derived category of
coherent sheafs also tied deeply with similar structures arising in
D-brane technologies as advocated by Aspinwall, Douglas et al.
Stringy orbifolds thus became a
topic of pursuit by such noted mathematicians as Batyrev, Kontsevich
and Reid.

Intimately tied thereto, were applications of the construction of
certain hyper-K\"ahler quotients, which
are themselves moduli spaces of certain gauge theories, as
gravitational instantons. The works by Kronheimer-Nakajima placed
the McKay Correspondence under the light of representation theory of
{\em quivers}. Douglas-Moore's construction mentioned above for the
orbifold gauge theories thus brought these quivers into a string
theoretic arena.

With the technology of D-branes to probe sub-stringy
distance scales, Aspinwall-Greene-Douglas-Morrison-Plesser 
made space-time
a derived concept as moduli space of world-volume theories. 
Consequently, novel perspectives arose, in the understanding of the
field known as
Geometric Invariant Theory (GIT), in the light of gauge invariant
operators 
in the gauge theories on the D-brane. Of great significance, was the
realisation that the Landau-Ginzberg/Calabi-Yau correspondence in the
linear sigma model of Witten, could be used to translate
between the gauge theory as a world-volume theory
and the moduli space as a {\bf GIT quotient}. 

In the case of toric varieties,
the sigma-model fields corresponded nicely to 
generators of the homogeneous co\"ordinate ring in the language of
Cox. This provided us with a alternative and computationally feasible
view from the traditional approaches to toric varieties. We shall take
advantage of this fact when we deal with toric duality later on.

This work will focus on how the
above construction of gauge theories leads to various intricacies in
algebraic geometry, representation theory and finite graphs, and vice
versa, how we could borrow techniques from the latter to address the
physics of the former.
In order to refresh
the reader's mind on the requisite mathematics, 
Book I is devoted to a review on the
relevant topics. Chapter 2 will be an overview of the geometry,
especially algebraic singularities and Picard-Lefschetz theory. Also
included will be a discussion on symplectic quotients as well as the
special case of toric varieties. Chapter 3 then prepares the reader
for the orbifolds, by reviewing the pertinent concepts from
representation theory of finite groups. Finally in Chapter 4, a
unified outlook is taken by studying quivers as well as the
constructions of Kronheimer and Nakajima.

Thus prepared with the review of the
mathematics in Book I and the physics in II, 
we shall then take the reader to Books III and IV, 
consisting of some of the author's 
work in the last four years at the Centre for Theoretical Physics at
MIT.

We begin with the D-brane probe picture. In Chapters
\ref{chap:9811183} and \ref{chap:9905212} we classify and study the
singularities of the orbifold type by discrete subgroups of $SU(3)$
and $SU(4)$ \cite{9811183,9905212}. The resulting physics consists of
catalogues of finite four dimensional Yang-Mills theories with 1 or 0
supersymmetry. These theories are nicely encoded by certain finite
graphs known as {\bf quiver diagrams}.
This generalises the work of
Douglas and Moore for abelian ALE spaces and subsequent work by
Johnson-Meyers for all ALE spaces as orbifolds of $SU(2)$. Indeed
McKay's Correspondence facilitates the ALE case; moreover the
ubiquitous ADE meta-pattern, emerging in so many seemingly unrelated
fields of mathematics and physics greatly aids our understanding.
 
In our work, as we move from two-dimensional quotients to three and
four dimensions, interesting
observations were made in relation to generalised McKay's
Correspondences. Connections to Wess-Zumino-Witten
models that are conformal field theories on the world-sheet,
especially the remarkable resemblance of the McKay graphs from the
former and fusion graphs from the latter were conjectured in
\cite{9811183}. Subsequently, a series of activities were initiated in
\cite{9903056,9911114,0009077} to attempt to address why weaker
versions of the complex of dualities which exists in dimension two may
persist in higher dimensions. Diverse subject matters such as
symmetries of the modular invariant
partition functions, graph algebras of the conformal field theory,
matter content of the probe gauge theory and crepant resolution of
quotient singularities all contribute to an intricate web of
inter-relations. Axiomatic approaches such as the quiver and ribbon
categories were also attempted.
We will discuss these issues in Chapters
\ref{chap:9903056}, \ref{chap:9911114} and \ref{chap:0009077}.

Next we proceed to address the T-dual versions of these D-brane probe
theories in terms of Hanany-Witten configurations. As mentioned
earlier, understanding these would greatly enlighten the understanding
of how these gauge theories embed into string theory. With the help of
orientifold planes, we construct the
first examples of non-Abelian configurations for $\IC^3$ orbifolds
\cite{9906031,9909125}. These are direct generalisations of the
well-known elliptic models and brane box models, which are a widely
studied class of conformal theories. These constructions will be the
theme for Chapters \ref{chap:9906031} and
\ref{chap:9909125}. 

Furthermore, we discuss the steps towards a
general method \cite{0012078}, which we dubbed as ``stepwise
projection,'' of finding Hanany-Witten setups for
arbitrary orbifolds in Chapter \ref{chap:0012078}. With the help of
Fr{\o}benius' induced representation theory, the stepwise procedure of
systematically obtaining non-Abelian gauge theories from the Abelian
theories,
stands as a non-trivial step towards solving the general problem of
T-dualising pure geometry into Hanany-Witten setups.

Ever since Seiberg and Witten's  realisation that the NS-NS B-field of
string theory, turned on along world-volumes of D-branes, leads to
non-commutative field theories, a host of activity ensued. In our
context, Vafa generalised the DHVW closed sector orbifold partition
function to include phases associated with the B-field. Subsequently,
Douglas and Fiol found that the open sector analogue lead to projective
representation of the orbifold group.

This inclusion of the background B-field has come to be known as
turning on {\bf discrete torsion}. Indeed a corollary of a theorem due to
Schur tells us that orbifolds of dimension two, i.e., the ALE spaces
do not admit such turning on. This is in perfect congruence with the
rigidity of the ${\cal N}=2$ superpotential. For ${\cal N}=0,1$
theories however, 
we can deform the superpotential consistently and arrive at
yet another wide class of field theories.

With the aid of such elegant mathematics as the Schur multiplier,
covering groups and the Cartan-Leray spectral sequence, 
we systematically study how and when it is
possible to arrive at these theories with discrete torsion
by studying the projective
representations of orbifold groups \cite{0010023,0011192} in Chapters
\ref{chap:0010023} and \ref{chap:dis2}.

Of course orbifolds, the next best objects to flat
(complex-dimensional) space, are but one class of local Calabi-Yau
singularities. Another intensively studied class of algebraic
varieties are the so-called toric varieties. As finite group
representation theory is key to the former, combinatorial geometry of
convex bodies is key to the latter. It is pleasing to have such
powerful interplay between such esoteric mathematics and our
gauge theories.

We address the problem of constructing
gauge theories of a D-brane probe on toric
singularities \cite{0003085} in Chapter \ref{chap:0003085}.
Using the technique of partial resolutions pioneered by Douglas,
Greene and Morrison, we formalise a so-called ``Inverse Algorithm'' to
Witten's gauged linear sigma model approach and carefully investigate
the type of theories which arise given the type of toric singularity.

Harnessing the degree of freedom in the toric data in the above
method, we will encounter a surprising phenomenon which we call
{\bf Toric Duality}. \cite{0104259}. This in fact gives us an
algorithmic technique to engineer gauge theories which flow to the
same fixed point in the infra-red moduli space. The manifestation of
this duality as Seiberg Duality for ${\cal N}=1$ \cite{0109063} came
as an additional bonus. Using a combination of field theory
calculations, Hanany-Witten-type of brane configurations and the
intersection theory of the mirror geometry \cite{Amer2}, we check that
all the cases produced by our algorithm do indeed give Seiberg duals
and conjecture the validity in general \cite{multi}.
These topics will constitute
Chapters \ref{chap:0104259} and \ref{chap:0109063}.

All these intricately tied and inter-dependent themes of D-brane
dynamics, construction of four-dimensional gauge theories, 
algebraic singularities and quiver graphs, will be the subject of this
present writing.
\part{LIBER PRIMUS: Invocatio Mathematic\ae}
\chapter{Algebraic and Differential Geometry}
\section*{Nomenclature}
Unless otherwise stated, we shall adhere to the following notations
throughout the writing:

\begin{tabular}{ll}
$X$			&Complex analytic variety\\
$T_pX$, $T_p^*X$	& Tangent and cotangent bundles (sheafs) of $X$
				at point $p$\\
${\cal O}(X)$		&Sheaf of analytic functions on $X$\\
${\cal O}^*(X)$		&Sheaf of non-zero analytic functions on $X$\\
$\Gamma(X, {\cal O})$	&Sections of the sheaf (bundle) ${\cal O}$
			over $X$\\ 
$\Omega^{p,q}(X)$	&Dolbeault $(p,q)$-forms on $X$\\
$\omega_X$		&The canonical sheaf of $X$\\
$f : \tilde{X} \rightarrow X$ &Resolution of the singularity $X$\\
$\mathfrak{g} = Lie(G)$	&The Lie Algebra of the Lie group $G$\\
$\widetilde{\mathfrak{g}}$	&The Affine extension of $\mathfrak{g}$\\
$\mu : M \rightarrow Lie(G)^*$
			& Moment map associated with the group $G$\\
$\mu^{-1}(c) // G$	&Symplectic quotient associated with the moment
			map $\mu$\\
$|G|$			& The order of the finite group $G$\\
$\chi_\gamma^{(i)}(G)$	& Character for the $i$-th irrep in the $\gamma$-th
			conjugacy class of $G$
\end{tabular}

{\hspace{-0.5in}{ A}}s the subject matter of this work is on
algebraic singularities and
their applications to string theory, what better place to
commence our mathematical invocations indeed,
than a brief review on some rudiments of the vast field of
singularities in algebraic varieties. The material contained herein
shall be a collage from such canonical texts as
\cite{Griffith,Hartshorne,Reid1,Reid2}, to which the
reader is highly recommended to refer. 
\section{Singularities on Algebraic Varieties}
Let $M$ be an $m$-dimensional 
complex algebraic variety; we shall usually deal with projective
varieties and shall take $M$ to be $\IP^m$, the complex projective
$m$-space, with projective
co\"ordinates  $(z_1,\ldots, z_m) = [Z_0: Z_2: \ldots: Z_m] \in
\IC^{m+1}$. In general, by Chow's Theorem,
any analytic subvariety $X$ of $M$ 
can be locally given as the zeores of a finite collection of
holomorphic functions $g_i(z_1,\ldots, z_m)$.
Our protagonist shall then be the variety $X := \left\{z |
g_i(z_1,\ldots, z_m) = 0~~\forall~i=1, \ldots, k \right\}$, especially
the singular points thereof. The following definition shall
distinguish such points for us:
\begin{definition}
A point $p \in X$ is called a smooth point of $X$ if $X$ is a
submanifold of $M$ near $p$, i.e.,
the Jacobian ${\cal J}(X) := \left(\frac{\partial g_i}{\partial z_j}
\right)_p$ has maximal rank, namely $k$.
\end{definition}
Denoting the locus of smooth points as $X^*$, then if $X = X^*$, $X$
is called a smooth variety. Otherwise, a point $s \in V\setminus V^*$ is
called a {\bf singular point}\index{Singularity!definition}.

Given such a singularity $s$ on a $X$, the first exercise one could
perform is of course its {\bf resolution}, defined to be a birational
morphism $f: \tilde{X} \rightarrow X$ from a nonsingular variety
$\tilde{X}$ \index{Resolution!definition}. The
preimage $f^{-1}(s) \subset \tilde{X}$ of the singular point is called the
{\bf exceptional divisor} in $\tilde{X}$. Indeed
if $X$ is a projective variety, then if we require the resolution $f$ to be
projective (i.e., it can be composed as $\tilde{X} \rightarrow X
\times \IP^N 
\rightarrow X$), then $\tilde{X}$ is a projective variety.

The singular variety $X$, of (complex) dimension $n$,
is called {\em normal} \index{Singularity!normal}
if the structure sheafs obey
${\cal O}_X = f^* {\cal O}_{\tilde{X}}$. We henceforth restrict our
attention to 
normal varieties. The point is that as a topological
space the normal variety $X$ is simply the quotient
\[
X = \tilde{X} / \sim,
\]
where $\sim$ is the equivalence which collapses the exceptional
divisor to a point\footnote{And so $X$ has the structure sheaf $f^*
	{\cal O}_{\tilde{X}}$, the set of regular functions on
	$\tilde{X}$ which are 
	constant on $f^{-1}(s)$.}, the so-called process of blowing
down. Indeed the
reverse, where we replace the singularity $s$ by a set of directions
(i.e., a projective space), is called blowing up
\index{Resolution!blow up}. As we shall mostly
concern ourselves with Calabi-Yau manifolds (CY) of dimensions 2 and 3, 
of the uttermost importance will be exceptional divisors of dimension
1, to these we usually refer as {\bf $\IP^1$-blowups}.

Now consider the {\bf canonical divisors} of $\tilde{X}$ and $X$. 
\index{Divisor!canonical}
We recall
that the canonical divisor $K_X$ of $X$ is any divisor in the linear
equivalence (differing by principal divisors) class as the canonical
sheaf $\omega_X$, the $n$-th (hence maximal) 
exterior power of the sheaf of
differentials. Indeed for $X$ Calabi-Yau, $K_X$ is trivial. In general
the canonical sheaf of the singular variety and that of its resolution
$\tilde{X}$ are
not so na\"{\i}vely related but differ by a term depending on the
exceptional divisors $E_i$:
\[
K_{\tilde{X}} = f^*(K_X) + \sum\limits_i a_i E_i.
\]

The term $\sum\limits_i a_i E_i$ is a formal sum over the exceptional
divisors \index{Divisor!exceptional}
and is called the {\em discrepancy} of
the resolution and the
values of the numbers $a_i$ categorise some commonly encountered
subtypes of singularities characterising $X$, which we tabulate below:
\[
\begin{array}{|c|c|c|c|} 
\hline
a_i \ge 0	& \mbox{canonical} & a_i > 0	& \mbox{terminal} \\
	\hline 
a_i \ge -1	& \mbox{log canonical} &  a_i > -1	
	& \mbox{log terminal} \\ \hline
\end{array}
\]

The type which shall be pervasive throughout this work will be the
canonical singularities. In the particular case when all $a_i = 0$,
and the discrepancy term vanishes, we have what is known as a {\bf
crepant resolution}. \index{Resolution!crepant}
In this case the canonical sheaf of the
resolution is simply the pullback of that of the singularity, when the
latter is trivial, as in the cases of orbifolds which we shall soon
see, the former remains trivial and hence Calabi-Yau. Indeed crepant
resolutions always exists for dimensions 2 and 3, the situations of
our interest, and are related by flops. Although in dimension 3, the
resolution may not be unique (q.v. e.g. \cite{Ruan}).
On the other hand, for
terminal singularities, any resolution will change the canonical sheaf
and such singular Calabi-Yau's will no longer have resolutions to
Calabi-Yau manifolds.

In this vein of discussion on Calabi-Yau's, of the greatest
relevance to us are the so-called\footnote{The definition more
	familiar to algebraists is that a singularity is Gorenstein if
	the local ring is a Gorenstein ring, i.e., a local Artinian
	ring with maximal ideal $m$ such that the annihilator of $m$
	has dimension 1 over $A/m$. Another commonly
	encountered terminology is the $\mathbb{Q}$-Gorenstein singularity;
	these have $\Gamma(X\setminus p,K_X^{\otimes n})$ a free ${\cal
	O}(X)$-module for some finite $n$ and are cyclic quotients of
	Gorenstein singularities.}
{\bf Gorenstein} singularities \index{Singularity!Gorenstein},
which admit a nowhere vanishing global holomorphic
$n$-form on $X \setminus s$; these are then precisely those
singularities whose
resolutions have the canonical sheaf as a trivial line bundle, or in
other words, these are the local Calabi-Yau singularities.

Gorenstein canonical singularities which admit crepant resolutions to
smooth Calabi-Yau varieties are therefore the subject matter of this
work.
\subsection{Picard-Lefschetz Theory}
\index{Picard-Lefschetz Theory!definition}
We have discussed blowups of singularities in the above, in
particular $\IP^1$-blowups. A most useful study is when we consider
the vanishing behaviour of these $S^2$-cycles. Upon this we now focus.
Much of the following is based on \cite{Arnold}; 
The reader is also encouraged to consult e.g. \cite{Cecotti,Amer} for
aspects of Picard-Lefschetz monodromy in string theory.

Let $X$ be an $n$-fold, and $f : X \rightarrow U \subset \IC$
a holomorphic function thereupon. For our purposes, we take $f$ to
be the embedding equation of $X$ as a complex algebraic variety (for
simplicity we here study a hypersurface rather than complete
intersections). The
singularities of the variety are then, in accordance with Definition
2.1.1,
$\left\{\vec{x} | f'(\vec{x}) = 0\right\}$ with $\vec{x} =
(x_1,...,x_n) \in M$. $f$ evaluated at these critical points $\vec{x}$
is called a critical value of $f$.

We have level sets $F_z := f^{-1}(z)$ for complex numbers $z$; these
are $n-1$ dimensional varieties. For any
non-critical value $z_0$ one can construct a loop $\gamma$ beginning
and ending at $z_0$ and encircling no critical value. The map
$h_{\gamma} : F_{z_0} \rightarrow  F_{z_0}$, which generates the
monodromy as one cycles the loop, the main theme of Picard-Lefschetz
Theory. In particular, we are concerned with the induced action
$h_{\gamma *}$ on the homology cycles of $F_{z_0}$.

When $f$ is Morse\footnote{That is to say, at all critical points
	$x_i$, the Hessian 
	$\frac{\partial f}{\partial x_i \partial x_j}$ has
	non-zero determinant and all critical values $z_i = f(x_i)$ 
	are distinct.}, in the neighbourhood of each critical point
$p_i$ , $f$ affords the Taylor series $f(x_1,\ldots,x_n) = z_i +
\sum\limits_{j=1}^n (x_j - p_j)^2$ in some coordinate system. Now adjoin a
critical value $z_i = f(p_i)$ with a non-critical value $z_0$ by a
path $u(t) : t \in [0,1]$ which does not pass through any other
critical value.
Then in the level set $F_{u(t)}$ we fix sphere $S(t) = \sqrt{u(t) -
z_i} S^{n-1}$ (with $S^{n-1}$ the standard $(n-1)$-sphere $\left\{
(x_1, \ldots, x_n) : |x|^2 = 1, \mbox{Im} x_i = 0 \right\}$. In
particular $S(0)$ is precisely the critical point $p_i$.
Under these premises, we call
the homology class $\Delta \in H_{n-1}(F_{z_0})$ in the non-singular
level set $F_{z_0}$ represented by the sphere $S(1)$ the {\bf
Picard-Lefschetz vanishing cycle}.

Fixing $z_0$, we have a set of such cycles, one from each of the
critical values $z_i$. Let us consider what are known as {\it simple
loops}. These are elements of $\pi_1\left( U \backslash 
\{z_i\}, z_0 \right)$,
the fundamental group of loops based at $z_0$ and going around the
critical values. For these simple loops $\tau_i$ 
we have the corresponding
{\it Picard-Lefschetz monodromy operator}
\[
h_i = h_{\tau_i *} : H_{\bullet}(F_{z_0}) \rightarrow
H_{\bullet}(F_{z_0}).
\]
On the other hand if $\pi_1\left( U \backslash \{z_i\}, z_0 \right)$
is a free 
group then the cycles $\{ \Delta_i \}$ are {\it weakly
distinguished}.

The {\em point d'appui} is the Picard-Lefschetz Theorem which
determines the monodromy of $f$ under the above setup:
\begin{theorem}
\label{PL}
The monodromy group of the singularity is generated by the
Picard-Lefschetz operators $h_i$, corresponding to a weakly
distinguished basis $\{ \Delta_i \} \subset H_{n-1}$ of the
non-singular level set of $f$ near a critical point. In particular for
any cycle $a \in H_{n-1}$ (no summation in $i$)
\[
h_i(a) = a + (-1)^{\frac{n(n+1)}{2}} (a \circ \Delta_i) \Delta_i.
\]
\end{theorem}
\section{Symplectic Quotients and Moment Maps}
We have thus far introduced canonical algebraic singularities and
monodromy actions on exceptional $\IP^1$-cycles. The spaces we shall
be concerned are K\"ahler (Calabi-Yau) manifolds 
and therefore naturally we have more structure. Of uttermost
importance, especially when we encounter moduli spaces of certain
gauge theories, is the symplectic structure.
\begin{definition}
Let $M$ be a complex algebraic variety, a {\em symplectic form}
$\omega$ on $M$ is a holomorphic 2-form, i.e. $\omega \in \Omega^2(M)
= \Gamma(M,\bigwedge^2 T^* M)$, such that
\begin{itemize}
\item $\omega$ is closed: $d\omega=0$;
\item $\omega$ is non-degenerate: $\omega(X,Y)=0$ for
	any $Y\in T_p M \Rightarrow X =0$ .
\end{itemize}
\end{definition}

\index{Symplectic!manifold}
Therefore on the {\em symplectic manifold} $(M,\omega)$ (which by the
above definition is locally a complex symplectic vector space, 
implying that $\dim_{\IC}M$ is
even) $\omega$ induces an isomorphism between the tangent and cotangent
bundles by taking $X \in TM$ to $i_X(\omega) := \omega(X,\cdot) 
\in \Omega^1(M)$.
Indeed for any global analytic function $f \in {\cal O}(M)$ we can
obtain its differential $df \in \Omega^1(M)$. However by the (inverse
map of the) above isomorphism, we can define a vector field $X_f$,
which we shall call the {\em Hamiltonian} vector field associated to
$f$ (a scalar called the Hamiltonian).
In the language of classical mechanics, this vector field is the
generator of infinitesimal canonical transformations\footnote{If we
	were to write local co\"{o}rdinates $(p_i,q_i)$ for $M$, 
	then $\omega = \sum_i dq_i \wedge dq_i$ and the Hamiltonian
	vector field is $X_f = \sum_i \frac{\partial f}{\partial p_i}
	\frac{\partial}{\partial q_i} - (p_i \leftrightarrow q_i)$ and
	our familiar Hamilton's Equations of motion are
	$i_{X_f}(\omega) = \omega(X_f,\cdot)
	= df.$
	\label{ft:hamilton}}.
In fact, $[X_f, X_g]$, the commutator between two Hamiltonian vector
fields is simply $X_{\{f,g\}}$, where $\{f,g\}$ is the familiar
Poisson bracket.

The vector field $X_f$ is actually symplectic in the sense that
\[
L_{X_f} \omega = 0,
\]
where $L_X$ is the Lie derivative with respect to the vector field
$X$. This is so since $L_{X_f} \omega = (d \circ i_{X_f} + i_{X_f} \circ
d)\omega = d^2f + i_{X_f}d\omega = 0$.
Let $H(M)$ be the Lie subalgebra of Hamiltonian vector fields (of the
tangent space at the identity), then we have an obvious exact sequence
of Lie algebras (essentially since energy is defined up to a constant),
\[
0 \rightarrow \IC \rightarrow {\cal O}(M) \rightarrow H(M) \rightarrow
0,
\]
where the Lie bracket in ${\cal O}(M)$ is the Poisson bracket.

Having presented some basic properties of symplectic manifolds, we
proceed to consider quotients of such spaces by certain equivariant
actions. We let $G$ be some algebraic group which acts symplectically
on $M$. In other words, for the action $g^*$ on $\Omega^2(M)$, induced
from the action $m \rightarrow gm$ on the manifold for $g\in G$, we
have $g^*\omega = \omega$ and so the symplectic structure is preserved.
The infinitesimal action of $G$ is prescribed by its Lie algebra,
acting as symplectic vector fields; this gives homomorphisms $k:
Lie(G) \rightarrow H(M)$ and $\tilde{k} : Lie(G) \rightarrow {\cal
O}(M)$. The action of $G$ on $M$ is called Hamiltonian if the following
modification to the above exact sequence commutes
\[
\ba{rcccl}
0 \rightarrow \IC \rightarrow & 
{\cal O}(M) & \rightarrow H(M) & \rightarrow & 0\\
& \tilde{k}\nwarrow & \uparrow k  & \\
&&Lie(G)& &
\ea
\]

\begin{definition}
Any such Hamiltonian $G$-action on $M$ gives rise to a $G$-equivariant
{\bf Moment Map} $\mu : M \rightarrow Lie(G)^*$ which
corresponds\footnote{Because  $\hom(Lie(G),\hom(M,\IC)) = 
	\hom(M,Lie(G)^*)$.} to the map $\tilde{k}$ and satisfies
\[
k(A) = X_{A \circ \mu} \quad \mbox{for any~~} A \in Lie(G),
\]
i.e., $d(A \circ \mu) = i_{k(A)} \omega$.
\end{definition}
\index{Symplectic!moment map}
Such a definition is clearly inspired by the Hamilton equations of
motion as presented in Footnote \ref{ft:hamilton}. We shall not delve
into many of the beautiful properties of the moment map, such as when
$G$ is translation in Euclidean space, it is nothing more
than momentum, or when $G$ is rotation, it is simply angular momentum;
for what we shall interest ourselves in the forthcoming, we are
concerned with a crucial property of the moment map, namely the
ability to form certain smooth quotients.

Let $\mu : M \rightarrow Lie(G)^*$ be a moment map and $c \in
[Lie(G)^*]^G$ be the $G$-invariant subalgebra of $Lie(G)^*$ (in other
words the co-centre), then the equivariance of $\mu$ says that $G$
acts on the fibre $\mu^{-1}(c)$ and we can form the quotient of the
fibre by the group action. This procedure is called the {\em symplectic
quotient} and the subsequent space is denoted $\mu^{-1}(c) // G$.
The following theorem guarantees that the result still
lies in the category of algebraic varieties.
\begin{theorem}
Assume that $G$ acts freely on $\mu^{-1}(c)$, then the {\bf symplectic
quotient} $\mu^{-1}(c) // G$ is a symplectic manifold, with a unique
symplectic form $\bar{\omega}$, which is the pullback of the
restriction of the symplectic form on $M$
$\omega|_{\mu^{-1}(c)}$; i.e., $\omega|_{\mu^{-1}(c)} = q^*
\bar{\omega}$ if
$q : \mu^{-1}(c) \rightarrow \mu^{-1}(c) // G$ is the quotient map.
\end{theorem}
\index{Symplectic!quotient}
A most important class of symplectic quotient varieties are the
so-called toric varieties. These shall be the subject matter of the
next section.
\section{Toric Varieties}
\index{Toric Variety!definition}
The types of algebraic singularities with which we are most concerned
in the ensuing chapters in Physics are quotient and toric
singularities. The former are the next best thing to flat spaces and
will constitute the topic of the Chapter on finite groups. For now,
having prepared ourselves with symplectic quotients from the above
section, we give a
lightening review on the vast subject matter of toric varieties, which
are the next best thing to tori. The reader is encouraged to consult
\cite{Fulton,Oda,Sturmfels,Ewald,Cox-rev} as canonical mathematical texts as
well as \cite{GLSM,Greene-Lec,Beasley-The} for nice discussions in the
context of string theory.

As a holomorphic quotient, a toric variety is simply a generalisation
of the complex projective space $\IP^d := (\IC^{d+1} \ \{0\})/\IC^*$
with the $\IC^*$-action being the identification $x \sim \lambda x$. A
toric variety of complex dimension $d$ is then the quotient
\[
(\IC^n \setminus F)/\IC^{*(n-d)}.
\]
Here the $\IC^{*(n-d)}$-action is given by $x_i \sim
\lambda_a^{Q_i^a} x_i$ ($i=1,\ldots,n; a=1,\ldots,n-d$) for some
integer matrix (of charges) $Q_i^a$. Moreover, $F \in \IC^n \setminus
\IC^{*n}$ is a closed set of points one must remove to make the quotient
well-defined (Hausdorff).

In the language of symplectic quotients, we can reduce the geometry of
such varieties to the combinatorics of certain convex sets.
\subsection{The Classical Construction}
Before discussing the quotient, let us first outline the standard
construction of a toric variety. What we shall describe is the
classical construction of a toric 
variety from its defining fan, due originally to MacPherson.
Let $N \simeq \IZ^n$ be an integer lattice and let $M =
\hom_{\IZ}(N,\IZ) \simeq \IZ^n$ be its dual. Moreover let $N_{\IR} :=
N \otimes_{\IZ} \IR \simeq \IR^n$ (and similarly for $M_{\IR}$). Then
\begin{definition}
A (strongly convex) polyhedral cone $\sigma$
is the positive hull of a finitely many vectors $v_1, \ldots, v_k$ in
$N$, namely
\[
\sigma = {\rm pos}\{v_{i=1, \ldots, k}\} := \sum\limits_{i=1}^k \IR_{
\ge 0} v_i.
\]
\end{definition}
From $\sigma$ we can compute its {\bf dual cone} $\sigma^{\vee}$ as
\[
\sigma^{\vee} := \left\{ u \in M_{\IR} | u \cdot v \ge 0 \forall v \in
\sigma \right\}.
\]

Subsequently we have a finitely generated monoid
\[
S_{\sigma} := \sigma^{\vee} \cap M = 
\left\{ u \in M | u \cdot \sigma \ge 0 \right\}.
\]
We can finally associate maximal ideals of the monoid algebra of the
polynomial ring adjoint $S_{\sigma}$ to points in an algebraic
(variety) scheme. This is the affine toric variety $X_\sigma$
associated with the cone $\sigma$:
\[
X_\sigma := {\rm Spec}( \IC[ S_{\sigma} ] ).
\]

To go beyond affine toric varieties, we simply paste together, as
co\"ordinate patches,
various $X_{\sigma_i}$ for a collection of cones $\sigma_i$; such a
collection is called a {\bf fan} $\Sigma = \bigsqcup_i \sigma_i$ and
we finally arrive at the general toric variety $X_\Sigma$.

As we are concerned with the singular behaviour of our varieties, the
following definition and theorem shall serve us greatly.
\begin{definition}
A cone $\sigma = {\rm pos}\{v_i\}$ is {\bf simplicial} is all the
vectors $v_i$ are linearly independent; it is {\bf regular} if
$\{v_i\}$ is a $\IZ$-basis for $N$. The fan $\Sigma$ is {\bf complete}
if its cones span the entirety of $\IR^n$ and it is regular if all its
cones are regular and simplicial.
\end{definition}
Subsequently, we have
\begin{theorem}
$X_\Sigma$ is compact iff $\Sigma$ is complete; it is {\bf
non-singular} iff $\Sigma$ is regular.
\end{theorem}
\index{Toric Variety!Calabi-Yau}
Finally we are concerned with Calabi-Yau toric varieties, these are
associated with what is know (recalling Section 1.1
regarding Gorenstein resolutions) as {\bf Gorenstein cones}. It turns
out that an $n$-dimensional toric variety satisfies the Ricci-flatness
condition if all the endpoints of the vectors of its cones lie on a
single $n-1$-dimensional hypersurface, in other words,
\begin{theorem}
The cone $\sigma$ is called {\em Gorenstein} if there exists a vector
$w \in N$ such that $\langle v_i,w \rangle=1$ for all the
generators $v_i$ of $\sigma$. Such cones give rise to toric Calabi-Yau
varieties.
\end{theorem}
We refer the reader to \cite{dais} for conditions when Gorenstein
cones admit crepant resolutions.

The name {\em toric} may not be clear from the above construction but
we shall see now that it is crucial. Consider each point $t$ the
algebraic torus $T^n 
:= (\IC^*)^n \simeq N \otimes_{\IZ} \IC^* \simeq \hom (M, \IC^*)
\simeq {\rm spec}(\IC[M])$ as a group homomorphism $t : M \rightarrow
\IC^*$ and each point $x \in X_\sigma$ as a monoid homomorphism $x :
S_\sigma \rightarrow \IC$. Then we see that there is a {\em natural
torus action} on the toric variety by the algebraic torus $T^n$ as
$x \rightarrow t \cdot x$ such that $(t \cdot x)(u) := t(u)x(u)$ for
$u \in S_\sigma$. For $\sigma = \{0\}$, this action is nothing other
than the group multiplication in $T^n = X_{\sigma = \{0\}}$.
\subsection{The Delzant Polytope and Moment Map}
\index{Symplectic!moment map}
How does the above tie in together with what we have discussed on
symplectic quotients? We shall elucidate here. It turns out such a
construction is canonically done for compact toric varieties embedded
into projective spaces, so we shall deal more with {\em polytopes}
rather than {\em polyhedral cones}. The former is simply a compact
version of the latter and is a bounded set of points instead of
extending as a cone. The argument below can be easily
extended for fans and
non-compact (affine) toric varieties. For now
our toric variety $X_\Delta$ is encoded in a polytope $\Delta$. 

Let $(X,\omega)$ be a symplectic manifold of real dimension $2n$.  Let
$\tau : T^n \rightarrow {\rm Diff}(X,\omega)$ be a Hamiltonian action
from the $n$-torus to vector fields on $X$. This immediately gives us
a moment map $\mu : X \rightarrow \IR^n$,
where $\IR^n$ is the dual of the Lie algebra for $T^n$ considered as
the Lie group $U(1)^n$. The image of $\mu$ is a polytope $\Delta$,
called a moment or {\bf Delzant Polytope}. The inverse image, up to
equivalence of the $T^n$-action, is then
nothing but our toric variety $X_\Delta$. But this is precisely the
statement that
\[
X_\Delta := \mu^{-1}(\Delta) //T^n
\]
and the toric
variety is thus naturally a symplectic quotient.

In general, given a convex polytope, Delzant's theorem guarantees that
if the following conditions are satisfied, then the polytope is
Delzant and can be used to construct a toric variety:
\begin{theorem}[Delzant] 
A convex polytope $\Delta \subset \IR^n$ is Delzant if:
\begin{enumerate}
\item There are $n$ edges meeting at each vertex $p_i$;
\item Each edge is of the form $p_i + \IR_{\ge 0} v_i$ with $v_{i=1,
\ldots, n}$ a basis of $\IZ^n$.
\end{enumerate}
\end{theorem}

We shall see in Liber II and III, that the moduli space of certain
gauge theories arise as toric singularities. In Chapter 5, we shall in
fact see a third, physically motivated construction for the toric
variety. For now, let us introduce another class of Gorenstein
singularities.  
\chapter{Representation Theory of Finite Groups}
{ A} wide class of Gorenstein canonical singularities are of course
quotients of flat spaces by appropriate discrete groups. When the
groups are chosen to be discrete subgroups of special unitary groups,
i.e., the holonomy groups of Calabi-Yau's, and when crepant
resolutions are admissible, these quotients are singular limits of
CY's and provide excellent local models thereof. Such quotients
of flat spaces by discrete finite subgroups of certain Lie actions,
are called {\bf orbifolds} (or V-manifolds, in their original guise in
\cite{Satake}). It is therefore a natural {\it point de d{\'e}part}
for us to go from algebraic geometry to a brief discussion on finite
group representations (q.v. e.g. \cite{Fulton-Rep} for more details of
which much of the following is a condensation).
\section{Preliminaries}
We recall that a {\bf representation} of a finite group $G$ on a finite
dimensional (complex) vector space $V$ is a homomorphism $\rho : G
\rightarrow GL(V)$ to the group of automorphisms $GL(V)$ of $V$. Of
great importance to us is the {\bf regular} representation, where $V$
is the vector space with basis $\{e_g | g\in G\}$ and $G$ acts on $V$
as $h \cdot \sum a_g e_g = \sum a_g e_{hg}$ for $h \in G$.

Certainly the corner-stone of representation theory is Schur's Lemma:
\begin{theorem}[Schur's Lemma]
If $V$ and $W$ are irreducible representations of $G$ and $\phi : V
\rightarrow W$ is a $G$-module homomorphism, then (a) either $\phi$ is
an isomorphism or $\phi=0$. If $V=W$, then $\phi$ is a homothety
(i.e., a multiple of the identity).
\end{theorem}

The lemma allows us to uniquely decompose any representation $R$ into
irreducibles $\{R_i\}$ as $R = R_1^{\oplus a_1} \oplus \ldots \oplus
R_n^{\oplus a_n}$. The three concepts of regular representations,
Schur's lemma and unique decomposition we shall extensively use
later in Liber III. 
Another crucial technique is that of character theory into which we
now delve.
\section{Characters}
\index{Finite Groups!characters}
If $V$ is a representation of $G$, we define its character $\chi_V$ to
be the $\IC$-function on $g \in G$:
\[
\chi_V(g) = \Tr(V(g)).
\]
Indeed the character is a {\bf class function}, constant on each
conjugacy class of $G$; this is due to the cyclicity of the trace:
$\chi_V(h g h^{-1}) = \chi_V(g)$. Moreover $\chi$ is a
homomorphism from vector spaces to $\IC$ as
\[
\chi_{V \oplus W} = \chi_V + \chi_W \qquad
\chi_{V \otimes W} = \chi_V \chi_W.
\]

From the following theorem
\begin{theorem}
There are precisely the same number of conjugacy classes are there are
irreducible representations of a finite group $G$,
\end{theorem}
and the above fact that $\chi$ is a class function,
we can construct a square matrix, the so-called {\bf character table},
whose entries are the characters
$\chi_{\gamma}^{(i)}:=\Tr(R_i(\gamma))$, as $i$ goes
through 
the irreducibles $R_i$ 
and $\gamma$, through the conjugacy classes. This table
will be of tremendous computational use for us in Liber III.

The most important important properties of the character table are its
two 
{\em orthogonality conditions}, the first of which is for the rows,
where we sum over conjugacy classes:
\[
\sum_{g \in G} \chi_g^{(i)*} \chi_g^{(j)*} = \sum_{\gamma = 1}^n
r_\gamma \chi_\gamma^{(i)*} \chi_\gamma^{(j)*} = |G| \delta_{ij},
\]
where $n$ is the number of conjugacy classes (and hence irreps) and
$r_\gamma$ the size of the $\gamma$-th conjugacy class.
The other orthogonality is for the columns, where we sum over irreps:
\[
\sum_{i = 1}^n \chi_k^{(i)*} \chi_l^{(i)*} = \frac{|G|}{r_k}
	\delta_{kl}.
\]

We summarise these relations as
\begin{theorem}
With respect to the inner product $(\alpha,\beta) := \frac{1}{|G|}
\sum\limits_{g \in G} \alpha^*(g) \beta(g) = \frac{1}{|G|}
\sum\limits_{\gamma = 1}^n r_\gamma \alpha^*(\gamma) \beta(\gamma)$,
the characters of the irreducible representations (i.e. the character
table) are orthonormal.
\end{theorem}
Many interesting corollaries follow. Of the most useful are the
following. Any representation $R$ is irreducible iff $(\chi_R,\chi_R)=1$
and if not, then $(\chi_R,\chi_{R_i})$ gives the multiplicity of the
decomposition of $R$ into the $i$-th irrep. 

For the {\em regular representation} $R_r$, the character is simply
$\chi(g) 
= 0$ if $g \ne \II$ and it is $|G|$ when $g = \II$ (this is simply
because any group element $h$ other than the identity will permute $g
\in G$ and in the vector basis $e_g$ correspond to a non-diagonal
element and hence do not contribute to the trace). Therefore if we
were to decompose the $R_r$ in to irreducibles, the
$i$-th would receive a multiplicity of $(R_r,R_i) = \frac{1}{|G|}
\chi_{R_i}(\II) |G| = \dim R_i$. Therefore any irrep $R_i$ appears in
the regular representation precisely $\dim R_i$ times.
\subsection{Computation of the Character Table}
There are some standard techniques for computing the character table
given a finite group $G$; the reader is referred to
\cite{Lomont,Lederman,Hill} for details.

For the $j$-th conjugacy class $c_j$, define a {\em class operator}
$C_j := \sum\limits_{g \in c_j} g$, as a formal sum of group elements
in the conjugacy class. This gives us a {\bf class multiplication}:
\[
C_j C_k = \sum\limits_{g \in c_j; h \in c_k} gh = \sum_k c_{jkl} C_l,
\]
where $c_{jkl}$ are ``fusion coefficients'' for the class
multiplication and can be determined from the multiplication table of
the group $G$. Subsequently one has, by taking characters,
\[
r_j r_k \chi_j^{(i)} \chi_k^{(i)} = \dim R_i \sum\limits_{l=1}^n
c_{jkl} r_l \chi_k^{(l)}.
\]
These are $n^2$ equations in $n^2+n$ variables $\{\chi_j^{(i)}; \dim
R_i\}$. We have another $n$ equations
from the orthonormality $\frac{1}{|G|}
\sum\limits_{j=1}^n r_j |\chi_j^{(i)}|^2=1$; these then suffice to
determine the characters and the dimensions of the irreps.
\section{Classification of Lie Algebras}
\index{Lie Algebras}
In Book the Third we shall encounter other aspects of representation
theory such as induced and projective representation; we shall deal
therewith accordingly.
For now let us turn to the representation
of Lie Algebras. It may indeed seem to the reader rather discontinuous
to include a discussion on the the classification of Lie Algebras in a
chapter touching upon finite groups. However the reader's patience
shall soon be rewarded in Chapter 4 as well as Liber III 
when we learn that certain
classifications of finite groups are intimately related, by what has
become known as McKay's Correspondence, to that of Lie
Algebras. Without further ado then let us simply present, for the sake
of refreshing the reader's memory, the classification of complex Lie
algebras.

Given a complex Lie algebra $\mathfrak{g}$, it has the {\em Levi
Decomposition}
\[
\mathfrak{g} = {\rm Rad}(\mathfrak{g}) \oplus \tilde{\mathfrak{g}} =
{\rm Rad}(\mathfrak{g}) \oplus 
\bigoplus\limits_i \mathfrak{g}_i,
\]
where Rad$(\mathfrak{g})$ is the radical, or the maximal solvable
ideal, of 
$\mathfrak{g}$. The representation of such solvable algebras is
trivial and can 
always be brought to $n \times n$ upper-triangular matrices by a basis
change.
On the other hand $\tilde{\mathfrak{g}}$ is semisimple and contains no
nonzero 
solvable ideals. We can decompose $\tilde{\mathfrak{g}}$ further into
a direct 
sum of {\bf simple} Lie algebras $\mathfrak{g}_i$ which contain no
nontrivial 
ideals. The $\mathfrak{g}_i$'s are then the nontrivial pieces of
$\mathfrak{g}$.

\begin{figure}
	\centerline{\psfig{figure=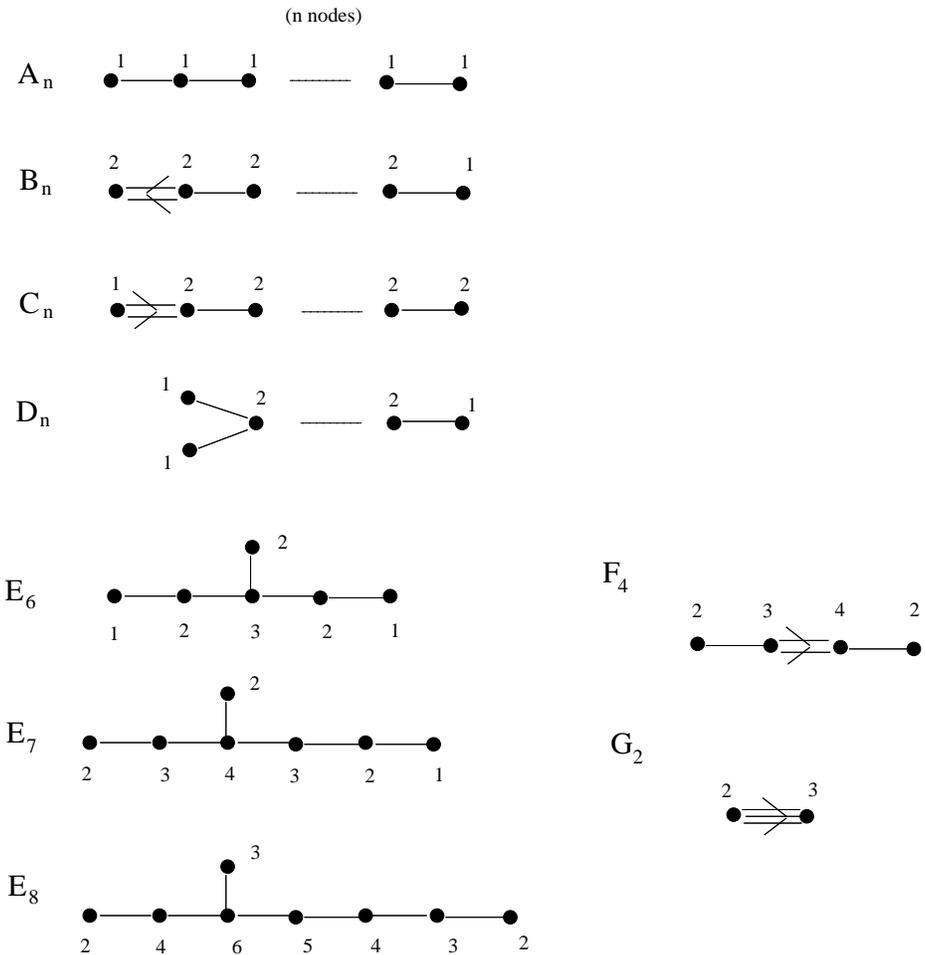,width=5in}}
\caption{The Dynkin diagrams of the simple complex Lie Algebras;
	the nodes are labelled with the comarks.}
\label{f:liealg}
\end{figure}

The great theorem is then the complete classification of the complex
simple Lie algebras due to Cartan, Dynkin and Weyl. These are the 
\begin{itemize}
\item {\bf Classical Algebras:} 
	$A_n := \mathfrak{sl}_{n+1}(\IC)$, $B_n :=
	\mathfrak{so}_{2n+1}(\IC)$, $C_n := \mathfrak{sp}_{2n}(\IC)$
	and $D_n := \mathfrak{so}_{2n}(\IC)$ for $n=1,2,3\ldots$;
\item {\bf Exceptional Algebras:} $E_{6,7,8}$, $F_{4}$ and $G_2$.
\end{itemize}
The Dynkin diagrams for these are given in \fref{f:liealg}. The nodes
are marked with the so-called comarks $a_i^\vee$ which we recall to be
the expansion coefficients of the highest root $\theta$ into the
simple coroots $\alpha_i^\vee := 2\alpha_i/|\alpha_i|^2$ 
($\alpha_i$ are the simple roots)
\[
\theta = \sum\limits_i^r a_i^\vee
\alpha_i^\vee,
\]
where $r$ is the rank of the algebra (or the number of nodes).

The dual Coxeter numbers are defined to be
\[
c := \sum\limits_i^r a_i^\vee + 1
\]
and the {\bf Cartan Matrix} is
\[
C_{ij} := (\alpha_i, \alpha_j^\vee).
\]

We are actually concerned more with {\bf Affine} counterparts of the
above simple algebras. These are central extensions of the above in 
the sense that if the commutation relation in the simple 
$\mathfrak{g}$ is
$[T^a,T^b] = f^{ab}_cT^c$, then that in the affine
$\widehat{\mathfrak{g}}$ is $[T^a_m,T^b_m] = f^{ab}_cT^c_{m+n} + k n
\delta_{ab}\delta_{m,-n}$. The generators $T^a$ of $\mathfrak{g}$ are
seen to be generalised to $T^a_m := T^a \otimes t^m$ of
$\widehat{\mathfrak{g}}$ by Laurent polynomials in $t$.
The above concepts of roots etc.~are directly generalised with the
inclusion of the affine root. The Dynkin diagrams are as in
\fref{f:liealg} but augmented with an extra affine node.

We shall see in Liber III
that the comarks and the dual Coxeter numbers will
actually show up in the dimensions of the irreducible representations
of certain finite groups. Moreover, the Cartan matrices will
correspond to certain graphs constructable from the latter.
\chapter{Finite Graphs, Quivers, and Resolution of Singularities}
{ W}e have addressed algebraic singularities, symplectic quotients and
orbifolds in relation to finite group representations. It is now time
to embark on a journey which would ultimately give a unified
outlook. To do so we must involve ourselves with yet another field of
mathematics, namely the theory of graphs.
\section{Some Rudiments on Graphs and Quivers}
\index{Quivers!definition}
As we shall be dealing extensively with algorithms on finite graphs in
our later work on toric singularities, let us first begin with the
fundamental concepts in graph theory. The reader is encouraged to
consult such classic texts as \cite{Graph1,Graph2}.
\begin{definition}
A finite graph is a triple $(V,E,I)$ such that $V,E$ are disjoint
finite sets (respectively the set of vertices and edges) with members
of $E$ joining those of $V$ according to the incidence relations $I$.
\end{definition}

The graph is {\em undirected} if for each edge $e$ joining vertex $i$
to $j$ there is another edge $e'$ joining $j$ to $i$; it is directed
otherwise. The graph is {\em simple} if there exists no {\em loops}
(i.e., edges joining a vertex to itself). The graph is
connected if any two vertices can be linked a series of edges, a
so-called {\em walk}. Two more commonly encountered concepts are the
{\em Euler} and {\em 
Hamilton} cycles, the first of which is walk returning to the
beginning vertex which traverses each edge only once and each vertex
at least once, while the latter, the vertices only once. 
Finally we call two graphs isomorphic if they are topologically
homeomorphic; we emphasise the unfortunate fact that the graph
isomorphism problem (of determining whether two graphs are isomorphic)
is thus far unsolved; it is believed to be neither P nor
NP-complete. This will place certain restrictions on our computations
later. 

We can represent a graph with $n$ vertices and $m$ edges by an $n
\times n$ matrix, the so-named {\bf adjacency matrix} $a_{ij}$ whose
$ij$-th entry is the number of edges from $i$ to $j$. If the graph is
simple, then we can also represent the graph by an {\bf incidence
matrix}, an $n \times m$ matrix $d_{ia}$ in whose $a$-th columns there
is a $-1$ (resp. 1) in row $i$ (resp. row $j$) if there an $a$-th edge
going from $i$ to $j$. We emphasise that the graph must be {\em
simple} for the incidence matrix to fully encapture its
information. Later on in Liber III we will see this is a shortcoming
when we are concerned with gauged linear sigma models.
\subsection{Quivers}
Now let us move onto a specific type of directed graphs, which we
shall call a {\bf quiver}. To any such a quiver $(V,E,I)$ is associated
the abelian category Rep$(V,E,I)$, of its representations (over say,
$\IC$). A (complex) {\em representation} of a quiver associates to every
vertex $i \in V$ a vector space $V_i$ and to any edge $i
\stackrel{a}{\rightarrow} j$ a linear map $f_a : V_i \rightarrow
V_j$. The vector $\vec{d} = (d_i := \dim_{\IC} V_i)$ is called the
dimension of the representation.

Together with its representation
dimension, we can identify a quiver as a {\em labelled graph} (i.e., a
graph with its nodes associate to integers)
$(V,E,I;\vec{d})$. Finally, as we shall encounter in the case of
gauge theories, one could attribute certain algebraic meaning
to the arrows by letting them be formal variables which satisfy
certain sets algebraic relations $R$; now we have to identify
the quiver as a quintuple $(V,E,I;\vec{d},R)$. These labelled directed
finite quivers with relations are what concern string theorist the
most.

In Liber III we shall delve further into the representation theory of
quivers in relation to gauge theories, for now let us introduce two
more preliminary concepts. We say a representation with dimension
$\vec{d'}$ is a sub-representation of that with $\vec{d}$ if 
$(V,E,I;\vec{d'}) \hookrightarrow (V,E,I;\vec{d})$ is an injective
morphism. In this case given a vector $\theta$ such that $\theta \cdot
d = 0$, we call a
representation with dimension $d$ {\bf $\theta$-semistable} if for any
subrepresentation with dimension $d'$, $\theta \cdot d' \ge 0$; we
call it $\theta$-stable for the strict inequality. King's beautiful
work \cite{King} has shown that $\theta$-stability essentially implies
existence of solutions to certain BPS equations in supersymmetric
gauge theories, the so-called F-D flatness conditions. But pray be
patient as this discussion would have to wait until Liber II.
\section{du Val-Kleinian Singularities}
\index{ADE Singularities}
Having digressed some elements of graph and quiver theories, let us
return to algebraic geometry. We shall see below a beautiful link
between the theory of quivers and that of orbifold of $\IC^2$.

First let us remind the reader of the classification of the quotient
singularities of $\IC^2$, these date as far back as F. Klein
\cite{Klein}. The affine equations of these so-called {\bf ALE}
(Asymptotically Locally Euclidean)
singularities can be written in $\IC[x,y,z]$ as 
\[
\begin{array}{|c|}
\hline
A_n: xy+z^n=0\\
D_n: x^2+y^2z+z^{n-1}=0\\
E_6: x^2+y^3+z^4=0\\
E_7: x^2+y^3+yz^3=0\\
E_8: x^2+y^3+z^5=0.\\
\hline
\end{array}
\]
We have not named these $ADE$ by coincidence. 
The resolutions of such singularities were studied extensively by
\cite{duVal} and one sees in fact that the $\IP^1$-blowups intersect
precisely in the fashion of the Dynkin diagrams of the simply-laced
Lie algebras $ADE$. For a illustrative review upon this elegant
subject, the reader is referred to \cite{Slodowy}.
\subsection{McKay's Correspondence}
\index{McKay Correspondence!definition}
Perhaps it is a good point here to introduce the famous McKay
correspondence, which will be a major part of Liber III. We shall be
brief now, promising to expound upon the matter later.

Due to the remarkable observation of McKay in \cite{McKay}, there is
yet another justification of naming the classification of the discrete
finite subgroups $\Gamma$ of $SU(2)$ as $ADE$. Take the defining
representation $R$ of $\Gamma$, and consider its tensor product with
all the irreducible representations $R_i$:
\[
R \otimes R_i = \bigoplus_j a_{ij} R_j.
\]
Now consider $a_{ij}$ as an adjacency matrix of a finite quiver with
labelling the dimensions of the irreps. Then McKay's Theorem states
that $a_{ij}$ of the $ADE$ finite group is precisely the Dynkin
diagram of the {\em affine} $ADE$ Lie algebra and the dimensions
correspond to the comarks of the algebra.
Of course for any finite group we can perform such a procedure, and we
shall call the quiver so-obtained the {\bf McKay Quiver}.
\section{ALE Instantons, hyper-K\"ahler Quotients and McKay Quivers}
It is the unique perspective of Kronheimer's work \cite{Kronheimer}
which uses the methods of certain symplectic quotients in conjunction
with quivers to study the resolution of the $\IC^2$ orbifolds. We must
digress one last time, to introduce instanton constructions.
\subsection{The ADHM Construction for the $E^4$ Instanton}
\index{ADHM construction}
For the Yang-Mills equation 
$D^aF_{ab} := \nabla^a F_{ab} + [A^a,F_{ab}] = 0$ 
obtained from the action
$L_{\rm YM} = -{1 \over 4} F_{ab} F^{ab}$ with connexion $A_a$ and
field strength 
$F_{ab} := \nabla_{[a} A_{b]} + [A_a,A_b]$, we seek {\em finite
action} solutions. 
These are known as {\bf instantons}.
A theorem due to Uhlenbeck \cite{Uhlenbeck} ensures that finding such
an instanton 
solution in Euclidean space $E^4$ amounts to investigating $G$-bundles
over $S^4$ since 
finite action requires the gauge field to be well-behaved at infinity
and hence the 
one-point compactification of $E^4$ to $S^4$.

Such $G$-bundles, at least for simple $G$, are classified by integers,
viz., the second 
Chern number of the bundle $E$,
$c_2(E) := {1 \over {8 \pi}} \int_{S^4} {\rm Tr} (F \wedge F)$;
this is known as the {\it instanton number} of the gauge field.
In finding the saddle points, so as to enable the evaluation
of the Feynman path integral for $L_{\rm YM}$, one can easily show
that only the 
self-dual and self-anti-dual solutions $F_{ab} = \* \pm F_{ab}$ give
rise to absolute 
minima in each topological class (i.e., for fixed instanton number).
Therefore we shall focus in particular on the self-dual instantons. We
note that 
self-duality implies solution to the Yang-Mills equation due to the
Bianchi identity. 
Hence we turn our attention to self-dual gauge fields.
There is a convenient theorem (see e.g. \cite{Twister}) which
translates the 
duality condition into the language of holomorphic bundles:
\begin{theorem}[Atiyah et al.] \label{Atiyah}
There is a natural 1-1 correspondence between
\begin{itemize}
\item Self-dual SU(n) gauge fields\footnote{Other classical groups
have also been done, but here 
	we shall exemplify with the unitary groups.}
	on $U$, an open set in $S^4$, and
\item Holomorphic rank $n$ vector bundles E over $\hat{U}$, an open
	set\footnote{ 
		There is a canonical mapping from $x \in U$ to
		$\hat{x} \in \hat{U}$  
		into which we shall not delve.} 
	in $\IP^3$, such that
	(a) $E|_{\hat{x}}$ is trivial $\forall x \in U$;
	(b) $\det E$ is trivial;
	(c) $E$ admits a positive real form.
\end{itemize}
\end{theorem}
Therefore the problem of constructing self-dual
instantons amounts to 
constructing a holomorphic vector bundle over $\IP^3$. 
The key technique is due to the monad 
concept of Horrocks \cite{Horrocks} where a sequence of vector bundles
$F \stackrel{A}{\rightarrow} G \stackrel{B}{\rightarrow} H$ is used to
produce the 
bundle $E$ as a quotient $E = {\rm ker} B / {\rm Im} A$.
Atiyah, Hitchin, Drinfeld and Manin then utilised this idea in their
celebrated paper 
\cite{ADHM} to reduce the self-dual Yang-Mills instanton problem from
partial differential 
equations to matrix equations; this is now known as the {\bf ADHM
construction}. 
Let $V$ and $W$ be complex vector spaces of dimensions $2k + n$ and
$k$ respectively 
and $A(Z)$ a linear map
\[
A(Z) : W \rightarrow V
\]
depending linearly on coordinates $\{Z^{a=0,1,2,3}\}$ of $\IP^3$ as $A(Z)
:= A_aZ^a$ with 
$A_a$ constant linear maps from $W$ to $V$.
For any subspace $U \subset V$, we define
\[
U^0 := \{v \in V | (u,v) = 0~~\forall u \in U \}
\]
with respect to the symplectic (nondegenerate skew bilinear) form $(~,~)$.
Moreover we introduce antilinear maps $\sigma : W \rightarrow W$ with
$\sigma^2 = 1$ 
and $\sigma : V \rightarrow V$ with $\sigma^2 = -1$ and impose the
conditions 
\beq
\label{conditions}
\begin{array}{ll}
(1) & \forall Z^a \ne 0, U_Z := A(Z)W~\mbox{has dimension}~k~\mbox{and
		is isotropic}~ 
		(U_Z \subset U_Z^0); \\
(2) & \forall w \in W, \sigma A(Z) w = A(\sigma Z) \sigma w.
\end{array}
\eeq
Then the quotient space $E_Z := U^0_Z / U_Z$ of dimension $(2k + n -
k) - k = n$ 
is precisely the rank $n$ $SU(n)$-bundle $E$ over $\IP^3$ which we
seek. One can further check that 
$E$ satisfies the 3 conditions in theorem \ref{Atiyah}, whereby giving
us the required self-dual instanton.
Therefore we see that the complicated task of solving the non-linear
partial differential 
equations for the self-dual instantons has been reduced to finding
$(2k + n) \times k$ 
matrices $A(Z)$ satisfying condition (\ref{conditions}), the second of
which is usually 
known - though perhaps here not presented in the standard way - as the
ADHM equation.
\subsection{Moment Maps and Hyper-K\"ahler Quotients}
The other ingredient we need is a generalisation of the symplectic
quotient discussed in Section 1.2, the so-called Hyper-K\"ahler
Quotients of Kronheimer \cite{Kronheimer} (see also the elucidation
in \cite{Bianchi}).
\index{Symplectic!hyper-Kahler quotient}
A Riemannian manifold $X$ with three covariantly constant complex
structures 
$i := I,J,K$ satisfying the quaternionic algebra is called {\bf
Hyper-K\"ahler}\footnote{ 
	In dimension 4, simply-connectedness and self-duality of the
	Ricci tensor 
	suffice to guarantee hyper-K\"ahlerity.}.
From these structures we can define closed (hyper-)K\"ahler 2-forms:
\[
\omega_i (V,W) := g(V,iW) \qquad {\rm for}~~~~i = I,J,K
\]
mapping tangent vectors $V,W \in T(X)$ to $\R$ with $g$ the metric
tensor. 

On a hyper-K\"ahler manifold with Killing vectors $V$ (i.e., ${\cal
L}_V g = 0$) 
we can impose {\bf triholomorphicity}:
${\cal L}_V \omega_i = V^\nu (d\omega_i)_\nu + d(V^\nu (\omega_i)_\nu)
= 0$ which 
together with closedness $d\omega_i = 0$ of the hyper-K\"ahler forms 
imply the existence of potentials $\mu_i$, such that $d\mu_i = V^\nu
(\omega_i)_\nu$. 
Since the dual of the Lie algebra $\mathfrak{g}$ of the group of
symmetries $G$ generated 
by the Killing vectors $V$ is canonically
identifiable with left-invariant forms, we have an induced map of such
potentials: 
\[
\mu_i : X \rightarrow \mu_i^a \in \R^3 \otimes \mathfrak{g}^* \qquad
i=1,2,3;~~a=1,...,\dim(G) 
\]
These maps are the (hyper-K\"ahler) {\bf moment maps} and
usually grouped as 
$\mu_\R = \mu_3$ and $\mu_{\mathbb{C}} = \mu_1 + i \mu_2$

Thus equipped, for any hyper-K\"ahler manifold $\Xi$ of dimension $4n$
admitting $k$ freely 
acting triholomorphic symmetries, we can construct
another, $X_\zeta$, of dimension $4n - 4k$ by the following two steps:
\begin{enumerate}
\item We have $3k$ moment maps and can thus define a level set of
dimension $4n - 3k$: 
	\[
	P_\zeta := \{ \xi \in \Xi | \mu_i^a(\xi)  = \zeta^a_i \};
	\]
\item When $\zeta \in \R^3 \otimes {\rm Centre}(\mathfrak{g}^*)$,
	$P_\zeta$ turns out to be 
	a principal $G$-bundle over a new hyper-K\"ahler manifold
	\[
	X_\zeta := P_\zeta / G \cong \{ \xi \in \Xi |
	\mu^a_{\mathbb{C}}(\xi) = 
	\zeta^a_{\mathbb{C}} \} / G^{\mathbb{C}}. 
	\]
\end{enumerate}
This above construction , where in fact the natural connection on the
bundle 
$P_\zeta \rightarrow X_\zeta$ is self-dual, is the celebrated
{\bf hyper-K\"ahler quotient} construction \cite{Kronheimer}.

Now we present a remarkable fact which connects these moment maps
to the previous section. If we write (\ref{conditions}) for $SU(n)$
groups 
into a (perhaps more standard) component form, we have the ADHM data
\[
M := \{ A,B; s,t^\dagger | A,B \in {\rm End}(V); 
	s,t^\dagger \in {\rm Hom}(V,W) \},
\]
with the ADHM equations
\[
\begin{array}{rcl}
[A,B] + ts &=& 0;\\
([A,A^\dagger] + [B,B^\dagger]) - s s^\dagger + t t^\dagger &=& 0.
\end{array}
\]
Comparing with the hyper-K\"ahler forms
$\omega_{\mathbb{C}} = {\rm Tr}(dA \wedge dB) + {\rm Tr}(dt\wedge ds)$ and
$\omega_\R = {\rm Tr}(dA \wedge dA^\dagger + dB \wedge dB^\dagger) -
	{\rm Tr}(ds^\dagger \wedge ds - dt \wedge dt^\dagger)$ which
are invariant under the action by $A,B,s,t^\dagger$, we immediately
arrive at the following fact:
\begin{proposition}
The moment maps for the triholomorphic $SU(n)$ isometries precisely
encode 
the ADHM equation for the $SU(n)$ self-dual instanton construction.
\end{proposition}
\subsection{ALE as a Hyper-K\"ahler Quotient}
\index{ADE Singularity!hyper-Kahler quotient}
\index{ALE Spaces}
Kronheimer subsequently used the above construction for the case of
$X$ being the ALE space, i.e. the orbifolds $\IC^2/(\Gamma \in
SU(2))$. Let us first clarify some notations:
$\Gamma \subset SU(2) :=$
		Finite discrete subgroup of $SU(2)$, i.e., $A_n$,
		$D_n$, or $E_{6,7,8}$;
		$Q :=$ The defining ${\mathbb{C}}^2$-representation;
		$R :=$ The regular $|\Gamma|$-dimensional complex
		representation;
		$R_{i = 0,..,r} := {\rm irreps}(\Gamma)$ of
		dimension $n_i$ with 0 
		corresponding to the affine node
		(the trivial irrep);
$(~)_\Gamma :=$	The $\Gamma$-invariant part;	
$a_{ij} :=$	The McKay quiver matrix for $\Gamma$, i.e.,
			$Q \otimes R_i = \bigoplus\limits_j a_{ij} R_j$;
$T :=$  	A one dimensional quaternion vector space
		$\qquad = \{x_0 + x_1 i + x_2 j + x_3 k | x_i \in \R\}$;
$\Lambda^+T^* :=$
		The self-dual part of the second exterior power of the
		dual space = span$\{\mbox{hyper-K\"ahler
		forms }\omega_{i=I,J,K}\}$;
$\left[ y \wedge y \right]:= (T^* \wedge T^*) \otimes [{\rm
End}(V),{\rm End}(V)], \mbox{ for } y \in T^* \otimes {\rm End}(V)$;
${\rm Endskew}(R) :=$ 
		The anti-Hermitian endomorphisms of $R$;
$Z :=$		Trace free part of Centre(Endskew$_\Gamma(R)$);
$G :=$		$\prod\limits_{i=1}^r U(n_i) =$ The group of unitary 
		automorphisms of $R$
		commuting with the action of $\Gamma$, modded
		out by $U(1)$ scaling\footnote{This is in the sense that the
		group $U(|\Gamma|)$ is broken down,  
		by $\Gamma$-invariance, to
		$\prod\limits_{i=0}^r U(n_i)$,  
		and then further reduced to $G$ by the modding out.}
$X_\zeta :=	\{y \in (T^* \otimes_\R {\rm Endskew}(R))_\Gamma | 
			[ y \wedge y ]^+ = \zeta \} / G \mbox{ for generic}
		 	\zeta \in \Lambda^+T^* \otimes Z$;
${\cal R} :=	\mbox{The natural bundle over $X_\zeta$, viz., }
		Y_\zeta \times_G R, 
		~{\rm with}~ Y_\zeta := \{y | [ y \wedge y ]^+ = \zeta
\}$; and finally
$\xi :=		\mbox{A tautological vector-bundle endormorphism}
		\mbox{ as an element in } T^* \otimes_\R {\rm
		Endskew}({\cal R})$.

We now apply the hyper-K\"ahler construction in the previous
subsection to the ALE  manifold 
\[
\begin{array}{|rl|}
\hline
\Xi := 	& (Q \otimes {\rm End}(R))_\Gamma = \{ \xi = \left(
	\begin{array}{c} \alpha \\ \beta \end{array} \right) \} \\
 =	& \bigoplus\limits_{ij} a_{ij}
 \hom({\mathbb{C}}^{n_i},{\mathbb{C}}^{n_j}) \\ 
 \cong	& \left( T^* \otimes_\R {\rm Endskew}(R) \right)_\Gamma = 
		\{ \xi = \left( \begin{array}{cc}
		\alpha	&	-\beta^\dagger \\
		\beta	&	\alpha^\dagger \\
		\end{array} \right) \} \\
\hline
\end{array}
\]
where $\alpha$ and $\beta$ are $|\Gamma| \times |\Gamma|$ matrices
satisfying $\left( \begin{array}{c} R_\gamma \alpha R_{\gamma^{-1}} \\
	R_\gamma \beta R_{\gamma^{-1}} \\ \end{array} \right) =
	Q_\gamma \left( \begin{array}{c} \alpha \\ \beta \\
\end{array} \right)$ 
for $\gamma \in \Gamma$. Of course this is simply the
$\Gamma$-invariance 
condition; or in a physical context, the projection of the matter
content 
on orbifolds. In the second line we have directly used the definition
of the McKay matrices\footnote{
	The steps are as follows: $({Q\otimes {\rm End}(R)})_\Gamma =
	(Q\otimes{\rm Hom}(\bigoplus\limits_i
R_i\otimes{\mathbb{C}}^{n_i},{\rm 
Hom}({\mathbb{C}}^{n_i},{\mathbb{C}}^{n_j})))_\Gamma = 
	(\bigoplus\limits_{ijk} a_{ik} {\rm
Hom}(R_k,R_j))_{\Gamma}\otimes{\rm
Hom}({\mathbb{C}}^{n_i},{\mathbb{C}}^{n_j})  = 
	\bigoplus\limits_{ij} a_{ij} {\rm
Hom}({\mathbb{C}}^{n_i},{\mathbb{C}}^{n_j})$ by
Schur's Lemma.} 
$a_{ij}$ and in the third, the canonical isomorphism
between ${\mathbb{C}}^4$ and the quaternions.

The hyper-K\"ahler forms are
$\omega_\R = {\rm Tr}(d\alpha \wedge d\alpha^\dagger) + {\rm
Tr}(d\beta \wedge d\beta^\dagger)$ 
and
$\omega_{\mathbb{C}} = {\rm Tr}(d\alpha \wedge d\beta)$, the moment maps,
$\mu_\R = [\alpha, \alpha^\dagger] + [\beta, \beta^\dagger]$ and
$\mu_{\mathbb{C}} = [\alpha,\beta]$. Moreover, the group of triholomorphic
isometries is 
$G = \prod\limits_{i=1}^r U(n_i)$ with a trivial $U(n_0) = U(1)$
modded out. 
It is then the celebrated theorem of Kronheimer \cite{Kronheimer} that
\begin{theorem}[Kronheimer]
The space
\[
X_\zeta := \{ \xi \in \Xi | \mu_i^a(\xi)  = \zeta^a_i \} / G
\]
is a smooth hyper-K\"ahler manifold of dimension\footnote{Since
	$\dim(X_\zeta) =  
	\dim(\Xi) - 4 \dim(G) = 2\sum\limits_{ij}a_{ij}n_i n_j -
	4(|\Gamma| - 1) = 
	 4|\Gamma| - 4|\Gamma| + 1 = 4$.}
four diffeomorphic to the resolution of the ALE orbifold
	${\mathbb{C}}^2 / \Gamma$. And conversely 
all ALE hyper-K\"ahler four-folds are obtained by such a resolution.
\end{theorem}
We remark that in the metric, $\zeta_{\mathbb{C}}$ corresponds to the
complex 
deformation while $\zeta_\R = 0$ corresponds to the singular limit
${\mathbb{C}}^2 / \Gamma$.
\subsection{Self-Dual Instantons on the ALE}
Kronheimer and Nakajima \cite{KN} subsequently applied the ADHM
construction 
on the ALE quotient constructed in the previous section.
In analogy to the usual ADHM construction, we begin with the data
$(V,W,{\cal A},\Psi)$ such 
that
\begin{center}
\begin{tabular}{rl}
$V,W$ := 	& A pair of unitary $\Gamma$-modules of complex
		dimensions \\  
		& $k$ and $n$ respectively; \\
$A,B$ :=	& $\Gamma$-equivariant endomorphisms of $V$; \\
${\cal A}$ := 	& $\left(\begin{array}{cc}
			A & -B^\dagger \\
			B & A^\dagger \end{array} \right)
		 \in (T^* \otimes_\R {\rm Endskew}(R))_\Gamma =	
		 \bigoplus\limits_{ij} a_{ij} {\rm Hom}(V_i,V_j)$; \\
$s,t^\dagger$ :=& homomorphisms from $V$ to $W$; \\
$\Psi$ :=	& $(s,t^\dagger) \in {\rm Hom}(S \otimes V,
		W)_\Gamma$.
\end{tabular}
\end{center}

Let us explain the terminology above. By $\Gamma$-module we simply
mean that 
$V$ and $W$ admit decompositions into the irreps of $\Gamma$ in the
canonical way: 
$V = \bigoplus\limits_i V_i \otimes R_i$ with $V_i \cong \C^{v_i}$
such that 
$k = \dim(V) = \sum\limits_i v_i n_i$ and similarly for $W$. By
$\Gamma$-equivariance 
we mean the operators as matrices can be block-decomposed (into $n_i
\times  n_j$) 
according to the decomposition of the modules $V$ and $W$.
In the definition of ${\cal A}$ we have used the McKay matrices in the 
reduction of $(T^* \otimes_\R {\rm Endskew}(R))_\Gamma$ in precisely
the same 
fashion as was in the definition of $\Xi$. For $\Phi$, we use
something analogous to 
the standard spin-bundle decomposition of tangent bundles
$T^* \otimes \C = S \otimes \bar{S}$, to positive and (dual) negative
spinors $S$ and $\bar{S}$.
We here should thus identify $S$ as the right-handed spinors and $Q$,
the left-handed. 

Finally we have an additional structure on $X_\zeta$. Now since
$X_\zeta$ is constructed 
as a quotient, with $P_\zeta$ as a principal $G$-bundle, we have an
induced natural bundle 
${\cal R} := P_\zeta \times_G R$ with trivial $R$  fibre. From this we
have a {\bf tautological 
bundle} ${\cal T}$ whose endomorphisms are furnished by 
$\xi \in T^* \otimes_\R {\rm Endskew}({\cal R})$. This is tautological
in the sense that 
$\xi \in \Xi$ and the points of the base $X_\zeta$ are precisely the
endomorphisms of the fibre $R$.

On $X_\zeta$ we define operators
${\cal A} \otimes {\rm Id}_{\cal T}, {\rm Id}_{V} \otimes \xi$ and
$\Psi \otimes {\rm Id}_{\cal T} : S \otimes V \otimes {\cal T}
\rightarrow W \otimes {\cal T}$. 
Finally we define the operator (which is a $(2k + n)|\Gamma| \times 2 k
|\Gamma|$ matrix because 
$S$ and $Q$ are of complex dimension 2, $V$, of dimension $k$ and
${\cal R}$ and ${\cal T}$, 
of dimension $|\Gamma|$)
\[
{\cal D} := ({\cal A} \otimes {\rm Id} - {\rm Id} \otimes \xi) \oplus
\Psi \otimes {\rm Id} 
\]
mapping $S \otimes V \otimes {\cal R} \rightarrow Q \otimes V \otimes
{\cal T} \oplus 
W \otimes {\cal R}$. We can restrict this operator to the
$\Gamma$-invariant part, 
viz., ${\cal D}_\Gamma$, which is now a $(2k + n) \times 2 k$
matrix. The adjoint is given by 
\[
{\cal D}^\dagger_\Gamma : \left( \bar{Q} \otimes \bar{V} \otimes {\cal
		  T} \right)_\Gamma \oplus 
		  \left( \bar{W} \otimes {\cal T} \right)_\Gamma \rightarrow
			S \otimes \left( \bar{V} \otimes {\cal T}
		  \right)_\Gamma, 
\]
where $\bar{V},\bar{W}$ and $\bar{Q}$ denote the trivial (Cartesian
product) bundle 
over $X_\zeta$ with fibres $V,W$ and $Q$.

Now as with the $\R^4$ case, the moment maps encode the ADHM equations,
except that instead of the right hand side being zero, we now have the 
deformation parametres $\zeta$. In other words, we have
$[ {\cal A} \wedge {\cal A} ]^+ + \{ \Psi^\dagger, \Psi \} = -\zeta_V$,
where $\{ \Psi^\dagger, \Psi \} \in \Lambda^+ T^* \otimes {\rm
Endskew}(V)$  
is the symmetrisation in the $S$ indices and
contracting in the $W$ indices of $\Psi^\dagger \otimes \Psi$, and
$\zeta_V$ is such that 
$\zeta_V \otimes {\rm Id} \in \Lambda^+ T^* \otimes {\rm
End}((V\otimes R)^\Gamma)$. 
In component form this reads
\beq \label{ADHM-ALE}
\begin{array}{rcl}
[A,B] + ts &=& -\zeta_\C;\\
([A,A^\dagger] + [B,B^\dagger]) - s s^\dagger + t t^\dagger &=& \zeta_\R,
\end{array}
\eeq
where as before $\zeta = \bigoplus\limits_{i=1}^{r} \zeta_i {\rm Id}_{v_i}
\in \R^3 \otimes Z$.

Thus equipped, the anti-self-dual\footnote{The self-dual 
	ones are obtained by reversing the
	orientation of the bundle.}
instantons can be constructed by the following theorem:
\begin{theorem}[Kronheimer-Nakajima]
For ${\cal A}$ and $\Psi$ satisfying injectivity of ${\cal D}_\Gamma$
and (\ref{ADHM-ALE}), 
all anti-self-dual $U(n)$ connections of instanton number $k$, 
on ALE can be obtained as the induced connection on the
bundle $E = {\rm Coker}({\cal D}_\Gamma)$.
\end{theorem}
More explicitly, we take an orthonormal frame $U$ of sections of 
Ker$({\cal D}_\Gamma^\dagger)$, i.e., a $(2k + n) \times n$ complex
matrix such that ${\cal D}_\Gamma^\dagger U = 0$ and $U^\dagger U =
{\rm Id}$. 
Then the required connection (gauge field) is given by
\[
A_\mu = U^\dagger \nabla_\mu U.
\]
\subsection{Quiver Varieties}
\index{Quivers!quiver variety}
We can finally take a unified perspective, combining what we have
explained concerning the construction of ALE-instantons as
Hyper-K\"ahler quotients and the quivers for th orbifolds of $\IC^2$.
Given an $SU(2)$ quiver (i.e., a McKay quiver constructed out of
$\Gamma$, a finite  
discrete subgroup of $SU(2)$) $Q$ with edges $H=\{h\}$, vertices
$\{1,2,...,r\}$,  
and beginning (resp. ends) of $h$ as $\alpha(h)$ (resp. $\beta(h)$),
we study the representation by associating 
vector spaces as follows: 
to each vertex $q$, we associate a pair of hermitian vector spaces
$V_q$ and $W_q$. 
We then define the complex vector space:
\[
\begin{array}{ccl}
M(v,w) & :=  & \left( \bigoplus\limits_{h \in H} {\rm
	Hom}(V_{\alpha(h)},V_{\beta(h)}) \right) \oplus 
	\left(\bigoplus\limits_{q=1}^r {\rm Hom}(W_q,V_q) \oplus {\rm
	Hom}(V_q,W_q)\right) \\ 
& := & \bigoplus\limits_{h,q} \left\{ B_h,i_q,j_q \right\}
\end{array}
\]
with $v := \left( \dim_\C V_1,~...~, \dim_\C V_n \right)$ and 
	$w := \left( \dim_\C W_1,~...~, \dim_\C W_n \right)$ being
	vectors of dimensions of the spaces associated with the nodes.

Upon $M(v,w)$ we can introduce the action by a group
\[
G := \prod\limits_q U(V_q) : \left\{ B_h,i_q,j_q \right\} \rightarrow 
		\left\{ g_{\alpha(h)} B_h g^{-1}_{\beta(h)}, g_q i_q,
		j_q g_q^{-1} \right\} 
\]
with each factor acting as the unitary group $U(V_q)$. We shall be
more concerned with 
$G' := G / U(1)$ where the trivial scalar action by an overall factor
of $U(1)$ has been modded out.

In $Q$ we can choose an orientation $\Omega$ and hence a signature for
each (directed) 
edge $h$, viz., $\epsilon(h) = 1$ if $h \in \Omega$ and $\epsilon(h) =
-1$ if $h \in \bar\Omega$. 
Hyper-K\"ahler moment maps are subsequently given by:
\beq
\label{moment}
\begin{array}{ccl}
\mu_\R(B,i,j) & := & {i \over 2}\left(\sum\limits_{h\in H,
q=\alpha(h)} B_h B_h^\dagger -  
		B_{\bar{h}}^\dagger B_{\bar{h}} + i_q i_q^\dagger -
		j_q^\dagger j_q \right) 
		\in \bigoplus\limits_q \mathfrak{u}(V_q) :=
		\mathfrak{g}, \\ 
\mu_\C(B,i,j) & := & \left(\sum\limits_{h\in H, q=\alpha(h)}
		\epsilon(h) B_h B_{\bar{h}}  
		+ i_q j_q \right)
		\in \bigoplus\limits_q \mathfrak{gl}(V_q) :=
		\mathfrak{g} \otimes \C.
\end{array}
\eeq
These maps (\ref{moment}) we recognise as precisely the ADHM equations
in a different 
guise. Moreover, the center $Z$ of $\mathfrak{g}$, being a set of
scalar $r \times r$ matrices, 
can be identified with $\R^n$. For Dynkin graphs\footnote{
	In general they are defined as $R_+ := \{\theta \in \Z_{\ge 0}^n | 
	\theta^t \cdot C \cdot \theta \le 2 \} \backslash \{ 0 \}$ for
	generalised Cartan matrix 
	$C := 2I - A$ with $A$ the adjacency matrix of the graph; 
	$R_+(v) := \{ \theta \in R_+ | \theta_q \le v_q = \dim_\C
	V_q~\forall q\}$ and 
	$D_\theta := \{ x \in \R^n | x \cdot \theta = 0 \}$.}
we can then define $R_+$, the set of positive
roots, $R_+(v)$, the positive roots bounded by $v$ and $D_\theta$, the
	wall defined 
by the root $\theta$. 

We rephrase Kronheimer's theorem as \cite{KN}:
\begin{theorem}
For the discrete subgroup $\Gamma \in SU(2)$, let $v =
(n_0,n_1,...,n_n)$, the vector of Dynkin labels 
of the Affine Dynkin graph associated with $\Gamma$ and let $w = 0$, then
for\footnote{$Z$ is the trace-free part of the centre and $\mu(B) =
\zeta$ means, component-wise 
	$\mu_\R = \zeta_\R$ and $\mu_\C = \zeta_\C$.}
$\zeta  := (\zeta_\R,\zeta_\C) \in \left\{ \R^3 \otimes Z \right\}
\backslash  
	\bigcup\limits_{\theta \in \R_+ \backslash \{n\}} \R^3 \otimes
D_\theta$, 
the manifold
\[
X_\zeta := \{ B \in M(v,0) | \mu(B) = \zeta \} // G'
\]
is the smooth resolution of $\C^2 / \Gamma$ with corresponding ALE
metric.
\end{theorem}
For our purposes this construction induces a natural bundle which will
give us the required 
instanton. In fact, we can identify $G' = \prod\limits_{q \ne 0}
U(V_q)$ as the gauge group 
over the non-Affine nodes and consider the bundle
\[
{\cal R}_l = \mu^{-1}(\zeta) \times_{G'} \C^{n_l}
\]
for $l = 1,...,r$ indexing the non-Affine nodes where $\C^{n_l}$ is
the space acted upon 
by the irreps of $\Gamma$ (whose dimensions, by the McKay
Correspondence, are precisely the 
Dynkin labels) such that $U(V_q)$ acts trivially (by Schur's Lemma)
unless $q=l$. For the affine 
node, we define ${\cal R}_0$ to be the trivial bundle (inspired by the
fact that this node 
corresponds to the trivial principal 1-dimensional irrep of $\Gamma$).
There is an obvious tautological bundle endomorphism:
\[
\xi := (\xi_h) \in \bigoplus\limits_{h \in H} {\rm Hom}({\cal
R}_{\alpha(h)},{\cal R}_{\beta(h)}). 
\]
We now re-phrase the Kronheimer-Nakajima theorem above as
\begin{theorem}
The following sequence of bundle endomorphisms
\[
\bigoplus\limits_q V_q \otimes {\cal R}_q
	\stackrel{\sigma}{\rightarrow}
\left( \bigoplus\limits_{h \in H} V_{\alpha(h)} \otimes {\cal
R}_{\beta(h)} \right) 
\oplus \left( \bigoplus\limits_q W_q \otimes {\cal R}_q \right)
	\stackrel{\tau}{\rightarrow}
\bigoplus\limits_q V_q \otimes {\cal R}_q,
\]
where
\[
\begin{array}{ccl}
\sigma & := & \left( B_{\bar{h}} \otimes {\rm Id}_{{\cal R}_{\beta(h)}}
		+ \epsilon(h) {\rm Id}_{V_{\alpha(h)}} \otimes \xi_h
		\right) \oplus 
		\left( j_q \otimes {\rm Id}_{{\cal R}_q} \right)\\
\tau & := & \left( \epsilon(h) B_{\bar{h}} \otimes {\rm Id}_{{\cal
		R}_{\beta(h)}} - 
		{\rm Id}_{V_{\alpha(h)}} \otimes \xi_{\bar{h}}, 
		i_q \otimes {\rm Id}_{V_q} \right)
\end{array}
\]
is a complex (since the ADHM equation $\mu_\C(B,i,j) = -\zeta_\C$
implies $\tau\sigma = 0$)  
and the induced connection $A$ on the bundle
\[
E := {\rm Coker}(\sigma,\tau^\dagger) \subset \left(
	\bigoplus\limits_{h \in H} V_{\alpha(h)}  
	\otimes {\cal R}_{\beta(h)} \right)
	\oplus \left( \bigoplus\limits_q W_q \otimes {\cal R}_q
	\right) 
\]
is anti-self-dual. And conversely all such connections are thus
obtained.
\end{theorem}

\begin{figure}
\centerline{\psfig{figure=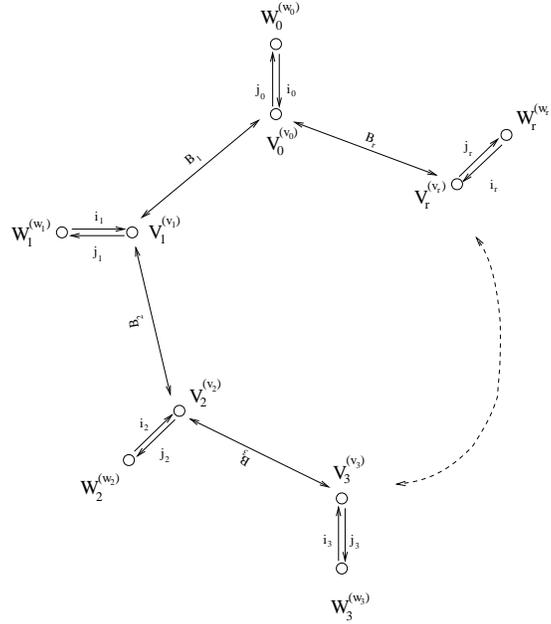,width=3.0in}}
\caption{
The Kronheimer-Nakajima quiver for $\IC^2/A_n$, extending the McKay
quiver to also encapture the information for the construction of the
ALE instanton.
}
\label{f:quivervar}
\end{figure}

We here illustrate the discussions above via explicit quiver diagrams;
though 
we shall use the $\widehat{A_2}$ as our diagrammatic example, the
generic structure should be captured.
The quiver is represented in \fref{f:quivervar} and the concepts
introduced
in the previous sections are elucidated therein. In the figure,
the vector space $V$ of dimension $k$ is decomposed into $V_0 \oplus
V_1 \oplus ... \oplus V_r$, 
each of dimension $v_i$ and
associated with the $i$-th node of Dynkin label $n_i = \dim(R_i)$ in
the affine Dynkin diagram of rank $r$. This is simply the usual McKay
quiver for  
$\Gamma \subset SU(2)$. Therefore we have $k = \sum\limits_i n_i
v_i$. 

To this we add the vector space $W$ of dimension $n$ decomposing
similarly as $W = W_0 \oplus W_1 \oplus ... \oplus W_r$, each of
dimension $w_i$ and 
$n = \sum\limits_i n_i w_i$. Now we have the McKay quiver with extra
legs. 
Between each pair of nodes $V_{q_1}$ and $V_{q_2}$ we have the map
$B_h$ with $h$ 
the edge between these two nodes. We note of course that due to McKay
$h$ is undirected 
and single-valence for $SU(2)$ thus making specifying merely one map
between two nodes sufficient. 
Between each pair $V_q$ and $W_q$ we have the maps 
$i_q : W_q \rightarrow V_q$ and $j_q$, in the other direction.
The group $U(k)$ is broken down to $(\prod\limits_{q=0}^r
U(v_q))/U(1)$. This is the 
group of $\Gamma$-compatible symplectic diffeomorphisms.
This latter gauge group is our required rank $n=\dim(W)$ unitary
bundle with anti-self-dual 
connection, i.e., an $U(n)$ instanton with instanton number $k =
\dim(V)$.

\newpage
\section*{\qquad \qquad \qquad \qquad { Epilogue}}
Thus we conclude Liber I, our preparatory journey into the requisite
mathematics. We have introduced canonical Gorenstein singularities
and monodromies thereon. Thereafter we have studied symplectic
structures one could impose, especially in the context of symplectic
quotients and moment maps. 
As a powerful example of such quotients we have
reviewed toric varieties.

We then digressed to the representation of finite groups, in
preparation of studying a wide class of Gorenstein singularities: the
orbifolds. We shall see in Liber III how all of the Abelian orbifolds
actually afford toric descriptions.
Subsequently we digressed again to the theory of finite graphs and
quiver, another key constituent of this writing. 

A unified outlook was
finally performed in the last sections of Chapter 4 where symplectic
quotients in conjunction with quivers were used to address orbifolds of
$\IC^2$, the so-called ALE spaces.
With all these tools in hand, let us now proceed to string theory.

\part{LIBER SECUNDUS: Invocatio Philosophi\ae~Naturalis}
\chapter{Calabi-Yau Sigma Models and ${\cal N}=2$ Superconformal
Theories}
\section*{Nomenclature}
We have by now prepared the reader, in the spirit of the Landau
School, with the requisite mathematics. Now let us move onto the theme
of this writing: string theory. The following 4 chapters will serve as an
introduction of the requisite background in physics. First, to
parallel Liber I, ket us clarify some notations:

\begin{tabular}{ll}
\\
$\alpha'$               & String tension\\
$l_s, g_s$              & String Length and Coupling\\
CY3                     & Calabi-Yau threefold \\
$Dp$			& Dirichlet $p$-brane \\
$NS5$			& Neveu-Schwarz 5-brane\\
$g_{YM}$		& Yang-Mills coupling \\
GLSM                    & Gauged linear $\sigma$-model\\
LG			& Landau-Ginsberg Theories\\
${\cal N}$              & Number of supersymmetries\\
VEV                     & Vacuum Expectation Value\\
$\zeta$			& Fayet-Illiopoulos Parametre
\end{tabular}

\newpage
{\hspace{-0.5in}{ A}} key feature of the type II superstring is
the 2-dimensional 
world-sheet ${\cal N}=2$ superconformal field theory with central
charge $c=15$. In compactification down to $\IR^4 \times CY3$, the
difficult part to study is the $c=9$ ${\cal N}=2$ theory internal 
to the Calabi-Yau, the properties of which determine the $c=6$ 
theory on the $\IR^4$ that is ultimately to give our real world.

A main theme therefore, is the construction of the various $c=9$
so-called ``internal'' ${\cal N}=2$ superconformal theories. Three
major subtypes have been widely studied (q.v. \cite{Greene-Lec} for
an excellent pedagogical review). These are
\begin{enumerate}
\item The non-linear sigma model, embedding the worldsheet, into
	the $CY3$ endowed with a metric $g_{\mu\nu}$ and
	anti-symmetric 2-form $B_{\mu\nu}$, with action
	\[
	\frac{1}{\alpha'}\int_{w.s} (g_{\mu\nu}+B_{\mu\nu})
		\partial^\mu X \partial^\nu X+ {\rm fermion};
	\]
\item The Landau-Ginsberg (LG) 
	theory, constructed from chiral superfields
	$\Psi_i,\bar{\Psi}_i$,
	and with a holomorphic polynomial superpotential 
	$W(\Psi_i)$ giving a unique vacuum.
	The action is an integral over the ${\cal N}=2$ superspace
	\[
	\int dz^2d\theta^4 K(\Psi_i,\bar{\Psi}_i) + (W(\Psi_i) +
		h.c.).
	\]
	We usually start with a non-conformal case and let it flow
	to a superconformal fixed point into IR;
\item The minimal models, being rational conformal field theories
	with a finite number of primary fields (and $c<1$ in the
	bosonic case or $c < 3/2$ in ${\cal N}=1$),
	furnishing unitary highest-weight representations of the
	(super)-Virasoro algebra. These can
	then be tensored together to achieve $c=9$.
\end{enumerate}
\index{LG Theory}
Now the LG theories can be seen as explicit Langrangian
realisation of tensor products of the minimal models \cite{LG-MM}.
On the other hand, the Gepner construction \cite{Gepner} relates
the chiral primaries in the minimal models with co\"{o}rdinates
in certain Calabi-Yau hypersurfaces, thereby relating 1 and 3.
Hence we shall focus on the inter-relation between 1 and 2.

Indeed this inter-relation between LG theories and Calabi-Yau
sigma models is what interests us most. The theme of this
writing is to study the behaviour of string theory on Calabi-Yau
varieties, modeled as algebraic singularities. The physics with
which we are concerned are supersymmetric gauge theories of
${\cal N}=0,1,2$ in 4 dimensions. These, with their matter content
and superpotential, can be written precisely in LG form. In
establishing the proposed correspondence, quantities in the
gauge theory can then be mapped to geometrical properties in the
Calabi-Yau. This correspondence was first provided by Witten in
\cite{GLSM}. With a brief review thereupon let us begin our
invocations in physics.
\section{The Gauged Linear Sigma Model}
\index{GLSM}
According to \cite{GLSM}, let us begin with neither the
Calabi-Yau sigma model nor the LG theory with superpotenetial,
let us begin instead with a {\bf linear sigma model} with gauge
group $U(1)$. The action is
\[
S = S_{kinetic} + S_D + \int d^2 z d^2 \theta \; W,
\]
where $W$ is our
superpotential in terms of the chiral super-fields 
$X=\{P,s_{1,...,5}\}$,
with $U(1)$ charges $Q:=(-5,1,...,1)$. We choose $W$ to be of the
form $W = P \cdot G(s_i)$ where $G$ is a homogeneous polynomial
of degree 5. On the other hand,
$S_D$ is the D-term of Fayet-Illiopoulos, of the form
\[
D = -e^2\left( \sum_i Q_i |X_i|^2 - r \right)
	= -e^2 \left( \sum_i |s_i|^2 - 5|p|^2 - r \right).
\]
The bosonic part of our potential then becomes
\[
U = |G(s_i)|^2 + |p|^2 \sum_i |\frac{\partial G}{\partial s_i}|^2
	+ \frac{1}{2e^2} + 2|\sigma|^2 \sum_i Q_i^2 |X_i|^2,
\]
with $\sigma$ a scalar field in the (twisted) chiral multiplet.
The vacuum of the theory, i.e., the moduli space, is then determined
by the minimum of $U$, which being a sum of squares, attains its
minimum when each of the terms does so.

What is crucial is the FI-parametre $r$ which we shall see as
an interpolator between phases.
\subsection*{The Phase $r > 0$}
When $r>0$, minimising the $D^2$ term in $U$ implies that at
least one $s_i$ is non-zero. This forces the second term in $U$
to attain its minimum at $p=0$, so too the argument applies to the
last term to force $\sigma=0$ and the first, to imply $G=0$.

Therefore our vacuum is parametrised by $\sum_i |s_i|^2 = r$,
together with the identification due to gauge symmetry, viz.,
$s_i \sim e^{i \theta} s_i$. In other words, the superfields live
in $\IC\IP^4$ (a toric variety).

The one more condition we obtained, namely $G=0$, implies that
for $r>0$ the fields actually live in a hypersurface in 
$\IC \IP^4$. Of course such hypersurface, the homogenenous quintic,
is a Calabi-Yau manifold.

We note therefore, in the limit of $r>0$, certain fields whose
masses in the original Lagrangian are determined by $r$, play no
r\^{o}le in recovering the Calabi-Yau and are effectively integrated
out. We have therefore obtained, in the IR,
a conformal non-linear sigma model on the CY as a hypersurface in
a toric variety.
\subsection*{The Phase $r<0$}
In the case of $r<0$, reasoning as above, we conclude that all $s_i$
vanish and $p = \sqrt{-r/5}$ which gives an unbroken $\IZ_5$
gauge symmetry because $p$ is of charge 5.
We actually arrive at a single point
for the vacuum and the $s_i$ act as fluctuations around it.
The configuration is thus $\IC^5/\IZ_5$ and is an orbifold
of a LG theory.

We conclude therefore that the gauged linear sigma model has
2 limits, a Calabi-Yau non-linear sigma model ($r>0$) and an 
(orbifolded) Landau-Ginsberg theory ($r<0$). In fact the 
complexified form of $r$, namely $e^{2 \pi i (b + i r)}$ serves
as the K\"ahler parametre of the moduli space.
\section{Generalisations to Toric Varieties}
The above approach of relating LG theories and Calabi-Yau sigma models
not only gave a physically enlightening way to intimately tie together
two methods of constructing ${\cal N}=2$ superconformal theories, but
also presented mathematicians with a novel perspective on toric
varieties. The construction was soon generalised to other toric
varieties as well as hypersurfaces therein
\cite{Monomial,SmallD,Aspin-Reso} (cf. also \cite{Beasley-The} and
\cite{Cox-rev}).

As we shall later describe the method in painstaking detail in Liber
III, where we shall construct gauge theories for D-brane
probes on arbitrary toric singularities, we shall be brief for the
moment. The idea is to generalise the charge vector $Q$ discussed
above to a product of $n-d$ $U(1)$ groups for $n$ superfields,
whereupon the charges become encoded by
an $n \times (n-d)$ integer matrix
$Q_{i=1,\ldots,n}^{a=1,\ldots,n-d}$ such that $\sum_i Q_i^a = 0$ so
that the D-term equations are written as
\[
\sum_i Q_i^a v_i = 0~\forall~a.
\]

It is with foresight in the above that we have written $v_i :=
|X_i|^2$ for the modulus-squared of the superfields. We identify $v_i$
as generators of a polyhedral cone (cf. Liber I, Section 1.3) and
define the toric variety accordingly, the toroidal $\IC^{*(n-d)}$ action
is prescribed exactly as
\[
\lambda_a : x_i \rightarrow \lambda_a^{Q_i^a} x_i
\]
for $x_i \in \IC^n$.

In this description therefore, the moment map defining the toric
variety is simply the D-term and the charge matrix of the linear sigma
model gives the relations among all the generators of the cone. In the
case of the toric variety being singular, the desingularisation
thereof simply corresponds to the acquisition of non-zero values of
the FI-parametre $r$.

In this way we can describe any toric variety as a gauged linear sigma
model with charge matrix $Q_i^a$ whose integer kernel has $\IZ$-span
$v_i$, which are the generators of the cone. The {\em homogeneous
co\"{o}rdinate ring} is given as the subring of $\IC[x_1,\ldots,x_n]$,
invariant under the above $\IC^*$ action by $Q_i^a$, namely
\[
\IC[x_1, \ldots, x_n]^Q = \{z_a = \prod_i x_i^{v_i^a} \}.
\]

Our above construction of the moduli space in the IR, will turn out to
be a crucial ingredient in the construction of gauge theories from
string theory. Indeed if we use D-branes to probe background
(Calabi-Yau) geometry, the IR moduli space of the world-volume theory
will precisely be the background.

This construction of gauge theories brings us to the motivation behind
all of our discussions.  Indeed if string theory promises to be Grand
Unified Theory, one must be able to construct the Standard Model gauge
theory therefrom. In the following 3 chapters we shall present 3
alternative methods towards this noble goal.
\chapter{Geometrical Engineering of Gauge Theories}
\index{Geometric Engineering}
{ A} natural approach to the construction of four dimensional
(supersymmetric) gauge theories is of course to consider the low
energy limit of String/M/F-theory in the context of compactifications
on Calabi-Yau spaces. Such an endeavour, of using the geometrical
properties of the underlying Calabi-Yau space to explain the
perturbative and non-perturbative effects of the field theory, was
pioneered in the beautiful papers \cite{geoeng1,geoeng2,geoeng3}.

Historical trends have shown that the more supersymmetry one has,
the easier the techniques become. The above papers initiated the study
of ${\cal N}=2$ theories; those with ${\cal N}=1$ came later
(q.v.~e.g \cite{Cachazo}). The
construction was based on the fabrication of ${\cal N}=2$  theories
by compactifying the heterotic $E_8 \times E_8$ or $Spin(32)/\IZ_2$ 
string theory on $K3 \times T^2$, which by string duality
\cite{Kachru-Vafa}, is equivalent to type IIA/B on a Calabi-Yau
threefold.
\section{Type II Compactifications}
Let us first briefly remind ourselves of some key facts in type II
compactifications (q.~v.~ \cite{typeII} for an excellent review). 
The spaces with which we are concerned are
Ricci-flat K\"ahler manifolds of $SU(3)$ holonomy with Hodge diamond
\[
h^{p,q} =
\ba{ccccccc}
&&&1&&&\\
&&0&&0&&\\
&0&&h^{1,1}&&0&\\
1&&h^{2,1}&&h^{2,1}&&1\\
&0&&h^{1,1}&&0&\\
&&0&&0&&\\
&&&1&&&\\
\ea
\]
There are hence two parametres serving to characterise such a
(restricted) Calabi-Yau 3-fold, namely $h^{2,1}$, the space of complex
structure and $h^{1,1}$, the space of K\"ahler structure.
Indeed string theory on such curved backgrounds gives rise to a
$(2,2)$ super-conformal sigma model, the spectrum of which is
therefore in one-to-one correspondence with the above cohomologies of
the Calabi-Yau. From the point of view of the resulting ${\cal N}=2$
theory in four dimensions, the aforementioned deformations of complex
and K\"ahler structures realise as the moduli space of vector ($M_V$) and
hyper-multiplets ($M_H$) of the supersymmetry algebra. Indeed for type IIA
compactifications, $M_V$ corresponds to the complexified\footnote{
	The K\"ahler form $J$ is complexified by the type II
	NS-NS B-field as $J + i B$.} K\"ahler
deformations and is of complex dimension $h^{2,1}$ while $M_H$
corresponds to complex deformations together with RR fields and the
dilation-axion, and has quaternionic dimension $h^{2,1}+1$. In other
words, the abelian gauge symmetry including the graviphoton
corresponding to the vector multiplets is $U(1)^{h^{1,1}+1}$. In
addition, there are $h^{2,1}+1$ massless hyper-multiplets.
One important fact to note is that since the dilaton lives in the
hypermultiplet, the vector couplings (gauge coupling and moduli space
metric) are purely classical.

The situation for type IIB is reversed, and the complex dimension 
$\dim_{\IC}(M_V) = h^{2,1}$ 
while the quaternionic dimension
$\dim_{Q}(M_H) = h^{1,1}+1$. Thus the vector (gauge) couplings
here are not affected by K\"ahler deformations which correspond to
world-sheet instantons and can be calculated purely geometrically.
This is of course a manifestation
of mirror symmetry upon which we shall touch lightly later in this
chapter.
\section{Non-Abelian Gauge Symmetry and Geometrical Engineering}
In the above, we have addressed the massless spectrum of type II
compactifications on Calabi-Yau 3-folds (CY3) where one could see the
emergence of an Abelian gauge symmetry. The construction of
non-Abelian gauge theories with adjoint matter fields was initiated in
\cite{geoeng1,geoeng2,geoeng3,matter}. As with all studies in
compactification, the method of attack was to start with the Calabi-Yau
2-fold, namely the K3 surface and consider complex fibrations of K3
to obtain the 3-fold.
The crucial realisation was that, due to the duality between heterotic
on $K3 \times T^2$ and type IIA on the CY3, itself as a $K3$-fibre bundle
\cite{Kachru-Vafa}, 
the relevant QFT moduli space comes from the $K3$ singularities so
that the gauge fields are obtained from wrapping type IIA D2 branes on
the vanishing 2-cycles thereof and that the matter comes from the
extra singularities of the base of the CY3. 

Let us digress a moment to remind the reader of the key features of K3
surfaces needed in the construction. We recall that a local
singularity of K3 can be modeled as an {\bf Asymptotically Locally
Euclidean} or ALE space. These are quotient spaces of $\IC^2$ by
discrete subgroups of the monodromy group $SU(2)$. We have learnt in
Liber I, Chapter 3 that such quotients are the 2-dimensional
orbifolds, or the du-Val-Klein singularities, with an
$ADE$-classification.
\index{ADE Singularities}
The steps of geometrical
engineering are therefore as follows: (i) specify the type of ADE
singularity of the K3 fibre; (ii) the gauge coupling is related to the
volume of the base as
\[
\frac{1}{g_{YM}} = \sqrt{V(B)};
\]
take the large $V(B)$ limit so that gravity decouples and 
so that only
the gauge dynamics becomes relevant; and (iii) consider the behaviour
of the string theory as D2-branes wrap the vanishing cycles
corresponding to the singularities of the fibre. In so doing, our
study of the vanishing cycles in the context of Picard-Lefschetz theory
in 2.1.1 will be of significance.

Let us illustrate with the canonical example of the $A_1$ singularity
corresponding to a $\IZ_2$ quotient of $\IC^2$, fibred over
$\IP^1$. The singularity is described by $xy=z^2$. We can set
$x=\phi_1^2\phi_2, y= \phi_3^2\phi_2$ and
$z=\phi_1\phi_2 \phi_3$ with $\phi_i$ the complex fields of a
two-dimensional SUSY gauged linear sigma model (GLSM); the D-term is
given by $U(\phi_1, \phi_2,\phi_3)=(\phi_1
\phi_1+\phi_3\phi_3-2\phi_2\phi_2-\zeta)^2$, with Fayet-Illiopoulos
parametre $\zeta$ serving as a K\"ahler resolution of the singularity
as a $\IP^1$-blowup.

Now let D2-branes wrap the $\IP^1$-blowups, which are the vanishing
cycles of the fibre. We obtain two vector particles $W^{\pm}_\mu$
depending on the orientation of wrapping, with masses proportional to
the volume of the blowup. These are charged under the $U(1)$ field
$Z^{\mu}_0$ obtained from decomposing the RR 3-form of IIA onto the
harmonic form of the $\IP^1$. As we shrink the size of the blow-up,
the $W$ and $Z$ become massless and form an adjoint of $SU(2)$ and we
obtain a 6D $SU(2)$ gauge theory. Further compactification upon the
base over which our type $A_1$ $K3$ is fibred to give the CY3 finally
gives us a 4D ${\cal N}=2$ pure $SU(2)$ Yang-Mills. The analysis
extends to all other $ADE$ groups and it is easy to remember that a
singularity of type $A$ (respectively $D$, $E$) gives a gauge group
which is the compact Lie groups under Dynkin classification type $A$
(respectively $D,E$)\footnote{The non-simply laced cases of $BCFG$ can
	be obtained as well after some modifications (q.~v.
	e.g. \cite{geoengrev})}.

To obtain matter, we consider {\em collisions of fibres}. For example,
letting an $A_{m-1}$ singularity of the K3 fibre meet with an
$A_{n-1}$ one
would give a gauge group $SU(m) \times  SU(n)$. The base geometry
would consist of two intersecting $\IP^1$'s whose volumes determine
the gauge couplings of each factor. Wrapping a linear combination of
the 2 vanishing cycles will give rise to {\bf bi-fundamental} fields
transforming as $(m,\bar{n})$ of the gauge group. Moreover, taking the
limit of one of the base volumes would make the gauge factor a flavour
symmetry and henceforth give rise to fundamental matter.

We can thus geometrically engineer 4 dimensional ${\cal N}=2$ Yang-Mills
theories with product gauge groups with (bi-)fundamental matter by the
pure classical geometry of CY3 modeled as $K3$-fibrations over $\IP^1$.
\subsection{Quantum Effects and Local Mirror Symmetry}
The above construction gave us classical aspects of the gauge theory
as one had to take the $\alpha' \rightarrow 0$ limit to decouple
gravity and consider only the low energy physics. Therefore we
consider only local geometry, or the non-compact
singularities which model the Calabi-Yau. This is why we discussed at
length the singularity behaviour of complex varieties in Liber
I and why we shall later make extensive usage of these local,
singular varieties. The large volume limits are suppressed by powers
of $\alpha'$.

However it is well-known that the classical moduli space of ${\cal
N}=2$ Super-Yang-Mills receives quantum corrections. The prepotential
of the pure $SU(2)$ case for example is of the form ${\cal F}(A) =
\frac12 \tau_0 A^2 +
\frac{i}{\pi}A^2\log\left(\frac{A}{\Lambda}\right)^2 + F_{inst}$ in
terms of the scalar in the ${\cal N}=2$ vector multiplet. The log-term
describes the 1-loop effects while $F_{inst}$ is the instanton
corrections as determined by the Seiberg-Witten curve
\cite{Seiberg-Witten}. The corresponding prepotential in type IIA has
the structure \cite{Candelas}
\[
{\cal F} = -\frac16 C_{ABC}t_A t_B t_C - \frac{\chi \zeta(3)}{4 \pi^3}
+ \frac{1}{8 \pi^3} \sum\_{d_1, \ldots, d_h} n_{d_1, \ldots, d_h}
Li_3(\exp( i \sum_A d_A t_A )),
\]
in terms of the K\"ahler moduli $t_{A = 1,\ldots, h^{1,1}}$, where
$n_{d_1, \ldots, d_h}$ are the rational curves in the Calabi-Yau
corresponding to the instantons.

To compute these instanton effects one evokes the mirror principle and
map the discussion to type IIB compactified on the mirror
Calabi-Yau. Now we need to consider $D3$ branes wrapping vanishing
3-cycles (conifold-type singularities). In the double-scaling limit as
we try to decouple gravity ($\alpha' \rightarrow 0$) and study low
energy dynamics (volume of cycles $\rightarrow 0$), we are finding
mirrors of non-compact Calabi-Yau's. Such a procedure, with the
prototypical example being the ALE-fibrations, is referred to as {\bf
local mirror} transformation as opposed to that for the compact manifolds
studied in the original context of mirror symmetry.

We shall not delve to much into the matter, a rich and beautiful field
in itself. Suffice to say that mathematicians and physicists alike
have made much progress in the local mirror phenomenon, especially in
the context of (our interested) toric varieties (cf
e.g.
\cite{Batyrev1,Batyrev2,SYZ,LocalMirror,Amer,Hori-Vafa,Leung-Vafa}).
The original conjecture was the statement in \cite{Batyrev2}, that
``every pair of $d$ dimensional dual reflexive Gorenstein $\sigma$ and
$\sigma^\vee$ of index $r$ gives rise to an ${\cal N}=2$
superconformal theory with central charge $c = 3(d-2(r-1))$. Moreover,
the superpotentials of the corresponding LG theories define two
families of generalised toric Calabi-Yau manifolds related by mirror
symmetry.'' We recall from Section 1.1 of Liber I the definitions of
dual and Gorenstein cones. Here we elucidate two more. By reflexive we
mean that the Gorenstein cone $\sigma$ has a dual cone $\sigma^\vee$
which is also Gorenstein. The index $r$ is the inner product of $w$
and $w^\vee$, the two vectors guaranteeing the Calabi-Yau conditions
($\langle w, \sigma \rangle = \langle w^\vee, \sigma^\vee \rangle =
0$).

In terms of the complex equations. If $M$ is the variety corresponding
to $\sigma$ generated by $v_i$ satisfying the charge relation
(cf. Section 4.2)
\[
\sum_{i=1}^{n+d} Q_i^a v_i =0 \qquad a=1, \ldots, n,
\]
then the mirror $W$ is defined by the equation
$\sum_i a_i m_i = 0$,
where $a_i$ are coefficients and $m_i$, monomials which satisfy
\[
\prod_{i=1}^{n+d} m_i^{Q_i^a}=1.
\]

Having addressed the method of geometrical engineering, we now move
onto a more physical realisation of gauge theories, involving certain
configurations of branes in the 10-dimensions of the superstring.
\chapter{Hanany-Witten Configurations of Branes}
\index{Hanany-Witten}
\section{Type II Branes}
{ I}t is well-known that type IIA (restively type IIB) superstring
theory has Dirichlet $p$-branes of world volume dimension $p+1$ for $p
= 0, 
2,4,6,8$ (resp. $-1,1,3,5,7,9$) which are coupled to the Ramond-Ramond
$p+1$-form electro-magnetically. They are of tension and hence
RR charge, in units of the fundamental string scale $l_s$,
\[
T_p = \frac{1}{g_s l_s^{p+1}}.
\]
The $Dp$-branes are BPS saturated objects preserving half of the 
32 supercharges of type II, namely those of the form
\[
\epsilon_L Q_L + \epsilon_R Q_R \qquad {\rm s.t.} \quad
\epsilon_L = \Gamma^0 \ldots \Gamma^p \epsilon_R,
\]
where $Q_{L,R}$ are the spacetime supercharges generated by left and
right moving worldsheet degree of freedom of opposite chirality.
\subsection{Low Energy Effective Theories}
The low-energy world-volume theory on an infinite $Dp$-brane is
a $p+1$-dimensional field theory with 16 SUSY's describing the
dynamics
of the ground state of the open string which end on the brane (for a
pedantic review upon this subject, q.v. \cite{branerev}, upon which
much of the ensuing in this section is based). 
The theory is obtained by dimensional reduction of the $9+1$-D
${\cal N}=1$ $U(1)$ super-Yang-Mills (SYM) with gauge coupling
$g_{YM}^2 = g_s l_s^{p-3}$ by dimensional analysis. Gravity can
thus be decoupled by holding $g_{YM}$ fixed while sending
$l_s \rightarrow 0$.
The
massless spectrum includes a $p+1$-D $U(1)$ gauge field $A_\mu$
whose world-volume degrees of freedom carry the {\em Chan-Paton}
factors of the open strings, as well as $9-p$ scalars $X^I$ described
by the transverse directions to the world-volume.

As BPS objects, parallel $Dp$-branes are shown in a celebrated
calculation of \cite{Polchinski} to exert zero force upon each other.
This subsequently inspired the famous result of \cite{Nbranes},
stating that the low-energy dynamics of $N_c$ parallel coincident
$Dp$-branes gives a $U(N_c)$ SYM in $p+1$ dimensions with 16
supercharges. With the addition of {\bf orientifold $p$-planes}, which
are fixed planes of a $\IZ_2$ action on the 10-D spacetime and
are of charge $\pm 2^{p-4}$ times the corresponding $Dp$-brane charge,
we can similarly fabricate SYM with $Sp$ and $SO$ gauge groups.

A last player upon our stage is of course the solitonic NS-NS
5-brane, of tension $T_{NS} = \frac{1}{g_s^2 l_s^6}$, which
couples magnetically to the NS-NS B-field. It too is BPS object
preserving 16 supercharges. The low-energy theory of a stack of
$k$ IIB NS-branes is a 6-D $(1,1)$ $U(k)$ SYM while that of the
IIA NS-brane is more exotic, being a non-Abelian
generalisation of a non-trivial $(2,0)$ tensor-multiplet theory in
6 dimensions. The most crucial fact with which we shall concern
ourselves is the above tension formula. Indeed in the low-energy
limit as $g_S \rightarrow 0$, the NS-brane is heavier than any
of the D-branes and can be considered as relatively non-dynamical.
\subsection{Webs of Branes and Chains of Dualities}
Having addressed stacks of $Dp$-branes, now consider $N_c$ $Dp$-branes
occupying $x^{0,\ldots,p}$ directions with $N_f$ $D(p+4)$-branes in the
$x^{0, \ldots, p+4}$ directions. The SUSY preservation conditions
become more constrained: $\epsilon_L = \Gamma^0 \ldots \Gamma^p
\epsilon_R = \Gamma^0 \ldots \Gamma^{p+4} \epsilon_R$, subsequently
another 1/2 SUSY is broken. This is the famous $Dp-D(p+4)$ system
where the $Dp$ probes the geometry of the latter and the relative
positions of the branes give various moduli of the gauge theory.

More precisely, the locations of the $D(p+4)$ give the masses of the
$N_f$ fundamentals, those of $Dp$, the VEV's in the adjoint of
$U(N_c)$ and parametrise the Coulomb branch, and finally the $Dp$
directions in the $D(p+4)$ are the VEV's of adjoint hypermultiplets.
The Higgs branch, parametrised by the VEV's of the fundamentals, is
then the moduli space of $N_c$ instantons with gauge group $U(N_f)$.

From the above setup, in conjunction with the usage of a chain of
dualities which we summarise below, we may arrive at a sequence of
other useful setups. Here then are the effects of $S$ and $T$
dualities on various configurations ($R_i$ is the compactification
radius):
\[
\ba{c}
\mbox{T-duality along}\\
\mbox{the $i$-th direction:}
\ea
\qquad
\ba{ccc}
R_i & \leftrightarrow & \frac{l_s^2}{R_i} \\
g_s & \leftrightarrow & \frac{g_s l_s}{R_i} \\
Dp \mbox{ wrapped on } x^i & \leftrightarrow &
	D(p-1) \mbox{ at a point on } x^i \\
NS5_{IIA} \mbox{ wrapped on } x^i & \leftrightarrow &
	NS5_{IIB} \mbox{ wrapped on } x^i \\
NS5 \mbox{ at a point on } x^i & \leftrightarrow &
	KK \mbox{ monopole}
\ea
\]
\[
\mbox{Type IIB S-duality:} \qquad
\ba{ccc}
g_s & \leftrightarrow & \frac{1}{g_s} \\
l_s^2 & \leftrightarrow & l_s^2 g_s \\
\mbox{Fundamental String} & \leftrightarrow & D1\\
D3 & \leftrightarrow & D3\\
NS5 & \leftrightarrow & D5\\
(p,q) 7~{\rm brane} &\leftrightarrow &(p',q') 7~{\rm brane}
\ea
\]
\section{Hanany-Witten Setups}
Equipped with the above chain of dualities, from the $Dp-D(p+4)$
system we can arrive at $Dp-D(p+2)$ by compactifying 2
directions as well its T-dual version  $D(p+1)-D(p+3)$. Notably one
has the $D3-D5$ system. Subsequent S-duality leads to $D3-NS5$
configuration as well as all $Dp-NS5$ for other values of $p$ by
repeated T-dualities.

\begin{figure}
        \centerline{\psfig{figure=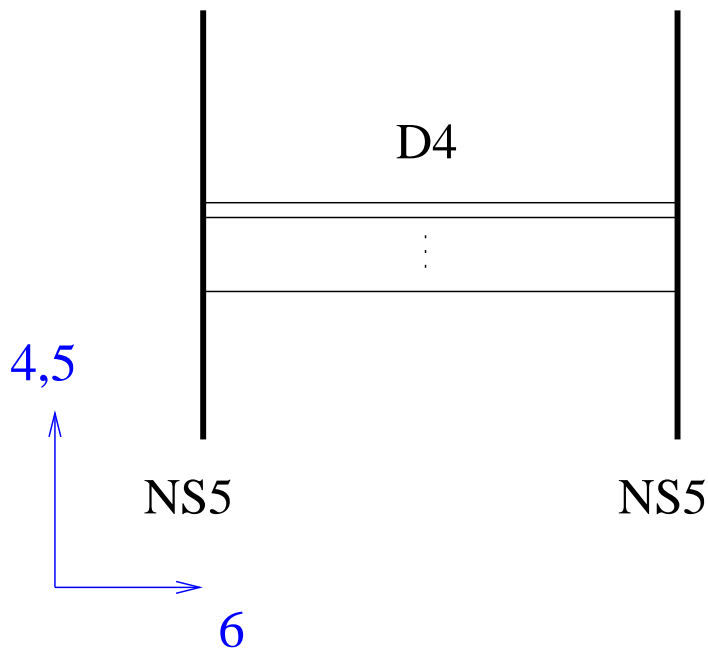,width=4in}}
\caption{The canonical example of the Hanany-Witten setup where a
	stack of D4-branes is stretched between 2 parallel NS5-branes.}
\label{f:hweg}
\end{figure}

Of particular interest is the type IIA setup, directly liftable to
M-theory, of a stack of $N_c$ $D4$-branes, stretched between 2 parallel
infinite NS5 branes (cf. \fref{f:hweg}). 
The D4-branes occupy directions $x^{0, \ldots 3,
6}$ and the two NS5 occupy $x^{0, \ldots 5}$, but at a distant $L_6$
apart. As discussed earlier the SUSY condition become more restricted and
the theory with 32 supercharges had the NS-brane been absent now
becomes one with a quarter as much, or 8. More precisely, the Lorentz
group breaks as $SO(1,9) \rightarrow SO(1,3) \times SO(2) \times
SO(3)$ respectively on $x^{0,1,2,3}$, $x^{4,5}$ and $x^{7,8,9}$. The
$SO(3)$ becomes a global $SU(2)$ R-symmetry of an ${\cal N}=2$ SYM
while the $SO(2)$, a $U(1)$ R-symmetry.
At low energies, to an 4-dimensional observer in $x^{0,1,2,3}$, bulk
10-D spacetime modes as well as those on the NS5 branes are higher
dimensional excitations and for length scales larger than $L_6$ the
excitations on the D4-branes essentially describe a 4-dimensional
(instead of 4+1) physics. What results is an ${\cal N}=2$ pure SYM
theory in 4 dimensions with gauge group $U(N_c)$. The gauge coupling
is $\frac{1}{g^2} = \frac{L_6}{g_s l_s}$ and in order to decouple gravity
effects and go to the low energy limit we need once again take the
double limit $g_s \rightarrow 0, L_6/l_s \rightarrow 0$.

What we have described above, is a prototypical example of the
celebrated {\bf Hanany-Witten} brane configuration where one
fabricates 4 dimensional supersymmetric gauge theories by suspending
D4-brane between NS-branes.
\subsection{Quantum Effects and M-Theory Solutions}
Of course the above discussion had been classical. Just as in the
geometrical engineering picture one has to use local mirror symmetry
to consider quantum effects, here too must one be careful. Indeed, in
type IIA the endpoints of the D4-branes on the NS5-branes are
singular and are governed by a Laplace-type of equation. An
approximate solution is to let the D4 exert a force and cause the two
NS5-branes to bend so that they are no longer strictly parallel with
respect to the 6th direction. In fact, the NS5-branes bend
logarithmically and the separation (which as we saw governs the gauge
coupling) varies and determines the logarithmic running of the
coupling.

The shapes of the branes thus incorporate the 1-loop effects. Now
since our theory is ${\cal N}=2$, there are no higher-loop
contributions due to non-renormalisation. Therefore what remains to be
considered are the non-perturbative instanton effects which we saw
above as the $Dp-D(p+4)$ system, or here, $D0$-branes in the $D4$.

The solution is the elegant ``lift to M-Theory'' \cite{M-Theory}. Of
course both the $D4$ and $NS5$ are different manifestation of the same
object in M-Theory, namely the $M5$-brane; the former is the $M5$
wrapped around the compact 11-th dimensional $S^1$ in going from
M-theory to type IIA while the latter is the $M5$ situated at a point
on the $S^1$.

The lift of the Hanany-Witten setup is then a Riemann surface $\Sigma$
in
11-dimensions. The bending condition from the 1-loop effects
determine the embedding equation of $\Sigma$ while the instantons are
automatically included since the $D0$-branes in M-theory are simply
Kaluza-Klein modes of the compactification. A most beautiful result of
\cite{M-Theory} is that $\Sigma$ is precisely the Seiberg-Witten curve
\cite{Seiberg-Witten}
describing the 4-dimensional field theory.

From geometrical engineering we have moved to configurations of
branes. Our next method will be D-branes at singular points in the
geometry. 
\chapter{Brane Probes and World Volume Theories}
\index{Brane Probes}
{ T}he third method of constructing gauge theories from string theory
which we shall now review in detail is the method of D-branes probing
background geometries. This is in some sense a mixture of the two
methods described above: it utilises both the geometry of local
Calabi-Yau as well as world-volume gauge theories living on D-branes. 

The pioneering work in this direction was initiated by Douglas and
Moore in \cite{DM}. Their technique is a physical realisation of the
mathematics which we described in Liber I, Chapter 3 and gives a
unifying application of such concepts as Hyper-K\"ahler quotients,
McKay quivers, Finite group representations and instanton moduli
spaces. 
\section{The Closed Sector}
Before we introduce D-branes and hence the open sector to our story
let us first briefly remind ourselves of the closed sector, in the
vein of the geometric engineering and compactifications presented
earlier in this Liber II as well as the mathematics of ALE spaces
introduced in Section 3.3 of Liber I. We recall that the ALE space
$M_{\Gamma}$ is the local model for K3 surfaces, being (resolutions
of) the orbifolds $\IC^2/(\Gamma \in SU(2))$. It is also known as a
{\bf gravitational instanton} in the sense that it is endowed with a
anti-self-dual hyper-K\"ahler metric (with $SU(2)$ holonomy).

For $\Gamma = A_{n-1}$, the metric is explicitly given as the
multi-centre Eguchi-Hanson metric \cite{Eguchi-Hanson}: 
\[
ds^2 = \left(\sum_{i=1}^n \frac{1}{|\vec{x} -\vec{x_i}|} \right)
^{-1}(dt  \vec{A} \cdot d \vec{x})^2 + V dx^2, 
\]
where $-\vec{\nabla}V = \vec{\nabla} \times \vec{A}$, $t$ is the
angular co\"ordinate, and $x_i$ the $n$ singular points. Choosing a
basis $\Sigma_i$ of $H^2(M_\Gamma;\IZ)$, the quantity
$\vec{\zeta}:=\vec{x_{i+1}} - \vec{x_i}$ is then equal to
$\int_{\Sigma_i} \vec{\omega}$, where $\vec{\omega}=\omega_{I,J,K}$ are
the three hyper-K\"ahler symplectic forms introduced in Section 1.2
of Liber I. The $\zeta_i$'s govern the size of the $\IP^1$-blowups and
are hence the K\"ahler parametres of the ALE space. 
The moduli space is of dimension $3n-6$, our familiar result for moduli
space of instantons.

When considering the ALE as the (two complex dimensional) target-space
for non-linear sigma models, we are left with ${\cal N}=(4,4)$
supersymmetry. On the other hand in the context of considering
superstrings propagating in the background $\IR^6 \times M_\Gamma$, we
have ${\cal N}=(0,1), (1,1)$ and $(0,2)$ respectively for types I, IIA
and IIB. The $SU(2)$ R-symmetry of the 6-dimensional gauge theory sits
as an unbroken subgroup of the $SO(4)$ isometry of the space. 
\section{The Open Sector}
Now let us add D5 branes to the picture. We do so for the obvious
reason that we shall consider D5 with its world-volume extending the
$\IR^6$ and transverse to the 4-dimensional $M_\Gamma$ (which together
constitute the 10-dimensions of type II superstring theory). Also
historically, Witten in \cite{Witten-small} considered the 5-brane
built as an instanton in the gauge theory of \cite{CHS}. The
6-dimensional ${\cal N}=1$ theory on the world-volume leads to a
hyper-K\"ahler quotient description of the vacuum moduli space. 

Consider a stack of $N$ D5-branes each filling the $\IR^6$ and at a
point in $\IC^2$. This gives us, as discussed in the previous chapter,
an $U(N)$ gauge theory in 6-dimensions with ${\cal N}=1$. Open strings
ending on the $i,j$-th D-brane carry Chan-Paton factors corresponding
to the gauge fields $A_\mu^{ij}$ as $N \times N$ Hermitian matrices;
we can write the states as 
\[
|A\rangle = A_\mu^{ij}\psi^\mu |ij\rangle,
\]
where $\psi^\mu$ are fermions; similarly we have scalars $X^i$ as $N
\times N$ matrices by dimensional reduction.

Thus prepared, let us move on to the configuration in question, viz.,
the stack of D5-branes situated at a point in the ALE orbifold of
$\IC^2$. The group $\Gamma$ has an induced action on the vectors as
well as scalars (and hence by supersymmetry the fermions), namely for
$g\in\Gamma$, $g: A_\mu(x) \rightarrow \gamma(g) A_\mu(x')
\gamma(g)^{-1}$ and $g: X^i(x) \rightarrow R(g)^i_j \gamma(g) X^j(x')
\gamma(g)^{-1}$ where $\gamma$ is a representation acting on the
Chan-Paton indices and $R$ is a representation that act additionally
on space-time. 

Due to this projection by the orbifold group, only a subsector of the
theory survives, namely  
\beq
\label{subsector}
A_\mu(x) = \gamma(g) A_\mu(x) \gamma(g)^{-1} \qquad
	X^i(x) = R(g)^i_j \gamma(g) X^j(x') \gamma(g)^{-1}.
\eeq

In Liber III, we shall present a detailed method of explicitly solving
these equations. For now we shall point out to the reader that such a
configuration of a stack of D-branes, placed transversely to a
singular point of the geometry, is called a {\bf brane probe}.
\subsection{Quiver Diagrams}
\index{Quivers!gauge theory}
We shall certainly delve into this matter further
in Liber III, a chief theme of which shall in fact be
the encoding of solutions to equations \eref{subsector}, 
namely those which describe the matter content
(and interaction) of the world-volume probe theory. Now,
let us here entice the reader with a few advertisements.

We shall learn that the world-volume super-Yang-Mills
(SYM) theory can be represented by a {\bf quiver diagram},
which we recall from Section 3.1 to be a labelled
directed finite graph together with a (complex)
representation.

To each vertex we associate the vector multiplet and to
each edge, the hypermultiplet. Generically we have product
$U(n_i)$'s for the gauge groups of the theory (with the
inclusion of orientifolds we can also obtain other groups).
Therefore we attribute a vector space $V_i$ as well as the
semisimple component (i.e., the $U(n_i)$ factor)
for the gauge group which acts on
$V_i$, to each vertex $v_i$. 

\begin{figure}
        \centerline{\psfig{figure=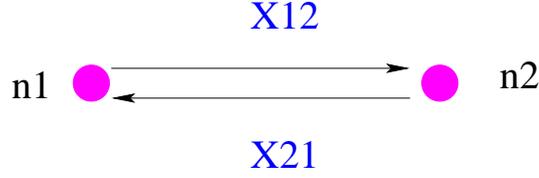,width=3in}}
\caption{An example of a quiver diagram encoding the matter
	content. Here theory has gauge group $U(n_1) \times U(n_2)$
	with hypermultiplet $(X_{12},X_{21})$.}
\label{f:quivereg}
\end{figure}

In other words, the vector multiplets
are seen as (Hermitian) matrices, representing 
adjoint gauge fields, acting on the space $V$. On the other
hand, an edge from vertex $v_i$ to $v_j$ is a complex
scalar transforming in the representation 
$\bar{V}_i \otimes V_j = \Hom(V_i,V_j)$ and hence
constitutes a mapping between the two vector spaces.
An undirected edge consisting of two oppositely directed
edges composes a single hypermultiplet. 
And so with this we can encode the matter content of 
a SYM theory on the D-brane probe as a quiver.
The example in \fref{f:quivereg} shall serve to clarify.
\subsection{The Lagrangian}
Having addressed the matter content for the theory discussed,
namely ${\cal N}=1$ in 6-dimensions (or ${\cal N}=2$ in 4), enough
supercharges exist to allow us to actually write down the Lagrangian,
rather conveniently in terms of hyper-K\"ahler geometry. The action
is of the form
\[
L = L_{BI} + L_{HM} + L_{CS} + \mbox{fermions},
\]
where $L_{BI}$ is the familiar Dirac-Born-Infeld action, $L_{HM}$,
the kinetic energy of the hypermultiplets, $L_{CS}$, a Chern-Simons
coupling term and the fermions form the SUSY completion. We 
concentrate on the first terms as they are purely in terms of the
scalars and shall provide the moduli space of the vacuum.

We recall from the previous subsection that the hypermultiplets
take values in $\Hom(V,V)$ for the vector space
$V := \{z^{a=1,\ldots,n}\}$ attributed to a vertex (and hence
a semisimple factor $G$ of the gauge group). 
Then letting the dual space
$V^*$ have co\"{o}rdinates $\{w_a\}$, the hypermultiplets can then
be written as Hermitian matrices 
$X^a := \mat{z^a & \bar{w}^a \cr -w_a & \bar{z}_a}$ which form
a quaternionic vector space with the Pauli matrices $\vec{\sigma}$
serving as the
3 complex structures. More generally, the $X^a$'s form a 
{\em hyper-K\"ahler manifold} with a triplet of symplectic forms:
$\omega^{\IR} = \frac{i}{2} dz^a d\bar{z}_a +dw_a d\bar{w}^a$ and
$\omega^{\IC} = d z^a \wedge d w_a$.

Finally $\mathfrak{g} := Lie(G)$ has a natural action on $X$ as
$\delta_{j=1,\ldots \dim(\mathfrak{g})} X^a = (t_j)^a_b X^b$. This
action is symplectic with respect to the above triplet of $\omega$'s
and we can write down a triplet of hyper-K\"ahler moment maps
\[
\vec{\mu}_j := \frac12 \mbox{tr} \vec{\sigma} 
		X^{\dagger}_a (t_j)^a_b X^b,
\]
being Noether charges of the symplectic action.

The scalar part of the action then reads
\[
L_{HM} + L_{BI} = \int_{D6} \sum_{j \in \mathfrak{g}}
	 \vec{D_j} \cdot (\vec{D_j} + \vec{\mu_j}),
\]
where $\vec{D_j}$ is the triple of auxiliary fields from the
D-fields in the vector-multiplet. For D3 branes (and hence
${\cal N}=2$ in 4-dimensions), $D^{\IR}$ is the FI D auxiliary field
while $D^{\IC}$ gives the F auxiliary field.
\subsection{The Vacuum Moduli Space}
\index{Vacuum Moduli Space}
Integrating out the $D$-fields from the Lagrangian above (we have
to include the $L_{CS}$ as well which we shall not discuss here),
we obtain an effective potential energy for the hypermultiplet:
$\sum\limits_j (\vec{\mu_j} - \vec{\phi_j})^2$. Here $\phi_j$'s are
scalars in the hypermultiplet
corresponding to the centre of $\mathfrak{g}$ acting as
$\Hom(V,V)$. Letting the VEV of $\phi_j$ be $\langle \vec{\phi_j}
\rangle = \vec{\zeta_j}$, the vacuum manifold is then
\[
\vec{\mu_j} = \vec{\zeta_j},
\]
modulo gauge transformations.
We remark that in fact for D5 and D4 branes, this
classical moduli space is the same the quantum one and for D3 probes,
the hyper-K\"ahler metric does not obtain quantum corrections.

We have of course seen this already in Section 3.3 of Liber I. This
moduli space is a hyper-K\"ahler quotient, with respect to the moment
maps. Such a space $M_{\zeta} = \mu^{-1}(\zeta) / G$ is precisely the
Kronheimer's {\bf ALE-instanton} \cite{Kronheimer}
as a resolution of the orbifold
$\IC^2/\Gamma$ in the case of the type IIB D5-brane probing the ALE
space as a local K3. More generically, when we include pairs of vector
spaces $(V,W)$ to each node in the spirit of \fref{f:quivervar}, 
we can actually obtain the {\bf quiver manifold} for the ALE space
\cite{KN}. This manifold is actually the moduli space of instantons on
$M_{\zeta}$ and we shall refer the reader to Section 3.3 for the
details.
\vspace{-0.2in}
\section*{\qquad \qquad \qquad \qquad { Epilogue}}
We have addressed three methods of constructing gauge theories in
4-dimensions, from which hopefully one day we can uniquely identify
our real world. It should be no surprise to us of course, that these
three prescriptions: geometrical engineering, Hanany-Witten setups and
D-brane probes, are all different guises of a single concept.

The key of course is T-duality, or for the mathematician, Mirror
Symmetry \cite{SYZ}. Using fractional branes and 3 consecutive
T-dualities, \cite{Karch:equiv} showed the equivalence between
Hanany-Witten and geometrical engineering. Furthermore, 
\cite{Ooguri-Vafa} showed that T-dual of NS-branes is precisely the
ALE instanton, whereby effectively establishing the equivalence between
NS-D-brane setups and D-brane probes.

Thus concludes our invocations. Prepared with some rudiments in the
mathematics and physics of a beautiful subfield of string theory, 
let us trudge on...

\part{LIBER TERTIUS: Sanguis, Sudor, et Larcrim\ae~Mei}
\vspace{0.5in}
\section*{\qquad \qquad \qquad \qquad { Prologue}}
\vspace{-0.2in}
{ H}aving hopefully by now conjured up the spirits of our gentle
readers, 
by these our invocations in mathematics and in physics, let us proceed
to the heart of this writing. I shall regret, to have enticed so much,
and yet shall soon provide so little. Though the ensuing pages will be
voluminous, my sheer want of wit shall render them uninspiring.

Yet I have laboured upon them and for some four years shed my blood,
sweat and tears upon these pages. I shall thus beg ye readers to open
your magnanimous hearts, to peruse and not to scoff, to criticise and
not to scorn.

Without further ado then allow me to summarise the contents of the
following chapters. This Book the Third itself divides into three
parts. The first, consists of chapters \ref{chap:9811183} till
\ref{chap:0009077}. They deal with gauge theory living on D-brane
probes transverse to quotient singularities of dimensions two, three
(Chap. \ref{chap:9811183}) and four (Chap. \ref{chap:9905212}). Certain
unified perspectives, from such diverse points of view as modular
invariants of WZW models, quiver categories and generalised McKay's
correspondences are discussed in Chapters \ref{chap:9903056},
\ref{chap:9911114} and \ref{chap:0009077}. Extensive use will be made
of the techniques of Chapters 2, 3 and 4 of Liber I.

The next part consists of Chapters \ref{chap:9906031} till
\ref{chap:0012078} where we address the more physical question of
realising the above probe theories as brane configurations of the
Hanany-Witten type. Thereafter, the two chapters \ref{chap:0010023}
and \ref{chap:dis2} consider the additional complication when there is
a background of the NS-NS B-field, which subsequently leads to the
study of projective representations of the orbifold group.

Finally the remaining chapters of the present Liber III are dedicated
to a detailed study of the IR moduli space of certain gauge theories,
in particular we venture beyond the orbifolds and study toric
singularities. Chapter 2 of Liber I and chapter 5 of Liber II will
therefore be of great use.
\chapter{Orbifolds I: $SU(2)$ and $SU(3)$}
\label{chap:9811183}
\section*{\center{{ Synopsis}}}
This is the first chapter on D-brane probes on orbifold
singularities where we study the world-volume 
${\cal N} = 4$ $U(n)$ super-Yang-Mills
theory orbifolded by discrete subgroups of $SU(2)$ and $SU(3)$.
We have reached many interesting observations that have
graph-theoretic interpretations. 

For the subgroups of $SU(2)$, we have 
McKay's correspondence to our aid. In the case of $SU(3)$ we have
constructed
a catalogue of candidates for finite (chiral) ${\cal N}=1$ theories, giving
the gauge group and matter content. 

To generalise the case of $SU(2)$, we
conjecture a McKay-type correspondence for Gorenstein singularities in
dimension 3 with modular invariants of WZW conformal models.
This implies a connection between a class of finite ${\cal N}=1$ supersymmetric
gauge theories in four dimensions and
the classification of affine $SU(3)$ modular invariant partition functions
in two dimensions \cite{9811183}.
\section{Introduction}
Recent advances on finite four dimensional gauge theories from string theory
constructions have been dichotomous: either from the geometrical perspective of
studying algebro-geometric singularities such as orbifolds
\cite{KS} \cite{LNV} \cite{BKV}, 
or from the intuitive perspective of studying various configurations of branes 
such as the so-called brane-box models \cite{HZ}.
(See \cite{HU} and references therein for a detailed description of these
models. A recent paper discusses the bending of non-finite
models in this context \cite{LR}.)
The two approaches lead to the realisation of finite, possibly chiral, ${\cal N}=1$
supersymmetric gauge theories, such as those discussed in
\cite{Leigh}.
Our ultimate dream is of course to have the flexibility of the equivalence and
completion of these approaches, allowing us to compute say, the duality group
acting on the moduli space of marginal gauge couplings \cite{HSU}.
(The duality groups for the ${\cal N}=2$ supersymmetric theories were
discussed in the context of these two approaches in \cite{geoeng3} and 
\cite{M-Theory}.)
The brane-box method has met great success in providing the intuitive picture
for orbifolds by Abelian groups: the elliptic model consisting of 
$k \times k'$ branes conveniently reproduces the theories on
orbifolds by $\IZ_k \times \IZ_{k'}$ \cite{HU}.
Orbifolds by $\IZ_k$ subgroups of $SU(3)$ are given by Brane Box Models with
non-trivial identification on the torus \cite{HSU} \cite{HU}.
Since by the structure
theorem that all finite Abelian groups are direct sums of cyclic ones, 
this procedure
can be presumably extended to all Abelian quotient singularities.
The non-Abelian groups however, present difficulties. By adding orientifold planes,
the dihedral groups have also been successfully attacked for theories with
${\cal N}=2$ supersymmetry \cite{Kapustin}. The
question still remains as to what could be done for the myriad of finite groups,
and thus to general Gorenstein singularities.

In this chapter we shall present a catalogue of these Gorenstein
singularities in dimensions 2 and 3, i.e., orbifolds constructed from 
discrete subgroups of $SU(2)$ and $SU(3)$ whose classification are complete.
In particular we shall concentrate on the gauge group, the fermionic and
bosonic matter content resulting from the orbifolding of an ${\cal N} = 4$
$U(n)$ super-Yang-Mills theory. In Section 2, we present the general arguments
that dictate the matter content for arbitrary finite group $\Gamma$. Then in
Section 3, we study the case of $\Gamma \subset SU(2)$ where we notice
interesting graph-theoretic descriptions of the matter matrices. 
We analogously analyse case by case, the discrete subgroups of $SU(3)$ in
Section 4, followed by a brief digression of possible mathematical 
interest in Section 5. This leads to a Mckay-type
connection between the classification
of two dimensional $SU(3)_k$ modular invariant partition functions and the
class of finite ${\cal N}=1$ supersymmetric gauge theories calculated in this
chapter. 
Finally we tabulate possible chiral theories
obtainable by such orbifolding techniques for these $SU(3)$ subgroups.

\section {The Orbifolding Technique}
\index{Brane Probes!Orbifolds}
\index{Singularity!Orbifolds}
\index{Orbifolds}
Prompted by works by Douglas, Greene, Moore and Morrison
on gauge theories which arise by placing D3 branes on orbifold singularities
\cite{DM} \cite{DG}, \cite{DGM},
Kachru and Silverstein \cite{KS} and subsequently
Lawrence, Nekrasov and Vafa \cite{LNV} noted that an orbifold theory 
involving the
projection of a supersymmetric ${\cal N}=4$ gauge theory on some discrete subgroup 
$\Gamma \subset SU(4)$ leads to a conformal field theory with 
${\cal N}\leq 4$ supersymmetry. 
We shall first briefly summarise their results here.

We begin with a $U(n)$ ${\cal N}=4$ super-Yang-Mills
theory which has an R-symmetry
of $Spin(6)\simeq SU(4)$. There are gauge bosons $A_{IJ}$ $(I,J=1,...,n)$
being singlets of $Spin(6)$, along with
adjoint Weyl fermions $\Psi _{IJ}^{\bf{4}}$
in the fundamental $\bf{4}$ of $SU(4)$ and adjoint scalars $\Phi _{IJ}^{%
\bf{6}}$ in the antisymmetric $\bf{6}$ of $SU(4)$. Then we choose a
discrete (finite) subgroup $\Gamma \subset SU(4)$ with the set of
irreducible representations $\left\{ {\bf r}_{i}\right\}$ acting on
the gauge group by breaking the $I$-indices up according to 
$\left\{ {\bf r}_{i}\right\} $, i.e., by 
$\bigoplus\limits_{i} {\bf r}_{i}=\bigoplus\limits_{i}%
 \IC^{N_{i}} {\bf r}_{i}$ such that $\IC^{N_{i}}$ 
accounts for the multiplicity of each ${\bf r}_{i}$ and 
$n=\sum\limits_{i=1}N_{i}\dim ({\bf r}_{i})$.
In the string theory picture, this decomposition of the gauge group
corresponds to permuting $n$ D3-branes and hence their Chan-Paton factors
which contain the $IJ$ indices,
on orbifolds of $\IR^{6}$. Subsequently by the Maldecena large $N$ conjecture
\cite{Maldecena}, we have
an orbifold theory on $AdS_{5}\times S^{5}$, with the R-symmetry
manifesting as the $SO(6)$ symmetry group of $S^{5}$ in which the 
branes now live \cite{KS}. The string perturbative calculation 
in this context, especially with respect to vanishing theorems for $\beta$-functions,
has been performed \cite{BKV}.

Having decomposed the gauge group, we must likewise do so for the matter 
fields: since an orbifold is invariant under the $\Gamma$-action,
we perform the so-called projection on the fields by keeping only the 
$\Gamma $-invariant fields in the theory. Subsequently we arrive at a
(superconformal) field theory with gauge group 
$G=\bigotimes\limits_{i}SU(N_{i})$ and Yukawa and quartic interaction
respectively as (in the notation of \cite{LNV}):
\[
\begin{array}{l}
Y = \sum_{ij k} \gamma_{ijk}^{f_{ij}, f_{jk}, f_{ki}} \Tr
\Psi_{f_{ij}}^{ij} \Phi_{f_{jk}}^{jk} \Psi_{f_{ki}}^{ki} \\ \\

V = \sum_{ijkl} \eta^{ijkl}_{f_{ij}, f_{jk}, f_{kl},
f_{li}} \Tr
\Phi_{f_{ij}}^{ij}\Phi_{f_{jk}}^{jk}\Phi_{f_{kl}}^{kl}\Phi_{f_{li}}^{li},
\end{array}
\]
where 
\[
\begin{array}{l}
\gamma_{ij k}^{f_{ij}, f_{jk}, f_{ki}} =
\Gamma_{\alpha\beta, m} \left( Y_{f_{ij}}
\right)^{\alpha}_{v_{i}{\bar v}_{j}}
\left( Y_{f_{jk}} \right)^{m}_{v_{j}{\bar v}_{k}} \left( Y_{f_{ki}}
\right)^{\beta}_{v_{k}{\bar v}_{i}} \\ \\

\eta^{ijkl}_{f_{ij}, f_{jk}, f_{kl}, f_{li}} =
\left(Y_{f_{ij}}\right)^{[ m}_{v_{i}{\bar v}_{j}}
\left(Y_{f_{jk}}\right)^{n]}_{v_{j}{\bar v}_{k}}
\left(Y_{f_{kl}}\right)^{[m}_{v_{k}{\bar v}_{l}}
\left(Y_{f_{li}}\right)^{n]}_{v_{l}{\bar v}_{i}},
\end{array}
\] 
such that $\left( Y_{f_{ij}} \right)^{\alpha}_{v_{i}{\bar v}_{j}}$,
$\left(Y_{f_{ij}}\right)^{m}_{v_{i}{\bar v}_{j}}$ are the $f_{ij}$'th
Clebsch-Gordan
coefficients corresponding to the projection of $4\otimes {\bf r}_i$
and $6\otimes {\bf r}_i$ onto
${\bf r}_j$, and $\Gamma_{\alpha\beta, m}$ is the invariant in
${\bf 4} \otimes {\bf 4} \otimes {\bf 6}$.

Furthermore, the matter content is as follows:
\begin{enumerate}
\item  Gauge bosons transforming as
\[
\hom \left( \IC^{n},\IC^{n}\right) ^{\Gamma}
=\bigoplus\limits_{i}\IC^{N_{i}} {\bf \otimes }\left( \IC^{N_{i}}\right) ^{*},
\]
which simply means that the original (R-singlet) adjoint $U(n)$ fields now
break up according to the action of $\Gamma$ to become the 
adjoints of the various $SU(N_i)$;

\item  $a_{ij}^{\bf{4}}$ Weyl fermions
 $\Psi _{f_{ij}}^{ij}$
($f_{ij}=1,...,a_{ij}^{\bf{4}}$ ) 
\[
\left( {\bf 4} \otimes \hom \left(\IC^{n},\IC^{n}\right) \right) ^{\Gamma
}=\bigoplus\limits_{ij}a_{ij}^{\bf 4}\IC^{N_{i}} {\bf \otimes }
\left( \IC^{N_{j}}\right) ^{*},
\]
which means that these fermions in the fundamental {\bf 4} of the 
original R-symmetry now become $\left(N_{i},\overline{N}_{j}\right)$
bi-fundamentals of $G$ and there are $a_{ij}^{\bf 4}$ copies of them;

\item  $a_{ij}^{\bf 6}$ scalars $\Phi _{f_{ij}}^{ij}$
($f_{ij}=1,...,a_{ij}^{\bf{6}}$ ) as
\[
\left( {\bf 6} \otimes \hom \left(\IC^{n},\IC^{n}\right) \right) ^{\Gamma}
= \bigoplus\limits_{ij}a_{ij}^{\bf{6}}\IC^{N_{i}} {\bf \otimes }
\left( \IC^{N_{j}}\right)^{*},
\]
similarly, these are $G$ bi-fundamental bosons, inherited from the
{\bf 6} of the original R-symmetry.
\end{enumerate}

For the above, we define $a_{ij}^{\cal R}$ (${\cal R} = {\bf 4}$ or {\bf 6}
for fermions and bosons respectively) as the composition coefficients
\begin{equation}
{\cal R}\otimes {\bf r}_{i}=\bigoplus\limits_{j}a_{ij}^{{\cal R}}
{\bf r}_{j}
\label{aij}
\end{equation}

Moreover, the supersymmetry of the projected theory must have its R-symmetry
in the commutant of $\Gamma \subset SU(4)$, which is $U(2)$ for 
$SU(2)$, $U(1)$ for $SU(3)$ and trivial for $SU(4)$, which means:
if $\Gamma \subset SU(2)$, we have an ${\cal N} = 2$ theory, 
if $\Gamma \subset SU(3)$, 
we have ${\cal N} = 1$, and finally for $\Gamma \subset $ the full $SU(4)$, 
we have a non-supersymmetric theory.

Taking the character $\chi $ for element $\gamma \in \Gamma $ on both sides
of (\ref{aij}) and recalling that $\chi $ is a $\left( \otimes
,\oplus \right) $-ring homomorphism, we have
\begin{equation}
\chi _{\gamma }^{{\cal R}}\chi _{\gamma
}^{(i)}=\sum\limits_{j=1}^{r}a_{ij}^{{\cal R}}\chi _{\gamma }^{(j)}
\label{aijchar}
\end{equation}
where $r=\left| \left\{ {\bf r}_{i}\right\} \right| $, the number of
irreducible representations, which by an elementary theorem on finite
characters, is equal to the number of inequivalent conjugacy classes of
$\Gamma$. We further recall the orthogonality theorem of finite characters,

\begin{equation}
\sum\limits_{\gamma =1}^{r}r_{\gamma }\chi_{\gamma }^{(i)*}
	\chi_{\gamma }^{(j)}=g\delta^{ij},
\label{ortho}
\end{equation}

where $g=\left| \Gamma \right| $ is the
order of the group and  $r_{\gamma }$ is the order of the conjugacy class
containing $\gamma$. Indeed, $\chi$ is a class function and is hence constant
for each conjugacy class; moreover, 
$\sum\limits_{\gamma =1}^{r}r_{\gamma }=g$ is the class equation for
$\Gamma$. This orthogonality
allows us to invert (\ref{aijchar}) to finally give the matrix $a_{ij}$ 
for the matter content
\begin{equation}
a_{ij}^{{\cal R}}=\frac{1}{g}\sum\limits_{\gamma
=1}^{r}r_{\gamma }\chi_{\gamma }^{{\cal R}}\chi_{\gamma }^{(i)}\chi
_{\gamma }^{(j)*}  \label{matter}
\end{equation}
where ${\cal R}$ $=\bf{4}\ $for Weyl fermions and  $\bf{6}$ for
adjoint scalars and the sum is effectively that over the columns of the
Character Table of $\Gamma $. Thus equipped, let us specialise to $\Gamma$
being finite discrete subgroups of $SU(2)$ and $SU((3)$.

\section{Checks for $SU(2)$}
\index{Finite Groups!$SU(2)$ subgroups}
The subgroups of $SU(2)$ have long been classified \cite{Klein};
discussions and applications thereof can be found in \cite{McKay}
\cite{Seiberg} \cite{Kronheimer} \cite{Blichfeldt}. 
To algebraic geometers they give rise to 
the so-called Klein singularities and are labeled by the first affine
extension of the simply-laced simple Lie groups $\widehat{A}\widehat{D}
\widehat{E}$ (whose associated Dynkin diagrams are those of $ADE$ adjointed
by an extra node), i.e., there are two infinite series and 3 exceptional
cases:

\begin{enumerate}
\item  $\widehat{A}_{n}=\IZ_{n+1}$, the cyclic group of order $n+1$;

\item  $\widehat{D}_{n}$, the binary lift of the ordinary dihedral group $d_n$;

\item  the three exceptional cases, $\widehat{E}_{6}$, $\widehat{E}_{7}$ and 
$\widehat{E}_{8}$, the so-called binary or double\footnote{
For $SO(3)\cong SU(2)/\IZ_2$ these would be the familiar 
symmetry groups of the respective
regular solids in $\IR^{3}$: the dihedron, tetrahedron, octahedron/cube and
icosahedron/dodecahedron. 
However since we are in the double cover
$SU(2),$ there is a non-trivial $\IZ_{2}$- lifting,\\
$
\begin{array}{ccccccccc}
0 & \rightarrow & \IZ_2 & \rightarrow & SU(2) & \rightarrow & SO(3) & \rightarrow
& 0, \\
  & & & & \bigcup  & & \bigcup  & &\\
  & & & & {\widehat D}, {\cal T}, {\cal O}, {\cal I} & \rightarrow &d, T, O, I & & \\
\end{array}
$ \\
hence the modifier ``binary''. Of course, the $A$-series, being abelian, receives no
lifting. Later on we shall briefly touch upon the
ordinary $d,T,I,O$ groups as well.} tetrahedral, 
octahedral and icosahedral groups ${\cal T, O, I}$.
\end{enumerate}

The character tables for these groups are known \cite{Lomont} \cite{Zaf} \cite{CFT} and
are included in Appendix \ref{append:9811183.I} for reference.
Therefore to obtain (\ref{matter}) the only difficulty remains in the
choice of ${\cal R}$. We know that whatever ${\cal R}$ is, it must be 4 
dimensional for the fermions and 6 dimensional for the bosons inherited from
the fundamental {\bf 4} and antisymmetric {\bf 6} of $SU(4)$.
Such an ${\cal R}$ must therefore be a 4 (or 6) dimensional irrep of 
$\Gamma$, or be the tensor sum of lower dimensional irreps (and hence be
reducible); for the character table, this means that the row
of characters for ${\cal R}$ (extending over the conjugacy 
classes of $\Gamma$) must be an existing row or the sum of
existing rows.
Now since the first column of the character table of any finite group
precisely gives
the dimension of the corresponding representation, it must therefore be that
$\dim({\cal R}) = 4,6$ should be partitioned into these numbers.
Out of these possibilities we must select the one(s)
consistent with the decomposition of the {\bf 4} and {\bf 6} of $SU(4)$ into the
$SU(2)$ subgroup\footnote{We note that even though 
this decomposition is that into irreducibles
for the full continuous Lie groups, such irreducibility may not be inherited by 
the discrete subgroup, i.e., the {\bf 2}'s {\it may} not be irreducible representations
of the finite $\Gamma$.}, namely:

\begin{equation}
\begin{array}{ccc}
SU(4) & \rightarrow & SU(2) \times SU(2) \times U(1) \\
{\bf 4} & \rightarrow & ({\bf 2,1})_{+1} \bigoplus ({\bf 1,2})_{-1} \\
{\bf 6} & \rightarrow & ({\bf 1,1})_{+2} \bigoplus ({\bf 1,1})_{-2} \bigoplus 
			({\bf 2,2})_0 \\
\end{array}
\label{break2}
\end{equation}

where the subscripts correspond to the $U(1)$ factors (i.e., the trace) and in
particular the $\pm$ forces the overall traceless condition.
 From (\ref{break2}) we know that $\Gamma \subset SU(2)$ inherits a {\bf 2}
while the complement is trivial. This means that the 4 dimensional represention
of $\Gamma$ must be decomposable into a nontrivial 2 dimensional one with a 
trivial 2 dimensional one. In the character language, this means that
${\cal R} = {\bf 4 = 2_{trivial} \oplus 2 } $ where
${\bf 2_{trivial} = 1_{trivial} \oplus 1_{trivial}}$, the tensor sum of two copies
of the (trivial) principal representation where all group elements are mapped to the
identity, i.e., corresponding to the first row in the character table. Whereas for the 
bosonic case we have
${\cal R} = {\bf 6 = 2_{trivial}
 \oplus 2 \oplus 2^{'} } $. We have denoted ${\bf 2^{'}}$ to signify that the
two {\bf 2}'s may not be the same, and correspond to inequivalent 
representations of $\Gamma$ with the same dimension. However we can restrict this 
further by recalling that the antisymmetrised tensor product 
$[{\bf 4} \otimes {\bf 4}]_A \rightarrow
1 \oplus 2  \oplus 2 \oplus [2 \otimes 2]_A$ must
in fact contain the {\bf 6}. Whence we conclude that ${\bf 2} = {\bf 2^{'}}$.
Now let us again exploit the additive property of the group character, i.e., a
homomorphism from a $\oplus$-ring to a +-subring of a number field (and indeed
much work has been done for the subgroups in the case of number fields of various
characteristics); this
means that we can simplify $\chi^{{\cal R}=x \oplus y}$ as $\chi^x + \chi^y$.
Consequently, our matter matrices become:
\[
\begin{array}{l}
a_{ij}^{{\bf 4}}=\frac{1}{g}\sum\limits_{\gamma=1}^{r}r_{\gamma} 
	\left(2 \chi_{\gamma}^{{\bf 1}} + \chi_{\gamma}^{{\bf 2}} \right)
	\chi_{\gamma }^{(i)}\chi_{\gamma }^{(j)*}
	= 2 \delta_{ij} + \frac{1}{g}\sum\limits_{\gamma=1}^{r}r_{\gamma} 
	\chi_{\gamma}^{{\bf 2}}
	\chi_{\gamma }^{(i)}\chi_{\gamma }^{(j)*}\\

a_{ij}^{{\bf 6}}=\frac{1}{g}\sum\limits_{\gamma=1}^{r}r_{\gamma} 
	\left(2 \chi_{\gamma}^{{\bf 1}} + \chi_{\gamma}^{{\bf 2}
	\oplus {\bf 2}} \right)
	\chi_{\gamma }^{(i)}\chi_{\gamma }^{(j)*}
	= 2 \delta_{ij} + \frac{2}{g}\sum\limits_{\gamma=1}^{r}r_{\gamma} 
	\chi_{\gamma}^{{\bf 2}} \chi_{\gamma }^{(i)}\chi_{\gamma }^{(j)*}\\
\end{array}
\label{matter2}
\]
where we have used the fact that $\chi$ of the trivial representation are
all equal to 1, thus giving by (\ref{ortho}), the $\delta_{ij}$'s. 
This simplification thus limits our attention to only 2 dimensional 
representations of $\Gamma$; however there still may remain many possibilities
since the {\bf 2} may be decomposed into nontrivial {\bf 1}'s or there may
exist many inequivalent irreducible {\bf 2}'s.

We now appeal to physics for further restriction. We know that the ${\cal N}
= 2$ theory (which we recall is the resulting case when $\Gamma \subset 
SU(2)$) is a non-chiral supersymmetric theory; this means our bifundamental
fields should not distinguish the left and right indices, i.e., the matter
matrix $a_{ij}$ must be {\it symmetric}.
Also we know that in the ${\cal N}=2$ vector multiplet there are 2 Weyl
fermions and 2 real scalars, thus the fermionic and bosonic matter matrices
have the same entries on the diagonal. Furthermore the hypermultiplet
has 2 scalars and 1 Weyl fermion in $(N_i, \bar{N}_j)$ and another 2 
scalars and 1 Weyl fermion in the complex conjugate representation, whence
we can restrict the off-diagonals as well, viz., $2 a_{ij}^{\bf 4} -
a_{ij}^{\bf 6}$ must be some multiple of the identity.
This supersymmetry
matching is of course consistent with (\ref{matter2}).

Enough said on generalities. Let us analyse the groups case by case.
For the cyclic group, the {\bf 2} must come from the tensor sum of two {\bf 1}'s.
Of all the possibilities, only the pairing of dual representations gives
symmetric $a_{ij}$. By dual we mean the two {\bf 1}'s which are
complex conjugates of each other (this of course includes when
${\bf 2} = {\bf 1_{trivial}^2}$,
which exist for all groups and gives us merely $\delta_{ij}$'s and can
henceforth be eliminated as uninteresting). We denote the nontrivial pairs
as ${\bf 1^{'}}$ and ${\bf 1^{''}}$. In this case we can easily perform
yet another consistency check. From (\ref{break2}), we have a traceless
condition seen as the cancelation of the $U(1)$ factors. That was on the
Lie algebra level; as groups, this is our familiar determinant unity 
condition. 
Since in the block decomposition (\ref{break2}) the 
${\bf 2_{trivial}} \subset$ the complement $SU(4)\backslash\Gamma$ 
clearly has determinant 1, this forces our {\bf 2} matrix
to have determinant 1 as well. However in this cyclic case, $\Gamma$ is
abelian, whence the characters are simply presentations of the group,
making the {\bf 2} to be in fact diagonal. Thus the determinant is simply the
product of the entries of the two rows in the character table. And indeed
we see for dual representations, being complex conjugate roots of unity,
the two rows do multiply to 1 for all members. Furthermore we note that
different dual pairs give $a_{ij}$'s that are mere permutations of each other.
We conclude that the fermion
matrix arises from ${\bf 1^2} \oplus {\bf 1^{'}} \oplus {\bf 1^{''}}$. For
the bosonic matrix, by (\ref{matter2}), we have  ${\bf 6} =
({\bf 1} \oplus {\bf 1^{'}} \oplus {\bf 1^{''}})^{\bf 2}$. These and ensuing
$a_{ij}$'s are included in Appendix \ref{append:9811183.II}.

For the dihedral case, the {\bf 1}'s are all dual to the principal, corresponding
to some $\IZ_2$ inner automorphism among the conjugacy classes and the
characters consist no more than $\pm 1$'s, giving us $a_{ij}$'s which are
block diagonal in $((1,0),(0,1))$ or $((0,1),(1,0))$ and are not terribly 
interesting. Let us rigorise this statement. Whenever we have the character
table consisting of a row that is composed of cycles of roots of unity,
which is a persistent theme for {\bf 1} irreps, this corresponds in general to
some $\IZ_k$ action on the conjugacy classes. This implies that our
$a_{ij}$ for this choice of {\bf 1} will be the Kronecker product of matrices
obtained from the cyclic groups which offer us nothing new. We shall refer to 
these cases as ``blocks''; they offer us another condition of
elimination whose virtues we shall exploit much. 
In light of this, for the dihedral the choice of the {\bf 2} comes 
from the irreducible {\bf 2}'s which again give symmetric $a_{ij}$'s that
are permutations among themselves.
Hence ${\cal R} = {\bf 4} = {\bf 1^2} \oplus {\bf 2}$ and
${\cal R} = {\bf 6} = {\bf 1^2} \oplus {\bf 2^2}$.
For reference we have done likewise for the dihedral series not in the full 
$SU(2)$, the choice for ${\cal R}$ is the same for them.

Finally for the exceptionals ${\cal T,O,I}$, the {\bf 1}'s again give 
uninteresting block diagonals and out choice of {\bf 2} is again unique
up to permutation. Whence still 
${\cal R} = {\bf 4} = {\bf 1^2} \oplus {\bf 2}$ and
${\cal R} = {\bf 6} = {\bf 1^2} \oplus {\bf 2^2}$.
For reference we have computed the ordinary exceptionals $T,O,I$ which live in
$SU(2)$ with its center removed, i.e., in $SU(2)/\IZ_2 \cong SO(3)$. For them
the {\bf 2} comes from the ${\bf 1^{'}} \oplus {\bf 1^{''}}$, the {\bf 2},
and the trivial ${\bf 1^2}$ respectively.

Of course we can perform an {\it a posteriori} check. In 
this case of $SU(2)$ we already know the matter content due
to the works on quiver diagrams \cite{DM} \cite{Seiberg} \cite{Kapustin}.
The theory dictates that the 
matter content $a_{ij}$ can be obtained by looking at the Dynkin diagram of the
$\widehat{A} \widehat{D} \widehat{E}$ group associated to $\Gamma $ whereby 
one assigns 2 for $a_{ij}$ on the diagonal as well as 1 for
every pair of connected nodes
$ i \rightarrow j$ and $ 0 $ otherwise, i.e., $a_{ij}$ is essentially the
\index{Quivers!adjacency matrix}
 adjacency matrix for the Dynkin diagrams treated as unoriented graphs.
Of course adjacency matrices for unoriented graphs are symmetric; this is
consistent with our nonchiral supersymmetry argument. Furthermore,
the dimension of $a_{ij}^{\bf 4}$ is required to be equal to the number of nodes 
in the associated affine Dynkin diagram (i.e., the rank).
This property is immediately seen to be satisfied by examining the character
tables in Appendix \ref{append:9811183.I} where we note 
that the number of conjugacy classes of the respective finite groups 
(which we recall is equal to the number of irreducible representations) and
hence the dimension of $a_{ij}$ is indeed
that for the ranks of the associated affine algebras, namely $n+1$ for
$\widehat{A_n}$ and $\widehat{D_n}$ and 7,8,9 for $\widehat{E_{6,7,8}}$
respectively.
We note in passing that the conformality condition $N_f = 2 N_c$ for 
this ${\cal N}=2$ \cite{KS} \cite{LNV} nicely translates to the graph language:
it demands that for the one loop $\beta$-function to vanish the label of each
node
(the gauge fields) must be $\frac12$ that of those connected thereto (the
bi-fundamentals).

Our results for $a_{ij}$ computed using (\ref {matter}),
Appendix \ref{append:9811183.I}, 
and the aforementioned decomposition of ${\cal R}$ are tabulated
in Appendix \ref{append:9811183.II}. They are 
precisely in accordance with the quiver theory and present themselves as the
relevant adjacency matrices. One interesting point to note is that for the
dihedral series, the ordinary $d_n$ (which are in $SO(3)$ and not $SU(2)$)
for even $n$ also gave the binary $\widehat{D_{n'=\frac{n+6}{2}}}$ 
Dynkin diagram while the odd $n$ case always gave the ordinary 
$D_{n'=\frac{n+3}{2}}$ diagram.

\index{McKay Correspondence!brane probes}
These results should be of no surprise to us, since a similar calculation was
in fact done by J. Mckay when he first noted his famous correspondence \cite
{McKay}. In the paper he computed the composition coefficients $m_{ij}$ in
$R \bigotimes R_j = \bigoplus\limits_{k} m_{jk} R_k$ for $\Gamma \subset
SU(2)$ with $R$ being a faithful representation thereof. He further noted that
for all these $\Gamma$'s there exists (unique up to automorphism) 
such $R$, which is precisely the 2 
dimensional irreducible representation for $\widehat{D}$ and $\widehat{E}$ 
whereas for $\widehat{A}$ it is the
direct sum of a pair of dual 1 dimensional representations. Indeed this is
exactly the decomposition of ${\cal R}$ which we have argued above from
supersymmetry. His {\it Theorema Egregium} was then\\
\index{Quivers!quiver category}
\noindent
{\large {\bf Theorem:}} The matrix $m_{ij}$ is $2 I$ minus
the cartan matrix, and is thus the adjacency matrix for the 
associated affine
Dynkin diagram treated as undirected $C_2$-graphs (i.e., maximal eigenvalue is 
2).\\

Whence $m_{ij}$ has 0 on the diagonal and 1 for connected nodes.
Now we note from our discussions above and results in Appendix
\ref{append:9811183.II}, 
that our $\cal R$ is precisely 
Mckay's $R$ (which we henceforth denote as $R_M$) plus two copies of the 
trivial representation for the {\bf 4} and
$R_M$ plus the two dimensional irreps in addition to the two copies of the 
trivial for the {\bf 6}. Therefore we conclude from (\ref{matter}):
\[
\begin{array}{l}
a_{ij}^{{\bf 4}}= \frac{1}{g}\sum\limits_{\gamma
	=1}^{r}r_{\gamma }\chi_{\gamma }^{R_M \oplus {\bf 1}^2 }
	\chi_{\gamma }^{(i)}
	\chi_{\gamma }^{(j)*} \\
a_{ij}^{{\bf 6}}=\frac{1}{g}\sum\limits_{\gamma
	=1}^{r}r_{\gamma }\chi_{\gamma }^{R_M \oplus R_M \oplus {\bf 1}^2 }
	\chi_{\gamma }^{(i)}\chi_{\gamma }^{(j)*} \\
\end{array}
\]
which implies of course, that our matter matrices should be
\[
\begin{array}{l}
a_{ij}^{{\bf 4}}= 2 \delta_{ij} + m_{ij} \\
a_{ij}^{{\bf 6}}= 2 \delta_{ij} + 2 m_{ij} \\
\end{array}
\]
with Mckay's $m_{ij}$ matrices. This is exactly the results
we have in Appendix \ref{append:9811183.II}.  
Having obtained such an elegant graph-theoretic
interpretation to our results, we remark that from this
point of view, oriented graphs means chiral gauge theory and connected means
interacting gauge theory. Hence we have the foresight that the ${\cal N} = 1$
case which we shall explore next will involve oriented graphs.

Now Mckay's theorem explains why the discrete subgroups of $SU(2)$ and hence
Klein singularities of algebraic surfaces (which our orbifolds essentially
are) as well as subsequent gauge theories thereupon afford this correspondence
with the affine simply-laced Lie groups. However they were originally proven
on a case by case basis, and we would like to know a deeper connection,
especially in light of quiver theories. We can partially answer this 
question by noting a beautiful
theorem due to Gabriel \cite{Gabriel} \cite{Bernstein} which forces the quiver
considerations by Douglas et al.\ \cite{DM}
to have the ADE results of Mckay.

It turns out to be convenient to formulate the theory axiomatically.
We define ${\cal L}(\gamma,\Lambda)$,
for a finite connected graph $\gamma$ with orientation $\Lambda$, vertices
$\gamma_0$ and edges $\gamma_1$, 
to be the category of quivers whose objects are any
collection $(V,f)$ of spaces $V_{\alpha \in \gamma_0}$ and mappings 
$f_{l \in \gamma_1}$ and whose morphisms are 
$\phi : (V,f) \rightarrow (V',f') $
a collection of linear mappings $\phi_{\alpha \in \Gamma_0} : V_{\alpha}
\rightarrow V'_{\alpha}$ compatible with $f$ by 
$\phi_{e(l)}f_l = f'_{l}\phi_{b(l)}$ where $b(l)$ and $e(l)$ are the beginning
and end of the directed edge $l$. Then we have\\

\noindent
{\large {\bf Theorem:}} If in the quiver category ${\cal L}(\gamma,\Lambda)$ 
there are only finitely many
non-isomorphic indecomposable objects, then $\gamma$ coincides with one of the
graphs $A_n,D_n,E_{6,7,8}$.\\

This theorem essentially compels any finite quiver theory to be constructible
only on graphs which are of the type of the Dynkin diagrams of $ADE$. And
indeed, the theories of Douglas, Moore et al.\ \cite{DM} \cite{Kapustin}
have explicitly made the physical realisations of these constructions.
We therefore see how McKay's calculations, quiver theory and 
our present calculations nicely fit together for the case of $\Gamma
\subset SU(2) $.

\index{Finite Groups!$SU(3)$ subgroups}
\section{The case for $SU(3)$}
We repeat the above analysis for $\Gamma = SU(3)$, though now we have no 
quiver-type theories to aid us. The discrete subgroups of $SU(3)$ have also 
been long classified \cite{Blichfeldt}. They include (the order of these groups are
given by the subscript), other than all those of $SU(2)$ since $SU(2) \subset SU(3)$, 
the following new cases. We point out that in addition to the cyclic group in $SU(2)$, 
there is now in fact another Abelian case $\IZ_k \times \IZ_{k'}$ for $SU(3)$ 
generated by the matrix $((e^{\frac{2 \pi i}{k}},0,0),(0,e^{\frac{2 \pi i}{k'}},0),
(0,0,e^{-\frac{2 \pi i}{k}-\frac{2 \pi i}{k'}}))$ much in the spirit that
$((e^{\frac{2 \pi i}{n}},0),(0,e^{-\frac{2 \pi i}{n}}))$ generates the $\IZ_n$ for
$SU(2)$. Much work has been done for this $\IZ_k \times \IZ_{k'}$ case, q.\ v.\, 
\cite{HU} and references therein.
\begin{enumerate}
\item Two infinite series $\Delta_{3n^2}$ and $\Delta_{6n^2}$ for $n \in \IZ$, 
which are analogues of the dihedral series in $SU(2)$:
\begin{enumerate}
\item $\Delta \subset$ only the full $SU(3)$: when $n = 0 \mbox{ mod } 3$ 
	where the number of classes for $\Delta(3n^2)$
	is $(8 + \frac{1}{3}n^2)$ and for $\Delta(6n^2)$, $\frac{1}{6}(24 + 9n + n^2)$;
\item $\Delta \subset$ both the full $SU(3)$ and $SU(3)/\IZ_3$: 
	when $n \ne 0\mbox{ mod } 3$ where the number 
	of classes for $\Delta(3n^2)$ is $\frac{1}{3}(8 + n^2)$ and 
	for $\Delta(6n^2)$, $\frac{1}{6}(8 + 9n + n^2)$;
\end{enumerate}
\item Analogues of the exceptional subgroups of $SU(2)$, and indeed
like the later, there are 
two series depending on whether the $\IZ_3$-center of $SU(3)$ has been 
modded out (we recall that
the binary ${\cal T,O,I}$ are subgroups of $SU(2)$, while the ordinary 
$T,O,I$ are subgroups of the center-removed $SU(2)$, i.e., $SO(3)$,
and not the full $SU(2)$):
\begin{enumerate}
\item For $SU(3)/\IZ_3$: \\ $\Sigma_{36}, \Sigma_{60} \cong A_5$, the alternating
	symmetric-5 group, which incidentally is precisely the ordinary icosahedral group
	 $I, \Sigma_{72}, 
	\Sigma_{168} \subset S_7$, the symmetric-7 group, $\Sigma_{216} \supset 
	\Sigma_{72} \supset \Sigma_{36},$ and 
	$\Sigma_{360} \cong A_6$, the alternating symmetric-6 group;
\item For the full\footnote {In his work on Gorenstein singularities \cite {Yau}, 
	Yau points out that since the cases of 
	$\Sigma_{60 \times 3}$ and $\Sigma_{168 \times 3}$ 
	are simply direct products of the respective cases in $SU(3)/\IZ_3$ 
	with $\IZ_3$, they are usually left out by most authors. 
	The direct product simply extends the class equation
	of these groups by 3 copies and acts as an inner automorphism 
	on each conjugacy class. Therefore the character table is that of the 
	respective center-removed cases, but with the entries 
	each multiplied by the matrix $((1,1,1),(1,w,w^2),(1,w^2,w))$ 
	where $w=\exp(2 \pi i / 3)$, i.e., the full character table is the Kronecker 
	product of that of the corresponding center-removed group with that of $\IZ_3$. 
	Subsequently, the matter matrices $a_{ij}$ become the Kronecker product of 
	$a_{ij}$ for the center-removed groups with that for $\Gamma = \IZ_3$ and 
	gives no interesting new results. In light of this, 
	we shall adhere to convention and call 
	$\Sigma_{60}$ and $\Sigma_{168}$ subgroups of both $SU(3)/\IZ_3$ 
	and the full $SU(3)$ and ignore
	$\Sigma_{60 \times 3}$ and $\Sigma_{168 \times 3}$.}
	$SU(3)$: \\ $\Sigma_{36 \times 3},
	\Sigma_{60 \times 3} \cong \Sigma_{60} \times \IZ_3,
	\Sigma_{168 \times 3} \cong \Sigma_{168} \times \IZ_3,
	\Sigma_{216 \times 3},$ and $\Sigma_{360 \times 3}$.
\end{enumerate}
\end{enumerate}

Up-to-date presentations of these
groups and some character tables may be found in \cite{Yau} \cite{Fairbairn}.
The rest have been computed with \cite {Prog}.
These are included in Appendix \ref{append:9811183.III} for reference.
As before we must narrow down our choices for ${\cal R}$. First we note
that it must be consistent with the decomposition:

\begin{equation}
\begin{array}{ccc}
SU(4) & \rightarrow & SU(3) \times U(1) \\
{\bf 4} & \rightarrow & {\bf 3}_{-1} \bigoplus {\bf 1}_{3} \\
{\bf 6} & \rightarrow & {\bf 3}_2 \bigoplus {\bf \bar{3}}_{-2} \\
\end{array}
\label{break3}
\end{equation}

This decomposition (\ref{break3}), as in the comments for (\ref{break2}),
forces us to consider only 3 dimensionals (possibly reducible) and for
the fermion case the remaining {\bf 1} must in fact be the trivial, giving
us a $\delta_{ij}$ in $a_{ij}^{\bf 4}$.

Now as far as the symmetry of $a_{ij}$ is concerned, since $SU(3)$ gives
rise to an ${\cal N} = 1$ chiral theory, the matter matrices are no longer
necessarily symmetric and we can no longer rely upon this property to guide us.
However we still have a matching condition between the bosons and the fermions.
In this ${\cal N} = 1$ chiral theory we have 2 scalars and a Weyl fermion in the 
chiral multiplet as well as a gauge field and a Weyl fermion in the vector multiplet.
If we denote the chiral and vector matrices as $C_{ij}$ and $V_{ij}$, and
recalling that there is only one adjoint field in the vector 
multiplet, then we should have:

\begin{equation}
\begin{array}{l}
a_{ij}^{\bf 4} = V_{ij} + C_{ij} = \delta_{ij} + C_{ij}\\
a_{ij}^{\bf 6} = C_{ij} + C_{ji}.\\
\end{array}
\label{break3mat}
\end{equation}

This decomposition is indeed consistent with (\ref{break3}); where the 
$\delta_{ij}$ comes from the principal {\bf 1} and the $C_{ij}$ and 
$C_{ji}$, from dual pairs of {\bf 3}; 
incidentally it also implies that the bosonic matrix should be symmetric
and that dual {\bf 3}'s should give matrices that are mutual transposes.
Finally as we have discussed in the $A_n$ case of $SU(2)$, if one is to
compose only from 1 dimensional representations, then the rows of characters for these
{\bf 1}'s must multiply identically to 1 over all conjugacy classes.
Our choices for ${\cal R}$ should thus be restricted by these general
properties.

Once again, let us analyse the groups case by case.
First the $\Sigma$ series. For the members which belong to the center-removed
$SU(2)$, as with the ordinary $T,O,I$ of $SU(2)/\IZ_2$, we expect nothing
particularly interesting (since these do not have non-trivial 3 dimensional
representations which in analogy to the non-trivial 2 dimensional irreps of
$\widehat{D_n}$ and $\widehat{E_{6,7,8}}$ should be the ones to give interesting results).
However, for completeness, we shall touch upon these groups, namely,
$\Sigma_{36,72,216,360}$. Now the {\bf 3} in (\ref{break3}) must be 
composed of {\bf 1} and {\bf 2}. The obvious choice is of course again 
the trivial one where we compose everything from only the principal {\bf 1}
giving $4 \delta_{ij}$ and $6 \delta_{ij}$ for the fermionic and bosonic 
$a_{ij}$ respectively.
We at once note that this is the only possibility for $\Sigma_{360}$, 
since its first non-trivial representation is 5 dimensional.
Hence this group is trivial for our purposes.
For $\Sigma_{36}$, the ${\bf 3}$ can come only from {\bf 1}'s for which case 
our condition that the rows must multiply to 1 implies that 
${\bf 3} = \Gamma_1 \oplus \Gamma_3 \oplus \Gamma_4$, or $\Gamma_1 \oplus \Gamma_2^2$,
both of which give uninteresting blocks, in the sense of what we have discussed
in Section 2. For $\Sigma_{72}$, we similarly
must have ${\bf 3} = \Gamma_2 \oplus \Gamma_3 \oplus \Gamma_4$ or 
${\bf 1} \oplus$ the self-dual ${\bf 2}$, both of which again give trivial
blocks. Finally for $\Sigma_{216}$, whose conjugacy classes consist 
essentially of $\IZ_3$-cycles in the 1 and 2 dimensional representations, 
the ${\bf 3}$ comes from ${\bf 1} \oplus {\bf 2}$
and the dual ${\bf 3}$, from ${\bf 1} \oplus {\bf 2^{'}}$.

For the groups belonging to the full $SU(3)$, 
namely $\Sigma_{168, 60, 36 \times 3, 216 \times 3, 360 \times 3}$, the situation is clear:
as to be expected in analogy to the $SU(2)$ case, there always exist dual 
pairs of {\bf 3} representations. The fermionic
matrix is thus obtained by tensoring the trivial representations with one member from a pair 
selected in turn out of the various pairs, i.e., ${\bf 1} \oplus {\bf 3}$; 
and indeed we have explicitly checked that 
the others (i.e., ${\bf 1} \oplus {\bf 3^{'}}$) are permutations thereof. On the
other hand, the bosonic matrix is obtained from tensoring any choice of a dual pair ${\bf 3} 
\oplus {\bf 3^{'}}$ and again we have explicitly checked that other dual pairs give rise to 
permutations. We may be tempted to construct the {\bf 3} out of the {\bf 1}'s and {\bf 2}'s
which do exist for $\Sigma_{36 \times 3, 216 \times 3}$, however we note that in 
these cases the {\bf 1} and {\bf 2} characters are all cycles of 
$\IZ_3$'s which would again give uninteresting blocks. Thus
we conclude still that for all these groups, ${\bf 4} = {\bf 1} \oplus {\bf 3}$ while
${\bf 6} = {\bf 3} \oplus$ dual ${\bf \bar{3}}$.
These choices are of course obviously in accordance with the 
decomposition (\ref{break3}) above.
Furthermore, for the $\Sigma$ groups that belong solely to the full $SU(3)$, the dual pair of 
{\bf 3}'s always gives matrices that are mutual transposes, consistent
with the requirement in (\ref{break3mat}) that the bosonic matrix be symmetric.

Moving on to the two $\Delta$ series. We note\footnote{
Though congruence in this case really means group isomorphisms, 
for our purposes since only the group characters concern us, in what follows
we might use the term loosely to mean identical character tables.}, that
for $n=1$, $\Delta_{3} \cong \IZ_3$ and $\Delta_{6} \cong d_6$ while for 
$n=2$, $\Delta_{12} \cong T := E_6$ and $\Delta_{24} \cong O := E_7$.
Again we note that for all $n > 1$ (we have already analysed the $n=1$ 
case\footnote{Of course for $\IZ_3$, we must have a different choice for ${\cal R}$,
in particular to get a good chiral model, we take the ${\bf 3} = {\bf 1'} \oplus
{\bf 1''} \oplus {\bf 1'''}$} for 
$\Gamma \subset SU(2)$), there exist
the dual {\bf 3} and ${\bf 3^{'}}$ representations as in the 
$\Sigma \subset$ full $SU(3)$ above; this is expected of course
since as noted before, all the $\Delta$ groups at least belong to the full $SU(3)$. 
Whence we again form the fermionic $a_{ij}$ from ${\bf 1} \oplus {\bf 3^{'}}$, giving
a generically nonsymmetric matrix (and hence a good chiral theory),
and the bosonic, from ${\bf 3} \oplus {\bf 3^{'}}$, giving us always a
symmetric matrix as required. We note in passing that when $n = 0 \mbox{ mod } 3$,
i.e,. when the group belongs to both the full and the center-removed
$SU(3)$, the $\Delta_{3n^2}$ matrices consist of a trivial diagonal block
and an L-shaped block. Moreover, all the $\Delta_{6n^2}$ matrices are
block decomposable. We shall discuss the significances of this observation
in the next section.
Our analysis of the discrete subgroups of $SU(3)$ is now complete; the
results are tabulated in Appendix \ref{append:9811183.IV}.

\section{Quiver Theory? Chiral Gauge Theories?}
\begin{figure}
	\centerline{\psfig{figure=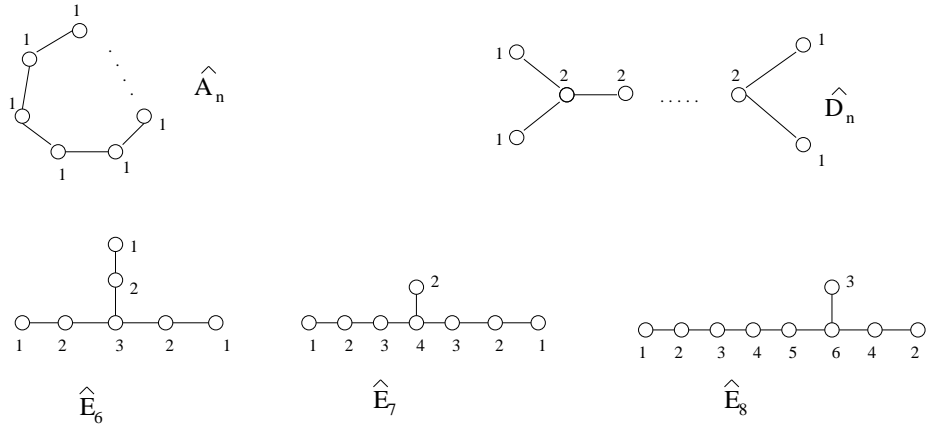,width=4.8in,angle=270}}
\caption{$\Gamma \subset$ full $SU(2)$ correspond to affine Dynkin diagrams
with the Dynkin labels $N_i$ on the nodes corresponding to the dimensions of the
irreps. In the quiver theory the nodes correspond to gauge groups and the
lines (or arrows for chiral theories), matter fields. For finite theories
each $N_i$ must be $\frac{1}{2}$ of the sum of neighbouring labels and the
gauge group is $\bigoplus\limits_i U(N_i)$. \label{su21}}
\end{figure}

\begin{figure}
	\centerline{\psfig{figure=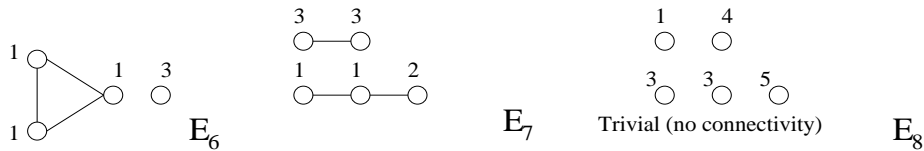,width=4.8in,angle=270}}
\caption{$\Gamma \subset SU(2)/\IZ_2$ give disconnected graphs \label{su22}}
\end{figure}

Let us digress briefly to make some mathematical observations.
We recall that in the $SU(2)$ case the matter matrices $a_{ij}$, 
due to McKay's
\index{Quivers!adjacency matrix}
theorem and Moore-Douglas quiver theories, are encoded as adjacency matrices of
affine Dynkin diagrams considered as unoriented graphs as given by Figures
\ref{su21} and \ref{su22}.


\begin{figure}
	\centerline{\psfig{figure=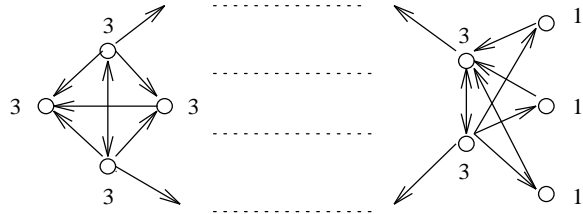,width=3.0in,angle=270}}
\caption{$\Delta_{3n^2} \subset SU(3)$ for $n \ne 0 $ mod 3. These belong
to both the full and center-removed $SU(3)$. \label{su3d}}
\end{figure}

\begin{figure}
	\centerline{\psfig{figure=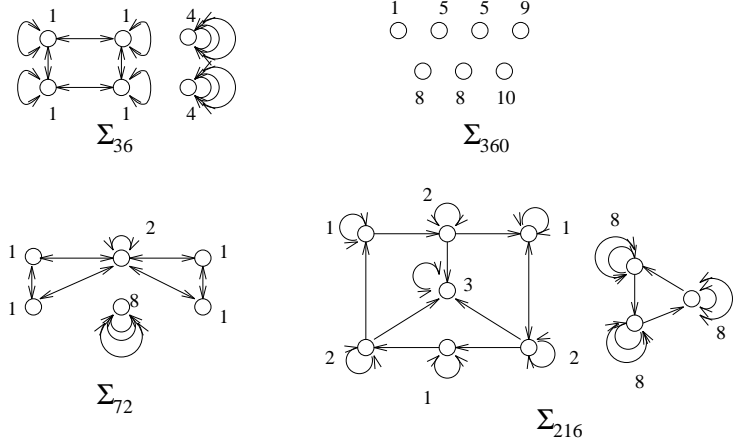,width=3.8in,angle=270}}
\caption{$\Sigma \subset SU(3)/\IZ_3$ gives unconnected graphs. \label{su3s1}}
\end{figure}

\begin{figure}
	\centerline{\psfig{figure=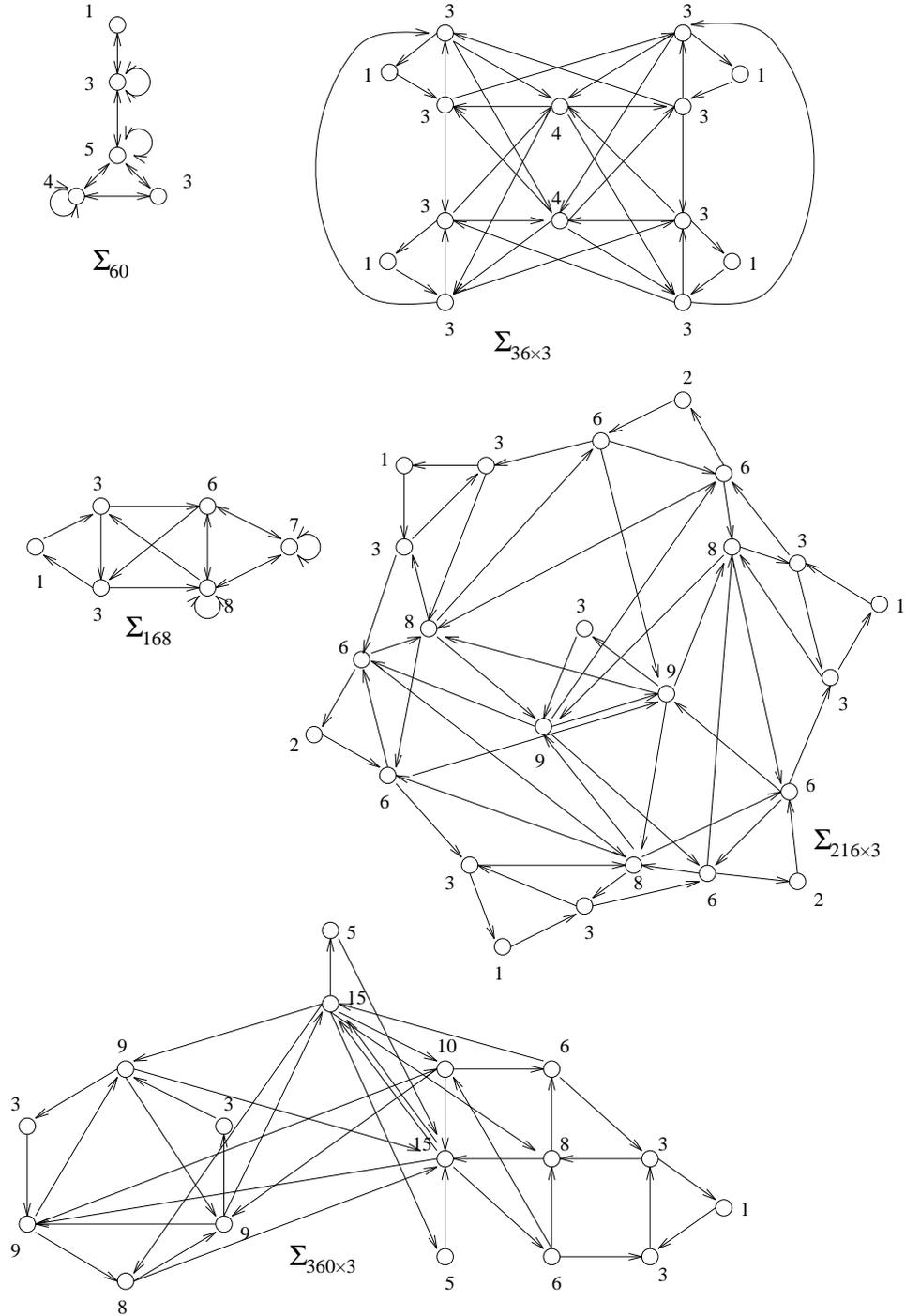,width=5.0in,angle=270}}
\caption{$\Sigma \subset$ full $SU(3)$. Only 
	$\Sigma_{36 \times 3, 216 \times 3, 360 \times 3}$ 
	belong only to the full $SU(3)$, for
	these we have the one loop $\beta$-function vanishing condition
        manifesting
	as the label of each node equaling to $\frac{1}{3}$ of that
	of the incoming and outgoing neighbours respectively.
        The matrix representation for these graphs are given in
	Appendix \ref{append:9811183.IV}.
        \label{su3s2}}
\end{figure}

We are of course led to wonder, whether in analogy, the $a_{ij}$ for $SU(3)$
present themselves as adjacency matrices for quiver diagrams associated to
some {\it oriented} graph theory because the theory is chiral.
This is very much in the spirit of 
recent works on extensions of Mckay correspondences by algebraic geometers
\cite{Ito} \cite{Roan}. We here present these quiver graphs in figures
\ref{su3d} \ref{su3s1} and \ref{su3s2}, hoping 
that it may be of academic interest.


Indeed we note that for the center-removed case, as with $SU(2)$, we get disconnected
(or trivial) graphs; this of course is the manifestation of the fact that
there are no non-trivial {\bf 3} representations for these groups (just
as there are no non-trivial {\bf 2}'s of $\Gamma \subset SU(2) / \IZ_2$).
On the other hand for $\Gamma \subset$ full $SU(3)$, we do get interesting connected and
oriented graphs, composed of various directed triangular cycles.

Do we recognise these graphs? The answer is sort of yes and the right
place to look for turns out to be in conformal field theory. 
In the work on general modular invariants in the WZW model for
$\widehat{su(n)_k}$ (which is equivalent to the study of the modular properties
of the characters for affine Lie algebras), an $ADE$ classification was noted
for $n=2$ \cite{CFT} \cite{Gepner1} \cite{Bernard}; 
this should somewhat be expected due to our 
earlier discussion on Gabriel's Theorem. For $n=3$, work has been done to
extract coefficients in the fusion rules and to treat them as entries of
adjacency matrices; this fundamentally is analogous to what we have done since
fusion rules are an affine version of finite group composition coefficients.
So-called generalised Dynkin diagrams have been constructed for 
$\widehat{su(3)}$ in analogy to the 5 simply-laced types corresponding to
$SU(2)$, they are: ${\cal A}_n, {\cal D}_{3n}, {\cal E}_5, {\cal E}_9$ and
${\cal E}_{21}$ where the subscripts denote the level in the representation
of the affine algebra \cite{CFT} \cite {DiFrancesco} \cite{Gannon}. 
We note a striking resemblance between these graphs
(they are some form of a dual and we hope to rigorise this similarity in
future work) with our quiver graphs: the ${\cal E}_5, {\cal E}_9$ and
${\cal E}_{21}$ correspond to $\Sigma_{216 \times 3}$, 
$\Sigma_{360 \times 3}$, and $\Sigma_{36 \times 3}$ respectively. Incidentally
these $\Sigma$ groups are the {\it only} ones that belong solely to the
full and not the center-removed $SU(3)$. The ${\cal D}_{3n}$ corresponds to
$\Delta_{3n^2}$ for $n \ne 0\mbox{ mod }3$, which are the non-trivial ones as
observed in the previous section and which again are those that belong
solely to the full $SU(3)$. The $\Delta_{6n^2}$ series, as noted above,
gave non-connected graphs, and hence do not have a correspondent. Finally
the ${\cal A}_n$, whose graph has complete $\IZ_3$ symmetry must come from
the Abelian subgroup of $SU(3)$, i.e., the $\widehat{A_n}$ case of $SU(2)$
but with ${\cal R} = {\bf 3}$ and not {\bf 2}. This beautiful relationship
prompts us to make the following conjecture upon which we may labour in
the near future:\\
\index{McKay Correspondence!and WZW}
{\large {\bf Conjecture:}} There exists a McKay-type correspondence
between Gorenstein singularities and the characters
of integrable representations of affine algebras $\widehat{su(n)}$
(and hence the modular invariants of the WZW model).\\

A physical connection between $\widehat{SU(2)}$ modular invariants and
quiver theories with 8 supercharges has been pointed out \cite{SD}. 
We remark that our conjecture is in the same spirit and a hint may come from
string theory. If we consider a D1 string on our orbifold, then this is
just our configuration of D3 branes after two T-dualities. In the strong
coupling limit, this is just an F1 string in such a background which
amounts to a non-linear sigma model and therefore some (super) conformal
field theory whose partition function gives rise to the modular invariants.
Moreover, connections between such modular forms and Fermat varieties
have also been pointed out \cite{Fermat}, this opens yet another door for
us and many elegant intricacies arise.

Enough digression on mathematics; let us return to physics. We would like
to conclude by giving a reference catalogue of chiral 
theories obtainable from
$SU(3)$ orbifolds. Indeed, though some of the matrices may not be terribly
interesting graph-theoretically, the non-symmetry of $a_{ij}^{\bf 4}$
is still an indication of a good chiral theory.

For the original $U(n)$ theory it is conventional to take a canonical
decomposition \cite{LNV}
as $n=N|\Gamma|$ \cite{LNV}, whence the (orbifolded) gauge group 
must be $\bigotimes\limits_{i}SU(N_i)$ as discussed in Section 3, such that 
$N|\Gamma|=n=\sum\limits_i N_i |{\bf r}_i|$. By an elementary theorem on
finite characters: $|\Gamma|=\sum \limits_i |{\bf r}_i|^2$, we see that the
solution is $N_i=N|{\bf r}_i|$. This thus immediately gives the form of 
the gauge group. Incidentally for $SU(2)$, the McKay
correspondence gives more information, it dictates that the dimensions
of the irreps of $\Gamma$ are actually the Dynkin labels for the diagrams.
This is why we have labeled the nodes in the graphs above. Similarly for
$SU(3)$, we have done so as well; these should be some form of generalised
Dynkin labels.

Now for the promised catalogue, we shall list below all the 
chiral theories obtainable from orbifolds of $\Gamma \subset SU(3)$ 
($\IZ_3$ center-removed or not). This is
done so by observing the graphs, connected or not, that contain unidirectional
arrows. For completeness, we also include the subgroups of $SU(2)$, which
are of course also in $SU(3)$, and which do give non-symmetric matter matrices
(which we eliminated in the ${\cal N}=2$ case) if we judiciously choose the
{\bf 3} from their representations. We use the short hand 
$(n_1^{k_1},n_2^{k_2},...,n_i^{k_i})$ to denote the gauge group
$\bigoplus\limits^{k_1} SU(n_1)...\bigoplus\limits^{k_i} SU(n_i)$. Analogous to
the discussion in Section 3, the conformality condition to one loop order
in this ${\cal N} = 1$
case, viz., $N_f = 3 N_c$ translates to the requirement that the label of
each node must be $\frac13$ of the sum of incoming and the sum of 
outgoing neighbours individually.  (Incidentally, the gauge anomaly cancelation 
condition has been pointed out as well \cite{LR}. In our language it demands the
restriction that $N_j a_{ij} = \bar{N}_j a_{ji}$.)
In the following table, the * shall
denote those groups for which this node condition is satisfied. We see that
many of these models contain the group $SU(3) \times SU(2) \times U(1)$ and hope
that some choice of orbifolds may thereby contain the Standard Model.
\[
\begin{array}{|c|c|}
\hline
\Gamma \subset SU(3)				&\mbox{Gauge Group} \\ \hline
\widehat{A_n} \cong \IZ_{n+1}			&(1^{n+1}) \\
\IZ_k \times \IZ_{k'}				&(1^{k k'}) *\\
\widehat{D_n}					&(1^4,2^{n-3}) \\
\widehat{E_6} \cong {\cal T}			&(1^3,2^3,3) \\
\widehat{E_7} \cong {\cal O}			&(1^2,2^2,3^2,4) \\
\widehat{E_8} \cong {\cal I}			&(1,2^2,3^2,4^2,5,6) \\
E_6 \cong T					&(1^3,3) \\
E_7 \cong O					&(1^2,2,3^2) \\
E_8 \cong I					&(1,3^2,4,5) \\
\Delta_{3n^2}	(n=0\mbox{ mod }3)	  	&(1^9,3^{\frac{n^2}{3}-1}) *\\
\Delta_{3n^2}	(n\ne0\mbox{ mod }3)		&(1^3,3^{\frac{n^2-1}{3}}) *\\
\Delta_{6n^2}	(n\ne0\mbox{ mod }3) 		&(1^2,2,3^{2(n-1)},6^{\frac{n^2-3n+2}{6}}) *\\
\Sigma_{168}					&(1,3^2,6,7,8) *\\
\Sigma_{216}					&(1^3,2^3,3,8^3) \\
\Sigma_{36\times 3}				&(1^4,3^8,4^2) *\\
\Sigma_{216\times 3}				&(1^3,2^3,3^7,6^6,8^3,9^2) *\\
\Sigma_{360\times 3}				&(1,3^4,5^2,6^2,8^2,9^3,10,15^2) *\\
\hline
\end{array}
\]

\section{Concluding Remarks}
By studying gauge theories constructed from orbifolding of an ${\cal N}=4$
$U(n)$ super-Yang-Mills theory in 4 dimensions, we have touched upon
many issues. We have presented the explicit matter content and gauge
group that result from such a procedure, for the cases of $SU(2)$ and
$SU(3)$. In the
first we have shown how our calculations agree with current quiver 
constructions and
in the second we have constructed possible candidates for chiral theories.
Furthermore we have noted beautiful graph-theoretic interpretations of these
results: in the $SU(2)$ we have used Gabriel's theorem to partially explain
the $ADE$ outcome and in the $SU(3)$ we have noted connections with 
generalised Dynkin diagrams and have conjectured the
existence of a McKay-type correspondence between these orbifold theories
and modular invariants of WZW conformal models.

Much work of course remains. In addition to proving this conjecture, we also
have numerous questions in physics. What about $SU(4)$, the full group?
These would give interesting non-supersymmetric theories. How do we
construct the brane box version of these theories? Roan has shown how
the Euler character of these orbifolds correspond to the class numbers
\cite{Roan}; we know the blow-up of these singularities correspond to
marginal operators. Can we extract the marginal couplings and thus the
duality group this way? We shall hope to address these problems in forth-coming
work. Perhaps after all, string orbifolds, gauge theories, modular
invariants of conformal field theories as well as Gorenstein singularities
and representations of affine Lie algebras, are all manifestations of
a fundamental truism.
\chapter{Orbifolds II: Avatars of McKay Correspondence}
\label{chap:9903056}
\section*{\center{{ Synopsis}}}
\index{ADE Meta-pattern}
Continuing with the conjecture from the previous chapter, we attempt
to view the ubiquitous ADE classification, manifesting as often
mysterious correspondences both in mathematics and physics, from a
string theoretic perspective.

On the mathematics side we delve into such matters as 
quiver theory, ribbon categories, 
and
the McKay Correspondence which relates finite group
representation theory to Lie algebras as well as
crepant resolutions of Gorenstein singularities.  On the
physics side, we investigate D-brane orbifold theories, 
the graph-theoretic classification of the WZW modular invariants,
as well as the relation between the string theory nonlinear
$\sigma$-models 
and Landau-Ginzburg orbifolds.

We here propose a unification scheme which naturally incorporates all 
these correspondences of the ADE type in two complex dimensions.
An intricate web of inter-relations is constructed, providing a
possible 
guideline to establish new directions of research 
or alternate pathways to the standing problems 
in higher dimensions \cite{9903056}.
\section{Introduction}
\begin{figure}
\centerline{\psfig{figure=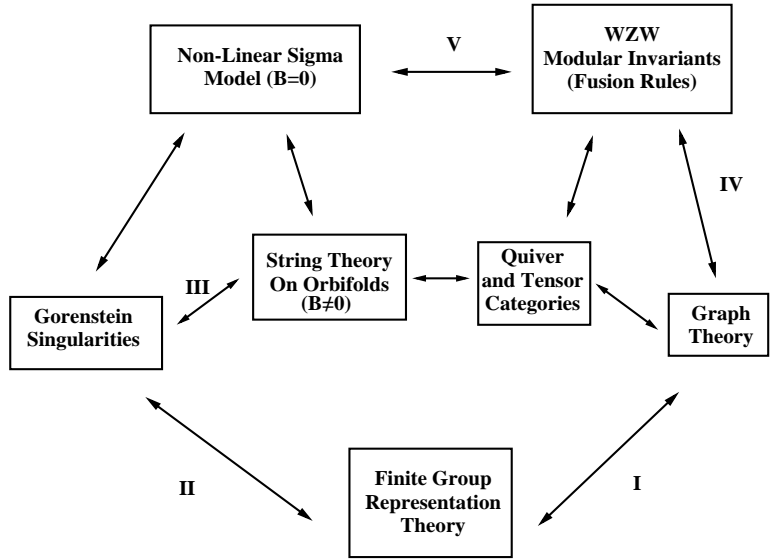,width=4.0in}}
\caption{The Myriad of Correspondences: it is the purpose of this
	chapter 
	to elucidate these inter-relations in 2-dimensions, so as to
	motivate a similar coherent picture in higher dimensions.
	Most of the subsectors in this picture have been studied
	separately by mathematicians and physicists, but they are in
	fact not as disparate as they are guised.}
\label{fig:mother}
\end{figure}

This chapter reviews the known facts about the various ADE
classifications that arise in mathematics and string theory and
organizes them into a unified picture.  This picture serves as a guide
for our on-going work in higher dimensions and naturally incorporates 
diverse concepts in mathematics.

In the course of their research on supersymmetric Yang-Mills theories 
resulting from the type IIB D-branes on orbifold singularities
(Chap. \ref{chap:9811183}), as 
prompted by collective works in constructing (conformal) gauge theories
in the physics literature (cf. previous chapter), 
it was conjectured that there may exist a McKay-type
correspondence between the bifundamental matter content and the modular 
invariant partition functions of the Wess-Zumino-Witten (WZW)
conformal field theory.  Phrased in another way, the correspondence,
if true, would
relate the Clebsch-Gordan  coefficients for tensor products of the 
irreducible representations
of finite subgroups of $SU(n)$ 
with the integrable characters for the affine algebras $\widehat{SU(n)}_k$
of some integral weight $k$.

Such a  relation has been well-studied in the case of $n=2$ and it
falls into 
an ADE classification scheme \cite{CFT,Gepner1,GW,Gannon2}.
Evidences for what might occur
in the case of $n=3$ were presented in Chap. \ref{chap:9811183} 
by computing the
Clebsch-Gordan coefficients extensively for  
the subgroups of $SU(3)$. Indications from the lattice
integrable model perspective were given in \cite{DiFrancesco}.

The natural question to pose is why there should be such correspondences.
Indeed, why should there be such an intricate chain of connections
among 
string theory on orbifolds, finite representation theory, graph
theory, affine 
characters and WZW modular invariants? In this chapter, we hope to 
propose a unified quest
to answer this question from the point of view of the conformal field
theory description of Gorenstein singularities.  We also observe that
category theory seems to prove a common basis for all these theories.

We begin in two dimensions, where there have been numerous independent 
works in the past few decades in both mathematics and physics to
establish 
various correspondences.
In this case, the all-permeating theme is the ADE classification.
In particular, there is the original McKay's correspondence between 
finite subgroups of $SU(2)$ and the ADE Dynkin diagrams \cite{McKay} 
to which we henceforth refer as the {\it Algebraic McKay Correspondence}.
On the geometry side, the representation rings of these groups 
were related to the Groethendieck (cohomology) rings of the resolved
manifolds 
constructed from the Gorenstein singularity of the respective groups
\cite{threedim,Roan}; we shall refer to this as the {\it Geometric McKay
Correspondence}. 
Now from physics, studies in conformal field theory (CFT) have prompted many
beautiful connections among graph theory, fusion algebra, and modular 
invariants \cite{CFT,Gepner1,GW,Gannon2,minimal,gepner}. The classification
of the modular invariant partition function associated with 
$\widehat{SU(2)}$ Wess-Zumino-Witten (WZW) models also mysteriously 
falls into an ADE type \cite{CIZ}. There have been some
recent 
attempts to explain this seeming accident from the
supersymmetric field theory and brane configurations perspective 
\cite{SD,ABKS}. In this chapter we push from the direction of the 
Geometric McKay Correspondence and see how Calabi-Yau (CY) non-linear
sigma models 
constructed on the Gorenstein singularities associated with the finite
groups may be related to Kazama-Suzuki coset models 
\cite{minimal,gepner,vafa1,vafa2,witten,GLSM}, which in turn can
 be related to 
the WZW models. This link would provide a natural setting for the emergence of
the ADE classification of the modular invariants.
In due course, we will review and establish a catalog of 
inter-relations, whereby 
forming a complex web and unifying many independently noted
correspondences.  Moreover, we find a common theme of categorical
axioms  that all of these
theories seem to satisfy and suggest why the ADE classification and its
extensions arise so naturally.
This web, presented in Figure~\ref{fig:mother}, is the central idea of
our chapter.  Most of the correspondences in \fref{fig:mother} actually
have been discussed in the string theory literature although not all
at once in a unified manner with a mathematical tint.

Our purpose is two-fold.  Firstly, we shall show that tracing 
through the arrows in \fref{fig:mother}
 may help to enlighten the links that may seem
accidental. Moreover, and perhaps more importantly, we propose that
this program may be extended beyond two dimensions and hence 
beyond $A$-$D$-$E$. Indeed, algebraic geometers have done extensive research 
in generalizing
McKay's original correspondence to Gorenstein singularities of dimension
greater than 2 (\cite{Gonzales} to \cite{Brylinski}); many standing
conjectures exist in this respect. On the other hand, there is the
conjecture mentioned above regarding the $\widehat{SU(n)}_k$ WZW
 and the subgroups of $SU(n)$ in general.
It is our hope that Figure~\ref{fig:mother}
remains valid for $n > 2$ so that these conjectures may be attacked
by the new pathways we propose. We require and sincerely hope for the 
collaborative effort of many experts in mathematics and physics 
who may take interest in this attempt to
unify these various connections.

\index{McKay Correspondence!in string theory}
The outline of the chapter follows the arrows drawn in
Figure~\ref{fig:mother}.  
We begin in \sref{sec:ubiquity} by
summarizing the ubiquitous ADE classifications, and
\sref{sec:arrows} will be devoted to clarifying these ADE links, while bearing
in mind how such ubiquity may permeate to
higher dimensions. It will be divided in to the following subsections:

\begin{itemize}
\item I. The link between representation
	 theory for finite groups and quiver graph theories (Algebraic
	 McKay); 
\item II. The link between finite groups and crepant 
	resolutions of Gorenstein singularities (Geometric McKay);
\item III. The link between resolved Gorenstein singularities, 
	Calabi-Yau manifolds and chiral rings for the associated
	non-linear sigma model 
        (Stringy Gorenstein resolution);
\item IV. The link between quiver graph theory for finite groups and WZW
	modular invariants in CFT, as discovered in the  study of
	of string orbifold theory (Conjecture in
	Chap. \ref{chap:9811183}); 
\end{itemize}
	and finally, to complete the cycle of correspondences,
\begin{itemize}
\item V. The link between the singular geometry and its conformal
	field theory description provided by an orbifoldized coset
	construction which contains the  WZW theory.
\end{itemize}

\vspace{2mm}
\noindent
In \sref{sec:CFT} we discuss arrow V which
fills the gap between mathematics and physics, explaining why WZW
models have the magical properties that are so closely related to the
discrete subgroups of the unitary groups as well as to geometry.
From all these links arises \sref{sec:conj} 
which consists of our conjecture that there exists a conformal
field theory description of the Gorenstein singularities for higher
 dimensions, encoding the relevant information about the discrete
 groups and the cohomology ring.
In \sref{sec:ribbons}, we hint at how these vastly different fields may have
similar structures by appealing to the so-called ribbon and quiver categories.

Finally in the concluding remarks of \sref{sec:conclusion},
 we discuss the projection for future labors.

We here transcribe our observations with the hope 
they would spark a renewed interest in the study of McKay 
correspondence under a possibly new light of CFT, and vice versa.
We hope that Figure~\ref{fig:mother} will open up many
interesting and exciting pathways of research and that not only some
existing 
conjectures may be solved by new methods of attack, but also further
beautiful 
observations could be made.

\vspace{1cm}
\noindent
{\large\bf  Notations and Nomenclatures}\\
\noindent
We put a \ $\widetilde{}$\ \/ over a singular variety
to denote its resolved geometry. By dimension we mean
complex dimension unless stated otherwise.
Also by ``representation ring of $\Gamma$,'' we mean the 
ring formed by the tensor product decompositions of the irreducible 
representations of $\Gamma$.  The capital Roman numerals, I--V, in front
of the section headings correspond to the arrows in
Figure~\ref{fig:mother}.

\section{Ubiquity of ADE Classifications} \label{sec:ubiquity}
\begin{figure}[ht]
\centerline{\psfig{figure=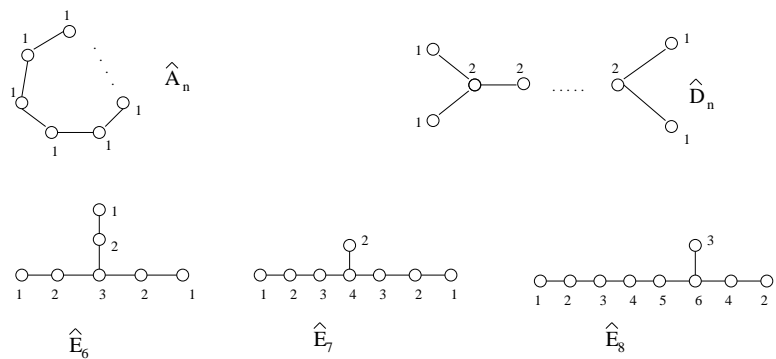,width=4.0in}}
\caption{The Affine Dynkin Diagrams and Labels.}
\end{figure}

\begin{table}[ht]
\vspace{5mm}
\begin{center}
\begin{tabular}{|p{1cm}|p{3cm}|p{4cm}|p{3cm}|}
\hline
 & Theory & Nodes & Matrices \\ 
\hline \hline
(a)& Finite Subgroup $\Gamma$ of $SU(2)$ & 
        Irreducible Representations &
        Clebsch-Gordan Coefficients\\
\hline
(b)& Simple Lie algebra of type $ADE$ & Simple Roots & 
        Extended Cartan matrix \\
\hline
(c)& Quiver Dynkin Diagrams & Dynkin Labels &
        Adjacency Matrix \\
\hline
(d)& Minimal Resolution $X\to\C^2/\Gamma$ &
        Irreducible Components of the Exceptional Divisor
        (Basis of $H_2(X,\Z)$) &
        Intersection Matrix \\
\hline
(e)& $\widehat{SU(2)_k}$ WZW Model &
        Modular Invariants / WZW Primary Operators &
        Fusion Coefficients \\
\hline
(f)& Landau-Ginzburg &
        Chiral Primary Operators & 
        Chiral Ring Coefficients \\
\hline
(g)& CY Nonlinear Sigma Model &
        Twisted Fields &
        Correlation Functions \\
\hline 
\end{tabular} 
\caption{ADE Correspondences in 2-dimensions. The same
	graphs and their affine extensions appear in
	different theories.}\label{table:ADE}
\end{center}
\end{table}

In this section, we summarize the appearance of the ADE
classifications in physics and mathematics and their commonalities.

It is now well-known that the complexity of particular algebraic and
geometric 
structures can often be organized into classification schemes of the 
ADE type. The first hint of this structure  began in the 1884 work
of F. Klein in which he classified the discrete subgroups $\Gamma$
of $SU(2)$ \cite{Klein}.
\index{Finite Groups!$SU(2)$ subgroups}
These were noted to be in 1-1 correspondence with the Platonic solids in 
$\R^3$, and with some foresight, we write them as:
\begin{itemize}
\item type A: the cyclic groups (the regular polygons);
\item type D: the binary dihedral groups (the regular dihedrons) and 
\item type E: the binary tetrahedral (the tetrahedron), octahedral
	(the cube and the octahedron) and icosahedral (the
	dodecahedron and 
	the icosahedron) groups,
\end{itemize}
where we have placed in parenthesis next to each group the geometrical shape
for which it is the double cover of the symmetry group.

The ubiquity of Klein's original hint has persisted till the present day.
The ADE scheme has manifested itself in such diverse fields as
representation theory of finite groups, quiver graph theory, Lie algebra
theory, algebraic geometry, conformal field theory, and string theory.
It will be the intent of the next section to explain the details of the
correspondences appearing in Table~\ref{table:ADE}, and we will subsequently
propose their extensions in the remainder of the chapter.
%

\section{The Arrows of Figure 1.} \label{sec:arrows}
In this section, we explain the arrows appearing in Figure 1.  We verify
that there are compelling evidences in favor of the picture for 
the case of $2_\C$-dimensions, and we will propose its generalization to
higher dimensions in the
subsequent sections, hoping that it will lead to new insights on the
McKay correspondence as well as conformal field theory.

\subsection{(I) The Algebraic McKay Correspondence} \label{sec:alg-mckay}
In the full spirit of the omnipresent ADE classification, it has been
noticed in 1980 by J. McKay that there exists a remarkable correspondence
between the discrete subgroups of $SU(2)$ and the affine Dynkin graphs
\cite{McKay}. 
Indeed, this is why we have labeled the subgroups in the
manner we have done.
\index{McKay Correspondence!definition}
\begin{definition}
For a finite group $\Gamma$, let $\left\{ r_i \right\}$ be its set of
irreducible representations (irreps), then we define the coefficients
$m_{ij}^k$ appearing in
\begin{equation}
r_k \otimes r_i = \bigoplus\limits_{j} m_{ij}^k r_j
\label{CG}
\end{equation}
to be the {\bf Clebsch-Gordan coefficients} of $\Gamma$.
\end{definition}
For $\Gamma \subset SU(2)$ McKay chose a fixed 
(not necessarily irreducible) representation
$R$ in lieu of general $k$ in \ref{CG} and defined matrices $m_{ij}^R$.
He has noted that up to automorphism, there always exists a
unique 2-dimensional representation,
which for type A is the tensor sum of 2 dual 1-dimensional irreps
and for all others the self-conjugate 2-dimensional irrep.
It is this $R=2$ which we choose and simply write the matrix as $m_{ij}$.
The remarkable observation of McKay can be summarized in the
following theorem;
the original proof was on a case-to-case basis and Steinberg gave a
unified proof in 1981 \cite{McKay}.
\begin{theorem}[{\sc McKay-Steinberg}]
For $\Gamma = A,D,E$, the matrix 
$m_{ij}$ is $2 I$ minus
the Cartan matrix of the 
affine extensions of the respective simply-laced simple Dynkin diagrams 
$\widehat{A}, \widehat{D}$ and $\widehat{E}$,
treated as undirected $C_2$-graphs (i.e., maximal eigenvalue of the
adjacency matrix is 2).
\end{theorem}
\index{Quivers!adjacency matrix}
Moreover, the Dynkin labels of the nodes of the affine
Dynkin diagrams are
precisely the dimensions of the irreps.  Given a discrete subgroup
$\Gamma \subset SU(2)$, 
there thus exists a Dynkin diagram that
encodes the essential information about the representation ring of $\Gamma$.
Indeed the
number of nodes should equal to the number of irreps and thus by a
rudimentary 
fact in finite representation theory, subsequently equals the number of 
conjugacy classes of $\Gamma$. Furthermore, if we remove the node
corresponding to the trivial 1-dimensional (principal) representation,
we obtain the regular ADE Dynkin diagrams. We present these facts in the
following diagram:
\index{Lie Algebras}
\vspace{3mm}
\bcenter
\parbox{1.2in}{Clebsch-Gordan Coefficients for $\Gamma=A,D,E$} \hs{2} 
$\longleftrightarrow$\hs{2} 
\parbox{1.2in}{Dynkin Diagram of
$\widehat{A},\widehat{D},\widehat{E}$}\hs{2}  
$\longleftrightarrow$\hs{2} \parbox{1.5in}{Cartan matrix and dual
Coxeter labels of  
	$\widehat{A},\widehat{D},\widehat{E}$}
\ecenter
\vspace{3mm}
This is Arrow I of Figure~\ref{fig:mother}.

Proofs and extension of McKay's results from geometric perspectives 
of this originally combinatorial/graph-theoretic 
theorem soon followed; they caused fervent activities in both algebraic
geometry and string theory (see e.g.,
\cite{orbifold,Gonzales,threedim,Roan}). Let us first turn to the former.

\subsection{(II) The Geometric McKay Correspondence}
In this section, we are interested in crepant resolutions of Gorenstein
quotient singularities. 
\begin{definition}
The singularities of
\index{Orbifolds}
\index{Singularity!Gorenstein}
$\C^n/\Gamma$ for $\Gamma\subset GL(n,\C)$ are called {\bf Gorenstein}
if there exists a nowhere-vanishing
holomorphic $n$-form\footnote{Gorenstein singularities thus provide
local 
models of singularities on Calabi-Yau manifolds.} 
on regular points.   
\end{definition}
Restricting $\Gamma$ to
$SU(n)$ would guarantee that the quotient singularities are
Gorenstein.  
\index{Resolution!crepant}
\begin{definition}
We say that a smooth variety
$\widetilde{M}$ is a {\bf crepant} resolution of a singular
variety
$M$ if there exists a birational morphism $\pi:\widetilde{M}\ra M$
such that the canonical sheaves $\cK_M$ and $\cK_{\widetilde{M}}$ are
the same, 
or more precisely, if $\pi^* (\cK_M) = \cK_{\widetilde{M}}$.
\end{definition}

For $n\leq 3$, Gorenstein singularities always admit crepant
resolutions \cite{threedim,Roan}. On the other hand, 
in dimensions greater than 3, there
are known examples of terminal Gorenstein singularities which do not
admit crepant resolutions.  It is believed, however, that when the
order of $\Gamma$ is sufficiently larger than $n$, there exist crepant
resolutions for most of the groups.

The traditional ADE classification is relevant in studying the
discrete subgroups of $SU(2)$ and resolutions of Gorenstein
singularities in two complex-dimensions.
Since we can choose an invariant Hermitian metric on $\C^2$, 
finite subgroups of $GL(2,\C )$ and $SL(2,\C)$ are conjugate to finite
subgroups of $U(2)$ and $SU(2)$, respectively.  Here,
motivated by the string compactification on manifolds of trivial
canonical bundle, we consider the linear
actions of non-trivial discrete subgroups $\Gamma$ of $SU(2)$ on $\C^2$.
Such quotient spaces $M=\ctg$, called {\it orbifolds},  have
fixed points which are
isolated Gorenstein singularities of the ADE type studied by Felix Klein.  

As discussed in the previous sub-section, McKay\cite{McKay} has observed
a 1-1 correspondence between the non-identity conjugacy 
classes of discrete subgroups of  
$SU(2)$ and the Dynkin diagrams
of $A$-$D$-$E$ simply-laced Lie algebras, and this relation
in turn provides an
indirect correspondence between the orbifold singularities of
 $M$ and the Dynkin
diagrams. 
In fact, there exists a direct geometric correspondence between
the crepant resolutions of $M$ and the Dynkin diagrams. 
 Classical theorems in algebraic geometry tell us
that there exists a unique crepant resolution $(\widetilde{M},\pi)$ of the
Gorenstein
singularity of $M$ for all
$\Gamma\subset SU(2)$.  Furthermore, the exceptional divisor
$E=\pi^{-1} (0)$ is a 
compact, connected union of irreducible $1_{\C}$-dimensional curves of
genus zero\footnote{We will refer to them as $\P^1$ blow-ups.} such that
their intersection matrix is represented by 
the simply-laced Dynkin diagram
associated to $\Gamma$.  More precisely,  each node of the diagram
corresponds to an irreducible $\P^1$, and the  intersection matrix
is negative of 
the Cartan matrix of the Dynkin diagram such that two $\P^1$'s intersect
transversely at one point if and only if the two nodes are connected
by a line in the diagram.
In particular, we see that
the curves have self-intersection numbers $-2$ which exhibits the
singular nature of the orbifold upon blowing them down.
Simple consideration shows that these curves form a basis of the 
homology group $H_2(\widetilde{M},\Z )$ which is seen to coincide with
the root 
lattice of the associated Dynkin diagram by the above identification.
Now, combined with the algebraic McKay correspondence, this crepant
resolution  
picture  yields a 1-1 correspondence
between the basis of $H_2(\widetilde{M},\Z )$ and the non-identity
conjugacy classes of $\Gamma$.
We recapitulate the above discussion in the following diagram:
\index{Resolution!blow up}
\vspace{3mm}
\bcenter
\parbox{1.2in}{$H_2(\widetilde{M},\Z)$ of the blow-up} \hs{2} 
$\longleftrightarrow$\hs{2} 
\parbox{1in}{Dynkin Diagram of $\Gamma$}\hs{2} 
$\longleftrightarrow$\hs{2} \parbox{1.5in}{Non-identity Conjugacy
Classes of $\Gamma$}
\ecenter
\vspace{3mm}
This is Arrow II in Figure~\ref{fig:mother}.
Note incidentally that one can think of irreducible representations as
being dual to 
conjugacy classes and hence as basis of $H^2(\widetilde{M},\Z)$.  This
poses 
a subtle question of which correspondence is more natural, but we will
ignore such issues in our discussions.

It turns out that $M$ is not only analytic but also
algebraic; that is, $M$ is isomorphic to $f^{-1} (0)$, where
$f:\C^3\rightarrow\C$
is one of the polynomials in \tref{rational} depending on $\Gamma$.
The orbifolds defined by the zero-loci of the polynomials are 
commonly referred to as the singular ALE spaces.
\index{ALE Spaces}
\index{ADE Singularities}
\begin{table}
\begin{center}
\begin{tabular}{||l|l|l||}  \hline
$f(x,y,z)$ & Subgroup $\Gamma$ & Order of $\Gamma$\\ \hline
$x^2 + y^2 +z^{k+1}$ & $A_k$ Cyclic & $k+1$\\ \hline
$x^2 + y^2z +z^{k-1}$ & $D_k$ Binary Dihedral & $4(k-2)$\\ \hline
$x^2 + y^4 +z^3$ & $E_6$ Binary Tetrahedral & 24 \\ \hline
$x^2 + y^3z +z^3$  & $E_7$ Binary Octahedral & 48\\ \hline
$x^2 + y^5 +z^3$ & $E_8$ Binary Icosahedral & 120\\ \hline
\end{tabular}
\caption{Algebraic Surfaces with Quotient Singularities \label{rational}}
\end{center}
\end{table}

\subsection{(II, III) McKay Correspondence and SCFT}

One of the first relevance of ADE series in conformal field
theory appeared in attempts to classify $N=2$ superconformal
field theories (SCFT) with central charge $c<3$ \cite{minimal}.  
Furthermore,  the exact forms of the ADE polynomials in
\tref{rational} appeared in a similar attempt to classify certain
classes of $N=2$ SCFT in terms of Landau-Ginzburg (LG) models.  The LG
super-potentials were precisely classified by the polynomials, and the
chiral ring and quantum numbers were computed with applications of
singularity theory \cite{vafa2}.  The LG theories which realize coset
models would appear again in this chapter to link the WZW to geometry.

In this subsection, we review how string theory, when the $B$-field is
non-vanishing, resolves the orbifold 
singularity and how it encodes the information about the cohomology of 
the resolved manifold.  
Subsequently, we will consider the singular limit
of the conformal field theory on orbifolds by turning off the $B$-field, and
we will argue  that, in this singular limit, 
the $\widehat{SU(2)}_k$ WZW fusion ring 
inherits the information about the cohomology ring from the smooth theory.

\index{Resolution!blow up}
\subsubsection{Orbifold Resolutions and Cohomology Classes} 
\label{sec:orbifold}
Our discussion here will be general and not restricted to $n=2$.
Many remarkable features of string theory stem from the fact that we can
``pull-back'' much of the physics on the target space to the
world-sheet, and as a result, the resulting
world-sheet conformal field theory somehow encodes the geometry of the
target space.  One example is that CFT is often\footnote{Not all CFT on
singular geometry are smooth.  For example, there are
examples of singular CFT's defined on singular backgrounds, such as
in the case of gauge symmetry enhancement of the Type IIA string theory
compactified on singular $K3$ where the $B$-field vanishes
\cite{aspinwall}.  Later, we will discuss a tensored coset model 
\cite{Ooguri-Vafa} 
describing this singular non-linear sigma model and relate it to the
algebraic McKay Correspondence.} insensitive to 
Gorenstein singularities and quantum effects revolve the singularity
so that the CFT is smooth.  More precisely, Aspinwall \cite{aspinwall2}
has shown that non-vanishing of the $NS$-$NS$ $B$-field makes the CFT
smooth.   In fact, string theory predicts the Euler characteristic
of the {\it resolved}\/ orbifold \cite{orbifold}; the local form of
the statement is

\begin{conjecture}[{\sc Stringy Euler characteristic}]\label{euler}  Let
$M=\C^n/\Gamma$ for $\Gamma\subset SU(n)$ a finite subgroup.  Then,
there exists a crepant resolution $\pi: \mt \ra M$ such that
\beq
	\chi (\mt) = | \{ \mbox{Conjugacy Classes of $\Gamma$}\}| \ .
\eeq
\end{conjecture}
Furthermore, the Hodge numbers of resolved orbifolds
were also predicted by Vafa for
CY manifolds realized as hypersurfaces in weighted projective
spaces  and by Zaslow for K\"{a}hler manifolds \cite{vafa4}.
In dimension three, it has been proved \cite{threedim,Roan,ito-reid} that
every  Gorenstein singularity
 admits a crepant resolution\footnote{In fact, a given Gorenstein
singularity generally admits many crepant resolutions \cite{Joyce}.
String theory so far has yielded two distinguished
desingularizations: the traditional CFT resolution without discrete
torsion and deformation with discrete torsion \cite{torsion}.  In this
chapter, we are concerned only with K\"{a}hler resolutions without
discrete torsion.}
and that every
crepant resolution satisfies the
Conjecture~\ref{euler} and the Vafa-Zaslow Hodge number formulae.  For
higher dimensions, there are compelling evidences that the formulae are
satisfied by all crepant resolutions, when they exist.

As the Euler Characteristic in mathematics is naturally defined by the
Hodge numbers of cohomology classes, motivated by the works of string
theorists and the fact that $\mt$ has no odd-dimensional
cohomology\footnote{See \cite{ito-reid} for a discussion on this point.}, 
mathematicians have
generalized 
the classical McKay Correspondence  \cite{threedim,Roan,ito-reid,batyrev}
to geometry.

The geometric McKay Correspondence in 2-dimensions
actually identifies the cohomology ring of
$\mt$ and the representation ring of $\Gamma$ not only as vector spaces
but as rings.
Given a finite subgroup $\Gamma\subset SU(2)$, the intersection matrix
of the irreducible components of the exceptional divisor of the
resolved manifold is given by
the negative of the Cartan matrix of the associated Dynkin diagram
which is specified by the algebraic McKay Correspondence.  Hence,
there exists 
an equivalence between the tensor product
decompositions of conjugacy classes and 
intersection pairings of homology classes.  Indeed in
\cite{nakajima}, Ito and Nakajima prove that for all $\Gamma\subset SU(2)$
and for  abelian
$\Gamma\subset SU(3)$, the Groethendieck (cohomology) ring
of $\mt$ is isomorphic as a $\Z$-module 
to the representation ring of $\Gamma$ and that the intersection
pairing on  its dual, the Groethendieck
group of coherent sheaves on $\pi^{-1} (0)$, can be expressed as the
Clebsch-Gordan coefficients.  Furthermore, string theory analysis also
predicts a similar relation between the two ring structures
\cite{Dijkgraaf}.

The geometric McKay Correspondence can thus be stated as
\begin{conjecture}[\sc Geometric McKay Correspondence] \label{gmc} Let
$\Gamma,M,$
and $\mt$ be as in Conjecture~\ref{euler}.  Then, there exist bijections
\barrayn
\mbox{\fbox{\rule[-3mm]{0mm}{9mm}  Basis of $H^* (\mt,\Z)$}} \ \
	&\longleftrightarrow&\ \  
	\mbox{\fbox{\rule[-3mm]{0mm}{9mm} $\{$Irreducible
Representations of  
$\Gamma \}$}}\\
\mbox{\fbox{\rule[-3mm]{0mm}{9mm}  Basis of $ H_* (\mt,\Z)$}} \ \
	&\longleftrightarrow&\ \  
	\mbox{\fbox{\rule[-3mm]{0mm}{9mm} $\{$Conjugacy Classes of
$\Gamma \}$}}\ , 
\earrayn
and there is an identification between the two ring structures.
\end{conjecture}
\index{Chiral Rings}
\subsubsection{Question of Ito and Reid and Chiral Ring}
In \cite{ito-reid}, Ito and Reid raised the question whether the
cohomology ring\footnote{Henceforth, $\dim M=n$
 is not restricted to 2.}
 $H^*(\widetilde{M})$ is generated by
$H^2(\widetilde{M})$.  In this subsection, we rephrase the question
in terms of $N=2$ SCFT on $M=\C^n/\Gamma$, $\Gamma\subset SU(n)$.
String theory provides a way\footnote{It is believed that string
theory somehow picks out a distinguished resolution of the orbifold,
and the following discussion pertains to such a resolution when it
exists.} of computing the cohomology of the resolved manifold $\mt$.
Let us briefly review the method for the present case \cite{orbifold}:  

The cohomology of $\mt$ consists of those elements of $H^*(\C^n)$ that
survive the projection under $\Gamma$ and new classes arising from the
blow-ups.  In this case, $H^0(\C^n)$ is a set of all constant functions
on $\C^n$ and survives\footnote{This cohomology class should 
correspond to the trivial
representation in the McKay correspondence.} 
the projection, while all other cohomology classes vanish.  Hence, all
other non-trivial elements of $H^*(\mt)$ arise from the blow-up
process; in string theory language, they correspond to the
twisted chiral primary operators, which are not necessarily all marginal.
In the $N=2$ SCFT of non-linear sigma-model on a compact CY manifold, 
the $U(1)$
spectral flow 
identifies the chiral ring of the SCFT with the cohomology ring of the
manifold, modulo
quantum corrections.  
For non-compact cases, by
considering a topological non-linear $\sigma$-model, 
the $A$-model chiral ring matches the cohomology ring and
the blow-ups still correspond to the twisted sectors.

 An $N=2$ non-linear sigma model on  a CY $n$-fold $X$
 has two topological
twists called the $A$ and $B$-models, of which the ``BRST'' non-trivial
observables \cite{tqft} encode the information about the K\"{a}hler and 
complex structures of $X$, respectively.
The correlation functions of the $A$-model receive instanton corrections
whereas the classical computations of the $B$-model give exact quantum
answers.  The most efficient way of computing the $A$-model
correlation functions is to map the theory to a $B$-model on another
manifold $Y$ which is a mirror\footnote{Mirror symmetry has been
intensely studied by both mathematicians and physicists for the past
decade, leading to many powerful tools in enumerative geometry.  A detailed
discussion of mirror symmetry is beyond the scope of this chapter, and we
refer the reader to \cite{mirror} for introductions to the subject and
for references.} of $X$ \cite{mirror}.  Then, the classical
computation of the $B$-model on $Y$ yields the full quantum answer
for the $A$-model on $X$. 

In this chapter, we are interested in K\"{a}hler resolutions of the Gorenstein
singularities and, hence, in the $A$-model whose chiral ring is a quantum 
deformation of the classical cohomology ring.
Since all non-trivial elements of the cohomology
ring, except for $H^0$, arise from the twisted sector or blow-up
contributions, we have
the following reformulation of the Geometric McKay
Correspondence which is well-established in string theory:
\begin{proposition}[{\sc String Theory McKay
Correspondence}]\label{conjecture1} 
Let $\Gamma$ be a
discrete subgroup of $SU(n)$ such that the Gorenstein singularities of
$M=\C^n/\Gamma$ has a crepant resolution $\pi:\mt\ra M$.  Then, there
exists a following bijection between the cohomology and $A$-model data:
\beq
	\mbox{\fbox{\rule[-3mm]{0mm}{9mm} Basis of $\bigoplus\limits_{i>0}
H^i (\mt)$} \ \ 
	$\longleftrightarrow$\ \  
	\fbox{\rule[-3mm]{0mm}{9mm} $\{ \mbox{Twisted Chiral Primary
	Operators}\}$}}\ ,
\eeq
or equivalently, by the Geometric McKay Correspondence,

\beq
	\mbox{\fbox{\rule[-3mm]{0mm}{9mm} \{Conjugacy classes of
	$\Gamma$\}}\ \ 
	$\longleftrightarrow$\ \  
	\fbox{\rule[-3mm]{0mm}{9mm} \{Twisted Elements of the Chiral
	Ring\}}}\ . 
\eeq
\end{proposition}
Thus, since all $H^i, i>0$ arise from the twisted chiral primary but
not necessarily marginal fields and since the marginal operators
correspond to $H^2$,
we can now reformulate the question of whether $H^2$ generates $H^*$ as
follows: 
\begin{quote}
{\it Do the marginal twisted
 chiral primary fields generate the entire twisted  chiral ring?}
\end{quote}
\noindent
This kind of string theory resolution of orbifold singularities
is Arrow III in Figure~\ref{fig:mother}.  In \sref{sec:CFT}, we will
see how a conformal field theory description of
the singular limit of these string theories naturally 
allows us to link geometry to representation theory.  In this way, we
hint why McKay correspondence and the discoveries of \cite{CIZ} 
are not mere happy flukes of nature, as it will become clearer as
we proceed.

\index{McKay Correspondence!and WZW}
\subsection{(I, IV) McKay Correspondence and WZW} \label{subsec:Mckay-WZW}
When we calculate the partition function for the WZW
model with its energy momentum tensor associated to an algebra
$\widehat{g_k}$ 
of level $k$, it will be of the form\footnote{we henceforth
use the notation in \cite{CFT}}:
\[
Z(\tau) = \sum_{\widehat{\lambda},\widehat{\xi} \in P^{(k)}_+}
	\chi_{\widehat{\lambda}}(\tau)
	{\cal M}_{\widehat{\lambda},\widehat{\xi}}
	\overline{\chi_{\widehat{\xi}}}(\overline{\tau})
\]
where $P^{(k)}_+$ is the set of dominant weights and 
$\chi_{\widehat{\lambda}}$
is the affine character of $\widehat{g_k}$. The matrix ${\cal M}$ gives the
multiplicity of the highest weight modules in the decomposition of the
Hilbert space and is usually referred to as the {\it mass
matrix}. Therefore 
the problem of classifying the modular invariant 
partition functions of WZW models is
essentially that of the integrable characters $\chi$ of affine Lie algebras.

In the case of $\widehat{g_k} = \widehat{SU(2)_k}$, all the modular 
invariant partition functions are classified, and they fall into an ADE 
scheme (\cite{CFT} to \cite{Gannon2}).
In particular, they are of the form of sums over
modulus-squared of combinations of the weight $k$ Weyl-Kac character
$\chi^k_\lambda$ for $\widehat{SU(2)}$ (which is in turn
expressible in term of Jacobi theta functions), where
the level $k$ is correlated with the rank of the ADE Dynkin diagrams as 
shown in Table~\ref{table:ADE-WZW} and $\lambda$ are the eigenvalues
for the 
adjacency matrices of the ADE Dynkin diagrams.
\begin{table}
\begin{center}
\begin{tabular}{||c|l||}  \hline
Dynkin Diagram of Modular Invariants &  Level of WZW\\ \hline
$A_n$ & $n-1$\\ \hline
$D_n$ & $2n-4$\\ \hline
$E_6$ & 10 \\ \hline
$E_7$ & 16 \\ \hline
$E_8$ & 28 \\ \hline
\end{tabular}
\caption{The $ADE$-Dynkin diagram representations of the 
modular invariants of the $\widehat{SU(2)}$ WZW. \label{table:ADE-WZW}}
\end{center}
\end{table}
Not only are the modular invariants classified by these graphs,
but some of the fusion ring algebra can be reconstructed from
the graphs.

Though still largely a mystery, the reason for this classification
can be somewhat traced to the so-called {\it fusion rules}.
In a rational conformal field theory, the fusion coefficient 
$N_{\phi_i \phi_j}^{\phi_k^*}$ is defined by

\begin{equation}
\phi_i \times \phi_j = \sum\limits_{\phi_k^*} 
{\cal N}_{\phi_i \phi_j}^{\phi_k^*}
	\phi_k^*
\label{fusion}
\end{equation}
\noindent
where $\phi_{i,j,k}$ are chiral\footnote{Chirality here means left- or 
right-handedness not chirality in the sense of $N=2$ superfields.} 
primary fields. This fusion rule provides 
such vital information as the number of independent coupling between the
fields and the multiplicity of the conjugate field $\phi_k^*$ appearing in
the operator product expansion (OPE) of $\phi_i$ and $\phi_j$.
In the case of the WZW model with the energy-momentum tensor taking values
in the algebra $\widehat{g_k}$ of level $k$, we can recall 
that the primary fields have integrable representations
$\widehat{\lambda}$ in the dominant weights of $\widehat{g_k}$, and
subsequently, (\ref{fusion}) reduces to
\[
\widehat{\lambda} \otimes \widehat{\mu} =
\bigoplus_{\widehat{\nu} \in P_+^k} 
{\cal N}_{\widehat{\lambda}\widehat{\mu}}^{\widehat{\nu}}\ \widehat{\nu}.
\]
Indeed now we see the resemblance of (\ref{fusion}) coming from
conformal field theory to (\ref{CG}) coming from finite representation
theory, hinting that there should be some underlying relation.
We can of course invert (\ref{CG}) using the properties of finite
characters, just as we can extract $\cal{N}$ by using the Weyl-Kac
character formula (or by the Verlinde equations).

Conformal field theorists, inspired by the ADE classification
of the minimal models, have devised similar methods to treat the fusion
coefficients. It turns out that in the simplest cases the fusion
rules can be generated entirely from one special case of 
$\widehat{\lambda} = f$, the so-called fundamental representation. This
is of course in analogy to the unique (fundamental) 2-dimensional 
representation $R$ in McKay's paper. In this case, all the information
about the fusion rule is encoded in a matrix $[N]_{ij} = {\cal
N}_{fi}^j$, 
to be treated as the adjacency matrix of a finite graph. Conversely we
can define a commutative algebra to any finite graph, whose
adjacency matrix is defined to reproduce the fusion rules for
the so-called {\it graph algebra}. It turns out that in the cases of
$A_n, D_{2n}, E_6$ and $E_8$ Dynkin diagrams, 
the resulting graph algebra has an
subalgebra which reproduces the (extended)
fusion algebra of the respective ADE $\widehat{SU(2)}$ WZW models.

From another point of view, we can study the WZW model by quotienting
it by discrete subgroups of $SU(2)$; this is analogous to the
twisted sectors in string theory where for the partition function we
sum over all states invariant under the action of the discrete 
subgroup. Of course in this case we also have an $A$-$D$-$E$-type
classification 
for the finite groups due to the McKay Correspondence, therefore
speculations have risen as to why both the discrete subgroups and
the partition functions are classified by the same graphs 
\cite{CFT,DiFrancesco}, which also reproduce the associated ring 
structures.
The reader may have noticed that this connection
is somewhat weaker than the others hitherto considered, in the sense
that the adjacency matrices do not correspond 1-1 to the fusion rules.
This subtlety will be addressed in \sref{sec:CFT} and \sref{sec:ribbons}.

Indeed, the graph algebra construction has been extended to $\widehat{SU(3)}$
and a similar classification of the modular invariants have in fact been done
and are shown to correspond to the so-called {\it generalized Dynkin
Diagrams} \cite{CFT,Gannon,DiFrancesco}. On the other hand,
the Clebsch-Gordan coefficients of the McKay type for the discrete subgroups
of $SU(3)$ have been recently computed in the context of studying 
D3-branes on orbifold singularities (Chap \ref{chap:9811183}). 
It was noted that the
adjacency 
graphs drawn in the two different cases are in some form of correspondence
and was conjectured that this relationship might extend
to $\widehat{SU(n)_k}$ model for $n$ other than 2 and 3 as well. It is
hoped that this problem may be attacked by going through the other
arrows.

We have now elucidated arrows I and IV in \fref{fig:mother}.

\section{The Arrow V: $\sigma$-model/LG/WZW Duality}  \label{sec:CFT}
We here summarize the link V in \fref{fig:mother} for ALE spaces, as has been
established in \cite{Ooguri-Vafa}.

It is well-known that
application of catastrophe theory leads to the ADE
classification of Landau-Ginzburg models \cite{vafa2}.  It has been
subsequently shown 
that the renormalization group fixed points of these theories
actually provide the Lagrangian formulations of
$N=2$ discrete minimal models \cite{witten}.  What is even more
surprising and beautiful is Gepner's another proposal \cite{gepner} that 
certain classes of
$N=2$ non-linear sigma-models on CY 3-folds are equivalent to 
tensor products of $N=2$ minimal models with the correct central charges and
$U(1)$ projections.  
Witten has successfully verified the claim  in
\cite{GLSM}  using a gauged linear-sigma model which interpolates
between Calabi-Yau compactifications and Landau-Ginzburg orbifolds.

In a similar spirit, Ooguri and Vafa have considered  LG orbifolds\footnote{
The universality classes of  the LG models are completely specified by their
superpotentials $W$, and such a simple characterization leads to very powerful 
methods of detailed computations \cite{vafa1,Vafa-LG}.  Generalizations
of these models have many important applications in string theory, and
the OPE coefficients of topological LG theories with judiciously chosen
non-conformal deformations yield the fusion algebra of rational 
conformal field theories.
In \cite{gepner2}, Gepner has shown 
that the topological
 LG models with deformed Grassmannian superpotentials yield the fusion
algebra of the $\widehat{SU(n)}_k$ WZW, illustrating
 that much information about
 non-supersymmetric RCFT can be extracted from
their $N=2$ supersymmetric counterparts.  Gepner's superpotential could be
viewed as a particular non-conformal deformation of the
superpotential appearing in Ooguri and Vafa's model.} of
the tensor product of $SL(2,\R)/U(1)$ and $SU(2)/U(1)$
Kazama-Suzuki models\footnote{The $SL(2,\R)/U(1)$ coset model describes the
two-dimensional black hole geometry \cite{Witten-black}, while
the $SU(2)/U(1)$ Kazama-Suzuki model is just
the $N=2$ minimal model.} \cite{Kazama} and have shown that the resulting
theory describes the singular conformal field theory of the non-linear
sigma-model with the $B$-field turned off.  In particular,  they have
shown that the singularity on $A_{n-1}$ ALE space is
described by the
\beq
	\frac{\frac{SL(2)_{n+2}}{U(1)} \times \frac{SU(2)_{n-2}}{U(1)}}{\Z_n}
\eeq
orbifold model which contains the $\widehat{SU(2)}_{n-2}$ WZW theory
at level $k=n-2$.  The coset descriptions of the non-linear $\sigma$-models on 
$D$ and $E$-type ALE spaces also contain the corresponding WZW theories whose
modular invariants are characterized by the $D$ and $E$-type resolution graphs
of the ALE spaces.  The full orbifoldized Kazama-Suzuki model has
fermions as well 
as an extra Feigin-Fuchs scalar, but we will be interested only in the
WZW sector 
of the theory, for this particular sector
contains the relevant information about the discrete group
$\Gamma$ and the cohomology of $\widetilde{\C^2/\Gamma}$.
We summarize the results in \tref{table:ALE-WZW}.

\begin{table}
\begin{center}
\begin{tabular}{||c|l||}  \hline
ALE Type &  Level of WZW\\ \hline
$A_n$ & $n-1$\\ \hline
$D_n$ & $2n-4$\\ \hline
$E_6$ & 10 \\ \hline
$E_7$ & 16 \\ \hline
$E_8$ & 28 \\ \hline
\end{tabular}
\caption{The WZW subsector of the Ooguri-Vafa conformal field theory
 description of the 
 singular non-linear sigma-model 
on ALE. \label{table:ALE-WZW}}
\end{center}
\end{table}

We now assert that
many amazing ADE-related properties of the $\widehat{SU(2)}$ WZW conformal 
field theory and the
McKay correspondence can be interpreted as consequences of the fact that the
conformal field theory description of the singularities of ALE spaces contains
the $\widehat{SU(2)}$ WZW.  That is, we argue that the WZW theory 
inherits most of the geometric information about the ALE spaces.

\index{Chiral Rings!fusion algebra}
\subsection{Fusion Algebra, Cohomology and Representation Rings}
Comparing the Table~\ref{table:ALE-WZW} with the Table~\ref{table:ADE-WZW},
we immediately see that the graphical representations of the homology intersections
of $H_2(\widetilde{\C^2/\Gamma},\Z)$
and the modular invariants of the associated
$\widehat{SU(2)}$ WZW subsector are identical.

Let us recall 
how $\widehat{SU(2)}_k$ WZW model has been historically 
related to the finite subgroups of
$SU(2)$. Meanwhile we shall recapitulate some of the key points in 
\sref{subsec:Mckay-WZW}.
The finite subgroups $\Gamma$ of $SU(2)$ 
have two infinite and one finite series.  The Algebraic McKay Correspondence
showed that the representation ring of each finite group admits a graphical 
representation such that the two infinite series have the precise $A$ and $D$
Dynkin diagrams while the finite series has the $E_{6,7,8}$ Dynkin diagrams.
Then, it was noticed that the same Dynkin diagrams classify 
the modular invariants of the $\widehat{SU(2)}_k$ WZW model, and this 
observation was interesting but there was no {\it a priori}\/ connection 
to the representation theory of finite subgroups.  It was later 
discovered that the Dynkin diagrams also encode the $\widehat{SU(2)}_k$ WZW
fusion rules or their extended versions\footnote{See \cite{CFT} for a more
complete discussion of this point.}.  Independently of the WZW models,
the Dynkin diagrams are also well-known to 
represent the homological intersection numbers on $\widetilde{\C^2/\Gamma}$, 
which are encoded the chiral ring structure of the sigma-model when $B\neq0$.
What Ooguri and Vafa have shown us is that when the $B$-field is set to zero,
the information about the chiral ring and the
discrete subgroup $\Gamma$ do not get destroyed but get transmitted to the
orbifoldized Kazama-Suzuki model which contains the $\widehat{SU(2)}_k$ WZW.

Let us demonstrate the fusion/cohomology correspondence for the
$A$-series.  Let $C_i$ be the basis of $H^2(\widetilde{\C^2/\Z_n},\Z)$ and
$Q_{ij}$ their intersection matrix inside the $A_{n-1}$ ALE space.  The
$\widehat{SU(2)}_k$ WZW at level $k=n-2$ has $k+1$ primary fields $\phi_a, 
a=0, 1,\ldots n-2$.  Then, the fusion of the fundamental field $\phi_1$ with
other primary fields 
\beq
	\phi_1 \times \phi_a = {{\cal N}_{1a}}^b\ \phi_b
\eeq
is precisely given by the intersection matrix, i.e. ${{\cal N}_{1a}}^b=Q_{ab}$.
Now, let $N_1$ be the matrix whose components are the fusion coefficients
$(N_1)_{ab}={{\cal N}_{1a}}^b$, and define
$k-1$ matrices $N_i, i=2,\ldots,k$ recursively by the following equations
\barrayn
	N_1N_1 &=& N_0 +N_2 \nonumber\\
	N_1N_2 &=& N_1+N_3\nonumber\\
	N_1N_3 &=& N_2 + N_4 \ \cdots\nonumber\\ 
	N_1N_{k-1} &=& N_{k-2} + N_{k}\nonumber\\
	N_1N_k  &=& N_{k-1} \label{eq:graph-alg}
\earrayn
where $N_0=\mbox{Id}_{(k+1)\times (k+1)}$.  That is, multiplication by $N_1$ with
$N_j$ just lists the neighboring nodes in the $A_{k+1}$ Dynkin diagram with a
sequential labeling.  Identifying the primary fields $\phi_i$ with the matrices
$N_i$, it is easy to see that the algebra of $N_i$ generated by 
the defining equations \eref{eq:graph-alg}
precisely reproduces the fusion algebra of the $\phi_i$ for the
$\widehat{SU(2)}_k$ WZW at level 
$k=n-2$.  This algebra is the aforementioned  graph algebra in conformal field 
theory.  The graph algebra has been known for many years, but what we
are proposing 
in this chapter is that the graph algebra is a 
consequence of the fact that the WZW 
contains the information about the cohomology of the corresponding ALE space.

Furthermore, recall from \sref{sec:alg-mckay} 
that the intersection matrix is identical to
the Clebsch-Gordan coefficients $m_{ij}$, ignoring the affine node.  This fact is in
accordance with the proof of Ito and Nakajima \cite{nakajima} that the
cohomology ring of $\widetilde{\C^2/\Gamma}$ is isomorphic to the representation
ring ${\cal R}(\Gamma)$  
of $\Gamma$.  At first sight, it appears that we have managed to reproduce
only a subset of Clebsch-Gordan coefficients of ${\cal R}(\Gamma)$ 
from the cohomology or equivalently the fusion
ring.  For the $A$-series, however, we can easily find all the Clebsch-Gordan 
coefficients of the irreps of $\Z_n$ from the fusion algebra by simply
relabeling the irreps and choosing a different
 self-dual 2-dimensional representation.  This is because the algebraic
McKay correspondence produces an $A_{n-1}$ Dynkin diagram 
for any self-dual 2-dimensional representation $R$ and choosing a different
$R$ amounts to relabeling the nodes with different irreps.  
The graph algebras of the $\widehat{SU(2)}_k$
WZW theory for the $D$ and $E$-series actually lead not to the fusion
algebra of the original theory but to that of the extended theories, and
these cases require further investigations.

\begin{figure}[ht]
\centerline{\psfig{figure=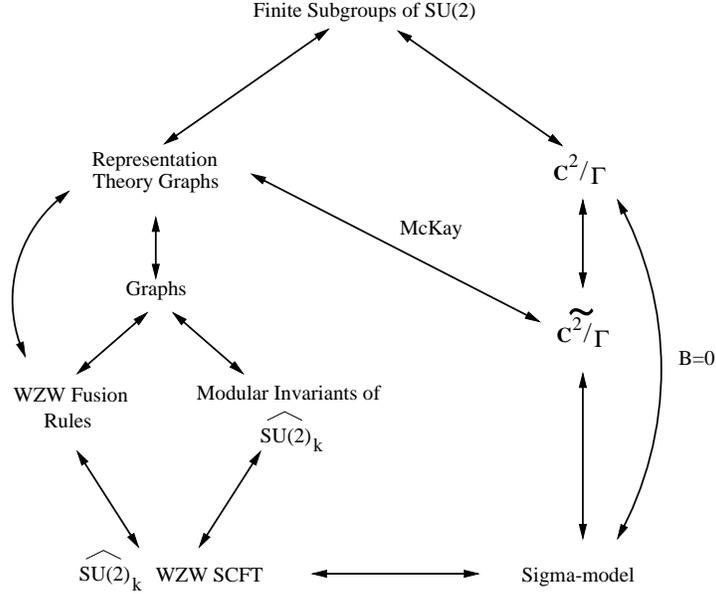,width=3.7in}}
\caption{Web of Correspondences:  \small Each finite group $\Gamma\subset SU(2)$ 
gives rise to an isolated  Gorenstein singularity as well as to its representation
ring ${\cal R}$.  The cohomology ring of the resolved manifold is
isomorphic to 
${\cal R}$.
The  $\widehat{SU(2)}_k$ WZW theory at level 
$k= \mbox{$\#$ Conjugacy classes of $\Gamma$}-2$ has a graphical
representation of its 
modular invariants and its fusion ring.  The resulting graph is
precisely the non-affine 
version of McKay's graph for $\Gamma$.  The WZW model arises as a
subsector of the  
conformal field theory description of the quotient singularity when
the $B$-field has 
been set to zero.  We further note that the three rings in the picture
are equivalent. 
\label{fig:2-dim}}
\end{figure}

String theory is thus telling us that the cohomology ring of
$\widetilde{\C^2/\Gamma}$, fusion ring of $\widehat{SU(2)}$ WZW and
the representation ring of $\Gamma$ are all equivalent.
We summarize the noted correspondences and our observations in 
Figure~\ref{fig:2-dim}.

%
\subsection{Quiver Varieties and WZW}
\index{Quivers!quiver variety}
In this subsection, we suggest how affine Lie algebras may be arising so
naturally in the study of two-dimensional quotient spaces.

Based on the previous studies of Yang-Mills instantons on ALE spaces as in
\cite{Kronheimer,KN},
Nakajima has introduced in \cite{Nakajima-quiver} the notion of a quiver
variety which is roughly a hyper-K\"{a}hler 
moduli space of representations of a quiver associated to a finite 
graph (We shall turn to quivers in the next section). There, he presents a beautiful
 geometric construction of representations of affine Lie algebras. 
In particular, he shows that when the graph is of the ADE type, 
the middle cohomology of the quiver variety is isomorphic to the weight space of
integrable highest-weight representations.  A famous example of a
quiver variety with this kind of affine Lie algebra symmetry is the
moduli space of instantons over ALE spaces.
 
In a separate paper \cite{nakajima}, Nakajima also shows that the quotient space
$\C^2/\Gamma$  admits a Hilbert scheme resolution $X$ which itself can be
identified with 
a quiver variety associated with the affine Dynkin diagram of $\Gamma$.
The analysis of \cite{Nakajima-quiver} thus seems to suggest that the second 
cohomology of the resolved space $X$ is isomorphic to the weight space of
some affine Lie algebra.  
We interpret Nakajima's work as telling physicists that the
$\widehat{SU(2)}_k$ WZW has every right to be present and
carries the geometric information about the second cohomology.
Let us demonstrate our thoughts when $\Gamma = \Z_n$.  In this case,
we have $\dim H^2 = n-1$, consisting of $n-1$ $\P^1$ blow-ups in a linear
chain.  We interpret the $H^2$ basis as furnishing a
representation  
of the $\widehat{SU(2)}_k$ WZW at level $k=n-2$, as
the basis matches the primary fields
of the WZW.  This interpretation agrees with the analysis of Ooguri and 
Vafa, but we are not certain how to reproduce the result directly from
Nakajima's work.

\subsection{T-duality and Branes}
In \cite{callan,SD,Gomis,ABKS}, the $\widehat{SU(2)}_k$ WZW theory arose in
a different 
but equivalent context of brane dynamics.  As shown in
\cite{Ooguri-Vafa}, the type IIA (IIB) string theory on an
$A_{n-1}$ ALE space is $T$-dual to
the type IIB (IIA) theory in the background of $n$ $NS5$-branes.
The world-sheet description of the
near-horizon geometry of the colliding $NS5$-branes is
 in terms of the $\widehat{SU(2)}_k$ WZW, a Feigin-Fuchs boson, and
their superpartners.  More precisely, the near-horizon geometry of $n$
$NS5$-branes is given by the WZW at level $n-2$, which is consistent with the
analysis of Ooguri and Vafa.

It was  conjectured in \cite{SD}, and further generalized in
\cite{Gomis}, that the string theory on the near horizon geometry of
the $NS5$-branes is dual to the decoupled theory on the world-volume
of the $NS5$-branes.  In this chapter, our main concern has been the
singularity structure of the ALE spaces, and  we
have  thus restricted ourselves only
to the transverse directions of the $NS5$-branes in the $T$-dual
picture.

\index{Quivers!quiver category}
\index{Quivers!ribbon category}
\section{Ribbons and Quivers at the Crux of Correspondences} \label{sec:ribbons}
There is a common theme in all the fields relevant to our observations
so far. In general we construct a theory and attempt to encode its
rules into 
some matrix, whether it be fusion matrices, Clebsch-Gordan coefficients, or
intersection numbers. Then we associate this matrix with some graph by
treating the former as the adjacency matrix of the latter and study the
properties of the original theory by analyzing the graphs\footnote{There is 
interesting work done to formalize to sub-factors and to investigate
the graphs generated \cite{Evans}.}.

Therefore there appears to be two steps in our program: firstly, we
need to study the commonalities in the minimal set of axioms in these
different fields, and secondly, we need to encode information
afforded by these axioms by certain graphical representations. It turns out
that there has been some work done in both of these steps, the first exemplified by
the so-called ribbon categories and the second, quiver categories.

\subsection{Ribbon Categories as Modular Tensor Categories}
Prominent work in the first step has been done by A. Kirillov
\cite{Kirillov} and we shall adhere to his notations. We are interested in
monoidal additive categories, in particular, we need the following:
\begin{definition}
A {\bf ribbon category} is an additive category $\cal C$ with the
following additional structures:
\begin{itemize}
\item BRAIDING: A bifunctor $\otimes:\cal C\times \cal C\to \cal C$
 along with 
 functorial associativity and commutativity isomorphisms for objects $V$
 and $W$:
\[\begin{array}{c}
a_{V_1,V_2,V_3}:(V_1\otimes V_2)\otimes V_3\to V_1\otimes(V_2\otimes V_3),\\
\check R_{V,W}:V\otimes W\to W\otimes V;
\end{array}\]
\item MONOIDALITY: A unit object 
${\bf 1} \in \mbox{Obj }\cal C$ along with isomorphisms
${\bf 1}\otimes V\to V, V\otimes {\bf 1}\to V$;
\item RIGIDITY of duals: for every object $V$ we have a (left) dual
$V^*$ and homomorphisms
\[\begin{array}{c}
e_V: V^*\otimes V\to {\bf 1}, \\
i_V:{\bf 1}\to V\otimes V^*;
\end{array}\]
\item BALANCING: functorial isomorphisms $\theta_V:V\to V$, 
        satisfying the compatibility condition
\[
\theta_{V\otimes W}= \check R_{W,V} \check R_{V, W}(\theta_V\otimes
\theta_W).
\]
\end{itemize}
\end{definition}
Of course we see that all the relevant rings in
Figure~\ref{fig:mother} fall under
such a category.  Namely, we see that the representation rings of
finite groups, chiral rings of non-linear $\sigma$-models, Groethendieck
rings of exceptional divisors or fusion rings of WZW, together with their
associated tensor products, are all different realizations of
a ribbon category \footnote{Of course they may possess additional
structures, e.g., these rings are all finite. We shall later see how
finiteness becomes an important constraint when going to step two.}.
This fact is perhaps obvious from the point of view of orbifold string
theory, in which the fusion ring naturally satisfies the
representation algebra of the finite group and the WZW arises as a
singular limit of the vanishing $B$-field.
The ingredients of each of these rings, respectively the irreps, chiral
operators and cohomology elements, thus manifest as the objects in $\cal C$.
Moreover, the arrows of \fref{fig:mother}, loosely speaking, 
become functors among these various representations of $\cal C$ 
whereby making our central diagram a (meta)graph associated to $\cal C$.
What this means is that as far as the ribbon category is concerned,
all of these theories discussed so far are axiomatically identical. 
Hence indeed any underlying correspondences will be natural.

What if we impose further constraints on $\cal C$?
\begin{definition} We define $\cal C$ to be {\bf semisimple} if
\begin{itemize}
\item It is defined over some field $\mathbb{K}$ and all the spaces of
        homomorphisms are finite-dimensional vector spaces over $mathbb{K}$;
\item Isomorphism classes of simple objects $X_i$ in $\cal C$
        are indexed by elements $i$ of some
        set $I$. This implies involution ${}^*:I\to I$ such that
        $X_i^*\simeq X_{i^*}$ (in particular, $0^*=0$);
\item ``Schur's Lemma'': $\hom (X_i, X_j) = \mathbb{K}\delta_{ij}$;
\item Complete Finite Reducibility: $\forall$ $V\in \mbox{Obj }\cal C$, 
        $V=\bigoplus\limits_{i\in I} N_i X_i,$ such that the sum is finite,
        i.e., almost all $N_i\in \Z_+$ are zero.
\end{itemize}
\end{definition}
Clearly we see that in fact our objects, whether they be WZW fields or
finite group irreps, actually live in a semisimple ribbon category.
It turns out that semisimplicity is enough to allow us to define composition 
coefficients of the ``Clebsch-Gordan'' type:
\[
X_i \oplus X_j = \bigoplus N_{ij}^k X_k,
\]
which are central to our discussion.

Let us introduce one more concept, namely the
matrix $s_{ij}$ mapping $X_i \to X_j$ represented graphically by the
simple ribbon 
tangle, i.e., a link of 2 closed directed cycles of maps from $X_i$ and $X_j$
respectively into themselves. The remarkable fact is that imposing that
\begin{itemize}
\item $s_{ij}$ be invertible and that
\item $\cal C$ have only a finite number of simple objects
        (i.e., the set $I$ introduced above is finite) 
\end{itemize}
naturally gives rise to modular properties. We define such semisimple ribbon
category equipped with these two more axioms as a {\bf Modular Tensor
Category}. 
If we define the matrix $t_{ij} = \delta_{ij} \theta_i$ with $\theta_i$ being
the functorial isomorphism introduced in the balancing axiom for $\cal C$, the
a key result is the following  \cite{Kirillov}:
\begin{theorem}
In the modular tensor category $\cal C$, the matrices $s$ and $t$ generate
precisely the modular group $SL(2,\Z)$.
\end{theorem}
Kirillov remarks in \cite{Kirillov} that it might seem mysterious that modular
properties automatically arise in the study of tensor categories and argues
in two ways why this may be so. Firstly, a projective action of $SL(2,\Z)$ 
may be defined for certain objects in $\cal C$. This is essentially
the construction 
of Moore and Seiberg \cite{MS} when they have found new modular invariants for
WZW, showing how WZW primary operators are objects in $\cal C$.
Secondly, he points out that geometrically one can associate a
topological quantum 
field theory (TQFT) to each tensor category, whereby the mapping
class group of the 
Riemann surface associated to the TQFT gives rise to the modular group.
If the theories in \fref{fig:mother} are indeed providing different
but equivalent
realizations of $\cal C$, we may be able to trace the 
origin of the $SL(2,\Z)$ action on the category to the WZW
modular invariant partition functions.  That is,  it seems that 
in two dimensions the ADE scheme, which also arises in other
representations of $\cal C$, naturally classifies some kind of 
modular invariants.  In a generic realization of the modular tensor
category, it may be difficult to identify such modular invariants, but
they are easily identified as the invariant partition functions 
in the WZW theories.

\subsection{Quiver Categories}
{Quivers!quiver category}
We now move onto the second step. Axiomatic studies of the encoding procedure 
(at least a version thereof)
have been done even before McKay's result. In fact, in 1972, Gabriel has
noticed that categorical studies of quivers lead to $A$-$D$-$E$-type
classifications \cite{Gabriel}.
\begin{definition}
We define the {\bf quiver category} ${\cal L}(\Gamma,\Lambda)$,
for a finite connected graph $\Gamma$ with orientation $\Lambda$, vertices
$\Gamma_0$ and edges $\Gamma_1$ as follows:
The objects in this category are any
collection $(V,f)$ of spaces $V_{\alpha}, \alpha \in \Gamma_0$ and mappings 
$f_{l}, l \in \Gamma_1$. The morphisms are 
$\phi : (V,f) \rightarrow (V',f') $
a collection of linear mappings $\phi_{\alpha}  : V_{\alpha}
\rightarrow V'_{\alpha}$ compatible with $f$ by 
$\phi_{e(l)}f_l = f'_{l}\phi_{b(l)}$ where $b(l)$ and $e(l)$ are the beginning
and the ending nodes of the directed edge $l$.
\end{definition}
Finally we define decomposability in the usual sense that
\begin{definition}
The object $(V,f)$ is {\bf indecomposable} iff there do not exist objects
$(V_1,f_1), (V_2,f_2) \in {\cal L}(\Gamma,\Lambda)$ such that
$V = V_1 \oplus V_2$ and $f = f_1 \oplus f_2$.
\end{definition}
Under these premises we have the remarkable result:
\begin{theorem}[{\sc Gabriel-Tits}] The graph
$\Gamma$ in ${\cal L}(\Gamma,\Lambda)$ coincides with one of the
graphs $A_n,D_n,E_{6,7,8}$, if and only if there are only finitely many
non-isomorphic indecomposable objects in the quiver category.
\end{theorem}
By this result, we can argue that the theories, which we have seen
to be different representations of the ribbon category $\cal C$ and
which all
have ADE classifications in two dimensions, each must in fact be
realizable as a finite quiver category $\cal L$ in dimension
two. Conversely, the finite quiver category has representations as
these theories in 2-dimensions. To formalize, we state
\begin{proposition}
In two dimensions, finite group representation ring, 
WZW fusion ring, Gorenstein
cohomology ring, and non-linear $\sigma$-model chiral ring, as
representations of a ribbon category $\cal C$, can be mapped to a
finite quiver category $\cal C$. In particular the ``Clebsch-Gordan''
coefficients ${\cal N}_{ij}^k$ of $\cal C$ realize as 
adjacency matrices of graphs in $\cal L$
\footnote{Here the graphs are ADE Dynkin diagrams. For higher
dimension we propose that there still is a mapping, though perhaps not
to a {\it finite} quiver category.}.
\end{proposition}
Now $\cal L$ has recently been given a concrete realization 
by the work of Douglas and Moore \ \cite{DM}, in the
context of investigating string theory on orbifolds. The objects in the
quiver category have found representations in the resulting ${\cal N}=2$
Super Yang-Mills theory. The modules $V$ (nodes) manifest themselves
as gauge groups
arising from the vector multiplet and the mappings $f$ (edges which in 
this case are really bidirectional arrows), as bifundamental matter.
This is the arrow from graph theory to string orbifold theory in the
center of \fref{fig:mother}.
Therefore it is not surprising that an ADE type of result in encoding
the physical content of the theory has been
obtained. Furthermore, attempts at brane configurations to construct these 
theories are well under way (e.g. \cite{Kapustin}).

Now, what makes ADE and two dimensions special?
A proof of the theorem due to Tits \cite{Gabriel} 
rests on the fact that the
problem can essentially be reduced to a Diophantine inequality in the
number of 
nodes and edges of $\Gamma$, of the general type:
\[
\sum\limits_i\frac{1}{p_i} \ge c
\]
where $c$ is some constant and $\left\{p_i\right\}$ is a set of integers
characterizing the problem at hand.
This inequality has a long history in mathematics \cite{Humphereys}.
In our context, we recall that the uniqueness of the five perfect solids
in $\R^3$ (and hence the discrete subgroups of $SU(2)$) 
relies essentially on the equation $1/p + 1/q \ge 1/2$ having only
5 pairs of integer solutions. Moreover we recall that Dynkin's classification
theorem of the simple Lie algebras depended on integer solutions of
$1/p + 1/q + 1/r \ge 1$.

Since Gabriel's theorem is so restrictive, extensions thereto have been done
to relax certain assumptions (e.g., see \cite{Nazarova}). This will
hopefully give us give more graphs and in particular those appearing
in finite group, WZW, orbifold theories or non-linear $\sigma$-models
at higher dimensions.
A vital step in the proof is that a certain quadratic form over the
$\mathbb{Q}$-module  
of indices on the nodes (effectively the Dynkin labels) 
must be positive-definite.
It was noted that if this condition is relaxed to positive semi-definity, then
$\Gamma$ would include the affine cases 
$\widehat{A},\widehat{D},\widehat{E}$ as well.
Indeed we hope that further relaxations of the condition would admit more graphs,
in particular those drawn for the $SU(3)$ subgroups.
This inclusion on the one hand would relate quiver graphs
to Gorenstein singularities in dimension three due to the 
link to string orbifolds\footnote{In this case we get ${\cal N} = 1$ 
Super-Yang-Mills theory in 4 dimension.} and on the other 
hand to the WZW graph algebras by the conjecture in
Chap. \ref{chap:9811183}.
Works in this direction are under way. 
It has been recently suggested that since the discrete subgroups 
of SU(4,5,6,7) have also been classified \cite{Gannon2}, graphs for
these could be 
constructed and possibly be matched to the modular invariants corresponding
to $\widehat{SU(n)}$ for $n=4,..,7$ respectively. Moreover, proposals
for unified 
schemes for the modular invariants by considering orbifolds by abelian $\Gamma$
in SU(2,3,..,6) have been made in \cite{AA}.

Let us summarize what we have found. We see that the representation ring
of finite groups with its associated $(\otimes,\oplus)$, the chiral ring of
nonlinear $\sigma$-model with its $(\otimes, \oplus)$, the fusion ring of
the WZW model with its $(\times,\oplus)$ and the Groethendieck ring of resolved
Gorenstein singularities with it $(\otimes,\oplus)$ manifest
themselves as
different realizations of a semisimple ribbon category $\cal C$. 
Furthermore, the requirement of finiteness and an
invertible $s$-matrix makes $\cal C$ into a  modular tensor category. The
ADE schemes in two dimensions, 
if they arise in one representation of $\cal C$, might naturally
appear in another. Furthermore, the quiver category $\cal L$ has a physical
realization as bifundamentals and gauge groups of SUSY Yang-Mills theories. The
mapping of the Clebsch-Gordan coefficients in $\cal C$ to the quivers in
$\cal L$ is therefore a natural origin for the graphical representations of the
diverse theories that are objects in $\cal C$.

\section{Conjectures}\label{sec:conj}
\index{McKay Correspondence!in string theory}
\begin{figure}[ht]
\centerline{\psfig{figure=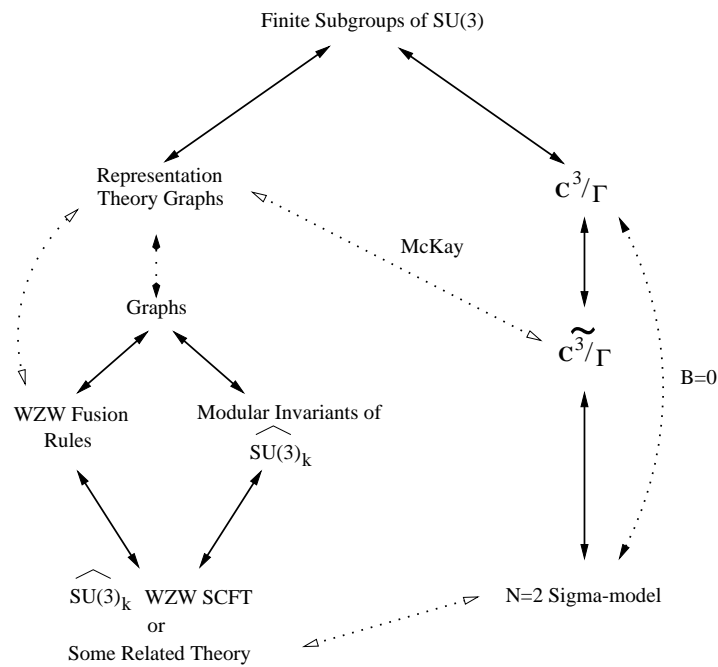,width=3.7in}}
\caption{Web of Conjectures: Recently, the graphs from the
representation theory 
side were constructed and were noted to resemble
those on WZW $\widehat{SU(3)}_k$ side (Chap. \ref{chap:9811183}).  
The solid lines have
been sufficiently 
well-established
while the dotted lines are either conjectural or ill-defined. 
\label{fig:3-dim}}
\end{figure}
We have seen  that there exists a remarkably coherent picture of
inter-relations  
in two dimensions among many different branches of mathematics and physics.
The organizing principle appears to be the mathematical 
theory of quivers and ribbon category, while the crucial bridge
between mathematics 
and physics is the conformal field theory description of the Gorenstein
singularities provided by the orbifoldized coset construction.

Surprisingly, similar features have been noted in three dimensions.
The Clebsch-Gordan coefficients for the tensor product of
irreducible representations for all discrete subgroups of $SU(3)$ were
computed 
in \cite{Muto,Greene} and Chap.~\ref{chap:9811183},
and a possible correspondence was noted, and conjectured for 
$n\geq 3$,
between the resulting Dynkin-like diagrams and the graphic representations of 
the fusion rules and modular invariants of $\widehat{SU(3)}_k$
WZW models.  Furthermore, 
as discussed previously, the Geometric McKay Correspondence between
the representation ring of the abelian
discrete subgroups $\Gamma\subset SU(3)$
and the cohomology ring of $\widetilde{\C^3/\Gamma}$ has been proved in
\cite{nakajima}. 
Hence, the situation in 3-dimensions as seen in \fref{fig:3-dim}
closely resembles that in 2-dimensions.

Now, one naturally inquires:

\begin{quotation}
\it
Are there graphical representations of the fusion rules and modular
invariants of the 
$\widehat{SU(n)}_k$ WZW model or some related theory
that contain the Clebsch-Gordan 
coefficients
for the representations of $\Gamma\subset SU(n)$?  And, in turn, are the 
Clebsch-Gordan coefficients related to  the (co)-homological 
intersections on the resolved geometry $\widetilde{\C^n/\Gamma}$ that
are contained in the chiral ring of the $N=2 \ \sigma$-model on $\C^n/\Gamma$ 
with a non-vanishing $B$-field?
{\bf Most importantly, what do 
these correspondences tell us about the two conformal field theories and their
singular limits?}
\end{quotation}
\noindent
As physicists, we believe that the
McKay correspondence and the classification of certain modular invariants in
terms of finite subgroups are
consequences of orbifolding and of some underlying 
quantum equivalence of the associated conformal field theories.

We thus believe that a picture similar to that seen in this chapter
for 2-dimensions 
persists in higher dimensions and 
 conjecture that there exists a conformal field theory description of the
Gorenstein singularities in higher dimensions.
If such a theory can be found, then it would explain the observation made
in Chap.~\ref{chap:9811183} of the 
resemblance of the graphical representations of
the representation 
ring of the finite subgroups of $SU(3)$ and the modular invariants of the 
$\widehat{SU(3)}_k$ WZW.  We have checked that the correspondence, if
any, between the 
finite subgroups of $SU(3)$ and the $\widehat{SU(3)}_k$ WZW theory is
not one-to-one. 
For example, the number of primary fields generically
does not match the number of conjugacy classes
of the discrete subgroups.  It has been observed in
Chap.~\ref{chap:9811183}, 
however, that some of 
the representation graphs appear to be subgraphs of the graphs encoding
the modular invariants.
We hope that the present chapter serves as a motivation for finding the 
correct conformal field
theory description in three dimensions which would tell us how to
``project'' the modular invariant graphs to retrieve the 
representation graphs of the finite graphs. 

Based on the above discussions, we summarize our speculations,
relating geometry, 
generalizations of the ADE classifications, 
representation theory, and string theory  in \fref{fig:3-dim}.  

\subsection{Relevance of Toric Geometry}
It is interesting to note that the toric resolution of
certain Gorenstein singularities also naturally admits graphical
representations of fans.  In fact, the exceptional divisors in 
the Geometric McKay Correspondence
for $\Gamma=\Z_n\subset SU(2)$ in 2-dimensions can be easily seen as
the vertices of new cones in the toric resolution, and these vertices
precisely form the $A_{n-1}$ Dynkin digram.   Thus, at least for the
abelian case in 2-dimensions, the McKay correspondence and the
classification of 
$\widehat{SU(2)}$ modular invariants seem to be most naturally
connected to geometry as toric diagrams of the resolved manifolds
$\widetilde{C^2/\Gamma}$.

Surprisingly---perhaps not so much so in retrospect---we 
have noticed a similar pattern in 3-dimensions.  That is, the toric
resolution diagrams of $\C^3/\Z_n\times \Z_n$ singularities reproduce
the graphs that classify the ${\cal A}$-type modular invariants of the
$\widehat{SU(3)}_k$ WZW models.  For which $k$?  It has been
previously observed in \cite{lattice} that there seems to be a
correspondence, up to some truncation, between the subgroups 
$\Z_n\times \Z_n\subset SU(3)$ and the ${\cal A}$-type 
$\widehat{SU(3)}_{n-1}$ modular invariants, which do appear as
subgraphs of the $\widetilde{\frac{\C^3}{\Z_n\times \Z_n}}$ toric diagrams.
On the other hand, a precise formulation of the correspondence with
geometry and the conformal field theory description of  Gorenstein
singularities still remains as an unsolved problem and will be
presented elsewhere \cite{Song}.

\section{Conclusion} \label{sec:conclusion}
Inspired by the ubiquity of ADE classification and prompted
by an observation of a mysterious 
relation between finite groups and WZW models,
we have proposed a possible unifying scheme.
Complex and intricate
webs of connections have been presented, the particulars of which
have either been hinted at
by collective works in the past few decades in mathematics and physics
or are conjectured to exist by arguments in this chapter.
These webs  include the 
McKay correspondences of various types 
as special cases and relate such seemingly disparate subjects as
finite group representation theory, graph theory, string orbifold
theory and sigma models, as well as
conformal field theory descriptions of Gorenstein singularities.
We note that the integrability of the theories that we are 
considering may play a role in understanding the deeper connections.

This chapter catalogs many observations which have been
put forth in the mathematics and physics literature and presents
them from a unified perspective. 
Many existing results and conjectures have been phrased under a new light.
We can summarize the contents of this chapter as follows:
\begin{enumerate}
\item In two dimensions, all of the
	correspondences mysteriously fall into an ADE type.
	We have provided, via \fref{fig:mother},
	a possible setting how these mysteries might arise
	naturally. Moreover, we have pointed out how axiomatic works
	done by 
	category theorists may demystify some of these links.  Namely,
	we have noted that the relevant rings of the theories can be
	mapped to 
	the quiver category. 
\item   We have also discussed the possible role played by the modular
 	tensor category in our picture, in which the modular invariants
	arise very naturally.  Together with the study of the
	 quiver category and
 	quiver variety, the ribbon category seems to provide the
 	reasons for the emergence of affine Lie algebra symmetry and
 	the ADE classification of the modular invariants.
\item	We propose the validity of our program to higher dimensions, 
	where the picture is far less clear since
	there are no ADE schemes, though some hints of generalized graphs
	have appeared.
\item  There are three standing conjectures:
\begin{itemize}\item  We propose that there exists a conformal field theory
description of the Gorenstein singularities in dimensions greater than
two.
\item  As noted in Chap.~\ref{chap:9811183}, we conjecture that the modular
invariants and the fusion rings of the $\widehat{SU(n)}, n>2$ WZW,
or their generalizations, may be related to the discrete
subgroups of the $SU(n)$.
\item  Then, there is the mathematicians' conjecture that there exits
a McKay correspondence between the cohomology ring
$H^*(\widetilde{C^n/\Gamma},\Z)$ and the representation ring of
$\Gamma$, for finite subgroup $\Gamma\subset SU(n)$.
\end{itemize}
We have combined these conjectures into a web so that proving one of
them would help proving the others.
\end{enumerate}
We hope
that \fref{fig:mother} essentially commutes and that the standing
conjectures 
represented by certain arrows therewithin may be solved by investigating
the other arrows.
In this way,
physics may provide us with a possible method of attack and
explanation for 
McKay's correspondence and many other related issues, and likewise
mathematical structures may help to clarify and rigorize some
observations made from string theory.

It is the purpose of this writing to
inform the physics and mathematics community of a possibly new
direction of research which could harmonize ostensibly different and
diverse branches of mathematics and physics into a unified picture.
\chapter{Orbifolds III: $SU(4)$}
\label{chap:9905212}
\index{Orbifolds}
\index{Finite Groups!$SU(4)$ subgroups}
\section*{\center{{ Synopsis}}}
Whereas chapter \ref{chap:9811183} studied the $SU(2)$ and $SU(3)$
orbifolds as local Calabi-Yau surfaces and threefolds, we here
present, 
in modern notation, the classification of 
the discrete finite
subgroups of $SU(4)$ as well as the character tables 
for the exceptional cases
thereof (Cf. http://pierre.mit.edu/$\sim$yhe/su4.ct).

We hope this catalogue will be useful to works on string orbifold
theories on Calabi-Yau fourfolds,
quiver theories, WZW modular invariants, Gorenstein resolutions, 
nonlinear sigma-models
as well as the inter-connections among them proposed in Chapter
\ref{chap:9903056}
\cite{9905212}.
\section{Introduction}
It is well known that the discrete finite subgroups of $SL(n=2,3;\C)$ have been
completely classified; works related to string orbifold theories and quiver
theories have of late used these results (see for example
\cite{9903056,DM,LNV,KS,Muto,Greene} and Chap.~\ref{chap:9811183}
as well as
references therein). Conjectures regarding higher $n$ have been raised and works
toward finite subgroups of $SU(4)$ are under way.
Recent works by physicists and mathematicians alike further beckon for a 
classification of the groups, conveniently presented,
in the case of $SU(4)$ \cite{Vafa-Frampton,Gannon}.
Compounded thereupon is the disparity of language under which the groups are discussed:
the classification problem in the past decades has chiefly been of interest to either
theoretical chemists or to pure mathematicians, the former of whom disguise them in
Bravais crystallographic notation (e.g.\ \cite{Chem}) while the latter abstract them in
fields of finite characteristic (e.g.\ \cite{Math}). Subsequently, there is
a need within the string theory community for a list of the finite subgroups of
$SU(4)$ tabulated in our standard nomenclature, complete with the generators
and some brief but not overly-indulgent digression on their properties.

The motivations for this need are manifold. There has recently been a host
of four dimensional finite gauge theories constructed by placing D3 branes
on orbifold singularities \cite{DM,LNV,KS}; brane setups have also been
achieved for some of the groups \cite{HU,HZ}. 
In particular, a theory with ${\cal N}=2,1,0$
supercharges respectively
is obtained from a $\C^N/\{\Gamma \subset SU(n=2,3,4)\}$ singularity with $N=2,3$
(see Chap.~\ref{chap:9811183},
\cite{DM,LNV,KS} and references therein). Now as
mentioned above $n=2,3$ have 
been discussed, and $n=4$ has yet to be fully attacked. This last case is of particular
interest because it gives rise to an ${\cal N}=0$, non-supersymmetric
theory. On the one hand these orbifold theories provide interesting
string backgrounds for checks on the AdS/CFT Correspondence \cite{BKV,AGMOO}.
On the other hand, toric descriptions for the Abelian cases of the
canonical Gorenstein singularities have been treated while the non-Abelian
still remain elusive \cite{Agata,Horizon}.
Moreover, the quiver theories arising from these string orbifold
theories (or equivalently, representation rings of finite subgroups of
$SU(n)$) have been hinted to be related to modular invariants of
$\widehat{su(n)}$-WZW models (or equivalently, affine characters of
$\widehat{su(n)}$) for arbitrary $n$ \cite{DiFrancesco}, and a generalised
McKay Correspondence, which would also relate non-linear sigma models, 
 has been suggested to provide a reason \cite{9903056}. Therefore a need for
the discrete subgroups of $SU(4)$ arises in all these areas.

Indeed the work has been done by Blichfeldt \cite{Blichfeldt} in 1917, or at least
 all the exceptional cases, though in an obviously outdated parlance and moreover with many   
infinite series being ``left to the reader as an exercise.'' It is therefore the
intent of the ensuing monograph to present the discrete subgroups $\Gamma$ of $SL(4,\C)$
in a concise fashion, hoping it to be of use to impending work, particularly 
non-supersymmetric conformal gauge theories from branes on orbifolds, resolution of
Gorenstein singularities in higher dimension, as well as
$\widehat{su(4)}$-WZW models.

\section*{Nomenclature}
Unless otherwise stated we shall adhere to the convention that
$\Gamma$ refers to a discrete subgroup of $SU(n)$ (i.e., a finite collineation group),
that $<x_1,..,x_n>$ is a finite group generated by $\{x_1,..,x_n\}$, that
$H \triangleleft G$ means $H$ is a normal subgroup of $G$, that
$S_n$ and $A_n$ are respectively the symmetric and alternating permutation groups on
$n$ elements, and that placing $*$ next to a group signifies that it belongs to
$SU(4) \subset SL(4;\C)$.

\section{Preliminary Definitions}
Let $\Gamma$ be a finite discrete subgroup of the general linear group, i.e.,
$\Gamma \subset GL(n,\C)$.
From a mathematical perspective, quotient varieties of the
form $\C^n / \Gamma$ may be constructed and by the theorem of Khinich and Watanabe
\cite{Khinich,Yau}, the quotient is Gorenstein\footnote{That is, if there 
exists a nowhere-vanishing holomorphic $n$-form. These varieties thus provide
local models of Calabi-Yau manifolds and are recently of great interest.}
if and only if 
$\Gamma$ is in fact in $SL(n,\C)$. Therefore we would like to focus on the
discrete subgroups of linear transformations {\it up to linear equivalence}, 
which are what has been dubbed in the old literature as 
{\bf finite collineation groups} \cite{Blichfeldt}. 
From a physics perspective, discrete subgroups of $SU(n) \subset SL(n;\C)$
have been subject to investigation in the early days of particle phenomenology 
\cite{Fairbairn} and have lately been of renewed interest in string theory, 
especially in the context of orbifolds (see for example
Chap.~\ref{chap:9811183} and
\cite{9903056,DM,LNV,KS,Vafa-Frampton,Gannon}).

There are some standard categorisations of finite collineation 
groups \cite{Blichfeldt,Yau}. They
first fall under the division of transitivity and intransitivity as follows:
\begin{definition}
If the $n$ variables upon which $\Gamma$ acts as a linear transformation
can be separated into 2 or more sets either directly or after a change of
variables, such that the variables of each set are transformed into linear
functions only of themselves, then $\Gamma$ is called {\bf Intransitive}; it is
called Transitive otherwise.
\end{definition}

The transitive $\Gamma$ can be further divided into the primitive and 
imprimitive cases:
\begin{definition}
If for the transitive $\Gamma$ the variables may be separated\footnote{
Again, either directly or after a change of variables.} into 2 or more
sets such that the variables of each are transformed into linear functions
of only those in any set according to the separation (either the same or
different), then $\Gamma$ is called {\bf Imprimitive}; it is called 
Primitive otherwise.
\end{definition}

Therefore in the  matrix representation of the groups, 
we may na\"{\i}vely construe
intransitivity as being block-diagonalisable and imprimitivity as being
block off-diagonalisable, whereby making primitive groups generically having
no {\it a priori} zero entries. We give examples of an intransitive,
a (transitive) imprimitive and a (transitive) primitive group, 
in their matrix forms, as follows:
\[
\begin{array}{ccc}
\left(\matrix{	\times &  \times &  0 &  0\\
		\times &  \times &  0 &  0 \\ 
		0 &  0 &  \times &  \times \\
		0 &  0 &  \times &  \times}
\right)
&
\left(\matrix{	0 &  0 &  \times &  \times \\
		0 &  0 &  \times &  \times \\
		\times &  \times &  0 &  0 \\
		\times &  \times &  0 &  0}
\right)
&
\left(\matrix{	\times &  \times &  \times &  \times \\
		\times &  \times &  \times &  \times \\
		\times &  \times &  \times &  \times \\
		\times &  \times &  \times &  \times}
\right)
\\
$Intransitive$	&	$Imprimitive$ 	& $Primitive$ \\
& \multicolumn{2}{c}{$Transitive$} \\
\end{array}
\]
\index{Finite Groups!colineation}
Let us diagrammatically summarise all these inter-relations as is done in \cite{Yau}:
\[
\Gamma \left\{\begin{array}{l}
	$Intransitive$ \\
	$Transitive$ \left\{\begin{array}{l}
		$Imprimitive$\\
		$Primitive$ \left\{\begin{array}{l}
			$Simple$ \\
			$Having Normal Primitive Subgroups$ \\
			$Having Normal Intransitive Subgroups$ \\
			$Having Normal Imprimitive Subgroups$ \\
		\end{array} \right.
	\end{array}  \right. 
\end{array}  \right.
\]

In some sense the primitive groups are the fundamental building 
blocks and
pose as the most difficult to be classified. It is those primitive groups
that Blichfeldt presented, as linear transformations, in \cite{Blichfeldt}.
These groups are what we might call {\it exceptionals} in the sense that they 
do not fall into infinite series, in analogy to the $E_{6,7,8}$ groups of
$SU(2)$. We present them as well as their sub-classifications first. Thereafter
we shall list the imprimitive and intransitives, which give rise to a host of
infinite series of groups, in analogy to the $A_n$ and $D_n$ of $SU(2)$.

Let us take a final digression to clarify the so-called 
{\bf Jordan Notation}, which is the
symbol $\phi$ commonly used in finite group theory. A linear
group $\Gamma$ often has its order
denoted as $|\Gamma| = g\phi$ for positive integers $g$ and $\phi$; 
the $\phi$ signifies the order of the subgroup of homotheties, or those
multiples of the identity which together form the center of the $SL(n;\C)$.
We know that $SU(n) \subset SL(n;\C)$, so a subgroup of the latter is not
necessarily that of the former. In the case of $SL(n=2,3;\C)$,
the situation is
simple\footnote{See \cite{Fairbairn} for a discussion on this point.}:
the finite subgroups belonged either to (A) $SU(n=2,3)$, or to (B) the 
center-modded\footnote{For $n=2$, this our familiar $SU(2)/\Z_2 \cong SO(3)$.}
$SU(n=2,3)/\Z_{2,3}$, or (C) to both. Of course a group with order $g$ 
in type (B) would have a natural lifting to type (A) and become a group
of order $g$ multiplied by $|\Z_2|=2$ or $|\Z_3=3|$ respectively, 
which is now a finite subgroup of the full $SU(2)$ or $SU(3)$, implying that
the Jordan $\phi$ is 2 or 3 respectively.

For the case at hand, the situation is slightly more complicated since 4 is
not a prime. Therefore $\phi$ can be either 2 or 4 depending how one lifts
with respect to the relation $SU(4)/\Z_2\times\Z_2 \cong SO(6)$ and we lose
a good discriminant of whether or not $\Gamma$ is in the full $SU(4)$. To
this end we have explicitly verified the unitarity condition for the group
elements and will place a star ($*$) next to those following groups which indeed
are in the full $SU(4)$. Moreover, from the viewpoint of string orbifold
theories which study for example the fermionic and bosonic matter content
of the resulting Yang-Mills theory, one naturally takes interest in
$Spin(6)$, or the full $\Z_2 \times \Z_2$ cover of $SO(6)$ which admits
spinor representations; for these we shall
look in particular at the groups that have $\phi = 4$ in the Jordan notation,
as will be indicated in the tables below.

\section{The Discrete Finite Subgroups of $SL(4;\C)$}
We shall henceforth let $\Gamma$ denote a finite subgroup of $SL(4;\C)$
unless otherwise stated.
\subsection{Primitive Subgroups}
There are in all 30 types of primitive cases for $\Gamma$. 
First we define the constants $w = e^{\frac{2\pi i}{3}}$, 
$\beta = e^{\frac{2\pi i}{7}}$,
$p = \beta + \beta^2 +\beta^4$, $q = \beta^3 + \beta^5 +\beta^6$,
$s = \beta^2 + \beta^5$, $t = \beta^3 + \beta^4$, and $u = \beta + \beta^6.$ 
Furthermore
we shall adhere to some standard notation and denote the permutation 
and the alternating permutation group on $n$ elements 
respectively as $S_n$ and $A_n$. Moreover, in what follows we shall use the
function $Lift$ to mean the lifting by (perhaps a subgroup) of the Abelian center
$C$ according to the exact sequence
$
\begin{array}{ccccccccc}
0 & \rightarrow & C & \rightarrow & SU(4) & \rightarrow & SU(4)/C & \rightarrow
& 0. \\
\end{array}
$

We present the relevant matrix generators as we proceed:
{\small
\[
\begin{array}{c}
F_1 = \left(
\matrix{1 & 0 & 0 & 0 \\ 
	0 & 1 & 0 & 0 \\
	0 & 0 & w & 0 \\
	0 & 0 & 0 & w^2 \\}
\right)

F_2 = \frac{1}{\sqrt{3}}\left(
\matrix{1 & 0 & 0 & \sqrt{2} \\ 
	0 & -1 & \sqrt{2} & 0 \\
	0 & \sqrt{2} & 1 & 0 \\
	\sqrt{2} & 0 & 0 & -1 \\}
\right)

F_3 = \left(
\matrix{\frac{\sqrt{3}}{2} & \frac{1}{2} & 0 & 0 \\ 
	\frac{1}{2} & -\frac{\sqrt{3}}{2} & 0 & 0 \\
	0 & 0 & 0 & 1 \\
	0 & 0 & 1 & 0 \\}
\right)

\\ \\

F'_2 = \frac{1}{3}\left(
\matrix{3 & 0 & 0 & 0 \\ 
	0 & -1 & 2 & 2 \\
	0 & 2 & -1 & 2 \\
	0 & 2 & 2 & -1 \\}
\right)

F'_3 = \frac{1}{4}\left(
\matrix{-1 & \sqrt{15} & 0 & 0 \\ 
	\sqrt{15} & 1 & 0 & 0 \\
	0 & 0 & 0 & 4 \\
	0 & 0 & 4 & 0 \\}
\right)

F_4 = \left(
\matrix{0 & 1 & 0 & 0 \\ 
	1 & 0 & 0 & 0 \\
	0 & 0 & 0 & -1 \\
	0 & 0 & -1 & 0 \\}
\right)

\\ \\

S = \left(
\matrix{1 & 0 & 0 & 0 \\ 
	0 & \beta & 0 & 0 \\
	0 & 0 & \beta^4 & 0 \\
	0 & 0 & 0 & \beta^2 \\}
\right)

T = \left(
\matrix{1 & 0 & 0 & 0 \\ 
	0 & 0 & 1 & 0 \\
	0 & 0 & 0 & 1 \\
	0 & 1 & 0 & 0 \\}
\right)

W = \frac{1}{i\sqrt{7}}\left(
\matrix{p^2 & 1 & 1 & 1 \\ 
	1 & -q & -p & -p \\
	1 & -p & -q & -p \\
	1 & -p & -p & -q \\}
\right)

\\ \\

R = \frac{1}{\sqrt{7}}\left(
\matrix{1 & 1 & 1 & 1 \\ 
	2 & s & t & u \\
	2 & t & u & s \\
	2 & u & s & t \\}
\right)

C = \left(
\matrix{1 & 0 & 0 & 0 \\ 
	0 & 1 & 0 & 0 \\
	0 & 0 & w & 0 \\
	0 & 0 & 0 & w^2 \\}
\right)

D = \left(
\matrix{w & 0 & 0 & 0 \\ 
	0 & w & 0 & 0 \\
	0 & 0 & w & 0 \\
	0 & 0 & 0 & 1 \\}
\right)

\\ \\

V = \frac{1}{i\sqrt{3}}\left(
\matrix{i\sqrt{3} & 0 & 0 & 0 \\ 
	0 & 1 & 1 & 1 \\
	0 & 1 & w & w^2 \\
	0 & 1 & w^2 & w \\}
\right)

F = \left(
\matrix{0 & 0 & -1 & 0 \\ 
	0 & 1 & 0 & 0 \\
	-1 & 0 & 0 & 0 \\
	0 & 0 & 0 & -1 \\}
\right)

\end{array}
\]
}
We see that all these matrix generators are unitary except $R$.

\subsubsection{Primitive Simple Groups}
There are 6 groups of this most fundamental type:
\[
\begin{array}{|c|c|c|c|}
\hline
$Group$	& $Order$	& $Generators$ 	& $Remarks$ \\
\hline \hline
$I$*	& 60\times 4	& F_1,F_2,F_3	& Lift(A_5)\\
$II$*	& 60		& F_1,F'_2,F'_3 & \cong A_5 \\
$III$*	& 360\times 4	& F_1,F_2,F_3	& Lift(A_6)\\
$IV$*	& \frac12 7!\times 2	& S,T,W		& Lift(A_7)\\
$V$	& 168\times 4	& S,T,R		& \\
$VI$*	& 2^6 3^4 5\times 2& T,C,D,E,F	& \\	
\hline
\end{array}
\]

\subsubsection{Groups Having Simple Normal Primitive Subgroups}
There are 3 such groups, generated by simple primitives
and the following 2 matrices:

{\small
\[
\begin{array}{c}
F' = \frac{1+i}{\sqrt{2}}\left(
\matrix{1 & 0 & 0 & 0 \\ 
	0 & 1 & 0 & 0 \\
	0 & 0 & 0 & 1 \\
	0 & 0 & 1 & 0 \\}
\right)

F'' = \left(
\matrix{0 & 1 & 0 & 0 \\ 
	-1 & 0 & 0 & 0 \\
	0 & 0 & 0 & 1 \\
	0 & 0 & -1 & 0 \\}
\right)
\end{array}
\]
}
The groups are then:
\[
\begin{array}{|c|c|c|c|}
\hline
$Group$	& $Order$	& $Generators$ 	& $Remarks$ \\
\hline \hline
$VII$*	& 120\times 4	& ($I$),F''	& Lift(S_5)\\
$VIII$*	& 120\times 4	& ($II$),F'	& Lift(S_5)\\
$IX$*	& 720\times 4	& ($III$),F''	& Lift(S_6)\\
\hline
\end{array}
\]

\subsubsection{Groups Having Normal Intransitive Subgroups}
There are seven types of $\Gamma$ in this case and their fundamental
representation matrices turn out to be Kronecker products of 
those of the exceptionals of $SU(2)$. In other words, 
for $M$, the matrix representation
of $\Gamma$, we have $M = A_1 \otimes_K A_2$ such that $A_i$ are the 
$2\times 2$ matrices representing $E_{6,7,8}$.
Indeed we know that $E_6 = \langle S_{SU(2)}, U^2_{SU(2)}\rangle,
E_7 = \langle S_{SU(2)}, U_{SU(2)}\rangle,
E_8 = \langle S_{SU(2)}, U^2_{SU(2)}, V_{SU(2)} \rangle$, where
{\small
\[
\begin{array}{c}
S_{SU(2)} = \frac12\left(
\matrix{-1 + i	& -1 + i \\
	1 + i	& -1 - i \\}
\right)

U_{SU(2)} = \frac{1}{\sqrt{2}}\left(
\matrix{1 + i	& 0 \\
	0	& 1 - i \\}
\right).

\\ \\

V_{SU(2)} = \left(
\matrix{\frac{i}{2}	& \frac{1-\sqrt{5}}{4} - i \frac{1+\sqrt{5}}{4}\\
	-\frac{1-\sqrt{5}}{4} - i \frac{1+\sqrt{5}}{4}	& -\frac{i}{2} \\}
\right)
\end{array}
\]
}

We use, for the generators, the notation $\langle A_i\rangle \otimes \langle B_j \rangle$
to mean that Kronecker products are to be formed between all combinations of $A_i$ with
$B_j$. Moreover the group (XI), a normal subgroup of (XIV), is formed by tensoring the
2-by-2 matrices 
$x_1 = \frac{1}{\sqrt{2}}\left(\matrix{1 & 1\cr i & -i}\right)$,
$x_2 = \frac{1}{\sqrt{2}}\left(\matrix{i & i\cr -1 & 1}\right)$,
$x_3 = \frac{1}{\sqrt{2}}\left(\matrix{-1 & -1\cr -1 & 1}\right)$,
$x_4 = \frac{1}{\sqrt{2}}\left(\matrix{i & 1\cr 1 & i}\right)$,
$x_5 = \frac{1}{\sqrt{2}}\left(\matrix{1 & -1\cr -i & -i}\right)$, and
$x_6 = \frac{1}{\sqrt{2}}\left(\matrix{i & -i\cr 1 & 1}\right)$.
The seven groups are:
\[
\begin{array}{|c|c|c|c|}
\hline
$Group$	& $Order$	& $Generators$ 	& $Remarks$ \\
\hline \hline
$X$*	& 144 \times 2	
	& \langle S_{SU(2)}, U^2_{SU(2)}\rangle \otimes \langle S_{SU(2)}, U^2_{SU(2)}\rangle
	& \cong E_6 \otimes_K E_6\\
$XI$*	& 288 \times 2
	& x_1 \otimes x_2, x_1 \otimes x^T_2, x_3 \otimes x_4, x_5 \otimes x_6
	& ($X$) \triangleleft \Gamma \triangleleft ($XIV$)\\
$XII$*	& 288 \times 2	
	& \langle S_{SU(2)}, U^2_{SU(2)}\rangle \otimes \langle S_{SU(2)}, U_{SU(2)}\rangle
	& \cong E_6 \otimes_K E_7\\
$XIII$*	& 720 \times 2
	& \langle S_{SU(2)}, U^2_{SU(2)}\rangle 
		\otimes \langle S_{SU(2)}, V_{SU(2)}, U^2_{SU(2)}\rangle		
	& \cong E_6 \otimes_K E_8\\
$XIV$*	& 576 \times 2	
	& \langle S_{SU(2)}, U_{SU(2)}\rangle \otimes \langle S_{SU(2)}, U_{SU(2)}\rangle
	& \cong E_7 \otimes_K E_7\\
$XV$*	& 1440 \times 2	
	& \langle S_{SU(2)}, U_{SU(2)}\rangle \otimes \langle S_{SU(2)}, V_{SU(2)}, U^2_{SU(2)}\rangle	
	& \cong E_7 \otimes_K E_8\\
$XVI$*	& 3600 \times 2	
	& \langle S_{SU(2)}, V_{SU(2)}, U^2_{SU(2)}\rangle 
		\otimes \langle S_{SU(2)}, V_{SU(2)}, U^2_{SU(2)}\rangle
	& \cong E_8 \otimes_K E_8\\
\hline
\end{array}
\]

\subsubsection{Groups Having X-XVI as Normal Primitive Subgroups}
There are in all 5 of these, generated by the above, together with
{\small
\[
\begin{array}{c}
T_1 = \frac{1+i}{\sqrt{2}}\left(
\matrix{1 & 0 & 0 & 0 \\ 
	0 & 0 & 1 & 0 \\
	0 & 1 & 0 & 0 \\
	0 & 0 & 0 & 1 \\}
\right)

T_2 = \left(
\matrix{1 & 0 & 0 & 0 \\ 
	0 & 0 & 1 & 0 \\
	0 & i & 0 & 0 \\
	0 & 0 & 0 & i \\}
\right)
\end{array}
\]
}
The group generated by (XIV) and $T_2$ is isomorphic to (XXI), generated by
(XIV) and $T_1$ so we need not consider it. The groups are:
\[
\begin{array}{|c|c|c|}
\hline
$Group$	& $Order$	& $Generators$ 	\\
\hline \hline
$XVII$*	&576\times 4 	& ($XI$),T_1 	\\
$XVIII$*&576\times 4  	& ($XI$),T_2	\\
$XIX$*	&288\times 4 	& ($X$),T_1	\\
$XX$*	&7200\times 4 	& ($XVI$),T_1	\\
$XXI$*	&1152\times 4	& ($XIV$),T_1	\\
\hline
\end{array}
\]

\subsubsection{Groups Having Normal Imprimitive Subgroups}
Finally these following 9 groups of order divisible by 5
complete our list of the primitive
$\Gamma$, for which we need the following generators:
{\small
\[
\begin{array}{c}
A = \frac{1+i}{\sqrt{2}}\left(
\matrix{1 & 0 & 0 & 0 \\ 
	0 & i & 0 & 0 \\
	0 & 0 & i & 0 \\
	0 & 0 & 0 & 1 \\}
\right)

B = \frac{1+i}{\sqrt{2}}\left(
\matrix{1 & 0 & 0 & 0 \\ 
	0 & 1 & 0 & 0 \\
	0 & 0 & 1 & 0 \\
	0 & 0 & 0 & -1 \\}
\right)

\\ \\

S' = \frac{1+i}{\sqrt{2}}\left(
\matrix{i & 0 & 0 & 0 \\ 
	0 & i & 0 & 0 \\
	0 & 0 & 1 & 0 \\
	0 & 0 & 0 & 1 \\}
\right)

T' = \frac{1+i}{2}\left(
\matrix{-i & 0 & 0 & i \\ 
	0 & 1 & 1 & 0 \\
	1 & 0 & 0 & 1 \\
	0 & -i & i & 0 \\}
\right)

R' = \frac{1}{\sqrt{2}}\left(
\matrix{1 & i & 0 & 0 \\ 
	i & 1 & 0 & 0 \\
	0 & 0 & i & 1 \\
	0 & 0 & -1 & -i \\}
\right)

\end{array}
\]
}

Moreover these following groups contain the group $K$ of order
$16\times 2$, generated by:
{\small
\[
\begin{array}{c}
A_1 = \left(
\matrix{1 & 0 & 0 & 0 \\ 
	0 & 1 & 0 & 0 \\
	0 & 0 & -1 & 0 \\
	0 & 0 & 0 & -1 \\}
\right)

A_2 = \left(
\matrix{1 & 0 & 0 & 0 \\ 
	0 & -1 & 0 & 0 \\
	0 & 0 & -1 & 0 \\
	0 & 0 & 0 & 1 \\}
\right)

\\ \\

A_3 = \left(
\matrix{0 & 1 & 0 & 0 \\ 
	1 & 0 & 0 & 0 \\
	0 & 0 & 0 & 1 \\
	0 & 0 & 1 & 0 \\}
\right)

A_4 = \left(
\matrix{0 & 0 & 1 & 0 \\ 
	0 & 0 & 0 & 1 \\
	1 & 0 & 0 & 0 \\
	0 & 1 & 0 & 0 \\}
\right)

\end{array}
\]
}
We tabulate the nine groups:
\[
\begin{array}{|c|c|c|}
\hline
$Group$	& $Order$		& $Generators$ 	\\
\hline \hline
$XXII$*	& 5\times 16 \times 4	& ($K$),T'	\\
$XXIII$*& 10\times 16 \times 4	& ($K$),T',R'^2	\\
$XXIV$*	& 20\times 16 \times 4	& ($K$),T,R	\\
$XXV$*	& 60\times 16 \times 4	& ($K$),T,S'B	\\
$XXVI$*	& 60\times 16 \times 4	& ($K$),T,BR'	\\
$XXVII$*& 120\times 16 \times 4	& ($K$),T,A	\\
$XXVIII$*& 120\times 16 \times 4& ($K$),T,B	\\
$XXIX$*	& 360\times 16 \times 4	& ($K$),T,AB	\\
$XXX$*	& 720\times 16 \times 4	& ($K$),T,S	\\
\hline
\end{array}
\]

\subsection{Intransitive Subgroups}
These cases are what could be constructed from the various combinations of the
discrete subgroups of $SL(2;\C)$ and $SL(3;\C)$ according to the various
possibilities of diagonal embeddings. Namely, they consist of those of the
form $(1,1,1,1)$ which represents the various possible Abelian groups with 
one-dimensional (cyclotomic) representation\footnote{These includes the
$\Z_m \times \Z_n \times \Z_p$ groups recently of interest in brane cube
constructions \cite{Uranga}.}, $(1,1,2)$, two Abelians and an $SL(2;\C)$ subgroup,
$(1,3)$, an Abelian and an $SL(3;\C)$ subgroup, and $(2,2)$, two $SL(2;\C)$ subgroups
as well as the various permutations thereupon. Since these embedded groups (as
collineation groups of lower dimension) have been well discussed in
Chap.~\ref{chap:9811183}, we
shall not delve too far into their account.

\subsection{Imprimitive Groups}
The analogues of the dihedral groups (in both $SL(2;\C)$ and $SL(3;\C)$), which
present themselves as infinite series, are to be found in these last cases of 
$\Gamma$. They are of two subtypes:
\begin{itemize}
\item (a) Generated by the canonical Abelian group of order $n^3$ for $n \in \Z^+$
	whose elements are
	\[
	\begin{array}{cc}
	\Delta =  \{\left(
	\matrix{\omega^i & 0 & 0 & 0 \\ 
		0 & \omega^j & 0 & 0 \\
		0 & 0 & \omega^k & 0 \\
		0 & 0 & 0 & \omega^{-i-j-k} \\}
	\right)\}
	&
	\begin{array}{c}
		\omega = e^{\frac{2 \pi i}{n}} \\
		i,j,k = 1,...,n
	\end{array}
	\end{array}
	\]
	as well as respectively the four groups $A_4$, $S_4$, the Sylow-8 subgroup
	$Sy \subset S_4$ (or the ordinary dihedral group of 8 elements)
	 and $\Z_2 \times \Z_2$;
\item (b) We define $H$ and $T''$ (where again $i = 1,...,n$) as:
	\[
	\begin{array}{c}
	H = \left(
	\matrix{a & b & 0 & 0 \\ 
		c & d & 0 & 0 \\
		0 & 0 & e & f \\
		0 & 0 & g & h \\}
	\right)

	T'' = \left(
	\matrix{0 & 0 & 1 & 0 \\ 
		0 & 0 & 0 & 1 \\
		\omega^i & 0 & 0 & 0 \\
		0 & \omega^{-i} & 0 & 0 \\}
	\right)
	\end{array}
	\]
	where the blocks of $H$ are $SL(2;\C)$ subgroups.
\end{itemize}
We tabulate these last cases of $\Gamma$ as follows:
\[
\begin{array}{|c|c|c|c|}
\hline
$Subtype$	& $Group$	& $Order$ & $Generators$ \\
\hline \hline
(a)		& $XXXI$*	& 12n^3	& \langle \Delta,A_4 \rangle\\
		& $XXXII$*	& 24n^3	& \langle \Delta,S_4 \rangle\\
		& $XXXIII$*	& 8n^3	& \langle \Delta,Sy \rangle\\
		& $XXXIII$*	& 4n^3	& \langle \Delta,\Z_2 \times \Z_2 \rangle\\
(b)		& $XXXIV$*	& 	& \langle H,T'' \rangle \\
\hline
\end{array}
\]

\section{Remarks}
We have presented, in modern notation, the classification of the discrete subgroups
of $SL(4,\C)$ and in particular, of $SU(4)$. The matrix generators and orders of these
groups have been tabulated, while bearing in mind how the latter fall into 
sub-categories of transitivity and primitivity standard to discussions on collineation
groups.

\begin{figure}
\centerline{\psfig{figure=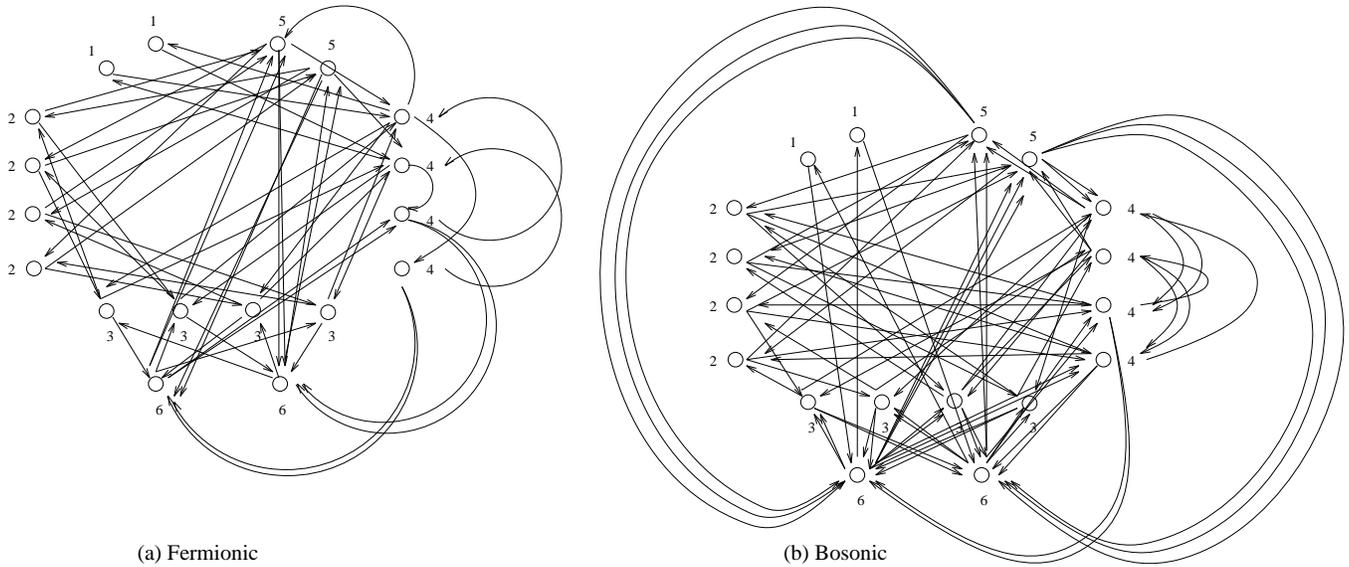,width=7in}}
\caption{
\label{fig:I}
The Quiver Diagram for Group (I), constructed for (a) the fermionic
$a_{ij}^4$ corresponding to the irreducible ${\bf 4}_3$ and (b) the bosonic
$a_{ij}^6$ corresponding to the irreducible ${\bf 6}_2$  (in the notation
of Chap.~\ref{chap:9811183}). We make this choice because we know that 
${\bf 4}_1 \otimes {\bf 4}_3 = {\bf 4}_3 \oplus {\bf 6}_1 \oplus {\bf 6}_2$
and that the two {\bf 6}'s are conjugates. 
The indices are the 
dimensions of the various
irreducible representations, a generalisation of Dynkin labels.
}
\end{figure}

Furthermore, we have computed the character table for the 30 exceptional 
cases \cite{Prog}; 
The interested reader may, at his or her convenience, find the character tables
at 
http://pierre.mit.edu/$\sim$yhe/su4.ct.
These tables will
be crucial to quiver theories.
As an example, we present in \fref{fig:I} the quiver for the irreducible {\bf 4} of
the group (I) of
order $60 \times 4$, which is the lift of the alternating 
permutation group on 5 elements.

Indeed such quiver diagrams may be constructed for all the groups using
the character tables mentioned above. We note in passing that since
$\Gamma \subset SU(4)$ gives rise to an ${\cal N}=0$ theory in 4 dimensions,
supersymmetry will not come to our aid in relating the fermionic $a_{ij}^{\bf 4}$
and the bosonic $a_{ij}^{\bf 6}$ as was done in Chap.~\ref{chap:9811183}.
However we can analyse the problem with a slight modification and place
a stack of M2 branes on the orbifold,
(which in the Maldacena picture corresponds to orbifolds on the $S^7$ factor
in $AdS_4 \times S^7$),
and obtain an ${\cal N}=2$ theory in 3 dimensions at least in the
IR limit as we lift from type IIA to M Theory
\cite{DM,LNV,KS,Agata,Horizon,Fabbri}. 
This supersymmetry would help us to impose the constraining relation between
the two matter matrices, and hence the two quiver diagrams. This would be
an interesting check which we leave to future work.

We see therefore a host of prospective research in various areas,
particularly in the context of string orbifold/gauge 
theories, WZW modular invariants, and singularity-resolutions in algebraic geometry.
It is hoped that this monograph, together with its companion tables on the web,
will provide a ready-reference to works in these directions.
\chapter{Finitude of Quiver Theories and Finiteness of Gauge Theories}
\section*{\center{{ Synopsis}}}
\label{chap:9911114}
The D-branes probe theories thusfar considered are all finite theories
with a conformal fixed point in the IR. Indeed,
asymptotic
freedom, finitude and IR freedom pose as a trichotomy of the
beta-function behaviour in quantum field theories.
Parallel thereto is a trichotomy in set theory of finite, tame and
wild representation types.
At the intersection of the above lies the theory of quivers.

We briefly review some of the terminology standard to the physics and to
the mathematics. Then we utilise certain results from graph theory and
axiomatic representation theory of path algebras to address physical
issues such as the implication of graph additivity to finiteness of
gauge theories, the impossibility of constructing completely IR free
string orbifold theories and the unclassifiability of ${\cal N}<2$
Yang-Mills theories in four dimensions. This perspective sheds a new
light on the speciality of $SU(2)$ ADE orbifolds \cite{9911114}. 
\section{Introduction}
In a quantum field theory (QFT), it has been known since the 70's (q.v. e.g. 
\cite{Peskin}), that
the behaviour of physical quantities such as mass and coupling constant are
sensitive to the renormalisation and evolve according to momentum scale
as dictated by the so-called {\it renormalisation flows}. In particular,
the correlation (Green's) functions, which encode the physical 
information relevant to Feymann's perturbative analysis of the theory and
hence unaffected by such flows, obey the famous Callan-Symanzik
Equations.
These equations assert the existence two universal functions $\beta(\lambda)$
and $\gamma(\lambda)$
shifting according to the coupling and field renormalisation
in such a way so as to compensate for the renormalisation scale.

A class of QFT's has lately received much attention, particularly among
the string theorists. These are the so-named {\bf finite} theories, 
characterised by the vanishing of the $\beta$-functions. These theories are
extremely well-behaved and no divergences can be associated with the coupling
in the ultraviolet; they were thus once embraced as the solution to ultraviolet
infinities of QFT's. Four-dimensional finite theories are restricted to 
supersymmetric gauge theories (or Super-Yang-Mills, SYM's), of which divergence
cancelation is a general feature, and have a wealth of interesting structure.
${\cal N} = 4$ SYM theories have been shown to be finite to all orders
(Cf. e.g. \cite{West,N4}) whereas for ${\cal N}=2$, the Adler-Bardeen
Theorem guarantees that no higher than 1-loop corrections exist for the 
$\beta$-function \cite{N2}. Finally, for the unextended ${\cal N}= 1$ 
theories, the vanishing at 1-loop implies that for 2-loops \cite{N1}.

When a {\bf conformal} field theory (CFT) with vanishing
$\beta$-function also has the anomalous dimensions vanishing, the
theory is in fact a finite theory.
This class of theories is without divergence and scale -- and here we enter
the realm of string theory. Recently much attempts have been undertaken in the
construction of such theories as low-energy limits of the world-volume theories
of D-brane probes on spacetime singularities (Chap.~\ref{chap:9811183},
\cite{DG,DGM,DM,Orb2,KS,LNV}) or
of brane setups of the Hanany-Witten type \cite{HW,HSU,HZ,HU}.
The construction of these theories not only supplies an excellent check for
string theoretic techniques but also, vice versa, facilitate the
incorporation of the Standard Model into string unifications.
These finite (super-)conformal theories in four dimensions
still remain a topic of fervent pursuit.

Almost exactly concurrent with these advances in physics was a host of 
activities in mathematics. Inspired by problems in linear representations
of partially ordered sets over a field \cite{Simson,Gabriel,Gabriel2,Sets,Dlab},
elegant and graphical methods have been
developed in attacking standing problems in algebra and combinatorics such
as the classification of representation types and indecomposables of 
finite-dimensional algebras.

In 1972, P. Gabriel introduced the concept of a ``K\"ocher''
in \cite{Gabriel}. This is what is known to our standard parlance today 
as a ``Quiver.'' What entailed
was a plethora of exciting and fruitful research in graph theory, axiomatic
set theory, linear algebra and category theory, among many other branches. 
In particular one result that has spurned interest is the great limitation
imposed on the shapes of the quivers once the concept of {\bf finite representation
type} has been introduced.

It may at first glance seem to the reader that these two disparate directions of
research in contemporary physics and mathematics may never share conjugal harmony.
However, following the works of
\cite{DM,Orb2,KS,LNV} those amusing quiver diagrams have 
surprisingly - or perhaps
not too much so, considering how that illustrious field of String Theory has
of late brought such enlightenment upon physics from seemingly most 
esoteric mathematics - taken a slight excursion from the reveries of the
abstract, and manifested themselves in SYM theories emerging
from D-branes probing orbifolds.
The gauge fields and matter content of
the said theories are conveniently encoded into quivers and further
elaborations upon relations to beyond orbifold theories have been suggested in
Chapters \ref{chap:9811183} and \ref{chap:9903056}.

It is therefore natural, for one to pause and step back awhile, and regard the
string orbifold theory from the perspective of a mathematician, and the quivers,
from that of a physicist. However, due to his inexpertise in both, the author
could call himself neither. Therefore we are compelled to peep at the
two fields as outsiders, and from afar attempt to make some observations
on similarities, obtain some vague notions of the beauty, and speculate upon
some underlying principles. This is then the purpose of this note:
to perceive, with a distant and weak eye; to inform, with a remote and feeble voice.

The organisation of this chapter is as follows. Though the main results
are given in \S 4, we begin with some preliminaries from
contemporary techniques in string theory on constructing four dimensional
super-Yang-Mills, focusing on what each interprets finitude to mean: 
\sref{ss:probe} on D-brane probes on orbifold singularities,
\sref{ss:HW} on Hanany-Witten setups and \sref{ss:geo} on geometrical
engineering. Then we move to the other direction and give preliminaries in
the mathematics, introducing quiver graphs and path 
algebras in \sref{ss:quiver},
classification of representation types in \sref{ss:type} and
how the latter imposes constraints on the former in \sref{ss:theorem}.
The physicist may thus liberally neglect \S 2 and the mathematician, 
\S 3. Finally in \sref{s:marriage} we shall see how those beautiful theorems
in graph theory and axiomatic set theory may be used to give surprising
results in constructing gauge theories from string theory.

\section*{Nomenclature}
Unless the contrary is stated, we shall throughout this chapter adhere to 
the convention that $k$ is a field of characteristic zero (and hence infinite),
that $Q$ denotes a quiver and $kQ$, the path algebra over the field $k$ associated
thereto, that rep$(X)$ refers to the representation of the object $X$, and that
irrep($\Gamma$) is the set of irreducible representations of the group $\Gamma$.
Moreover, {\sf San serif} type setting will be reserved for categories,
calligraphic ${\cal N}$ is used to denote the number of supersymmetries and
$\widehat{~~}$, to distinguish the Affine Lie Algebras or Dynkin graphs.

\section{Preliminaries from the Physics}
\index{Finiteness!gauge theory}
The Callan-Symanzik equation of a QFT dictates the behaviour, under
the renormalisation 
group flow, of the $n$-point correlator $G^{(n)}(\{\phi(x_i)\};M,\lambda)$ 
for the quantum fields $\phi(x)$,
according to the renormalisation of the coupling $\lambda$ and momentum
scale $M$ (see e.g. \cite{Peskin}, whose conventions we shall adopt):
\[
\left[ M \frac{\partial}{\partial M} + \beta(\lambda) \frac{\partial}{\partial \lambda}
+ n \gamma(\lambda) \right] G^{(n)}(\{\phi(x_i)\}; M,\lambda) = 0.
\]
The two universal dimensionless functions $\beta$ and $\gamma$ are known respectively
as the {\bf $\beta$-function} and the {\bf anomalous dimension}. They determine how the
shifts $\lambda \rightarrow \lambda + \delta \lambda$ in the coupling constant and
$\phi \rightarrow (1 + \delta \eta) \phi$ in the wave function compensate for the
shift in the renormalisation scale $M$:
\[
\beta(\lambda) := M \frac{\delta \lambda}{\delta M}~~~~~~
\gamma(\lambda) := -M \frac{\delta \eta}{\delta M}.
\]
Three behaviours are possible in the region of small $\lambda$:
(1) $\beta(\lambda) > 0$;
(2) $\beta(\lambda) < 0$; and
(3) $\beta(\lambda) = 0$.
The first has good IR behaviour and admits valid Feynmann perturbation at large-distance,
and the second possesses good perturbative behaviour at UV limits and are
asymptotically free. The third possibility is where the coupling constants do not
flow at all and the renormalised coupling is always equal to the bare coupling.
The only possible divergences in these theories are associated with field-rescaling
which cancel automatically in physical $S$-matrix computations. It seems that to
arrive at these well-tamed theories, some supersymmetry (SUSY) is needed so as to induce the
cancelation of boson-fermion loop effects\footnote{Proposals for non-supersymmetric
finite theories in four dimensions have been recently made in
\cite{KS,LNV,Vafa-Frampton,9905212}; to their
techniques we shall later turn briefly.}.
These theories are known as the {\bf finite theories} in QFT.

Of particular importance are the finite theories 
that arise from {\bf conformal} field theories which generically have
in addition to the vanishing $\beta$-functions, also zero anomalous
dimensions.
Often this subclass belongs to a continuous manifold of scale invariant
theories and is characterised by the existence of exactly marginal operators and whence
dimensionless coupling constants, the
set of mappings among which constitutes the {\it duality group} \`a la Mantonen-Olive
of ${\cal N} = 4$ SYM, a hotly pursued topic.

A remarkable phenomenon is that if there is a choice of coupling constants such that
all $\beta$-functions as well as the anomalous dimensions (which themselves do vanish
at leading order if the manifold of fixed points include the free theory) vanish at first 
order then the theory is finite to all orders (Cf. references in \cite{HSU}).
A host of finite theories arise as low energy effective theories of String Theory. It will
be under this light that our discussions proceed. There are three contemporary methods of
constructing (finite, super) gauge theories: (1) geometrical engineering; (2)
D-branes probing singularities and (3) Hanany-Witten brane setups. Discussions on the 
equivalence among and extensive reviews for them have been in wide circulation 
(q.v. e.g. \cite{Methods,Karch:equiv,ZD,9906031,9909125}). Therefore we
shall not delve too far into their account; 
we shall recollect from them what each interprets {\it finitude} to mean.

\subsection{D-brane Probes on Orbifolds} \label{ss:probe}
\index{Orbifolds}
\index{McKay Correspondence!brane probes}
\index{McKay Correspondence!in string theory}
When placing $n$ D3-branes on a space-time orbifold singularity $\C^m/\Gamma$, 
out of the parent
${\cal N} = 4$ $SU(n)$ SYM one can fabricate a $\prod\limits_{i} U(N_i)$ 
gauge theory with irrep$(\Gamma) := \{{\bf r}_i \}$ 
and $\sum\limits_{i} N_i \dim{\bf r}_i = n$ \cite{LNV}. The resulting SUSY in
the four-dimensional worldvolume is ${\cal N} = 2$ if the orbifold is
$\C^2 / \{ \Gamma \subset SU(2) \}$ as studied in \cite{DM,Orb2},
${\cal N} = 1$ if $\C^3 / \{ \Gamma \subset SU(3) \}$ as in
Chap.~\ref{chap:9811183}
and non-SUSY if $\C^3 / \{ \Gamma \subset SU(4) \}$ as in
Chap.~\ref{chap:9905212}.
The subsequent matter fields are
$a_{ij}^{\bf{4}}$ Weyl fermions $\Psi _{f_{ij}=1,...,a_{ij}^{\bf{4}}}^{ij}$ 
and $a_{ij}^{\bf 6}$ scalars $\Phi _{f_{ij}}^{ij}$ with $i,j = 1,...,n$ and
$a_{ij}^{\cal R}$ defined by
\beq
{\cal R}\otimes {\bf r}_i=\bigoplus\limits_{j}a_{ij}^{\cal R} {\bf r}_j
\label{aij2}
\eeq
respectively for ${\cal R} = 4,6$.
It is upon these matrices $a_{ij}$, which we call {\bf bifundamental
matter matrices} that we shall dwell. They dictate how many matter fields
transform under the $(N_i,\bar{N}_j)$ of the product gauge group.
It was originally pointed out in \cite{DM,Orb2} that one can encode this information
in {\bf quiver diagrams} where one indexes the vector multiplets (gauge) by nodes
and hypermultiplets (matter) by links in a (finite) graph so that the bifundamental
matter matrix defines the (possibly oriented) adjacency matrix for this graph.
In other words, one draws $a_{mn}$ number of arrows from node $m$ to $n$.
Therefore to each vertex $i$ is associated a vector space $V_i$ and a semisimple 
component $SU(N_i)$ of the gauge group acting on $V_i$. Moreover 
an oriented link from $V_1$ to $V_2$ represents a complex field transforming under
hom($V_1,V_2$). We shall see in section \sref{ss:quiver} what all this means.

When we take the dimension of both sides of (\ref{aij2}), we obtain the matrix
equation 
\beq
\label{dimaij}
\dim({\cal R}) r_i = a_{ij}^{\cal R} r_j
\eeq
where $r_i := \dim{\bf r}_i$.
As discussed in Chap.~\ref{chap:9811183}, the remaining SUSY must be in the 
commutant of $\Gamma$ in the $SU(4)$ R-symmetry of the parent
${\cal N} = 4$ theory. In the case of ${\cal N} = 2$ this means that 
$4 = 1 + 1 + 2$ and by SUSY, $6 = 1 + 1 + 2 + 2$ where the 1 is the principal
(trivial) irrep and 2, a two-dimensional irrep. Therefore due to the additivity and
orthogonality of group characters, it was thus pointed out ({\it cit. ibid.}) that
one only needs to investigate the fermion matrix $a_{ij}^4$, which is actually
reduced to $2 \delta_{ij} + a_{ij}^2$. Similarly for ${\cal N} = 1$, we have
$\delta_{ij} + a_{ij}^3$. It was subsequently shown that
(\ref{dimaij}) necessitates the vanishing of the $\beta$-function
to one loop. Summarising these points, we state the condition for {\it finitude}
from the orbifold perspective:
\beq
\label{orb_cond}
\begin{array}{c|c}
$SUSY$		&	$Finitude$ \\ \hline
{\cal N} = 2	&	2 r_i = a_{ij}^2 r_j \\
{\cal N} = 1	&	3 r_i = a_{ij}^3 r_j \\
{\cal N} = 0	&	4 r_i = a_{ij}^4 r_j \\
\end{array}
\eeq
In fact it was shown in \cite{LNV,geoeng3}, that the 1-loop $\beta$-function
is proportional to $d r_i - a_{ij}^d r_j$ for $d=4 - {\cal N}$ whereby
the vanishing thereof signifies finitude, exceeding zero signifies asymptotical freedom
and IR free otherwise\footnote{As a cautionary note, these conditions
	are necessary but may not be sufficient. In the cases
	of ${\cal N}<2$, one needs to check the superpotential.
	However, throughout the chapter we shall focus on the necessity
	of these conditions.}.
We shall call this expression $d \delta_{ij} - a_{ij}^d$
the {\bf discriminant function} since its relation with respective to zero
(once dotted with the vector of labels) discriminates the behaviour of the
QFT. This point shall arise once again in \sref{s:marriage}.

\subsection{Hanany-Witten} \label{ss:HW}
\index{Hanany-Witten}
In brane configurations of the Hanany-Witten type \cite{HW}, D-branes
are stretched between sets of NS-branes, the presence of which break the
SUSY afforded by the 32 supercharges of the type II theory. In particular,
parallel sets of NS-branes break one-half SUSY, giving rise to ${\cal N} = 2$ in
four dimensions \cite{HW} whereas rotated NS-branes \cite{Erlich} or grids of NS-branes
(the so-called Brane Box Models) \cite{HSU,HZ,HU} break one further half
SUSY and gives ${\cal N}=1$ in four dimensions.

The Brane Box Models (BBM) (and possible extensions to brane cubes) provide an intuitive
and visual realisation of SYM. They generically give rise to ${\cal N} = 1$, with
${\cal N} = 2$ as a degenerate case. Effectively, the D-branes placed in the
boxes of NS-branes furnish a geometrical way to encode the representation properties
of the finite group $\Gamma$ discussed in \sref{ss:probe}. The bi-fundamentals,
and hence the quiver diagram, are constructed from oriented open strings connecting
the D-branes according to the rule given in \cite{HZ}:
\[
{\bf 3} \otimes r_i = \bigoplus\limits_{j\in {\tiny \begin{array}{c}
{\rm N,~E,~SW}\\ {\rm Neighbours} \end{array}}} r_j.
\]
This is of course (\ref{aij2}) in a different guise and we clearly see the equivalence
between this and the orbifold methods of \sref{ss:probe}.

Now in \cite{HW}, for the classical setup of stretching a D-brane between two NS-branes,
the asymptotic bending of the NS-brane controls the evolution of the gauge coupling
(since the inverse of which is dictated by the distance between the NS-branes).
Whence NS-branes bending towards each other gives an IR free theory (case (1) defined
above for the $\beta$-functions), while bending away give an UV free (case (2))
theory. No bending thus indicates the non-evolution of the $\beta$-function and
thus finiteness; this is obviously true for any brane configurations, intervals,
boxes or cubes. We quote \cite{HSU} verbatim on this issue:
{\it Given a brane configuration which has no bending, the corresponding
field theory which is read off from the brane configuration by using the rules of
\cite{HZ} is a finite theory.}

Discussions on bending have been treated in \cite{LR,Randall} while works
towards the establishment of the complete correspondence between Hanany-Witten
methods and orbifold probes (to beyond the Abelian case) are well under
way \cite{ZD,9906031,9909125}. Under this light, we would like to lend
this opportunity to point 
out that the anomaly cancelation equations (2-4) of \cite{LR} which discusses the 
implication of tadpole-cancelation to BBM in excellent detail, are precisely
in accordance with (\ref{aij2}). In particular, what they referred as the Fourier
transform to extract the rank matrix for the $\Z_k \times \Z_{k'}$ BBM is
precisely the orthogonality relations for finite group characters (which in the
case of the Abelian groups conveniently reduce to roots of unity and hence
Fourier series). The generalisation of these equations for non-Abelian groups
should be immediate. We see indeed that there is a close intimacy between
the techniques of the current subsection with \sref{ss:probe}; let us now move
to a slightly different setting.

\subsection{Geometrical Engineering} \label{ss:geo}
\index{Geometric Engineering}
On compactifying Type IIA string theory on a non-compact Calabi-Yau 
threefold, we can geometrically engineer \cite{geoeng2,matter,geoeng3}
an ${\cal N}=2$ SYM.
More specifically when we compactify Type IIA on a K3 surface,
locally modeled by an ALE singularity, we arrive at an ${\cal N}=2$
SYM in 6 dimensions with gauge group $ADE$ depending on the singularity
about which D2-branes wrap in the zero-volume limit. However if we were
to further compactify on $T^2$, we would arrive at an ${\cal N} = 4$ SYM
in 4 dimensions. In order to kill the extraneous scalars we require
a 2-fold without cycles, namely {\bf P}$^1$, or the 2-sphere.
Therefore we are effectively compactifying our original 10 dimensional theory
on a (non-compact) Calabi-Yau threefold which is an ALE (K3) fibration
over {\bf P}$^1$, obtaining a pure ${\cal N}=2$ SYM in 4 dimensions with
coupling $\frac{1}{g^2}$ equaling to the volume of the base {\bf P}$^1$.

To incorporate matter \cite{matter,geoeng3} we let an $A_{n-1}$ ALE
fibre collide with an $A_{m-1}$ one to result in an $A_{m+n-1}$
singularity; this corresponds
to a Higgsing of $SU(m+n) \rightarrow SU(m) \times SU(n)$, giving
rise to a bi-fundamental matter $(n,\bar{m})$. Of course, by
colliding the $A$ singular fibres appropriately (i.e., in accordance with
Dynkin diagrams) this above idea can easily be generalised to fabricate
generic product $SU$ gauge groups.
Thus as opposed to \sref{ss:probe} where bi-fundamentals 
(and hence the quiver diagram) arise from linear maps between
irreducible modules of finite group representations, or \sref{ss:HW}
where they arise from open strings linking D-branes, in the context of
geometrical engineering, they originate from colliding fibres of the
Calabi-Yau.

The properties of the $\beta$-function from this geometrical perspective
were also investigated in \cite{geoeng3}. The remarkable fact, using the
Perron-Frobenius Theorem, is that the {\it possible} resulting SYM is highly
restricted. The essential classification is that {\it if the ${\cal
N}=2$ $\beta$-
function vanishes (and hence a {\rm finite} theory), then the
quiver diagram encoding the bi-fundamentals must be the affine 
$\widehat{ADE}$ Dynkin Diagrams} and when it is less than zero (and
thus an asymptotically free theory), the quiver must be the ordinary
$ADE$. We shall see later how one may graphically arrive at these results.

Having thus reviewed the contemporary trichotomy of the methods of
constructing SYM from string theory fashionable of late,
with special emphasis on what the word {\it finitude} means in each,
we are obliged, as prompted by the desire to unify, to ask ourselves
whether we could study these techniques axiomatically. After all, the
quiver diagram does manifest under all these circumstances. And it is
these quivers, as viewed by a graph or representation theorist, that
we discuss next.

\section{Preliminaries from the Mathematics}
We now formally study what a quiver is in a mathematical sense.
There are various approaches one could take, depending on whether one's
interest lies in category theory or in algebra. We shall commence with
P. Gabriel's definition, which was the genesis of the excitement which ensued.
Then we shall introduce the concept of path algebras and representation types
as well as a host of theorems that limit the shapes of quivers depending on
those type. As far as convention and nomenclature are concerned, \sref{ss:quiver} and
\sref{ss:type} will largely follow \cite{Rep,Mod,Simson}.

\subsection{Quivers and Path Algebras} \label{ss:quiver}
\index{Quivers!quiver category}
\index{Quivers!definition}
In his two monumental papers \cite{Gabriel,Gabriel2},
Gabriel introduced the following concept:
\begin{definition}
A {\bf quiver} is a pair $Q = (Q_0,Q_1)$, where $Q_0$ is a
set of vertices and $Q_1$, a set of arrows such that each element $\alpha \in Q_1$
has a beginning $s(\alpha)$ and an end $e(\alpha)$ which are vertices, i.e.,
$\{s(\alpha) \in Q_0\} \stackrel{\alpha}{\rightarrow} \{e(\alpha) \in Q_0$\}.
\end{definition}
In other words a quiver is a (generically) {\it directed graph}, 
possibly with multiple arrows and loops. 
We shall often denote a member $\gamma$ of $Q_1$ by the beginning and
ending vertices, as in $x \stackrel{\gamma}{\rightarrow} y$.

Given such a graph, we can generalise $Q_{0,1}$ by defining a {\bf path of length $m$} 
to be the formal composition
$\gamma = \gamma_1 \gamma_2 \dots \gamma_m := 
(i_0 \stackrel{\gamma_1}{\rightarrow} i_1 \dots \stackrel{\gamma_m}{\rightarrow} i_m)$
with $\gamma_j \in Q_1$ and $i_j \in Q_0$ such that $i_0 = s(\gamma_1)$ and 
$i_t = s(\gamma_{t-1}) = e(\gamma_t)$ for $t = 1,...,m$. This is to say that we
follow the arrows and trace through $m$ nodes. Subsequently we 
let $Q_m$ be the set of all paths of length $m$ and for the identity define,
for each node $x$, a trivial path of length zero, $e_x$, starting and ending at $x$.
This allows us to associate $Q_0 \sim \{e_x\}_{x \in Q_0}$ and 
$(i \stackrel{\alpha}{\rightarrow} j) \sim e_i \alpha = \alpha e_j$.
Now $Q_m$ is defined for all non-negative $m$, whereby giving a gradation in $Q$.

Objects\footnote{We could take this word literally and indeed we shall later
briefly define the objects in a Quiver Category.} may be assigned to the
nodes and edges of the quiver so as to make its conception more concrete. This is
done so in two closely-related ways:
\begin{enumerate}
\item By the {\bf representation of a quiver}, rep$(Q)$, we mean to 
	associate to each vertex $x \in Q_0$ of $Q$, a vector space $V_x$ 
	and to each arrow $x \rightarrow y$,
	a linear transformation between the corresponding vector spaces 
	$V_x \rightarrow V_y$.
\item Given a field $k$ and a quiver $Q$, a {\bf path algebra} $kQ$ is an algebra which 
	as a vector space over $k$ has its basis prescribed by the paths in $Q$.
\end{enumerate}

There is a 1-1 correspondence between $kQ$-modules and rep$(Q)$.
Given rep$(Q) = \{V_{x\in Q_0},(x \rightarrow y) \in Q_1 \}$, the associated
$kQ$ module is $\bigoplus\limits_x V_x$ whose basis is the set of paths $Q_m$.
Conversely, given a $kQ$-module $V$, we define $V_x = e_x V$ and the arrows to be
prescribed by the basis element $u$ such that $u \sim e_y u = u e_x$ whereby
making $u$ a map from $V_x$ to $V_y$.

On an algebraic level, due to the gradation of the quiver $Q$ by $Q_m$, the path
algebra is furnished by

\beq
\label{pathalg}
kQ := \bigoplus\limits_m kQ_m
~~~{\rm with}~~~
kQ_m := \bigoplus\limits_{\gamma \in Q_m} \gamma k
\eeq

As a $k$-algebra, the addition and multiplication axioms of $kQ$ are as
follows: given
$a = \sum\limits_{\alpha \in Q_m;~a_\alpha \in k} \alpha a_\alpha$
and
$b = \sum\limits_{\beta \in Q_n;~a_\beta \in k} \beta a_\beta$
as two elements in $kQ$,
$a+b = \sum\limits_\alpha \alpha (a_\alpha + b_\alpha)$ and
$a \cdot b = \sum\limits_{\alpha,\beta} \alpha \beta a_\alpha b_\beta$
with $\alpha \beta$ being the joining of paths (if the endpoint of
one is the beginning of another, otherwise it is defined to be 0).

\begin{figure}
\centerline{\psfig{figure=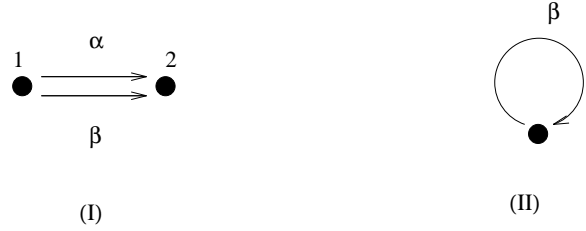,width=3.0in}}
\caption{Two examples of quivers with nodes and edges labeled.}
\label{f:ex}
\end{figure}

This correspondence between path algebras and quiver representations
gives us the flexibility of freely translating between the two, an advantage
we shall later graciously take.
As illustrative examples of concepts thus far introduced, we have
drawn two quivers in \fref{f:ex}.
In example (I), $Q_0 = \{1,2\}, Q_1 = \{\alpha,\beta\}$ and
$Q_{m > 1} = \{\}$. The path algebra is then the so-called Kronecker
Algebra: 
\[
kQ = e_1 k \oplus e_2 k \oplus \alpha k \oplus \beta k
= \left[ \begin{array}{cc} k & k^2 \\ 0 & k \\
	\end{array} \right].
\]
On the other hand, for example (II), 
$Q_{m \in \{0,1,2,...\}} = \{\beta^m\}$ and the
path algebra becomes $\bigoplus\limits_m \beta^m k = k[\beta]$,
the infinite dimensional free algebra of polynomials of one
variable over $k$.

In general, $kQ$ is finitely generated if there exists a finite
number of vertices and arrows in $Q$ and $kQ$ is finite-dimensional
if there does not exist any oriented cycles in $Q$.

\begin{figure}
\centerline{\psfig{figure=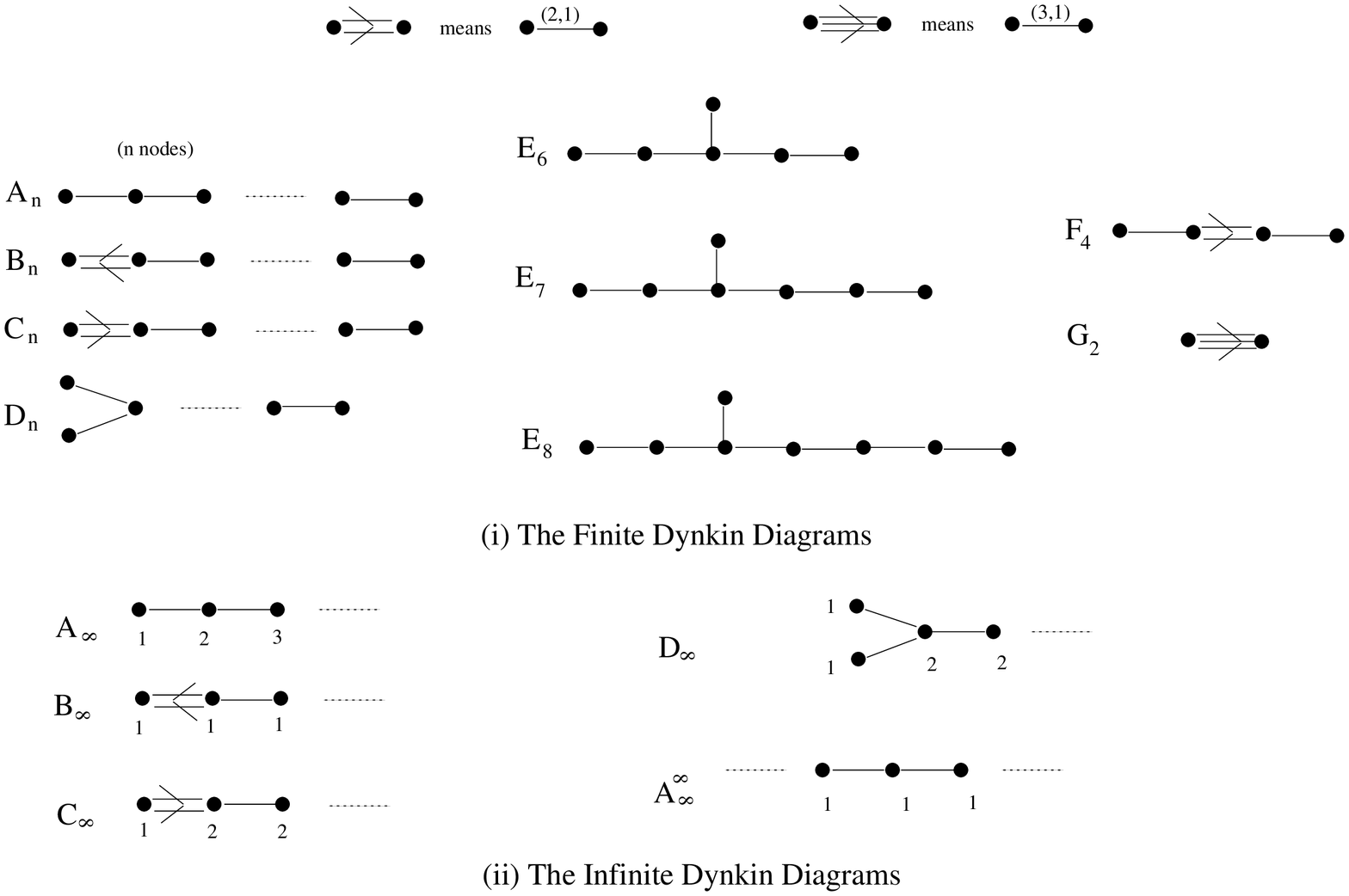,width=6.0in}}
\caption{The Finite and Infinite Dynkin Diagrams as labeled quivers.
	The finite cases are the well-known Dynkin-Coxeter graphs
	in Lie Algebras (from Chapter 4 of \cite{Rep}).}
\label{f:dynkin}
\end{figure}

\begin{figure}
\centerline{\psfig{figure=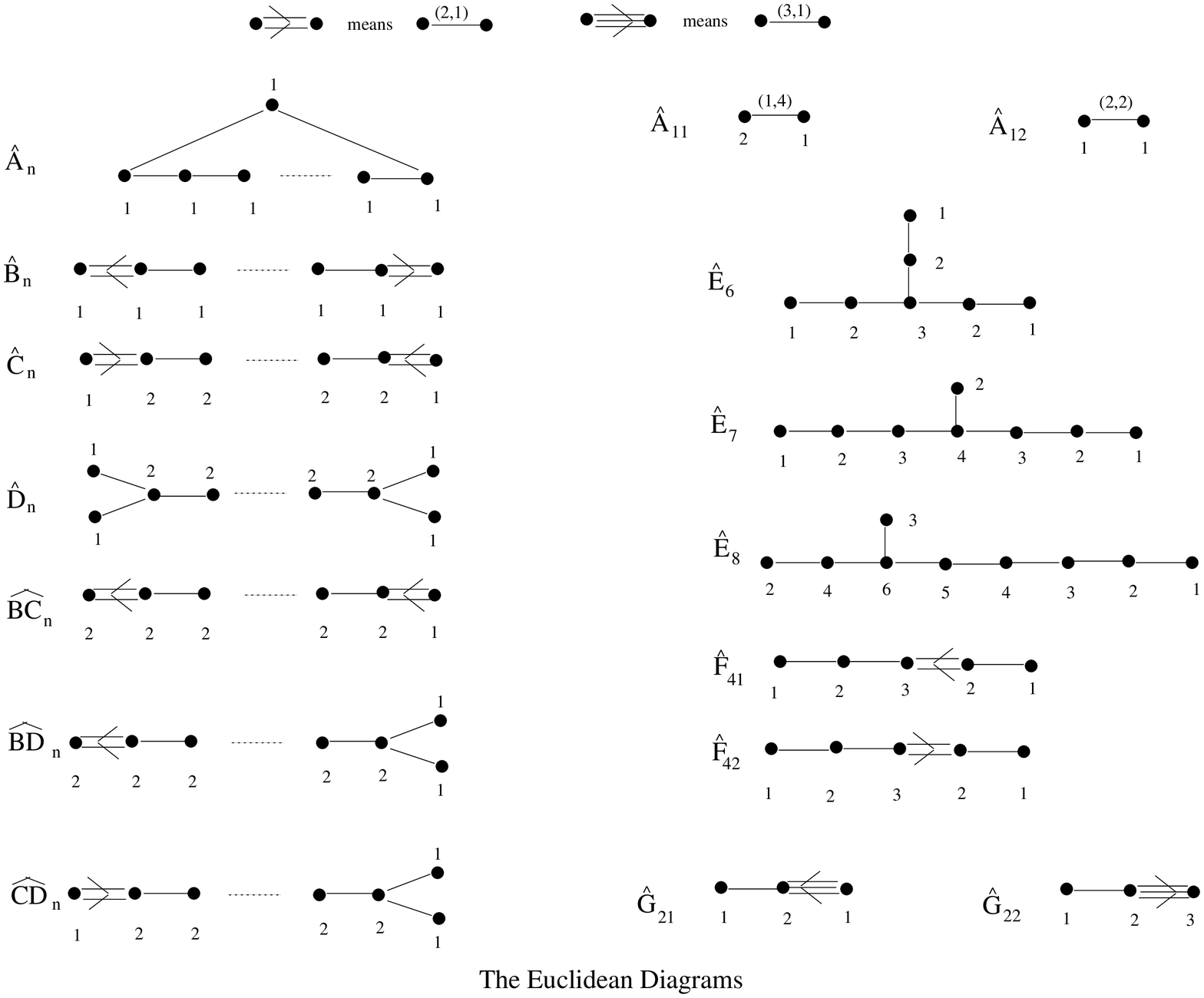,width=6.0in}}
\caption{The Euclidean Diagrams as labeled quivers; we recognise that
	this list contains the so-called Affine Dynkin Diagrams
	(from Chapter 4 of \cite{Rep}).}
\label{f:euclid}
\end{figure}

To specify the quiver even further one could introduce labeling
schemes for the nodes and edges; to do so we need a slight excursion
to clarify some standard terminology from graph theory.
\begin{definition} The following are common categorisations of graphs:
\begin{itemize}
\item A {\bf labeled graph} is a graph which has, for each of its edge
	$(i \stackrel{\gamma}{\rightarrow} j)$, a pair of positive
	integers $(a_{ij}^\gamma, a_{ji}^\gamma)$ associated thereto;
\item A {\bf valued graph} is a labeled graph for which there exists a
	positive integer $f_i$ for each node $i$, such that 
	$a_{ij}^\gamma f_j = a_{ji}^\gamma f_i$ for each arrow\footnote{
	Thus a labeled graph without any cycles is always a valued graph since
	we have enough degrees of freedom to solve for a consistent set of
	$f_i$ whereas cycles would introduce extra constraints.
	(Of course there is no implicit summation assumed in the equation.)}.
\item A {\bf modulation} of a valued graph consists of an assignment of
	a field $k_i$ to each node $i$, and a $k_i$-$k_j$
	bi-module $M_{ij}^\gamma$ to each arrow
	$(i \stackrel{\gamma}{\rightarrow} j)$ satisfying
	\begin{enumerate}
	\def\theenumi{\alph{enumi}}\def\labelenumi{(\theenumi)}
		\item $M_{ij}^\gamma \cong \hom_{k_i}(M_{ij}^\gamma,k_i)
			\cong \hom_{k_j}(M_{ij}^\gamma,k_j)$;
		\item $\dim_{k_i}(M_{ij}^\gamma) = a_{ij}^\gamma.$
	\end{enumerate}
\item A {\bf modulated quiver} is a valued graph with
	a modulation (and orientation).
\end{itemize}
\end{definition}
We shall further adopt the convention that we omit the label to edges
if it is $(1,1)$. We note that of course according to this labeling,
the matrices $a_{ij}$ are almost what we call {\bf adjacency matrices}.
In the case of unoriented single-valence edges between say nodes $i$ and
$j$, the adjacency matrix has $a_{ij} = a_{ji} = 1$, precisely the
label $(1,1)$. However, directed edges, as in \fref{f:dynkin}
and \fref{f:euclid}, are slightly more involved. This is
exemplified by $\bullet \Rightarrow \bullet$
which has the label $(2,1)$ whereas the conventional adjacency
matrix would have the entries $a_{ij} = 2$ and $a_{ji} = 0$. Such
a labeling scheme is of course so as to be consistent with the
entries of the Dynkin-Cartan Matrices of the semi-simple Lie Algebras.
To this subtlety we shall later turn.

The canonical examples of labeled (some of them are valued) graphs 
are what are known as the {\bf Dynkin} and {\bf Euclidean} graphs. 
The Dynkin graphs are further subdivided into the
finite and the infinite; the former are simply the Dynkin-Coxeter
Diagrams well-known in Lie Algebras while the latter are analogues
thereof but with infinite number of labeled nodes (note that 
the nodes are not labeled so as to make them valued graphs;
we shall shortly see what those numbers signify.)
The Euclidean graphs are the so-called Affine Coxeter-Dynkin Diagrams
(of the affine extensions of the semi-simple Lie algebras) but
with their multiple edges differentiated by oriented labeling schemes.
These diagrams are shown in \fref{f:dynkin} and \fref{f:euclid}.
\index{Lie Algebras}

How are these the {\it canonical} examples? We shall see the reason in 
\sref{ss:theorem}
why they are ubiquitous and atomic, constituting, when certain finiteness
conditions are imposed, the only elemental quivers.
Before doing so however, we need some facts from representation theory of
algebras; upon these we dwell next.

\subsection{Representation Type of Algebras} \label{ss:type}
\index{Finiteness!quivers}
Henceforth we restrict ourselves to infinite fields, as some of the
upcoming definitions make no sense over finite fields. This is of no loss 
of generality because in physics we are usually concerned with the field $\C$.
When given an algebra, we know its quintessential properties once we determine
its decomposables (or equivalently the irreducibles of the associated module).
Therefore classifying the behaviour of the indecomposables is the main goal
of classifying {\bf representation} types of the algebras.

The essential idea is that an algebra is of finite type if there are only
finitely many indecomposables; otherwise it is of infinite type.
Of the infinite type, there is one well-behaved subcategory, namely the
algebras of tame representation type, which has its indecomposables of
each dimension coming in finitely many one-parameter families with only
finitely many exceptions. Tameness in some sense still suggests
classifiability of the infinite indecomposables. On the other hand, an
algebra of wild type includes the free algebra on two variables, $k[X,Y]$,
(the path algebra of \fref{f:ex} (II), but with two self-adjoining arrows),
which indicates representations of arbitrary finite dimensional algebras,
and hence unclassifiability\footnote{
For precise statements of the unclassifiability of modules of
two-variable free algebras as Turing-machine undecidability,
cf. e.g. Thm 4.4.3 of \cite{Rep} and \cite{Turing}.}.

We formalise the above discussion into the following definitions:
\begin{definition}
Let $k$ be an infinite field and $A$, a finite dimensional algebra.
\begin{itemize}
\item $A$ is of {\bf finite representation type} if there are only finitely
	many isomorphism classes of indecomposable $A$-modules, otherwise
	it is of infinite type;
\item $A$ is of {\bf tame representation type} if it is of infinite type and
	for any dimension $n$, there is a finite set of 
	$A$-$k[X]$-bimodules\footnote{
		Therefore for the polynomial ring $k[X]$,
		the indeterminate $X$ furnishes the parameter for the one-parameter
		family mentioned in the first paragraph of this subsection. 
		Indeed the indecomposable $k[X]$-modules 
		are classified by powers of irreducible polynomials over $k$.}
	$M_i$ which obey the following:
	\begin{enumerate}
	\item $M_i$ are free as right $k[X]$-modules;
	\item For some $i$ and some indecomposable $k[X]$-module $M$,
		all but a finitely many indecomposable $A$-modules
		of dimension $n$ can be written as $M_i \otimes_{k[X]} M$.
	\end{enumerate}
	If the $M_i$ may be chosen independently of $n$, then we say $A$
	is of {\bf domestic representation type}.
\item $A$ is of {\bf wild representation type} if it is of infinite
	representation type and there is a finitely generated
	$A$-$k[X,Y]$-bimodule $M$ which is free as a right $k[X,Y]$-module
	such that the functor $M \otimes_{k[X,Y]}$ from finite-dimensional
	$k[X,Y]$-modules to finite-dimensional $A$-modules preserves 
	indecomposability and isomorphism classes.
\end{itemize}
\end{definition}
We are naturally led to question ourselves whether the above list is exhaustive.
This is indeed so: what is remarkable is the so-called trichotomy 
theorem which says that all finite dimensional algebras
must fall into one and only one of the above
classification of types\footnote{For a discussion on this theorem and how
	similar structures arises for finite groups, cf. e.g. \cite{Rep,Simson}
	and references therein.}:
\begin{theorem} {\rm (Trichotomy Theorem)}
For $k$ algebraically closed, every finite dimensional algebra $A$ is
of finite, tame or wild representation types, which are mutually exclusive.
\end{theorem}

To this pigeon-hole we may readily apply our path algebras of \sref{ss:quiver}.
Of course such definitions of representation types can be generalised to
additive categories with unique decomposition property. Here by an additive
category $\sf B$ we mean one with finite direct sums and an Abelian
structure on ${\sf B}(X,Y)$, the set of morphisms from object $X$ to $Y$
in $\sf B$ such that the composition map 
${\sf B}(Y,Z) \times {\sf B}(X,Y) \rightarrow {\sf B}(X,Z)$ 
is bilinear for $X,Y,Z$ objects
in $\sf B$. Indeed, that (a) each object in $\sf B$ can be finitely decomposed via
the direct sum into indecomposable objects and that (b) the ring of 
endomorphisms between objects has a unique maximal ideal guarantees that
$\sf B$ possesses unique decomposability as an additive category \cite{Simson}.

The category ${\sf rep}(Q)$, what \cite{Bernstein} calls the {\bf Quiver
Category}, has as its objects the pairs $(V,\alpha)$ with
linear spaces $V$ associated to the nodes and linear mappings $\alpha$, to
the arrows. The morphisms of the category are mappings
$\phi : (V,\alpha) \rightarrow (V',\alpha')$
compatible with $\alpha$ by 
$\phi_{e(l)}\alpha_l = \alpha'_{l}\phi_{s(l)}$.
In the sense of the correspondence between representation of quivers and path 
algebras as discussed in \sref{ss:quiver}, the category ${\sf rep}(Q)$ 
of finite dimensional representations of $Q$,
as an additive category, is equivalent to ${\sf mod}(kQ)$, the category of finite
dimensional (right) modules of the path algebra $kQ$ associated to $Q$.
This equivalence
\[
{\sf rep}(Q) \cong {\sf mod}(kQ)
\]
is the axiomatic statement of the correspondence and justifies why we can hereafter
translate freely between the concept of representation types of quivers
and associated path algebras.

\subsection{Restrictions on the Shapes of Quivers} \label{ss:theorem}
Now we return to our quivers and in particular combine \sref{ss:quiver} and 
\sref{ss:type} to address the problem of how the representation types of the path algebra
restricts the shapes of the quivers. Before doing so let us first justify, as advertised in
\sref{ss:quiver}, why \fref{f:dynkin} and \fref{f:euclid} are canonical.
We first need a preparatory definition: we say a labeled graph $T_1$ is {\bf smaller} 
than $T_2$ if there is an
injective morphism of graphs $\rho : T_1 \rightarrow T_2$ such that for
each edge $(i \stackrel{\gamma}{\rightarrow} j)$ in $T_1$,
$a_{ij} \le a_{\rho(i)\rho(j)}$ (and $T_1$ is said to be strictly
smaller if $\rho$ can not be chosen to be an isomorphism).
With this concept, we can see that the Dynkin and Euclidean graphs 
are indeed our archetypal examples of labeled graphs due to the following theorem:
\begin{theorem} {\rm \cite{Rep,Mod}}
Any connected labeled graph $T$ is one and only one of the 
following:
\begin{enumerate}
	\item $T$ is Dynkin (finite or infinite);
	\item There exists a Euclidean graph smaller than $T$.
\end{enumerate}
\end{theorem}
This is a truly remarkable fact which dictates that the atomic constituents of
all labeled graphs are those arising from semi-simple (ordinary and affine) 
Lie Algebras. The omni-presence of such meta-patterns is still largely
mysterious (see e.g. \cite{9903056,Gannon} for discussions on this point).

\index{ADE Meta-pattern}
Let us see another manifestation of the elementarity of the Dynkin and Euclidean
Graphs. Again, we need some rudimentary notions.
\begin{definition}
The {\bf Cartan Matrix} for a labeled graph $T$ with labels
$(a_{ij},a_{ji})$ for the edges is the matrix\footnote{This definition
	is inspired by, but should be confused with, Cartan matrices for semisimple
	Lie algebras; to the latter we shall refer as Dynkin-Cartan matrices.
	Also, in the definition we have summed over edges $\gamma$ adjoining
	$i$ and $j$ so as to accommodate multiple edges between the two nodes
	each with non-trivial labels.}
$c_{ij} := 2 \delta_{ij} - \sum\limits_\gamma a^\gamma_{ij}$
\end{definition}
We can symmetrise the Cartan matrix for valued graphs as
$\tilde{c}_{ij} = c_{ij}f_j$ with $\{f_j\}$ the valuation of the nodes
of the labeled graph.
With the Cartan matrix at hand, let us introduce an important 
function on labeled graphs:
\begin{definition} \label{additive}
A {\bf subadditive function} $n(x)$ on a labeled graph $T$ is a function
taking nodes $x \in T$ to $n \in \mathbb{Q}^+$ such that 
$\sum\limits_i n(i) c_{ij} \ge 0~\forall~j$.
A subadditive function is {\bf additive} if the equality holds.
\end{definition}
It turns out that imposing the existence such a function highly restricts
the possible shape of the graph; in fact we are again led back to our 
canonical constituents. This is dictated by the following
\begin{theorem} {\rm (Happel-Preiser-Ringel \cite{Rep})}
\label{thm:HPR}
Let $T$ be a labeled graph and $n(x)$ a subadditive function thereupon,
then the following holds:
\begin{enumerate}
	\item $T$ is either (finite or infinite) Dynkin or Euclidean;
	\item If $n(x)$ is not additive, then $T$ is finite Dynkin or $A_\infty$;
	\item If $n(x)$ is additive, then $T$ is infinite Dynkin or Euclidean;
	\item If $n(x)$ is unbounded then $T = A_\infty$
\end{enumerate}
\end{theorem}
We shall see in the next section what this notion of graph additivity
\cite{Rep,Gannon} signifies for super-Yang-Mills theories. For now,
let us turn to the {\it Theorema Egregium} of Gabriel that definitively
restricts the shape of the quiver diagram once the {\it finitude} of
the representation type of the corresponding path algebra is imposed.
\begin{theorem} {\rm (Gabriel \cite{Gabriel,Gabriel2,Simson})}
\label{thm:Gab}
A finite quiver $Q$ (and hence its associated path algebra over an
infinite field)
is of finite representation type if and only if it is a disjoint
union of Dynkin graphs of type $A_n$, $D_n$ and $E_{678}$, i.e.,
the ordinary simply-laced $ADE$ Coxeter-Dynkin diagrams.
\end{theorem}
In the language of categories \cite{Bernstein}, where a proof of the
theorem may be obtained
using Coxeter functors in the Quiver Category, the above proposes that
the quiver is (unions of) $ADE$ if and only if there are a 
finite number of non-isomorphic 
indecomposable objects in the category {\sf rep}$(Q)$.

Once again appears the graphs of \fref{f:dynkin}, and in fact only
the single-valence ones: that ubiquitous $ADE$ meta-pattern! We recall 
from discussions in \sref{ss:quiver}
that only for the simply-laced (and thus simply-valanced quivers)
cases, viz. $ADE$ and $\widehat{ADE}$, do the
labels $a^\gamma_{ij}$ precisely prescribe the adjacency matrices.
To what type of path algebras then, one may ask, do the affine 
$\widehat{ADE}$ Euclidean graphs correspond? The answer is given by
Nazarova as an extension to Gabriel's Theorem.
\begin{theorem} {\rm (Nazarova \cite{Poly2,Simson})}
Let $Q$ be a connected quiver without oriented cycles and let
$k$ be an algebraically closed field, then $kQ$ is of
tame (in fact domestic)
representation type if and only if $Q$ is the one of the Euclidean 
graphs of type $\hat{A}_n$, $\hat{D}_n$ and $\hat{E}_{678}$, i.e.,
the affine $ADE$ Coxeter-Dynkin diagrams.
\end{theorem}
Can we push further? What about the remaining quivers of in our canonical
list? Indeed, with the introduction of modulation on the quivers, 
as introduced in \sref{ss:quiver}, the results can be further
relaxed to include more graphs, in fact all the Dynkin and Euclidean
graphs:
\begin{theorem} {\rm (Tits, Bernstein-Gel'fand-Ponomarev,
	Dlab-Ringel, Nazarova-Ringel 
	\cite{Bernstein,Dlab,Poly1,Rep})}
\label{thm:gen}
Let $Q$ be a connected modulated quiver, then
\begin{enumerate}
\item If $Q$ is of finite representation type then $Q$ is Dynkin;
\item If $Q$ is of tame representation type, then $Q$ is Euclidean.
\end{enumerate}
\end{theorem}
This is then our dualism, on the one level of having finite graphs
encoding a (classifiability) infinite algebra and on another level
having the two canonical constituents of all labeled graphs being
partitioned by finitude versus infinitude\footnote{This is much in the
	spirit of that wise adage, ``Cette opposition nouvelle, 
	`le fini et l'infini', ou mieux `l'infini dans le fini',
	remplace le dualisme de l'\^etre et du para\^{\i}tre: ce qui
	para\^{\i}t, en effet, c'est seulement un {\it aspect} de 
	l'objet et l'objet est tout entier {\it dans} 
	cet aspect et tout entier hors de lui \cite{Sartre}.''}.

\section{Quivers in String Theory and Yang-Mills in Graph Theory} \label{s:marriage}
\index{McKay Correspondence!in string theory}
We are now equipped with a small arsenal of facts; it is now our duty to 
expound upon them. Therefrom we shall witness how
axiomatic studies of graphs and representations may shed light on current
developments in string theory.

Let us begin then, upon examining condition (\ref{orb_cond}) and Definition 
\ref{additive}, with the following

\begin{observation}
\label{ob:fin}
The condition for finitude of ${\cal N} = 2$ orbifold SYM theory is equivalent to
the introduction of an additive function on the corresponding quiver as a
labeled graph.
\end{observation}

This condition that for the label $n_i$ to each node $i$ and adjacency matrix
$A_{ij}$, $2 n_i = \sum\limits_j a_{ij} n_j$ is a very
interesting constraint to which we
shall return shortly. What we shall use now is Part 3 of 
Theorem \ref{thm:HPR} in conjunction with the above observation to deduce

\begin{corollary} \label{cor:N2}
All finite ${\cal N} = 2$ super-Yang-Mills Theories with bi-fundamental matter
have their quivers as (finite disjoint unions) of the single-valence (i.e.,
$(1,1)$-labeled edges) cases
of the Euclidean (\fref{f:euclid}) or Infinite Dynkin (\fref{f:dynkin}) graphs.
\end{corollary}

A few points to remark. This is slightly a more extended list than that given in
\cite{geoeng3} which is comprised solely of the $\widehat{ADE}$ quivers.
These latter cases are the ones of contemporary interest because they,
in addition to being geometrically constructable (Cf. \sref{ss:geo}),
are also obtainable from the string orbifold technique\footnote{
	And in the cases of $A$ and $D$ also from Hanany-Witten
	setups \cite{Kapustin,ZD,9906031,9909125}.} (Cf. \sref{ss:probe})
since after all the finite discrete subgroups of $SU(2)$ fall into
an $\widehat{ADE}$ classification due to McKay's Correspondence 
\cite{McKay,DG,DGM} and Chap.~\ref{chap:9811183}.
In addition to the above well-behaved cases, we also 
have the infinite simply-laced Dynkin graphs: $A_\infty,D_\infty$ and
$A_\infty^\infty$. The usage of the Perron-Frobenius Theorem in
\cite{geoeng3} restricts one's attention to finite matrices.
The allowance for infinite graphs of course implies an infinitude of nodes
and hence infinite products for the gauge group. One needs not exclude
these possibilities as after all in the study of D-brane probes,
Maldacena's large $N$ limit has been argued in \cite{KS,LNV,HSU}
to be required for conformality and finiteness. In this limit of
an infinite stack of D-branes, infinite gauge groups may well arise.
In the Hanany-Witten picture, $A_\infty^\infty$ for example would 
correspond to an infinite array of NS5-branes, and $A_\infty$,
a semi-infinite array with enough D-branes on the other side
to ensure the overall non-bending and parallelism of the NS.
Such cases had been considered in \cite{Erlich}.

Another comment is on what had been advertised earlier in \sref{ss:quiver}
regarding the adjacency matrices. Theorem \ref{thm:HPR} does not exclude
graphs with multiple-valanced oriented labels. This issue does not
arise in ${\cal N}=2$ which has only single-valanced and unoriented
quivers.
However, going beyond to ${\cal N} = 1,0$, requires generically oriented and 
multiply-valanced quivers (i.e., non-symmetric, non-binary matter matrices)
(Cf. Chapters \ref{chap:9811183} and \ref{chap:9905212}); 
or, it is conceivable that certain theories
not arising from orbifold procedures may also possess these generic traits.
Under this light we question ourselves how one may
identify the bi-fundamental matter matrices not with strict adjacency
matrices of graphs but with the graph-label matrices $a_{ij}^\gamma$
of \sref{ss:quiver} so as to accommodate multiple, chiral bi-fundamentals
(i.e. multi-valence, directed graphs). In other words, could 
Corollary \ref{cor:N2} {\it actually be relaxed to incorporate all of 
the Euclidean and infinite Dynkin graphs} as dictated by Theorem \ref{thm:HPR}?
Thoughts on this direction, viz., how to realise Hanany-Witten brane
configurations for non-simply-laced groups have been engaged but still
waits further clarification \cite{chat}.

Let us now turn to Gabriel's famous Theorem \ref{thm:Gab} and see its implications
in string theory and vice versa what information the latter provides for
graph theory. First we make a companion statement to Observation \ref{ob:fin}:

\begin{observation}
\label{ob:af}
The condition for asymptotically free ($\beta < 0$) ${\cal N} = 2$
SYM theory with bi-fundamentals
is equivalent to imposing a subadditive (but not additive) function
of the corresponding quiver.
\end{observation}

This may thus promptly be utilised together with Part 2 of Theorem
\ref{thm:HPR} to 
conclude that the only such theories are ones with $ADE$ quiver, or,
allowing infinite gauge groups, $A_\infty$ as well (and indeed all finite
Dynkin quivers once, as mentioned above, non-simply-laced groups have
been resolved). 
This is once again a slightly extended version of the results in
\cite{geoeng3}.

Let us digress, before trudging on, a moment to consider what is means to
encode SYM with quivers. Now we recall that for the quiver $Q$, 
the assignment of objects and morphisms to the category {\sf rep}$(Q)$,
or vector spaces and linear maps to nodes $Q_0$ and edges $Q_1$ in $Q$,
or bases to the path algebra $kQ$, are all equivalent procedures.
From the physics perspective, these assignments are precisely what
we do when we associate vector multiplets to nodes and hypermultiplets to
arrows as in the orbifold technique, or NS-branes to nodes and oriented
open strings between D-branes to arrows as in the Hanany-Witten configurations, or 
singularities in Calabi-Yau to nodes and colliding fibres to arrows as in
geometrical engineering. In other words the three methods, \sref{ss:quiver},
\sref{ss:probe} and \sref{ss:geo}, of constructing
gauge theories in four dimensions currently in vogue are different
representations of {\sf rep}$(Q)$ and are hence {\it axiomatically}
equivalent as far as quiver theories are concerned.

Bearing this in mind, and in conjunction with Observations \ref{ob:fin}
and \ref{ob:af}, as well as Theorem \ref{thm:Gab} together with its
generalisations, and in particular Theorem \ref{thm:gen}, we make the following

\index{Finiteness!gauge theory}
\index{Finiteness!quivers}
\begin{corollary} \label{cor:alg}
To an asymptotically free ${\cal N}= 2$ SYM with bi-fundamentals is associated a
finite path algebra and to a finite one, a tame path algebra. The association
is in the sense that these SYM theories (or some theory categorically equivalent thereto)
prescribe representations of the only quivers of such representation types.
\end{corollary}

What is even more remarkable perhaps is that due to the Trichotomy Theorem,
the path algebra associated to
{\bf all other quivers} must be of wild representation type. What this means,
as we recall the unclassifiability of algebras of wild representations, is
that these quivers are unclassifiable. In particular, if we assume that
SYM with ${\cal N}=0,1$ and arbitrary bi-fundamental matter content can
be constructed (either from orbifold techniques, Hanany-Witten, or geometrical 
engineering), then these theories {\it can not} be classified, in the strict
sense that they are Turing undecidable and there does not exist, in any finite
language, a finite scheme by which they could be listed. Since the set of SYM with
bi-fundamentals is a proper subset of all SYM, the like applies to general SYM.
What this signifies is that however ardently we may continue to provide
more examples of say finite ${\cal N}=1,0$ SYM, the list can never be finished
nor be described, unlike the ${\cal N}=2$ case where the above discussions
exhaust their classification. We summarise this amusing if not depressing
fact as follows:

\begin{corollary} \label{cor:un}
The generic ${\cal N} = 1,0$ SYM in four dimensions are unclassifiable 
in the sense of being Turing undecidable.
\end{corollary}

We emphasise again that by unclassifiable here we mean not {\it
completely} classifiable because we have given a subcategory (the
theories with bi-fundamentals) which is unclassifiable. Also, we rest upon
the assumption that for any bi-fundamental matter content an SYM could
be constructed. Works in the direction of classifying all possible
gauge invariant operators in an ${\cal N}=1$ SUSY Lagrangian have been
pursued \cite{Skenderis}. Our claim is much milder as no further
constraints than the possible na\"{\i}ve matter content are imposed; we
simply state that the complete generic problem of classifying the
${\cal N}<2$ matter content is untractable. In \cite{Skenderis}, the
problem has been reduced to manipulating a certain cohomological
algebra; it would be interesting to see for example, whether such BRST
techniques may be utilised in the classification of certain categories
of graphs.

Such an infinitude of gauge theories need not worry us as there certainly
is no shortage of say, Calabi-Yau threefolds which may be used to
geometrically engineer them.
This unclassifiability is rather in the spirit of that of, for example,
four-manifolds.
Indeed, though we may never exhaust the list, we are not precluded
from giving large exemplary subclasses which are themselves
classifiable, e.g., those prescribed by the orbifold theories. 
Determining these theories amounts to the classification of the finite
discrete subgroups of $SU(n)$. 

We recall from Corollary \ref{cor:N2} that
${\cal N}=2$ is given by the affine and infinite Coxeter-Dynkin graphs of
which the orbifold theories provide the $\widehat{ADE}$ cases. What
remarks could one make for ${\cal N} = 0,1$, i.e., $SU(3,4)$ 
McKay quivers (Cf. Chap. \ref{chap:9811183} and \ref{chap:9905212})?
Let us first see ${\cal N}=2$ from the graph-theoretic perspective, which
will induce a relationship between additivity (Theorem \ref{thm:HPR}) and
Gabriel-Nazarova (Theorems \ref{thm:Gab} and extensions).
The crucial step in Tit's proof of Gabriel's Theorem is the introduction
of the quadratic form on a graph \cite{Bernstein,Donovan}:
\begin{definition}
For a labeled quiver $Q = (Q_0,Q_1)$, one defines the (symmetric bilinear)
{\bf quadratic form} $B(x)$ on the set $x$ of the labels as follows:
\[
B(x) := \sum\limits_{i \in Q_0} x_i^2 - \sum\limits_{\alpha \in Q_1}
	x_{s(\alpha)} x_{e(\alpha)}.
\]
\end{definition}
\index{ADE Meta-pattern}
The subsequent work was then to show that finitude of representation is
equivalent to the positive-definity of $B(x)$, and in fact, as in
Nazarova's extension, that tameness is equivalent to positive-semi-definity.
In other words, finite or tame representation type can be translated,
in this context, to a Diophantine inequality which dictates the nodes and
connectivity of the quiver (incidentally the very same Inequality which
dictates the shapes of the Coxeter-Dynkin Diagrams or the vertices and
faces of the Platonic solids in $\R^3$):
\[
B(x) \ge 0 \Leftrightarrow \widehat{ADE},ADE~~~~~~~B(x) > 0 \Leftrightarrow ADE.
\]
Now we note that $B(x)$ can be written as 
$\frac 12 x^T \cdot c \cdot x$ where $(c)_{ij}$ is {\it de facto} 
the Cartan Matrix for graphs as
defined in \sref{ss:theorem}. The classification problem thus, because
$c := 2I - a$, becomes
that of classifying graphs whose adjacency matrix $a$ has maximal eigenvalue
2, or what McKay calls $C_2$-graphs in \cite{McKay}. This issue was addressed
in \cite{Smith} and indeed the $\widehat{ADE}$ graphs emerge.
Furthermore the additivity condition $\sum\limits_j c_{ij} x_j \ge 0~\forall~i$
clearly implies the constraint $\sum\limits_{ij} c_{ij} x_i x_j \ge 0$ 
(since all labels are positive) and thereby the like on the quadratic form.
Hence we see how to arrive at the vital step in Gabriel-Nazarova through
graph subadditivity.

The above discussions relied upon the specialty of the number 2.
Indeed one could translate between the graph quadratic form $B(x)$
and the graph Cartan matrix precisely because the latter is defined by
$2I - a$. From a physical perspective this is precisely the discriminant function
for ${\cal N}=2$ orbifold SYM (i.e. $d = 2$) as discussed at the
end of \sref{ss:probe}. This is why $\widehat{ADE}$ arises in all these contexts.
We are naturally led to question ourselves, what about general\footnote{
	In the arena of orbifold SYM, $d=1,2,3$, but in a broader settings,
	as in generalisation of McKay's Correspondence, $d$ could be
	any natural number.} $d$?
This compels us to consider a {\bf generalised Cartan matrix} for graphs
(Cf. Definition in \sref{ss:theorem}), given
by $c_{ij} := d \delta_{ij} - a_{ij}$, our discriminant function of \sref{ss:probe}.
Indeed such a matrix was considered in \cite{Steinberg} for general McKay quivers.
As a side remarks, due to such an extension, Theorem \ref{thm:HPR} must likewise
be adjusted to accommodate more graphs; a recent paper \cite{McKay2} shows
an example, the so-dubbed semi-Affine Dynkin Diagrams, where a new class of
labeled graphs with additivity with respect to the extended $c_{ij}$ emerge.

Returning to the generalised Cartan matrix, in \cite{Steinberg},
the McKay matrices $a_{ij}$ were obtained, for an arbitrary
finite group $G$, by tensoring a faithful $d$-dimensional representation with
the set of irreps: $r_d \otimes r_i = \oplus_j a_{ij} r_j$.
What was noticed was that the scalar product defined with respect
to the matrix $d \delta_{ij} - a_{ij}$ (precisely our generalised Cartan) was
positive semi-definite in the vector space $V = \{x_i\}$ of labels.
In other words, $\sum\limits_{ij} c_{ij} x_i x_j \ge 0$.
We briefly transcribe his proof in Appendix \ref{append:9911114}.
What this means for us is that is the following
\index{Orbifolds}
\index{Brane Probes!Orbifolds}
\begin{corollary}
\label{cor:d}
String orbifold theories can not produce a completely IR free 
(i.e., with respect to all semisimple components of the gauge group) 
QFT (i.e., Type (1), $\beta > 0$).
\end{corollary}

To see this suppose there existed such a theory. Then $\beta > 0$, implying
for our discriminant function that $\sum\limits_j c_{ij} x_j < 0~\forall~i$ for
some finite group.
This would then imply, since all labels are positive, that 
$\sum\limits_{ij} c_{ij} x_i x_j < 0$, violating the positive semidefinity
condition that it should always be nonnegative for any finite group according to
\cite{Steinberg}. Therefore by {\it reductio ad absurdum}, we conclude 
Corollary \ref{cor:d}.

On a more general setting, if we were to consider using the
generalised Cartan matrix $d \delta_{ij} - a_{ij}$
to define a generalised subadditive function (as opposed to merely $d=2$),
could we perhaps have an extended classification scheme?
To our knowledge this is so far an unsolved problem for indeed take
the subset of these graphs with all labels being 1 and 
$d n_i = \sum\limits_j a_{ij} n_j$, these are known as $d$-regular graphs
(the only 2-regular one is the $\widehat{A}$-series) and these are
already unclassified for $d>2$. We await input from mathematicians on this point.

\section{Concluding Remarks and Prospects}
The approach of this writing has been bilateral. On the one hand, we
have briefly reviewed the three contemporary techniques of obtaining 
four dimensional gauge theories from string theory, namely
Hanany-Witten, D-brane probes and geometrical engineering. In
particular, we focus on what finitude signifies for these theories and
how interests in quiver diagrams arises. Subsequently, we approach
from the mathematical direction and have taken a promenade in the
field of axiomatic representation theory of algebras associated to
quivers. The common ground rests upon the language of graph theory,
some results from which we have used to address certain issues in 
string theory.

From the expression of the one-loop $\beta$-function, we have defined
a discriminant function $f := d \delta_{ij} - a_{ij}^d$ for the quiver with
adjacency matrix $a_{ij}$ which encodes the bi-fundamental matter
content of the gauge theory. The nullity (resp. negativity/positivity)
of this function gives a necessary condition for the finitude
(resp. IR freedom/asymptotic freedom) of the associated gauge theory.
We recognise this function to be precisely the generalised Cartan
matrix of a (not necessarily finite) graph and the nullity (resp.
negativity) thereof, the additivity (resp. strict subadditivity) of
the graph. In the case of $d=2$, such graphs are completely
classified: infinite Dynkin or Euclidean if $f = 0$ and finite Dynkin
or $A_\infty$ if $f < 0$. In physical terms, this means that these
are the {\it only} ${\cal N}=2$ theories with bi-fundamental matter
(Corollary \ref{cor:N2} and Observation \ref{ob:af}).
This slightly generalises the results of \cite{geoeng3} by the
inclusion of infinite graphs, i.e., theories with infinite product
gauge groups. From the mathematics alone, also included are the
non-simply-laced diagrams, however we still await progress in the
physics to clarify how these gauge theories may be fabricated.

For $d>2$, the mathematical problem of their classification is so far
unsolved. A subclass of these, namely the orbifold theories coming
from discrete subgroups of $SU(n)$ have been addressed upto $n=4$
\cite{DM,Orb2,9811183,9905212}. A general remark we can make about these
theories is that, due to a theorem of Steinberg, D-brane probes on
orbifolds can never produce a completely IR free QFT (Corollary
\ref{cor:d}).

From a more axiomatic stand, we have also investigated possible
finite quivers that may arise. In particular we have reviewed the
correspondence between a quiver and its associated path algebra. Using
the Trichotomy theorem of representation theory, that all finite
dimensional algebras over an algebraically closed field are of either
finite, tame or wild type, we have seen that all quivers are
respectively either $ADE$, $\widehat{ADE}$ or unclassifiable. In
physical terms, this means that asymptotically free and finite 
${\cal N}=2$ SYM in four dimensions respectively exhaust the {\it
only} quiver theories of respectively finite and tame type (Corollary
\ref{cor:alg}). What these particular path algebras mean in a physical
context however, is yet to be ascertained. For the last type, we have
drawn a melancholy note that all other theories, and in particular,
${\cal N}<2$ in four dimensions, are in general Turing
unclassifiable (Corollary \ref{cor:un}).

Much work remains to be accomplished. It is the main purpose of this
note, through the eyes of a neophyte, to inform readers in each of two
hitherto disparate fields of gauge theories and axiomatic
representations, of certain results from the other. It is hoped that
future activity may be prompted.
\index{McKay Correspondence!and WZW}
\index{McKay Correspondence!in string theory}
\chapter{Orbifolds IV: Finite Groups and WZW Modular Invariants, Case
Studies for $SU(2)$ and $SU(3)$}
\label{chap:0009077}
\section*{\center{{ Synopsis}}}
Inspired by Chapters \ref{chap:9811183} and \ref{chap:9903056} 
which contained some
attempts to formulate various correspondences between the
classification of affine $SU(k)$ WZW modular-invariant partition
functions and that of discrete finite subgroups of $SU(k)$, we present
a small and perhaps interesting observation in this light. 

In particular we show
how the groups generated by the permutation of the terms in the
exceptional $\widehat{SU(2)}$-WZW invariants encode the corresponding
exceptional $SU(2)$ subgroups. We also address a weaker analogue for
$SU(3)$ \cite{0009077}.
\newpage
\section{Introduction}
\index{ADE Meta-pattern}
The ubiquitous $ADE$ meta-pattern of mathematics makes her mysterious emergence in
the classification of the modular invariant partition functions in Wess-Zumino-Witten
(WZW) models of rational conformal field theory (RCFT). Though this fact is by
now common knowledge, little is known about why {\it a fortiori} these invariants
should fall under such classification schemes \cite{CFT}.
Ever since the original work in the completion of the classification for $\widehat{su(2)}$
WZW invariants by Cappelli-Itzykson-Zuber \cite{su21,su22} as well as the subsequent
case for $\widehat{su(3)}$ by Gannon \cite{su31,su32}, many efforts have been
made to attempt to clarify the reasons behind the said emergence. These
include perspectives from lattice integrable systems where the invariants are
related to finite groups \cite{DiFrancesco}, and from generalised root systems and
$N$-colourability of graphs \cite{Reflection,Folding}. Furthermore, there has been a 
recent revival of interest in the matter as viewed from string theory
where sigma models and orbifold constructions are suggested to provide a link
\cite{9811183,9903056,Song}.

Let us first briefly review the situation at hand (much shall follow the conventions of
\cite{CFT} where a thorough treatment may be found). The $\widehat{g}_k$-WZW model
(i.e., associated to an affine Lie algebra $g$ at level $k$) is a non-linear
sigma model on the group manifold $G$ corresponding to the algebra $g$. Its
action is
\[
S^{\rm WZW} = \frac{k}{16\pi} \int_G \frac{d^2 x}{X_{\rm rep}} {\rm Tr}
	(\partial^\mu g^{-1} \partial_\mu g) + k \Gamma
\]
where $k \in \Z$ is called the level, $g(x)$, a matrix bosonic field
with target space\footnote{We are really integrating over the
	pull-back to the world sheet.}
 $G$ and $X_{\rm rep}$ the Dynkin index for the representation
of $g$. The first term is our familiar pull back in sigma models while the second
\[
\Gamma = \frac{-i}{24 \pi} \int_B \frac{d^3 y}{X_{\rm rep}} 
	\epsilon_{\alpha \beta \gamma} {\rm Tr}
	(\tilde{g}^{-1} \partial^\alpha\tilde{g} \tilde{g}^{-1} 
	\partial^\beta\tilde{g} \tilde{g}^{-1} \partial^\gamma\tilde{g} )
\]
is the WZW term added to ensure conformal symmetry. $B$ is a manifold such that
$\partial B = G$ and $\tilde{g}$ is the subsequent embedding of $g$ into $B$.
The conserved currents $J(z) := \sum\limits_a J^a t^a$ and
$J^a := \sum\limits_{n \in \Z} J^a_n z^{-n-1}$ 
(together with an independent anti-holomorphic copy)
form a {\bf current algebra} which is precisely the level $k$ affine algebra 
$\widehat{g}$:
\[
\left[ J^a_n, J^b_m \right] = i \sum\limits_c f_{abc} J^c_{n+m} + 
	k n \delta_{a b} \delta_{n+m,0}.
\]
The energy momentum tensor $T(z) = \frac{1}{d+k} \sum\limits_a J^a J^a$
with $d$ the dual Coxeter number of $g$ furnishes a Virasoro algebra
with central charge
\[
c(\widehat{g}_k) = \frac{k \dim g}{k+d}.
\]

Moreover, the primary fields are in 1-1 correspondence with the heighest weights
$\widehat{\lambda} \in P^k_+$ of $\widehat{g}$, which, being of a finite number,
constrains the number of primaries to be finite, thereby making WZW a RCFT.
The {\bf fusion algebra} of the primaries $\phi$ for this RCFT is consequently given by
$\phi_i \times \phi_j = \sum\limits_{\phi_k^*} 
{\cal N}_{\phi_i \phi_j}^{\phi_k^*}\phi_k^*$, or in the integrable representation
language of the affine algebra:
\[
\widehat{\lambda} \otimes \widehat{\mu} =
\bigoplus_{\widehat{\nu} \in P_+^k} 
{\cal N}_{\widehat{\lambda}\widehat{\mu}}^{\widehat{\nu}}\ \widehat{\nu}.
\]

The Hilbert Space of states decomposes into holomorphic and anti-holomorphic parts
as
${\cal H} = \bigoplus\limits_{\widehat{\lambda},\widehat{\xi} \in P^{(k)}_+}
{\cal M}_{\widehat{\lambda},\widehat{\xi}} H_{\widehat{\lambda}} 
\otimes H_{\widehat{\xi}}$ 
with the {\bf mass matrix} 
${\cal M}_{\widehat{\lambda},\widehat{\xi}}$
counting the multiplicity
of the $H$-modules in the decomposition. Subsequently, the partition function
over the torus, 
$Z(q) := {\rm Tr}_{\cal H} q^{L_0 - \frac{c}{24}} 
\bar{q}^{\bar{L}_0 - \frac{c}{24}}$
with $q := e^{2 \pi i \tau}$ reduces to
\beq
\label{Z}
Z(\tau) = \sum_{\widehat{\lambda},\widehat{\xi} \in P^k_+}
        \chi_{\widehat{\lambda}}(\tau)
        {\cal M}_{\widehat{\lambda},\widehat{\xi}}
        \bar{\chi}_{\widehat{\xi}}(\bar{\tau})
\eeq
with $\chi$ being the affine characters of $\widehat{g}_k$.
Being a partition function on the torus, (\ref{Z}) must obey the $SL(2;\Z)$
symmetry of $T^2$, i.e., it must be invariant under the {\bf modular group}
generated by $S : \tau \rightarrow -1/\tau$ and $T : \tau \rightarrow \tau + 1$.
Recalling the modular transformation properties of the affine characters, viz.,
\[
\begin{array}{lll}
T : \chi_{\widehat{\lambda}} & \rightarrow & \sum\limits_{\widehat{\mu} \in P^k_+}
	{\cal T}_{\widehat{\lambda}\widehat{\mu}} \chi_{\widehat{\mu}} \\
S : \chi_{\widehat{\lambda}} & \rightarrow & \sum\limits_{\widehat{\mu} \in P^k_+}
	{\cal S}_{\widehat{\lambda}\widehat{\mu}} \chi_{\widehat{\mu}}
\end{array}
\]
with
\[
\begin{array}{lll}
{\cal T}_{\widehat{\lambda} \widehat{\mu}} & = & \delta_{\widehat{\lambda}\widehat{\mu}}
	e^{\pi i (\frac{| \widehat{\lambda} + \widehat{\rho} |^2}{k+d} - 
			\frac{|\widehat{\rho}|^2}{d})} \\
{\cal S}_{\widehat{\lambda} \widehat{\mu}} & = & K\sum\limits_{w \in W} \epsilon(w)
	e^{-\frac{2 \pi i}{k + d} (w(\lambda + \rho), \mu + \rho)}
\end{array}
\]
where $\widehat{\rho}$ is the sum of the fundamental weights, $W$, the Weyl group
and $K$, some proportionality constant.
Modular invariance of (\ref{Z}) then implies $\left[{\cal M}, {\cal S} \right] =
\left[{\cal M}, {\cal T} \right] = 0$. 
The problem of classfication of the physical
modular invariants of $\widehat{g}_k$-WZW then amounts to solving for
{\em all nonnegative
integer matrices ${\cal M}$ such that ${\cal M}_{00} = 1$} (so as to guarantee uniqueness
of vacuum) and {\em satisfying these commutant relations.}

The fusion coefficients ${\cal N}$ can be, as it is with modular tensor categories
(q.v. e.g. \cite{9903056}), related to the matrix ${\cal S}$ by the celebrated
Verlinde Formula:
\beq
\label{Verlinde}
{\cal N}^t_{rs} = \sum\limits_m 
	\frac{{\cal S}_{rm}{\cal S}_{sm}{\cal S}^{-1}_{mt}}{{\cal S}_{0m}}.
\eeq
Furthermore, in light of the famous McKay Correspondence 
(Cf. e.g. \cite{9811183,9903056} 
for discussions of the said correspondence in this context), to establish 
correlations between modular invariants and graph theory, one can
chose a fundamental representation $f$ and regard $(N)_{st} := {\cal N}^t_{fs}$ as
an adjacency matrix of a finite graph. Conversely out of the adjacency
matrix $(G)_{st}$ for some finite graph, one can extract a set of matrices 
$\{(N)_{st}\}_i$ such that $N_0 = \I$ and $N_f = G$. We diagonalise $G$ as
${\cal S} \Delta {\cal S}^{-1}$ and define, as inspired by (\ref{Verlinde}),
the set of matrices 
$N_r := \{(N)_{st}\}_r = \sum\limits_m 
\frac{{\cal S}_{rm}{\cal S}_{sm}{\cal S}^{-1}_{mt}}{{\cal S}_{0m}}$, which clearly
satisfy the constriants on $N_{0,f}$.
This set of matrices $\{N_i\}$, each associated to a vertex in the judiciously chosen graph,
give rise to a {\bf graph algebra} and appropriate subalgebras thereof, by virtue of
matrix multiplication, constitute a representation for the fusion algebra, i.e.,
$N_i \cdot N_j = \sum\limits_k {\cal N}_{ij}^k N_k$. In a more axiomatic language,
the Verlinde equation (\ref{Verlinde}) is essentially the inversion of the McKay
composition
\beq
\label{McKay}
R_r \otimes R_s = \bigoplus\limits_t {\cal N}^t_{rs} R_t
\eeq
of objects $\{R_i\}$ in a (modular) tensor category. 
The ${\cal S}$ matrices are then the characters
of these objects and hence the matrix of eigenvectors of $G = {\cal N}^t_{rs}$ once
fixing some $r$ by definition (\ref{McKay}). The graph algebra is essentially the
set of these matrices ${\cal N}^t_{rs}$ as we extrapolate $r$ from 0 (giving $\I$)
to some fixed value giving the graph adjacency matrix $G$.

Thus concludes our brief review on the current affair of things. Let
us now proceed to present our small observation.
\section*{Nomenclature}
Throughout the chapter, unless otherwise stated, we shall adhere to the
folloing conventions:
$G_n$ is group $G$ of order $n$.
$\gen{x_i}$ is the group generated by the (matrix) elements $\{x_i\}$.
$k$ is the level of the WZW modular invariant partition function $Z$.
$\chi$ is the affine character of the algebra $\widehat{g}$.
${\cal S},{\cal T}$ are the generators of the modular group
$SL(2;\Z)$ whereas $S,T$ will be these matrices in a new basis, to be
used to generate a finite group. $E_{6,7,8}$ are the ordinary
tetrahedral, octahedral and icosahedral groups while
$\widehat{E_{6,7,8}}$ are their binary counterparts. Calligraphic font
(${\cal A,D,E}$) shall be reserved for the names of the modular invariants.

\section{$\widehat{su(2)}$-WZW}
\index{McKay Correspondence!and WZW}
The modular invariants of $\widehat{su(2)}$-WZW were originally classified in
the celebrated works of \cite{su21,su22}.
The only solutions of the abovementioned conditions for 
$k,{\cal S}, {\cal T}$ and ${\cal M}$ give rise to the following:
\beq
\label{su2ST}
{\cal S}_{ab}=\sqrt{{2\over k+2}}\,\sin(\pi\,{(a+1)(b+1)\over {k+2}}),\qquad 
{\cal T}_{ab}=\exp[ \pi i ({(a+1)^2\over 2(k+2)} - {1 \over 4})]\,\delta_{a,b}
\qquad a,b = 0,...,k
\eeq
with the partition functions
\beq
\label{su2Z}
\begin{array}{lll}
k & {\cal A}_{k+1} & Z = \sum\limits_{n=0}^{k} |\chi_n|^2 \\

k = 4m & {\cal D}_{2m+2} & Z = \sum\limits_{n=0,\even}^{2m-2} 
	|\chi_n + \chi_{k - n}|^2 + 2 |\chi_{2m}|^2 \\

k = 4m-2 & {\cal D}_{2m+1} & Z = |\chi_{k \over 2}|^2 + 
	\sum\limits_{n=0,\even}^{4m-2} |\chi_n|^2 + 
	\sum\limits_{n=1,\odd}^{2m-1}(\chi_n \bar{\chi}_{k-n} + c.c.) \\

k = 10 & {\cal E}_6 & Z = |\chi_0+\chi_6|^2+|\chi_3+\chi_7|^2+|\chi_4+\chi_{10}|^2 \\

k = 16 & {\cal E}_7 & Z = |\chi_0+\chi_{16}|^2+|\chi_4+\chi_{12}|^2+|\chi_6+\chi_{10}|^2
	+(\bar{\chi}_8(\chi_2+\chi_{14}) + c.c.) \\

k = 28 & {\cal E}_8 & Z = |\chi_0+\chi_{10}+\chi_{18}+\chi_{28}|^2+
	|\chi_6+\chi_{12}+\chi_{16}+\chi_{22}|^2 \\
\end{array}
\eeq

We know of course that the simply-laced simple Lie algebras, as well as the
discrete subgroups of $SU(2)$ fall precisely under such a classification. The now
standard method is to associate the modular invariants to subalgebras of the
graph algebras constructed out of the respective ADE-Dynkin Diagram. This is done in
the sense that the adjacency matrices of these diagrams\footnote{These are the well-known 
	symmetric matrices of eigenvalues $\le 2$, or equivalently, the McKay 
	matrices for $SU(2)$; for a discussion on this point q.v. e.g. 
\cite{9911114}.}
are to define $N_1$ and subsets of $N_i$ determine the fusion rules. The correspondence
is rather weak, for in addition to the necessity of the truncation to
subalgebras, only $A_k$, $D_{2k}$
and $E_{6,8}$ have been thus related to the graphs while $D_{2k+1}$ and $E_7$
give rise to negative entries in ${\cal N}_{ij}^k$. However as an encoding process,
the above correspondences has been very efficient, especially in generalising to WZW
of other algebras.

The first attempt to explain the ADE scheme in the $\widehat{su(2)}$ modular invariants
was certainly not in the sophistry of the above context. It was in fact done in the
original work of \cite{su22}, where the authors sought to relate their invariants
to the discrete subgroups of $SO(3) \cong SU(2)/\Z_2$. It is under the inspiration of
this idea, though initially abandoned ({\it cit. ibid.}), that the current
writing has its birth. We do not promise to find a stronger correspondence, yet we
shall raise some observations of interest.

The basic idea is simple. To ourselves we pose the obvious question: what, algebraically
does it mean for our partition functions (\ref{su2Z}) to be modular invariant?
It signifies that the action by ${\cal S}$ and ${\cal T}$ thereupon must permute the
terms thereof in such a way so as not to, by virtue of the transformation properties
of the characters (typically theta-functions), introduce extraneous terms.
In the end of the monumental work \cite{su22}, the authors, as a diversion,
used complicated identities of theta and eta functions to rewrite the
${\cal E}_{6,7,8}$ cases of (\ref{su2Z}) into sum of
terms on whose powers certain combinations of ${\cal S}$ and ${\cal T}$ act. These
combinations were then used to generate finite groups which in the
case of ${\cal E}_6$, did
give the ordinary tetrahedral group $E_6$ and ${\cal E}_8$, the
ordinary icosahedral group $E_8$, which are
indeed the finite groups associated to these Lie algebras, a fact which dates back to
F. Klein. As a postlude, \cite{su22} then speculated upon the reasons for this correspondence
between modular invariants and these finite groups, as being attributable to the
representation of the modular groups over finite fields, since afterall
$E_6 \cong PSL(2;\Z_3)$ and $E_8 \cong PSL(2;\Z_4) \cong PSL(2;\Z_5)$.

We shall not take recourse to the complexity of manipulation of theta functions
and shall adhere to a pure group theoretic perspective. We translate the aforementioned
concept of the permutation of terms into a vector space language.
First we interpret the characters appearing in (\ref{su2Z}) as basis upon which
${\cal S}$ and ${\cal T}$ act. For the $k$-th level they are defined as the canonical
bases for $\C^{k+1}$:
\[
\chi_0 := (1,0,...,0); \quad ... \quad \chi_i := (\I)_{i+1}; 
\quad ... \quad \chi_k := (0,0,...,1).
\]
Now ${\cal T}$ being diagonal clearly maps these vectors to multiples of themselves
(which after squaring the modulus remain uneffected); the interesting
permutations are performed by ${\cal S}$.
\subsection{The $E_6$ Invariant}
Let us first turn to the illustrative example of ${\cal E}_6$. 
From $Z$ in (\ref{su2Z}), we see that we are clearly interested in the vectors
$v_1 := \chi_0 + \chi_6 = (1, 0, 0, 0, 0, 0, 1, 0, 0, 0, 0),
 v_2 := \chi_4 + \chi_{10} = (0, 0, 0, 0, 1, 0, 0, 0, 0, 0, 1)$ and
$v_3 := \chi_3 + \chi_7 = (0, 0, 0, 1, 0, 0, 0, 1, 0, 0, 0)$.
Hence (\ref{su2ST}) gives
${\cal T} : v_1 \rightarrow e^{{-{5 \pi i} \over {24}} } v_1$,
${\cal T} : v_2 \rightarrow e^{{{19 \pi i} \over {24}} } v_2$ and
${\cal T} : v_3 \rightarrow e^{{{ 5 \pi i} \over {12}} } v_3$.
Or, in other words in the subspace spanned by $v_{1,2,3}$, ${\cal T}$ acts
as the matrix $T := {\rm Diag}(e^{{-{5 \pi i} \over {24}}},
	e^{{{19 \pi i} \over {24}}}, e^{{{ 5 \pi i} \over {12}}})$.
Likewise, ${\cal S}$ becomes a 3 by 3 matrix; we present them below:
\beq
\label{e6newST}
S =
\left( 
\matrix{ {1\over 2} & {1\over 2} & {1\over {{\sqrt{2}}}} \cr
    {1\over 2} & {1\over 2} & -{1\over {{\sqrt{2}}}} \cr
    {1\over {{\sqrt{2}}}} & -{1\over {{\sqrt{2}}}} & 0 }
\right)
\qquad
T =
\left(
\matrix{{e^{-{{5 \pi i}\over {24}}}} & 0 & 0
   \cr 0 & {e^{{{19 \pi i}\over {24}}}} & 0 ,
   \cr 0 & 0 & {e^{{{5 \pi i}\over {12}}}}}
\right)
\eeq

Indeed no extraneous vectors are involved, i.e., of the 11 vectors  $\chi_i$ 
and all combinations of sums thereof, only the combinations $v_{1,2,3}$ appear
after actions by ${\cal S}$ and ${\cal T}$. This closure of course is what is needed
for modular invariance. What is worth of note, is that we have collapsed an
11-dimensional representation of the modular group acting on $\{\chi_i\}$, to
a (non-faithful) 3-dimensional representation which corresponds the
subspace of interest (of the initial $\C^{11}$) by virtue of the appearance of the
terms in the associated modular invariant. Moreover the new matrices
$S$ and $T$, being of finite order (i.e., $\exists m,n \in \Z_+$ s.t. $S^m = T^n = \I$),
actually generate a {\it finite group}. It is this finite group that we shall compare to
the ADE-subgroups of $SU(2)$.

The issue of the finiteness of the initial group generated 
by ${\cal S}$ and ${\cal T}$ was addressed in a recent work by Coste and Gannon \cite{Coste}.
Specifically, the group
\beq
\label{poly}
P := \{S,T | T^N = S^2 = (ST)^3 = \I \},
\eeq
generically known as the {\em polyhedral (2,3,N) group}, is infinite
for $N > 5$.
On the other hand, for $N=2,3,4,5$, $G \cong \Gamma / \Gamma(N) := SL(2;\Z/N\Z)$,
which, interestingly enough, for these small values are, the symmetric-3, the tetrahedral,
the octahedral and icosahedral groups respectively.

We see of course that our matrices in \eref{e6newST} satisfy
the relations of \eref{poly} with $N=48$ (along with additional
relations of course) and hence generates a subgroup of $P$. Indeed,
$P$ is the modular group in a field of finite characteristic $N$ and
since we are dealing with nonfaithful representations of the modular
group, the groups generated by $S,T$, as we shall later see, in the
cases of other modular invariants are all finite subgroups of $P$.

In our present case, $G=\gen{S,T}$ is of order 1152. Though $G$ itself
may seem unenlightening, upon closer inspection we find that it has
12 normal subgroups $H \triangleleft G$ and {\em only one} of which is of
order 48. In fact this $H_{48}$ is $\IZ_4 \times \IZ_4 \times
\IZ_3$. The observation is that the quotient group formed between $G$
and $H$ is precisely the binary tetrahedral group $\widehat{E_6}$, i.e.,
\beq
\label{e6res}
G_{1152} / H_{48} \cong \widehat{E_6}.
\eeq

We emphasize again the {\em uniqueness} of this procedure: as will be
with later examples, given $G({\cal E}_6)$, there exists a unique
normal subgroup which can be quotiented to give $\widehat{E_6}$, and
moreover there does not exist a normal subgroup which could be used to
generate the other exceptional groups, viz.,  $\widehat{E_{7,8}}$. We
shall later see that such a 1-1 correpondence between the exceptional modular
invariants and the exceptional discrete groups persists.

This is a pleasant surprise; it dictates that the symmetry group
generated by the permutation of the terms in the ${\cal E}_6$ modular
invariant partition function of $\widetilde{SU(2)}$-WZW, upon
appropriate identification, is exactly the symmetry group assocaited
to the $\widehat{E_6}$ discrete subgroup of $SU(2)$. Such a correspondence may
{\it a priori} seem rather unexpected.
\subsection{Other Invariants}
It is natural to ask whether similar circumstances arise for the
remaining invariants. Let us move first to the the case of $E_8$. By
procedures completely analogous to \eref{e6newST} as applied to the
partition function in \eref{su2Z}, we see that the basis is composed
of $v_1 =
\chi_0 + \chi_{10} + \chi_{18} + \chi_{28} = \{1, 0, 0, 0, 0, 0, 0, 0,
0, 0, 1, 0, 0, 0, 0, 0, 0, 0, 1, 0, 0, 0, 0, 0, 0, 0, 0, 0, 1\}$ and
$v_2 = \chi_6 + \chi_{12} + \chi_{16} + \chi_{22} = \{0, 0, 0, 0, 0,
0, 1, 0, 0, 0, 0, 0, 1, 0, 0, 0, 1, 0, 0, 0, 0, 0, 1, 0, 0, 0, 0, 0, 0
\}$, under which $S$ and $T$ assume the forms as summarised in
Table \ref{Esummary}.

This time $G = \gen{S,T}$ is of order 720, with one {\it unique}
normal subgroup of order 6 (in fact $\IZ_6$). Moreover we find that
\beq\label{e8res}
G_{720} / H_6 \cong \widehat{E_8},
\eeq
in complete analogy with \eref{e6res}. Thus once again, the symmetry
due to the permutation of the terms inherently encode the associated
discrete $SU(2)$ subgroup.

What about the remaining exceptional invariant, $E_7$? The basis as
well as the matrix forms of $S,T$ thereunder are again presented in
Table \ref{Esummary}. The group generated thereby is of order 324,
with 2 non-trivial normal subgroups of orders 27 and
108. Unfortunately, no direct quotienting could possibly give the
binary octahedral group here. However $G / H_{27}$ gives a group of
order 12 which is in fact the {\it ordinary} octahedral group $E_7=A_4$, which
is in turn isomorphic to $\widehat{E_7} / \IZ_2$. Therefore for our present
case the situation is a little more involved:
\beq\label{e7res}
G_{324} / H_{27} \cong \widehat{E_7} / \IZ_2 \cong E_7.
\eeq

We recall \cite{CFT} that a graph algebra \eref{Verlinde} based on the
Dynkin diaram of $E_7$ has actually not been succesully constructed
for the $E_7$ modular invariant. Could we speculate that the slight
complication of \eref{e7res} in comparison with \eref{e6res} and
\eref{e8res} be related to this failure?

We shall pause here with the exceptional series as for the infinite
series the quotient of the polyhedral $(2,3,N)$ will never give any
abelian group other than $\IZ_{1,2,3,4,6}$ or any dihedral group other
than $D_{1,3}$ \cite{Com}. More complicated procedures are called for
which are yet to be ascertained \cite{Prog}, though we remark here
briefly that for the $A_{k+1}$ series, since $Z$
is what is known as the {\it diagonal invariant}, i.e., it includes all
possible $\chi_n$-bases, we need not perform any basis change and
whence $S,T$ are simply the original ${\cal S,T}$ and there is an
obvious relationship that $G := \gen{T^8} \cong \IZ_{k+2}
:= A_{k+1}$.

Incidentally, we can ask ourselves whether any such correspondences
could possibly hold for the {\em ordinary} exceptional groups. From
\eref{e7res} we see that $G({\cal E}_7)/H_{27}$ does indeed correspond
to the ordinary octahedral group. Upon further investigation, we find
that $G({\cal E}_6)$ could not be quotiented to give the ordinary $E_6$
while $G({\cal E}_8)$ does have a normal subgroup of order 12 which
could be quotiented to give the ordinary $E_8$.
Without much further
ado for now, let us summarise these results:
\[
\ba{|c|c|c|c|c|}
\hline
	& G := \gen{S,T} & \mbox{Normal Subgroups} &
		\multicolumn{2}{|c|}{\mbox{Relations}} \\ \hline
{\cal E}_6 & G_{1152} & H_{3,4,12,16,48,64,192,192',384,576} &
	G_{1152} / H_{48} \cong \widehat{E_6} & - \\
{\cal E}_7 & G_{324} & H_{27,108} & G_{324} / H_{27} \cong \widehat{E_7} / \IZ_2
	& G_{324} / H_{27} \cong E_7 \\
{\cal E}_8 & G_{720} & H_{2,3,4,6,12,120,240,360} & G_{720} / H_6
	\cong \widehat{E_8} & G_{720} / H_{12} \cong E_8 \\
\hline
\ea
\]
\subsubsection*{Table of $SU(2)$ Exceptional Invariants}
\beq
\label{Esummary}
{\scriptsize \hspace{-0.5cm}
\begin{array}{|c|c|c|}
\hline
	& \mbox{\normalsize Matrix Generators}	&  \mbox{\normalsize
Basis} \\ \hline 
E_6	& S = \left( 
	\matrix{ {1\over 2} & {1\over 2} & {1\over {{\sqrt{2}}}} \cr
    		{1\over 2} & {1\over 2} & -{1\over {{\sqrt{2}}}} \cr
    		{1\over {{\sqrt{2}}}} & -{1\over {{\sqrt{2}}}} & 0 }
	\right)
	  T = \left(\matrix{{e^{{{-5 \pi i}\over {24}}}} & 0 & 0
   		\cr 0 & {e^{{{19 \pi i}\over {24}}}} & 0 ,
   		\cr 0 & 0 & {e^{{{5 \pi i}\over {12}}}}}
	\right)
	&
	\begin{array}{l}
		v_1 = \chi_0 + \chi_6 \\
		v_2 = \chi_4 + \chi_{10} \\
		v_3 = \chi_3 + \chi_7 \\
	\end{array}
	\\ &&\\
\hline
E_7	& 
	\begin{array}{c}
	S = {1 \over 3} \left(
 	\matrix{ \sin ({{\pi }\over {18}}) + \sin ({{17\,\pi }\over {18}}) & 
   		\sin ({{5\,\pi }\over {18}}) + \sin ({{85\,\pi }\over {18}}) & 
   		\sin ({{7\,\pi }\over {18}}) + \sin ({{119\,\pi }\over {18}}) & 2 & 1 \cr 
   		\sin ({{5\,\pi }\over {18}}) + \sin ({{13\,\pi }\over {18}}) & 
   		\sin ({{25\,\pi }\over {18}}) + \sin ({{65\,\pi }\over {18}}) & 
   		\sin ({{35\,\pi }\over {18}}) + \sin ({{91\,\pi }\over {18}}) & 2 & 1 \cr 
   		\sin ({{7\,\pi }\over {18}}) + \sin ({{11\,\pi }\over {18}}) & 
   		\sin ({{35\,\pi }\over {18}}) + \sin ({{55\,\pi }\over {18}}) & 
   		\sin ({{49\,\pi }\over {18}}) + \sin ({{77\,\pi }\over {18}}) & -2 & -1
    		\cr 1 & 1 & -1 & 1 & -1 \cr 1 & 1 & -1 & -2 & 2 \cr  } 
	\right) \\ \\
	T = \left(\matrix{ {e^{{{-2\,i}\over 9}\,\pi }} & 0 & 0 & 0 & 0 \cr 0 & 
	   	{e^{{{4\,i}\over 9}\,\pi }} & 0 & 0 & 0 \cr 0 & 0 & 
   		{e^{{{-8\,i}\over 9}\,\pi }} & 0 & 0 \cr 0 & 0 & 0 & 1 & 0 \cr 0 & 0 & 0 & 
   		0 & 1 \cr  }
	\right)
	\end{array}
	&
	\begin{array}{l}
		v_1 = \chi_0+\chi_{16} \\
		v_2 = \chi_4+\chi_{12} \\
		v_3 = \chi_6+\chi_{10} \\
		v_4 = \chi_8 \\
		v_5 = \chi_2+\chi_{14} \\
	\end{array} \\
\hline &&\\
E_8	&
	\begin{array}{c}
	S = {1 \over \sqrt{15}} \left(
\matrix{ \matrix{\sin ({{\pi }\over {30}}) + \sin ({{11\,\pi
	}\over {30}}) + \cr 
     \quad + \sin ({{19\,\pi }\over {30}}) + \sin ({{29\,\pi }\over {30}}) 
	} & 
   \matrix{ \sin ({{7\,\pi }\over {30}}) + \sin ({{77\,\pi }\over {30}}) + \cr
    \quad + \sin ({{133\,\pi }\over {30}}) + \sin ({{203\,\pi }\over {30}})} \cr 
   \matrix{\sin ({{7\,\pi }\over {30}}) + \sin ({{13\,\pi }\over {30}}) + \cr
    \quad + \sin ({{17\,\pi }\over {30}}) + \sin ({{23\,\pi }\over {30}})} & 
   \matrix{\sin ({{49\,\pi }\over {30}}) + \sin ({{91\,\pi }\over {30}}) + \cr
     \quad +\sin ({{119\,\pi }\over {30}}) + \sin ({{161\,\pi }\over {30}})}
	\cr  
	} \right)\\ \\
	T = \left(\matrix{ {e^{{{-7\,i}\over {30}}\,\pi }} & 0 \cr 0 & 
   	{e^{{{17\,i}\over {30}}\,\pi }} \cr  } \right)
	\end{array}
	&
	\begin{array}{l}
		v_1 = \chi_0+\chi_{10}+\chi_{18}+\chi_{28} \\
		v_2 = \chi_6+\chi_{12}+\chi_{16}+\chi_{22} \\
	\end{array} \\
\hline
\end{array}
}
\eeq
\section{Prospects: $\widehat{su(3)}$-WZW and Beyond?}
\index{McKay Correspondence!in string theory}
There has been some recent activity \cite{DiFrancesco,9811183,9903056,Song}
in attempting to explain the patterns emerging in the modular invariants
beyond $\widehat{su(2)}$. Whether from the perspective of integrable systems,
string orbifolds or non-linear sigma models, proposals of the invariants
being related to subgroups of $SU(n)$ have been made.
It is natural therefore for us to inquire whether the correspondences from the
previous subsection between $\widehat{su(n)}$-WZW and the discrete subgroups of
$SU(n)$ for $n=2$ extend to $n=3$.

We recall from \cite{su31,su32} that the modular invariant partition functions
for $\widehat{su(3)}$-WZW have been classified to be the following:
\beq
\label{su3Z}
\begin{array}{lll}
{\cal A}_k & := &  \sum\limits_{\lambda \in P^k} |\chi_\lambda^k|^2,\qquad
	\forall k\ge 1; \cr

{\cal D}_k & := & \sum\limits_{(m,n)\in P^k} \chi^k_{m,n} \chi^{k*}_{\omega^{k(m-n)}(m,n)}, 
	\quad {\rm for}~k \not\equiv 0 \bmod 3~{\rm and}~k \ge 4; \cr

{\cal D}_k & := & {1\over 3}
	\sum\limits_{{(m,n)\in P^k} \atop {m \equiv n \bmod 3}}
	|\chi^k_{m,n}+\chi_{\omega(m,n)}^k+\chi^k_{\omega^2(m,n)}|^2; \cr

{\cal E}_5 & := & |\chi^5_{1,1}+\chi^5_{3,3}|^2+|\chi^5_{1,3}+\chi^5_{4,3}|^2+
	|\chi^5_{3,1}+\chi^5_{3,4}|^2 + \\&&
	|\chi^5_{3,2}+\chi^5_{1,6}|^2+|\chi^5_{4,1}+
	\chi^5_{1,4}|^2+|\chi^5_{2,3}+\chi^5_{6,1}|^2; \cr

{\cal E}_9^{(1)} & := & |\chi^9_{1,1}+\chi^9_{1,10}+\chi^9_{10,1}+\chi^9_{5,5}+
	\chi^9_{5,2}+\chi^9_{2,5}|^2+2|\chi^9_{3,3}+\chi^9_{3,6}+\chi^9_{6,3}|^2; \cr 

{\cal E}_9^{(2)} & := &  |\chi^9_{1,1}+\chi^9_{10,1}+\chi^9_{1,10}|^2+|\chi^9_{3,3}+
	\chi^9_{3,6}+\chi^9_{6,3}|^2+ 2|\chi^9_{4,4}|^2 \\&&
	+ |\chi^9_{1,4}+\chi^9_{7,1}+ \chi^9_{4,7}|^2+|\chi^9_{4,1}+\chi^9_{1,7}+
	\chi^9_{7,4}|^2+|\chi^9_{5,5}+\chi^9_{5,2}+\chi^9_{2,5}|^2 \\&&
	+ (\chi^9_{2,2}+\chi^9_{2,8}+\chi^9_{8,2})\chi^{9*}_{4,4}+
	\chi^9_{4,4}(\chi^{9*}_{2,2}+\chi^{9*}_{2,8}+\chi^{9*}_{8,2}); \cr

{\cal E}_{21} & := & |\chi^{21}_{1,1}+\chi^{21}_{5,5}+\chi^{21}_{7,7}+
	\chi^{21}_{11,11}+\chi^{21}_{22,1}+\chi^{21}_{1,22}
	+\chi^{21}_{14,5}+\chi^{21}_{5,14}+\chi^{21}_{11,2}+\chi^{21}_{2,11}+\chi^{21}_{10,7}
	+\chi^{21}_{7,10}|^2 \\&&
	+|\chi^{21}_{16,7}+\chi^{21}_{7,16}+\chi^{21}_{16,1}+\chi^{21}_{1,16}+
		\chi^{21}_{11,8}+\chi^{21}_{8,11}
	+\chi^{21}_{11,5}+\chi^{21}_{5,11}+
		\chi^{21}_{8,5}+\chi^{21}_{5,8}+\chi^{21}_{7,1}+\chi^{21}_{1,7}|^2; \cr
\end{array}
\eeq
where we have labeled the level $k$ explicitly as subscripts. Here the highest weights
are labeled by two integers $\lambda = (m,n)$ as in the set
\[	
P^k := \{\lambda = m \beta_1 + n\beta_2\,|\,m,n\in\Z,~ 0 <  m, n, m+n < k+3 \}
\]
and $\omega$ is the operator $\omega :(m,n) \rightarrow (k+3-m-n,n)$. The modular
matrices are simplified to
\beq
\label{su3ST}
\begin{array}{lll}
{\cal S}_{\lambda \lambda'} & = & {-i \over {\sqrt{3}(k+3)}} \bigr\{ e_k(2mm'+mn'+nm'+2nn')
	+e_k(-mm'-2mn'-nn'+nm') \\&& +e_k(-mm'+mn'-2nm'-nn')-e_k(-2mn'-mm'-nn'-2nm') \\&&
	-e_k(2mm'+mn'+nm'-nn')-e_k(-mm'+mn'+nm'+2nn')\} \cr
{\cal T}_{\lambda \lambda'} & = & e_k(-m^2-mn-n^2+k+ 3)\,\delta_{m,m'}\,\delta_{n,n'}
\end{array}
\eeq
with $e_k(x) := \exp[{-2\pi i x\over 3(k+3)}]$.

We imitate the above section and attempt to generate various finite
groups by $S,T$ under appropriate transformations from \eref{su3ST} to
new bases. We summarise the results below:
\[
\ba{|c|c|c|}
\hline
& {\rm Basis} & G:=\gen{S,T}\\
\hline
{\cal E}_5 & \{\chi_{1,1}+\chi_{3,3}; \chi_{1,3}+\chi_{4,3};
         \chi_{3,1}+\chi_{3,4}; \chi_{3,2}+\chi_{1,6};
         \chi_{4,1}+\chi_{1,4}; \chi_{2,3}+\chi_{6,1}\}
& G_{1152}\\ \hline
{\cal E}_9^{(1)} &
\{\chi_{1,1}+\chi_{1,10}+\chi_{10,1}+\chi_{5,5}+\chi_{5,2}+\chi_{2,5}];
  \chi_{3,3}+\chi_{3,6}+\chi_{6,3}\}
& G_{48}\\ \hline
{\cal E}_9^{(2)} & 
\ba{c}
\{\chi_{1,1}+\chi_{1,10}+\chi_{10,1};
\chi_{5,5}+\chi_{5,2}+\chi_{2,5}; \\
\chi_{3,3}+\chi_{3,6}+\chi_{6,3}; \chi_{4,4};
\chi_{4,1}+\chi_{1,7}+\chi_{7,4}; \\
\chi_{1,4}+\chi_{7,1}+\chi_{4,7};
\chi_{2,2}+\chi_{2,8}+\chi_{8,2}\}
\ea
& G_{1152}\\ \hline
{\cal E}_{21} &
\ba{c}
\{\chi_{1,1}+\chi_{5,5}+\chi_{7,7}+\chi_{11,11}+\chi_{22,1}+
 \chi_{1,22}+\chi_{14,5}+\chi_{5,14}+ \\
 \chi_{11,2}+\chi_{2,11}+ \chi_{10,7}+\chi_{7,10}; \\
 \chi_{16,7}+\chi_{7,16}+\chi_{16,1}+\chi_{1,16}+\chi_{11,8}+
 \chi_{8,11} +\\
 \chi_{11,5}+\chi_{5,11}+\chi_{8,5}+\chi_{5,8}+
 \chi_{1,7}+\chi_{7,1}\}
\ea
& G_{144} \\ \hline
\ea
\]

We must confess that unfortunately the direct application of our
technique in the previous section has yielded no favourable results,
i.e., no quotients groups of $G$ gave any of the exceptional $SU(3)$
subgroups $\Sigma_{36\times3, 72\times3, 216\times 3, 360\times3}$ or
nontrivial quotients thereof (and {\it vice versa}), even though the
fusion graphs for the former and the McKay quiver for the latter have
been pointed out to have certain similarities
\cite{DiFrancesco,Reflection,9811183}. These similarities are a little
less direct than the Mckay Correspondence for $SU(2)$ and involve
truncation of the graphs, the above failure of a na\"{\i}ve
correspondence by quotients may be related to this complexity.

Therefore much work yet remains for us. Correspondences
for the infinite series in the $SU(2)$ case still needs be formulated
whereas a method of attack is still pending for $SU(3)$ (and
beyond). It is the main purpose of this short note to inform the
reader of an intriguing correspondence between WZW modular invariants
and finite groups which may hint at some deeper mechanism yet to be
uncovered.

\index{Hanany-Witten!brane box}
\chapter{Orbifolds V: The Brane Box Model for $\IC^3 / Z_k \times D_{k'}$}
\label{chap:9906031}
\section*{\center{{ Synopsis}}}

In the next four chapters we shall study the T-dual aspects of
what had been discussed in the previous chapters;
these are the so-called
Hanany-Witten brane setups. 

In this chapter,
an example of a non-Abelian Brane Box Model, namely 
one corresponding to a $Z_k \times D_{k'}$ orbifold singularity of
$\IC^3$, is constructed. Its self-consistency and hence equivalence to
geometrical methods are subsequently shown. 
It is demonstrated how a group-theoretic twist of the
non-Abelian group circumvents the problem of inconsistency that arise
from na\"{\i}ve attempts at the construction \cite{9906031}.
\newpage
\section{Introduction}
Brane setups \cite{HW} have been widely attempted to 
provide an alternative to algebro-geometric methods in the
construction of gauge  
theories (see \cite{branerev} and references therein). The advantages
of the latter  
include the enlightening of important properties of manifolds such as mirror
symmetry, the provision of convenient supergravity descriptions and in instances
of pure geometrical engineering, the absence of non-perturbative objects.
The former on the other hand, give
intuitive and direct treatments of the gauge theory. One can conveniently
read out much information concerning the
gauge theory from the brane setups, such as the dimension of the Coulomb and Higgs 
branches \cite{HW}, the mirror symmetry \cite{HW,Boer,Kapustin,
P-Zaf} in 3 
dimensions first shown in \cite{Seiberg}, 
the Seiberg-duality in 4 
dimensions \cite{Methods}, and exact solutions
when we lift the setups from Type IIA to M Theory \cite{M-Theory}.

In particular, when discussing ${\cal N}=2$ supersymmetric gauge theories in 
4 dimensions, there are three known methods currently in favour.
The first method is {\it geometrical engineering}
exemplified by works in \cite{geoeng3}; 
the second uses D3 branes as {\it probes} on orbifold singularities of 
the type 
$\C^2/\Gamma$ with $\Gamma$ being a finite discrete subgroup of $SU(2)$
\cite{DM}, and the third, the usage of {\it brane setups}.
These three approaches are related to each other by
proper T or S Dualities \cite{Karch:equiv,Karch2}.
For example,
the configuration of stretching Type IIA D4 branes between $n+1$ NS5 
branes placed in a 
circular fashion, the so-called {\bf elliptic model}\footnote{We call it elliptic 
even though
there is only an $S^1$ upon which we place the D4 branes; this is because from the
M Theory perspective, there is another direction: an $S^1$ on which we compactify to 
obtain type Type IIA. The presence of two $S^1$'s makes the theory toroidal, or
elliptic.
Later we shall see how to make use of $T^2=S^1 \times S^1$ in Type IIB.
For clarity we shall refer to the former as the ${\cal N}=2$ elliptic model and the latter,
the ${\cal N}=1$ elliptic model.}, 
is precisely T-dual to D3 branes stacked upon ALE\footnote{ 
Asymptotically Locally Euclidean, i.e., Gorenstein singularities that locally represent
Calabi-Yau manifolds.} singularities of type $\widehat{A_n}$ (see
\cite{M-Theory,Karl,B-Karch,Park,Erlich} for detailed discussions).

The above constructions can be easily generalised to ${\cal N}=1$ 
supersymmetric field theories in 4 dimensions. 
Methods from geometric engineering as well as D3 branes as probes
now dictate the usage of orbifold singularities of the type 
$\C^3/\Gamma$ with $\Gamma$ being a finite discrete subgroup of 
$SU(3)$ \cite{KS,LNV,BKV,DG,DGM}.
A catalogue of all the discrete subgroups of $SU(3)$ in this context
is given in \cite{9811183,Muto}.
Now from the brane-setup point of view, there are two ways to arrive at 
the theory. The first is to rotate certain branes in the configuration to
break the supersymmetry from ${\cal N}=2$ to ${\cal N}=1$ \cite{Methods}. 
The alternative is to add
another type of NS5 branes, viz., a set of NS5$'$ branes placed
perpendicularly to the original NS5, whereby
constructing the so-called {\bf Brane Box Model} \cite{HZ,HU}. 
Each of these two different approaches has its own merits. 
While the former (rotating branes) facilitates the deduction of Seiberg Duality, 
for the latter (Brane Box Models), it is easier to construct a class of new,
finite, chiral 
field theories \cite{HSU}.  By finite we mean that in the field theory the 
divergences may be cancelable.
From the perspective of branes on geometrical singularities, 
this finiteness corresponds to the
cancelation of tadpoles in the orbifold background and from that of
brane setups, it corresponds to the no-bending requirement of 
the branes \cite{Karch:equiv,Karch2,HSU,LR}. 
Indeed, as with the ${\cal N}=2$ case, we can still show 
the equivalence among these
different perspectives by suitable S or T Duality transformations. 
This equivalence is explicitly shown in \cite{HU} for the case of 
the Abelian finite subgroups of $SU(3)$.
More precisely, for the group $Z_k \times Z_{k'}$ or $Z_k$ 
and a chosen decomposition
of ${\bf 3}$ into appropriate irreducible representations thereof
one can construct 
the corresponding Brane Box Model that gives the
same quiver diagram as the one 
obtained directly from the geometrical methods of
attack; this is what we mean by equivalence \cite{KS,LNV,BKV}. 

Indeed, we are not satisfied with the fact that this abovementioned equivalence
so far exists only for Abelian singularities and would like to see how it may be
extended to non-Abelian cases.
The aim for constructing Brane Box Models of non-Abelian finite 
groups is twofold: firstly we would generate a new category of finite supersymmetric
field theories and secondly we would demonstrate how the equivalence between the 
Brane Box Model and D3 branes as probes is true beyond the Abelian case and hence
give an interesting physical perspective on non-Abelian groups.
More specifically, the problem we wish to tackle is that given any finite
discrete subgroup $\Gamma$ of $SU(2)$ or $SU(3)$,
what is the brane setup (in the T-dual picture)
that corresponds to D3 branes as probes on orbifold singularities afforded by
$\Gamma$?
For the $SU(2)$ case, the answer for the $\widehat{A}$ series
was given in \cite{M-Theory} and that for the $\widehat{D}$ series,
in \cite{Kapustin}, yet $\widehat{E_{6,7,8}}$ are still unsolved.
For the $SU(3)$ case, the situation is even worse.
While \cite{HZ,HU} have given solutions to the Abelian groups
$Z_k$ and $Z_k\times Z_{k'}$, the non-Abelian $\Delta$ and $\Sigma$ series
have yet to be treated.
Though it is not clear how the generalisation can be done for
arbitrary non-Abelian singularities,
it is the purpose of this writing to take one further step from \cite{HZ,HU},
and address the next simplest series of dimension three
orbifold theories, viz., those of $\C^3/Z_k \times D_{k'}$ and
construct the corresponding
Brane Box Model and show its equivalence to geometrical methods. In addition to
equivalence we demonstrate how the two pictures are bijectively related for the
group of interest and that given one there exists a unique description in the other.
The key input is given by Kutasov, Sen and Kapustin in \cite{Kapustin,Kutasov,Sen}.
Moreover \cite{Han-Zaf2} 
has briefly pointed out how his results may be used, but
without showing the consistency and equivalence.

The chapter is organised as follows.
In section \sref{sec:review} we shall briefly review some techniques of brane setups
and orbifold projections in the context of finite quiver theories. Section
\sref{sec:group} is then devoted to a crucial digression on the mathematical properties
of the group of our interest, or what we call $G:=Z_k \times D_{k'}$. In section
\sref{sec:BB} we construct the Brane Box Model for $G$, followed by concluding
remarks in section \sref{sec:conc}.

\section*{Nomenclature}
Unless otherwise stated, we shall, throughout our chapter, 
adhere to the notation
that $\omega_n = e^{\frac{2 \pi i}{n}}$, the $n$th root of unity,
that $G$ refers to the group \GZD, that without ambiguity
$Z_k$ denotes $\Z_k$, the cyclic group of $k$ elements, that $D_k$ is
the binary dihedral group of order $4k$ and gives the affine Dynkin
diagram of $\widehat{D}_{k+2}$,
and that $d_k$ denotes
the ordinary dihedral group of order $2k$. Moreover $\delta$ will be defined as
$(k,2k')$, the greatest common divisor (GCD) of $k$ and $2k'$. 

\section{A Brief Review of $D_n$ Quivers, Brane Boxes, 
	and Brane Probes on Orbifolds} \label{sec:review}
The aim of this chapter is to construct the Brane Box Model of the 
non-Abelian finite group \GZD and to show its consistency as well as
equivalence to geometric methods. To do so,
we need to know how to read out the gauge groups and matter content
from quiver diagrams which describe a particular 
field theory from the geometry side. 
The knowledge for such a task is supplied in \sref{subsec:Quiver}.
Next, as mentioned in the introduction, to construct field theories
which could be encoded in the $D_k$ quiver diagram, 
we need an important result from \cite{Kapustin,Kutasov,Sen}.
A brief review befitting
our aim is given in \sref{subsec:Kapustin}. 
Finally in \sref{subsec:BBZZ} we present the rudiments of the
Brane Box Model.

\subsection{Branes on Orbifolds and Quiver Diagrams} \label{subsec:Quiver}
\index{Orbifolds}
\index{Brane Probes!Orbifolds}
It is well-known that a stack of coincident $n$ D3 branes gives rise to an ${\cal N}=4$ 
$U(n)$ super-Yang-Mills theory on the four dimensional world volume. The $U(1)$ factor
of the $U(n)$ gauge group decouples when we discuss the low energy dynamics of 
the field theory and can be ignored, therefore giving us an effective $SU(n)$ theory.
For ${\cal N}=4$ in 4 dimensions the R-symmetry is $SU(4)$. Under such an R-symmetry, 
the fermions in the vector multiplet transform in
the spinor representation of $SU(4) \simeq Spin(6)$ and the scalars,
in the vector representation of $Spin(6)$, the universal cover of $SO(6)$. 
In the brane picture we can identify the R-symmetry as the $SO(6)$ 
isometry group which acts on the six transverse directions of the D3-branes.
Furthermore, in the AdS/CFT picture, 
this $SU(4)$ simply manifests as the $SO(6)$ isometry group of the
5-sphere in $AdS_{5}\times S^{5}$ \cite{KS,LNV,BKV}.

We shall refer to this gauge theory of the D3 branes as the parent theory and
consider the consequences of putting the stack on geometric singularities.
A wide class of finite Yang-Mills theories of various
gauge groups and supersymmetries is obtained when the parent theory is placed on orbifold
singularities of the type $\C^m/\Gamma$ where $m=2,3$.
What this means is that we select a discrete finite group $\Gamma \subset SU(4)$
and let its irreducible representations $\{{\bf r}_i\}$ act on the Chan-Paton 
indices $I,J=1,...,n$ 
of the D3 branes by permutation. Only those matter fields of the parent theory
that are invariant under the group action of $\Gamma$ remain, the rest are
eliminated by this so-called ``orbifold projection''.
We present the properties of the parent and the orbifolded theory in the following
diagram:
\[
\begin{array}{|l|lll|}
\hline
&$Parent Theory$	& \stackrel{\Gamma,{\rm~irreps~}=\{{\bf r}_i\}}{\longrightarrow}
			&$Orbifold Theory$\\
\hline
$SUSY$			&{\cal N}=4	&	&
	\begin{array}{l}
	{\cal N}=2, {\rm~for~} \C^2/\{\Gamma\subset SU(2)\}	\\
	{\cal N}=1, {\rm~for~} \C^3/\{\Gamma\subset SU(3)\}	\\
	{\cal N}=0, {\rm~for~} (\C^3\simeq\R^6)/\{\Gamma\subset \{SU(4)\simeq SO(6)\}\}	\\
	\end{array}\\
\hline
\begin{array}{c}
	$Gauge$ \\
	$Group$
\end{array}		&U(n)		&	& \prod\limits_{i} SU(N_i), 
		{\rm~~~~~~~where~}\sum\limits_{i} N_i \dim{\bf r}_i = n\\
\hline
$Fermion$	&\Psi_{IJ}^{\bf{4}} &	& \Psi_{f_{ij}}^{ij} \\
$Boson$		&\Phi_{IJ}^{\bf{6}} &	& \Phi_{f_{ij}}^{ij} 
		{\rm~~~~~~~where~} I,J=1,...,n; f_{ij}=1,...,a_{ij}^{{\cal R}={\bf 4},{\bf 6}}\\
			&&&~~~~~~~~~~~~~~~~~~~~~~~~~~~~~~~~~~~~~~{\cal R}\otimes 
			{\bf r}_{i}=\bigoplus\limits_{j}a_{ij}^{{\cal R}}\\
\hline
\end{array}
\]
Let us briefly explain what the above table summarises.
In the parent theory,
there are, as mentioned above, gauge bosons $A_{IJ=1,...,n}$ as singlets of $Spin(6)$,
adjoint Weyl fermions $\Psi_{IJ}^{\bf{4}}$
in the fundamental $\bf{4}$ of $SU(4)$ and adjoint scalars 
$\Phi_{IJ}^{\bf{6}}$ in the antisymmetric $\bf{6}$ of $SU(4)$.
The projection is the condition that
\[
A = \gamma(\Gamma) \cdot A \cdot \gamma(\Gamma)^{-1}
\]
for the
gauge bosons and 
\[
\Psi({\rm~or~}\Phi) = R(\Gamma) \cdot \gamma(\Gamma) \cdot 
\Psi({\rm~or~}\Phi) \cdot \gamma(\Gamma)^{-1}
\]
for the fermions and bosons respectively
($\gamma$ and $R$ are appropriate representations of $\Gamma$).

Solving these relations by using Schur's Lemma gives the information on
the orbifold theory.
The equation for $A$ tell us that the original $U(n)$ 
gauge group is broken to
$\prod\limits_{i} SU(N_i)$ where $N_i$ are positive integers such that
$\sum\limits_{i} N_i \dim{\bf r}_i = n$. We point out here that henceforth
we shall use the {\it regular representation} where $n = N|\Gamma|$ for some
integer $N$ and $n_i = N \dim{\bf r}_i$. Indeed other choices are possible
and they give rise to {\it Fractional Branes}, which not only provide interesting
dynamics but are also crucial in showing the equivalence between brane setups
and geometrical engineering \cite{Dog1,Karch:equiv}.
The equations for $\Psi$ and $\Phi$ dictate that they become bi-fundamentals
which transform
under various pairs $(N_i,\bar{N_j})$ within the product gauge group. We have
a total of
$a_{ij}^{\bf{4}}$ Weyl fermions $\Psi _{f_{ij}=1,...,a_{ij}^{\bf{4}}}^{ij}$ 
and $a_{ij}^{\bf 6}$ scalars $\Phi _{f_{ij}}^{ij}$
where $a_{ij}^{\cal R}$ is defined by
\begin{equation}
{\cal R}\otimes {\bf r}_i=\bigoplus\limits_{j}a_{ij}^{\cal R} {\bf r}_j
\label{aij3}
\end{equation}
respectively for ${\cal R} = 4,6$.

The supersymmetry of the orbifold theory is determined by analysing the 
commutant of $\Gamma$ as it embeds into the parent $SU(4)$ R-symmetry.
For $\Gamma$ belonging to $SU(2)$, $SU(3)$ or the full $SU(4)$,
we respectively obtain ${\cal N}=2,1,0$. The corresponding geometric
singularities are as presented in the table.
Furthermore, the action of $\Gamma$ clearly differs for $\Gamma \subset
SU(2,3,$~or~$4)$ and the {\bf 4} and {\bf 6} that give rise to
the bi-fundamentals must be decomposed appropriately.
Generically, the number of trivial (principal) 1-dimensional irreducible representations 
corresponds to the co-dimension of the singularity. 
For the matter matrices $a_{ij}$, these irreducible representations give
a contribution of $\delta_{ij}$ and therefore to guaranteed adjoints.
For example, in the case of ${\cal N}=2$, there are
2 trivial {\bf 1}'s in the {\bf 4} and for ${\cal N}=1$, 
${\bf 4} = {\bf 1}_{\rm trivial} \oplus {\bf 3}$.
In this chapter, we focus on the latter case since \GZD
is in $SU(3)$ and gives rise to ${\cal N}=1$. Furthermore we acknowledge the
inherent existence of the trivial 1-dimensional irrep and focus on the decomposition
of the {\bf 3}.

The matrices $a_{ij}^{{\cal R}={\bf 4,6}}$ in (\ref{aij3})
and the numbers $\dim{\bf r}_i$ contain
all the information about the matter fields and gauge groups of the orbifold theory.
They can be conveniently encoded into so-called {\bf quiver diagrams}.
Each node of such a diagram treated as a finite graph represents 
a factor in the product gauge group
and is labeled by $\dim{\bf r}_i$. The (possibly oriented) 
adjacency matrix for the graph is prescribed precisely by $a_{ij}$. 
The cases of ${\cal N} = 2,3$ are done \cite{DM,9811183,Muto,Greene}
and works toward the (non-supersymmetric) ${\cal N} =0$ case are
underway \cite{9905212}.
In the ${\cal N} = 2$ case, the quivers must 
coincide with $ADE$ Dynkin diagrams treated
as unoriented graphs in order that the orbifold theory be finite
\cite{geoeng3}.
The quiver diagrams in general are suggested to be related to WZW modular 
invariants \cite{9811183,9903056}.

This is a brief review of the construction via geometric methods and it is our
intent now to see how brane configurations reproduce examples thereof.

\subsection{$D_k$ Quivers from Branes}\label{subsec:Kapustin}
\index{Hanany-Witten!elliptic model}
Let us first digress briefly to $A_k$ quivers from branes.
In the case of $SU(2) \supset \Gamma = \widehat{A_k} \simeq Z_{k+1}$, the quiver
theory should be represented by an affine $A_k$ Dynkin diagram, i.e., a regular
polygon with $k+1$ vertices. The gauge group is 
$\prod\limits_{i} SU(N_i) \times U(1)$ 
with $N_i$ being a $k+1$-partition of $n$ since ${\bf r}_i$ are all 
one-dimensional\footnote{The $U(1)$ corresponds to the
centre-of-mass motion and decouples from other parts of the theory 
so that when we discuss
the dynamical properties, it does not contribute.}.
However, we point out that on a classical level we expect
$U(N_i)$'s from the brane perspective rather than $SU(N_i)$. It is only after 
considering the one-loop quantum corrections in the field theory
(or bending in the brane picture) that we realise that
the $U(1)$ factors are frozen. This is explained in \cite{M-Theory}.
On the other hand, from the point of view of D-branes
as probes on the orbifold singularity, 
associated to the anomalous $U(1)$'s are field-dependent
Fayet-Illiopoulos terms generating which freezes the $U(1)$ factors.
Thess two prespectives are T-dual to each other.
Further details can be found in \cite{LRA}.

\begin{figure}
\centerline{\psfig{figure=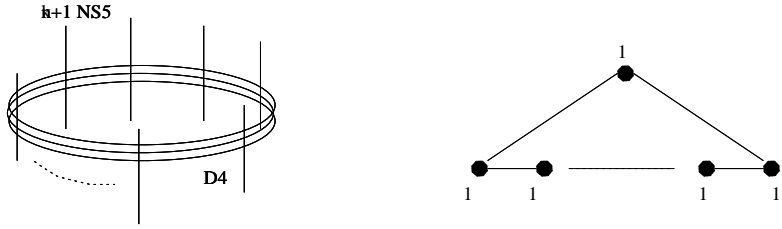,width=4.0in}}
\caption{The ${\cal N}=2$ elliptic model of D4 branes stretched between NS5 branes
	to give quiver theories of the $\widehat{A_k}$ type.}
\label{fig:A}
\end{figure}

Now, placing $k+1$ NS5 branes on a circle with $N_i$ stacked D4 branes 
stretched between the
$i$th and $i+1$st NS5 reproduces precisely this gauge group
with the correct bifundamentals provided by open strings ending on the adjacent 
D4 branes (in the compact direction). This circular model thus furnishes the brane
configuration of an $A_n$-type orbifold theory and is summarised in \fref{fig:A}.
Indeed T-duality in the compact direction transforms the $k+1$ NS5 branes into
a nontrivial metric, viz., the $k+1$-centered Taub-NUT, precisely that expected
from the orbifold picture.
Since both the NS5 and the D4 are offsprings of the M5 brane, in the
M-Theory context, 
the circular configuration becomes $\R^4 \times \bar\Sigma$ in
$\R^{10,1}$, where 
$\bar\Sigma$ is a $k+1$-point compactification of a the Riemann
surface $\Sigma$ 
swept out by the worldvolume of the fivebrane \cite{M-Theory}. The
duality group, which is the 
group of automorphisms among the marginal couplings that arise in the
resulting field theory, 
whence becomes the fundamental group of ${\cal M}_{k+1}$, 
the moduli space of an elliptic curve with $k+1$ marked points.

The introduction of ON$^0$ planes facilitates the next type of ${\cal N}=2,d=4$
quiver theories, namely those encoded by affine $\widehat{D_k}$ Dynkin 
diagrams \cite{Kapustin}. 
The gauge group is now
$SU(2N)^{k-3} \times SU(N)^4 \times U(1)$ (here $U(1)$ decouples also, as explained before) 
dictated by the Dynkin indices of
the $\widehat{D_k}$ diagrams.

\begin{figure}
\centerline{\psfig{figure=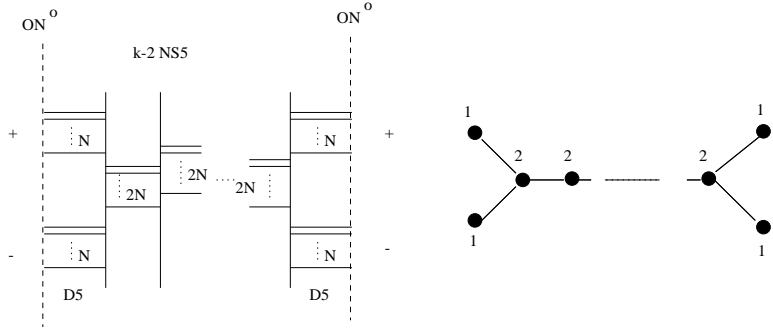,width=4.0in}}
\caption{D5 branes stretched between ON$^0$ branes, interrupted by NS5 branes
	to give quiver theories of the $\widehat{D_k}$ type.}
\label{fig:D}
\end{figure}

There are two ways to see the $\widehat{D_k}$ quiver in the brane picture: one in 
Type IIA theory and the other, in Type IIB. 
Because later on in the construction of the Brane Box 
Model we will use D5 branes which are in Type IIB, we will focus on Type IIB only (for 
a complete description and how the two descriptions are related by T-duality, 
see \cite{Kapustin}).
In this case, what we need is the ON$^0$-plane 
which is the S-dual of a peculiar pair: a D5 brane on top of an O5$^-$-plane.
The one
important property of the ON$^0$-plane is that it has an orbifold description
$\R^6 \times \R^4/{\cal I}$ where
${\cal I}$ is a product of world sheet fermion operator $(-1)^{F_L}$ with the
parity inversion of the $\R^4$ \cite{Sen}.
Let us place 2 parallel vertical ON$^0$ planes and $k-2$ NS5 branes in between and 
parallel to both as in \fref{fig:D}. Between the ON$^0$ and its immediately
adjacent NS5, we stretch $2N$ D5 branes; $N$ of positive charge
on the top and $N$ of negative charge below. 
Now due to the projection of the ON$^0$ plane, $N$ D5 branes of positive charge give
one $SU(N)$ gauge group and $N$ D5 branes of negative charge give
another. Furthermore, these D5 branes end on NS5 branes and the
boundary condition on the NS5 projects out the bi-fundamental hypermultiplets
of these two $SU(N)$ gauge groups 
(for the rules of such projections see \cite{Kapustin}). 
Moreover, between the two adjacent interior
NS5's we stretch $2N$ D5 branes, giving $SU(2N)$'s for the gauge group.
From this brane setup we immediately see 
that the gauge theory is encoded in the affine Quiver diagram of
$\widehat{D_k}$.

\subsection{Brane Boxes}\label{subsec:BBZZ}
\index{Hanany-Witten!brane box}
We have seen in the last section, that positioning appropriate branes 
according to Dynkin diagrams - which for $\Gamma \subset SU(2)$ have
their adjacency matrices determined by the representation of $\Gamma$,
due to the McKay Correspondence \cite{9811183} - branengineers some
orbifold theories that can be geometrically engineered. The exceptional
groups however, have so far been elusive
\cite{Kapustin}. For $\Gamma \subset SU(3)$, perhaps related to the fact
that there is not yet a general McKay Correspondence\footnote{For
Gorenstein singularities of dimension 3, only those of the Abelian
type such that 1 is not an eigenvalue of $g$ $\forall g\in \Gamma$ are isolated. 
This restriction perhaps limits na\"{\i}ve brane box
constructions to Abelian orbifold groups \cite{HU}. For a discussion on
the McKay Correspondence as a ubiquitous thread, see \cite{9903056}.}
above dimension 2, the problem becomes more subtle; brane setups have 
been achieved for orbifolds of the Abelian type, a restriction that has been
argued to be necessary for consistency \cite{HZ,HU}. It is thus
the purpose of this writing to show how a group-theoretic ``twisting''
can relax this condition and move beyond Abelian theories; to this we shall
turn later.

We here briefly review the so-called $Z_k \times Z_{k'}$ elliptic brane
box model. The orbifold theory corresponds to 
$\C^3 / \{ \Gamma = Z_k \times Z_{k'} \subset SU(3) \}$
and hence by arguments before we are
in the realm of ${\cal N} = 1$ super-Yang-Mills. The generators for $\Gamma$
are given, in its fundamental 3-dimensional representation\footnote{We
have chosen the directions in the transverse spacetime upon which
each cyclic factor acts; the choice is arbitrary. In the language of
finite groups, we have chosen the transitivity of the collineation sets.
The group at hand, $Z_k \times Z_{k'}$, is in fact the first example of
an intransitive subgroup of $SU(3)$.
For a discussion of finite subgroups of unitary groups, see \cite{9905212} and
references therein.}, by diagonal
matrices $diag(e^{\frac{2\pi i}{k}},e^{\frac{-2\pi i}{k}},1)$ corresponding
to the $Z_k$ which act non-trivially on the first two coordinates of $\C^3$
and $diag(1,e^{\frac{2\pi i}{k'}},e^{\frac{-2\pi i}{k'}})$
corresponding to the $Z_{k'}$ which act non-trivially on the 
last two coordinates of $\C^3$.

\begin{figure}
\centerline{\psfig{figure=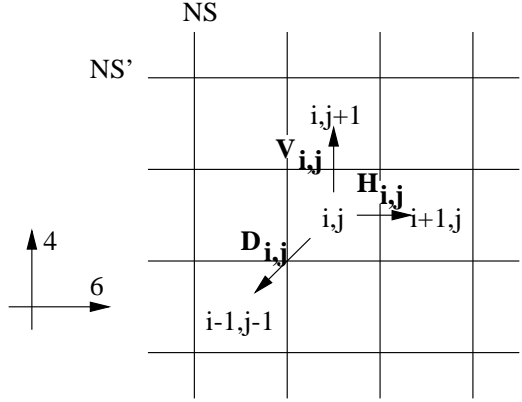,width=2.7in}}
\caption{Bi-fundamentals arising from D5 branes stretched between grids of
	NS5 and NS5$'$ branes in the elliptic brane box model.}
\label{fig:BBZZ}
\end{figure}

Since $\Gamma$ is a direct product of Abelian groups, the representation
thereof is simply a Kronecker tensor product of the two cyclic groups.
Or, from the branes perspective, we should in a sense take a Cartesian 
product or sewing between two ${\cal N}=2$ elliptic $A_{k-1}$ and $A_{k'-1}$ 
models discussed above,
resulting in a brane configuration on $S^1 \times S^1 = T^2$. This is
the essence of the (${\cal N}=1$ elliptic) Brane Box Model \cite{HZ,HU}.
Indeed the placement of a perpendicular set of branes breaks the supersymmetry
of the ${\cal N} = 2$ model by one more half, thereby giving the desired
${\cal N}=1$. More specifically, we place $k$ NS5 branes in the $012345$
and $k'$ NS5$'$ branes in the $012367$ directions, whereby forming
a grid of $kk'$ boxes as in \fref{fig:BBZZ}.
We then stretch $n_{ij}$ D5 branes in the $012346$ directions
within the $i,j$-th box and compactify the $46$ directions (thus making
the low-energy theory on the D5 brane to be 4 dimensional).
The bi-fundamental fields are then given according to adjacent boxes
horizontally, vertically and diagonally and the gauge groups is
$(\bigotimes\limits_{i,j}SU(N))\times U(1)  = SU(N)^{kk'}\times U(1)$ (here
again the $U(1)$ decouples) as expected from geometric
methods. Essentially we construct one box for each irreducible representation of 
$\Gamma=Z_k \times Z_{k'}$ such that going in the 3 directions as shown
in \fref{fig:BBZZ} corresponds to tensor decomposition of the irreducible
representation in that grid and a special ${\bf 3}$-dimension representation
which we choose when we construct the Brane Box Model.

We therefore see the realisation of Abelian orbifold theories in dimension
3 as brane box configurations; twisted identifications of the grid can
in fact lead to more exotic groups such as $Z_k \times Z_{kk'/l}$.
More details can be found in \cite{HU}.

\section{The Group $G=Z_k \times D_{k'}$} \label{sec:group}
It is our intent now to investigate the next simplest example of intransitive
subgroups of $SU(3)$, i.e., the next infinite series of orbifold theories
in dimension 3 (For definitions on the classification of collineation groups,
see for example \cite{9905212}). This will give us a first example of a Brane Box Model
that corresponds to non-Abelian singularities.

Motivated by the $Z_k \times Z_{k'}$ treated in section
\sref{sec:review}, we let the second factor be the binary dihedral group
of $SU(2)$, or the $D_{k'}$ series (we must point out that in our notation, 
the $D_{k'}$ group gives the $\widehat{D}_{k'+2}$ Dynkin diagram). 
Therefore $\Gamma$ is the group
$G = Z_k \times D_{k'}$, generated by
\[
\alpha = \left(  \begin{array}{ccc}
                        \omega_{k} & 0 & 0  \\
                         0 & \omega_{k}^{-1} & 0\\
                         0  &  0 & 1
                \end{array}
        \right)
~~~~~~~~
\beta = \left(  \begin{array}{ccc}
                        1  & 0  & 0  \\
                        0 & \omega_{2k'} & 0  \\
                        0 &  0 & \omega_{2k'}^{-1}  
                \end{array}
        \right)
~~~~~~~~
\gamma =\left(  \begin{array}{ccc} 
                1  &  0  &  0  \\
                0  &  0  &  i \\
                0  &  i &   0 
                \end{array}
        \right)
\]
where $w_x := e^{\frac{2 \pi i}{x}}$. We observe that indeed $\alpha$ generates
the $Z_k$ acting on the first two directions in $\C^3$ while $\beta$ and $\gamma$
generate the $D_{k'}$ acting on the second two.

We now present some crucial properties of this group $G$ which shall be used in
the next section. First we remark that the $\times$ in $G$ is really an abuse of
notation, since $G$ is certainly not a direct product of these two groups. This
is the cause why na\"{\i}ve constructions of the Brane Box Model fail and to this
point we shall turn later.
What we really mean is that the actions on the first two and last two coordinates
in the transverse directions by these subgroups are to be construed as separate.
Abstractly, we can write the presentation of $G$ as
\begin{equation}
\alpha \beta = \beta \alpha,~~~~\beta \gamma =\gamma \beta^{-1},~~~~
\alpha^{m} \gamma \alpha^{n} \gamma =\gamma \alpha^{n} \gamma \alpha^{m}
~~~~\forall m,n \in \Z
\label{relations}
\end{equation}
These relations compel all elements in $G$ to be writable in the form
$\alpha^{m} \gamma \alpha^{\tilde{m}} \gamma^{n} \beta^{p}$. However, before discussing
the whole group, we find it very useful to discuss the subgroup generated by $\beta$ and
$\gamma$, i.e the binary dihedral group $D_{k'}$ as a degenerate ($k=1$) case of $G$,
because the properties of the binary dihedral group turn out to be crucial for the structure
of the Brane Box Model and the meaning of ``twisting'' which we shall clarify later.

\subsection{The Binary Dihedral $D_{k'} \subset G$}\label{subsec:D}
All the elements of $D_{k'}$ can be written as
$ \beta^{p}\gamma^{n}$ with $n=0,1$ and $p=0,1,...,2k'-1$, giving the
order of the group as $4k'$. We now move onto Frobenius characters.
It is easy to work out the structure of conjugate classes. We have two conjugate
classes $(1), (\beta^{k'})$ which have only one element, $(k'-1)$ conjugate classes
$(\beta^p,\beta^{-p}),p=1,..,k'-1$ which have two elements and two conjugate classes
$( \beta^{p~\even}\gamma),( \beta^{p~\odd}\gamma)$ which have $k'$ elements.
The class equation is thus as follows:
\[
4k' = 1 + 1 + (k' - 1)\cdot 2 + 2 \cdot k'.
\]
Moreover there are 4 1-dimensional and $k'-1$ 2-dimensional irreducible
representations
such that the characters for the 1-dimensionals depend on the parity of $k'$.
Now we have enough facts to clarify our notation: the group $D_{k'}$ gives
$k'+3$ nodes (irreducible representations) which corresponds to the Dynkin diagram of 
$\widehat{D_{k'+2}}$.

We summarise the character table as follows:\\
\[
\doublerulesep 0.7pt
\begin{array}{cc}
k' \even &
\begin{array}{|c|c|c|c|c|c|c|}
\hline \hline
        & C_{n=0}^{p=0} &  C_{n=0}^{p=k'} &  C_{n=0}^{\pm \even~p} &
	  C_{n=0}^{\pm \odd~p} & C_{n=1}^{\even~p} & C_{n=1}^{\odd~p} \\ \hline
|C|  	& 1 & 1 & 2 & 2  & k' & k' \\ \hline
\#C   	& 1 & 1 & \frac{k'-1}{2} & \frac{k'-1}{2} 
	& 1 & 1 \\ \hline \hline
\Gamma_1	& 1 &  1 &  1 &  1  & 1 & 1 \\ \hline
\Gamma_2	& 1 &  -1 &  1 &  -1 & 1 & -1 \\ \hline
\Gamma_3	& 1 & 1 & 1 & 1 & -1 & -1 \\ \hline
\Gamma_4	& 1 & -1 & 1 & -1 & -1 & 1 \\ \hline
\Gamma_l	& \multicolumn{4}{|c|}{(\omega_{2k'}^{lp}+
			\omega_{2k'}^{-lp})~~~~l=1,..,k'-1} & 0 & 0 \\ \hline
\end{array}
\\ \\
k' \odd &
\doublerulesep 0.7pt
\begin{array}{|c|c|c|c|c|c|c|}
\hline \hline
        & C_{n=0}^{p=0} &  C_{n=0}^{p=k'} &  C_{n=0}^{\pm \even~p} &
	  C_{n=0}^{\pm \odd~p} & C_{n=1}^{\even~p} & C_{n=1}^{\odd~p} \\ \hline
|C|  	& 1 & 1 & 2 & 2  & k' & k' \\ \hline
\#C   	& 1 & 1 & \frac{k'-2}{2} & \frac{k'}{2} 
	& 1 & 1 \\ \hline \hline
\Gamma_1	& 1 & 1 &  1 &  1  & 1 & 1 \\ \hline
\Gamma_2	& 1 & 1 &  1 &  -1 & \omega_4 & -\omega_4 \\ \hline
\Gamma_3	& 1 & 1 & 1 & 1 & -1 & -1 \\ \hline
\Gamma_4	& 1 & 1 & 1 & -1 & -\omega_4 & \omega_4 \\ \hline
\Gamma_l	& \multicolumn{4}{|c|}{(\omega_{2k'}^{lp}+
			\omega_{2k'}^{-lp})~~~~l=1,..,k'-1} & 0 & 0 \\ \hline
\end{array}
\\
\end{array}
\]

In the above tables, $|C|$ denotes the number of group elements in conjugate class
$C$ and $\#C$, the number of conjugate classes belonging to this type. Therefore
$\sum\limits_C \#C\cdot|C|$ should equal to order of the group.
When we try to
look for the character of the 1-dimensional irreps, we find it to be the same as the
character of the factor group $D_{k'}/N$ where $N$ is the normal subgroup generated by 
$\beta$. This factor group is Abelian of order 4 and is different depending
on the parity of $k'$. When $k'=\even$, it
is $Z_2\times Z_2$ and when $k'=\odd$ it is $Z_4$. Furthermore, the conjugate class
$(\beta^p,\beta^{-p})$ corresponds to different elements in this factor group 
depending on the parity of $p$, and we distinguish the two different cases in the 
table as $C_{n=0}^{\pm \odd~p}$ and $C_{n=0}^{\pm \even~p}$.

\index{Finite Groups!$SU(3)$ subgroups!$Z$-$D$ type}
\subsection{The whole group $G = Z_k \times D_{k'}$}
Now from (\ref{relations}) we see that all elements of G can be written in the form
$\alpha^{m} \gamma \alpha^{\tilde{m}} \gamma^{n} \beta^{p}$ with 
$m,\tilde{m}=0,..,k-1$, $n=0,1$ and $p=0,..2k'-1$, which we abbreviate as
$(m,\tilde{m},n,p)$. In the matrix form of our fundamental representation, they become
\[
\begin{array}{ll}
(m,\tilde{m},n=0,p)= & (m,\tilde{m},n=1,p)= \\

\left(  \begin{array}{ccc}
                        \omega_{k}^{m+\tilde{m}} & 0 & 0\\
                        0 & 0 & i\omega_{k}^{-m} \omega_{2k'}^{-p}  \\
                        0 & i\omega_{k}^{-\tilde{m}}\omega_{2k'}^{p} & 0
                        \end{array}
                \right),
&
\left(  \begin{array}{ccc}
                        \omega_{k}^{m+\tilde{m}} & 0 & 0 \\
                         0 & -\omega_{k}^{-m} \omega_{2k'}^{p} & 0 \\
                        0 & 0 & -\omega_{k}^{-\tilde{m}}\omega_{2k'}^{-p} 
                        \end{array}
                \right). \\
\end{array}
\]
Of course this representation is not faithful and there is a non-trivial orbit; we can
easily check the repeats:
\begin{equation}
\begin{array}{l}
(m,\tilde{m},n=0,p)=(m+\frac{k}{(k,2k')},\tilde{m}-\frac{k}{(k,2k')},n=0,p-\frac{2k'}
{(k,2k')}), \\
(m,\tilde{m},n=1,p)=(m+\frac{k}{(k,2k')},\tilde{m}-\frac{k}{(k,2k')},n=1,p+\frac{2k'}
{(k,2k')}) 
\end{array}
\label{orbit}
\end{equation}
where $(k,2k')$ denotes the largest common divisor between them. Dividing by the
factor of this repeat immediately gives the 
order of $G$ to be $\frac{4k'k^2}{(k,2k')}$.

We now move on to the study of the characters of the group. The details of the
conjugation automorphism, class equation and irreducible
representations we shall leave to the appendix \ref{append:9906031} 
and the character tables we shall present below; again we have two cases, 
depending on the parity of $\frac{2k'}{(k,2k')}$. First however we start with
some preliminary definitions. We define $\eta$ as a function of $n$, $p$ and
$h = 1,2,3,4$.
\begin{equation}
\begin{array}{cc}
k' = \even &
\begin{array}{ccccc}
     &  (n=1,p=\even) & (n=1,p=\odd) & (n=0,p=\even) & (n=0,p=\odd) \\
\eta^{1} & 1 & 1 & 1 & 1 \\
\eta^{2} & 1 & -1 & 1 & -1 \\
\eta^{3} & 1 & 1 & -1 & -1 \\
\eta^{4} & 1 & -1 & -1 & 1 
\end{array}
\\
k' = \odd &
\begin{array}{ccccc}
        & (n=1,p=\odd) & (n=1,p=\even) & (n=0,p=\even) & (n=0,p=\odd)\\
\eta^{1} & 1 & 1 & 1 & 1 \\
\eta^{2} & 1 & -1 & \omega_4 & -\omega_4 \\
\eta^{3} & 1 & 1 & -1 & -1 \\
\eta^{4} & 1 & -1 & - \omega_4 &  \omega_4
\end{array}
\\
\end{array}
\label{eta}
\end{equation}
Those two tables simply give the character tables of $Z_2\times Z_2$ and $Z_4$
which we saw in the last section.

Henceforth we define $\delta := (k,2k')$. 
Furthermore, we shall let 
$\Gamma^n_x$ denote an $n$-dimensional irreducible representation indexed by some (multi-index) $x$.
For $\frac{2k'}{\delta} = \even$,
there are $4k$ 1-dimensional irreducible representations indexed by $(l,h)$ with $l=0,1,..,k-1$ and 
$h=1,2,3,4$ and
$k(\frac{k'k}{(k,2k')}-1)$ 2-dimensionals indexed by $(d,l)$ with
$d=1,..,\frac{k'k}{(k,2k')}-1;l=0,..,k-1$.
For $\frac{2k'}{\delta} = \odd$,
there are $2k$ 1-dimensional irreducible representations indexed by $(l,h)$ with $l=0,..,k-1;h=1,3$ and
$k(\frac{k'k}{(k,2k')}-\frac{1}{2})$ 2-dimensionals indexed by $(d,l)$
$d=1,..,\frac{k'k}{(k,2k')}-1;l=0,..,k-1$ and
$d=\frac{k'k}{(k,2k')};l=0,..,\frac{k}{2}-1$.
Now we present the character tables.\\

{\large $\frac{2k'}{\delta} = \even$}
\[
{\hspace{-1cm}}
\doublerulesep 0.7pt
\begin{array}{|c|c|c|c|}
\hline \hline
|C| & 1 & 2 & \frac{k'k}{(k,2k')} \\ \hline
\#C & 2k & k(\frac{k'k}{(k,2k')}-1) & 2k \\
\hline \hline
	&
	\begin{array}{c}
		m = 0,..,\frac{k}{\delta}-1; ~i = 0,..,\delta-1; \\
			~\tilde{m} = m + \frac{i k}{\delta}; ~n=1; \\
			~p=k'-\frac{ik'}{(k,2k')},2k'-\frac{ik'}{(k,2k')}
	\end{array}
	&
	\begin{array}{c}
		m = 0,..,\frac{k}{\delta}-1; ~i = 0,..,\delta-1; ~n=1 \\
		\left\{
		\begin{array}{l}
	  	s = 0,..,m-1; ~p = 0,..2k'-1;\\
			~~~~~~\tilde{m} = s + \frac{i k}{\delta};\\
	  	s = m; \mbox{and require further that} \\
			~~~~~~p < (-p - \frac{2 i k'}{\delta}) \bmod (2k')\\
		\end{array}
		\right.
	\end{array}
	&
	\begin{array}{c}
		m = 0;\\
		\tilde{m} = 0,..,k-1;\\
		p = 0,1;\\
		n = 0\\
	\end{array} \\ \hline
\Gamma^1_{(l,h)} & \multicolumn{3}{|c|}{
	\omega_{k}^{(m+\tilde{m})l} \eta^{h},~~~~~~l=0,1,..,k-1;~~h=1,..,4}
	\\ \hline
\Gamma^2_{(d,l)} & \multicolumn{2}{|c|}{
	\begin{array}{c}
	(-1)^{d}(\omega_{k}^{-dm}\omega_{2k'}^{dp}+\omega_{k}^{-d\tilde{m}}
	\omega_{2k'}^{-dp}) \omega_{k}^{(m+\tilde{m})l}\\
	~~~~~~~~~~~~~d\in[1,\frac{k'k}{(k,2k')}-1];~~l\in [0,k)\\
	\end{array}
	}
	& 0 \\ \hline
\end{array}
\]

{\large $\frac{2k'}{\delta} = \odd$}
\[
\doublerulesep 0.7pt
{\hspace{-1cm}}
\begin{array}{|c|c|c|c|}
\hline \hline
|C| & 1 & 2 & \frac{k'k}{(k,2k')} \\ \hline
\#C & k & k(\frac{k'k}{(k,2k')}-\frac12) & k \\
\hline \hline
	&
	\begin{array}{c}
		m = 0,..,\frac{k}{\delta}-1;\\
		i = 0,..,\delta-1 \mbox{ and even}; \\
			~\tilde{m} = m + \frac{i k}{\delta}; ~n=1; \\
			~p=k'-\frac{ik'}{(k,2k')},\\~~~~~~2k'-\frac{ik'}{(k,2k')}
	\end{array}
	&
	\begin{array}{c}
		m = 0,..,\frac{k}{\delta}-1; ~i = 0,..,\delta-1; ~n=1 \\
		\left\{
		\begin{array}{l}
	  	s = 0,..,m-1; ~p = 0,..2k'-1;\\
			~~~~~~\tilde{m} = s + \frac{i k}{\delta};\\
	  	s = m; \mbox{and require further that} \\
			~~~~~~p < (-p - \frac{2 i k'}{\delta}) \bmod (2k') 
				\mbox{ for even } i\\
			~~~~~~p \le (-p - \frac{2 i k'}{\delta}) \bmod (2k') 
				\mbox{ for odd } i\\	
		\end{array}
		\right.
	\end{array}
	&
	\begin{array}{c}
		m = 0;\\
		\tilde{m} = 0,..,k-1;\\
		p = 0;\\
		n = 0\\
	\end{array} \\ \hline
\Gamma^1_{(l,h)} & \multicolumn{3}{|c|}{
	\omega_{k}^{(m+\tilde{m})l} \eta^{h},~~~~~~l=0,1,..,k-1;~~h=1,3}
	\\ \hline
\Gamma^2_{(d,l)} & \multicolumn{2}{|c|}{
	\begin{array}{c}
	(-1)^{d}(\omega_{k}^{-dm}\omega_{2k'}^{dp}+\omega_{k}^{-d\tilde{m}}
	\omega_{2k'}^{-dp}) \omega_{k}^{(m+\tilde{m})l}\\
	~~~~~~~~~~~~d\in[1,\frac{k'k}{(k,2k')}-1];~~l\in [0,k)\\
	\end{array}
	}
	& 0 \\ \hline
\Gamma^2_{(d,l)} & \multicolumn{2}{|c|}{
	\begin{array}{c}
	(-1)^{d}(\omega_{k}^{-dm}\omega_{2k'}^{dp}+\omega_{k}^{-d\tilde{m}}
	\omega_{2k'}^{-dp}) \omega_{k}^{(m+\tilde{m})l}\\
	~~~~~~~~~~~~d=\frac{k'k}{(k,2k')};~~l\in [0,\frac{k}{2})\\
	\end{array}
	}
	& 0 \\ \hline
\end{array}
\]

Let us explain the above tables in more detail. The third row of each table
give the representative elements of the various conjugate classes. 
The detailed description of the group elements in 
each conjugacy class is given in appendix \ref{append:9906031}.
It is easy to see, by using the above character tables, that given two
elements $(m_i,\tilde{m}_i,n_i,p_i)~~i=1,2$, if they share the same 
characters (as given in the last two rows), they belong to 
same conjugate class as to be expected since the character is a class
function.

We can be more precise and actually write down the 2 dimensional irreducible representation
indexed by $(d,l)$ as
\begin{equation}
\begin{array}{l}
(m,\tilde{m},n=0,p)= \omega_{k}^{(m+\tilde{m})l} 
\left(   \begin{array}{cc}
0 & i^d \omega_{k}^{-dm} \omega_{2k'}^{-dp} \\
i^d \omega_{k}^{-d\tilde{m}} \omega_{2k'}^{dp} & 0
\end{array} \right)
\\
(m,\tilde{m},n=1,p)=\omega_{k}^{(m+\tilde{m})l} 
\left(   \begin{array}{cc}
 (-1)^d \omega_{k}^{-dm} \omega_{2k'}^{dp} & 0 \\
0 & (-1)^d \omega_{k}^{-d\tilde{m}} \omega_{2k'}^{-dp}
\end{array} \right)
\end{array}
\label{2d}
\end{equation}

\subsection{The Tensor Product Decomposition in $G$}\label{subsec:decomp}
A concept crucial to character theory and representations 
is the decomposition of tensor products into tensor sums among the
various irreducible representations, namely the equation
\[
{\bf r}_k \otimes {\bf r}_i = \bigoplus\limits_j a_{ij}^k {\bf r}_j.
\]
Not only will such an equation enlighten us as to the structure of the
group, it will also provide quintessential information to the brane box
construction to which we shall turn later. Indeed the ${\cal R}$ in
(\ref{aij3}) is decomposed into direct sums of irreducible
representations ${\bf r}_k$, which
by the additive property of the characters, makes the fermionic and bosonic
matter matrices $a_{ij}^{\cal R}$ ordinary sums of matrices $a_{ij}^k$.
In particular, knowing the specific decomposition of the {\bf 3},
we can immediately construct the quiver diagram prescribed by 
$a_{ij}^{\bf 3}$ as discussed in \sref{subsec:Quiver}.

We summarise the decomposition laws as follows (using the multi-index notation
for the irreducible representations introduced in the previous
section), with the case of $\frac{2k'}{\delta} = \even$ in
\eref{evencase} and odd, in \eref{oddcase}.
{\vspace{-2cm}}
\beq
\label{evencase}
\begin{array}{|c|c|c|}
\hline
{\bf 1} \otimes {\bf 1}' & (l_1,h_1)_1 \otimes (l_2,h_2)_1 = (l_1+l_2,h_3)_1\\
		& \mbox{where $h_3$ is such that $\eta^{h_1}\eta^{h_2} = \eta^{h_3}$
		according to (\ref{eta}).} \\ \hline
{\bf 2} \otimes {\bf 1}  & (d,l_1)_2 \otimes (l_2,h_2)_1=\left\{
	\begin{array}{l}
        	(d,l_1+l_2)_2~~{\rm when}~~h_2=1,3.  \\
        	(\frac{k'k}{(k,2k')}-d,l_1+l_2-d)_2~~{\rm when}~~h_2=2,4
        \end{array}
        \right.
	\\ \hline
{\bf 2} \otimes {\bf 2}' &
	\begin{array}{l}
	(d_1,l_1)_2 \otimes (d_2 \le d_1,l_2)_2 = \\
	~~~~~~~~~~(d_1+d_2,l_1+l_2)_2 \oplus (d_1-d_2,l_1+l_2-d_2)_2 \\
	{\rm where} \\
	(d_1-d_2,l_1+l_2-d_2)_2 := \\
	~~~~~~~~~~(l_1+l_2-d_2,h=1)_1 \oplus (l_1+l_2-d_2,h=3)_1 
	{\rm~~if~~} d_1 = d_2 \\
	(d_1+d_2,l_1+l_2)_2 := \\
	~~~~~~~~~~(l_1+l_2,h=2)_1 \oplus (l_1+l_2,h=4)_1
	{\rm~~if~~} d_1+d_2=\frac{k'k}{\delta} \\
	(d_1+d_2,l_1+l_2)_2 := \\
	~~~~~~~~~~(\frac{2k'k}{(k,2k')}-(d_1+d_2),(l_1+l_2)-(d_1+d_2))_2
	{\rm~~if~~} d_1+d_2>\frac{k'k}{\delta} \\
	\end{array}
	\\ \hline
\end{array}
\eeq

\beq
\label{oddcase}
{\vspace{-1cm}}
\begin{array}{|c|c|}
\hline
{\bf 1} \otimes {\bf 1}' & (l_1,h_1)_1 \otimes (l_2,h_2)_1=\left\{
	\begin{array}{l}
	        (l_1+l_2,h=1)_1~~{\rm if}~~h_1=h_2  \\
        	(l_1+l_2,h=3)_1~~{\rm if}~~h_1\neq h_2
	\end{array}   \right.
		\\ \hline
{\bf 2} \otimes {\bf 1}  & (d,l_1)_2 \otimes (l_2,h_2)_1=\left\{
	\begin{array}{l}
		(d,l_1+l_2)_2 \\
		(d,l_1+l_2-\frac{k}{2})_2 {\rm~~if~~} 
		d=\frac{k'k}{(k,2k')} {\rm~and~} l_1+l_2 \ge \frac{k}{2}
	\end{array} \right.
	\\ \hline
{\bf 2} \otimes {\bf 2}' &
	\begin{array}{l}
	(d_1,l_1)_2 \otimes (d_2 \le d_1,l_2)_2 = \\
	~~~~~~~~~~(d_1+d_2,l_1+l_2)_2 \oplus (d_1-d_2,l_1+l_2-d_2)_2 \\
	{\rm where} \\
	(d_1-d_2,l_1+l_2-d_2)_2 := \\
	~~~~~~~~~~(l_1+l_2-d_2,h=1)_1 \oplus (l_1+l_2-d_2,h=3)_1
	{\rm~~if~~} d_1 = d_2 \\
	(d_1+d_2,l_1+l_2)_2 := \\
	~~~~~~~~~~(d_1 + d_2, l_1 + l_2 - \frac{k}{2})_2
	{\rm~~if~~} d_1+d_2=\frac{k'k}{\delta}{\rm~and~} l_1 + l_2 \ge \frac{k}{2}\\
	(d_1+d_2,l_1+l_2)_2 := \\
	~~~~~~~~~~(\frac{2k'k}{(k,2k')}-(d_1+d_2),(l_1+l_2)-(d_1+d_2))_2
	{\rm~~if~~} d_1+d_2>\frac{k'k}{\delta} \\
	\end{array}
	\\ \hline
\end{array}
\eeq

\subsection{$D_{\frac{kk'}{\delta}}$, an Important Normal Subgroup}\label{subsec:H}
We now investigate a crucial normal subgroup $H \triangleleft G$. The purpose
is to write $G$ as a canonical product of $H$ with the factor group formed by
quotienting $G$ thereby, i.e., as $G \simeq G/H \times H$. 
The need for this rewriting of the group will 
become clear in \sref{sec:BB} on the brane box construction.
The subgroup we desire is the one presented in the following:
\begin{lemma}
The subgroup
\[
H := \{(m,-m,n,p)|m=0,..,k-1;n=0,1;p=0,...,2k'-1 \}
\]
is normal in $G$ and is isomorphic to $D_{\frac{kk'}{\delta}}$.
\end{lemma}
To prove normality we use the multiplication and conjugation rules in $G$ 
given in appendix \ref{append:9906031} as (\ref{conj}) and (\ref{multi}).
Moreover, let $D_{\frac{kk'}{\delta}}$ be generated by $\tilde{\beta}$ and 
$\tilde{\gamma}$ using the notation of \sref{subsec:D}, then isomorphism
can be shown by the following bijection:
\[
\begin{array}{l}
(m,-m,1,p) \longleftrightarrow \tilde{\beta}^{\frac{2k'}{\delta}m-
\frac{k}{\delta}(p-k')},  \\
(m,-m,0,p) \longleftrightarrow \tilde{\beta}^{\frac{2k'}{\delta}m+
\frac{k}{\delta}p} \tilde{\gamma}.
\end{array}
\]
Another useful fact is the following:
\begin{lemma}
The factor group $G/H$ is isomorphic to $Z_k$.
\end{lemma}
This is seen by noting that $\alpha^l,l=0,1,...k-1$ can be used as representatives
of the cosets. We summarise these results into the following

\begin{proposition}
There exists another representation of $G$, namely
$Z_k \times D_{k'} \simeq Z_k \rtimes D_{\frac{kk'}{\delta}}$, generated by the same
$\alpha$ together with
\[
\begin{array}{cc}
\tilde{\beta}^{\frac{2k'}{\delta}m- \frac{k}{\delta}p} := (m,-m,1,p+k') = ~~~~~~~~~~
 & \tilde{\gamma} := \gamma = (0,0,0,0)= ~~~~~~~~~~\\
\left(  \begin{array}{ccc}
1  &  0  &  0  \\
0 & \omega_{k}^{-m} \omega_{2k'}^{p}  & 0  \\
0 & 0 & \omega_{k}^{m} \omega_{2k'}^{-p}   
\end{array}  \right), &
\left(  \begin{array}{ccc}
1 & 0 & 0 \\
0 & 0 & i \\
0 & i & 0 \\
\end{array}  \right).
\end{array}
\]
The elements of the group can now be written as
$\alpha^a \tilde{\beta}^b \tilde{\gamma}^n$ with $a\in [0,k)$, 
$b\in [0,\frac{2kk'}{\delta})$ and $n=0,1$, constrained by the presentation
\[
\{\alpha^k=\tilde{\beta}^{\frac{2kk'}{\delta}}=1,
\tilde{\beta}^{\frac{kk'}{\delta}}=\tilde{\gamma}^2=-1,
\alpha \tilde{\beta} = \tilde{\beta} \alpha , \tilde{\beta} \tilde{\gamma} =
\tilde{\gamma} \tilde{\beta}^{-1},
\alpha  \tilde{\gamma} =\tilde{\beta}^{\frac{2k'}{\delta}} \tilde{\gamma} \alpha\}
\]
\end{proposition}
In the proposition, by $\rtimes$ we do mean the internal semi-direct product between
$Z_k$ and $H := D_{\tilde{k}} := D_{\frac{kk'}{\delta}}$, in the sense \cite{Alperin} that
(I) $G=HZ_k$ as cosets, (II) $H$ is normal in $G$ and $Z_k$ is another subgroup, 
and (III) $H \cap Z_k = 1$. Now we no longer abuse the symbol $\times$ and
unambiguously use $\rtimes$ to show the true structure of $G$.
We remark that this representation is in some sense more natural (later we shall see
that this naturality is not only mathematical but also physical). The mathematical
natuality is seen by the lift from the normal subgroup
$H$. We will see what is the exact meaning 
of the ``twist'' we have mentioned before. When we include the generator 
$\alpha$ and lift the normal subgroup $D_{\frac{kk'}{\delta}}$ 
to the whole group $G$, the structure of
conjugacy classes will generically change as well. For example, from
\begin{equation}
\alpha (\tilde{\beta}^b \tilde{\gamma}) \alpha^{-1} =
(\tilde{\beta}^{b+\frac{2k'}{\delta}} \tilde{\gamma}),
\label{add1}
\end{equation}
we see that the two different conjugacy classes $(\tilde{\beta}^{\even~b} \tilde{\gamma})$
and $(\tilde{\beta}^{\odd~b} \tilde{\gamma})$ 
will remain distinct if $\frac{2k'}{\delta}=\even$
and collapse into one single conjugacy class 
if $\frac{2k'}{\delta}=\odd$. We formally call the latter case
{\bf twisted}. Further clarifications regarding the 
structure of the conjugacy classes
of $G$ from the new points of view, especially physical, shall be most welcome.

After some algebraic manipulation, we can
write down all the conjugacy classes of $G$ in this new description.
For fixed $a$ and $\frac{2k'}{\delta}=\even$, 
we have the following classes: $(\alpha^a \tilde{\beta}^{-\frac{k'}{\delta}a})$,
$(\alpha^a \tilde{\beta}^{\frac{kk'}{\delta}-\frac{k'}{\delta}a}),$
$(\alpha^a \tilde{\beta}^b, \alpha^a \tilde{\beta}^{-b-\frac{2k'}{\delta}a})$
(with $b \neq -\frac{k'}{\delta}a$ and $\frac{kk'}{\delta}-\frac{k'}{\delta}a$), 
$(\alpha^b \tilde{\beta}^{p~\even} \tilde{\gamma})$ and 
$(\alpha^b \tilde{\beta}^{p~\odd} \tilde{\gamma})$. The crucial point here is that,
for every value of $a$, the structure of conjugacy classes is almost the same as that
of $D_{\frac{kk'}{\delta}}$. There is a 1-1
correspondence (or the lifting without the ``twist'') as we go from the
conjugacy classes of $H$ to $G$, making it possible to use
the idea of \cite{Han-Zaf2} to construct the corresponding
Brane Box Model. We will see this point more clearly later.
On the other hand, when
$\frac{2k'}{\delta}=\odd$, for fixed $a$, the conjugacy classes are 
no longer in 1-1 correspondence
between $H$ and $G$. Firstly, the last two 
classes of $H$ will combine into only one of $G$. Secondly,
the classes which contain only one element (the first two in $H$) will remain
so only for $a=\even$; for $a=\odd$, the they will combine into
one single class of $G$ which has two elements. 

So far the case of $\frac{2k'}{\delta}=\odd$ befuddles us and we do not know 
how the twist obstructs the construction of the Brane Box Model. This
twist seems to suggest quiver theories on non-affine $D_k$ diagrams because
the bifurcation on one side collapses into a single node, a phenomenon
hinted before in \cite{9811183,Han-Zaf2}.
It is a very interesting problem which we leave to further work.

\section{The Brane Box for \GZD} \label{sec:BB}
\subsection{The Puzzle}
The astute readers may have by now questioned themselves why such a long digression on
the esoterica of $G$ was done; indeed is it not enough to straightforwardly combine the
$D_{k'}$ quiver technique with the elliptic model and stack $k$ copies of Kapustin's
configuration on a circle to give the \GZD brane boxes?
Let us investigate where this na\"{\i}vet\'{e} fails. According to the discussions in
\sref{subsec:BBZZ}, one must construct one box for each irreducible representation of $G$. Let us place 2
ON$^0$ planes with $k'$ parallel NS5 branes in between as in \sref{subsec:Kapustin},
and then copy this $k$ times in the direction of the ON$^0$ and compactify that direction.
This would give us $k + k$ boxes each containing 2 1-dimensional irreducible representations corresponding
to the boxes bounded by one ON$^0$ and one NS5 on the two ends. And in the middle
we would have $k(k'-1)$ boxes each containing 1 2-dimensional irreducible representation.

Therefrom arises a paradox already! From the discussion of the group 
$G=Z_k \times D_{k'}$ in \sref{sec:group}, we recall that there are
$4k$ 1-dimensional irreducible representations and $k(\frac{k'k}{(k,2k')}-1)$ 2-dimensionals if
$\frac{2k'}{\delta} = \even$ and for $\frac{2k'}{\delta} = \odd$, $2k$ 
1-dimensionals and $k(\frac{k'k}{(k,2k')}-\frac12)$ 2-dimensionals.
Our attempt above gives a mismatch of the number the 2-dimensionals
by a factor of as large as $k$; there are far too many 2-dimensionals
for $G$ to be placed into the required $kk'$ boxes.
This mismatch tells us that such na\"{\i}ve constructions of the Brane Box Model fails.
The reason is that in this case what we are dealing with is a non-Abelian group
and the noncommutative property thereof twists the na\"{\i}ve structure
of the singularity. To correctly account for the property of the singularity 
after the non-Abelian twisting, we should attack in a new direction.
In fact, the discussion of the normal
subgroup $H$ in \sref{subsec:H} is precisely the way to see the 
structure of singularity more properly.
Indeed we have hinted, at least for $\frac{2k'}{\delta}=\even$,
that the na\"{\i}ve structure of the Brane Box Model can be applied again with a little 
modification, i.e., with the replacement of $D_{k'}$ by $D_{\frac{kk'}{\delta}}$.
Here again we have the generator of $Z_k$ acting on the first two coordinates of
$\C^3$ and the generators of $D_{\frac{kk'}{\delta}}$ acting on the last two.
This is the subject of the next sub section where we will 
give a consistent Brane Box Model for $G = Z_k \times D_{k'}$.

\subsection{The Construction of Brane Box Model}
Let us first discuss the decomposition of the fermionic {\bf 4} for which we
shall construct the brane box (indeed the model will dictate the fermion 
bi-fundamentals, bosonic matter fields will be given therefrom by supersymmetry).
As discussed in \cite{9811183} and \sref{subsec:Quiver}, 
since we are in an ${\cal N}=1$ (i.e., a co-dimension
one theory in the orbifold picture), the {\bf 4} must decompose into 
${\bf 1} \oplus {\bf 3}$ with the {\bf 1} being trivial. More precisely, 
since $G$ has only 1-dimensional or 2-dimensional irreducible representations, for giving the correct quiver diagram which corresponds to the Brane Box Model
the
{\bf 4} should  go into one trivial 1-dimensional, one non-trivial 1-dimensional
and one 2-dimensional according to
\[
{\bf 4} \longrightarrow (0,1)_1 \oplus (l',h')_1 \oplus (d,l)_2.
\]
Of course we need a constraint so as to ensure that such a decomposition
is consistent with the unity-determinant condition of the matrix representation
of the groups. Since from (\ref{2d}) we can compute the determinant of the $(d,l)_2$ to be
$(-1)^{(n+1)(d+1)}\omega_{k}^{(m+\tilde{m})(2l-d)}$, the constraining condition
is  $l'+2l-d\equiv 0(\bmod k)$.
In particular we choose
\begin{equation}
{\bf 3} \longrightarrow (l'=1,h'=1)_1+(d=1,l=0)_2;
\label{decomp}
\end{equation}
indeed this choice is precisely in accordance with the defining matrices of
$G$ in \sref{sec:group} and we will give the Brane Box Model corresponding to this
decomposition and check consistency.

\begin{figure}
\centerline{\psfig{figure=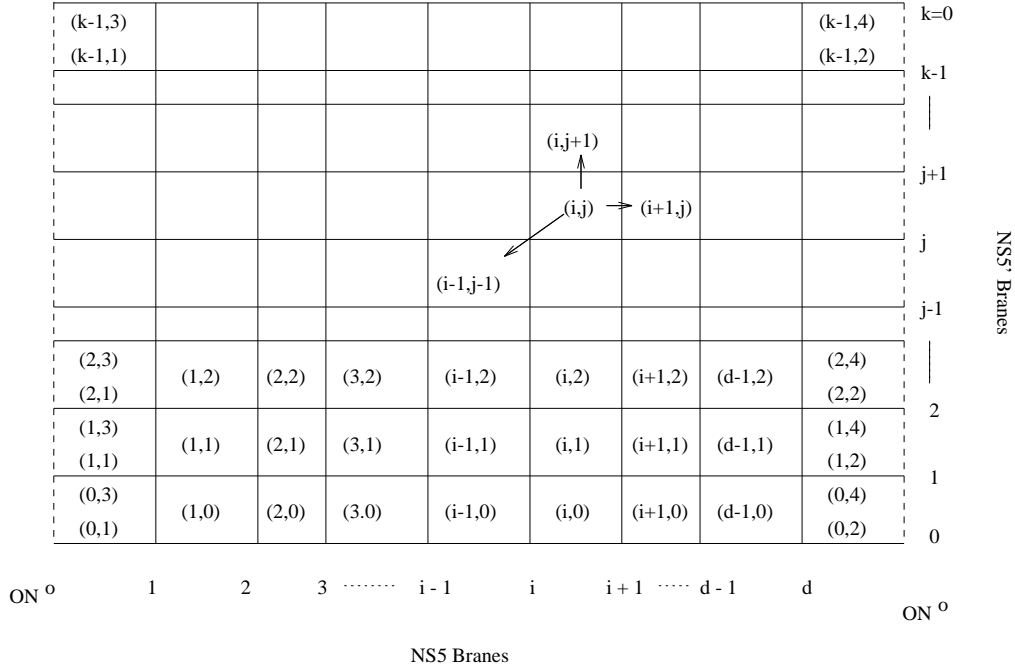,width=5.4in}}
\caption{The Brane Box Model for \GZD. We place $d := \frac{kk'}{\delta}$
NS5 branes in between 2 parallel ON$^0$-planes and $k$ NS5$'$ branes perpendicularly
while identifying the 0th and the $k$th circularly. Within the boxes of this grid, we
stretch D5 branes, furnishing bi-fundamental as indicated by the arrows shown.}
\label{fig:BBZD}
\end{figure}

Now we construct the brane box using the basic idea in \cite{Han-Zaf2} .
Let us focus on the case of $\delta := (k,2k')$ being even where we have
$4k$ 1-dimensional irreducible representations and $k(\frac{k'k}{(k,2k')}-1)$ 2-dimensionals.
We place 2 ON$^0$ planes vertically at two sides. Between them we place 
$\frac{kk'}{\delta}$ vertically parallel NS5 branes 
(which give the structure of $D_{\frac{kk'}{\delta}}$). 
Next we place $k$ NS5$'$ branes horizontally (which give the
structure of $Z_k$) and identify the $k$th with the zeroth. 
This gives us a grid of $k(\frac{kk'}{\delta}+1)$ boxes. Next we put $N$ D5 branes with
positive charge and $N$ with negative charge in those grids.
Under the decomposition (\ref{decomp}), we can connect the structure of singularity to
the structure of Brane Box Model by placing  the irreducible representations into the grid of boxes
\`{a} la \cite{HZ,HU} as follows (the setup is shown in
\fref{fig:BBZD}).

First we place the $4k$ 1-dimensionals at the two sides such that those boxes each 
contains two: at the left we have $(l'=0,h'=1)_1$ and $(l'=0,h'=3)_1$ 
at the lowest box and with the upper boxes containing subsequent increments on $l'$.
Therefore we have the list, in going up the boxes,
$\{ (0,1)_1~\&~(0,3)_1; 
(1,1)_1 ~\&~ (1,3)_1; (2,1)_1 ~\&~ (2,3)_1; ... (k-1,1)_1 ~\&~ (k-1,3)_1\}$. 
The right side has a similar list: 
$\{ (0,2)_1 ~\&~ (0,4)_1; 
(1,2)_1 ~\&~ (1,4)_1;$ 
$(2,2)_1 ~\&~ (2,4)_1; ... (k-1,2)_1 ~\&~ (k-1,4)_1\}$. 
Into the middle grids we place the 2-dimensionals, one to a box, such that the bottom
row consists of 
$\{(d=1,l=0)_2,(2,0)_2,(3,0)_2,...(\frac{kk'}{\delta}-1,0)_2 \}$ 
from left to right. And as we go up we increment $l$ until $l=k-1$ ($l=k$ is
identified with $l=0$ due to our compactification).
Now we must check the consistency condition. We choose the bi-fundamental directions
according to the conventions in \cite{HZ,HU}, i.e., East, North and Southwest.
The consistency condition is that for the irreducible representation
in box $i$, forming the tensor product with 
the {\bf 3} chosen in (\ref{decomp}) should be the tensor sum of the irreducible representations
of the neighbours in the 3 chosen directions, i.e.,
\begin{equation}
{\bf 3} \otimes R_i = \bigoplus\limits_{j\in{\rm Neighbours}} R_j
\label{consistency}
\end{equation}
Of course this consistency condition is precisely (\ref{aij3}) in a different
guise and
checking it amounts to seeing whether the Brane Box Model gives the
same quiver theory as does the geometry, whereby showing the equivalence 
between the two methods.
Now the elaborate tabulation in \sref{subsec:decomp} is seen to be not in vain;
let us check (\ref{consistency}) by column in the brane box as in 
\fref{fig:BBZD}.
For the $i$th entry in the leftmost column, containing $R_i=(l',1~{\rm or}~3)$,
we have $R_i \otimes {\bf 3} = (l',1~{\rm or}~3)_1 \otimes ((1,1)_1 \oplus
(1,0)_2) = (l'+1,1~{\rm or}~3)_1 \oplus (1,l')_2$. The righthand side is
precisely given by the neighbour of $i$ to the East and to the North and since
there is no Southwest neighbour, consistency (\ref{consistency}) holds for
the leftmost column. A similar situation holds for the rightmost column,
where we have ${\bf 3} \otimes (l',2~{\rm or}~4) = (l'+1,2~{\rm or}~4)_1
\oplus (\frac{kk'}{\delta}-1,l'-1)_2$, the neighbour to the North and the 
Southwest.

Now we check the second column, i.e., one between the first and second NS5-branes.
For the $i$th entry $R_i = (1,l)_2$, after tensoring with the {\bf 3},
we obtain $(1,l+1)_2  \oplus (l+1,l+0)_2 \oplus ((l+0-1,1)_1 \oplus
(l+0-1,3)_1)$, which are the irreducible representations precisely in the 3 neighbours: respectively
East, North and the two 1-dimensional in the Southwest. Whence 
(\ref{consistency}) is checked. Of course a similar situation occurs for the
second column from the right where we have ${\bf 3} \otimes 
(R_i = (\frac{kk'}{\delta}-1,l)_2) = (\frac{kk'}{\delta}-1,l+2)_2 \oplus
(\frac{kk'}{\delta}-1-1,l-1)_2 \oplus ((l,2)_1 \oplus (l,4)_1)$, or
respectively the neighbours to the North, Southwest and the East.

The final check is required of the interior box, say $R_i = (d,l)_2$.
Its tensor with {\bf 3} gives $(d,l+1)_2 \oplus (d-1,l-1)_2 \oplus
(d+1,l)_2$, precisely the neighbours to the North, Southwest and East.

\subsection{The Inverse Problem}
A natural question arises from our quest for the correspondence
between brane box constructions and branes as probes: is such a
correspondence bijective? Indeed if the two are to be related by some
T Duality or generalisations thereof, this bijection would be necessary.
Our discussions above have addressed one direction: given a \GZD singularity,
we have constructed a consistent Brane Box Model. Now we must ask
whether given  such a configuration with $m$ NS5 branes between two
ON$^0$  planes and $k$ NS5$'$ branes according to \fref{fig:BBZD},
could we find a unique \GZD orbifold which corresponds thereto?
The answer fortunately is in the affirmative and is summarised in the following:
\begin{proposition}
For $\frac{2k'}{(k,2k')}$ being even\footnote{Which is the case upon which
we focus.}, there exists a bijection\footnote{Bijection in the sense that given
a quiver theory produced from one picture there exists a unique method in the
other picture which gives the same quiver.} between the Brane Box Model and
the D3 brane-probes on the orbifold for the
group $G := Z_k \times D_{k'} \cong Z_k \rtimes D_{m:=\frac{kk'}{(k,2k')}}$.
In particular
\begin{itemize}
\item (I) Given $k$ and $k'$, whereby determining $G$ and hence the orbifold theory,
one can construct a unique Brane Box Model;
\item (II) Given $k$ and $m$ with the condition that $k$ is a divisor of $m$,
where $k$ is  the number of NS5 branes perpendicular to $ON^0$ planes and 
$m$ the number of NS5 branes between two  $ON^0$ planes
as in \fref{fig:BBZD}, one can extract a unique orbifold theory.
\end{itemize}
\end{proposition}

Now we have already shown (I) by our extensive discussion in the previous
sections. Indeed, given integers $k$ and
$k'$, we have twisted $G$ such that it is characterised by $k$ and
\begin{equation}
\label{addm}
m:=\frac{kk'}{(k,2k')},
\end{equation}
two numbers that uniquely fix the brane configuration.
The crux of the remaining direction (II) seems to be the issue whether we could,
given $k$ and $m$, ascertain the values
of $k$ and $k'$ uniquely? For if so then our Brane Box Model, which is solely
determined by $k$ and $m$, would be uniquely mapped to a \GZD orbifold, characterised
by $k$ and $k'$. We will show below that though this is not so and $k$ and $k'$ cannot
be uniquely solved, it is still true that $G$ remains unique. Furthermore, we will
outline the procedure by which we can find convenient choices of $k$ and $k'$ that
describe $G$.

Let us analyse this problem in more detail.
First we see that $k$, which determines the $Z_k$ in $G$, remains unchanged.
Therefore our problem is further reduced to: given $m$, is there a unique
solution of $k'$ at fixed $k$? We write $k,k',m$ as:

\begin{equation}
\label{kk}
\begin{array}{c}
k=2^{q}l f_2  \\
k'=2^{p} l f_1 \\
m=2^{n} f_3
\end{array}
\end{equation}

where with the extraction of all even factors, $l,f_1$ and $f_2$
are all odd integers and $l$ is the greatest common divisor of $k$ and $k'$ so that
$f_1,f_2$ are coprime. What we need to know are $l,f_1$ and $p$ given $k,q,n$ and $f_3$.
The first constraint is that $\frac{2k'}{(k,2k')}=\even$, a condition
on which this chapter 
focuses. This immediately yields the inequality $p\geq q$. The definition of
$m$ (\ref{addm}) above further gives
\[
2^{n} f_3=m=2^p l f_1 f_2= 2^{p-q}k f_1.
\]
From this equation, we can solve
\begin{equation}
\label{OI1}
p=n,~~~~~~~f_1=\frac{m}{2^{p-q}k}
\end{equation}
Now it remains to determine $l$. However, the solution for
$l$ is not unique. For example, if we take $l=l_1 l_2$ and $(l_2,f_1)=1$, then
the following set $\{\tilde{k},\tilde{k'}\}$ will give same $k,m$:
\[
\begin{array}{c}
\tilde{k}=k=2^{q}l_1 l_2 f_2  \\
\tilde{k'}=2^{p} l_1 f_1 \\
m=2^{n} f_3
\end{array}
\]
This non-uniqueness in determining $k,k'$ from $k,m$ may at first seem discouraging.
However we shall see below that different pairs of $\{k,k'\}$ that give the
same $\{k,m\}$ must give the same group $G$.

We first recall that $G$ can be written as
$Z_k \rtimes D_{m=\frac{kk'}{(k,2k')}}$.
For fixed $k,m$ the two subgroups $Z_k$ and $D_m$ are same. For the whole group
$Z_k \rtimes D_{m=\frac{kk'}{(k,2k')}}$ be unique no matter which $k'$ we choose we just need to show that the algebraic relation which generate 
$Z_k \rtimes D_{m=\frac{kk'}{(k,2k')}}$ from $Z_k$ and $D_m$ is same. For that,
 we recall from the proposition
in section \sref{subsec:H}, that in twisting $G$ into its internal semi-direct form,
the crucial relation is
\[
\alpha \tilde{\gamma}=\tilde{\beta}^{\frac{2k'}{(k,2k')}} \tilde{\gamma} \alpha
\]
Indeed we observe that $\frac{k'}{(k,2k')}= \frac{m}{k}$ where the condition
that $k$ is a divisor of $m$ makes the expression having meaning. Whence given $m$ and $k$,
the presentation of $G$ as $Z_k \rtimes D_m$ is uniquely fixed, and hence $G$
is uniquely determined. This concludes our demonstration for the above proposition.

Now the question arises as to what values of $k$ and $k'$ result in the
same $G$ and how the smallest pair (or rather, the smallest $k'$ since
$k$ is fixed) may be selected. In fact our discussion
above prescribes a technique of finding such a pair. First we
solve $p,f_1$ using (\ref{OI1}), then we find the largest factor $h$ of $k$ which
satisfies $(h,f_1)=1$. The smallest value of $k'$ is then such that
$l=\frac{k}{h}$ in (\ref{kk}).
Finally, we wish to emphasize
that the bijection we have discussed is not true for arbitrary $\{m,k\}$ and we
require that $k$ be a divisor of $m$ as is needed in demonstration of the
proposition. Indeed, given $m$ and $k$ which do not satisfy
this condition, the 1-1 correspondence between the Brane Box Model and the 
orbifold singularity is still an enigma and will be 
left for future labours.

\section{Conclusions and Prospects} \label{sec:conc}
We have briefly reviewed some techniques in two contemporary 
directions in the construction
of gauge theories from branes, namely branes as geometrical probes on orbifold
singularities or as constituents of configurations of D branes stretched between
NS branes. 
Some rudiments in the orbifold procedure,
in the brane setup of
${\cal N}=2$ quiver theories of the $\widehat{D_k}$ type as well as in the
${\cal N}=1$ $Z_k \times Z_{k'}$ Brane Box Model have been introduced.
Thus inspired, we have
constructed the Brane Box Model for an infinite series of
non-Abelian finite subgroups of $SU(3)$, by combining some methodology
of the aforementioned brane setups.

In particular, we have extensively studied the properties, especially the
representation and character theory of the intransitive collineation group 
$G := Z_k \times D_{k'} \subset SU(3)$,
the next simplest group after $Z_k \times Z_{k'}$ and a natural extension thereof.
From the geometrical perspective, this amounts to the study of Gorenstein
singularities of the type $\C^3 / G$ with the $Z_k$ acting on the first two
complex coordinates of $\C^3$ and $D_{k'}$, the last two.

We have shown why na\"{\i}ve Brane Box constructions for $G$ fail (and indeed
why non-Abelian groups in general may present difficulties). It is only after
a ``twist'' of $G$ into a semi-direct product form $Z_k \rtimes D_{\frac{kk'}{(k,2k')}}$,
an issue which only arises because of the non-Abelian nature of $G$, that
the problem may be attacked. For $\frac{2k'}{(k,2k')}$ even, we have successfully
established a consistent Brane Box Model. The resulting gauge theory is that of
$k$ copies of $\widehat{D}$-type quivers circularly arranged (see \fref{fig:BBZD}).
However for $\frac{2k'}{(k,2k')}$ odd, a degeneracy occurs and we seem to arrive at
ordinary (non-Affine) $D$ quivers, a phenomenon hinted at by some previous 
works \cite{Han-Zaf2,9811183} but still remains elusive. Furthermore, we have
discussed the inverse problem, i.e., whether given a configuration of the
Brane Box Model we could find the corresponding branes as probes on orbifolds. 
We have shown that when $k$ is a divisor of $m$
the two perspectives are bijectively related
and thus the inverse problem can be solved.
For general $\{m,k\}$, the answer of the inverse problem is still not clear.

Many interesting problems arise and are open. Apart from clarifying the physical
meaning of ``twisting'' and hence perhaps treat the $\frac{2k'}{(k,2k')}$ odd case,
we can try to construct Brane Boxes for more generic non-Abelian groups.
Moreover, marginal couplings and duality groups thereupon may be extracted
and interpreted as brane motions; this is of particular interest because
toric methods from geometry so far have been restricted to Abelian singularities.
Also, recently proposed brane diamond models \cite{Aganagic} may be combined with
our techniques to shed new insight. There is a parallel
work that deals with brane configurations for
$\C^3/\Gamma$ 
singularities for non-Abelian $\Gamma$ (i.e the $\Delta$ series in $SU(3)$)
by $(p,q)$5-brane webs \cite{ZD}.
We hope that our construction, as the  Brane Box Model realisation of
a non-Abelian 
orbifold theory in dimension 3, may lead to insight in these various
directions.
\index{Hanany-Witten!brane box}
\index{Finite Groups!$SU(3)$ subgroups!$Z$-$D$ type}
\chapter{Orbifolds VI: $Z$-$D$ Brane Box Models}
\label{chap:9909125}
\section*{\center{{ Synopsis}}}
Generalising the ideas of the previous chapter, 
we address the problem of
constructing Brane Box
Models of what we call the $Z$-$D$ Type from a new point of view, so as
to establish the complete correspondence between these brane setups
and orbifold singularities of the non-Abelian
$G$ generated by $Z_k$ and $D_d$ under certain 
group-theoretic constraints to
which we refer as the BBM conditions.
Moreover, we present
a new class of ${\cal N}=1$ quiver theories of the ordinary 
dihedral group $d_k$ as well as the ordinary exceptionals $E_{6,7,8}$
which have
non-chiral matter content and discuss issues related to
brane setups thereof \cite{9909125}.
\section{Introduction}
Configurations of branes \cite{HW} have been proven to be a 
very useful method to
study the gauge field theory which emerges as the low energy limit of string
theory (for a complete reference, see Giveon and Kutasov \cite{branerev}).
The advantage of 
such setups is that they provide an intuitive picture so that we can very easily
deduce many properties of the gauge theory.
For example, brane setups have been used to study mirror symmetry in 3 
dimensions \cite{HW,Seiberg,P-Zaf,Boer,Kapustin}, Seiberg Duality in 4
dimensions \cite{Methods},
and exact solutions when lifting Type IIA setups to M-theory
\cite{M-Theory,Karl}.
After proper T- or S-dualities, we can
transform the above brane setups to D-brane as probes on some target space with
orbifold singularities \cite{DM,Orb2,LNV}.

For example, the brane setup
of streching Type IIA D4-branes between $n+1$ NS5-branes placed in a circular
fashion (the ``elliptic model'' \cite{M-Theory}) is precisely T-dual
to D3-branes 
stacked upon ALE singularities of the type $\widehat{A_n}$, or in other words 
orbifold singularities of the form $\C^2/Z_{n+1}$,  where $Z_{n+1}$ is
the cyclic group on $n+1$ elements and is a finite discrete 
subgroup of $SU(2)$.
As another example, the Brane Box Model \cite{HZ,HU,HSU}
is T-dual to D3-branes as probes on orbifold singularities of the
type $\C^3/\Gamma$ with $\Gamma=Z_k$ or $Z_k \times Z_{k'}$ now being a finite
discrete subgroup of $SU(3)$ \cite{HU}. A brief review of some of these 
contemporary techniques can be found in the previous chapter.
In fact, it is a very interesting problem to see how in general 
the two different methods, viz., brane setups and D3-branes
as probes on geometrical singularities, are connected to each other
by proper duality transformations \cite{Karch:equiv}.

The general construction and methodology for D3-branes as probes
on orbifold singularities has been given \cite{LNV}. However, the complete
list of the
corresponding brane setups is not yet fully known. For orbifolds 
$\C^2/\{\Gamma \in SU(2)\}$, we have the answer
for the $\widehat{A_n}$ series (i.e., $\Gamma=Z_{n+1}$) and the 
$\widehat{D_n}$ series (i.e., $\Gamma=D_{n-2}$, the binary dihedral groups) 
\cite{Kapustin}, but not for the exceptional cases $\widehat{E_{6,7,8}}$.
At higher dimensions, the situation is even more disappointing: for orbifolds of 
$\C^3/\{\Gamma \in SU(3)\}$, brane setups are until recently limited to only
Abelian singularities, namely $\Gamma=Z_k$ or $Z_{k} \times Z_{k'}$ \cite{HU}.

In the previous chapter, we went beyond the Abelian restriction in
three dimensions and gave 
a new result concerning the correspondence of the two methods.
Indeed we showed that\footnote{In that chapter we used the notation
$Z_k \times D_{k'}$ and pointed out that the symbol $\times$ was
really an abuse. We shall here use the symbol $*$ and throughout this
chapter reserve $\times$ to mean strict direct product of groups and
$\rtimes$, the semi-direct product.} for $\Gamma := G = Z_k * D_{k'}$ 
a finite 
discrete subgroup of $SU(3)$, the corresponding brane setup (a Brane Box Model)
T-dual to the orbifold discription can be obtained.
More explicitly, the group $G \in SU(3)$ is defined as being generated by the
following matrices that act on $\C^3$:

\begin{equation}
\label{gen1}
\alpha = \left(  \begin{array}{ccc}
                        \omega_{k} & 0 & 0  \\
                         0 & \omega_{k}^{-1} & 0\\
                         0  &  0 & 1
                \end{array}
        \right)
~~~~~~~~
\beta = \left(  \begin{array}{ccc}
                        1  & 0  & 0  \\
                        0 & \omega_{2k'} & 0  \\
                        0 &  0 & \omega_{2k'}^{-1}  
                \end{array}
        \right)
~~~~~~~~
\gamma =\left(  \begin{array}{ccc} 
                1  &  0  &  0  \\
                0  &  0  &  i \\
                0  &  i &   0 
                \end{array}
        \right)
\end{equation}
where $\omega_n := e^{\frac{2 \pi i}{n}}$, the $n$th root of unity.

The abstract presentation of the groups is as follows:
\begin{equation}
\alpha \beta = \beta \alpha,~~~~\beta \gamma =\gamma \beta^{-1},~~~~
\alpha^{m} \gamma \alpha^{n} \gamma =\gamma \alpha^{n} \gamma \alpha^{m}
~~~~\forall m,n \in \Z
\label{rel1}
\end{equation}

Because of the non-Abelian property of $G$,
the preliminary attempts at the corresponding Brane Box Model by using the idea
in a previous work \cite{Han-Zaf2} met great difficulty.
However, via careful analysis, 
we found that the group $G$ can be written as the semidirect product of $Z_{k}$ and
$D_{\frac{kk'}{\gcd(k,2k')}}$. Furthermore, when $\frac{2k'}{\gcd(k,2k')}=\even$,
the character table of $G$ as the semidirect product 
$Z_k \rtimes D_{\frac{kk'}{\gcd(k,2k')}}$
preserves the structure of that of $D_{\frac{kk'}{\gcd(k,2k')}}$,
in the sense that it seems to be composed of $k$ copies of the latter.
Indeed, it was noted \cite{9906031} that only under this parity condition 
of $\frac{2k'}{\gcd(k,2k')}=\even$, can we construct, with
the two group factors $Z_{k}$ and $D_{\frac{kk'}{\gcd(k,2k')}}$, a consistent
Brane Box Model with the ideas in the abovementioned paper \cite{Han-Zaf2}.

The success of the above construction, constrained by certain 
conditions, hints that
something fundamental is acting as a key r\^ole in the construction of
non-Abelian brane setups above two (complex) dimensions.
By careful consideration, it seems that the following three conditions
presented themselves to be crucial in the study of $Z_k * D_{k'}$
which we here summarize:
\begin{enumerate}
\label{condition}
\item The whole group $G$ can be written as a semidirect product:
	$Z_k \rtimes D_{d}$;

\item The semidirect product of $G$ preserves the structure of the 
	irreducible representations of $D_{d}$, i.e., it appears
	that the irreps of $G$
	consist of $k$ copies of those of the subgroup $D_{d}$;

\item There exists a (possibly reducible) representation of $G$ 
	in 3 dimensions such that the representation matrices belong to $SU(3)$.
	Henceforth, we shall call such a representation,
	consistent with the $SU(3)$ requirement (see more discussions
	\cite{9906031,9811183} on decompositions), as 
	``{\it the chosen decomposition of {\bf 3}}''.
\end{enumerate}
We will show in this chapter that these conditions are sufficient for
constructing Brane Box Model of the $Z$-$D$ type. Here we will call the
Brane Box Model in the previous chapter as 
{\bf Type $Z$-$D$} and similarly, 
that in earlier works \cite{HZ,HU} we shall call the
$Z$-$Z$ Type. We shall see this definition more clearly in subsection
\sref{subsec:BBMZD}.
It is amusing to notice that Brane Box Models of Type $Z$-$Z$ also 
satisfy the above three conditions since they correspond to the
group $Z_k \times Z_{k'}$, which is a direct product.

Furthermore, we shall answer a mysterious question posed
at the end of the previous chapter. 
There, we discussed the so-called
{\it Inverse Problem}, i.e., given a consistent Brane Box Model, how
may one determine, from the structure of the setup 
(the number and the positioning
of the branes), the corresponding group $\Gamma$ in the orbifold 
structure of $\C^3/\Gamma$.
We found there that only when $k$ is the divisor of $d$ can we
find the corresponding group defined in (\ref{gen1}) with proper
$k,k'$. This was very unsatisfying. However, the structure of the
Brane Box Model of Type $Z$-$D$ was highly suggestive of the solution for
general $k,d$. We shall here mend that short-coming and for
arbitrary $k,d$ we shall construct the
corresponding group $\Gamma$ which satisfies above three conditions. 
With this result, we establish the complete correspondence 
between the Brane Box Model of Type $Z$-$D$ and 
D3-branes as probes on orbifold singularities
of $\C^3/\Gamma$ with properly determined $\Gamma$.

The three conditions which are used for solving the inverse problem can be 
divided into two conceptual levels. The first two are at the level of pure
mathematics, i.e., we can consider it from the point of view of abstract group theory
without reference to representations or to finite discrete subgroup of
$SU(n)$. The third condition is at the level of physical applications. From
the general structure \cite{LNV} we see that for constructing 
${\cal N}=2$
or ${\cal N}=1$ theories we respectively need the group $\Gamma$ to be a finite 
subgroup of $SU(2)$ or $SU(3)$. This requirement subsequently means that
we can find a faithful (but possibly reducible) 2-dimensional or 3-dimensional 
representation with the matrices satisfying the determinant 1 and unitarity
conditions. 
In other words, what supersymmetry (${\cal N}=2$ or 1) we 
will have in the orbifold theory by the standard procedure 
\cite{LNV} depends only on the chosen representation (i.e., the
decomposition of ${\bf 2}$ or ${\bf 3}$). Such distinctions were already
shown before \cite{HU,9811183}. The 
group $Z_3$ had been considered \cite{HU}.
If we choose its action on $\C^3$ as 
$(z_1,z_2,z_3) \longrightarrow
(e^{\frac{2\pi i}{3}} z_1,e^{\frac{-2\pi i}{3}} z_2,z_3)$
we will have ${\cal N}=2$ supersymmetry, but if we choose
the action to be 
$(z_1,z_2,z_3)\longrightarrow
(e^{\frac{2\pi i}{3}} z_1,e^{\frac{2\pi i}{3}} z_2,e^{\frac{2\pi i}{3}} z_3)$
we have only ${\cal N}=1$. This phenomenon mathematically corresponds to 
what are called sets of transitivity of collineation group actions
\cite{9905212,Yau}.

Moreover, we notice that the ordinary dihedral group $d_k$ which is
excluded from the classification of finite subgroup of $SU(2)$ can be
imbedded\footnote{Since it is in fact a subgroup of $SU(2)/\Z_2 \cong SO(3)$,
the embedding is naturally induced from $SO(3) \hookrightarrow SU(3)$.
In fact the 3-dimensional representation in $SU(3)$ is faithful.} into $SU(3)$.
Therefore we expect that $d_k$ should be useful in constructing some
${\cal N}=1$ gauge field theories by the standard 
procedures \cite{LNV,9811183}.
We show in this chapter that this is so.
With the proper decompositions, we can obtain new types of gauge theories by 
choosing $\C^3$ orbifolds to be of the type $d_k$.
For completeness, we also give the quiver 
diagrams of ordinary tetrahedral, octahedral and icosahedral groups 
($E_{6,7,8}$), which by a similar token, can be imbedded into $SU(3)$.

The organisation of the chapter is as follows. In \sref{sec:dir} we
give a simple 
and illustrative example of constructing a Brane Box Model for the direct
product $Z_k \times D_{k'}$, whereby initiating the study of brane setups
of what we call Type $Z$-$D$. In \sref{sec:twist} we deal
with the twisted case which we encountered earlier in the previous chapter.
We show that we can imbed the latter into the direct product (untwisted) case of
\sref{sec:dir} and arrive at another member of Brane Box Models of the $Z$-$D$ 
type. 
In \sref{sec:new} we give a new class of $SU(3)$ quiver
which are connected to the ordinary dihedral group $d_k$. Also, we give 
an interesting brane configuration that will give matter matter content as the 
$d_{k=\even}$ quiver but a different superpotential on the gauge theory level.
Finally in \sref{sec:con} we give concluding remarks and suggest future prospects.

\section*{Nomenclature}
Unless otherwise specified, we shall throughout the chapter adhere to the
notation that the group binary operator $\times$ refers to the 
strict direct product,
$\rtimes$, the semi-direct product, and $*$, a general (twisted)
product by combining the generators of the 
two operands\footnote{Therefore in the previous chapter, the
group $G := Z_k \times D_{k'}$ in this convention should be written as
$Z_k*D_{k'}$, {\it q.v. Ibid.} for discussions on how these different group
compositions affect brane constructions.}.
Furthermore, $\omega_n$ is defined to be $e^{\frac{2 \pi i}{n}}$, the $n$th
root of unity; $H \triangleleft G$ mean that $H$ is a normal subgroup of $G$; and
a group generated by the set $\{x_i\}$ under relations 
$f_i(\{x_j\}) = 1$ is denoted as $\langle x_i | f_j \rangle$.

\section{A Simple Example: The Direct Product $Z_{k} \times D_{k'}$}
\label{sec:dir}
\index{Finite Groups!$SU(3)$ subgroups!$Z$-$D$ type}
We recall that in a preceeding chapter,
we constructed the Brane Box Model (BBM) for
the group $Z_k * D_{k'}$ as generated by (\ref{gen1}), satifying 
the three conditions mentioned above, which we shall dub as the 
{\bf BBM condition}
for groups. However, as we argued in the introduction, there may exist in
general, groups not isomorphic to the one addressed \cite{9906031} but still
obey these conditions.
As an illustrative example, we start with the simplest member of the family of
$Z*D$ groups that satisfies the BBM condition, namely the
direct product $G=Z_k \times D_{k'}$. We define $\alpha$
as the generator for the $Z_{k}$ factor and $\gamma,\beta$, those for
the $D_{k'}$. Of course by definition of the direct product $\alpha$ must
commute with both $\beta$ and $\gamma$.
The presentation of the group is clearly as follows:
\[
\begin{array}{ll}
\alpha^k=1; & 	$The Cyclic Group $Z_k \\
\beta^{2k'} =1,~~~~~\beta^{k'}=\gamma^2,~~~\beta \gamma=\gamma \beta^{-1}; &
	$The Binary Dihedral Group $D_{k'} \\
\alpha \beta =\beta \alpha,~~~\alpha \gamma =\gamma \alpha &
	$Mutual commutation$ \\
\end{array}
\]

We see that the first two of the BBM conditions are trivially satisfied.
To satisfy the third, we need a 3-dimensional matrix 
represenation of the group.
More explicitly, as discussed \cite{9906031}, to construct
the BBM of the $Z$-$D$ type, one needs the decomposition of ${\bf 3}$
into one nontrivial 1-dimensional irrep and one 2-dimensional 
irrep. In light of this, we can write down the $SU(3)$ matrix
generators of the group as

\begin{equation}
\label{gen_dir}
\alpha = \left(  \begin{array}{ccc}
			\omega_{k}^{2} & 0 & 0  \\
			 0 & \omega_{k}^{-1} & 0\\
			 0  &  0 & \omega_{k}^{-1}
		\end{array}
	\right)
~~~~~~~~
\beta = \left(  \begin{array}{ccc}
			1  & 0  & 0  \\
			0 & \omega_{2k'} & 0  \\
			0 &  0 & \omega_{2k'}^{-1}  
		\end{array}
	\right)
~~~~~~~
\gamma =\left(  \begin{array}{ccc} 
		1  &  0  &  0  \\
		0  &  0  &  i \\
		0  &  i &   0 
		\end{array}
	\right)
\end{equation}

Here, we notice a subtle point. When $k=\even$, $\alpha^{\frac{k}{2}}$
and $\beta^{k'}$ give the same matrix form. In other words,
(\ref{gen_dir}) generates a {\it non-faithful} representation. We will
come back to this problem later, but before diving into a detailed 
discussion on the whole group $Z_k \times D_{k'}$, let us first give 
the necessary properties of the factor $D_{k'}$.

\subsection{The Group $D_{k'}$}
\label{subsec:Dk'}
One can easily check that all the elements of the binary dihedral
$D_{k'}=\langle \beta, \gamma \rangle$ group can be written, 
because $\gamma^2=\beta^{k'}$, as
\[
\gamma^n \beta^p,~~~{\rm with}~~~n=0,1~~p=0,1,...,2k'-1.
\]

From this constraint and the conjugation relation 
\[
(\gamma^{n_1} \beta^{p_1})^{-1} (\gamma^{n} \beta^{p}) 
(\gamma^{n_1} \beta^{p_1}) = 
\gamma^{n} \beta^{p_1(1-(-1)^{n})+(-1)^{n_1}p},
\]
we can see that the group is of order $4k'$ and moreover affords
4 1-dimensional irreps and $(k'-1)$ 2-dimensional irreps. The
classes of the group are:
\[
\begin{array}{ccccc}
        & C_{n=0}^{p=0} & C_{n=0}^{p=k'} & C_{n=0}^{\pm p} & C_{n=1}^{p \bmod 2} \\
|C|  & 1 & 1 & 2 & k' \\
\#C   & 1 & 1 & k'-1 & 2
\end{array}
\]

To study the character theory of $G := D_{k'}$, we recognise that
$H := \{\beta^{p}\}$ for $p$ even is a normal subgroup of $G$. Whence we can use
the Frobenius-Clifford theory of induced characters
to obtain the irreps of $G$ from the
factor group $\widetilde{G} :=G/H={1,\beta,\gamma,\gamma \beta}$.
For $k'$ even, $\widetilde{G}$ is $Z_2\times Z_2$ and for $k'$ odd, it is simply $Z_4$.
these then furnish the 1-dimensional irreps.
We summarise the characters of these 4 one dimensionals as follows:
\[
\begin{array}{c|c}
k' = \even
&
k' = \odd \\ \hline
\begin{array}{c|cccc}
     &	\beta^{p=\even} & \beta(\beta^{\odd}) & \gamma(\gamma \beta^{\even}) &
	\gamma \beta (\gamma \beta^{\odd}) \\
\chi^{1} & 1 & 1 & 1 & 1 \\
\chi^{2} & 1 & -1 & 1 & -1 \\
\chi^{3} & 1 & 1 & -1 & -1 \\
\chi^{4} & 1 & -1 & -1 & 1 
\end{array}
&
\begin{array}{cccc}
	\beta^{\even} & \beta(\beta^{\odd}) & \gamma(\gamma \beta^{\even}) &
	\gamma \beta (\gamma \beta^{\odd}) \\
1 & 1 & 1 & 1 \\
1 & -1 & \omega_4 & -\omega_4 \\
1 & 1 & -1 & -1 \\
1 & -1 & - \omega_4 &  \omega_4
\end{array} \\
\end{array}
\]

The 2-dimensional irreps can be directly obtained from the
definitions; they are indexed by a single integer $l$:

\begin{equation}
\label{2dreps}
\chi_{2}^{l}(C_{n=1})=0,~~~~\chi_{2}^{l}(C_{n=0}^{p})= 
(\omega_{2k'}^{lp}+\omega_{2k'}^{-lp}),~~l=1,..,k'-1.
\end{equation}

The matrix representations of these 2-dimensionals are given below:
\[
\beta^{p} = \left(  \begin{array}{cc}  \omega_{2k'}^{lp} & 0 \\
					0 & \omega_{2k'}^{-lp}
			\end{array}
		\right)
~~~~~~~
\gamma \beta^{p} = \left(  \begin{array}{cc}  0 & i^l\omega_{2k'}^{-lp} \\
					i^l \omega_{2k'}^{lp} & 0 
			\end{array}
		\right)
\]
 From (\ref{2dreps}) we immediately see that 
$\chi_2^{l}=\chi_2^{-l}=\chi_2^{2k'-l}$ which we use to restrict the index
$l$ in $\chi_2^l$ into the region $[1,k'-1]$. 

Now for the purposes of the construction of the BBM, we aboveall need to 
know the tensor decompositions of the group; these we summarise below.
\[
\begin{array}{|l|l|}
\hline
{\bf 1} \otimes {\bf 1}'
&
\begin{array}{c|c}
	k' = \even	& k' = \odd \\
	\begin{array}{ccc}
	\chi_1^2\chi_1^2=\chi_1^1  & \chi_1^3\chi_1^3=\chi_1^1 &
	\chi_1^4\chi_1^4=\chi_1^1  \\
	\chi_1^2 \chi_1^3=\chi_1^4 & \chi_1^2\chi_1^4=\chi_1^3 &
	\chi_1^3\chi_1^4=\chi_1^2
	\end{array}
	&
	\begin{array}{ccc}
	\chi_1^2\chi_1^2=\chi_1^3  & \chi_1^3\chi_1^3=\chi_1^1 &
	\chi_1^4\chi_1^4=\chi_1^3  \\
	\chi_1^2 \chi_1^3=\chi_1^4 & \chi_1^2\chi_1^4=\chi_1^1 &
	\chi_1^3\chi_1^4=\chi_1^2
	\end{array}
\end{array}
\\ \hline
{\bf 1} \otimes {\bf 2}
&
\chi_1^{h} \chi_2^l = \left\{ \begin{array}{l}
\chi_2^l~~~~h=1,3  \\
\chi_2^{k'-l}~~~~h=2,4 
\end{array}
\right.
\\ \hline
{\bf 2} \otimes {\bf 2'}
&
\chi_2^{l_1} \chi_2^{l_2}=\chi_2^{(l_1+l_2)}+\chi_2^{(l_1-l_2)}
{\rm ~where~}
\begin{array}{l}
	\chi_2^{(l_1+l_2)}= \left\{ \begin{array}{l}
	\chi_2^{(l_1+l_2)}~~~~{\rm if}~~~l_1+l_2<k',  \\
	\chi_2^{2k'-(l_1+l_2)}~~~~{\rm if}~~~l_1+l_2>k', \\
	\chi_1^2+\chi_1^4~~~~{\rm if}~~~l_1+l_2=k'.
	\end{array}
	\right.
	\\
	\chi_2^{(l_1-l_2)}= \left\{ \begin{array}{l}
	\chi_2^{(l_1-l_2)}~~~~{\rm if}~~~l_1>l_2,  \\
	\chi_2^{(l_2-l_1)}~~~~{\rm if}~~~l_1<l_2, \\
	\chi_1^1+\chi_1^3~~~~{\rm if}~~~l_1=l_2.
	\end{array}
	\right.
\end{array}
\\
\hline
\end{array}
\]

\subsection{The Quiver Diagram}
\index{Orbifolds}
\index{Brane Probes!Orbifolds}
The general method of constructing gauge field theories from orbifold 
singularities of $C^3/\Gamma \subset SU(3)$ has been given \cite{LNV,9811183}.
Let us first review briefly the essential results.
Given a finite discrete subgroup $\Gamma \subset SU(3)$ with irreducible
representations $\{ r_i \}$,
we obtain, under the orbifold projection, an ${\cal N}=1$ super Yang-Mills 
theory with gauge group
\[
\bigotimes_{i} SU(N|r_i|),~~~~~~|r_i|=\dim(r_i),N \in \Z
\]
To determine the matter content we need to choose the decomposition of ${\bf 3}$
(i.e., the $3 \times 3$ matrix form) of $\Gamma$ which describes how it acts upon
$\C^3$. 
We use $R$ to denote the representation of chosen ${\bf 3}$ and
calculate the tensor decomposition 

\begin{equation}
\label{decomp2}
R \otimes r_i= \bigoplus_{j} a_{ij}^{R} r_{j}
\end{equation}

The matrix $a_{ij}^{R}$ then tells us how many bifundamental chiral multiplets
of $SU(N_i) \times SU(N_j)$ there are which transform under the representation 
$(N_i,\bar{N_j})$, where $N_i := N|r_i|$.
Furthermore, knowing this matter content we can also write down
the superpotential whose explicit form is given in (2.7) and (2.8) of
Lawrence, Nekrasov and Vafa \cite{LNV}.
We do not need the detailed form thereof but we emphasize that all terms in 
the superpotential are cubic and there are no quatic term. This condition
is necessary for finiteness \cite{HSU,LNV}
and we will turn to this fact later.

We can encode the above information into a ``{\bf quiver diagram}''.
Every node $i$ with index $\dim{r_i}$ in the 
quiver denotes the gauge group $SU(N_i)$. 
Then we connect $a_{ij}^{R}$ arrows from node $i$ to $j$ in order
to denote the correpsonding bifundamental chiral multiplet $(N_i,\bar{N_j})$.
When we say that a BBM construction is {\bf consistent} we mean that
it gives the same quiver diagram as one obtains from the geometrical
probe methods \cite{LNV}.

Now going back to our example $Z_k \times D_{k'}$, its
character table is easily written: it is simply the 
Kronecker product of the
character tables of $Z_k$ and $D_{k'}$ (as matrices).
We promote (\ref{2dreps}) to a double index
\[
(a,\chi_{i}^{l})
\]
to denote the charaters, where $a = 0, ... , k-1$ and
are characters of $Z_k$ (which are simply various $k$th roots of unity)
and $\chi$ are the characters of $D_{k'}$ as presented in the previous
subsection. We recall that $i = 1$ or 2 and for the former, there
are 4 1-dimensional irreps indexed by $l=1,..,4$; and for the latter, 
there are $k'-1$ 2-dimensional irreps indexed by $l=1,..,k'-1$.
It is not difficult to see from (\ref{gen_dir}) that the chosen decomposition 
should be:
\[
{\bf 3} \longrightarrow (2,\chi_{1}^{1}) \oplus (-1,\chi_{2}^{1})
\]
The relevant tensor decomposition which gives the quiver is then
\begin{equation}
\label{tensor}
[(2,\chi_{1}^{1}) \oplus (-1,\chi_{2}^{1})]\otimes (a,\chi_{i}^{l})
=(a+2,\chi_{i}^{l}) \oplus (a-1,\chi_{i}^{l} \otimes \chi_{2}^{1}),
\end{equation}
which is thus reduced to the decompositions as tabulated in the previous
subsection.

\subsection{The Brane Box Model of $Z_k \times D_{k'}$}
\label{subsec:BBMZD}
\index{Hanany-Witten!brane box}
Now we can use the standard methodology \cite{HU,9906031,Han-Zaf2}
to construct the BBM. The general 
idea is that for the BBM corresponding to the singularity $\C^3/\Gamma$,
we put D-branes whose number is determined by the irreps of $\Gamma$ 
into proper grids in Brane Boxes constructed out of NS5-branes.
Then the genetal rule of the resulting BBM is that we have gauge group
$SU(N_i)$ in every grid and bifundamental chiral multiplets going along
the North, East and SouthWest directions.
The superpotential can also
be read by closing upper or lower triangles in the grids \cite{HU}.
The quiver diagram is also readily readable from the structure of the
BBM (the number and the positions of the branes).

Indeed, in comparison with geometrical methods, because the two quivers
(the orbifold quiver and the BBM quiver) seem to arise from two 
vastly disparate contexts, they need not match a priori.
However, by judicious choice of irreps in each grid
we can make these two quiver exactly the same; this is what is meant
by the {\bf equivalence} between the BBM and orbifold methods.
The consistency condition we impose on the BBM for such equivalence is

\begin{equation}
\label{consistency2}
{\bf 3} \otimes r_i = \bigoplus_{j \in \{{\rm North,East,SouthWest}\}} r_j.
\end{equation}

Of course we observe this to be precisely (\ref{decomp2}) in a different
guise.

\begin{figure}
\centerline{\psfig{figure=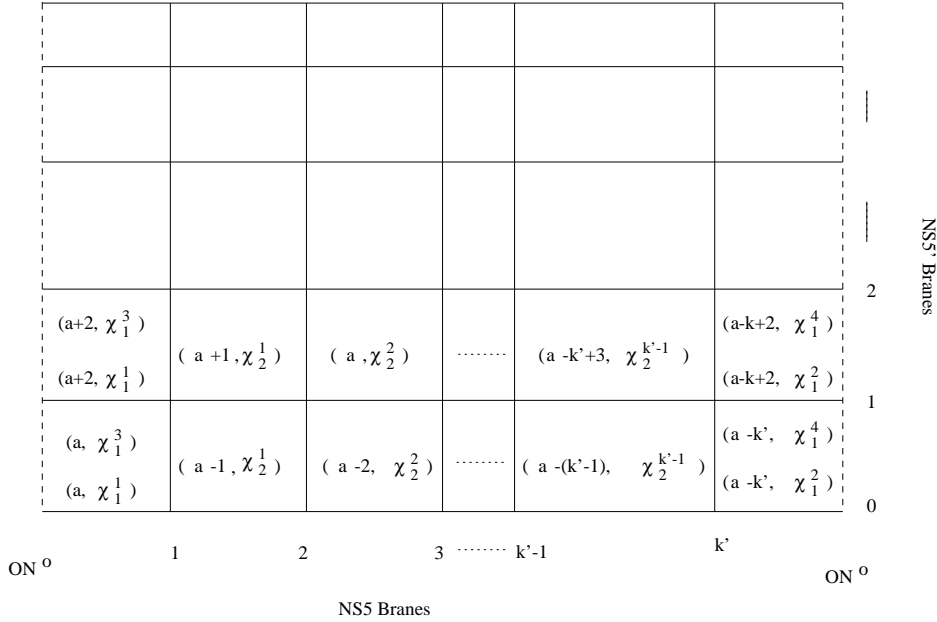,width=5.0in}}
\caption{The Brane Box Model for $Z_k \times D_{k'}$. Notice that 
	for every step along
	the vertical direction from the bottom to top, 
	the first index has increment 2, while along the 
	horizontal direction from left to right, 
	the first index has decrement 1 and
	the second index, increment 1. The vertical direction
	is also periodically identified so that $k(~$or$~\frac{k}{2}) \equiv 0$.}
\label{fig:BBM1}
\end{figure}

Now we return to our toy group $Z_k \times D_{k'}$. The grids are furnished
by a parallel set of $k'$ NS5-branes with 2 ON$^0$ planes intersected by
$k$ (or $\frac{k}{2}$ when $k$ is even; see explanation below) 
NS5$'$-branes perpendicular thereto and periodically identified such
that $k($or$~\frac{k}{2}) \equiv 0$ as before \cite{9906031}. This is shown in 
\fref{fig:BBM1}. The general brane setup of this form involving
2 sets of NS5-branes and 2 ON-planes we shall call, as mentioned in
the introduction, the BBM of {\bf the $Z$-$D$ Type}.

The irreps are placed in the grids as follows.
First we consider the leftmost column.
We place a pair of  irreps $\{(a,\chi_1^1),(a,\chi_1^3)\}$ at the bottom
(here $a$ is some constant initial index), 
then for each incremental grid going up we increase the index $a$ by 2.
Now we notice the fact that when $k$ is odd, such an indexing makes one
return to the bottom grid after $k$ steps whereas if $k$ is even, it suffices
to only make $\frac{k}{2}$ steps before one returns.
This means that when $k$ is odd, the periodicity of $a$ is
precisely the same as that required by our circular
identification of the NS5$'$-branes. However, when $k$ is even it seems
that we can not put all irreps into a single BBM. We can circumvent the
problem by dividing the irreps $(a,\chi)$ into 2 classes depending on
the parity of $a$, each of which gives a BBM consisting of $k\over 2$
NS5$'$-branes.
We should not be surprised at this phenomenon. As we mentioned at
the beginning of this section, the matrices (\ref{gen_dir}) generate a
non-faithful representation of the group when $k$ is even
(i.e., $\alpha^{\frac{k}{2}}$ gives the same matrix as $\beta^{k'}$).
This non-faithful decomposition of ${\bf 3}$ is what is responsible for
breaking the BBM into 2 disjunct parts.

The same phenomenon appears in the $Z_k \times Z_{k'}$ BBM as well.
For $k$ even, if we choose the decomposition as 
${\bf 3} \longrightarrow(1,0)+(0,1)+(-1,-1)$ 
we can put all irreps into $kk'$ grids, however
if we choose
${\bf 3} \longrightarrow (2,0)+(0,1)+(-2,-1)$
we can only construct two BBM's each with $\frac{kk'}{2}$ grids and
consisting of one half of the total irreps.
Indeed this a general phenomenon which we shall use later:

\begin{proposition}
\label{nonfaith}
Non-faithful matrix representations of $\Gamma$ give rise to
corresponding Quiver Graphs which are disconnected.
\end{proposition}

Having clarified this subtlety we continue to construct the BBM.
We have fixed the content for the leftmost column.
Now we turn to the bottom row.
Starting from the second column (counting from the
left side) we place the irreps
$(a-1,\chi_2^{1}), (a-2,\chi_2^{2}) ,..., (a-(k'-1),\chi_2^{k'-1})$
until we reach the right side (i.e., $(a-j,\chi_2^{j})$ with $j=1,...k'-1$)
just prior to the rightmost column; there we place
the pair $\{(a-k',\chi_1^2),(a-k',\chi_1^4)\}$.
For the remaining rows we imitate what we did for the leftmost column
and increment, for the $i$-th column, the first index by 2 each time
we ascend one row, i.e., $(b,\chi_i^{j}) \rightarrow (b+2,\chi_i^{j})$.
The periodicity issues are as discussed above.

Our task now is to check the consistency of the BBM, namely 
(\ref{consistency2}).
Let us do so case by case. First we check the grid at the first 
(leftmost) column at the $i$-th row; the content there is
$\{(a+2i,\chi_1^1),(a+2i,\chi_1^3)\}$.
Then (\ref{consistency2}) dictates that
\[
\begin{array}{c}
[(2,\chi_{1}^{1}) \oplus (-1,\chi_{2}^{1})] 
	\otimes(a+2i,\chi_1^1 ~$or$~ \chi_1^3) \\
=(a+2(i+1),\chi_1^1 ~$or$~ \chi_1^3) \oplus ((a+2i)-1,\chi_{2}^{1})
\end{array}
\]
by using the table of tensor decompositions in subsection \sref{subsec:Dk'}
and our chosen {\bf 3} from (\ref{tensor}). Notice that the
first term on the right is exactly the content of the box to the
North and second term, the content of the East. Therefore consistency
is satisfied.
Next we check the grid in the second column at the $i$-th row where
$((a+2i)-1,\chi_{2}^{1})$ lives. As above we require
\[
\begin{array}{c}
[(2,\chi_{1}^{1}) \oplus (-1,\chi_{2}^{1})]\otimes ((a+2i)-1,\chi_{2}^{1}) \\
=((a+2(i+1))-1,\chi_{2}^{1}) \oplus ((a+2i)-2,\chi_{2}^{2})
	\oplus (a+2(i-1),\chi_1^1) \oplus (a+2(i-1),\chi_1^3)
\end{array}
\]
whence we see that the first term corresponds to the grid to the North, 
and second, East, and the last two, SouthWest. 
We proceed to check the grid in the $j+1$-th column ($2\leq j\leq k'-2$) at
the $i$-th row where $((a+2i)-j,\chi_2^{j})$ resides.
Again (\ref{consistency2}) requires that
\[
\begin{array}{c}
[(2,\chi_{1}^{1}) \oplus (-1,\chi_{2}^{1})]\otimes ((a+2i)-j,\chi_2^{j}) \\
=((a+2(i+1))-j,\chi_2^{j}) \oplus ((a+2i)-(j+1),\chi_2^{j+1})
	\oplus ((a+2(i-1))-(j-1),\chi_2^{j-1})
\end{array}
\]
where again the first term gives the irrep the grid to the North, 
the second, East and the third, SouthWest. 
Next we check the grid in the $k'$-th column and $i$-th row, where the
irrep is $((a+2i)-(k'-1),\chi_2^{k'-1})$. Likewise the requirement is
\[
\begin{array}{c}
[(2,\chi_{1}^{1}) \oplus (-1,\chi_{2}^{1})]\otimes ((a+2i)-(k'-1),\chi_2^{k'-1})\\
=((a+2(i+1))-(k'-1),\chi_2^{k'-1}) \oplus ((a+2i)-k',\chi_1^{2}) \\ \oplus
	((a+2i)-k',\chi_1^{4}) \oplus ((a+2(i-1))-(k'-2),\chi_2^{k'-2})
\end{array}
\]
whence we see again the first term gives the grid to the North, the second 
and third, East and the fourth, SouthWest. 
Finally, for the last (rightmost) column, the grid in the $i$-th row has
$((a+2i)-k',\chi_1^{2})$ and $((a+2i)-k',\chi_1^{4})$. We demand
\[
\begin{array}{c}
[(2,\chi_{1}^{1}) \oplus (-1,\chi_{2}^{1})]\otimes 
	((a+2i)-k',\chi_1^{2} ~$or$~ \chi_1^{4}) \\
=((a+2(i+1))-k',\chi_1^{2} ~$or$~ \chi_1^{4}) \oplus 
	((a+2(i-1))-(k'-1),\chi_{2}^{k'-1}))
\end{array}
\]
where the first term gives the grid to the North and the second term,
Southwest. So we have finished all checks and our BBM is consistent.

From the structure of this BBM it is very clear that each row 
gives a $D_{k'}$ quiver and the different rows simply copies it $k$ times
according to the $Z_k$. This repetition hints that
there should be some kind of direct product, which is precisely what
we have.

\subsection{The Inverse Problem}
\label{subsec:inverse}
Now we address the inverse problem: given a BBM of type 
$Z$-$D$, with $k'$ 
vertical NS5-branes bounded by 2 ON$^0$-planes and $k$ 
horizontal NS5$'$-branes, what is the
corresponding orbifold, i.e., the group which acts on $\C^3$? 
The answer is now very clear:
if $k$ is odd we can choose the group $Z_{k} \times D_{k'}$ or 
$Z_{2k} \times D_{k'}$ with the action as defined in (\ref{gen_dir});
if $k$ is even, then we can choose the group to be 
$Z_{2k} \times D_{k'}$ with the same action.

In this above answer, we have two candidates when $k$ is odd since we recall
from discussions in \sref{subsec:BBMZD} the vertical direction of the BBM
for the group $Z_{2k}\times D_{k'}$
only has periodicity $\frac{k}{2}$ and the BBM separates into two pieces.
We must ask ourselves, what is the relation between these two 
candidates?
We notice that (\ref{gen_dir}) gives an non-faithful representation
of the group $Z_{2k}\times D_{k'}$. 
In fact, it defines a new group of which has the
faithful representation given by above matrix form 
and  is a factor group of
$Z_{2k} \times D_{k'}$ given by

\begin{equation}
\label{modH}
G:=(Z_{2k} \times D_{k'})/H,~~~{\rm with}~~~
	H=\langle1,\alpha^k \beta^{k'} \rangle
\end{equation}

In fact $G$ is isomorphic to $Z_{k} \times D_{k'}$. We can see this by
the following arguments.
denote the generators of $Z_{2k} \times D_{k'}$ as $\alpha,\beta,\gamma$ 
and those of $Z_{k} \times D_{k'}$ as 
$\tilde{\alpha},\tilde{\beta},\tilde{\gamma}$. 
An element of $G$ can be expressed as 
$[\alpha^a \beta^b \gamma^n] \equiv
[\alpha^{a+k} \beta^{b+k'} \gamma^n]$. We then see the homomorphism
from $G$ to $Z_{k} \times D_{k'}$ defined by
\[
[\alpha^a \beta^b \gamma^n] \longrightarrow 
\tilde{\alpha}^a \tilde{\beta}^{ak'+b} \tilde{\gamma}^n
\]
is in fact an isomorphism (we see that $[\alpha^a \beta^b \gamma^n]$ and
$[\alpha^{a+k} \beta^{b+k'} \gamma^n]$ are mapped to same element as
required; in proving this the $k=\odd$ condition is crucial).

We see therefore that given the data from the BBM, viz., $k$ and $k'$,
we can uniquely find the $\C^3$ orbifold singularity and our inverse
problem is well-defined.

\section{The General Twisted Case}
\label{sec:twist}
We have found in the previous chapter that the group $Z_{k} * D_{k'}$
(in which we called $Z_{k} \times D_{k'}$) defined by
(\ref{gen1}) can be written in another form as 
$Z_k \rtimes D_{\frac{kk'}{\gcd(k,2k')}}$ where it becomes an
(internal) semidirect product.
We would like to know how the former, which is a special
case of what we shall call a
{\bf twisted} group\footnote{As mentioned in the
Nomenclature section, $*$ generically denotes twisted products
of groups.} is related to the direct
product example, which we shall call the {\bf untwisted} case,
upon which we expounded in the previous section.

The key relation which describes the semidirect product structure
was shown \cite{9906031} to be
$\alpha \gamma =\beta^{\frac{2k'}{\gcd(k,2k')}} \gamma \alpha$.
This is highly suggestive and hints us to 
define a one-parameter family of groups\footnote{We note that
this is unambiguously the semi-direct product $\rtimes$: defining
the two subgroups $D := \langle \beta, \gamma \rangle$ and
$Z := \langle \alpha \rangle$, we see that $G(a) = DZ$ as cosets,
that $D \triangleleft G(a)$ and $D \cap Z = 1$, whereby all
the axioms of semi-directness are obeyed.}
$G(a) := \{Z_k \rtimes D_d\}$
whose presentations are

\begin{equation}
\label{gen_twist}
\alpha \beta=\beta \alpha,~~~~\alpha \gamma=\beta^a \gamma \alpha.
\end{equation}

When the parameter $a=0$, we have $G(0) = Z_k \times D_{k'}$ as discussed
extensively in the previous section. Also, when 
$a = \frac{kk'}{\gcd(k,2k')}$, $G(a)$
is the group $Z*D$ treated in the previous chapter.
We are concerned with
members of $\{G(a)\}$ that satisfy the BBM conditions and though indeed
this family may not exhaust the list of all groups that satisfy
those conditions they do provide an illustrative subclass.

\subsection{Preserving the Irreps of $D_d$}
We see that the first of the BBM conditions is trivially
satisfied by our definition (\ref{gen_twist} of 
$G(a) := Z_k \rtimes D_d$. Therefore
we now move onto the second condition.
We propose that $G(a)$ preserves the 
structure of the irreps of the $D_d$ factor if $a$ is even.
The analysis had been given in detail \cite{9906031} so
here we only review briefly.
Deducing from (\ref{gen_twist}) the relation, for $b \in \Z$,
\[
\alpha (\beta^b \gamma) \alpha^{-1}= \beta^{b+a} \gamma,
\]
we see that $\beta^b \gamma$ and $\beta^{b+a} \gamma$ belong to the same
conjugacy class after promoting $D_d$ to the semidirect product
$Z_k \rtimes D_d$. Now we recall from subsection \sref{subsec:Dk'} that
the conjugacy classes of $D_d$ are $\beta^0,\beta^d$,
$\beta^{\pm p}(p\neq 0,d)$, $\gamma \beta^{\even}$ and
$\gamma \beta^{\odd}$. Therefore we see that
when $a=\even$, the conjugacy structure of $D_d$ is
preserved since therein $\beta^b \gamma$ and $\beta^{b+a} \gamma$,
which we saw above belong to same conjugate class in $D_d$,
are also in the same conjugacy class in $G(a)$ and everything is fine.
However, when $a=\odd$, they live in two
different conjugacy classes at the level of $D_d$ but in the
same conjugacy class in $G(a)$ whence violating
the second condition. Therefore $a$ has to be even.

\subsection{The Three Dimensional Representation}
Now we come to the most important part of finding the 3-dimensional 
representations for $G(a)$, i.e., condition 3.
We start with the following form for the generators

\begin{equation}
\label{betagamma}
\beta = \left(  \begin{array}{ccc}
			1  & 0  & 0  \\
			0 & \omega_{2d} & 0  \\
			0 &  0 & \omega_{2d}^{-1}  
		\end{array}
	\right)
~~~~~~~
\gamma =\left(  \begin{array}{ccc} 
		1  &  0  &  0  \\
		0  &  0  &  i \\
		0  &  i &   0 
		\end{array}
	\right)
\end{equation}

and 

\begin{equation}
\label{alpha}
\alpha=\left( \begin{array}{ccc} 
\omega_{k}^{-(x+y)} &  0  & 0 \\
0 & \omega_{k}^{x} & 0  \\
0 & 0 & \omega_{k}^{y} 
		\end{array}
	\right)
\end{equation}
where $x,y \in \Z$ are yet undetermined integers
(notice that the form (\ref{alpha}) is fixed by the matrix
(\ref{betagamma}) of $\beta$ and the algebraic 
relation $\alpha \beta=\beta \alpha$). 
Using the defining relations (\ref{gen_twist}) of $G(a)$,
i.e relation $\alpha \gamma=\beta^a \gamma \alpha$, we 
immediately have the following constraint on $x$ and $y$:

\begin{equation}
\label{xy}
\omega_k ^{x-y} =\omega_{2d}^a
\end{equation}

which has integer solutions \footnote{Since (\ref{xy}) implies 
$\frac{2\pi (x-y)}{k} - \frac{2\pi a}{2d} = 2 \pi \Z$,
we are concerned with Diophantine equations of the form 
$\frac pq - \frac mn \in \Z$. This in turn requires that
$n p = m q \Rightarrow q = \frac{nl}{\gcd(m,n)},~l \in \Z$ by diving
through by the greatest common divisor of $m$ and $n$.
Upon back-substitution, we arrive at $p = \frac{m l}{\gcd(m,n)}$.}
only when

\begin{equation}
\label{ksol}
k=(\frac{2d}{\delta})l~~~~~l \in \Z~~{\rm and}~~\delta := \gcd (a,2d)
\end{equation}

with the actual solution being
\[
x-y=\frac{a}{\delta}l.
\]
Equation (\ref{ksol}) is a nontrivial condition which signifiess that for
arbitrary $k,2d,a$, the third of the BBM conditions may be violated,
and the solution, not found.
This shows that even though $G(a=\even)$
satisfies the first two of the BBM conditions, they can not
in general be applied to construct BBM's of Type $Z$-$D$ unless
(\ref{ksol}) is also respected. However, before starting
the general discussion of those cases of $Z*D$ where (\ref{ksol})
is satisfied, let us first see how the group treated before \cite{9906031}
indeed satisfies this condition.

For $Z_k * D_{k'}$ in the previous chapter and defined by (\ref{gen1}),
let $\delta_1 := \gcd(k,2k')$. We have
$d=\frac{kk'}{\delta_1}$, $a=\frac{2k'}{\delta_1}$ from 
Proposition (3.1) in that
chapter. Therefore $\delta=\gcd(a,2d)=a$ and
$k=\frac{2d}{\delta}$ so that (\ref{ksol}) is
satisfied with $l=1$ and we have the solution $x-y=1$.
Now if we choose $y=0$, then we have 

\begin{equation}
\label{alpha2}
\alpha=\left( \begin{array}{ccc} 
\omega_{k}^{-1} &  0  & 0 \\
0 & \omega_{k}^{1} & 0  \\
0 & 0 & 1 
		\end{array}
	\right).
\end{equation}

Combining with the matrices in (\ref{betagamma}), we see that they generate
a faithful 3-dimensional representation of $Z_k * D_{k'}$. It is easy to see
that what they generate is in fact isomorphic to a group with
matrix generators, as given in (\ref{gen_dir}),

\begin{equation}
\label{gen2}
\alpha^{-1} = \left(  \begin{array}{ccc}
			\omega_{2k}^{-2} & 0 & 0  \\
			 0 & \omega_{2k}^{1} & 0\\
			 0  &  0 & \omega_{2k}^{1}
		\end{array}
	\right)
~~~~~~~~
\beta = \left(  \begin{array}{ccc}
			1  & 0  & 0  \\
			0 & \omega_{2d} & 0  \\
			0 &  0 & \omega_{2d}^{-1}  
		\end{array}
	\right)
~~~~~~~
\gamma =\left(  \begin{array}{ccc} 
		1  &  0  &  0  \\
		0  &  0  &  i \\
		0  &  i &   0 
		\end{array}
	\right)
\end{equation}

by noticing that $\alpha^{-1} \beta^{\frac{k'}{\delta}}$ in
(\ref{gen2}) is precisely (\ref{alpha2}).
But this is simply a non-faithful representation of 
$Z_{2k} \times D_{d=\frac{kk'}{\gcd(k,2k')}}$, our direct
product example! 
Furthermore, when $k=odd$,
by recalling the results of \sref{subsec:inverse} we conclude
in fact that the group $Z_k * D_{k'}$
is isomorphic to $Z_k \times D_d$.
However, for $k=\even$, although $Z_k * D_{k'}$ is 
still embeddable into $Z_{2k} \times D_{d=\frac{kk'}{\gcd(k,2k')}}$
with a non-faithful representation (\ref{gen_dir}), it is 
not isomorphic to $Z_k \times D_d$ and the BBM thereof
corresponds to an intrintically twisted case (and unlike when
$k=\odd$ where it is actually isomorphic to a direct product group). 
We emphasize here an obvious but crucial fact exemplified by
(\ref{modH}): {\it non-faithful representations of a group $A$
can be considered as the faithful representation of a new
group $B$ obtained by quotienting an appropriate normal subgroup
of $A$.}
This is what is happening above.
This explains also why we have succeeded \cite{9906031} in 
constructing the BBM only when we wrote $Z_k * D_{k'}$ in the form 
$Z_{k} \rtimes D_{d=\frac{kk'}{\gcd(k,2k')}}$.

Now let us discuss the general case. We recall from the
previous subsection that $a$ has to be even; we thus let $a:=2m$.
With this definition, putting (\ref{xy}) into (\ref{alpha},)
we obtain for the quantity $\alpha \beta^{-m}$:

\begin{equation}
\label{atilde}
\tilde{\alpha}=\alpha \beta^{-m}=\left( \begin{array}{ccc} 
\omega_{k}^{-2y}\omega_{2d}^{2m} &  0  & 0 \\
0 & \omega_{k}^{y}\omega_{2d}^{-m} & 0  \\
0 & 0 & \omega_{k}^{y} \omega_{2d}^{-m}
		\end{array}
	\right)
\end{equation}

This $\tilde{\alpha}$ generates a cyclic group $Z_{\tilde{k}}$ 
and combined with (\ref{betagamma}) gives the direct 
product group of $Z_{\tilde{k}} \times D_d$, but
with a non-faithful representation as in (\ref{gen_dir}).
Therefore for the general twisted case, we can obtain 
the BBM of $Z$-$D$ type of $G(a)$ by imbedding $G(a)$ into a 
larger group $Z_{\tilde{k}} \times D_d$
which is a direct product 
just like we did  for $Z_k * D_{k'}$ embeding to
$Z_{k} \rtimes D_{d=\frac{kk'}{\gcd(k,2k')}}$ two
paragraphs before,  and for which, by our
etude in \sref{sec:dir}, a consistent BBM can always be
established.
However, we need to emphasize that in general such an
embedding (\ref{atilde}) gives non-faithful representations 
so that the quiver diagram of the twisted group
will be a union of disconnected pieces, as demanded by
Proposition \ref{nonfaith}, each of which
corresponds to a Type $Z$-$D$ BBM.
We summarise these results by stating
\begin{proposition}
\label{embed}
The group $G(a):=Z_k*D_d$ satisfies the BBM conditions if $a$ is even and
the relation (\ref{ksol}) is obeyed. In this case its matrices
actually furnish a non-faithful representation of a direct
product $\tilde{G} := Z_{\tilde{k}} \times D_d$ and hence 
affords a BBM\footnote{Though possibly disconnected with
the number of components 
depending on the order of an Abelian subgroup $H\triangleleft \tilde{G}$.}
of Type $Z$-$D$.
\end{proposition}
This action of $G(a) \hookrightarrow \tilde{G}$ is what we mean by embedding.
We conclude by saying that the simple example of \sref{sec:dir}
where the BBM is constructed for untwisted (direct-product)
groups is in fact general and Type $Z$-$D$ BBM's can be
obtained for twisted groups by imbedding into such direct-product
structures.

\section{A New Class of $SU(3)$ Quivers}
\label{sec:new}
\index{Hanany-Witten!brane box}
\index{Quivers}
\index{Finite Groups!$SU(3)$ subgroups}
It would be nice to see whether the ideas presented in the above sections can be 
generalised to give the BBM of other types such as Type $Z$-$E$,
$Z$-$d$ or $D$-$E$ 
whose definitions are obvious. Moreover, $E$ refers to the exceptional
groups $\widehat{E_{6,7,8}}$ and $d$ the ordinary dihedral group.
Indeed, we must first have the brane setups for these groups.
Unfortunately as of yet the $E$ groups still remain elusive.
However we will give an account of the
ordinary dihedral groups and the quiver theory thereof, as well as
the ordinary $E$ groups from a new perspectively from an earlier 
work \cite{9811183}.
These shall give us a new class of $SU(3)$ quivers.

We note that, as pointed out \cite{9811183}, the ordinary
di-, tetra-, octa- and iscosa-hedral groups (or $d$, $E_6,7,8$ 
respectively) are excluded from the
classification of the discrete finite subgroups of $SU(2)$ because
they in fact belong to the centre-modded group $SO(3) \cong SU(2)/\Z_2$.
However due to the obvious embedding $SO(3) \hookrightarrow SU(3)$,
these are all actually $SU(3)$ subgroups. Now the $d$-groups were not
discussed before \cite{9811183} because they did not have non-trivial
3-dimensional irreps and are not considered as non-trivial (i.e.,
they are fundamentally 2-dimensional collineation groups) in the
standard classification of $SU(3)$ subgroups; or in a mathemtical language
\cite{9905212,Yau}, they are transitives. Moreover, $E_6$ is precisely
what was called $\Delta(3 \times 2^2)$ earlier \cite{9811183}, $E_7$, 
$\Delta(6 \times 2^2)$ and $E_8$, $\Sigma_{60}$ and were discussed
there. However we shall here see all these groups together under a new
light, especially the ordinary dihedral group to which we now turn.

\subsection{The Group $d_{k'}$}
The group is defined as
\[
\beta^{k'} = \gamma^2 = 1,~~~~~~\beta \gamma=\gamma \beta^{-1},
\]
and differs from its binary cousin $D_{k'}$ in subsection
\sref{subsec:Dk'} only by having the orders of $\beta, \gamma$
being one half of the latter. Indeed, defining the normal
subgroup $H := \{1,\beta^{k'}\} \triangleleft D_{k'}$ we have
\[
d_{k'} \cong D_{k'}/H.
\]
We can subsequently obtain the character table of $d_{k'}$ from
that of $D_{k'}$ by using the theory of subduced representations,
or simply by keeping all the irreps of $D_{k'}$ which are 
invariant under the equivalence by $H$.
The action of $H$ depends on the parity of $k'$. When
it is even, the two conjugacy classes $(\gamma \beta^{even})$ and 
$(\gamma \beta^{odd})$ remain separate. Furthermore, 
the four 1-dimensional irreps are invariant while for the
2-dimensionals we must restrict the index $l$ as defined in
subsection \sref{subsec:Dk'} to $l=2,4,6,...,k'-2$ so as to
observe the fact that the two conjugacy classes 
$\{\beta^a,\beta^{-a}\}$ and $\{\beta^{k-a},\beta^{a-k}\}$
combine into a single one. All in all, we have 4 1-dimensional 
irreps and $\frac{k'-2}{2}$ 2-dimensionals.
On the other hand, for $k'$ odd, we have the two classes
$(\gamma \beta^{even})$ and $(\gamma \beta^{odd})$ collapsing
into a single one, whereby we can only keep $\chi^1,\chi^3$
in the 1-dimensionals and restrict $l=2,4,6,...,k'-1$ for the
2-dimensionals. Here we have a total of 2 1-dimensional
irreps and $\frac{k'-1}{2}$ 2-dimensionals.

In summary then, the character tables are as follows:
\[
\begin{array}{ll}
\begin{array}{|c|c|c|c|c|c|c|}
\hline
        & 1 & 2 & 2 & \cdots  & 2 & n \\ \hline
\Gamma_{1} & 1 & 1 & 1 & \cdots  & 1 & 1 \\ \hline
\Gamma_{2} & 1 & 1 & 1 & \cdots  & 1 & -1 \\ \hline
\Gamma_{3} & 2 & 2\cos \phi  & 2\cos 2\phi  & \cdots  & 2\cos m\phi  & 0 \\ \hline
\Gamma_{4} & 2 & 2\cos 2\phi  & 2\cos 4\phi  & \cdots  & 2\cos 2m\phi  & 0 \\ \hline
        \vdots  & \vdots  & \vdots  & \vdots  & \cdots  & \vdots  & \vdots  \\ \hline
\Gamma_{\frac{k'+3}{2}} & 2 & 2\cos m\phi  & 2\cos 2m\phi  & \cdots 
        & 2\cos m^{2}\phi  & 0 \\ \hline
\end{array}
& 
\begin{array}{l}
k' $ odd$  \\ 
m=\frac{k'-1}{2} \\ 
\phi =\frac{2\pi }{k'}
\end{array}
\end{array}
\]

\[
\begin{array}{|c|c|c|c|c|c|c|c|c|}
\hline
        & 1 & 2 & 2 & \cdots  & 2 & 1 & n/2 & n/2 \\ \hline
\Gamma_{1} & 1 & 1 & 1 & \cdots  & 1 & 1 & 1 & 1 \\ \hline
\Gamma_{2} & 1 & 1 & 1 & \cdots  & 1 & 1 & -1 & -1 \\ \hline
\Gamma_{3} & 1 & -1 & 1 & \cdots  & (-1)^{m-1} & (-1)^{m} & 1 & -1 \\ \hline
\Gamma_{4} & 1 & -1 & 1 & \cdots  & (-1)^{m-1} & (-1)^{m} & -1& 1 \\ \hline
\Gamma_{5} & 2 & 2 \cos \phi  & 2\cos 2\phi  & \cdots  & 
        2\cos(m-1)\phi  & 2\cos m\phi  & 0 & 0 \\ \hline
\Gamma_{6} & 2 & 2\cos 2\phi  & 2\cos 4\phi  & \cdots  & 2\cos
        2(m-1)\phi  & 2\cos 2m\phi  & 0 & 0 \\ \hline
\vdots  &\vdots  &\vdots  &\vdots  &\cdots  &\vdots  &\vdots  &\vdots  
	&\vdots \\ \hline
\Gamma_{\frac{k'+6}{2}} & 2 & 2\cos (m-1)\phi  & 2\cos 2(m-1)\phi  &
        \cdots  & 2\cos (m-1)^{2}\phi  & 2\cos m(m-1)\phi  & 0 & 0 \\ \hline
\end{array}
\]
for $k'$ even, $m=\frac{k'}{2}$ and $\phi =\frac{2\pi }{k'}$.
\subsection{A New Set of Quivers}
\begin{figure}
\centerline{\psfig{figure=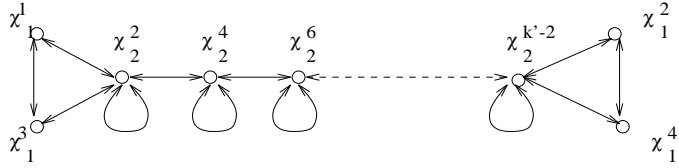,width=3.5in}}
\caption{The quiver diagram for $d_{k=\even}$. Here the notation of the irreps
	placed on the nodes is
 	borrowed from $D_k$ in \sref{subsec:Dk'}. Notice that it
	gives a finite theory with non-chiral matter content.}
\label{fig:dkeven}
\end{figure}

\begin{figure}
\centerline{\psfig{figure=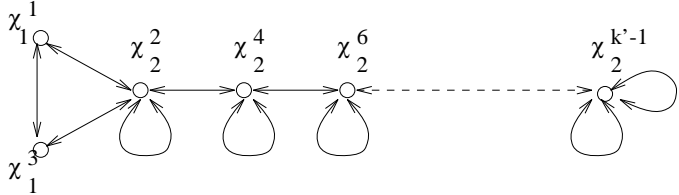,width=3.5in}}
\caption{The quiver diagram for $d_{k=\odd}$. Here again we use the notation 
	of the irreps of $D_k$ to index the nodes. 
	Notice that the theory is again finite and
	non-chiral.}
\label{fig:dkodd}
\end{figure}

Now we must choose an appropriate $SU(3)$ decomposition of the
{\bf 3} for our group in order to make physical sense for
the bifundamentals. The choice is
\[
{\bf 3} \longrightarrow \chi_1^3+\chi_2^2.
\]
Here, we borrow the notation of the irreps of $d_k$ from 
$D_k$ in \sref{subsec:Dk'}. 
The relationship between the irreps of the
two is discussed in the previous subsection. The advantage of using this
notation is that we can readily use the tabulated tensor decompositions
of $D_k$ in \sref{subsec:Dk'}.
With this chosen decomposition, we can immediately arrive at 
the matter matrices $a_{ij}$ and subsequent quiver diagrams.
The $k' = \even$ case gives a quiver which is very much like the
affine $\widehat{D_{k'+2}}$ Dynkin
Diagram, differing only at the two ends, where the nodes corresponding
to the 1-dimensionals are joined, as well as the existence of
self-joined nodes.
This is of course almost what one would expect from an ${\cal N}=2$
theory obtained from the binary dihedral group as a finite
subgroup of $SU(2)$; this clearly reflects the intimate relationship
between the ordinary and binary dihedral groups. 
The quiver is shown in \fref{fig:dkeven}.
On the other hand, for $k'$ odd, we have a quiver which looks like
an ordinary $D_{k'+1}$ Dynkin Diagram with 1 extra line
joining the 1-nodes as well as self-adjoints.
This issue of the dichotomous appearance of affine
and ordinary Dynkin graphs of the D-series in brane setups
has been raised before \cite{Han-Zaf2,9906031}. The diagram for $k'$ odd
is shown in \fref{fig:dkodd}.

\begin{figure}
\centerline{\psfig{figure=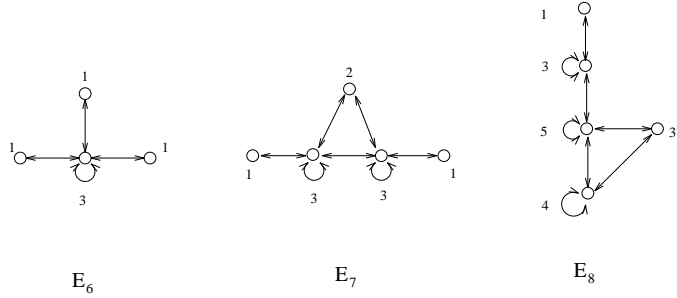,width=3.5in}}
\caption{The quiver diagrams for
	$E_6 = A_4 = \Delta(3 \times 2^2)$,
	$E_7 = S_4 = \Delta(6 \times 2^2)$ and
	$E_8 = A_5 = \Sigma_{60}$. The theories are
	finite and non-chiral.}
\label{fig:excep}
\end{figure}

For completeness and comparison we hereby also include
the 3 exceptional groups of $SO(3) \subset SU(3)$. For these,
we must choose the {\bf 3} to be the unique (up to 
automorphism among the conjugacy classes) 3-dimensional irrep.
Any other decompostion leads to non-faithful representations of
the action and subsequently, by our rule discussed earlier,
to disconnected quivers. This is why when they were considered
as $SU(2)/\Z_2$ groups with 
${\bf 3} \rightarrow {\bf 1} \oplus {\bf 2}$ chosen, 
uninteresting and disconnected quivers were obtained
\cite{9811183}. Now under this new light, we present
the quivers for these 3 groups in \fref{fig:excep}.

There are two points worth emphasising. All the above quivers
correspond to theories which are finite and non-chiral.
By {\bf finite} we mean the condition \cite{LNV}
for anomaly cancelation, that the matter matrix $a_{ij}^R$ must
satisfy
\[
\sum\limits_j a_{ij}^R \dim(r_j) = \sum\limits_j a_{ji}^R \dim(r_j)
\]
What this mean graphically is that for each node, the sum of the
indices of all the neighbouring nodes flowing thereto (i.e., having
arrows pointing to it) must equal to the sum of those flowing
therefrom, and must in fact, for an ${\cal N} = 1$ theory, be
equal to 3 times the index for the node itself.
We observe that this condition is satisfied for all the quivers
presented in \fref{fig:dkodd} to \fref{fig:excep}.

On the other hand by {\bf non-chiral} we mean that for every
bi-fundamental chiral multiplet $(N_i,\bar{N_j})$ there exists
a companion $(N_j,\bar{N_i})$ (such that the two combine together to give
a bi-fundamental hypermultiplet in the sense of ${\cal N}=2$). 
Graphically, this dictates that
for each arrow between two nodes there exists another in the
opposite direction, i.e., the quiver graph is unoriented.
Strangely enough, non-chiral matter content is a trademark
for ${\cal N}=2$ theories, obtained from 
$\C^2 / \Gamma \subset SU(2)$ singularities, while ${\cal N}=1$
usually affords chiral (i.e., oriented quivers) theories.
We have thus arrived at a class of finite, non-chiral
${\cal N} = 1$ super Yang-Mills theories. 
This is not that peculiar because all these groups belong to $SO(3)$ and
thus have real representations; the reality compel the existence
of complex conjugate pairs.
The more interesting fact is that these groups give quivers that
are in some sense in between the generic non-chiral
$SU(2)$ and chiral $SU(3)$ quiver theories. Therefore
we expect that the corresposnding gauge theory will have better properties,
or have more control, under the evolution along some energy scale.

\subsection{An Interesting Observation}
Having obtained a new quiver, for the group $d_k$, it is natural to ask 
what is the corresponding brane setup. 
Furthermore, if we can realize such a brane setup,
can we apply the ideas in the previous sections to realize the
BBM of Type $Z$-$d$? We regrettably have no answers at this stage as
attempts at the brane setup have met great difficulty.
We do, however, have an
interesting brane configuration which gives the correct matter content of 
$d_k$ but has a different superpotential.
The subtle point is that $d_k$ gives only ${\cal N}=1$ 
supersymmetry and unlike ${\cal N}=2$, one must specify both the matter 
content and the superpotential. Two theories with the same matter content but 
different superpotential usually have different low-energy behavior.

\begin{figure}
\centerline{\psfig{figure=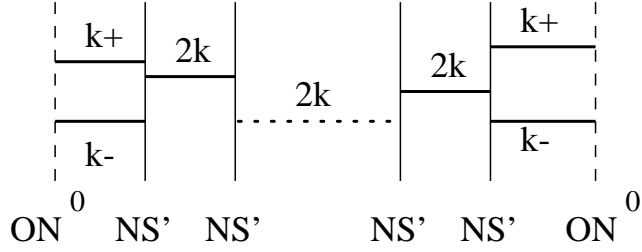,width=3.3in}}
\caption{The brane configuration which gives the same matter content as 
	the $d_{k=\even}$ quiver.}
\label{fig:d_k}
\end{figure}

We now discuss the brane configuration connected with
$d_k$, which turns out to be
a rotated version of the configuration for $D_k$ as given by 
Kapustin \cite{Kapustin}
(related examples \cite{9906031,Erlich} on how rotating branes
breaks supersymmetry further may be found).
In particular we rotate all NS5-branes (along direction
(12345)) between the two ON$^0$-plane as drawn in Figure 1 of Kapustin
\cite{Kapustin} 
to NS5$'$-branes (along direction (12389)). The setup is shown in
\fref{fig:d_k}.
Let us analyse this brane setup more carefully. 
First when we end D4-branes (extended along direction
(1236)) on the ON$^0$-plane, they can have two different charges: positive 
or negtive. With the definition of the matrix
\[
\Omega=\left(  \begin{array}{cc}
	1_{k+\times k+}  &  0  \\
	0  &  -1_{k-\times k-}
	\end{array}  \right),
\]
the projection on the Chan-Paton matrix of the D4-branes is as follows.
The scalar fields in the D4-worldvolume are projected as
\[
\phi^{\alpha}= \Omega \phi^{\alpha}\Omega^{-1}
~~~{\rm and}~~~
\phi^{i}=- \Omega \phi^{i}\Omega^{-1}
\]
where $\alpha$ runs from $4$ to $5$ and describes the oscillations of the
D4-branes in the directions parallel to the ON$^0$-plane while $i$ runs from
$7$ to $9$ and describes the transverse oscillations.
If we write the scalars as matrice in block form, 
the remaining scalars that survive the projection are
\[
\phi^{\alpha}=\left(  \begin{array}{cc}
	U_{k+\times k+}  &  0  \\
	0  &  U_{k-\times k-} \end{array} \right)
~~~{\rm and}~~~
\phi^{i}=\left(   \begin{array}{cc}
	0  &  U_{k+ \times k-}  \\
	U_{k-\times k+}  &  0  
\end{array}  \right).
\]

\begin{figure}
\centerline{\psfig{figure=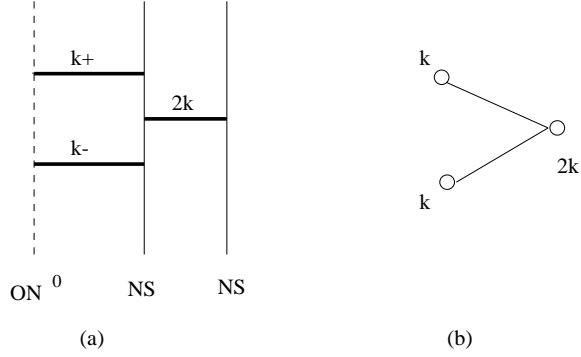,width=3.0in}}
\caption{(a). The brane configuration of the projection using NS5-branes.
	(b). The quiver diagram for the brane configuration in (a).}
\label{fig:NSp}
\end{figure}

From these we immediately see that $\phi^{\alpha}$ give scalars
in the adjoint representation and $\phi^{i}$, in the bifundamental representation.
Next we consider the projection conditions when we end the other side of 
our D4-brane on the NS-brane.
If we choose the NS5-brane to extend along (12345), then the scalars $\phi^{\alpha}$
will be kept while $\phi^{i}$ will be projected out and we would have an
${\cal N}=2$ $D_k$ quiver (see \fref{fig:NSp}).

\begin{figure}
\centerline{\psfig{figure=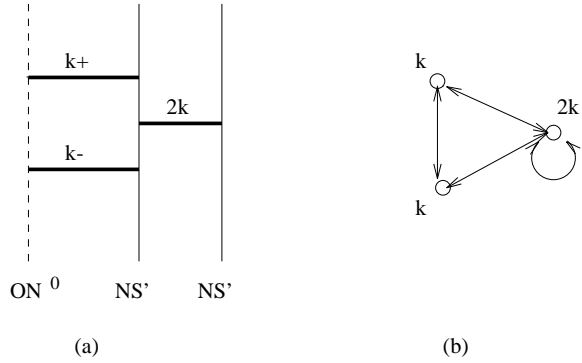,width=3.0in}}
\caption{(a). The brane configuration of projection using NS5$'$-branes.
	(b). The quiver diagram for the brane configuration in (a).}
\label{fig:NSrp}
\end{figure}

However, if we choose the NS5-branes to extend along (12389), then
$\phi^{\alpha}$ and $\phi^{i=7}$ will be projected out while
$\phi^{i=8,9}$ will be kept. It is in this case that
we see immediately that we obtain the same matter content 
as one would have from a $d_{k=\even}$ orbifold discussed in the
previous subsection (see \fref{fig:NSrp}).

Now we explain why the above brane configuration, though giving the
same matter content as
the $d_{k=\even}$,
is insufficient to describe the full theory.
The setup in \fref{fig:d_k} is obtained by the
rotation of NS-branes to NS$'$-branes; in this process of rotation, 
in general we change the geometry from an orbifold to a conifold. 
In other words, by rotating,
we break the ${\cal N}=2$ theory to ${\cal N}=1$ by giving masses to 
scalars in the ${\cal N}=2$ vector-multiplet. 
After integrating out the massive adjoint scalar in low
energies, we usually get quartic terms in the superpotential (for more 
detailed discussion of rotation see Erlich et al. \cite{Erlich}). 
Indeed Klebanov and Witten \cite{Kle-Wit}
have explained this point carefully and shows that the quartic terms will exist
even at the limiting case when the angle of rotation is $\frac{\pi}{2}$ and the
NS5-branes become NS5$'$-branes. On the other hand, the superpotential for
the orbifold singularity of $d_k$ contains only cubic terms as required by
Lawrence et al. \cite{LNV} and as we emphasized in \sref{sec:dir}.
It still remains an intersting problem to construct consistent brane
setups for $d_k$ that also has the right superpotential; this would
give us one further stride toward attacking non-Abelian brane configurations.

\section{Conclusions and Prospects}
\label{sec:con}
As inspired by the Brane Box Model (BBM) constructions \cite{9906031}
for the group $Z_k * D_{k'}$ generated by (\ref{gen1}), we have discussed
in this chapter a class of groups which are generalisations thereof.
These groups we have called the twisted groups 
(that satisfy BBM conditions).
In particular we have analysed at great length, the simplest memeber of this
class, namely the direct product $Z_k \times D_d$, focusing on how the
quiver theory, the BBM construction as well as the inverse problem (of
recovering the group by reading the brane setup) may be established.
The brane configuration for such an example, as in \fref{fig:BBM1}, 
we have called a {\bf BBM of Type $Z$-$D$}; consisting generically of a
grid of NS5-branes with the horizontal direction bounded by 2 ON-planes
and the vertical direction periodically identified.
We have also addressed, as given in Proposition \ref{nonfaith} the
issue of how non-faithful representations lead to disconnected
quivers graphs, or in other words several disjunct pieces of the
BBM setup.

What is remarkable is that the twisted groups, of which the one 
in the previous chapter
is a special case, can under certain circumstances
be embedded into a direct product structure (by actually furnishing
a non-faithful representation thereof). This makes our na\"{\i}ve
example of $Z_k \times D_d$ actually possess great generality as
the twisted cases untwist themselve by embedding into this, in a sense,
universal cover in the fashion of Proposition \ref{embed}.
What we hope is that this technique may be extended to address
more non-Abelian singularities of $\C^3$, whereby the generic finite 
discrete group $G \subset SU(3)$
maybe untwisted into a direct-product cover. In order to do so,
it seems that $G$ needs to obey a set of what we call {\bf BBM conditions}.
We state these in a daring generality: (1) That $G$ maybe written as
a semi-direct product $A \rtimes B$, (2) that the structure of the irreps
of $G$ preserves those of the factors $A$ and $B$ and (3) that there
exists a decomposition into the irreps of $G$ consistent with the unitarity
and determinant 1 constraints of $SU(3)$.

Indeed it is projected and hoped, with reserved optimism,
that if $A,B$ are $SU(2)$ subgroups for
which a brane setup is known, the techniques presented above may inductively
promote the setup to a BBM (or perhaps even brane cube for $SU(4)$ 
singularities). Bearing this in mind, we proceeded further to study
more examples, hoping to attack for example, BBM's of the $Z$-$d$
type where $d$ is the ordinary dihedral group. Therefrom arose our interest
in the ordinary groups $d,E_{6,7,8}$ as finite subgroups of
$SO(3) \subset SU(3)$. These gave us a new class of quiver
theories which have ${\cal N}=1$ but non-chiral matter content.
Brane setups that reproduce the matter content, but unfortunately not the
superpotential, have been established for the ordinary dihedral groups.
These give an interesting brane configuration involving rotating NS5-brane
with respect to ON-planes.

Of course much work remains to be done. In addition to finding the
complete brane setups that reproduce the ordinary dihedral quiver as well as
superpotential, we have yet to clarify the BBM conditions for groups
in general and head toward that beacon of brane realisations of non-Abelian
orbifold theories.
\index{Hanany-Witten!stepwise projection}
\chapter{Orbifolds VII: Stepwise Projection, or Towards Brane Setups
for Generic Orbifold Singularities}
\label{chap:0012078}
\section*{\center{{ Synopsis}}}
Having addressed, in the previous two chapters,
a wide class of non-Abelian orbifolds in
dimension 3, let us see how much further can we go. 

The construction of brane setups for the
exceptional series $E_{6,7,8}$ 
of $SU(2)$ orbifolds remains an ever-haunting
conundrum. Motivated by techniques in some works by Muto  
on non-Abelian $SU(3)$ orbifolds, we here provide an algorithmic
outlook, a method which we call stepwise projection, 
that may shed some light on this puzzle. We exemplify this method,
consisting of transformation rules for obtaining complex quivers and
brane setups
from more elementary ones, to the cases of the $D$-series and $E_6$
finite subgroups of $SU(2)$. Furthermore, we demonstrate the
generality of the stepwise procedure by appealing to Fr{\o}benius'
theory of Induced Representations.
Our algorithm suggests the existence of
generalisations of the orientifold plane in string theory
\cite{0012078}.
\section{Introduction}
It is by now a well-known fact that a stack of $n$ parallel coincident
D3-branes has on its world-volume, an ${\cal N}=4$, four-dimensional
supersymmetric $U(n)$ gauge theory. Placing such a stack at an orbifold
singularity of the form $\C^k/\{ \Gamma \subset SU(k) \}$ reduces
the supersymmetry to ${\cal N}=2,1$ and 0, respectively
for $k=2,3$ and $4$, and the gauge group is broken down to a product
of $U(n_i)$'s \cite{DM,Orb2,LNV}.

Alternatively, one could realize the gauge theory living on D-branes
by the so-called Brane Setups \cite{HW,branerev} (or ``Comic
Strips'' as dubbed by Rabinovici \cite{Comic}) where D-branes are
stretched between NS5-branes and orientifold planes. Since these two
methods of orbifold projections and brane setups provide the same gauge
theory living on D-branes, there should exist some kind of duality to
explain the connection between them.

Indeed, we know now that by
T-duality one can map D-branes probing certain classes of orbifolds to 
brane configurations. For example, the two-dimensional orbifold
$\C^2/\{\IZ_k \subset SU(2) \}$, 
also known as an ALE singularity of type $A_{k-1}$, is mapped into a circle
of $k$ NS-branes (the so-called elliptic model) after proper T-duality
transformations. Such a mapping is easily generalized to some other
cases, such as the three-dimensional orbifold $\C^3/\{ \IZ_k\times \IZ_l
\subset SU(3) \}$ being mapped to the so-named Brane Box 
Model \cite{HZ,HU} or the four-dimensional case of $\C^4/\{ \IZ_k \times
\IZ_l \times \IZ_m\subset SU(4) \}$ being mapped to the brane cube model
\cite{Uranga}. With the help of orientifold planes, we can T-dualise
$\C^2/\{ D_k \subset SU(2) \}$ to a brane configuration with
$ON$-planes \cite{Sen,Kapustin}, or $\C^3/\{ \IZ_k \times D_l
\subset SU(3) \}$ to brane-box-like models with $ON$-planes
\cite{9906031,9909125}.

A further step was undertaken by Muto \cite{Muto,ZD,Muto3} where an
attempt was made to establish the brane setup which corresponds to the
three-dimensional non-Abelian orbifolds $\C^3/\{ \Gamma \subset SU(3)
\}$ with $\Gamma=\Delta(3n^2)$ and $\Delta(6n^2)$. The key idea was
to arrive at these theories by judiciously quotienting the well-known
orbifold $\C^3/\{ \IZ_k\times \IZ_l \subset SU(3) \}$ whose brane
configuration is the Brane Box Model. 
In the process of this quotienting, a non-trivial $\IZ_3$ action on the
brane box is required. Though mathematically obtaining the quivers of
the former from those of the latter seems perfectly sound, such a
$\IZ_3$ action appears to be an unfamiliar symmetry in string theory. 
We shall briefly address this point later.

Now, with the exception of the above list of examples, there have been
no other successful brane setups for the myriad of orbifolds in
dimension two, three and four. Since we believe that the 
methods of orbifold projection and brane configurations are
equivalent to each other in giving D-brane world-volume gauge
theories, finding the T-duality mappings for arbitrary orbifolds is of 
great interest.

The present chapter is a small step toward such an aim. In particular, we
will present a so-called {\bf stepwise projection} algorithm which
attempts to systematize the quotienting idea of Muto, and, as we hope,
to give hints on the brane construction of generic orbifolds.

We shall chiefly focus on the orbifold projections by the $SU(2)$
discrete subgroups $D_k$ and $E_6$ in relation to $\IZ_n$. Thereafter,
we shall
evoke some theorems on induced representations which justify why our
algorithm of stepwise 
projection should at least work in general mathematically.
In particular, we will first demonstrate
how the algorithm gives the quiver of $D_k$ from that of $Z_{2k}$. We
then interpret this mathematical projection physically as precisely
the orientifold projection, whereby arriving at the brane setup of $D_k$ from
that of $\IZ_{2k}$, both of which are well-known and hence giving us a
consistency check.

Next we apply the same idea to $E_6$. We find that one can
construct its quiver from that of $\IZ_6$ or $D_2$ by an appropriate
$\IZ_3$ action. This is slightly mysterious to us physically as it
requires a $\IZ_3$ symmetry in string theory which we could use to
quotient out the $\IZ_6$ brane setup; such a symmetry we do not
know at this moment. However, in comparison with Muto's work, our
$\IZ_3$ action and the $\IZ_3$ investigated by Muto in light of the
$\Delta$ series of $SU(3)$, hint that there might be some objects in
string theory which provide a $\IZ_3$ action, analogous to the orientifold
giving a $\IZ_2$, and which we could use on the known brane setups to
establish those yet unknown, such as those corresponding to the
orbifolds of the exceptional series.

The organisation of the chapter is as follows. In \S 2 we review the
technique of orbifold projections in an explicit matrix language
before moving on to \S 3 to present our stepwise projection
algorithm. In particular, \S 3.1 will demonstrate how to obtain the
$D_k$ quiver from the $\IZ_{2k}$ quiver, \S 3.2 and \S 3.3 will show how
to get that of $E_6$ from those of $D_2$ and $\IZ_6$ respectively. We
finish with comments on the algorithm in \S 4. We will use induced
representation theory in \S 4.1 to prove the validity of our methods
and in \S 4.2 we will address how the present work may be used as a
step toward the illustrious goal of obtaining brane setups for the
generic orbifold singularity.

During the preparation of the manuscript, it has come to our attention
that independent and variant forms of the method have been in
germination \cite{Uranga2,Berenstein}; we sincerely hope that our
systematic treatment of the procedure may be of some utility thereto.

\section*{Nomenclature}

Unless otherwise stated we shall adhere to the convention that
$\Gamma$ refers to a discrete subgroup of $SU(n)$ (i.e., a finite
collineation group), that $\gen{x_1,..,x_n}$ is a finite group generated
by $\{x_1,..,x_n\}$, that $|\Gamma|$ is the order of the group
$\Gamma$, that $D_k$ is the binary dihedral group of order $4k$,
that $E_{6,7,8}$ are the binary exceptional subgroups of $SU(2)$, and that
$R_{G(n)}^{\bullet}(x)$ is a representation of the element $x \in G$ of
dimension $n$ with $\bullet$ denoting properties such as regularity,
irreducibility, etc., and/or simply a label. Moreover, $S^T$ shall denote
the transpose of the matrix $S$ and $A \otimes B$ is the tensor
product of matrices $A$ and $B$ with block matrix elements $A_{ij} B$.
Finally we frequently use the Pauli matrices $\{\sigma_i, i=1,2,3\}$
as well as $\I_N$ for the $N \times N$ identity matrix.
We emphasise here that
the notation for the binary groups differs from the previous chapters
in the exclusion of~$\widehat{~}$~and in the convention for the
sub-index of the binary dihedral group.
\section{A Review on Orbifold Projections}
\index{Orbifolds}
\index{Brane Probes!Orbifolds}
The general methodology of how the finite group structure of the
orbifold projects the gauge theory has been formulated in
\cite{LNV}. The complete lists of two and three dimensional cases have
been treated respectively in \cite{DM,Orb2} and \cite{9811183,Muto} as
well as the four dimensional case in \cite{9905212}.
For the sake of our forth-coming discussion, we shall not use the
nomenclature in \cite{LNV,9811183,9906031,9909125} where recourse to McKay's
Theorem and abstractions to representation theory are taken. Instead,
we shall adhere to the notations in \cite{Orb2} and explicitly indicate
what physical fields survive the orbifold projection.

Throughout we shall focus on two dimensional orbifolds $\C^2/\{\Gamma
\subset SU(2)\}$. The parent theory has an $SU(4)\cong Spin(6)$
R-symmetry from the ${\cal N}=4$ SUSY. The $U(n)$ gauge bosons $A^\mu_{IJ}$
with $I,J=1,...,n$ are R-singlets. Furthermore, there are Weyl
fermions $\Psi_{IJ}^{i=1,2,3,4}$ in the fundamental $\bf{4}$ of $SU(4)$ and
scalars $\Phi_{IJ}^{i=1,..,6}$ in the antisymmetric $\bf{6}$.

The orbifold imposes a projection condition upon these fields due to
the finite group $\Gamma$. Let $R^{reg}_\Gamma(g)$ be the regular
representation of $g \in \Gamma$, by which we mean
\[
R^{reg}_{\Gamma}(g) := \bigoplus\limits_i \Gamma_i(g) \otimes
	\I_{\dim(\Gamma_i)}
\]
where $\{\Gamma_i\}$ are the irreducible representations of
$\Gamma$. In matrix form, $R^{reg}_{\Gamma}(g)$ is composed of blocks of irreps,
with each of dimension $j$ repeated $j$ times. Therefore it is a
matrix of size $\sum\limits_i \dim(\Gamma_i)^2 = |\Gamma|$.

Let Irreps$(\Gamma) = \{\Gamma_1^{(1)},\ldots,\Gamma_{m_1}^{(1)};
\Gamma_1^{(2)},\ldots,\Gamma_{m_2}^{(2)};\ldots \ldots;
\Gamma_{1}^{(n)},\ldots,\Gamma_{m_n}^{(n)}\}$, 
consisting of $m_j$ irreps of dimension $j$, then $R^{reg}_\Gamma := $
\beq
\label{reg}
{\tiny
\left(
\ba{cccccccccc}
\Gamma_1^{(1)} &  &  &  &  &  &  &  &  &  \\
 & \ddots &  &  &  &  &  &  &  &  \\
 &  & \Gamma_{m_1}^{(1)} &  &  &  &  &  &  &  \\
	 &  &  & \left(\matrix{\Gamma_1^{(2)} & ~ \cr
	~ & \Gamma_1^{(2)}}\right) &  &  &  &  &  &  \\
 &  &  &  & \ddots &  &  &  &  &  \\
 &  &  &  &  & \left(\matrix{\Gamma_{m_2}^{(2)} & ~ \cr
	~ & \Gamma_{m_2}^{(2)}}\right) &  &  &  &  \\
 &  &  &  &  &  & \ddots &  &  &  \\
 &  &  &  &  &  &  &
	\left(\ba{ccc}
	\Gamma_1^{(n)} & & \\
	& \ddots & \\
	& & \Gamma_1^{(n)}
	\ea\right)_{n \times n} &  &  \\
 &  &  &  &  &  &  &  & \ddots &  \\
 &  &  &  &  &  &  &  &  & 
	\left(\ba{ccc}
	\Gamma_{m_n}^{(n)} & & \\
	& \ddots & \\
	& & \Gamma_{m_n}^{(n)} \\
	\ea\right)_{n \times n}
\ea
\right)
}.
\eeq
Of the parent fields $A^\mu, \Psi, \Phi$, only those invariant under
the group action will remain in the orbifolded theory; this imposition
is what we mean by {\em surviving the projection}:
\beq
\label{proj}
\ba{l}
A^\mu = R^{reg}_{\Gamma}(g)^{-1} \cdot A^\mu \cdot R^{reg}_{\Gamma}(g)\\
\Psi^i = \rho(g)^i_j ~R^{reg}_{\Gamma}(g)^{-1} \cdot \Psi^j 
	\cdot R^{reg}_{\Gamma}(g)\\
\Phi^i = \rho'(g)^i_j ~R^{reg}_{\Gamma}(g)^{-1} \cdot \Phi^j \cdot
	R^{reg}_{\Gamma}(g) 
\qquad
\forall~~g\in\Gamma,
\ea
\eeq
where $\rho$ and $\rho'$ are induced actions because the matter fields
carry R-charge (while the gauge bosons are R-singlets).
Clearly if $\Gamma = \gen{x_1, ...,  x_n}$, it suffices to
impose \eref{proj} for the generators $\{x_i\}$ in order to find the
matter content of the orbifold gauge theory; this observation we shall
liberally use henceforth.

Letting $n = N |\Gamma|$ for some large $N$ and $n_i = \dim(\Gamma_i)$, 
the subsequent gauge group becomes $\prod\limits_i U(n_iN)$ with
$a_{ij}^4$ Weyl fermions as bifundamentals
$\left({\bf n_iN},{\bf \overline{n_jN}}\right)$ as well as $a_{ij}^6$ scalar
bifundamentals. These bifundamentals are pictorially summarised in
quiver diagrams whose adjacency matrices are the $a_{ij}$'s.

Since we shall henceforth be dealing primarily with $\C^2$
orbifolds, we have ${\cal N}=2$ gauge theory in four dimensions
\cite{LNV}. In particular we choose the induced group action on the
R-symmetry to be ${\bf 4} = {\bf1}_{trivial}^2 \oplus {\bf 2}$ and
${\bf 6} = 
{\bf 1}_{trivial}^2 \oplus {\bf 2}^2$ in order to preserve the
supersymmetry. For 
this reason we can specify the final fermion and scalar matter
matrices by a single quiver characterised by the ${\bf 2}$ of $SU(2)$
as the trivial ${\bf 1}$'s give diagonal 1's. These issues are 
addressed at length in \cite{9811183}.
\section{Stepwise Projection}
Equipped with the clarification of notations of the previous section
we shall now illustrate a technique which we shall call {\bf stepwise
projection}, originally inspired
by \cite{Muto,ZD,Muto3}, who attempted brane realisations of certain
non-Abelian orbifolds of $\C^3$, an issue to which we shall later turn.

The philosophy of the technique is straight-forward\footnote{A recent
work \cite{Berenstein} appeared during
the final preparations of this draft; it beautifully
addresses issues along a similar vein. In particular, cases where
$\Gamma_1$ is normal in $\Gamma_2$ are discussed in detail.
However, our stepwise method is not restricted by normality.}:
say we are given a group $\Gamma_1 = \gen{x_1,...,x_n}$ with
quiver diagram $Q_1$ and $\Gamma_2 = \gen{x_1,...,x_{n+1}} \supset
\Gamma_1$  with quiver $Q_2$, we wish to determine $Q_2$ from $Q_1$ by
the projection \eref{proj} by $\{x_1,...,x_n\}$ followed by another
projection by $x_{n+1}$.

We now proceed to analyse the well-known examples of the
cyclic and binary dihedral quivers under this new light.
\subsection{$D_k$ Quivers from $A_k$ Quivers}
We shall concern ourselves with orbifold theories of $\C^2/\IZ_k$ and
$\C^2/D_k$.
Let us first recall that the cyclic group $A_{k-1} \cong \IZ_k$
has a single generator
\[
\beta_k := \left(
\matrix{\omega_k & 0 \cr 0 & \omega_k^{-1}} \right),
\qquad \mbox{with} \quad \omega_n := e^{2 \pi i \over n}
\]
and that the generators for the binary dihedral group $D_{k}$ are
\[
\beta_{2k} = \left(
\matrix{\omega_{2k} & 0 \cr 0 & \omega_{2k}^{-1}} \right),
\qquad
\gamma := \left( \matrix{0 & i \cr i & 0} \right).
\]
We further recall from \cite{9906031,9909125} that 
$D_k / \IZ_{2k} \cong \IZ_2$.

Now all irreps for $\IZ_k$ are 1-dimensional (the $k^{th}$ roots of
unity), and \eref{reg} for the generator reads
\[
R^{reg}_{\IZ_k}(\beta_k) = \left(
\begin{array}{ccccc}
1 & 0 & 0 & 0 & 0 \\
0 & \omega_k & 0 & 0 & 0 \\
0 & 0 & \omega^2_k & 0 & 0 \\
0 & 0 & 0 & \ddots & 0 \\
0 & 0 & 0 & 0 & \omega_k^{k-1}
\end{array}
\right).
\]
On the other hand, $D_{k}$ has 1 and 2-dimensional irreps and
\eref{reg} for the two generators become
\[
R^{reg}_{D_k}(\beta_{2k}) = 
{\tiny
\left(
\begin{array}{ccccccc}
\left( \matrix{1 & 0 \cr 0 & -1} \right) & 0 & 0 & 0 & 0 & 0 & 0 \\
0 & \left( \matrix{1 & 0 \cr 0 & -1} \right) & 0 & 0 & 0 & 0 & 0 \\
0 & 0 & \left( \matrix{\omega_{2k} & 0 \cr 0 & \omega_{2k}^{-1}}
	\right) & 0 & 0 & 0 & 0 \\
0 & 0 & 0 & \left( \matrix{\omega_{2k} & 0 \cr 0 & \omega_{2k}^{-1}}
	\right) & 0 & 0 & 0\\
\vdots & \vdots & \vdots & \vdots & \ddots & \vdots & \vdots \\
0 & 0 & 0 & 0 & 0 & 
	\left( \matrix{\omega^{k-1}_{2k} & 0 \cr 0 & \omega_{2k}^{-(k-1)}}
	\right)& 0 \\
0 & 0 & 0 & 0 & 0 & 0 &
	\left( \matrix{\omega^{k-1}_{2k} & 0 \cr 0 & \omega_{2k}^{-(k-1)}}
	\right)
\end{array}
\right)
}
\]
and
\[
R^{reg}_{D_k}(\gamma) = 
{\tiny
\left(
\begin{array}{ccccccc}
\left(
\matrix{1 & 0 \cr 0 & i^{k \bmod 2}} \right) & 0 & 0 & 
0 & 0 & 0 & 0 \\ 
0 & \left( \matrix{-1 & 0 \cr 0 & -i^{k \bmod 2}} \right)
& 0 & 0 & 0 & 0 & 0 \\
0 & 0 & \left( \matrix{0 & i \cr i & 0} \right) & 0 & 0 & 0 & 0 \\
0 & 0 & 0 & \left( \matrix{0 & i \cr i & 0} \right) & 0 & 0 & 0\\
\vdots & \vdots & \vdots & \vdots & \ddots & \vdots & \vdots \\
0 & 0 & 0 & 0 & 0 & \left( \matrix{0 & i^{k-1} \cr i^{k-1} &
	0}\right)& 0 \\ 
0 & 0 & 0 & 0 & 0 & 0 & \left( \matrix{0 & i^{k-1} \cr i^{k-1} &
	0}\right)
\end{array}
\right)
}.
\]
In order to see the structural similarities between the regular
representation of $\beta_{2k}$ in $\Gamma_1 = \IZ_{2k}$ and $\Gamma_2 =
D_k$, we need to perform a change of basis. We do so such that each
pair (say the $j^{th}$) of the 2-dimensional irreps of $D_2$ becomes
as follows:
\[
\Gamma^{(2)}(\beta_{2k}) =
\left(
\begin{array}{cc}
\left( \matrix{\omega^{j}_{2k} & 0 \cr 0 & \omega_{2k}^{-j}} \right) &
	0 \\
0 & \left( \matrix{\omega^{j}_{2k} & 0 \cr 0 & \omega_{2k}^{-j}}
	\right)
\end{array}
\right)
\rightarrow
\left(
\begin{array}{cc}
\omega^{j}_{2k} \left( \matrix{1 & 0 \cr 0 & 1} \right) & 0 \\
0 & \omega_{2k}^{-j} \left( \matrix{1 & 0 \cr 0 & 1} \right)
\end{array}
\right)
\]
where $j = 1, 2, \ldots, k - 1.$
In this basis, the 2-dimensionals of $\gamma$ become
\[
\Gamma^{(2)}(\gamma) =
\left(
\begin{array}{cc}
\left( \matrix{0 & i^j \cr i^j & 0} \right) & 0 \\
0 & \left( \matrix{0 & i^j \cr i^j & 0} \right)
\end{array}
\right)
\rightarrow
\left(
\begin{array}{cc}
0 & i^j \left(\matrix{1 & 0 \cr 0 & 1}\right) \\
i^j \left(\matrix{1 & 0 \cr 0 & 1}\right) & 0
\end{array}
\right).
\]

Now for the 1-dimensionals, we also permute the basis:
{\scriptsize
\[
\ba{c}
\Gamma^{(1)}(\beta_{2k}) =
\left(
\matrix{ 1 & 0 & 0 & 0 \cr 0 & 
     -1 & 0 & 0 \cr 0 & 0 & 1 & 0 \cr 0 & 0 & 0 & -1 \cr} 
\right)
\rightarrow
\left(
\matrix{ 1 & 0 & 0 & 0 \cr 0 & 1 & 0 & 0 \cr
	0 & 0 & -1 & 0 \cr 0 & 0 & 0 & -1 \cr  }
\right)
\\
\Gamma^{(1)}(\gamma) =
\left(
\matrix{ 1 & 0 & 0 & 0 \cr 0 & i^{k \bmod 2} & 0 & 0 \cr
	0 & 0 & -1 & 0 \cr 0 & 0 & 0 & -i^{k \bmod 2} \cr  }
\right)
\rightarrow
\left(
\matrix{ 1 & 0 & 0 & 0 \cr 0 & -1 & 0 & 0 \cr
	0 & 0 & i^{k \bmod 2} & 0 \cr
	0 & 0 & 0 & -i^{k \bmod 2} \cr  }
\right).
\ea
\]
}
Therefore, we have
{\scriptsize
\[
{\hspace{-0.7in}
R^{reg}_{D_k}(\beta_{2k}) =
\left(
\begin{array}{ccccccc}
1 & 0 & 0 & 0 &  & 0 & 0 \\
0 & -1 & 0 & 0 &  & 0 & 0 \\
0 & 0 & \omega_{2k} & 0 &  & 0 & 0 \\
0 & 0 & 0 & \omega_{2k}^{-1} &  & 0 & 0 \\
\vdots & \vdots & \vdots & \vdots & \ddots & \vdots & \vdots \\
0 & 0 & 0 & 0 & & \omega_{2k}^{k-1} & 0 \\
0 & 0 & 0 & 0 & & 0 & \omega_{2k}^{-(k-1)}
\end{array}
\right)
\otimes
\left( \matrix{1 & 0 \cr 0 & 1} \right),
}
\]
}
which by now has a great resemblance to the regular representation of
$\beta_{2k} \in \IZ_{2k}$; indeed, after one final change of basis, 
by ordering the powers of $\omega_{2k}$ in an ascending fashion while
writing $\omega_{2k}^{-j} = \omega_{2k}^{2k-j}$ to ensure only
positive exponents, we arrive at
\beq
\label{betafinDk}
\ba{rcl}
R^{reg}_{D_k}(\beta_{2k}) & = & 
\left(
\begin{array}{ccccc}
1 & 0 & 0 & & 0 \cr
0 & \omega_{2k} & 0 & & 0\cr
0 & 0 & \omega_{2k}^2 & & 0 \cr
\vdots & \vdots & \vdots & \ddots & \vdots \\
0 & 0 & 0 & & \omega_{2k}^{2k-1}
\end{array}
\right)
\bigotimes
\left( \matrix{1 & 0 \cr 0 & 1} \right)
\\
& = & R^{reg}_{\IZ_{2k}}(\beta_{2k}) \otimes \I_2,
\ea
\eeq
the key relation which we need.

Under this final change of basis,
{\scriptsize
\beq
\label{gammafin}
{\hspace{-0.55in}
R^{reg}_{D_k}(\gamma) =
\left(
\matrix{
\left( \matrix{1 & 0 \cr 0 & -1}\right)
	& 0 & 0 & 0 & 0 & 0 & 0 & 0 \cr
0 & 0 & 0 & 0 & 0 & m_{k-3} & 0 & 0 \cr
0 & 0 & 0 & 0 & 0 & 0 & m_{k-2} & 0 \cr
0 & \vdots & \vdots & \ddots & \vdots & \vdots & 0 & m_{k-1} \cr 
\vdots & & & & \left(\matrix{i^{k \bmod 2} & 0 \cr 0 & -i^{k \bmod 2}}
	\right) & & & \vdots \cr
0 & m_{k-3} & 0 & 0 & 0 & 0 & 0 & 0 \cr
0 & 0 & m_{k-2} & 0 & 0 & 0 & 0 & 0 \cr
0 & 0 & 0 & m_{k-1} & 0 & 0 & 0 & 0 \cr} 
\right)
},
\eeq
}
where $m_{n} := \bee{n}$.

Our strategy is now obvious. We shall first project according to
\eref{proj}, using \eref{betafinDk}, which is {\em equivalent to} a
projection by $\IZ_{2k}$, except with two identical copies (physically, this
simply means we place twice as many D3-brane probes). Thereafter we
shall project once again using \eref{gammafin} and the resulting theory
should be that of the $D_k$ orbifold.

\subsection*{An Illustrative Example}
Let us turn to a concrete example, namely $\IZ_4 \rightarrow D_2$. The
key points to note are that $D_2 :=\gen{\beta_4,\gamma}$
and $\IZ_4 \cong \gen{\beta_4}$. We shall therefore
perform stepwise projection by $\beta_4$ followed by $\gamma$.

Equation \eref{betafinDk} now reads
\beq
R^{reg}_{D_2}(\beta_4) = 
R^{reg}_{\IZ_{4}}(\beta_4) \otimes \I_2 = 
\left(\matrix{ 1 & 0 & 0 & 0 \cr 0 & i & 0 & 0 \cr
	0 & 0 & i^2 & 0 \cr 0 & 0 & 0 & i^3 \cr  } 
\right)
\otimes \I_2 .
\label{betafinD2}
\eeq
We have the following matter content in the parent (pre-orbifold)
theory: gauge field $A^\mu$, fermions $\Psi^{1,2,3,4}$ and scalars
$\Phi^{1,2,3,4,5,6}$ (suppressing gauge indices $IJ$).
Projection by $R^{reg}_{D_2}(\beta_4)$ in \eref{betafinD2} according
to \eref{proj} gives a $\IZ_4$ orbifold theory, which restricts the
form of the fields to be as follows:
\beq
\label{D2step1}
\ba{l}
A^\mu, \Psi^{1,2}, \Phi^{1,2} =
\left(
\begin{array}{cccc}
\sq & & & \\ & \sq & & \\ & & \sq & \\ & & & \sq
\end{array}
\right);
\\
\Psi^3, \Phi^{3,5} =
\left(
\begin{array}{cccc}
& \sq & & \\ & & \sq & \\ & & & \sq \\ \sq & & &
\end{array}
\right);
\quad
\Psi^4, \Phi^{4,6} =
\left(
\begin{array}{cccc}
& & & \sq \\ \sq & & & \\ & \sq & & \\ & & \sq &
\end{array}
\right)
\ea
\eeq
where $\sq$ are $2 \times 2$ blocks. We recall from the
previous section that we have chosen the R-symmetry decomposition as
${\bf 4} = {\bf 1}_{trivial}^2 \oplus {\bf 2}$ and
${\bf 6} = {\bf 1}_{trivial}^2 \oplus {\bf 2}^2$. The fields in
\eref{D2step1} are defined in accordance thereto: the fermions
$\Psi^{1,2}$ and scalars $\Phi^{1,2}$ are respectively in the two
trivial {\bf 1}'s of the {\bf 4} and {\bf 6}; $(\Psi^3,\Psi^4)$,
$(\Phi^3,\Phi^4)$ and $(\Phi^5,\Phi^6)$ are in the doublet {\bf 2} of $\Gamma$
inherited from $SU(2)$.
\EPSFIGURE[ht]{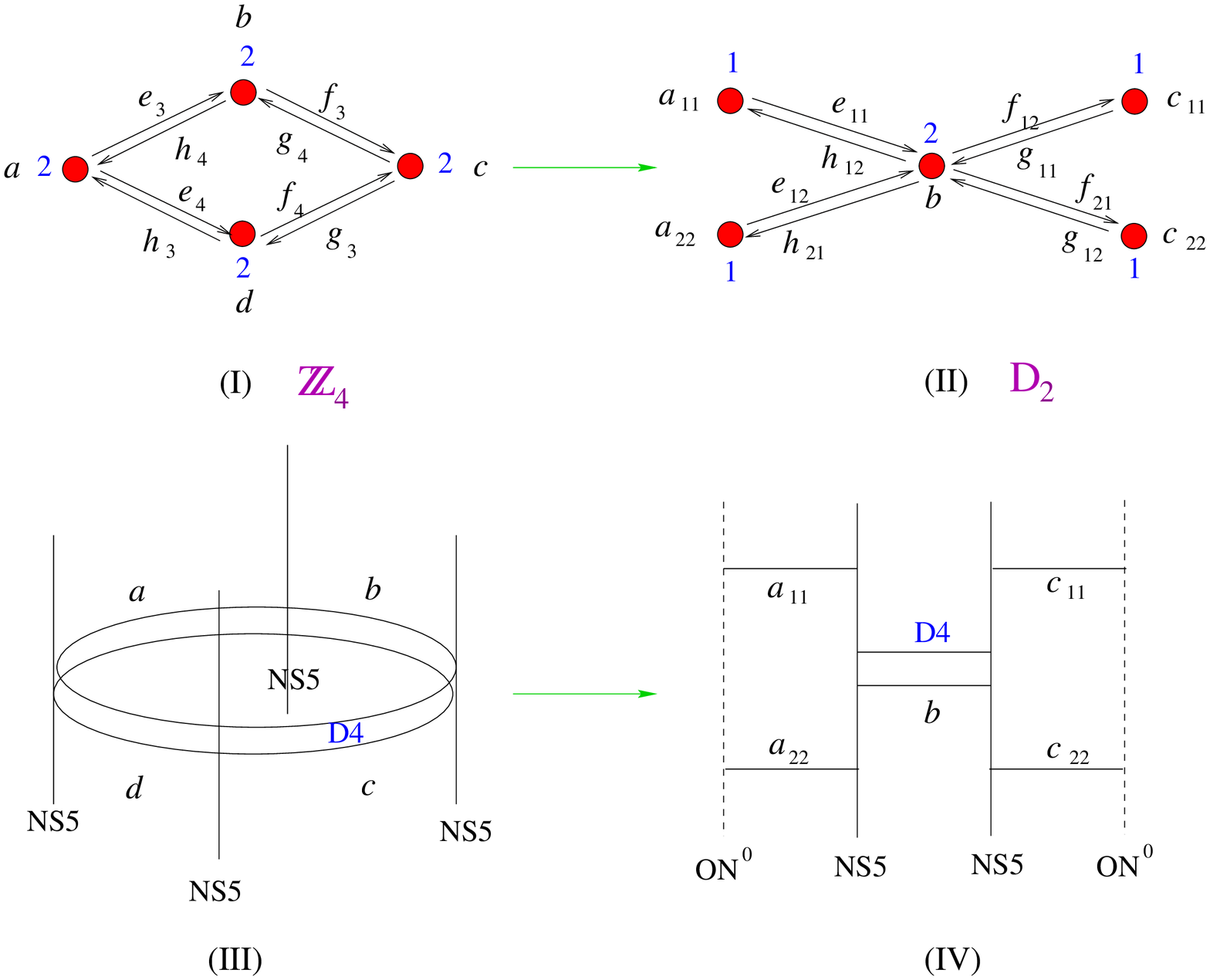,width=6.5in}
{
From the fact that $D_2 :=\gen{\beta_4,\gamma}$ is generated by $\IZ_4
= \beta_4$ together with $\gamma$, our stepwise projection, first by
$\beta_4$, and then by $\gamma$, gives 2 copies of the $\IZ_4$ quiver
in Part (I) and then the $D_2$ quiver in Part (II) by appropriate
joining/splitting of the nodes and arrows. The brane configurations
for these theories are given in Parts (III) and (IV).
\label{f:D2}
}
Indeed, the $R^{reg}_{\IZ_4}(\beta_4)$
projection would force $\sq$ to be numbers and not matrices as we do
not have the extra $\I_2$ tensored to the group action, in which case
\eref{D2step1} would be $4\times 4$ matrices prescribing the adjacency
matrices of the $\IZ_4$ quiver. For this
reason, the quiver diagram for the $\IZ_4$ theory as drawn in
part (I) of \fref{f:D2} has the nodes labelled 2's instead of the
usual Dynkin labels of 1's for the $A$-series. In physical terms we
have placed twice as many image D-brane probes. The key point is that because
$\sq$ are now matrices (and \eref{D2step1} are $8 \times 8$), further
projection internal thereto may change the number and structure of the
product gauge groups and matter fields.

Having done the first step by the $\beta_4$ projection,
next we project with the regular representation of $\gamma$:
\beq
\label{gammafin2}
R^{reg}_{D_2}(\gamma) = 
{\scriptsize
\left(
\begin{array}{cccc}
\left( \matrix{ 1 & 0 \cr 0 & -1 \cr  }\right) & 0 & 0 & 0 \cr
0 & 0 & 0 & \left( \matrix{ i & 0 \cr 0 & i \cr  }\right) \cr 
0 & 0 & \left( \matrix{ 1 & 0 \cr 0 & -1 \cr} \right) & 0 \cr
0 & \left( \matrix{ i & 0 \cr 0 & i \cr} \right) & 0 & 0 \cr
\end{array}
\right)
}
:=
\left(
\matrix{\sigma_3 & 0 & 0 & 0 \cr 0 & 0 & 0 & i\I_2 \cr
0 & 0 & \sigma_3 & 0 \cr 0 & i\I_2 & 0 & 0 \cr}
\right).
\eeq
In accordance with \eref{D2step1}, let the gauge field be
\[
A^\mu := \left(\matrix{a & 0 & 0 & 0 \cr 0 & b & 0 & 0 \cr
0 & 0 & c & 0 \cr 0 & 0 & 0 & d \cr} 
\right),
\]
with $a,b,c,d$ denoting the $2 \times 2$ blocks $\sq$, \eref{proj} for
\eref{gammafin2} now reads
\[
A^\mu = R^{reg}_{D_2}(\gamma)^{-1} \cdot A^\mu \cdot R^{reg}_{D_2}(\gamma)
\Rightarrow
\]
{\small
\[
\left(\matrix{a & 0 & 0 & 0 \cr 0 & b & 0 & 0 \cr
0 & 0 & c & 0 \cr 0 & 0 & 0 & d \cr} \right)
=
\left(
\matrix{\sigma_3 & 0 & 0 & 0 \cr 0 & 0 & 0 & -i\I_2 \cr
0 & 0 & \sigma_3 & 0 \cr 0 & -i\I_2 & 0 & 0 \cr}
\right)
\left(\matrix{a & 0 & 0 & 0 \cr 0 & b & 0 & 0 \cr
0 & 0 & c & 0 \cr 0 & 0 & 0 & d \cr} \right)
\left(
\matrix{\sigma_3 & 0 & 0 & 0 \cr 0 & 0 & 0 & i\I_2 \cr
0 & 0 & \sigma_3 & 0 \cr 0 & i\I_2 & 0 & 0 \cr}
\right),
\]
}
giving us a set of constraining equations for the blocks:
\beq
\label{D2cons1}
\sigma_3 \cdot a \cdot \sigma_3 = a; \qquad d = b; \qquad 
\sigma_3 \cdot c \cdot \sigma_3 = c.
\eeq
Similarly, for the fermions in the {\bf 2}, viz.,
\[
\Psi^3 = \left( \matrix{0 & e_3 & 0 & 0 \cr 0 & 0 & f_3 & 0\cr
0 & 0 & 0 & g_3 \cr h_3 & 0 & 0 & 0 \cr} \right),
\quad
\Psi^4 = \left( \matrix{0 & 0 & 0 & e_4 \cr f_4 & 0 & 0 & 0 \cr
0 & g_4 & 0 & 0 \cr 0 & 0 & h_4 & 0 \cr} \right),
\]
the projection \eref{proj} is
\[
\gamma \cdot \left( \begin{array}{cc} \Psi^3 \\ \Psi^4 \end{array}\right)
= R^{reg}_{D_2}(\gamma)^{-1} \cdot
 \left( \begin{array}{cc} \Psi^3 \\ \Psi^4 \end{array}\right)
 \cdot R^{reg}_{D_2}(\gamma).
\]
We have used the fact that the induced action $\rho(\gamma)$, having
to act upon a doublet, is simply the $2 \times 2$ matrix
$\gamma$ herself. Therefore, writing it out explicitly, we have
{\small
\[
i \left( \matrix{0 & 0 & 0 & e_4 \cr f_4 & 0 & 0 & 0 \cr
0 & g_4 & 0 & 0 \cr 0 & 0 & h_4 & 0 \cr} \right)
= \left(
\matrix{\sigma_3 & 0 & 0 & 0 \cr 0 & 0 & 0 & -i\I_2 \cr
0 & 0 & \sigma_3 & 0 \cr 0 & -i\I_2 & 0 & 0 \cr}
\right)
\left( \matrix{0 & e_3 & 0 & 0 \cr 0 & 0 & f_3 & 0\cr
0 & 0 & 0 & g_3 \cr h_3 & 0 & 0 & 0 \cr} \right)
\left(
\matrix{\sigma_3 & 0 & 0 & 0 \cr 0 & 0 & 0 & i\I_2 \cr
0 & 0 & \sigma_3 & 0 \cr 0 & i\I_2 & 0 & 0 \cr}
\right)
\]
}
and
{\small
\[
i \left( \matrix{0 & e_3 & 0 & 0 \cr 0 & 0 & f_3 & 0\cr
0 & 0 & 0 & g_3 \cr h_3 & 0 & 0 & 0 \cr} \right)
= \left(
\matrix{\sigma_3 & 0 & 0 & 0 \cr 0 & 0 & 0 & -i\I_2 \cr
0 & 0 & \sigma_3 & 0 \cr 0 & -i\I_2 & 0 & 0 \cr}
\right)
\left( \matrix{0 & 0 & 0 & e_4 \cr f_4 & 0 & 0 & 0 \cr
0 & g_4 & 0 & 0 \cr 0 & 0 & h_4 & 0 \cr} \right)
\left(
\matrix{\sigma_3 & 0 & 0 & 0 \cr 0 & 0 & 0 & i\I_2 \cr
0 & 0 & \sigma_3 & 0 \cr 0 & i\I_2 & 0 & 0 \cr}
\right),
\]
}
which gives the constraints
\beq
\label{D2cons2}
f_4 = -h_3 \cdot \sigma_3; \qquad g_4 = \sigma_3 \cdot g_3; \qquad
h_4 = -f_3 \cdot \sigma_3; \qquad e_4 = \sigma_3 \cdot e_3.
\eeq
The doublet scalars $(\Phi^{3,5},\Phi^{4,6})$ of course give the same
results, as should be expected from supersymmetry.

In summary then, the final fields which survive both $\beta_4$ and
$\gamma$ projections (and thus the entire group $D_2$) are
{\small
\beq
\label{D2step2}
\ba{l}
A^\mu = \left(
\begin{array}{cccc}
\left( \matrix{a_{11} & 0 \cr 0 & a_{22}} \right) & & &\\
& b & &\\
& & \left( \matrix{c_{11} & 0 \cr 0 & c_{22}} \right) & \\
& & & b
\end{array}
\right);
\quad
\left\{
\ba{c}
e_3 = \left( \matrix{e_{11} & e_{12} \cr 0 & 0} \right), \quad
f_3 = \left( \matrix{0 & f_{12} \cr 0 & f_{22}} \right), \\ \\
g_3 = \left( \matrix{g_{11} & g_{12} \cr 0 & 0} \right), \quad
h_3 = \left( \matrix{0 & h_{12} \cr 0 & h_{22}} \right),
\ea
\right.
\\
\Psi^3 = \left( \matrix{0 & e_3 & 0 & 0 \cr 0 & 0 & f_3 & 0\cr
0 & 0 & 0 & g_3 \cr h_3 & 0 & 0 & 0 \cr} \right),
\quad
\Psi^4 = \left(\matrix{0 & 0 & 0 & \sigma_3 \cdot e_3 \cr 
-h_3 \cdot \sigma_3 & 0 & 0 & 0 \cr
0 & \sigma_3 \cdot g_3 & 0 & 0 \cr
0 & 0 & -f_3 \cdot \sigma_3 & 0 \cr}
\right).
\ea
\eeq
}
The key features to be noticed are now apparent in the structure of
these matrices in \eref{D2step2}.
We see that the 4 blocks of $A^\mu$ in \eref{D2step1}, which give
the four nodes of the $\IZ_4$ quiver, now undergo a metamorphosis:
we have written out the components of $a,c$ explicitly and have used
\eref{D2cons1} to restrict both to diagonal matrices, while $b$ and
$d$ are identified, but still remain blocks without internal structure
of interest. Thus we have a total of 5 non-trivial constituents
$a_{11},a_{22},c_{11},c_{22}$ and $b$, precisely the 5
nodes of the $D_2$ quiver (see parts (I) and (II) of \fref{f:D2}).
Thus {\em nodes of the quiver merge and
split as we impose further projections}, as we mentioned a few
paragraphs ago.

As for the bifundamentals, i.e., the arrows of the quiver,
\eref{D2step1} prescribes the blocks $e_{3,4}, f_{3,4}, g_{3,4}$ and
$h_{3,4}$ as the 8 arrows of Part (I) of \fref{f:D2}. After the
projection by $\gamma$, and imposing the constraint \eref{D2cons2} as
well as the fact that all entries of matter matrices must be
non-negative, we are left with the 8 fields $e_{11,12}, f_{12,22},
g_{11,12}$ and $h_{12,22}$, precisely the 8 arrows in the $D_2$ quiver
(see Part (II) of \fref{f:D2}).

\subsection*{The General Case}
The generic situation of obtaining the $D_k$ quiver from that of
$\IZ_{2k}$ is completely analogous. We would always have two end nodes of
the $\IZ_{2k}$ quiver each splitting into two while the middle ones
coalesce pair-wise, as is shown in \fref{f:Dk}.
\EPSFIGURE[ht]{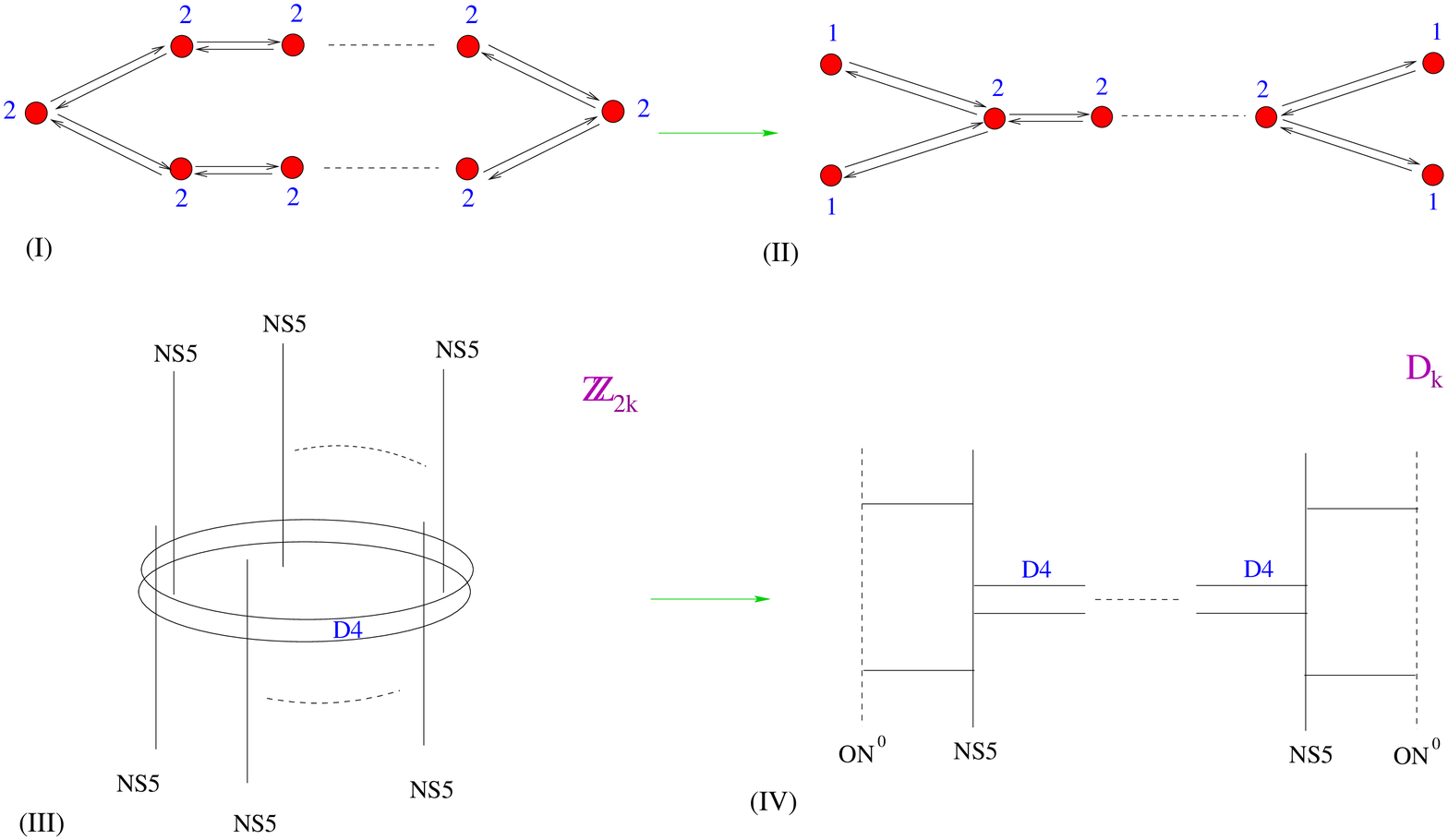,width=6in}
{
Obtaining the $D_k$ quiver (II) from the $\IZ_{2k}$ quiver (I) by the
stepwise projection algorithm. The brane setups are given respectively
in (IV) and (III).
\label{f:Dk}
}
\subsection{The $E_6$ Quiver from $D_2$}
We now move on to tackle the binary tetrahedral group $E_6$ (with the
relation that $E_6 / D_2 \cong \IZ_3$),
whose generators are
\[
\beta_4 = \left( \matrix{i & 0 \cr 0 & -i} \right), \quad 
\gamma=\left( \matrix{0 & i \cr i & 0} \right), \quad
\delta := {1 \over 2} \left( \matrix{ 1 - i & 1 - i  \cr -1 - i  & 
	1 + i  \cr  }\right).
\]
We observe therefore that it has yet one more generator $\delta$ than
$D_2$, hence we need to continue our stepwise projection from the
previous subsection, with the exception that we should begin with more
copies of $\IZ_4$. To see this let us first present
the irreducible matrix representations of the three generators of
$E_6$:
\[
\ba{c|c|c|c}
& \beta_4 & \gamma & \delta \\ \hline \hline
\Gamma^{(1)}_1 & 1 & 1 & 1 \\
\Gamma^{(1)}_2 & 1 & 1 & \omega_3 \\
\Gamma^{(1)}_3 & 1 & 1 & \omega_3^2 \\
\Gamma^{(2)}_4 & \beta_4 & \gamma & \delta \\
\Gamma^{(2)}_5 & \beta_4 & \gamma & \omega_3 \delta \\
\Gamma^{(2)}_6 & \beta_4 & \gamma & \omega_3^2 \delta \\
\Gamma^{(3)}_7 & \left( \matrix{ -1 & 0 & 0 \cr 0 & 1 & 0 \cr 0 & 0 &
		-1 \cr} \right) 
 		& \left(  \matrix{ 0 & 0 & -1 \cr 0 & -1 & 0 \cr -1 &
		0 & 0 \cr  } \right)
		& \left( \matrix{ -\frac{i}{2} & \frac{i}{\sqrt{2}} &  
     -\frac{i}{2} \cr -\frac{1}{\sqrt{2}} & 0 & \frac{1}
    {\sqrt{2}} \cr \frac{i}{2} & -\frac{i}{\sqrt{2}} & \frac{i}{2} \cr  } 
\right)
\ea
\]
The regular representation for these generators is therefore a matrix
of size $3 \cdot 1^2 + 3 \cdot 2^2 + 3^3 = 24$, in accordance with
\eref{reg}.

Our first step is as with the case of $D_2$, namely to change to a
convenient basis wherein $\beta_4$ becomes diagonal:
\beq
\label{betafinE6}
R^{reg}_{E_6}(\beta_4) = R^{reg}_{\IZ_4}(\beta_4) \otimes \I_6.
\eeq
The only difference between the above and \eref{betafinD2} is that we
have the tensor product with $\I_6$ instead of $\I_2$,
therefore at this stage we have a $\IZ_4$
quiver with the nodes labeled 6 as opposed to 2 as in Part (I) of
\fref{f:D2}. In other words we have 6 times the usual number of
D-brane probes.

Under the basis of \eref{betafinE6},
\beq
\label{gammafinE6}
R^{reg}_{E_6}(\gamma) = \left(
\matrix{\Sigma_3 & 0 & 0 & 0 \cr 0 & 0 & 0 & i\I_6 \cr
0 & 0 & \Sigma_3 & 0 \cr 0 & i\I_6 & 0 & 0 \cr}
\right)
\quad
{\rm where}
\quad
\Sigma_3 := \sigma_3 \otimes \I_3 = 
{\scriptsize
\left(\matrix{
1 & 0 & 0 & 0 & 0 & 0 \cr 0 & 1 & 0 & 0 & 0 & 0 \cr 0 & 0 & 1 & 0 & 0 & 
0 \cr 0 & 0 & 0 & -1 & 0 & 0 \cr 0 & 0 & 0 & 0 & 
-1 & 0 \cr 0 & 0 & 0 & 0 & 0 & -1 \cr}
\right).
}
\eeq
Subsequent projection gives a $D_2$ quiver as in part (II) of
\fref{f:D2}, but with the nodes labeled as $3,3,6,3,3$, three times
the usual. Note incidentally that \eref{betafinE6} and
\eref{gammafinE6} can be re-written in terms of regular
representations of $D_2$ directly: $R^{reg}_{E_6}(\beta_4) =
R^{reg}_{D_2}(\beta_4) \otimes \I_3$ and $R^{reg}_{E_6}(\gamma) =
R^{reg}_{D_2}(\gamma) \otimes \I_3$. To this fact we shall later turn.

To arrive at $E_6$, we proceed with one more projection,
by the last generator
$\delta$, the regular representation of which, observing the table
above, has the form (in the basis of \eref{betafinE6})
\beq
\label{deltafinE6}
R^{reg}_{E_6}(\delta) = \left(
\matrix{
S_1 & 0 & S_2 & 0 \cr 0 & \omega_8^{-1}P & 0 & \omega_8^{-1}P \cr
S_3 & 0 & S_4 & 0 \cr 0 & -\omega^8 P & 0 & \omega_8 P \cr}
\right)
\eeq
where
\[
S_1 := \left(\matrix{ 1 & 0 \cr 0 & 0
	\cr } \right) \otimes R^{reg}_{\IZ_3}(\beta_3), \quad
S_2 := \left(\matrix{ 0 & 0 \cr 1 & 0 \cr} \right) \otimes
	\left( \matrix{ 0 & 0 & 1 \cr 0 & 1 & 0 \cr 1 & 0 & 0 \cr  }
	\right),
\]
\[
S_3 := -i \left(\matrix{ 0 & 0 \cr 0 & 1 \cr} \right) \otimes
	\left( \matrix{ 0 & 0 & 1 \cr 0 & 1 & 0 \cr 1 & 0 & 0 \cr}
	\right),
\quad
S_4 := i \left(\matrix{ 0 & 1 \cr 0 & 0 \cr} \right) \otimes \I_3
\]
and
\[
P := R^{reg}_{\IZ_3}(\beta_3) \otimes \frac{1}{\sqrt{2}} \I_2;
\quad
\mbox{recalling that}
\quad
R^{reg}_{\IZ_3}(\beta_3) := \left(
\matrix{ 1 & 0 & 0 \cr 0 & \omega_3 & 0 \cr 0 & 0 & \omega_3^2 \cr  } 
\right).
\]
The inverse of \eref{deltafinE6} is readily determined to be
\[
R^{reg}_{E_6}(\delta)^{-1} = \left(
\matrix{
\tilde{S_1} & 0 & -S_3 & 0 \cr 0 & \frac12\omega_8 P^{-1} & 0 & 
	-\frac12\omega_8^{-1}P^{-1} \cr
S_2^T & 0 & -S_4^T & 0 \cr 0 & \frac12\omega_8 P^{-1} & 0 & 
	 \frac12\omega_8^{-1}P^{-1}\cr}
\right),
\qquad
\tilde{S_1} := \left(\matrix{ 1 & 0 \cr 0 & 0
	\cr } \right) \otimes R^{reg}_{\IZ_3}(\beta_3)^{-1}.
\]
Thus equipped, we must use \eref{proj} with \eref{deltafinE6} on the
matrix forms obtained in \eref{D2step2} (other fields can of course be
checked to have the same projection), with of course each number
therein now being $3 \times 3$ matrices.
The final matrix for $A^\mu$ is as in \eref{D2step2}, but with
\[
a_{11} = \left(
\matrix{a_{11(1)} & 0 & 0 \cr 0 & a_{11(2)} & 0 \cr 0 & 0 & a_{11(3)}}
\right)_{3 \times 3};
\quad
c_{11} = c_{22} = a_{22};
\quad
b = \left(
\matrix{b_{11} & 0 & 0 \cr 0 & b_{22} & 0 \cr 0 & 0 & b_{33} \cr }
\right)_{6 \times 6}
\]
where $a_{22}$, $c_{ii}$ are $3 \times 3$ while $b_{ii}$ are $2 \times
2$ blocks.
We observe therefore, that there are 7 distinct gauge group factors of
interest, namely $a_{11(1)}, a_{11(2)}, a_{11(3)}, a_{22}, b_{11},
b_{22}$ and $b_{33}$, with Dynkin labels $1,1,1,3,2,2,2$ respectively.
What we have
now is the $E_6$ quiver and the bifundamentals split and join
accordingly; the reader is referred to Part (I) of \fref{f:E6}.

\subsection{The $E_6$ Quiver from $\IZ_6$}
Let us make use of an interesting fact, that actually
$E_6 = \langle \beta_4, \gamma, \delta \rangle = \langle
\beta_4, \delta \rangle = \langle \gamma, \delta \rangle$. Therefore,
alternative to the previous subsection wherein we exploited the
sequence $\IZ_4 = \gen{\beta_4} \stackrel{+ \gamma}{\longrightarrow} D_2
\stackrel{+ \delta}{\longrightarrow} E_6$, we could equivalently apply
our stepwise projection on $\IZ_6 = \gen{\delta}
\stackrel{+ \beta_4}{\longrightarrow} E_6$. 

Let us first project with $\delta$, an element of order 6 and the
regular representation of which, after appropriate rotation is
\beq
\label{deltaE6}
R^{reg}_{E_6}(\delta) = R^{reg}_{\IZ_6}(\delta) \otimes \I_4.
\eeq
Therefore at this stage we have a $\IZ_6$ quiver with labels of
six 4's due to the $\I_4$; this is drawn in Part (II) of
\fref{f:E6}. The gauge group we shall denote as
$A^\mu := {\rm Diag}(a,b,c,d,e,f)_{24 \times 24}$, 
with $a, b, \cdots,f$ being $4 \times 4$ blocks.
\EPSFIGURE[ht]{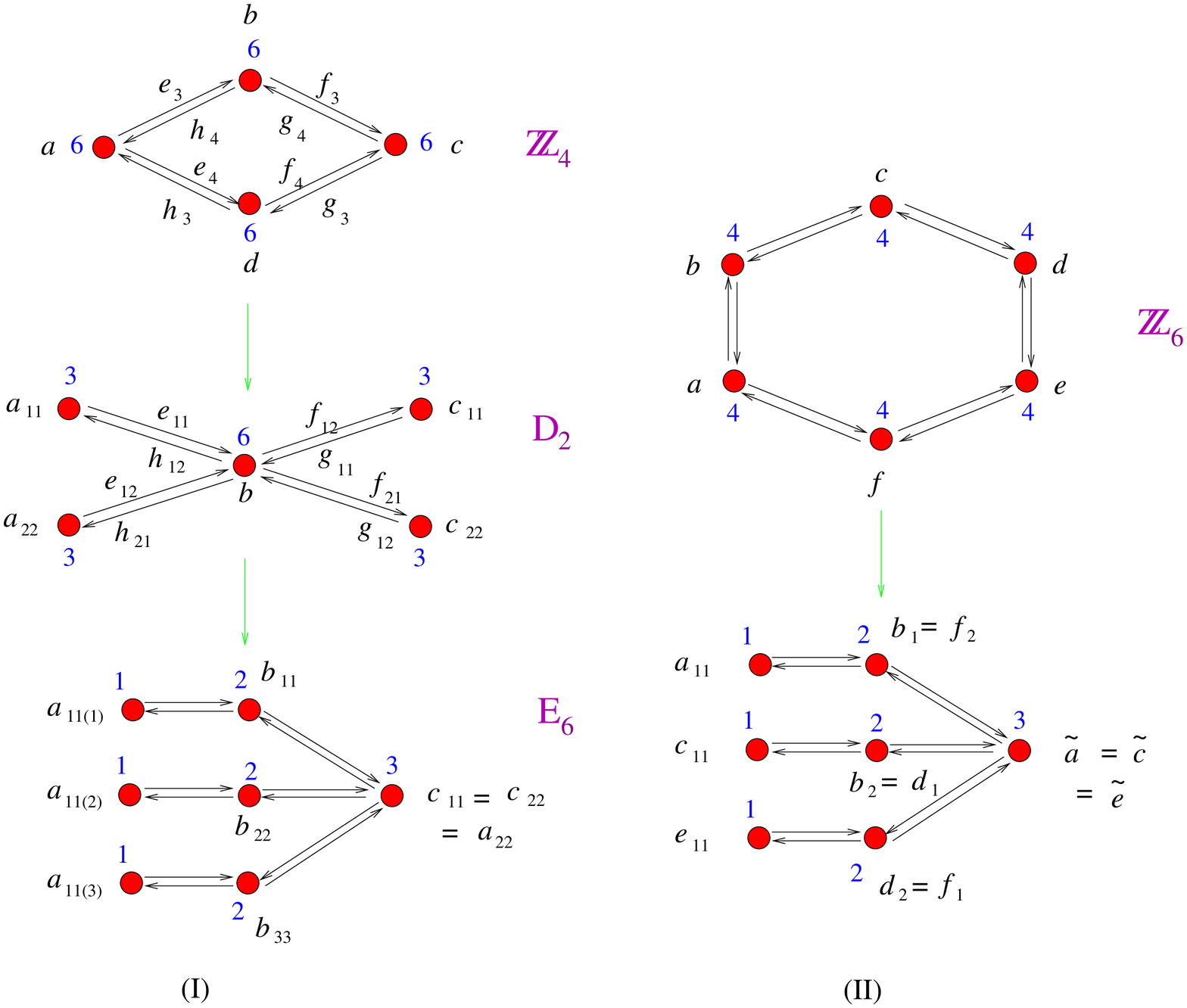,width=6in}
{
Obtaining the quiver diagram for the binary tetrahedral group $E_6$.
We compare the two alternative stepwise projections: (I) $\IZ_4 =
\gen{\beta_4} \rightarrow D_2 = \gen{\beta_4,\gamma} \rightarrow
E_6 = \gen{\beta_4,\gamma,\delta}$ and (II) $\IZ_6 = \gen{\delta}
\rightarrow E_6 = \gen{\delta,\beta_4}$.
\label{f:E6}
}

Next we perform projection by $R^{reg}_{E_6}(\beta_4)$
in the rotated basis, splitting and joining the gauge groups (nodes)
as follows
\[
A^\mu = 
{\scriptsize
\left(
\ba{cccccc}
\left(\matrix{a_{11} & 0 \cr 0 & \tilde{a}}\right) & 0 & 0 & 0 & 0 & 0 \\ 
0 & \left(\matrix{b_1 & 0 \cr 0 & b_2}\right) & 0 & 0 & 0 & 0 \\
0 & 0 & \left(\matrix{c_{11} & 0 \cr 0 & \tilde{c}}\right) & 0 & 0 & 0 \\ 
0 & 0 & 0 & \left(\matrix{d_1 & 0 \cr 0 & d_2} \right) & 0 & 0 \\
0 & 0 & 0 & 0 & \left(\matrix{e_{11} & 0 \cr 0 & \tilde{e}}\right) & 0 \\
0 & 0 & 0 & 0 & 0 & \left(\matrix{f_1 & 0 \cr 0 & f_2} \right)
\ea
\right)};
~~s.~t.~~
\ba{l}
\tilde{a} = \tilde{c} = \tilde{e},\\
b_2 = d_1,\\
d_2 = f_1,\\
f_2 = b_1,\\
\ea
\]
which upon substitution of the relations, gives us 7 independent factors:
$a_{11}, c_{11}$ and $e_{11}$ are numbers, giving 1
as Dynkin labels in the quiver; $b_1, b_2$ and
$d_2$ are $2 \times 2$ blocks, giving the 2 labels; while
$\tilde{a}$ is $3 \times 3$, giving the 3. We refer the reader to Part
(II) of \fref{f:E6} for the diagrammatical representation.
\section{Comments and Discussions}
Our procedure outlined above is originally inspired by a series of
papers \cite{Muto,ZD,Muto3}, where the quivers for the $\Delta$
series of $\Gamma \subset SU(3)$ were observed to be obtainable from
the $\IZ_n \times \IZ_n$ series after an appropriate identification.
In particular, it was noted that 

$\Delta(3n^2) = \gen{\left\{ \IZ_n \times
\IZ_n := {\scriptsize \left( \matrix{\omega_n^i & 0 & 0 \cr 0 &
\omega_n^j & 0 \cr 0 
& 0 & \omega_n^{-i-j}}\right)_{i,j=0,\cdots,n-1} }\right\}, 
{\scriptsize \left(\matrix{ 0
& 0 & 1 \cr 1 & 0 & 0 \cr 
0 & 1 & 0 \cr}\right), \left(\matrix{ 0 & 1 & 0 \cr 0 & 0 & 1 \cr 1 &
0 & 0 \cr} \right)}}$ and subsequently the quiver for $\Delta(3n^2)$ is that of
$\IZ_n \times \IZ_n$ modded out by a certain $\IZ_3$ quotient. Similarly,
the quiver for
\[
\Delta(6n^2) = \langle \IZ_n \times \IZ_n, {\scriptsize
\left(\matrix{ 0 & 0 & 1 \cr 1 & 0 & 0 \cr 
0 & 1 & 0 \cr}\right), \left(\matrix{ 0 & 1 & 0 \cr 0 & 0 & 1 \cr 1 &
0 & 0 \cr} \right), \left(\matrix{ -1 & 0 & 0 \cr 0 & 0 & -1 \cr 0 &
-1 & 0 \cr  } \right), \left(  \matrix{ 0 & -1 & 0 \cr
-1 & 0 & 0 \cr 
0 & 0 & -1 \cr  }\right), \left( \matrix{ 0 & 0 & -1 \cr 0 & -1 & 0
\cr -1 & 0 & 0 \cr  }\right) } \rangle
\]
is that
of $\IZ_n \times \IZ_n$ modded out by a certain $S_3$ quotient.
In \cite{Muto3}, it was further commented that the $\Sigma$ series
could be likewise treated.

The motivation for those studies was to realise a brane-setup for the
non-Abelian $SU(3)$ orbifolds as geometrical quotients of the
well-known Abelian case of $\IZ_m \times \IZ_n$, viz., the Brane Box
Models. The key idea was to recognise that the irreducible
representations of these groups could be labelled by a double index
$(l_1,l_2) \in \IZ_n \times \IZ_n$ up to identifications.

Our purpose here is to establish an algorithmic treatment along
similar lines, which would be generalisable to arbitrary finite
groups. Indeed, since any finite group $\Gamma$ is finitely generated,
starting from the cyclic subgroup (with one single generator), our stepwise
projection would give the quiver for $\Gamma$ as appropriate splitting
and joining of nodes, i.e., as a certain geometrical action, of the
$\IZ_n$ quiver.
\subsection{A Mathematical Viewpoint}
\index{Finite Groups!Frobenius induction}
To see why our stepwise projection works on a more axiomatic level, we
need to turn to a brief review of the Theory of Induced Representations.

It was a fundamental observation of Fr{\o}benius that
the representations of a group could be constructed from an arbitrary
subgroup. The aforementioned chain of groups, where we tried to relate
the regular representations, is precisely in this vein. Though we
shall largely follow the nomenclature of \cite{Lederman}, we shall now
briefly review this theory in the spirit of the above discussions.

Let $\Gamma_1 = \gen{x_1,...,x_n}$ and $\Gamma_2 =
\gen{x_1,...,x_{n+1}}$. We see thus that $\Gamma_1 \subset
\Gamma_2$. Now let $R_{\Gamma_1}(x)$ be a representation (not
necessarily irreducible) of the element $x \in \Gamma_1$. Extending it to
$\Gamma_2$ gives
\[
R_{\Gamma_2}(y) = \left\{
\ba{ll}
R_{\Gamma_1}(x) & {\rm if} \quad y = x \in \Gamma_1 \\
0 & {\rm if} \quad y \not\in \Gamma_1
\ea
\right.
\]
It follows then that if we decompose $\Gamma_2$ as (right) cosets of
$\Gamma_1$,
\[
\Gamma_2 = \Gamma_1 t_1 \cup \Gamma_1 t_2 \cup \cdots \cup \Gamma_1 t_m
\]
we have an {\bf Induced Representation} of $\Gamma_2$ as
\beq
\label{induced}
R_{\Gamma_2}(y) = R_{\Gamma_1}(t_i y t_j^{-1}) =
\left(\ba{cccc}
R_{\Gamma_1}(t_1 y t_1^{-1}) & R_{\Gamma_1}(t_1 y t_2^{-1}) & \cdots &
	R_{\Gamma_1}(t_1 y t_m^{-1}) \\ 
R_{\Gamma_1}(t_2 y t_1^{-1}) & R_{\Gamma_1}(t_2 y t_2^{-1}) & \cdots &
	R_{\Gamma_1}(t_2 y t_m^{-1}) \\ 
\vdots & \vdots & & \vdots \\
R_{\Gamma_1}(t_m y t_1^{-1}) & R_{\Gamma_1}(t_m y t_2^{-1}) & \cdots &
	R_{\Gamma_1}(t_m y t_m^{-1}) 
\ea\right).
\eeq
A beautiful property of \eref{induced} is that it has only one member
of each row or column non-zero and whereby it is essentially a
generalised permutation (see e.g., 3.1 of \cite{Lederman}) matrix acting
on the $\Gamma_1$-stable submodules of the $\Gamma_2$-module.

Now, for the case at hand the coset decomposition is simple due to the
addition of a single new generator: the (right) transversals
$t_1, \cdots, t_m$ are simply powers of the extra generator $x_{n+1}$
and $m$ is simply the index of $\Gamma_1 \subset \Gamma_2$, namely
$|\Gamma_2|/|\Gamma_1|$, i.e.,
\beq
\label{ti}
t_i = x_{n+1}^{i-1} \qquad i = 1,2,\cdots,m; \quad 
	m=\frac{|\Gamma_2|}{|\Gamma_1|}.
\eeq

Now let us define an important concept for an element $x \in \Gamma_2$
\begin{definition}
We call a representation $R_{\Gamma_2}(x)$ {\bf factorisable} if it
can be written, up to possible change of bases, as a tensor product
$R_{\Gamma_2}(x) = R_{\Gamma_1}(x) \otimes \I_k$ for some integer $k$.
\end{definition}
Factorisability of the element, in the physical
sense, corresponds to the ability to initialise our stepwise
projection algorithm, by which we mean that the orbifold projection by
this element is performed on $k$ copies as in the usual sense, i.e.,
a stack of $k$ copies of the quiver. Subsequently we could continue
with the stepwise algorithm to demonstrate how the nodes of these
copies merge or split. In the corresponding D-brane picture this simply
means that we should consider $k$ copies of each image
D-brane probe in the covering space.

The natural question to ask is of course why our examples in the
previous section permitted factorisable generators so as to in turn
permit the performance of the stepwise projection.
The following claim shall be of great assurance to us:
\begin{proposition}
Let $H$ be a subgroup of $G$, then the representation $R_G(x)$ for an
element $x \in H \subset G$ induced
from $R_H(x)$ according to \eref{induced} is factorisable and $k$ is
equal to $|G|/|H|$, the index of $H$ in $G$.
\end{proposition}
Proof: 
Take $R_H(x \in H)$, and tensor it with $\I_{k=|G|/|H|}$; this remains
of course a representation for $x\in H$. It then remains to find the
representations of $x \not\in H$, which we supplement by the
permutation actions of these elements on the $H$-cosets. At the end of
the day we arrive at a representation $R'_G(x)$ of dimension $k$, such
that it is factorisable for $x\in H$
and a general permutation for $x \not\in H$. However by the uniqueness
theorem of induced representations (q.v. e.g. \cite{Serre} Thm 11)
such a linear representation $R'_G(x)$ must in fact be
isomorphic to $R_G(x)$. Thus by
explicit construction we have shown that $R_G(x\in H) = R_H(x) \otimes
\I_k$. $\stackrel{~}{\tiny \sq}$

We can be more specific and apply Proposition 4.1 to our case of the
two groups the second of which is generated by the first with one
additional generator. Using the elegant property that the induction
of a regular representation remains regular (q.v. e.g., 3.3 of
\cite{Serre}), we have:
\begin{corollary}
Let $\Gamma_1$ and $\Gamma_2$ be as defined above, then
\[
R^{reg}_{\Gamma_2}(x_i) = R^{reg}_{\Gamma_1}(x_i) \otimes
\I_{|\Gamma_2|/|\Gamma_1|} \qquad \mbox{for common generators} \qquad
i = 1,2,\ldots,n.
\]
\end{corollary}
In particular, since any $G = \gen{x_1,\ldots,x_n}$ contains a cyclic
subgroup generated by, say $x_1$ of order $m$, i.e., $\IZ_m =
\gen{x_1}$, we conclude that
\begin{corollary}
$R^{reg}_G(x_1) = R^{reg}_{\IZ_m}(x_1) \otimes \I_{|G|/m}$, and hence
the quiver for $G$ can always be obtained by starting with the $\IZ_m$
quiver using the stepwise projection.
\end{corollary}

Let us revisit the examples in the previous section equipped with the
above knowledge.
For the case of $\Gamma_1 = \IZ_4 = \gen{\beta_4}$ and $\Gamma_2
= D_2$ with the extra generator $\gamma$, \eref{ti} becomes 
$t_1 = \I$ and $t_2 = \gamma$ as the index of $\IZ_4$ in $D_2$ is
$\frac{|D_2|=8}{|\IZ_4|=4}=2$.
The induced representation of $\beta_4$ according to \eref{induced}
reads
\[
R_{D_2}(\beta_4) =
\left(
\matrix{R^{reg}_{\IZ_4}(\I\beta_4\I^{-1}) &
		R^{reg}_{\IZ_4}(\I\beta_4\gamma^{-1}) \cr 
	R^{reg}_{\IZ_4}(\gamma\beta_4\I^{-1}) &
		R^{reg}_{\IZ_4}(\gamma\beta_4\gamma^{-1} )}
\right)
=
\left(
\matrix{R^{reg}_{\IZ_4}(\beta_4) & 0 \cr
	0 & R^{reg}_{\IZ_4}(\beta_4^{-1})}
\right)
\]
using the fact that $\gamma \beta_k \gamma^{-1} = \beta_k^{-1}$ in
$D_k$ for the last entry.  Recalling that $R^{reg}_{\IZ_4}(\beta_4)
= {\scriptsize \left( \matrix{ 1 & 0 & 0 & 0 \cr 0 & i & 0 & 0 \cr
0 & 0 & i^2 & 0 \cr 0 & 0 & 0 & i^3 \cr  }\right)}$, this is
subsequently equal to  
$R^{reg}_{\IZ_4} \otimes \I_2$ after appropriate permutation of basis.
Thus Corollary 4.1 manifests her validity as we see that the $R_{D_2}$
obtained by Fr{\o}benius induction of $R^{reg}_{\IZ_4}$ is indeed
regular and moreover factorisable, as \eref{betafinD2} dictates.

Similarly with the case of $\IZ_6 \rightarrow E_6$, we see that
Corollary 4.1 demands that for the common generator $\delta$,
$R^{reg}_{E_6}(\delta)$ should be factorisable, as is indeed
indicated by \eref{deltaE6}. So too is it with $\IZ_4 \rightarrow E_6$,
where $R^{reg}_{E_6}(\beta_4)$ should factorise, precisely as shown by
\eref{betafinE6}.

The above have actually been special cases of Corollary 4.2, where we
started with a cyclic subgroup; in fact we have also presented an
example demonstrating the general truism of Proposition 4.1. In the
case of $D_2 \rightarrow E_6$, we mentioned earlier that
$R^{reg}_{E_6}(\beta_4) =
R^{reg}_{D_2}(\beta_4) \otimes \I_3$ and $R^{reg}_{E_6}(\gamma) =
R^{reg}_{D_2}(\gamma) \otimes \I_3$ for the common generators as was
seen from \eref{betafinE6} and \eref{gammafinE6}; this is exactly as
expected by the Proposition.
\subsection{A Physical Viewpoint: Brane Setups?}
\index{Brane Probes!Orbifolds}
\index{Hanany-Witten}
Now mathematically it is clear what is happening to the quiver as we
apply stepwise projection. However this is only half of the story; as
we mentioned in the introduction, we expect T-duality to take D-branes
at generic orbifold singularities to brane setups.
It is a well-known fact that the brane setups for the $A$ and $D$-type
orbifolds $\C^2/\IZ_n$ and $\C^2/D_n$ have been realised (see
\cite{HZ,HU} and \cite{Kapustin} respectively). It has been the main
intent of a collective 
of works (e.g \cite{9906031,9909125,ZD,Muto3}) to establish such setups for
the generic singularity. 

In particular, the problem of finding a consistent brane-setup for the
remaining case of the exceptional groups
$E_{6,7,8}$ of the $ADE$ orbifold singularities of $\C^2$ (and indeed
analogues thereof for $SU(3)$ and $SU(4)$ subgroups) so far has been
proven to be stubbornly intractable. An original motivation for the
present work is to attempt to formulate an algorithmic outlook wherein
such a problem, with the insight of the algebraic structure of an
appropriate chain of certain relevant groups, may be addressed
systematically.
\subsubsection{The $\IZ_2$ Action on the Brane Setup}
Let us attempt to recast our discussion in Subsection 3.1 into a
physical language. First we try to interpret the action by
$R^{reg}_{D_k}(\gamma)$ in \eref{gammafin} on the $\IZ_{2k}$ quiver as
a string-theoretic action on brane setups to get the corresponding
brane setup of $D_k$ from that of $\IZ_{2k}$.

Now the brane configuration for the $\IZ_{2k}$ orbifold is the
well-known {\em elliptic model} consisting of $2k$ NS5-branes arranged
in a circle with D4-branes stretched in between as shown in Part (III)
of \fref{f:D2}. After stepwise projection by $\gamma$, the quiver in
Part (I) becomes that in Part(II) (see \fref{f:Dk} also). There is an
obvious $\IZ_2$ quotienting involved, where the nodes $i$ and $2k-i$
for $i=1,2,...,k-1$ are identified while each of the nodes $0$ and $k$ 
splits into two parts. Of course, this symmetry is not immediately
apparent from the properties of $\gamma$, which is a group element of
order 4. This phenomenon is true in general: {\sf the order of the
generator used in the stepwise projection does not necessarily
determine what symmetry the parent quiver undergoes to arrive at the
resulting quiver; instead we must observe {\it a posteriori} the shapes of the
respective quivers.}

Let us digress a moment to formulate the above results in the language
used in \cite{Muto,ZD}. Recalling from the brief comments in the
beginning of Section 4, we adopt their idea of labelling 
the irreducible representations of $\Delta$ by $\IZ_n \times \IZ_n$ up
to appropriate identifications, which in our terminology is simply the
by-now familiar stepwise projection of the parent $\IZ_n \times \IZ_n$ quiver.
As a comparison, we apply this idea to the case of $\IZ_{2k}
\rightarrow D_k$. Therefore we need to label the irreps of $D_k$ or
appropriate tensor sums thereof, in terms of certain (reducible)
2-dimensional representations of $\IZ_{2k}$.
Motivated by the factorization property \eref{betafinD2}, we chose
these representations to be
\beq
\label{z2kindex}
R_{\IZ_{2k}(2)}^l := R_{\IZ_{2k}(1)}^{l,irrep} 
\oplus R_{\IZ_{2k}(1)}^{l,irrep}
\eeq
where $l \in \IZ_{2k}$, and amounts to precisely a $\IZ_{2k}$-valued index
on the representations of $D_k$ 
(since $\IZ_{2k}$ is Abelian), which with foresight, we shall later use
on $D_k$.
We observe that such a labelling scheme has a symmetry
\[
R_{\IZ_{2k}(2)}^l \cong R_{\IZ_{2k}(2)}^{-l},
\]
which is obviously a $\IZ_2$ action. Note that $l=0$ and $l=k$ are
fixed points of this $\IZ_2$.

We can now associate the 2-dimensional irreps of $D_k$ with the non-trivial
equivalence classes of the $\IZ_{2k}$ representations \eref{z2kindex},
i.e., for $l=1,2,\ldots,k-1$ we have 
\beq
R_{\IZ_{2k}(2)}^l \cong R_{\IZ_{2k}(2)}^{-l} 
\rightarrow R_{D_{k}(2)}^{l,irrep}. 
\label{map1}
\eeq
These identifications correspond
to the merging nodes in the associated quiver diagram. 
As for the fixed points, we need to map
\beq
\ba{l}
R_{\IZ_{2k}(2)}^0 \rightarrow R_{D_{k}(1)}^{1,irrep} 
\oplus R_{D_{k}(1)}^{2,irrep} \\
R_{\IZ_{2k}(2)}^k \rightarrow R_{D_{k}(1)}^{3,irrep} 
\oplus R_{D_{k}(1)}^{4,irrep}.
\ea
\label{map2}
\eeq 
These fixed points are associated precisely with the nodes that split.

This construction shows clearly how, in the labelling scheme of
\cite{Muto,ZD}, our stepwise algorithm derives
the $D_k$ quiver as a $\IZ_2$ projection of the $\IZ_{2k}$ quiver. 
The consistency
of this description is verified by substituting the 
representations $R_{\IZ_{2k}(2)}^l$ in the $\IZ_{2k}$ quiver 
relations ${\cal R} \otimes R_{\IZ_{2k}(2)}^l = \bigoplus\limits_{\bar l}
a_{l {\bar l}}^{\IZ_{2k} ({\cal R})}  R_{\IZ_{2k}(2)}^{\bar l}$ using
\eref{map1} and \eref{map2}, which results 
exactly in the $D_k$ quiver relations. 
We can of course apply the stepwise projection for the case of 
$\IZ_n \times \IZ_n \rightarrow \Delta$, and would arrive at the results
in \cite{Muto,ZD}.

In the brane setup picture, the identification of the
nodes $i$ and $2k-i$ for $i=1,2,...,k-1$ corresponds to the
identification of these intervals of NS5-branes as well as the
D4-branes in between in the $X^{6789}$
directions (with direction-6 compact). Thus the $\IZ_2$ action on the
$\IZ_{2k}$ quiver should include a space-time action which identifies
$X^{6789}=-X^{6789}$. Similarly, the splitting of gauge fields in
intervals $0$ and $k$ hints the existence of a $\IZ_2$ action on the
string world-sheet. Thus the overall $\IZ_2$ action should include two
parts: a space-time symmetry which identifies and a world-sheet
symmetry which splits respective gauge groups.

What then is this action physically? What object in string theory
performs the tasks in the above paragraph? Fortunately, the
space-time parity and string world-sheet $(-1)^{F_L}$ actions
\cite{Sen,Kapustin} are precisely the aforementioned symmetries. In
other words, the {\em ON-plane} is that which we seek.
This is of great assurance to us, because the brane setup for $D_k$
theories, as given in \cite{Kapustin}, is indeed a configuration which
uses the ON-plane to project out or identify fields in a manner
consistent with our discussions.
\subsubsection{The General Action on the Brane Setup?}
It seems therefore, that we could now be boosted with much confidence:
since we have proven in the previous subsection that our stepwise
projection algorithm is a constructive method of arriving at {\em any}
orbifold quiver by appropriate quotient of the $\IZ_n$ quiver, could we
not simply find the appropriate object in string theory which would
perform such a quotient, much in the spirit of the orientifold
prescribing $\IZ_2$ in the above example, on the well-known $\IZ_n$
brane setup, in order to solve our problem?

Such a confidence, as is with most in life, is overly optimistic. Let
us pause a moment to consider the $E_6$ example. The action by
$\delta$ in the case of $D_2 \rightarrow E_6$ in \S 3.2 and
that of $\beta_4$ in the case of $\IZ_6 \rightarrow E_6$ in \S 3.3
can be visualised in Parts (I) 
and (II) of \fref{f:E6} to be an $\IZ_3$ action on the respective
parent quivers. In particular, the identifications $c_{11} \sim c_{22}
\sim a_{22}$ and $\tilde{a} \sim \tilde{c} \sim \tilde{e}; b_1 \sim
f_2, b_2 \sim d_1, d_2 \sim f_1$ respectively for Parts (I) and (II) 
are suggestive of a $\IZ_3$ action on $X^{6789}$. The tripartite
splittings for $b, a_{11}$ and $a,b,d$ respectively also hint at a
$\IZ_3$ action on the string world-sheet.

Again let us phrase the above results in the scheme of
\cite{Muto,ZD}, and manifestly show how
the $E_6$ quiver results from a $\IZ_3$ projection of the $D_2$ quiver.
We define the following representations of $D_2$: 
$R_{D_{2}(6)}^0 = R_{D_{2}(2)}^{irrep} \oplus  R_{D_{2}(2)}^{irrep}
\oplus  R_{D_{2}(2)}^{irrep}$ and $R_{D_{2}(3)}^l = 
R_{D_{2}(1)}^{l,irrep}  \oplus R_{D_{2}(1)}^{l,irrep}  \oplus
R_{D_{2}(1)}^{l,irrep}$ where $l \in \IZ_4$ labels the four 1-dimensional
irreducible representations of $D_2$. There is an identification
\[
R_{D_{2}}^l \cong  R_{D_{2}}^{f(l)}
\]
where
\[
f(l) = \left\{\begin{array}{c}
0, \;\; l=0 \\
2, \;\; l=1 \\
3, \;\; l=2 \\
1, \;\; l=3 \\
\end{array}\right.
\]
Clearly this is a $\IZ_3$ action on the index $l$.
Note that we have two representations labelled with $l=0$ 
which are fixed points of this action. 
In the quiver diagram of $D_2$ these correspond to the middle node
and another one arbitrarily selected from the
remaining four, both of which split into three. 
The remaining three nodes are consequently merged into a single one 
(see \fref{f:E6}). 
To derive the $E_6$ quiver we need to map the nodes of the parent $D_2$
quiver as
\[
\ba{c}
R_{D_{2}(6)}^0 \rightarrow 
R_{E_{6}(2)}^{1,irrep} \oplus R_{E_{6}(2)}^{2,irrep} 
\oplus R_{E_{6}(2)}^{3,irrep}\\
R_{D_{2}(3)}^0 \rightarrow 
R_{E_{6}(1)}^{1,irrep} \oplus R_{E_{6}(1)}^{2,irrep} 
\oplus R_{E_{6}(1)}^{3,irrep}\\
R_{D_{2}(3)}^l \cong R_{D_{2}(3)}^{f(l)} \rightarrow  R_{E_{6}(3)}^{irrep},
\qquad l \in
\IZ_{4}-\{0\}. \\
\ea
\]
Consistency requires that if we replace $R_{D_{2}}$ in the $D_2$ quiver 
defining relations and then use the above mappings, we get the $E_6$ quiver
relations for $R_{E_{6}}^{irrep}$.

The origin of this $\IZ_3$ analogue of the orientifold
$\IZ_2$-projection is
thus far unknown to us. If an object with this property is to exist,
then the brane setup for the $E_6$ theory could be implemented; on the
other hand if it does not, then we would be suggested at why the
attempt for $E_6$ has been prohibitively difficult.

The $\IZ_3$ action has been noted to arise in \cite{ZD} in the
context of quotienting the $\IZ_n \times \IZ_n$ quiver to arrive at the
quiver for the $\Delta$-series. Indeed from our comparative study in Section
4.2.1, we see that in general, labelling the irreps by a
multi-index is precisely our stepwise algorithm in disguise, as
applied to a product Abelian group: the $\IZ_n \times \cdots \times
\IZ_n$ orbifold. Therefore in a sense we have explained why the
labelling scheme of \cite{Muto,ZD} should work.

And the same goes with $E_7$ and $E_8$: we could perform stepwise
projection thereupon and mathematically obtain their quivers as
appropriate quotients of the $\IZ_n$ quiver by the symmetry $S$ of the
identification and splitting of nodes. To find a physical brane setup,
we would then need to find an object in string theory which has an
$S$ action on space-time and the string world-sheet. Note that the above
are cases of the $\C^2$ orbifolds; for the $\C^k$-orbifold we should
initialise our algorithm with, and perform stepwise projection on the
quiver of $\IZ_n \times \cdots \times \IZ_n$ ($k-1$ times), i.e., the
brane box and cube ($k=2,3$).

Though mathematically we have found a systematic treatment of
constructing quivers under a new light, namely the ``stepwise
projection'' from the Abelian quiver, much work remains. In the field
of brane setups for singularities, our algorithm is intended to be a
small step for an old standing problem. We must now diligently seek
a generalisation of the orientifold plane with symmetry $S$ in
string theory, that could perform the physical task which our
mathematical methodology demands.

\chapter{Orbifolds VIII: Orbifolds with Discrete Torsion and the Schur Multiplier}
\index{Brane Probes!discrete torsion}
\index{Orbifolds!discrete torsion}
\label{chap:0010023}
\section*{\center{{ Synopsis}}}
{ L}et us now proceed with another aspect of D-brane probes on
singularities, namely with the presence of background B-fields, i.e.,
to allow discrete torsion.
Armed with the explicit computation of Schur
Multipliers, we offer a classification of $SU(n)$ orbifolds for
$n=2,3,4$ which
permit the turning on of discrete torsion. 

As a
by-product, we find a hitherto unknown class of ${\cal N}=1$ 
orbifolds with non-cyclic discrete torsion group.
Furthermore,
we supplement the {\it status quo ante} by investigating a first example of
a non-Abelian orbifold admitting discrete torsion, namely the ordinary
dihedral group as a subgroup of $SU(3)$. A comparison of the quiver
theory thereof with that of its
covering group, the binary dihedral group, without discrete torsion,
is also performed \cite{0010023}.
\section{Introduction}
The study of string theory in non-trivial NS-NS B-field backgrounds has of
late become one of the most pursued directions of research. Ever since
the landmark papers \cite{SW}, where it was shown
that in the presence of such non-trivial B-fields along the
world-volume 
directions of the D-brane, the gauge theory living thereupon
assumes a non-commutative guise in the large-B-limit, most works were
done in this direction of space-time non-commutativity. However, there
is an alternative approach in the investigation of the effects of the
B-field, namely {\bf discrete torsion}, which is of great
interest in this respect. On the other hand, as discrete torsion
presents itself to be a natural generalisation to the study of orbifold
projections of D-brane probes at space-time singularities, a topic
under much research over the past few years, it is also
mathematically and physically worthy of pursuit under this light.

A brief review of the development of the matter from a historical
perspective shall serve to guide the reader. Discrete torsion first
appeared in \cite{torsion} in the study of the
closed string partition function $Z(q,\bar{q})$ on the orbifold $G$. 
And shortly thereafter, it effects on the geometry of space-time were
pointed out \cite{VafaWit}.
In particular, \cite{torsion} noticed that
$Z(q,\bar{q})$ could contain therein, phases $\epsilon(g,h) \in U(1)$
for $g,h \in G$, coming from the twisted sectors of the theory, as
long as
\beq\label{epsilon}
\ba{l}
\epsilon(g_1g_2,g_3) = \epsilon(g_1,g_3)\epsilon(g_2,g_3)\\
\epsilon(g,h) = 1/\epsilon(h,g)\\
\epsilon(g,g) = 1,
\ea
\eeq
so as to ensure modular invariance.

Reviving interests along this line, Douglas and Fiol
\cite{Doug-tor,DougFiol} extended discrete torsion to the open string
sector by showing that the usual procedure of projection by orbifolds
on D-brane probes \cite{DM,LNV}, applied to {\bf projective
representations} instead of the ordinary {\em linear representations}
of the orbifold group $G$, gives exactly the gauge theory with discrete
torsion turned on. In other words, for the invariant matter fields which
survive the orbifold, $\Phi$ such that $\gamma^{-1}(g) \Phi \gamma(g)
= r(g) \Phi,\quad\forall~g \in G$, we now need the representation
\beq\label{proj1}
\ba{l}
\gamma(g) \gamma(h) = \alpha(g,h) \gamma(gh),\quad g,h \in G~~$with$\\
\alpha(x,y)\alpha(xy,z) = \alpha(x,yz)\alpha(y,z), \qquad
        \alpha(x,\II_G) = 1 = \alpha(\II_G,x) \quad
        \forall x,y,z \in G, \qquad \\
\ea
\eeq
where $\alpha(g,h)$ is known as a cocycle. These cocycles constitute,
up to the equivalence
\beq\label{proj2}
\alpha(g,h) \sim \frac{c(g)c(h)}{c(gh)} \alpha(g,h),
\eeq
the so-called second cohomology group $H^2(G,U(1))$
of $G$, where $c$ is any function (not necessarily a homomorphism)
mapping $G$ to $U(1)$; this is what we usually mean by 
{\em discrete torsion being classified by $H^2(G,U(1))$}.
We shall formalise all these definitions in the subsequent sections.

In fact, one can show \cite{torsion}, that the choice
\[
\epsilon(g,h) = \frac{\alpha(g,h)}{\alpha(h,g)},
\]
for $\alpha$ obeying \eref{proj1} actually satisfies \eref{epsilon}, whereby
linking the concepts of discrete torsion in the closed and open string
sectors.
We point this out as one could be easily confused as to the
precise parametre called discrete torsion and which is actually
classified by the second group cohomology.

Along the line of \cite{Doug-tor,DougFiol}, a series of papers by
Berenstein, Leigh and Jejjala \cite{BL,BJL} developed the technique
to study the {\em non-commutative} moduli space of the ${\cal N}=1$ 
gauge theory living on $\IC^3/\IZ_m \times \IZ_n$ parametrised as
an algebraic variety. A host of activities followed in the
generalisation of this abelian orbifold, notably to $\C^4/\IZ_2
\times \IZ_2 \times \IZ_2$ by \cite{Ray}, to the inclusion of
orientifolds by \cite{Klein-tor}, and to the orbifolded conifold by
\cite{Tatar}.

Along the mathematical thread, Sharpe has presented a prolific series
of works to relate discrete torsion with connection on gerbes
\cite{Sharpe}, which could allow generalisations of the concept to
beyond the 2-form B-field. Moreover, in relation to twisted K-theory
and attempts to unify space-time cohomology with group cohomology in
the vein of the McKay Correspondence (see e.g. \cite{9903056}), works 
by Gomis \cite{Gomis2} and Aspinwall-Plesser \cite{AspinPles,Aspin-tor}
have given some guiding light.

Before we end this review of the current studies, we would like to
mention the work by Gaberdiel \cite{Gab}. He pointed out that there
exists a different choice, such that the original intimate relationship
between discrete torsion in the closed string sector and the
non-trivial cocycle in the open sector can be loosened. It would be
interesting to investigate further in this spirit.

We see however, that during these last three years of renewed
activity, the focus has mainly been on Abelian orbifolds. It is one of
the main intentions of this chapter to initiate the study of non-Abelian
orbifolds with discrete torsion, which, to the best of our knowledge,
have not been discussed so far in the literature\footnote{In the
context of conformal field theory on orbifolds, there has been a 
recent work addressing some non-Abelian cases \cite{Katrin}.}. 
We shall classify
the general orbifold theories with ${\cal N}=0,1,2$ supersymmetry which
{\em could allow discrete torsion} by exhaustively computing the
second cohomology of the discrete subgroups of $SU(n)$ for $n=4,3,2$.

Thus rests the current state of affairs. Our main objectives are
two-fold: to both supplement the past, by presenting and studying
a first example of a non-Abelian orbifold which affords discrete
torsion, and to presage the future, by classifying the orbifold
theories which could allow discrete torsion being turned on.
\newpage
\section*{Nomenclature}
Throughout this chapter, unless otherwise specified, we shall adhere to
the following conventions for notation:\\
\begin{tabular}{ll}
$\omega_n$ & $n$-th root of unity;\\
$G$ & finite group of order $|G|$; \\
$\IF$ & (algebraically closed) number field; \\
$\IF^*$ & multiplicative subgroup of $\IF$; \\
$\gen{x_i|y_j}$ & the group generated by elements $\{x_i\}$ with
	relations $y_j$;\\ 
$<G_1,G_2,\ldots,G_n>$ & group generated by the generators of groups
$G_1,G_2,\ldots,G_n$; \\
$\gcd(m,n)$ & the greatest common divisor of $m$ and $n$; \\
$D_{2n},E_{6,7,8}$ & ordinary dihedral, tetrahedral, octahedral and
	icosahedral groups;\\ 
$\widehat{D_{2n}},\widehat{E_{6,7,8}}$ & 
	the binary counterparts of the above; \\
$A_n$ and $S_n$ & alternating and symmetric groups on $n$ elements; \\
$H \triangleleft G$ & $H$ is a normal subgroup of $G$; \\
$A \rtimes B$ & semi-direct product of $A$ and $B$;\\
$Z(G)$ & centre of  $G$;\\
$N_G(H)$ & the normaliser of $H \subset G$;\\
$G':=[G,G]$ & the derived (commutator) group of $G$;\\
$\exp(G)$ & exponent of group $G$.\\
\end{tabular}
\section{Some Mathematical Preliminaries}
\index{Finite Groups!projective representation}
\subsection{Projective Representations of Groups}
We begin by first formalising \eref{proj1}, the group representation
of our interest:
\begin{definition}
A {\bf projective} representation of $G$ over a field $\IF$ 
(throughout we let $\IF$ be an algebraically closed field with
 characteristic $p\geq 0$) is a
mapping $\rho : G \rightarrow GL(V)$ such that
\[
(A)~~~ \rho(x)\rho(y) = \alpha(x,y)\rho(xy)~~\forall~~x,y \in G;
\qquad
(B)~~~ \rho(\II_G) = \II_V.
\]
\end{definition}
Here $\alpha : G \times G \rightarrow \IF^*$ is a mapping whose
meaning we shall clarify later.
Of course we see that if $\alpha=1$ trivially, then we have our familiar
ordinary representation of $G$ to which we shall refer as {\em
linear}. Indeed, the mapping $\rho$ into $GL(V)$ defined above is
naturally equivalent to a homomorphism into the projective linear group
$PGL(V) \cong GL(V)/\IF^*\II_V$, and hence the name ``projective.''
In particular we shall be concerned with projective {\em
matrix} representations of $G$ where we take $GL(V)$ to be
$GL(n,\IF)$.

The function $\alpha$ can not be arbitrary and two immediate
restrictions can be placed thereupon purely from the structure of the group:
\beq
\label{alpha-tor}
\ba{clcl}
(a)& \mbox{Group Associativity} & \Rightarrow &
	\alpha(x,y)\alpha(xy,z) = \alpha(x,yz)\alpha(y,z), \quad\forall
	x,y,z \in G\\
(b)& \mbox{Group Identity} & \Rightarrow &
	\alpha(x,\II_G) = 1 = \alpha(\II_G,x), \quad \forall x \in G.
\ea
\eeq
These conditions on $\alpha$ naturally leads to another discipline of
mathematics.
\subsection{Group Cohomology and the Schur Multiplier}
The study of such functions on a group satisfying \eref{alpha-tor}
is precisely the subject of the
theory of {\bf Group Cohomology}. In general we let $\alpha$ to take
values in $A$, an abelian coefficient group ($\IF^*$ is certainly a
simple example of such an $A$) 
and call them {\bf cocycles}. The set of all cocycles we
shall name $Z^2(G,A)$. Indeed it is straight-forward to see that
$Z^2(G,A)$ is an abelian group. We subsequently define a set of
functions satisfying
\beq
\label{cobound}
B^2(G,A) := \{(\delta g)(x,y) := g(x) g(y) g(xy)^{-1}\}
\quad\mbox{for any~}g :G \rightarrow A \mbox{~such that~} g(\II_G)=1,
\eeq
and call them {\em coboundaries}.
It is then obvious that $B^2(G,A)$ is a (normal) subgroup of
$Z^2(G,A)$ and in fact constitutes an equivalence relation on the
latter in the manner of \eref{proj2}. Thus it becomes 
a routine exercise in cohomology to define
\[
H^2(G,A) := Z^2(G,A) / B^2(G,A),
\]
the {\em second cohomology} group of $G$.

Summarising what we have so far, we see that the projective
representations of $G$ are
classified by its second cohomology $H^2(G,\IF^*)$. To facilitate the
computation thereof, we shall come to an important concept:
\begin{definition}
The {\bf Schur Multiplier} $M(G)$ of the group $G$ is the second
cohomology group with respect to the trivial action of $G$ on $\IC^*$:
\[
M(G) := H^2(G,\IC^*).
\]
\end{definition}
\index{Finite Groups!Schur multiplier}
\index{Finite Groups!cohomology}
Since we shall be mostly concerned with the field $\IF = \IC$, the
Schur multiplier is exactly what we need. However, the properties
thereof are more general. In fact, for any algebraically closed field
$\IF$ of zero characteristic,
$M(G)\cong H^2(G,\IF^*)$. In our case of $\IF =
\IC$, it can be shown that \cite{Klein-tor},
\[
H^2(G,\IC^*) \cong H^2(G,U(1)).
\]
This terminology is the more frequently encountered one
in the physics literature.

One task is thus self-evident: the calculation of the Schur Multiplier
of a given group $G$ shall indicate possibilities of projective
representations of the said group, or in a physical language, the
possibilities of turning on discrete torsion in string theory on the
orbifold group $G$. In particular, if $M(G) \cong \II$, then the
second cohomology of $G$ is trivial and no non-trivial discrete
torsion is allowed. We summarise this
\[
\mbox{KEY POINT:}\quad \mbox{Calculate }M(G) \Rightarrow 
\mbox{Information on Discrete Torsion.}
\]
\subsection{The Covering Group}
\index{Finite Groups!covering group}
The study of the actual projective representation of $G$ is very
involved and what is usually done in fact is to ``lift to an ordinary
representation.'' What this means is that for a central 
extension\footnote{i.e., $A$ in the centre
	$Z(G^*)$ and $G^*/A \cong G$ according to the exact sequence
$1 \rightarrow A \rightarrow G^* \rightarrow G \rightarrow 1$.} 
$A$ of $G$ to $G^*$, we say a projective
representation $\rho$ of $G$ {\bf lifts} to a linear
representation $\rho^*$ of $G^*$ if (i) $\rho^*(a \in A)$ is
proportional to $\II$ and (ii) there is a section\footnote{i.e., for the
	projection $f:G^*\rightarrow G$, $\mu \circ f = \II_G$.}
$\mu : G \rightarrow G^*$ such that $\rho(g) = \rho^*(\mu(g)),\quad
\forall g \in G$. Likewise it {\em lifts projectively} if $\rho(g) =
t(g) \rho^*(\mu(g))$ for a map $t:G \rightarrow \IF^*$. Now we are
ready to give the following:
\begin{definition}\label{defcover}
We call $G^*$ a {\bf covering group}\footnote{
	Sometimes is also known as {\bf representation group}.}
of $G$ over $\IF$ if the following are satisfied:\\
(i) $\exists$ a central extension $1 \rightarrow A \rightarrow G^*
\rightarrow G \rightarrow 1$ such that any projective representation
of $G$ lifts projectively to an ordinary representation of $G^*$;\\
(ii) $|A| = |H^2(G,\IF^*)|$.
\end{definition}
The following theorem, initially due to Schur, characterises covering groups.
\begin{theorem}\label{cover}{\rm (\cite{Karp} p143)}
$G^{\star}$ is a covering
group of $G$ over $\IF$ if and only if the following conditions hold:\\
(i) $G^{\star}$ has a finite subgroup $A$ with $A\subseteq
Z(G^{\star}) \cap [G^{\star},G^{\star}]$;\\
(ii) $G \cong G^{\star}/A$;\\
(iii) $|A|=|H^2(G,F^{\star})|$\\
where $[G^{\star},G^{\star}]$ is the derived group\footnote{For a
	group $G$, $G' := [G,G]$ is the group generated by elements of
	the form $xyx^{-1}y^{-1}$ for $x,y \in G$.}
$G^{*'}$ of $G^*$.
\end{theorem}

Thus concludes our prelude on the mathematical rudiments, the utility
of the above results shall present themselves in the ensuing.
%
\section{Schur Multipliers and String Theory Orbifolds}
The game is thus afoot. Orbifolds of the form $\IC^k/\{G \in SU(k)\}$
have been widely studied in the context of gauge theories living on
D-branes probing the singularities. We need only to compute $M(G)$ for
the discrete finite groups of $SU(n)$ for $n=2,3,4$ 
to know the discrete torsion
afforded by the said orbifold theories.
\subsection{The Schur Multiplier of the Discrete Subgroups of
$SU(2)$}
\index{Finite Groups!$SU(2)$ subgroups!Schur multiplier}
Let us first remind the reader of the well-known $ADE$
classification of the discrete finite subgroups of $SU(2)$.
Here are the presentations of these groups:
\beq
\label{SU2}
\ba{|c|c|c|c|}
\hline
G	& $Name$ & $Order$	& $Presentation$ \\
\hline
\widehat{A_n} & $Cyclic$, \cong \IZ_{n+1} & n & \gen{a|a^n=\II} \\ \hline
\widehat{D_{2n}} & \mbox{Binary Dihedral} & 4n &
		\gen{a,b|b^2=a^n, abab^{-1}=\II} \\ \hline
\widehat{E_6} & \mbox{Binary Tetrahedral} & 24 &
		\gen{a,b|a^3=b^3=(ab)^3} \\ \hline
\widehat{E_7} & \mbox{Binary Octahedral} & 48 &
		\gen{a,b|a^4=b^3=(ab)^2} \\ \hline
\widehat{E_8} & \mbox{Binary Icosahedral} & 120 &
		\gen{a,b|a^5=b^3=(ab)^2} \\ \hline
\ea
\eeq
We here present a powerful result due to Schur (1907) (q.v. Cor. 2.5,
Chap. 11 of \cite{Karp2}) which aids us to explicitly compute large
classes of Schur multipliers for finite groups:
\begin{theorem} {\rm (\cite{Karp} p383)}
\label{Schur}
Let $G$ be generated by $n$ elements with (minimally)
$r$ defining relations and let the Schur multiplier $M(G)$ have a
minimum of $s$ generators, then
\[
r \ge n+s.
\]
In particular, $r=n$ implies that $M(G)$ is trivial and $r=n+1$, that
$M(G)$ is cyclic.
\end{theorem}
Theorem \ref{Schur} could be immediately applied to $G\in SU(2)$.

Let us proceed with the computation case-wise. The $\widehat{A_n}$
series has 1 generator with 1 relation, thus $r=n=1$ and
$M(\widehat{A_n})$ is trivial. Now for the $\widehat{D_{2n}}$ series, we
note briefly that the usual presentation is $\widehat{D_{2n}} :=
\gen{a,b|a^{2n}=\II, b^2=a^n, bab^{-1}=a^{-1}}$ as in \cite{9909125};
however, we can see easily that the last two relations imply the
first, or explicitly: $a^{-n} := (bab^{-1})^n=b a^n b^{-1} = a^n$,
(q.v. \cite{Karp2} Example 3.1, Chap. 11), whence making $r=n=2$,
i.e., 2 generators and 2 relations, and further making
$M(\widehat{D_{2n}})$ trivial. Thus too are the cases of the 3
exceptional groups, each having 2 generators with 2 relations. In
summary then we have the following corollary of Theorem \ref{Schur},
the well-known \cite{AspinPles} result that
\begin{corollary}
All discrete finite subgroups of $SU(2)$ have second
cohomology $H^2(G,\IC^*) = \II$, and hence afford no non-trivial
discrete torsion.
\end{corollary}

It is intriguing that the above result can actually be hinted from
physical considerations without recourse to heavy mathematical
machinery. The orbifold theory for $G \subset SU(2)$
preserves an ${\cal N}=2$ supersymmetry on the world-volume of the
D3-Brane probe.
Inclusion of discrete torsion would deform the coefficients
of the superpotential. However, ${\cal N}=2$ supersymmetry is highly
restrictive and in general does not permit the existence of such
deformations. This is in perfect harmony with the triviality of the
Schur Multiplier of $G \subset SU(2)$ as presented in the above Corollary.

To address more complicated groups we need a methodology to compute
the Schur Multiplier, and we have many to our aid, for after all the
computation of $M(G)$ is a vast subject entirely by itself. We quote
one such method below, a result originally due to Schur:
\begin{theorem}
{\rm (\cite{Karp3} p54)} Let $G = F/R$ be the defining finite
presentation of $G$ with $F$ the free group of rank $n$ and $R$ is
(the normal closure of) the set of relations. Suppose $R/[F,R]$ has
the presentation $\gen{x_1,\ldots,x_m;y_1,\ldots,y_n}$ with all $x_i$ of
finite order, then
\[
M(G) \cong \gen{x_1,\ldots,x_n}.
\]
\end{theorem}

Two more theorems of great usage are the following:
\begin{theorem}{\rm (\cite{Karp3} p17)}
\label{exponent}
Let the exponent\footnote{i.e., the lowest common multiple of the orders
of the elements.} of $M(G)$ be $\exp(M(G))$, then
\[
\exp(M(G))^2 \mbox{ divides } |G|.
\]
\end{theorem}
And for direct products, another fact due to Schur,
\begin{theorem}{\rm (\cite{Karp3} p37)} \label{directprod}
\[
M(G_1 \times G_2) \cong M(G_1) \times M(G_2) \times (G_1 \otimes G_2),
\]
where $G_1 \otimes G_2$ is defined to be $\Hom_{\IZ}(G_1/G_1',G_2 /G_2')$.
\end{theorem}

With the above and a myriad of useful results (such as the Schur
Multiplier for semi-direct products), and especially with the aid of the
Computer Algebra package {\sf GAP} \cite{Prog} 
using the algorithm developed for the $p$-Sylow
subgroups of Schur Multiplier \cite{Holt}, 
we have engaged in the formidable task of giving
the explicit Schur Multiplier of the list of groups of our interest.

Most of the details of the computation we shall
leave to the appendix, to
give the reader a flavour of the calculation but not distracting him
or her from the main course of our writing.
Without much further ado then, we now proceed with the list of Schur
Multipliers for the
discrete subgroups of $SU(n)$ for $n=3,4$, i.e., the ${\cal N}=1,0$
orbifold theories.
\subsection{The Schur Multiplier of the Discrete Subgroups of $SU(3)$}
\index{Finite Groups!$SU(3)$ subgroups!Schur multiplier}
The classification of the discrete finite groups of $SU(3)$ is
well-known (see e.g. \cite{Fairbairn,9811183,Muto} 
for a discussion thereof in the
context of string theory). It was pointed out in \cite{9909125} that the usual
classification of these groups does not include the so-called {\em
intransitive} groups (see \cite{9905212} for definitions), which are perhaps
of less mathematical interest. Of course from
a physical stand-point, they all give well-defined orbifolds. More
specifically \cite{9909125}, all the ordinary polyhedral subgroups of
$SO(3)$, namely 
the ordinary dihedral group $D_{2n}$ and the ordinary $E_6 \cong 
A_4 \cong \Delta(3 \times 2^2), E_7 \cong S_4 \cong \Delta(6 \times
2^2), E_8 \cong \Sigma_{60}$, due to
the embedding $SO(3) \hookrightarrow SU(3)$, are obviously
(intransitive) subgroups thereof and thus we
shall include these as well in what follows. We discuss some aspects
of the intransitives in Appendix \ref{append:0010023.B} 
and are grateful to D. Berenstein
for pointing out some subtleties involved \cite{Berenstein}.
We insert one more cautionary note. The $\Delta(6n^2)$ series does not
actually include the cases for $n$ odd \cite{Muto}; therefore $n$
shall be restricted to be even.

Here then are the Schur Multipliers of the $SU(3)$ discrete subgroups
(I stands for Intransitives and T, intransitives).
\beq
\hspace{-0.5in}
\label{SU3}
\ba{|c|c|c|c|}
\hline
	& $G$ & $Order$ &  \mbox{Schur Multiplier } M(G) \\
\hline
\mbox{I}	& \IZ_n \times \IZ_m	& n \times m & \IZ_{\gcd(n,m)} \\ \hline
			& <\IZ_n, \widehat{D_{2m}}> & 
        \left\{\ba{lc}  n \times 4m & n~\odd \\ \frac{n}{2} \times 4m
        & n~\even \ea \right. &
	\left\{\ba{lc}  \II &  n~{\rm mod}~4 \neq 1 \\ 
        \IZ_2 &  n~{\rm mod}~4 = 0, m~\odd \\ 
        \IZ_2 \times \IZ_2 &  n~{\rm mod}~4 = 0, m~\even 
        \ea \right. \\ \hline

			& <\IZ_n, \widehat{E_6}> & 
        \left\{\ba{lc}  n \times 24 & n~\odd \\ \frac{n}{2} \times 24
        & n~\even \ea \right.  &
	\IZ_{\gcd(n,3)} 
	 \\ \hline

			& <\IZ_n, \widehat{E_7}>  &  
        \left\{\ba{lc}  n \times 48 & n~\odd \\ \frac{n}{2} \times 48
        & n~\even \ea \right.  &
	\left\{\ba{lc} \II & n~{\rm mod}~4 \neq 0 \\ \IZ_2 & n~{\rm mod}~4 = 0         \ea \right. \\ \hline

			& <\IZ_n, \widehat{E_8}>  &  
        \left\{\ba{lc}  n \times 120 & n~\odd \\ \frac{n}{2} \times 120
        & n~\even \ea \right.    &\II \\ \hline
			& \mbox{Ordinary Dihedral } D_{2n} & 2n &
	\IZ_{\gcd(n,2)}
	\\ \hline

			& <\IZ_n, D_{2m}> & 
        \left\{\ba{lc}  n \times 2m & m~\odd \\  
                        n \times 2m & m~\even, n~\odd \\
                        \frac{n}{2} \times 2m & m~\even, n~\even \\ \ea \right.  &
	\left\{\ba{lc}\ 
			\IZ_{\gcd(n,2)} & m~\odd \\ 
			\IZ_2  & m~\even, n~{\rm mod}~4 = 1,2,3 \\
			\IZ_2  &  m~{\rm mod}~4 \neq 0,  n~{\rm mod}~4 = 0 \\
			\IZ_2 \times \IZ_2  & m~{\rm mod}~4 = 0,  n~{\rm mod}~4 = 0 \\
	\ea\right. \\ \hline
\mbox{T} 	& \Delta_{3n^2} & 3n^2 &
		\left\{\ba{cc}\IZ_n \times \IZ_3, &\gcd(n,3)\neq 1 \\
		\IZ_n, & \gcd(n,3)=1 \ea \right. \\ \hline
		 	& \Delta_{6n^2}~(n~\even) & 6n^2 & \IZ_2 \\ \hline
	& \Sigma_{60}\cong A_5 & 60 & \IZ_2 \\ \hline
	& \Sigma_{168} & 168 & \IZ_2 \\ \hline
	& \Sigma_{108} & 36 \times 3 & \II \\ \hline
	& \Sigma_{216} & 72 \times 3 & \II \\ \hline
	& \Sigma_{648} & 216 \times 3 & \II \\ \hline
	& \Sigma_{1080}& 360 \times 3 & \IZ_2 \\ \hline
\ea
\eeq
\newpage
Some immediate comments are at hand.
The question of whether any discrete subgroup of $SU(3)$ admits
non-cyclic discrete torsion was posed in \cite{AspinPles}. From our
results in table \eref{SU3}, we have shown by explicit construction
that the answer is in the affirmative: not only the various
intransitives give rise to product cyclic Schur Multipliers, so too
does the transitive $\Delta(3n^2)$ series for $n$ a multiple of 3.

In Appendix \ref{append:0010023.A} we shall present the calculation for
$M(\Delta_{3n^2})$ and $M(\Delta_{6n^2})$ for illustrative
purposes. Furthermore, as an example of non-Abelian orbifolds with
discrete torsion, we shall investigate the series of the ordinary
dihedral group in detail with applications to physics in mind. For
now, for the reader's edification or amusement, let us continue
with the $SU(4)$ subgroups.
\newpage
\subsection{The Schur Multiplier of the Discrete Subgroups of $SU(4)$}
\index{Finite Groups!$SU(4)$ subgroups!Schur multiplier}
The discrete finite subgroups of $SL(4,\IC)$, which give rise
to non-supersymmetric orbifold theories, are presented in modern
notation in \cite{9905212}. Using the notation therein, and recalling that
the group names in $SU(4) \subset SL(4,\IC)$ were accompanied with a
star ({\it cit. ibid.}), let us tabulate in \eref{SU4}
the Schur Multiplier of the
exceptional cases of these particulars (cases XXIX$*$ and XXX$*$ were
computed by Prof. H. Pahlings to whom we are grateful).
\beq
\label{SU4}
\ba{|c|c|c|}
\hline
	$G$ & $Order$ & \mbox{Schur Mult. } M(G) \\
\hline
	 $I$* & 60\times 4 &  \II \\ \hline
	 $II$* \cong \Sigma_{60} & 60 & \IZ_2 \\ \hline
	 $III$* & 360 \times 4 & \IZ_3  \\ \hline
	 $IV$* & \frac12 7!\times 2  & \IZ_3 \\ \hline
	 $VI$*   & 2^6 3^4 5\times 2 &  \II \\ \hline
	 $VII$*   & 120\times 4 & \IZ_2 \\ \hline
	 $VIII$* & 120\times 4	& \IZ_2	\\ \hline	
	 $IX$*   & 720\times 4	& \IZ_2	\\ \hline
	 $X$*    & 144 \times 2 & \IZ_2 \times \IZ_3 \\ \hline
	 $XI$*   & 288 \times 2 & \IZ_2 \times \IZ_3 \\ \hline
	 $XII$*  & 288 \times 2 & \IZ_2 \\ \hline
	 $XIII$* & 720 \times 2 & \IZ_2 \\ \hline
	 $XIV$*  & 576 \times 2 & \IZ_2 \times \IZ_2 \\ \hline
	 $XV$*   & 1440 \times 2 & \IZ_2 \\ \hline
\ea
\quad
\ba{|c|c|c|}
\hline
	$G$ & $Order$ & \mbox{Schur Mult. } M(G) \\
\hline
	 $XVI$*  & 3600 \times 2 & \IZ_2 \\ \hline
	 $XVII$* &576\times 4 & \IZ_2	\\ \hline
	 $XVIII$*&576\times 4 & \IZ_2 \times \IZ_3 \\ \hline
	 $XIX$*  &288\times 4 & \II	\\ \hline
	 $XX$*   &7200\times 4 & \II	\\ \hline
	 $XXI$*  &1152\times 4 & \IZ_2 \times \IZ_2 \\ \hline
	 $XXII$* & 5\times 16 \times 4   & \IZ_2 \\ \hline
	 $XXIII$*& 10\times 16 \times 4  & \IZ_2 \times \IZ_2 \\ \hline
	 $XXIV$* & 20\times 16 \times 4  & \IZ_2 \\ \hline
	 $XXV$*  & 60\times 16 \times 4  & \IZ_2 \\ \hline
	 $XXVI$* & 60\times 16 \times 4  & \IZ_2 \times \IZ_4 \\ \hline
	 $XXVII$*& 120\times 16 \times 4 & \IZ_2 \times \IZ_2 \\ \hline
	 $XXVIII$*& 120\times 16 \times 4& \IZ_2 \\ \hline
	 $XXIX$* & 360\times 16 \times 4 & \IZ_2 \times \IZ_3\\ \hline
	 $XXX$*  & 720\times 16 \times 4 & \IZ_2 \\ \hline
\ea
\eeq
\section{$D_{2n}$ Orbifolds: Discrete Torsion for a
non-Abelian Example}
As advertised earlier at the end of subsection 3.2, 
we now investigate in depth the discrete
torsion for a non-Abelian orbifold.
The ordinary dihedral group $D_{2n} \cong \IZ_n \rtimes \IZ_2$ of order
$2n$, has the presentation
\[
D_{2n}=\gen{a,b|a^n=1, b^2=1, bab^{-1}=a^{-1}}.
\]
As tabulated in \eref{SU3}, the Schur Multiplier is $M(D_{2n})=\II$ 
for $n$ odd and $\IZ_2$ for $n$ even \cite{Karp}.
Therefore the $n$ odd cases are no different from the ordinary linear
representations as studied in \cite{9909125} since they have trivial Schur
Multiplier and hence trivial discrete torsion. On the other hand, for
the $n$ even case, we will demonstrate the following result:
\begin{proposition}
The binary dihedral group $\widehat{D_{2n}}$ of the
$D$-series of the discrete subgroups of $SU(2)$ (otherwise
called the generalised quaternion group) 
is the covering group of $D_{2n}$ when $n$ is even. 
\end{proposition}
Proof: The definition of the binary dihedral group $\widehat{D_{2n}}$, of
order $4n$, is
\[
\widehat{D_{2n}}=\gen{a,b|a^{2n}=1, b^2=a^n, bab^{-1}=a^{-1}},
\]
as we saw in subsection 3.1.
Let us check against the conditions of Theorem \ref{cover}.
It is a famous result that $\widehat{D_{2n}}$ is the double cover of
$D_{2n}$ and whence an $\IZ_2$ central extension.
First we can see that $A=Z(\widehat{D_{2n}})=\{1,a^n\}\cong \IZ_2$ and
condition (ii) is satisfied.

Second we find that the commutators are $[a^x,a^y] :=
(a^x)^{-1}(a^y)^{-1} a^x a^y=1$, $[a^xb, a^yb]=a^{2(x-y)}$ 
and $[a^x b, a^y]=a^{2y}$. From these we see that the derived group
$[\widehat{D_{2n}},\widehat{D_{2n}}]$ is generated by $a^2$ and is thus
equal to $\IZ_n$ (since $a$ is of order $2n$). An important point is that
only when $n$ is even does $A$ belong to $Z(\widehat{D_{2n}}) \cap
[\widehat{D_{2n}},\widehat{D_{2n}}]$. This result is consistent with the fact that
for odd $n$, $D_{2n}$ has trivial Schur Multiplier. Finally of course, 
$|A| = |H^2(G,\IF^*)| = 2$. Thus conditions (i)
and (iii) are also satisfied.
We therefore conclude that for even $n$, 
$\widehat{D_{2n}}$ is the covering group of $D_{2n}$.

\subsection{The Irreducible Representations}
With the above Proposition, we know by the very definition of the
covering group, that the projective representation of $D_{2n}$ should be
encoded in the linear representation of $\widehat{D_{2n}}$, which is a
standard result that we can recall from \cite{9909125}.
The latter has four 1-dimensional and $n-1$ 2-dimensional irreps.
The matrix representations of these 2-dimensionals for the generic
elements $a^p, b a^p$ ($p=0,...,2n-1$) are given below:
\beq 
\label{binarydihedral_2}
a^{p} = \mat{\ba{cc}  \omega_{2n}^{lp} & 0 \\
				0 & \omega_{2n}^{-lp}
		\ea}
\qquad
b a^{p} = \mat{\ba{cc}  0 & i^l\omega_{2n}^{-lp} \\
				i^l \omega_{2n}^{lp} & 0 
		\ea},
\eeq
with $l=1,...,n-1$; these are denoted as $\chi^l_2$. 
On the other hand, the four 1-dimensionals are
\beq
{\small
\label{binarydihedral_1}
\ba{c|c}
n = \even
&
n = \odd \\ \hline
\ba{c|cccc}
     &	a^{\even} & a(a^{\odd}) & b(b a^{\even}) &
	b a (b a^{\odd}) \\
\chi^{1}_1 & 1 & 1 & 1 & 1 \\
\chi^{2}_1 & 1 & -1 & 1 & -1 \\
\chi^{3}_1 & 1 & 1 & -1 & -1 \\
\chi^{4}_1 & 1 & -1 & -1 & 1 
\ea
&
\ba{cccc}
	a^{\even} & a(a^{\odd}) & b(b a^{\even}) &
	b a (b a^{\odd}) \\
1 & 1 & 1 & 1 \\
1 & -1 & \omega_4 & -\omega_4 \\
1 & 1 & -1 & -1 \\
1 & -1 & - \omega_4 &  \omega_4
\ea
\ea}
\eeq

We can subsequently obtain all irreducible projective
representations of $D_{2n}$ from the above (henceforth $n$ will be 
even).
Recalling that $\widehat{D_{2n}}/\{1,a^n\}\cong D_{2n}$ from property (ii) of
Theorem \ref{cover}, we can choose one element of each of the
transversals of $\widehat{D_{2n}}$ with respect to the $\IZ_2$ to be
mapped to $D_{2n}$.
For convenience we choose $b^{x} a^{y}$ with $x=0,1$ and
$y=0,1,...,n-1$, a total of $4n/2=2n$ elements. Thus we are
effectively expressing $D_{2n}$ in terms of $\widehat{D_{2n}}$ elements.

For the matrix representation of $a^n \in \widehat{D_{2n}}$, there are two cases. 
In the first, we have $a^n=1\times I_{d\times d}$ where $d$ is the dimension
of the representation. This case includes all four 1-dimensional representations
and $(n/2-1)$ 2-dimensional representations in \eref{binarydihedral_2}
for $l=2,4,...,n-2$. Because $a^n$ has the same matrix form as $\II$, we see that
the elements $b^{x} a^{y}$ and $b^{x} a^{y+n}$ also have the same
matrix form. Consequently, when we map them to $D_{2n}$, they automatically give the
irreducible linear representations of $D_{2n}$.

In the other case, we have $a^n=-1\times I_{d\times d}$ and this happens
when $l=1,3,...,n-1$. It is precisely these cases\footnote{Sometimes
	also called {\bf negative representations} in such cases.}
which give the {\em irreducible projective representations} of
$D_{2n}$. Now, because $a^n$ has a different matrix form
from $\II$, the matrices for $b^{x} a^{y}$ and $b^{x} a^{y+n}$ differ.
Therefore, when we map $\widehat{D_{2n}}$ to $D_{2n}$, there
is an ambiguity as to which of the matrix forms,
$b^{x} a^{y}$ or $b^{x} a^{y+n}$,
to choose as those of $D_{2n}$.

This ambiguity is exactly a feature of projective representations.
Preserving the notations of Theorem \ref{cover}, we let $G^* =
\bigcup\limits_{g_i \in G} A g_i$ be the decomposition into
transversals of $G$ for the normal subgroup $A$. Then choosing 
one element in every transversal, say $A_q g_i$ for some fixed $q$, we
have the ordinary (linear) representation thereof being precisely the
projective representation of $g_i$. Of course different choices of
$A_q$ give different but projectively equivalent (projective)
representations of $G$.

By this above method, we can construct all irreducible projective
representations of $D_{2n}$ from \eref{binarydihedral_2}.
We can verify this by matching dimensions:
we end up with $n/2$ 2-dimensional
representations inherited from $\widehat{D_{2n}}$ and
$2^2 \times (n/2)=2n$, which of course is the order of $D_{2n}$ as it
should.
\subsection{The Quiver Diagram and the Matter Content}
The projection for the matter content $\Phi$ is well-known (see
e.g., \cite{LNV,9811183}):
\begin{equation}
\label{projection}
\gamma^{-1}(g) \Phi \gamma(g)= r(g) \Phi,
\end{equation}
for $g \in G$ and $r, \gamma$ appropriate (projective) representations.
The case of $D_{2n}$ without torsion was discussed as a new class of
non-chiral ${\cal N}=1$ theories in \cite{9909125}. We recall that
for the group $D_{2n}$ we choose the generators (with action on
$\IC^3$) as
\begin{equation} 
\label{dihedral_2}
a = \left(  \begin{array}{ccc} 1 & 0 & 0 \\0 &  \omega_{n} & 0 \\
				0 &	0 & \omega_{n}^{-1}
			\end{array}
		\right)
\qquad
b  = \left(  \begin{array}{ccc}  -1 & 0 & 0\\ 0 & 0 & -1 \\
				0 & -1 & 0 
			\end{array}
		\right).
\end{equation}
Now we can use these explicit forms to work out the matter content
(the quiver diagram) and superpotential. For the regular
representation, we
choose $\gamma(g)$ as block-diagonal in which every 2-dimensional irreducible 
representation repeats twice with labels $l=1,1,3,3,..
,n-1,n-1$ (as we have shown in the previous section that the even
labels correspond to the linear representation of $D_{2n}$).
With this $\gamma(g)$, we calculate the matter content below.

For simplicity, in the actual calculation we would not use 
\eref{projection} but rather the standard method given by Lawrence,
Nekrasov and Vafa \cite{LNV}, generalised appropriately
to the projective case by
\cite{AspinPles}. We can do so because we are armed with
Definition \ref{defcover} and results from the previous subsection,
and directly use the linear representation of the covering group:
we lift the action of $D_{2n}$ into the action of its covering group
$\widehat{D_{2n}}$. It is easy to see that we get the 
same matter content either by using the projective representations of
the former or the linear representations of the latter.

From the point of view of the covering group, the representation $r(g)$ in 
\eref{projection} is given by
\beq\label{decomp-tor}
{\bf 3} \longrightarrow \chi^3_1+\chi^2_2
\eeq
and the representation $\gamma(g)$ is given by
$\gamma \longrightarrow \sum\limits_{l=0}^{n/2-1} 2 \chi^{2l+1}_2$.
We remind ourselves that the ${\bf 3}$ must in fact be a {\em linear}
representation of $D_{2n}$ while $\gamma(g)$ is the one that has to be {\em
projective} when we include discrete torsion \cite{Doug-tor}.

For the purpose of tensor decompositions we recall the result for the binary
dihedral group \cite{9909125}:\\
\begin{equation}
\begin{array}{|l|l|}
\hline
{\bf 1} \otimes {\bf 1}'
&
\begin{array}{c|c}
	n = \even	& n = \odd \\
	\begin{array}{ccc}
	\chi_1^2\chi_1^2=\chi_1^1  & \chi_1^3\chi_1^3=\chi_1^1 &
	\chi_1^4\chi_1^4=\chi_1^1  \\
	\chi_1^2 \chi_1^3=\chi_1^4 & \chi_1^2\chi_1^4=\chi_1^3 &
	\chi_1^3\chi_1^4=\chi_1^2
	\end{array}
	&
	\begin{array}{ccc}
	\chi_1^2\chi_1^2=\chi_1^3  & \chi_1^3\chi_1^3=\chi_1^1 &
	\chi_1^4\chi_1^4=\chi_1^3  \\
	\chi_1^2 \chi_1^3=\chi_1^4 & \chi_1^2\chi_1^4=\chi_1^1 &
	\chi_1^3\chi_1^4=\chi_1^2
	\end{array}
\end{array}
\\ \hline
{\bf 1} \otimes {\bf 2}
&
\chi_1^{h} \chi_2^l = \left\{ \begin{array}{l}
\chi_2^l~~~~h=1,3  \\
\chi_2^{n-l}~~~~h=2,4 
\end{array}
\right.
\\ \hline
{\bf 2} \otimes {\bf 2'}
&
\chi_2^{l_1} \chi_2^{l_2}=\chi_2^{(l_1+l_2)}+\chi_2^{(l_1-l_2)}
{\rm ~where~}
\begin{array}{l}
	\chi_2^{(l_1+l_2)}= \left\{ \begin{array}{l}
	\chi_2^{(l_1+l_2)}~~~~{\rm if}~~~l_1+l_2<n,  \\
	\chi_2^{2n-(l_1+l_2)}~~~~{\rm if}~~~l_1+l_2>n, \\
	\chi_1^2+\chi_1^4~~~~{\rm if}~~~l_1+l_2=n.
	\end{array}
	\right.
	\\
	\chi_2^{(l_1-l_2)}= \left\{ \begin{array}{l}
	\chi_2^{(l_1-l_2)}~~~~{\rm if}~~~l_1>l_2,  \\
	\chi_2^{(l_2-l_1)}~~~~{\rm if}~~~l_1<l_2, \\
	\chi_1^1+\chi_1^3~~~~{\rm if}~~~l_1=l_2.
	\end{array}
	\right.
\end{array}
\\
\hline
\end{array}
\end{equation}

From these relations we immediately obtain the matter content.
Firstly, there are $n/2$ $U(2)$ gauge groups ($n/2$ nodes in the quiver). 
Secondly, because $\chi^3_1 \chi_2^l=\chi_2^l$ we have one adjoint 
scalar for every gauge group. Thirdly, since $\chi_2^{2} \chi_2^{2l+1}=
\chi_2^{2l-1}+\chi_2^{2l+3}$ (where for $l=0$, 
$\chi_2^{2l-1}$ is understood to
be $\chi_2^{1}$ and for $l=n/2-1$, $\chi_2^{2l+3}$ is understood to be
$\chi_2^{n-1}$), we have two bi-fundamental chiral supermultiplets. 
We summarise these results in \fref{fig:dihedral}.
\EPSFIGURE[ht]{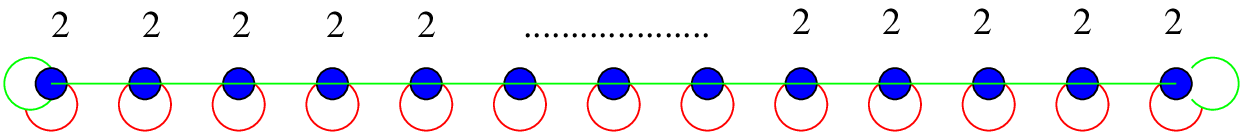,width=6.2in}
{
The quiver diagram of the ordinary dihedral group $D_{2n}$
with non-trivial projective representation. In this case of
discrete torsion being turned on, we have a product of $n/2$ $U(2)$
gauge groups (nodes).
The line connecting two nodes without arrows means
that there is one chiral multiplet in each direction. Therefore we
have a non-chiral theory.
\label{fig:dihedral}
}

We want to emphasize that by lifting to the covering group, in
general we not only find the matter content (quiver diagram) as we
have done above, but also the superpotential as well. The formula is
given in (2.7) of \cite{LNV}, which could be applied here without
any modification (of course, one can use the matrix form of the group
elements to obtain the superpotential directly as
done in \cite{Doug-tor,DougFiol,DM,BL,BJL,Ray,Klein-tor}, 
but (2.7), expressed in terms of the Clebsh-Gordan coefficients, is more 
convenient).

Knowing the above quiver (cf. \fref{fig:dihedral}) of the ordinary dihedral
group $D_{2n}$ {\em with} discrete torsion, we wish to question
ourselves as to the relationships between this quiver and that of
its covering group, the
binary dihedral group $\widehat{D_{2n}}$ {\em without} discrete
torsion (as well as that of $D_{2n}$ without discrete 
torsion).
The usual quiver of $\widehat{D_{2n}}$ is well-known \cite{Orb2,9811183};
we give an example for $n=4$ in part (a) of \fref{fig:comparing}.
The quiver is obtained 
by choosing the decomposition of ${\bf 3}\longrightarrow 
\chi_1^1+\chi_2^1$ (as opposed to \eref{decomp-tor} because this is the
linear representation of $\widehat{D_{2n}}$); also
$\gamma(g)$ is in the regular representation of dimension $4n$.
A total of $(n-1)+4=n+3$ nodes results.
We recall that when getting the quiver of $D_{2n}$
with discrete torsion in the above, we chose the decomposition of 
${\bf 3} \longrightarrow\chi_1^3+\chi_2^2$ in \eref{decomp-tor} which
provided a linear representation of $D_{2n}$.
Had we made this same choice for $\widehat{D_{2n}}$, our familiar quiver 
of $\widehat{D_{2n}}$ would have split into two parts: 
one being precisely the quiver of  $D_{2n}$ 
without discrete torsion as discussed in \cite{9909125} and the other, 
that of $D_{2n}$ with discrete
torsion as presented in \fref{fig:dihedral}. 
These are given respectively in
parts (b) and (c) of \fref{fig:comparing}.

From this discussion, we see that in some sense discrete torsion is
connected with different choices of decomposition in the usual orbifold
projection. We want to emphasize that the example of $D_{2n}$ is very
special because its covering group $\widehat{D}_{2n}$ belongs to $SU(2)$.
In general, the covering group does not even belong to $SU(3)$ and the meaning
of the usual orbifold projection of the covering group in string theory is
vague.
\EPSFIGURE[ht]{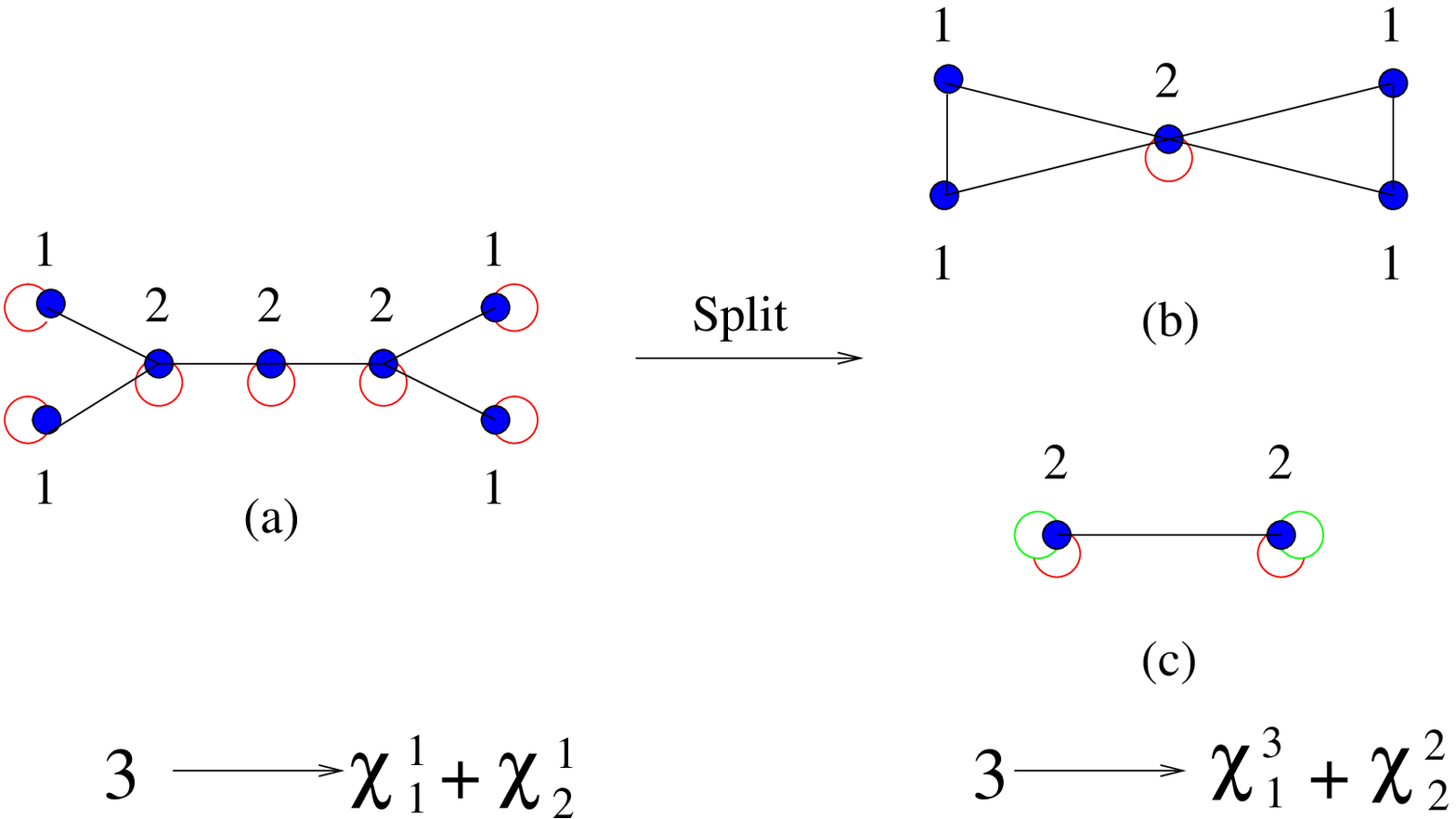,width=4.5in}
{
(a) The quiver diagram of the binary dihedral group 
$\widehat{D}_4$ {\em without}
discrete torsion; (b) the quiver of the ordinary dihedral group 
$D_4$ {\em without}
discrete torsion; (c) the quiver of the ordinary dihedral group 
$D_4$ {\em with} discrete torsion.  
\label{fig:comparing}
}
\section{Conclusions and Prospects}
Let us pause here awhile for reflection. A key purpose of this writing
is to initiate the investigation of discrete torsion for the generic
D-brane orbifold theories. Inspired by this goal, we have shown that
computing the Schur Multiplier $M(G)$ for the finite group $G$
serves as a beacon in our quest.

In particular, using the fact that $M(G)$ is an indicator of when we
can turn on a non-trivial 
NS-NS background in the orbifold geometry and when we cannot: only when
$M(G)$, as an Abelian group is not trivially $\II$ can the former be
executed. As a guide for future investigations, we have computed
$M(G)$ for the discrete subgroups $G$ in $SU(n)$ with $n=2,3,4$, which
amounts to a {\em classification of which D-brane orbifolds afford
non-trivial discrete torsion}.

As an explicit example, in supplementing the present lack of studies
of non-Abelian orbifolds with discrete torsion in the current literature,
we have pursued in detail the ${\cal N}=1$ gauge
theory living on the D3-Brane probe on the orbifold singularity
$\IC^3/D_{2n}$, corresponding to the ordinary dihedral group of order $2n$
as a subgroup of $SU(3)$. As the group has Schur Multiplier $\IZ_2$
for even $n$, we have turned on the discrete torsion and arrived at an
interesting class of non-chiral theories.

The prospects are as manifold as the interests are diverse and much
work remains to be done. An immediate task is
to examine the gauge theory living on the world-volume of D-brane probes
when we turn on the discrete torsion of a given
orbifold wherever allowed by our classification.
This investigation is currently in progress.

Our results of the Schur Multipliers could also be interesting to the
study of K-theory in connexion to string theory.
Recent works \cite{Gomis2,AspinPles,Kapustin2} have noticed an
intimate relation between twisted K-theory and discrete torsion. 
More specifically, the Schur Multiplier of an orbifold group may
in fact supply information about the torsion subgroup of the cohomology
group of space-time in the light of a generalised McKay Correspondence
\cite{AspinPles,9903056}.

It is also tempting to further study the non-commutative moduli
space of non-Abelian orbifolds in the spirit of \cite{DougFiol,BL,BJL}
which treated Abelian cases at great length. How the framework
developed therein extends to the non-Abelian groups should be interesting.
Works on discrete torsion in relation to permutation orbifolds and
symmetric products \cite{Sym} have also been initiated, we hope that our
methodologies could be helpful thereto.

Finally, there is another direction of future study. 
The boundary state formalism was used in
\cite{Gab} where it was suggested that the ties between close and
open string sectors maybe softened with regard to discrete torsion.
It is thus natural to ask if such ambiguities may exist also
for non-Abelian orbifolds.

All these open issues, of concern to the physicist and the
mathematician alike, present themselves to the intrigue of the
reader.
\chapter{Orbifolds IX: Discrete Torsion, Covering Groups and Quiver Diagrams}
\index{Finite Groups!Schur multiplier}
\index{Finite Groups!covering group}
\index{Brane Probes!discrete torsion}
\label{chap:dis2}
\section*{\center{{ Synopsis}}}
Extending the previous chapter and
without recourse to the sophisticated machinery
of twisted group algebras,
projective character tables and explicit values of 2-cocycles, we here
present a simple algorithm to study the gauge theory data of D-brane probes
on a generic orbifold $G$ with discrete torsion turned on.

We show in particular
that the gauge theory can be obtained with the knowledge of no more
than the {\em ordinary} character tables of $G$ and its covering group
$G^*$. Subsequently we present the quiver diagrams of certain
illustrative examples of
$SU(3)$-orbifolds which have non-trivial Schur
Multipliers. This chapter continues with the preceeding
and aims to initiate a systematic and computationally
convenient study of discrete torsion \cite{0011192}.
\section{Introduction}
Discrete torsion \cite{torsion,VafaWit} has become a meeting ground for
many interesting
sub-fields of string theory; its intimate relation with background
B-fields and non-commutative geometry is one of its many salient
features. In the context of D-brane probes on orbifolds with discrete
torsion turned on, new classes of gauge theories may be fabricated and
their (non-commutative) moduli spaces, investigated (see from
\cite{Doug-tor,DougFiol} to \cite{Sym}). Indeed, as it was
pointed out in \cite{Doug-tor,DougFiol}, projection on the matter
spectrum in the 
gauge theory by an orbifold $G$ with non-trivial discrete torsion
is performed by the {\em projective representations} of $G$, 
rather than the mere linear (ordinary) represenations as in the case
without.

In the previous chapter, to which the present shall be a
companion, we offered a classification of the orbifolds with ${\cal
N}=0,1,2$ supersymmetry which permit the turning on of discrete torsion.
We have pointed there that for the orbifold group $G$, the discriminant
agent is the Abelian group known as the {\bf Schur Multiplier} $M(G) :=
H^2(G,\IC^*)$; only if $M(G)$ were non-trivial could $G$ afford a
projective representation and thereby discrete torsion.

In fact one can do more and for actual physical computations one needs
to do more. The standard procedure of calculating the matter content
and superpotential of the orbifold gauge theory as developed in
\cite{LNV} can, as demonstrated in \cite{AspinPles}, be directly
generalised to the case with discrete torsion. Formulae given in terms
of the ordinary characters have their immediate counterparts in terms of
the projective characters, the {\it point d'appui} being that the
crucial properties of ordinary characters, notably orthogonality,
carry over without modification, to the projective case.

And thus our task would be done if we had a method of computing the
projective characters. Upon first glance, this perhaps seems
formidable: one seemingly is required to know the values of the cocycle
representatives $\alpha(x,y)$ in $M(G)$ for all $x,y \in G$. In
actuality, one can dispense with such a need. There exists a canonical
method to arrive at the projective characters, namely by recourse to
the {\bf covering group} of $G$. We shall show in this writing
the methodology standard in the mathematics
literature \cite{Karp,Humph} by which one, once
armed with the Schur Multiplier, arrives
at the cover. Moreover, in light of the physics, we will show how,
equipped with no more than the knowledge of the character table of $G$
and that of its cover $G^*$, one obtains the matter content of
the orbifold theory with discrete torsion.

The chapter is organised as follows. Section 2 introduces the
necessary mathematical background for our work. Due to the
technicality of the details, we present a paragraph at the beginning 
of the section to summarise the useful facts; the reader may then
freely skip the rest of Section 2 without any loss.
In Section 3, we commence with an explicit example, viz., the
ordinary dihedral group, to demonstrate the method to construct the
covering group. Then we present all the covering groups for transitive
and intransitive discrete subgroups of $SU(3)$.
In Section 4, we use these covering groups
to calculate the corresponding gauge 
theories (i.e., the quiver diagrams) for all exceptional subgroups
of $SU(3)$ admitting discrete torsion as well as some examples for the
Delta series. In
particular we demonstrate the algorithm of extracting the quivers from
the ordinary character tables of the group and its cover.
As a by-product, in Section 5 we present a method to calculate the
{\em cocycles} directly which will be useful for future reference.
The advantage of our methods for the quivers and the cocycles is
their simplicity and generality. Finally, in Section 6 we give some
conclusions and further directions for research.
\newpage
\section*{Nomenclature}
Throughout this chapter, unless otherwise specified, we shall adhere to
the following conventions for notation:

\begin{tabular}{rl}
$\omega_n$ & $n$-th root of unity;\\
$G$ & a finite group of order $|G|$;\\
$[x,y]$ &  $:= xyx^{-1}y^{-1}$, the group commutator of $x,y$;\\
$\gen{x_i|y_j}$ & the group generated by elements $\{x_i\}$ with
        relations $y_j$;\\ 
$\gcd(m,n)$ & the greatest common divisor of $m$ and $n$; \\
$Z(G)$ & centre of  $G$;\\
$G':=[G,G]$ & the derived (commutator) group of $G$;\\
$G^*$ & the covering group of $G$;\\
$A=M(G)$ & the Schur Muliplier of $G$;\\
char$(G)$ & ordinary (linear) character table of $G$, given as an
$(r + 1) \times r$ matrix \\
 & with $r$ the \# of conjugacy classes and
	the extra row for class numbers;\\
$Q_\alpha(G,{\cal R})$ & $\alpha$-projective quiver for $G$ associated to
	the chosen representation ${\cal R}$.
\end{tabular}
\section{Mathematical Preliminaries}
We first remind the reader of some properties of the the theory of
projective representations; in what follows we adhere to the notation
used in our previous work \cite{0010023}.

Due to the technicalities in the ensuing, the audience might be distracted
upon the first reading. Thus as promised in the introduction, we here
summarise the keypoints in the next fews paragraphs, so that the
remainder of this section may be loosely perused without any loss.

Our aim of this work is to attempt to construct the gauge theory
living on a D-brane probing an orbifold $G$ when
``discrete torsion'' is turned on. To accomplish such a goal, we need
to know the projective representations of the finite group $G$, which
may not be immediately available.
However, mathematicians have shown that there exists (for
representations in $GL(\IC)$) a
group $G^*$ called the {\em covering group} of $G$, such that
there is a one-to-one correspondence between the projective
representations of $G$ and the linear (ordinary) representations of
$G^*$.
Thus the method is clear: we simply need to find the covering group and
then calculate the ordinary characters of its (linear) representations.

More specificaly, we first introduce the concept of the
covering group in Definition 2.2. Then in Theorem 2.1, we introduce
the necessary and sufficient conditions for $G^*$ to be a covering
group; these conditions are very important and we use them
extensively during actual computations.

However, $G^*$ for any given $G$ is not unique and there
exist non-isomorphic groups which all serve as covering groups.
To deal with this, we introduce {\em isoclinism} and show that these
non-isomorphic covering groups must be isoclinic to each other in
Theorem 2.2. Subsequently, in Theorem 2.3, we give an upper-limit on 
the number of non-isomorphic covering groups of $G$. 
Finally in Thereom 2.4 we present
the one-to-one correspondence of all projective representations of
$G$ and all linear representations of its covering group $G^*$.

Thus is the summary for this section. The uninterested reader
may now freely proceed to Section 3.
\subsection{The Covering Group}
\index{Finite Groups!covering group}
Recall that a {\bf projective representation} of $G$ over
$\IC$ is a mapping $\rho : G \rightarrow GL(V)$ such that 
$\rho(\II_G) = \II_V$
and $\rho(x) \rho(y) = \alpha(x,y) \rho(x y)$
for any elements $x,y
\in G$. The function $\alpha$, known as the {\em cocycle}, is a map 
$G \times G \rightarrow \IC^*$ which is classified by $H^2(G,\IC^*)$,
the second $\IC^*$-valued cohomology of $G$.
This case of $\alpha=1$ trivially is of course our familiar ordinary
(non-projective) representation, which will be called {\bf linear}.

The Abelian group $H^2(G,\IC^*)$ is known as the {\bf Schur
Multiplier} of $G$ and will be
denoted by $M(G)$. Its triviality or otherwise is a discriminant of
whether $G$ admits projective representation. In a physical context,
knowledge of $M(G)$ provides immediate information as to the
possibility of turning on discrete torsion in the orbifold model under
study. A classification of $M(G)$ for all discrete finite 
subgroups of $SU(3)$ and the exceptional subgroups of $SU(4)$
was given in the companion work \cite{0010023}. 

The study of the projective representations of a given group $G$
is greatly facilitated by introducing an auxilliary object $G^*$, the
{\bf covering group} of $G$, which ``lifts projective representations
to linear ones.'' Let us refresh our memory what this means.
Let there be a {\bf central extension} according to the exact sequence
$1 \rightarrow A \rightarrow G^* \rightarrow G \rightarrow 1$ such
that $A$ is in the centre of $G^*$. Thus we have $G^*/A \cong G$. Now
we say 
\begin{definition}
\index{Finite Groups!projective representation}
A projective representation $\rho$ of $G$ {\bf lifts} to a linear
representation $\rho^*$ of $G^*$ if\\ 
(i) $\rho^*(a \in A)$ is proportional to $\II$ and \\
(ii) there is a section\footnote{i.e., for the
        projection $f:G^*\rightarrow G$, $\mu \circ f = \II_G$.}
$\mu : G \rightarrow G^*$ such that $\rho(g) = \rho^*(\mu(g)),~
\forall g \in G$.
\end{definition}
Likewise it {\em lifts projectively} if $\rho(g) =
t(g) \rho^*(\mu(g))$ for a map (not necessarily a homomorphism)
$t:G \rightarrow \IC^*$.
\begin{definition}
$G^*$ is called a {\bf covering group} (or otherwise 
known as the {\bf representation group}, Darstellungsgruppe)
of $G$ over $\IC$ if the following are satisfied:
\begin{description}
\item[(i)] $\exists$ a central extension $1 \rightarrow A \rightarrow G^*
\rightarrow G \rightarrow 1$ such that any projective representation
of $G$ lifts projectively to an ordinary representation of $G^*$;
\item[(ii)] $|A| = |M(G)| = |H^2(G,\IC^*)|$.
\end{description}
\end{definition}

The covering group will play a central r\^{o}le in our work; as we will
show in a moment, {\em the matter content of an orbifold theory
with group $G$ having discrete torsion switched-on is
encoded in the quiver diagram of $G^*$}.

For actual computational purposes, the following theorem, initially
due to Schur, is of extreme importance:
\begin{theorem}\label{cover2}{\rm (\cite{Karp} p143)}
$G^{\star}$ is a covering
group of $G$ over $\IC$ if and only if the following conditions hold:\\
(i) $G^{\star}$ has a finite subgroup $A$ with $A\subseteq
	Z(G^{\star}) \cap [G^{\star},G^{\star}]$;\\
(ii) $G \cong G^{\star}/A$;\\
(iii) $|A|=|M(G)|$.
\end{theorem}
In the above, $[G^{\star},G^{\star}]$ is the {\bf derived group}
$G^{*'}$ of $G^*$. We remind ourselves that for a
group $H$, $H' := [H,H]$ is the group generated by elements of
the form $[x,y] := xyx^{-1}y^{-1}$ for $x,y \in H$.
We stress that conditions (ii) and (iii) are easily satisfied while
(i) is the more stringent imposition.

The solution of the problem of finding covering groups
is certainly {\em not} unique:
$G$ in general may have more than one covering groups (e.g.,
the quaternion and the dihedral group of order 8 are both covering
groups of $\IZ_2 \times \IZ_2$). The problem of finding the necessary
conditions which two groups $G_1$ and $G_2$ must satisfy in order for
both to be covering groups of the same group $G$ is a classical one.

The well-known solution starts with the following
\begin{definition}
\label{isoclinic}
Two groups $G$ and $H$ are said to be {\bf isoclinic} if there exist two 
isomorphisms
\[
\alpha : G/Z(G) \stackrel{\cong}{\rightarrow} H/Z(H) \quad {\rm and} \quad
\beta : G' \stackrel{\cong}{\rightarrow} H'
\]
such that
$\alpha(x_1 Z(G))=x_2 Z(H)~~{\rm and}~~\alpha(y_1 Z(G))=y_2 Z(H)
\Rightarrow \beta([x_1,y_1])= [x_2,y_2],$
\end{definition}
where we have used the standard notation that $x Z(G)$ is a coset
representative in $G/Z(G)$.
We note in passing that every Abelian group is obviously 
isoclinic to the trivial group $\gen{\II}$.

We introduce this concept of isoclinism because of the following
important Theorem of Hall:
\begin{theorem}{\rm (\cite{Karp} p441)}\label{Hall}
Any two covering groups of a given finite group $G$ are isoclinic.
\end{theorem}

Knowing that the covering groups of $G$ are not isomorphic
to each other, but isoclinic, a natural question to ask is how many 
non-isomorphic covering groups can one have. Here a theorem due to
Schur shall be useful:
\begin{theorem}{\rm (\cite{Karp} p149)}
\label{num_cover}
For a finite group $G$, let
\[
G/G' = \IZ_{e_1}\times ...\times \IZ_{e_r}
\]
and
\[
M(G)= \IZ_{f_1}\times ...\times \IZ_{f_s}
\]
be decompositions of these Abelian groups into cyclic factors. 
Then the number of 
non-isomorphic covering groups of $G$ is less than or 
equal to
\[
\prod_{1\leq i\leq r,1\leq j \leq s} \gcd(e_i,f_j).
\]
\end{theorem}
\subsection{Projective Characters}
\index{Finite Groups!characters}
\index{Finite Groups!projective representation}
With the preparatory remarks in the previous subsection, we now delve
headlong into the heart of the matter. By virtue of the construction
of the covering group $G^*$ of $G$ , we have the following 1-1
correpondence which will enable us to compute $\alpha$-projective
representations of $G$ in terms of the linear representations of
$G^*$:
\begin{theorem}{\sf [Theorema Egregium]} 
{\rm (\cite{Karp} p139; \cite{Hoff-Hum} p8)}
Let $G^*$ be the covering group of $G$ and $\lambda :
A \rightarrow \IC^*$ a homomorphism. Then
\begin{description}
\item[(i)] For every linear representation $L : G^* \rightarrow GL(V)$ 
of $G^*$ such that $L(a)=\lambda(a) \II_V~\forall a \in A$, 
there is an induced projective representation $P$ on $G$ defined by
\[
P(g) := L(r(g)), ~\forall~ g \in G,
\]
with $ r : G \rightarrow G^*$ the map that associates to each coset $g \in
G \cong G^*/A $ a representative element\footnote{i.e., $r(g)A
	\rightarrow g$  is the isomorphism $G^*/A
	\stackrel{\cong}{\rightarrow} G$.} in $G^*$; and
vice versa,
\item[(ii)] Every $\alpha$-projective representation for $\alpha \in
M(G)$ lifts to an ordinary representation of $G^*$.
\end{description}
\end{theorem}
An immediate consequence of the above is the fact that knowing the
linear characters of $G^*$ suffices to establish
the projective characters of $G$ for all $\alpha$ \cite{Humph}. 
This should ease
our initial fear in that {\em one does not need to know a priori the
specific values of the cocycles $\alpha(x,y)$ for all $x,y \in G$ (a
stupendous task indeed) in order to construct the $\alpha$-projective
character table for $G$.}

We shall leave the uses of this crucial observation to later
discussions. For now, let us focus on some explicit computations of
covering groups.
\index{Finite Groups!covering group}
\section{Explicit Calculation of Covering Groups}
To theory we must supplant examples and to abstraction, concreteness.
We have prepared ourselves in the previous section the rudiments of
the theory of covering groups; in the present section we will
demonstrate these covers for the discrete finite subgroups of
$SU(3)$. First we shall illustrate our techniques with the case of
$D_{2n}$, the ordinary dihedral group, before tabulating the complete
results.
\subsection{The Covering Group of The Ordinary Dihedral Group}
The presentation of the ordinary dihedral group of order $2n$ is
standard (the notation is different from some of our earlier chapters
where the following would be called $D_n$):
\[
D_{2n}= \gen{\tilde{\alpha},\tilde{\beta} | \tilde{\alpha}^n=1, \tilde{\beta}^2=1, 
	\tilde{\beta} \tilde{\alpha} \tilde{\beta}^{-1} = \tilde{\alpha}^{-1}}.
\]
We recall from \cite{0010023} that the Schur Multiplier for $G = D_{2n}$ is
$\IZ_2$ when $n$ is even and trivial otherwise, thus we restrict
ourselves only to the case of $n$ even.
We let $M(D_{2n})$ be $A = \IZ_2$ generated by $\{ a |
 a^2=\II\}$. We let the covering group be $G^* =
\gen{\alpha,\beta, a}$.

Now having defined the generators we proceed to constrain relations
thereamong. Of course, $A \subset Z(G^*)$ immediately implies that
$\alpha a= a \alpha$ and
$\beta a= a \beta$. Moreover, $\alpha,\beta$ must map
to $\tilde{\alpha},\tilde{\beta}$ when we identify $G^{\star}/A \cong
D_{2n}$  (by part (ii) of Theorem \ref{cover2}). This means that $\II_G$
must have a preimage in $A \subset G^*$, giving us: $\alpha^n \in
A, \beta^2 \in A$ and $\beta \alpha \beta^{-1} \alpha \in A$ by virtue
of the presentation of $G$. And hence we have 8 possibilities, each
being a central extension of $D_{2n}$ by $A$:
\begin{equation}
\label{extension}
\ba{l}
G^*_1=\gen{\alpha,\beta, a |\alpha a= a\alpha,~~
\beta a= a\beta,~~  a^2=1,~~ \alpha^n=1,~~
\beta^2=1,~~ \beta \alpha \beta^{-1}= \alpha^{-1}}
\\
G^*_2=\gen{\alpha,\beta, a |\alpha a= a\alpha,~~
\beta a= a\beta,~~  a^2=1,~~ \alpha^n=1,~~
\beta^2=1,~~ \beta \alpha \beta^{-1}= \alpha^{-1} a}
\\
G^*_3=\gen{\alpha,\beta, a |\alpha a= a\alpha,~~
\beta a= a\beta,~~  a^2=1,~~ \alpha^n=1,~~
\beta^2= a,~~ \beta \alpha \beta^{-1}= \alpha^{-1}}
\\
G^*_4=\gen{\alpha,\beta, a |\alpha a= a\alpha,~~
\beta a= a\beta,~~  a^2=1,~~ \alpha^n=1,~~
\beta^2= a,~~ \beta \alpha \beta^{-1}= \alpha^{-1}
 a}
\\
G^*_5=\gen{\alpha,\beta, a |\alpha a= a\alpha,~~
\beta a= a\beta,~~  a^2=1,~~ \alpha^n= a,~~
\beta^2=1,~~ \beta \alpha \beta^{-1}= \alpha^{-1}}
\\
G^*_6=\gen{\alpha,\beta, a |\alpha a= a\alpha,~~
\beta a= a\beta,~~  a^2=1,~~ \alpha^n= a,~~
\beta^2=1,~~ \beta \alpha \beta^{-1}= \alpha^{-1} a}
\\
G^*_7=\gen{\alpha,\beta, a |\alpha a= a\alpha,~~
\beta a= a\beta,~~  a^2=1,~~ \alpha^n= a,~~
\beta^2= a,~~ \beta \alpha \beta^{-1}= \alpha^{-1}}
\\
G^*_8=\gen{\alpha,\beta, a |\alpha a= a\alpha,~~
\beta a= a\beta,~~  a^2=1,~~ \alpha^n= a,~~
\beta^2= a,~~ \beta \alpha \beta^{-1}= \alpha^{-1} a}
\ea
\end{equation}

Therefore, conditions (ii) and (iii) of Theorem \ref{cover2} are
satified. One must check (i) to distinguish the covering group among
these 8 central extensions in \eref{extension}. Now since $A$ is
actually the centre, it suffices to check whether $A \subset
G^{*'}_i = [G^*_i,G^*_i]$.

We observe $G^*_1$ to be precisely $D_{2n}\times \IZ_2$, from which we
have $G^{*'}_1 = \IZ_{n/2}$, generated by $\alpha^2$. 
Because $A = \{\II,a\}$ clearly is not included in this $\IZ_{n/2}$
we conclude that $G^*_1$ is not the covering group. 
For $G^*_2$, we have $G^{*'}_2 = \gen{\alpha^2 a}$, which means that
when $n/2=\odd$ (recall that $n=\even$), $G^{*'}_2$ can contain $a$
and hence $A \subset G^{*'}_2$, whereby making $G^*_2$ a covering group. 
By the same token we find that
$G^{*'}_3 = \gen{\alpha^2}$,
$G^{*'}_4 = \gen{\alpha^2 a}$,
$G^{*'}_5 = \gen{\alpha^2}$,
$G^{*'}_6 = \gen{\alpha^2 a}$, and
$G^{*'}_7 = \gen{\alpha^2}$.
We summarise these results in the following table:
\[
\ba{c|c|c|c|c}
{\rm Group}  & G^{*'} & Z(G^*) & G^*/Z(G^*) & \mbox{Covering Group?}\\ \hline
G^*_1  & \IZ_{n/2}=\gen{\alpha^2 } & \IZ_2\times \IZ_2=\gen{a,\alpha^{n/2} }
& D_{n} & $no$ \\
G^*_2(n=4k+2) & \IZ_n=\gen{\alpha^2a } & \IZ_2=\gen{ a } & D_{2n} &
$yes$ \\
G^*_2(n=4k)  & \IZ_{n/2}=\gen{ \alpha^2a } & 
\IZ_2\times \IZ_2=\gen{ a,\alpha^{n/2} }& D_{n}& $no$ \\
G^*_3  & \IZ_{n/2}=\gen{ \alpha^2} & \IZ_2\times
\IZ_2=\gen{a,\alpha^{n/2} }& D_{n} & $no$ \\
G^*_4(n=4k+2)  & \IZ_n=\gen{ \alpha^2a} & \IZ_2=\gen{ a }& D_{2n} & $yes$
\\
G^*_4(n=4k)  & \IZ_{n/2}=\gen{ \alpha^2a} & \IZ_2\times \IZ_2=\gen
{ a,\alpha^{n/2} } & D_{n} & $no$ \\
G^*_5  & \IZ_n=\gen{\alpha^2 } &  \IZ_2=\gen{ a }& D_{2n} & $yes$ \\
G^*_6(n=4k+2)  & \IZ_{n/2}=\gen{\alpha^2a} & \IZ_4=\gen{\alpha^{n/2} }
& D_{n} & $no$ \\
G^*_6(n=4k)  & \IZ_n=\gen{\alpha^2a} & \IZ_2=\gen{ a } & D_{2n} & $yes$\\
G^*_7  & \IZ_n=\gen{\alpha^2 } &  \IZ_2=\gen{ a }& D_{2n} & $yes$\\
G^*_8(n=4k+2)  & \IZ_{n/2}=\gen{ \alpha^2a }&  \IZ_4=\gen{\alpha^{n/2} }
 & D_{n}  & $no$ \\
G^*_8(n=4k)  & \IZ_n=\gen{ \alpha^2a }&  \IZ_2=\gen{ a }& D_{2n} & $yes$\\
\ea
\]
Whence we see that $G^*_1$ and $G^*_3$ are not
covering groups, while for $n/2 = \odd$ $G^*_{2,4}$ are covers, for
$n/2 = \even$ $G^*_{6,8}$ are covers as well 
and finally $G^*_{5,7}$ are always
covers. Incidentally, $G^*_7$ is actually the binary dihedral group and
we know that it is indeed the (double) covering group from \cite{0010023}.
Of course in accordance with Theorem \ref{Hall}, these different
covers must be isoclinic to each other. Checking against Definition
\ref{isoclinic}, we see that for $G^*$ being $G^*_{2,4}$ with $n=4k+2$,
$G^*_{6,8}$ with $n = 4k$ and $G^*_{5,7}$ for all even $n$, $G^{*'} \cong
\IZ_n$ and $G^*/Z(G^*) \cong D_{2n}$; furthermore the isomorphisms
$\alpha$ and $\beta$ in the Definition are easily seen to satisfy the
prescribed conditions. Therefore all these groups are indeed isoclinic.
We make one further remark, for both the 
cases of $n=4k$ and $n=4k+2$, we have found 4
non-isomorphic covering groups. Recall Theorem \ref{num_cover},
here we have $f_1=2$ and $G/G'=\IZ_2\times \IZ_2$ (note that
$n$ is even) and so $e_1=e_2=2$, whence the upper limit is exactly
$2\times 2=4$ which is saturated here. This demonstrates that our method
is general enough to find all possible covering groups.
\newpage
\enlargethispage*{1000pt}
\subsection{Covering Groups for the Discrete Finite Subgroups of
	$SU(3)$}
By methods entirely analogous to the one presented in the above
subsection, we can arrive at the covering groups for the discrete
finite groups of $SU(3)$ as tabulated in \cite{0010023}. 
Let us list the results (of course in comparison with Table 3.2
in \cite{0010023}, those with trivial Schur Multipliers have no
covering groups and will not be included here).
Of course, as mentioned earlier, the covering group is not unique. The
particular ones we have chosen in the following table are the same as
generated by the computer package GAP using the Holt algorithm \cite{Prog}.
\subsection*{Intransitives}
We used the shorthand notation $(x/y/\ldots/z)$ to mean the
relation to be applied to each of the elements $x,y,\ldots,z$.
\beq 
\label{ZZ}
\hspace{-1.8in}
\ba{ll}
\bullet \quad G = & \IZ_m \times \IZ_n = 
	\gen{\tilde{\alpha},\tilde{\beta} | 
	\tilde{\alpha}^n=1, \tilde{\beta}^m=1,
	\tilde{\alpha}\tilde{\beta}=\tilde{\beta}\tilde{\alpha}};\\
	& M(G) = \IZ_{p=\gcd(m,n)} = \gen{a | a^p=\II}; \\
	&  G^* = \gen{\alpha, \beta,  a |\alpha  a= a\alpha,
	\beta a=   a\beta,    a^p=1, \alpha^n=1,
	\beta^m=1, \alpha \beta =\beta \alpha  a}\\
\ea
\eeq
\beq \label{ZbinD}
\hspace{-0.6in}
\ba{ll}
\bullet \quad  G = & \gen{\IZ_{n=4k} , \widehat{D_{2m}}} = 
	\gen{\tilde{\alpha},\tilde{\beta},\tilde{\gamma}|
	\tilde{\alpha}\tilde{\beta}=\tilde{\beta}\tilde{\alpha},
	\tilde{\alpha}\tilde{\gamma}=\tilde{\gamma} \tilde{\alpha},
	\tilde{\alpha}^{n/2}=\tilde{\beta}^m, 
	\tilde{\beta}^{2m}=1, \tilde{\beta}^m=\tilde{\gamma}^2,
	\tilde{\gamma}\tilde{\beta}\tilde{\gamma}^{-1}=\tilde{\beta}^{-1}};\\
	& \left\{
	\ba{ll}
	m~\even \quad & M(G) = \IZ_2 \times \IZ_2=\gen{a,b|a^2=1=b^2, ab=ba};\\
		& \ba{ll}
		  G^* = & \langle 
			\alpha, \beta, \gamma,a,b| ab=ba,
			\alpha a=a \alpha, \alpha b=b \alpha, 
			\beta a= a \beta, \beta b= b \beta,\\
			& \gamma a=a \gamma, \gamma b=b \gamma, a^2=1=b^2,
				\alpha\beta =\beta\alpha a, 
				\alpha \gamma=\gamma  \alpha b,\\
			& \alpha^{n/2}=\beta^m, \beta^{2m}=1, 
				\beta^m=\gamma^2, \gamma \beta
				\gamma^{-1}=\beta^{-1} \rangle
		 \ea \\
	m~\odd,  \quad & M(G) = \IZ_2=\gen{a | a^2=1 };\\
		& \ba{ll}
		G^* = & \langle
		\alpha, \beta, \gamma,a | a^2=1,\alpha a=a \alpha,
		\beta a= a \beta,\gamma a=a \gamma, \alpha\beta
		=\beta\alpha, \\ 
		& \alpha \gamma=\gamma  \alpha a, 
		 \alpha^{n/2}=\beta^m, \beta^{2m}=1, \beta^m=\gamma^2, 
		\gamma \beta \gamma^{-1}=\beta^{-1} \rangle
		\ea \\
	\ea
	\right.
\ea
\eeq
\beq \label{ZE7}
\hspace{-1in}
\ba{ll}
\bullet \quad G = & \gen{\IZ_{n=4k},\widehat{E_7}} =
\gen{
\tilde{\alpha},\tilde{\beta},\tilde{\gamma}|
\tilde{\alpha}\tilde{\beta}=\tilde{\beta}\tilde{\alpha},
\tilde{\alpha}\tilde{\gamma}=\tilde{\gamma} \tilde{\alpha},
\tilde{\alpha}^{n/2}=\tilde{\beta}^4, \tilde{\beta}^4=\tilde{\gamma}^3=
(\tilde{\beta}\tilde{\gamma})^2
};\\
& M(G) = \IZ_2 = \gen{a | a^2=\II}; \\
& \ba{ll}
G^* = & \langle \alpha, \beta, \gamma,a |  a^2=1,\alpha a=a \alpha,
\beta a= a \beta,\gamma a=a \gamma, \alpha^{n/2}=\beta^4, \\ 
& \alpha \beta =\beta \alpha a, \alpha \gamma=\gamma \alpha, 
\beta^4=\gamma^3 = (\beta \gamma)^2  \rangle
\ea \ea \vspace{-1.0in} \eeq
\newpage
\enlargethispage*{1000pt}
\beq \label{ZE6}
\hspace{-0.5in}
\ba{lll}
\bullet \quad G = & \gen{\IZ_{n=3k},\widehat{E_6}} & \\
& k~\odd
&
G \cong \IZ_n \times \widehat{E_6} = 
\gen{\tilde{\alpha},\tilde{\beta},\tilde{\gamma}|
\tilde{\alpha}\tilde{\beta}=\tilde{\beta}\tilde{\alpha},
\tilde{\alpha}\tilde{\gamma}=\tilde{\gamma} \tilde{\alpha},
\tilde{\alpha}^n=1, \tilde{\beta}^3=\tilde{\gamma}^3=
(\tilde{\beta}\tilde{\gamma})^2}; \\
& & M(G) = \IZ_3 = \gen{a | a^3=\II}; \\
& & \ba{ll}
  G^* = & \langle
\alpha, \beta, \gamma,a | a^3=1,\alpha a=a \alpha,
\beta a= a \beta,\gamma a=a \gamma, \alpha^n=1, \\ 
& \alpha\beta =\beta\alpha a^{-1}, \alpha \gamma=\gamma  \alpha a,
\beta^3=\gamma^3=(\beta\gamma)^2 \rangle
\ea \\
& k = 2(2p+1) & G \cong \IZ_{n/2} \times \widehat{E_6} \\
& k = 4p & G \cong (\IZ_{n} \times \widehat{E_6})/\IZ_2 = \\
& &
\gen{\tilde{\alpha},\tilde{\beta},\tilde{\gamma}|
\tilde{\alpha}\tilde{\beta}=\tilde{\beta}\tilde{\alpha},
\tilde{\alpha}\tilde{\gamma}=\tilde{\gamma} \tilde{\alpha},
\tilde{\alpha}^{n/2}=\tilde{\beta}^3, \tilde{\beta}^3=\tilde{\gamma}^3=
(\tilde{\beta}\tilde{\gamma})^2}; \\
& & M(G) = \IZ_3 = \gen{a | a^3=\II}; \\
& & \ba{ll}
  G^* = & \langle
\alpha, \beta, \gamma,a | a^3=1,\alpha a=a \alpha,
\beta a= a \beta,\gamma a=a \gamma, \alpha^{n/2}=\beta^3, \\ 
& \alpha\beta =\beta\alpha a^{-1}, \alpha \gamma=\gamma  \alpha a,
\beta^3=\gamma^3=(\beta\gamma)^2 \rangle
\ea \\
\ea\eeq
\beq \label{ZD}
\hspace{-0.3in}
\ba{lll}
\bullet \quad  G = & \gen{\IZ_n,  D_{2m}} & \\  
& 
n~\odd, m~\even  & G = \IZ_n \times   D_{2m}=
\langle
\tilde{\alpha},\tilde{\beta},\tilde{\gamma}| \tilde{\alpha}^n
=1,
\tilde{\alpha}\tilde{\beta}=\tilde{\beta}\tilde{\alpha},
\tilde{\alpha}\tilde{\gamma}=\tilde{\gamma}\tilde{\alpha},
\tilde{\beta}^m=1, \\
&&\qquad \qquad
\tilde{\gamma}^2=1, 
\tilde{\gamma} \tilde{\beta}\tilde{\gamma}^{-1}=\tilde{\beta}^{-1}
\rangle;\\
&  & M(G) = \IZ_2 = \gen{a|a^2= 1};\\
& &\ba{ll}
G^*= & \langle \alpha, \beta, \gamma,a |  a^2=1,
a (\alpha/\beta/\gamma)=(\alpha/\beta/\gamma) a, 
\alpha (\beta/\gamma)= \\ &
(\beta/\gamma) \alpha, \alpha^n=1,
\beta^m=a, \gamma^2=1, \gamma \beta \gamma^{-1}=\beta^{-1} \rangle
\ea \\
& n~\even,m~\odd & G = \IZ_n \times   D_{2m} \\
& & M(G) = \IZ_2 = \gen{a|a^2= 1};\\
& & \ba{ll}
G^* = & \langle \alpha, \beta, \gamma,a |  a^2=1,
a (\alpha/\beta/\gamma)=(\alpha/\beta/\gamma) a, 
\alpha \beta=\beta \alpha, \\ &
\alpha \gamma =\gamma \alpha a, \alpha^n=1,
\beta^m=1, \gamma^2=1, \gamma \beta \gamma^{-1}=\beta^{-1}
\rangle \ea \\
& m~\even, n=2(2l+1) & G = \IZ_{n/2} \times   D_{2m} \\
& n=4k, m=2(2l+1)  & G = (\IZ_{n} \times   D_{2m})/\IZ_2=
\langle
\tilde{\alpha},\tilde{\beta},\tilde{\gamma}| \tilde{\alpha}^{n/2}
=\beta^{m/2},
\tilde{\alpha}\tilde{\beta}=\tilde{\beta}\tilde{\alpha},
\tilde{\alpha}\tilde{\gamma}=\tilde{\gamma}\tilde{\alpha},\\
&&\qquad \qquad
\tilde{\beta}^m=1, \tilde{\gamma}^2=1, 
\tilde{\gamma} \tilde{\beta}\tilde{\gamma}^{-1}=\tilde{\beta}^{-1}
\rangle;\\
&  & M(G) = \IZ_2 = \gen{a|a^2= 1};\\
& &\ba{ll}
G^*= & \langle \alpha, \beta, \gamma,a |  a^2=1,
a (\alpha/\beta/\gamma)=(\alpha/\beta/\gamma) a, 
\alpha \beta=\beta \alpha, \\ &
\alpha \gamma=\gamma \alpha a,
\alpha^{n/2}=\beta^{m/2},
 \beta^m=1, \gamma^2=1, \gamma \beta \gamma^{-1}=\beta^{-1} \rangle
\ea \\
& n=4k, m=4l  & G = (\IZ_{n} \times   D_{2m})/\IZ_2 \\
&  & M(G) = \IZ_2 \times \IZ_2 = \gen{a,b|a^2= 1, b^2=1, ab=ba};\\
& &\ba{ll}
G^*= & \langle \alpha, \beta, \gamma,a,b |  a^2=1,
a (\alpha/\beta/\gamma)=(\alpha/\beta/\gamma) a, 
\alpha \beta=\beta \alpha b, \\ &
\alpha \gamma=\gamma \alpha a,
\alpha^{n/2}=\beta^{m/2},
 \beta^m=1, \gamma^2=1, \gamma \beta \gamma^{-1}=\beta^{-1} \rangle
\ea \\
\ea \eeq
\newpage
\enlargethispage*{1000pt}
{\vspace{-0.5in}}
{\bf Transitives}

We first have the two infinite series.
\beq \label{del3}
\ba{ll}
\bullet \quad G = & \Delta(3n^2) = \gen{\alpha,\beta,\gamma |
	\alpha^n = \beta^n = \gamma^3 = 1,
        \alpha \beta = \beta \alpha,
        \alpha \gamma =  \gamma \alpha^{-1} \beta,
        \beta \gamma \alpha = \gamma};\\
& \left\{ \ba{ll}
\gcd(n,3) = 1, n~\even \quad & M(G) = \IZ_n = \gen{a | a^n=1};\\
	& \ba{ll} G^* = & \langle
	\alpha, \beta, \gamma, a | 
	(\alpha/\beta/\gamma) a = a(\alpha/\beta/\gamma),\\
	&
         a^n = \alpha^n a^{n/2} = \beta^n a^{n/2} = \gamma^3 = 1,\\
	&
        \alpha \beta = \beta \alpha a,
        \alpha \gamma = \gamma \alpha^{-1} \beta,
        \beta \gamma \alpha = \gamma
		\rangle; \ea \\
\gcd(n,3) = 1, n~\odd \quad & M(G) = \IZ_n;\\
	& \ba{ll} G^* = & \langle
	\alpha, \beta, \gamma, a | 
	(\alpha/\beta/\gamma) a = a(\alpha/\beta/\gamma),\\
	&
         a^n = \alpha^n = \beta^n = \gamma^3 = 1,\\
	&
        \alpha \beta = \beta \alpha a,
        \alpha \gamma = \gamma \alpha^{-1} \beta,
        \beta \gamma \alpha = \gamma
		\rangle; \ea \\
\gcd(n,3) \ne 1, n~\even \quad & M(G) = \IZ_n \times \IZ_3 = 
			\gen{a,b|a^n=1,b^3=1};\\
	& \ba{ll} G^* = & \langle
	\alpha, \beta, \gamma, a, b | 
	(\alpha/\beta/\gamma)(a/b) = (a/b)(\alpha/\beta/\gamma),\\
	&
        ab=ba, a^n=b^3 = \gamma^3 =\alpha^n a^{n/2} b = 1, \\
	&
        \beta^n a^{n/2}= b,
        \alpha \beta = \beta \alpha a b,
        \alpha \gamma = \gamma \alpha^{-1} \beta,
        \beta \gamma \alpha = \gamma
		\rangle; \ea \\
\gcd(n,3) \ne 1, n~\odd \quad & M(G) = \IZ_n \times \IZ_3; \\
	& \ba{ll} G^* = & \langle
	\alpha, \beta, \gamma, a, b |
	(\alpha/\beta/\gamma)(a/b) = (a/b)(\alpha/\beta/\gamma),\\
	&
	a^n=b^3 = \gamma^3 =\alpha^n b = \beta^n b^{-1}= 1, \\
	&
	ab=ba,
        \alpha \beta = \beta \alpha a b,
        \alpha \gamma = \gamma \alpha^{-1} \beta,
        \beta \gamma \alpha = \gamma
		\rangle; \ea \\
\ea
\right.
\ea
\eeq

\beq \label{del6}
\ba{ll}
\bullet \quad G = & \Delta(6n^2) = 
	\langle \alpha,\beta,\gamma,\delta |
	\alpha^n = \beta^n = \gamma^3 = \delta^2 = 1,
        \alpha \beta = \beta \alpha,
        \alpha \gamma = \gamma \alpha^{-1} \beta,
        \beta \gamma \alpha = \gamma,\\
	&
	\qquad \qquad \qquad
	\alpha \delta \alpha = \delta,
        \beta \delta = \delta \alpha^{-1} \beta,
        \gamma \delta \gamma = \delta
	\rangle;\\
	& M(G) = \IZ_2 = \gen{a | a^2 = 1};\\
& \left\{ \ba{ll}
\ba{ll} \gcd(n,4) = 4 \quad G^* = & \langle
	\alpha, \beta, \gamma, \delta, a |
	\alpha^n = \beta^n = \gamma^3 = \delta^2 = a^2 = 1,\\
	&
        ~~(\alpha/\beta/\gamma/\delta) a = a (\alpha/\beta/\gamma/\delta),
	\alpha \beta = \beta \alpha a,
        \alpha \gamma = \gamma \alpha^{-1} \beta,\\
	&
        ~~\beta \gamma \alpha = \gamma,
        \alpha \delta \alpha = \delta,
        \beta \delta = \delta \alpha^{-1} \beta,
        \gamma \delta \gamma = \delta
	\rangle; \ea \\
\ba{ll} \gcd(n,4) = 2 \quad G^* = & \langle
	\alpha, \beta, \gamma, \delta, a |
	\alpha^n a = \beta^n a = \gamma^3 = \delta^2 = a^2 = 1,\\
	&
        ~~(\alpha/\beta/\gamma/\delta) a = a (\alpha/\beta/\gamma/\delta),
	\alpha \beta = \beta \alpha a,
        \alpha \gamma = \gamma \alpha^{-1} \beta,\\
	&
	~~\beta \gamma \alpha = \gamma,
        \alpha \delta \alpha = \delta,
        \beta \delta = \delta \alpha^{-1} \beta,
        \gamma \delta \gamma = \delta
	\rangle; \ea \\
\ea
\right.
\ea
\eeq
\newpage
Next we present the three exceptionals that admit discrete torsion.
\beq \label{sig60}
\hspace{-2.2in}
\ba{ll}
\bullet \quad G = & \Sigma(60) \cong A_5 =
	\langle \alpha, \beta |
	\alpha^5 = \beta^3 = (\alpha \beta^{-1})^3 = (\alpha^2
	\beta)^2 = 1\\
	&
	\qquad \qquad \qquad \qquad \alpha \beta \alpha \beta \alpha \beta =  
	\alpha \gamma \alpha^{-1} \beta \alpha^2 \beta \alpha^{-2}
	\beta = 1
	\rangle;\\
	& M(G) = \IZ_2; \\
	& G^* = \langle \alpha, \beta, a|
	 \alpha^5 = a, \beta^3 = a^2 =1, (\alpha/\beta)a =
	a(\alpha/\beta)\\
	&
	\qquad \qquad (\alpha \beta^{-1})^3 = 1, (\alpha^2
	\beta)^2 = a
	\rangle; \\

\ea
\eeq

\beq \label{sig168}
\hspace{-0.5in}
\ba{ll}
\bullet \quad G = & \Sigma(168) = \gen{ \alpha, \beta, \gamma |
	\gamma^2 = \beta^3 = \beta \gamma \beta \gamma =
	(\alpha\gamma)^4 = 1,
	\alpha^2 \beta  = \beta \alpha, 
	\alpha^3 \gamma \alpha^{-1} \beta = \gamma \alpha \gamma 
	};\\
	& M(G) = \IZ_2; \\
	& G^*  = \langle \alpha, \beta, \gamma, \delta |
	\delta^2 = \gamma^2 \delta = \beta^3 \delta = (\beta \alpha)^3
	= (\alpha \gamma)^3 = 1, \\
	& \qquad \qquad
	\beta \gamma \beta = \gamma, \alpha \delta = \delta \alpha, 
	\beta^2 \alpha^2 \beta = \alpha, 
	\beta^{-1} \alpha^{-1} \beta \gamma \alpha^{-1} \gamma =
	\gamma \alpha \beta
	\rangle;\\
\ea
\eeq

\beq \label{sig1080}
\hspace{-0.5in}
\ba{ll}
\bullet \quad G = & \Sigma(1080) = \langle \alpha, \beta, \gamma, \delta |
	\alpha^5 = \beta^2 = \gamma^2 = \delta^2 = (\alpha \beta)^2 
	(\beta \gamma)^2 = (\beta \delta)^2 = 1,\\
	& \qquad \qquad \qquad \qquad 
	(\alpha \gamma)^3 = (\alpha \delta)^3 = 1, 
	\gamma \beta = \delta \gamma \delta, 
	\alpha^2 \gamma \beta \alpha^2 = \gamma \alpha^2 \gamma
	\rangle; \\
	& M(G) = \IZ_2; \\
	& G^* = \langle \alpha, \beta, \gamma, \delta, \epsilon |
	\alpha^5 = \epsilon^2 = \gamma^2 \epsilon^{-1} = \beta^2 \epsilon^{-1} 
	= \delta^2 \epsilon^{-1} = (\alpha \delta)^3 = 1, \\
	& \qquad \qquad \qquad \qquad
	\alpha^{-1} \epsilon \alpha = \beta^{-1} \epsilon \beta
	= \gamma^{-1} \epsilon \gamma
	= \delta^{-1} \epsilon \delta = \epsilon, \\
	& \qquad \qquad \qquad \qquad
	(\alpha \beta)^2 = (\beta \gamma)^2 = (\beta \delta)^2 =
	\gamma \beta \delta \gamma \delta =
	(\alpha \gamma)^3 = \epsilon,\\
	& \qquad \qquad \qquad \qquad
	\alpha^2 \gamma \beta \alpha^2 \gamma \alpha^{-2} \gamma=1
	\rangle;\\
\ea
\eeq
\section{Covering Groups, Discrete Torsion and Quiver Diagrams}
\index{Brane Probes!discrete torsion}
\index{Finite Groups!covering group}
\subsection{The Method}
The introduction of the host of the above concepts is not without a
cause. In this section we shall provide an {\bf algorithm} which
permits the construction of the quiver $Q_{\alpha}(G,{\cal R})$ of an
orbifold theory with group $G$ having discrete torsion $\alpha$
turned-on, and with a linear representation ${\cal R}$ of $G$ acting 
on the transverse space. 

Our method dispenses of the need of the knowledge of the cocycles
$\alpha(x,y)$, which in general is a formidable task from the
viewpoint of cohomology, but which the
current literature may lead the reader to believe to be required for
finding the projective representations. We shall demonstrate that the
problem of finding these $\alpha$-representations is reducible to the far
more manageable duty of finding the covering group, constructing its character
table (which is of course straightforward) and then applying the usual
prodecure of extracting the quiver therefrom.
One advantage of this method is that we immediately obtain 
the quiver for all cocycles (including the trivial cocycle
which corresponds to having no discrete torsion at all) and in fact
the values of $\alpha(x,y)$ (which we shall address in the next
section) in a unified framework. 

All the data we require are\\
(i) $G$ and its (ordinary) character table char$(G)$;\\
(ii) The covering group $G^*$ of $G$ and its (ordinary) character
	table char$(G^*)$.

We first recall from \cite{Doug-tor,DougFiol} that turning on discrete torsion
$\alpha$ in an orbifold projection amounts to using an
$\alpha$-projective representation\footnote{More rigorously, this
	statement holds only when the D-brane probe is pointlike in the
        orbifold directions. More generally, when D-brane probes
	extend along the orbifold directions, one may need to use 
	ordinary as well as projective 
	representations. For further details, please refer to
	\cite{Gab} as well as \cite{Craps1}.} 
 $\Gamma_{\alpha}$ of $g \in G$
\beq
\label{proj-tor}
\Gamma_{\alpha}(g) \cdot A \cdot \Gamma^{-1}_{\alpha}(g) = A,
\qquad
\Gamma_{\alpha}(g) \cdot \Phi \cdot \Gamma^{-1}_{\alpha}(g) 
= {\cal R}(g) \cdot \Phi
\eeq
on the gauge field $A$ and matter fields $\Phi$.

The above equations have been phrased in a more axiomatic setting (in
the language of \cite{LNV}), by virtue of the fact that much of
ordinary representation theory of finite group extends in direct
analogy to the projective case, in \cite{AspinPles}.
{\em However, we hereby emphasize that with the aid of the
linear representation of the covering
group, one can perform orbifold projection with discrete torsion
entirely in the setting of \cite{LNV} without usage of the
formulae in \cite{AspinPles} generalised to twisted group algebras and
modules.}
In other words, if we use the matrix of the linear representation of $G^*$
instead of that of the corresponding projective representation of $G$, we will
arrive at the same gauge group and matter contents in the orbifold
theory. This can be alternatively shown as follows.

When we lift the
projective matrix representation of $G$ into the linear one of
$G^*$, every matrix $\rho(g)$ will 
map to $\rho(ga_i)$ for every $a_i\in A$. The crucial 
fact is that $\rho(ga_i)=\lambda_i \rho(g)$ 
where $\lambda_i$ is simply a phase factor. 
Now in \eref{proj-tor} (cf. Sections 4.2 and 5 for more details),
$\Gamma_{\alpha}(g)$ and $\Gamma^{-1}_{\alpha}(g)$
always appear in pairs, when we replace them by
$\Gamma(ga_i)$ and $\Gamma^{-1}(ga_i)$,
the phase factor $\lambda_i$ will cancel out and leave the
result invariant. This shows that the two results,
the one given by projective matrix representations of $G$
and the other by linear matrix representations of $G^*$,
will give identical answers in orbifold projections.
\subsection{An Illustrative Example: $\Delta(3\times3^2)$}
Without much further ado, an illustrative example of the group
$\Delta(3\times3^2) \in SU(3)$ shall serve to enlighten the reader. We
recall from \eref{del3} that this group of order 27 has presentation
$\gen{\alpha,\beta,\gamma | \alpha^3 = \beta^3 = \gamma^3 = 1,
\alpha \beta = \beta \alpha,\alpha \gamma = \gamma \alpha^{-1} \beta,\beta
\gamma \alpha = \gamma}$ and its covering group of order 243 (since
the Schur Multiplier is $\IZ_3 \times \IZ_3$) is $G^* = \gen{\alpha,
\beta, \gamma, a, b | (\alpha/\beta/\gamma)(a/b) =
(a/b)(\alpha/\beta/\gamma), a^3=b^3 = \gamma^3 =\alpha^3 b = 
\beta^3 b^{-1}= 1,
ab=ba, \alpha \beta = \beta \alpha a b, \alpha \gamma = \gamma 
\alpha^{-1} \beta,
\beta \gamma \alpha = \gamma}$.

Next we require the two (ordinary) character tables. As pointed out in
the Nomenclatures section, character tables are given as $(r+1) \times
r$ matrices with $r$ being the number of conjugacy classes (and equivalently
the number of irreps), and the first row giving the conjugacy class numbers.
\beq 
\label{del3n3}
{\rm char}(\Delta(3\times3^2)) = 
{\tiny
\ba{|c|c|c|c|c|c|c|c|c|c|c|}
\hline
1 & 1 & 1 & 3 & 3 & 3 & 3 & 3 & 3 & 3 & 3 \\ \hline 1 & 1 & 1 & 1 & 1 & 1 & 1 & 1 & 1 & 1 & 1 \\ \hline 1 & 1 & 1 & 1 & 1 & \omega_3 & \omega_3 & \omega_3 & {\bar{\omega}_3} & 
   {\bar{\omega}_3} & {\bar{\omega}_3} \\ \hline 1 & 1 & 1 & 1 & 1 & {\bar{\omega}_3} & {\bar{\omega}_3} & {\bar{\omega}_3} & \omega_3 & \omega_3 & \omega_3 \\ \hline 1 & 1 & 1 & \omega_3 & {\bar{\omega}_3
   } & 1 & \omega_3 & {\bar{\omega}_3} & 1 & \omega_3 & {\bar{\omega}_3} \\ \hline 1 & 1 & 1 & \omega_3 & {\bar{\omega}_3} & \omega_3 & {\bar{\omega}_3} & 1 & {\bar{\omega}_3
   } & 1 & \omega_3 \\ \hline 1 & 1 & 1 & \omega_3 & {\bar{\omega}_3} & {\bar{\omega}_3} & 1 & \omega_3 & \omega_3 & {\bar{\omega}_3} & 1 \\ \hline 1 & 1 & 1 & {\bar{\omega}_3} & \omega_3 & 1 & {\bar{\omega}_3
   } & \omega_3 & 1 & {\bar{\omega}_3} & \omega_3 \\ \hline 1 & 1 & 1 & {\bar{\omega}_3} & \omega_3 & \omega_3 & 1 & {\bar{\omega}_3} & {\bar{\omega}_3} & \omega_3 & 1 \\ \hline 1 & 1 & 1 & {\bar{\omega}_3
   } & \omega_3 & {\bar{\omega}_3} & \omega_3 & 1 & \omega_3 & 1 & {\bar{\omega}_3} \\ \hline 3 & 3{\bar{\omega}_3} & 3\omega_3 & 0 & 0 & 0 & 0 & 0 & 0 & 0 & 0 \\ \hline 3 & 3\omega_3 & 3
   {\bar{\omega}_3} & 0 & 0 & 0 & 0 & 0 & 0 & 0 & 0 \\ \hline
\ea
}
\eeq
\beq
\label{cgdel3n3}
\setlength{\arraycolsep}{0.2mm}
\ba{l}
{\rm char}(\Delta(3\times3^2)^*) = \\
{\tiny
\ba{|c|c|c|c|c|c|c|c|c|c|c|c|c|c|c|c|c|c|c|c|c|c|c|c|c|c|c|c|c|c|c|c|c|c|c|}
\hline
   1 & 1 & 1 & 1 & 1 & 1 & 1 & 1 & 1 & 9 & 9 & 9 & 9 & 9 & 9 & 9 & 9 & 9 & 9 & 9 & 9 & 9 & 9 & 9 & 9 & 9 & 9 & 9 & 9 & 9 & 9 & 9 & 9 & 9 & 9 \\ \hline 1 & 1 & 1 & 1 & 
   1 & 1 & 1 & 1 & 1 & 1 & 1 & 1 & 1 & 1 & 1 & 1 & 1 & 1 & 1 & 1 & 1 & 1 & 1 & 1 & 1 & 1 & 1 & 1 & 1 & 1 & 1 & 1 & 1 & 1 & 1 \\ \hline 1 & 1 & 1 & 1 & 1 & 1 & 1 & 1 & 
   1 & 1 & 1 & 1 & 1 & 1 & 1 & 1 & 1 & \omega_3 & \omega_3 & \omega_3 & \omega_3 & \omega_3 & \omega_3 & \omega_3 & \omega_3 & \omega_3 & {\bar{\omega}_3} & {\bar{\omega}_3} & {\bar{\omega}_3} & {\bar{\omega}_3} & {\bar{\omega}_3
   } & {\bar{\omega}_3} & {\bar{\omega}_3} & {\bar{\omega}_3} & {\bar{\omega}_3} \\ \hline 1 & 1 & 1 & 1 & 1 & 1 & 1 & 1 & 1 & 1 & 1 & 1 & 1 & 1 & 1 & 1 & 1 & {\bar{\omega}_3
   } & {\bar{\omega}_3} & {\bar{\omega}_3} & {\bar{\omega}_3} & {\bar{\omega}_3} & {\bar{\omega}_3} & {\bar{\omega}_3} & {\bar{\omega}_3} & {\bar{\omega}_3
   } & \omega_3 & \omega_3 & \omega_3 & \omega_3 & \omega_3 & \omega_3 & \omega_3 & \omega_3 & \omega_3 \\ \hline 1 & 1 & 1 & 1 & 1 & 1 & 1 & 1 & 1 & 1 & 1 & \omega_3 & \omega_3 & \omega_3 & {\bar{\omega}_3} & {\bar{\omega}_3} & {\bar{\omega}_3
   } & 1 & 1 & 1 & \omega_3 & \omega_3 & \omega_3 & {\bar{\omega}_3} & {\bar{\omega}_3} & {\bar{\omega}_3} & 1 & 1 & 1 & \omega_3 & \omega_3 & \omega_3 & {\bar{\omega}_3} & {\bar{\omega}_3} & {\bar{\omega}_3
   } \\ \hline 1 & 1 & 1 & 1 & 1 & 1 & 1 & 1 & 1 & 1 & 1 & \omega_3 & \omega_3 & \omega_3 & {\bar{\omega}_3} & {\bar{\omega}_3} & {\bar{\omega}_3} & \omega_3 & \omega_3 & \omega_3 & {\bar{\omega}_3} & {\bar{\omega}_3
   } & {\bar{\omega}_3} & 1 & 1 & 1 & {\bar{\omega}_3} & {\bar{\omega}_3} & {\bar{\omega}_3
   } & 1 & 1 & 1 & \omega_3 & \omega_3 & \omega_3 \\ \hline 1 & 1 & 1 & 1 & 1 & 1 & 1 & 1 & 1 & 1 & 1 & \omega_3 & \omega_3 & \omega_3 & {\bar{\omega}_3} & {\bar{\omega}_3} & {\bar{\omega}_3} & {\bar{\omega}_3} & 
   {\bar{\omega}_3} & {\bar{\omega}_3} & 1 & 1 & 1 & \omega_3 & \omega_3 & \omega_3 & \omega_3 & \omega_3 & \omega_3 & {\bar{\omega}_3} & {\bar{\omega}_3} & {\bar{\omega}_3
   } & 1 & 1 & 1 \\ \hline 1 & 1 & 1 & 1 & 1 & 1 & 1 & 1 & 1 & 1 & 1 & {\bar{\omega}_3} & {\bar{\omega}_3} & {\bar{\omega}_3} & \omega_3 & \omega_3 & \omega_3 & 1 & 1 & 1 & {\bar{\omega}_3} & 
   {\bar{\omega}_3} & {\bar{\omega}_3} & \omega_3 & \omega_3 & \omega_3 & 1 & 1 & 1 & {\bar{\omega}_3} & {\bar{\omega}_3} & {\bar{\omega}_3
   } & \omega_3 & \omega_3 & \omega_3 \\ \hline 1 & 1 & 1 & 1 & 1 & 1 & 1 & 1 & 1 & 1 & 1 & {\bar{\omega}_3} & {\bar{\omega}_3} & {\bar{\omega}_3} & \omega_3 & \omega_3 & \omega_3 & \omega_3 & \omega_3 & \omega_3 & 1 & 1 & 1 & \bar{\omega}_3 & {\bar{\omega}_3} & {\bar{\omega}_3} & {\bar{\omega}_3} & {\bar{\omega}_3} & {\bar{\omega}_3
   } & \omega_3 & \omega_3 & \omega_3 & 1 & 1 & 1 \\ \hline 1 & 1 & 1 & 1 & 1 & 1 & 1 & 1 & 1 & 1 & 1 & {\bar{\omega}_3} & {\bar{\omega}_3} & {\bar{\omega}_3} & \omega_3 & \omega_3 & \omega_3 & {\bar{\omega}_3} & 
   {\bar{\omega}_3} & {\bar{\omega}_3} & \omega_3 & \omega_3 & \omega_3 & 1 & 1 & 1 & \omega_3 & \omega_3 & \omega_3 & 1 & 1 & 1 & {\bar{\omega}_3} & {\bar{\omega}_3} & {\bar{\omega}_3
   } \\ \hline 3 & 3 & 3 & 3 & 3 & 3 & 3 & 3 & 3 & 3{\bar{\omega}_3} & 3
   \omega_3 & 0 & 0 & 0 & 0 & 0 & 0 & 0 & 0 & 0 & 0 & 0 & 0 & 0 & 0 & 0 & 0 & 0 & 0 & 0 & 0 & 0 & 0 & 0 & 0 \\ \hline 3 & 3 & 3 & 3 & 3 & 3 & 3 & 3 & 3 & 3\omega_3 & 3
   {\bar{\omega}_3} & 0 & 0 & 0 & 0 & 0 & 0 & 0 & 0 & 0 & 0 & 0 & 0 & 0 & 0 & 0 & 0 & 0 & 0 & 0 & 0 & 0 & 0 & 0 & 0 \\ \hline 3 & 3\omega_3 & 3{\bar{\omega}_3} & 3 & 3
   \omega_3 & 3{\bar{\omega}_3} & 3 & 3\omega_3 & 3
   {\bar{\omega}_3} & 0 & 0 & 0 & 0 & 0 & 0 & 0 & 0 & 0 & 0 & 0 & X & Y & Z & 0 & 0 & 0 & 0 & 0 & 0 & 0 & 0 & 0 & P & M & N \\ \hline 3 & 3\omega_3 & 3
   {\bar{\omega}_3} & 3 & 3\omega_3 & 3{\bar{\omega}_3} & 3 & 3\omega_3 & 3
   {\bar{\omega}_3} & 0 & 0 & 0 & 0 & 0 & 0 & 0 & 0 & 0 & 0 & 0 & Z & X & Y & 0 & 0 & 0 & 0 & 0 & 0 & 0 & 0 & 0 & M & N & P \\ \hline 3 & 3\omega_3 & 3
   {\bar{\omega}_3} & 3 & 3\omega_3 & 3{\bar{\omega}_3} & 3 & 3\omega_3 & 3
   {\bar{\omega}_3} & 0 & 0 & 0 & 0 & 0 & 0 & 0 & 0 & 0 & 0 & 0 & Y & Z & X & 0 & 0 & 0 & 0 & 0 & 0 & 0 & 0 & 0 & N & P & M \\ \hline 3 & 3{\bar{\omega}_3} & 3
   \omega_3 & 3 & 3{\bar{\omega}_3} & 3\omega_3 & 3 & 3{\bar{\omega}_3} & 3
   \omega_3 & 0 & 0 & 0 & 0 & 0 & 0 & 0 & 0 & 0 & 0 & 0 & M & P & N & 0 & 0 & 0 & 0 & 0 & 0 & 0 & 0 & 0 & Z & Y & X \\ \hline 3 & 3{\bar{\omega}_3} & 3\omega_3 & 3 & 3
   {\bar{\omega}_3} & 3\omega_3 & 3 & 3{\bar{\omega}_3} & 3
   \omega_3 & 0 & 0 & 0 & 0 & 0 & 0 & 0 & 0 & 0 & 0 & 0 & N & M & P & 0 & 0 & 0 & 0 & 0 & 0 & 0 & 0 & 0 & Y & X & Z \\ \hline 3 & 3{\bar{\omega}_3} & 3\omega_3 & 3 & 3
   {\bar{\omega}_3} & 3\omega_3 & 3 & 3{\bar{\omega}_3} & 3
   \omega_3 & 0 & 0 & 0 & 0 & 0 & 0 & 0 & 0 & 0 & 0 & 0 & P & N & M & 0 & 0 & 0 & 0 & 0 & 0 & 0 & 0 & 0 & X & Z & Y \\ \hline 3 & 3 & 3 & 3\omega_3 & 3\omega_3 & 3\omega_3 & 3
   {\bar{\omega}_3} & 3{\bar{\omega}_3} & 3{\bar{\omega}_3} & 0 & 0 & A & -B & C & -A & 
    -C & B & 0 & 0 & 0 & 0 & 0 & 0 & 0 & 0 & 0 & 0 & 0 & 0 & 0 & 0 & 0 & 0 & 0 & 0 \\ \hline 3 & 3 & 3 & 3\omega_3 & 3\omega_3 & 3\omega_3 & 3{\bar{\omega}_3} & 3
   {\bar{\omega}_3} & 3{\bar{\omega}_3} & 0 & 0 & -B & C & A & -C & B & 
    -A & 0 & 0 & 0 & 0 & 0 & 0 & 0 & 0 & 0 & 0 & 0 & 0 & 0 & 0 & 0 & 0 & 0 & 0 \\ \hline 3 & 3 & 3 & 3\omega_3 & 3\omega_3 & 3\omega_3 & 3{\bar{\omega}_3} & 3{\bar{\omega}_3} & 3
   {\bar{\omega}_3} & 0 & 0 & C & A & -B & B & -A & -C & 0 & 0 & 0 & 0 & 0 & 0 & 0 & 0 & 0 & 0 & 0 & 0 & 0 & 0 & 0 & 0 & 0 & 0 \\ \hline 3 & 3\omega_3 & 3
   {\bar{\omega}_3} & 3\omega_3 & 3{\bar{\omega}_3} & 3 & 3{\bar{\omega}_3} & 3 & 3\omega_3 & 0 & 0 & 0 & 0 & 0 & 0 & 0 & 0 & -C & -A & B & 0 & 0 & 0 & 0 & 0 & 0 & 
    -B & C & A & 0 & 0 & 0 & 0 & 0 & 0 \\ \hline 3 & 3\omega_3 & 3{\bar{\omega}_3} & 3\omega_3 & 3{\bar{\omega}_3} & 3 & 3{\bar{\omega}_3} & 3 & 3
   \omega_3 & 0 & 0 & 0 & 0 & 0 & 0 & 0 & 0 & -A & B & -C & 0 & 0 & 0 & 0 & 0 & 0 & A & -B & C & 0 & 0 & 0 & 0 & 0 & 0 \\ \hline 3 & 3\omega_3 & 3{\bar{\omega}_3} & 3\omega_3 & 3
   {\bar{\omega}_3} & 3 & 3{\bar{\omega}_3} & 3 & 3\omega_3 & 0 & 0 & 0 & 0 & 0 & 0 & 0 & 0 & B & -C & -A & 0 & 0 & 0 & 0 & 0 & 0 & C & A & 
    -B & 0 & 0 & 0 & 0 & 0 & 0 \\ \hline 3 & 3{\bar{\omega}_3} & 3\omega_3 & 3\omega_3 & 3 & 3{\bar{\omega}_3} & 3{\bar{\omega}_3} & 3
   \omega_3 & 3 & 0 & 0 & 0 & 0 & 0 & 0 & 0 & 0 & 0 & 0 & 0 & 0 & 0 & 0 & M & N & P & 0 & 0 & 0 & Z & X & Y & 0 & 0 & 0 \\ \hline 3 & 3{\bar{\omega}_3} & 3\omega_3 & 3
   \omega_3 & 3 & 3{\bar{\omega}_3} & 3{\bar{\omega}_3} & 3
   \omega_3 & 3 & 0 & 0 & 0 & 0 & 0 & 0 & 0 & 0 & 0 & 0 & 0 & 0 & 0 & 0 & P & M & N & 0 & 0 & 0 & X & Y & Z & 0 & 0 & 0 \\ \hline 3 & 3{\bar{\omega}_3} & 3\omega_3 & 3
   \omega_3 & 3 & 3{\bar{\omega}_3} & 3{\bar{\omega}_3} & 3
   \omega_3 & 3 & 0 & 0 & 0 & 0 & 0 & 0 & 0 & 0 & 0 & 0 & 0 & 0 & 0 & 0 & N & P & M & 0 & 0 & 0 & Y & Z & X & 0 & 0 & 0 \\ \hline 3 & 3 & 3 & 3{\bar{\omega}_3} & 3
   {\bar{\omega}_3} & 3{\bar{\omega}_3} & 3\omega_3 & 3\omega_3 & 3\omega_3 & 0 & 0 & -A & -C & B & A & 
    -B & C & 0 & 0 & 0 & 0 & 0 & 0 & 0 & 0 & 0 & 0 & 0 & 0 & 0 & 0 & 0 & 0 & 0 & 0 \\ \hline 3 & 3 & 3 & 3{\bar{\omega}_3} & 3{\bar{\omega}_3} & 3{\bar{\omega}_3} & 3
   \omega_3 & 3\omega_3 & 3\omega_3 & 0 & 0 & -C & B & -A & -B & C & A & 0 & 0 & 0 & 0 & 0 & 0 & 0 & 0 & 0 & 0 & 0 & 0 & 0 & 0 & 0 & 0 & 0 & 0 \\ \hline 3 & 3 & 3 & 3
   {\bar{\omega}_3} & 3{\bar{\omega}_3} & 3{\bar{\omega}_3} & 3\omega_3 & 3\omega_3 & 3\omega_3 & 0 & 0 & B & -A & -C & C & A & 
    -B & 0 & 0 & 0 & 0 & 0 & 0 & 0 & 0 & 0 & 0 & 0 & 0 & 0 & 0 & 0 & 0 & 0 & 0 \\ \hline 3 & 3\omega_3 & 3{\bar{\omega}_3} & 3{\bar{\omega}_3} & 3 & 3\omega_3 & 3\omega_3 & 3
   {\bar{\omega}_3} & 3 & 0 & 0 & 0 & 0 & 0 & 0 & 0 & 0 & 0 & 0 & 0 & 0 & 0 & 0 & X & Z & Y & 0 & 0 & 0 & P & N & M & 0 & 0 & 0 \\ \hline 3 & 3\omega_3 & 3
   {\bar{\omega}_3} & 3{\bar{\omega}_3} & 3 & 3\omega_3 & 3\omega_3 & 3
   {\bar{\omega}_3} & 3 & 0 & 0 & 0 & 0 & 0 & 0 & 0 & 0 & 0 & 0 & 0 & 0 & 0 & 0 & Y & X & Z & 0 & 0 & 0 & N & M & P & 0 & 0 & 0 \\ \hline 3 & 3\omega_3 & 3
   {\bar{\omega}_3} & 3{\bar{\omega}_3} & 3 & 3\omega_3 & 3\omega_3 & 3
   {\bar{\omega}_3} & 3 & 0 & 0 & 0 & 0 & 0 & 0 & 0 & 0 & 0 & 0 & 0 & 0 & 0 & 0 & Z & Y & X & 0 & 0 & 0 & M & P & N & 0 & 0 & 0 \\ \hline 3 & 3{\bar{\omega}_3} & 3
   \omega_3 & 3{\bar{\omega}_3} & 3\omega_3 & 3 & 3\omega_3 & 3 & 3{\bar{\omega}_3} & 0 & 0 & 0 & 0 & 0 & 0 & 0 & 0 & -B & A & C & 0 & 0 & 0 & 0 & 0 & 0 & -C & B & 
    -A & 0 & 0 & 0 & 0 & 0 & 0 \\ \hline 3 & 3{\bar{\omega}_3} & 3\omega_3 & 3{\bar{\omega}_3} & 3\omega_3 & 3 & 3\omega_3 & 3 & 3
   {\bar{\omega}_3} & 0 & 0 & 0 & 0 & 0 & 0 & 0 & 0 & A & C & -B & 0 & 0 & 0 & 0 & 0 & 0 & -A & -C & B & 0 & 0 & 0 & 0 & 0 & 0 \\ \hline 3 & 3{\bar{\omega}_3} & 3
   \omega_3 & 3{\bar{\omega}_3} & 3\omega_3 & 3 & 3\omega_3 & 3 & 3{\bar{\omega}_3} & 0 & 0 & 0 & 0 & 0 & 0 & 0 & 0 & C & -B & A & 0 & 0 & 0 & 0 & 0 & 0 & B & -A & 
    -C & 0 & 0 & 0 & 0 & 0 & 0 \\ \hline
\ea
}\\
\ea
\eeq
with $A := -\omega_3 + \bar{\omega}_3, B := \omega_3 +
2\bar{\omega}_3, C := 2\omega_3 + \bar{\omega}_3; M := -\omega_9^2 - 2
\bar{\omega}_9^4, N := \omega_9^2 + \bar{\omega}_9^4, P := -\omega_9^2
+ \bar{\omega}_9^4; X := \omega_9^4 - \bar{\omega}_9^2, Y :=
\omega_9^4 + 2 \bar{\omega}_9^2, Z := -2 \omega_9^4 - \bar{\omega}_9^2$.

A comparative study of these two tables shall suffice to demonstrate
the method. We have taken extreme pains to re-arrange the columns and
rows of char$(G^*)$ for the sake of perspicuity; whence we immediately
observe that char$(G)$ and char$(G^*)$ are unrelated but that the
latter is organised in terms of ``cohorts'' \cite{Hum-Chat} of the
former. What this means is as follows: columns 1 through 9 of
char$(G^*)$ have their first 11 rows (not counting the row of class
numbers) identical to the first column of char$(G)$, so too is
column 10 of char$(G^*)$ with column 2 of char$(G)$, {\it et cetera}
with $\{11\} \rightarrow \{3\}$, $\{12,13,14\} \rightarrow \{4\}$, 
$\{15,16,17\} \rightarrow \{5\}$, $\{18,19,20\} \rightarrow \{6\}$, 
$\{21,22,23\} \rightarrow \{7\}$, $\{24,25,26\} \rightarrow \{8\}$, 
$\{27,28,29\} \rightarrow \{9\}$, $\{30,31,32\} \rightarrow \{10\}$, 
and $\{33,34,35\} \rightarrow \{11\}$; using the notation that $\{X\}
\rightarrow \{Y\}$ for the first 11 rows of columns $\{X\} \subset$
 char$(G^*)$ are mapped to column $\{Y\} \subset$ char$(G)$. These are
the so-called ``splitting conjugacy classes'' in $G^*$ which give the
(linear) char$(G)$ \cite{Hoff-Hum}.
In other words, (though the conjugacy class numbers may differ),
up to repetition char$(G) \subset$ char$(G^*)$. This of course is in the
spirit of the technique of Fr{\o}benius Induction of finding
the character table of a group from that of its subgroup; for a
discussion of this in the context of orbifolds, the reader
is referred to \cite{0012078}. 
Thus the first 11 rows of char$(G^*)$ corresponds exactly to the
{\em linear irreps} of $G$. The rest of the rows
we shall shortly observe to correspond to the
projective representations. 

To understand these above remarks, 
let $A := \Z_3 \times \Z_3$ so that $G^* / A \cong G$ as
in the notation of Section 2. Now $A \subseteq Z(G^*)$, hence the
matrix forms of all of its elements must be $\lambda \II_{d\times d}$,
where $d$ is the dimension of the irreducible representation and 
$\lambda$ some phase factor. Indeed the first 9 columns of char$(G^*)$
have conjugacy class number 1 and hence correspond to elements of this
centre. Bearing this in mind, if we only tabulated the phases
$\lambda$ (by suppressing the factor $d=$ 1 or 3 coming from $\II_{d\times
d}$) of these first 9 columns, we arrive at the following table
(removing the first row of conjugacy class numbers):
\[
\ba{|c||c|c|c|c|c|c|c|c|c|}
\hline
\mbox{rows} & \II & a  & a^2 & b  & ab  & a^2 b & b^2 & a b^2 & a^2 b^2 \\ \hline
2 - 12	& 1   & 1    &1      & 1   & 1   & 1      & 1   & 1   & 1  \\ \hline
13-15	& 1   &\omega_3&\bar{\omega}_3  & 1   &\omega_3 &\bar{\omega}_3 
                     & 1   &\omega_3 &\bar{\omega}_3\\ \hline
16-18	& 1 &\bar{\omega}_3 & \omega_3 & 1 &\bar{\omega}_3 & \omega_3 
                       & 1 &\bar{\omega}_3 & \omega_3 \\ \hline
19-21	& 1   & 1    &1 &\omega_3&\omega_3&\omega_3  &\bar{\omega}_3&\bar{\omega}_3
                        &\bar{\omega}_3 \\ \hline
22-24	& 1   &\omega_3&\bar{\omega}_3 &\omega_3 &\bar{\omega}_3 &1 
                     &\bar{\omega}_3  & 1  &\omega_3 \\ \hline
25-27	& 1 &\bar{\omega}_3 & \omega_3  &\omega_3  & 1  &\bar{\omega}_3 
                     &\bar{\omega}_3   &\omega_3 & 1\\ \hline
28-30	& 1   & 1    &1  &\bar{\omega}_3&\bar{\omega}_3&\bar{\omega}_3
                        &\omega_3&\omega_3&\omega_3 \\ \hline
31-33	& 1   &\omega_3&\bar{\omega}_3  &\bar{\omega}_3  & 1  &\omega_3
                        &\omega_3 &\bar{\omega}_3 &1 \\ \hline
34-36	& 1 &\bar{\omega}_3 & \omega_3  &\bar{\omega}_3  &\omega_3& 1
                        &\omega_3 & 1 & \bar{\omega}_3 \\ \hline
\ea
\]
The astute reader would instantly recognise this to be the character
table of $\IZ_3 \times \IZ_3 = A$ (and with foresight we have labelled
the elements of the group in the above table). This certainly is to be
expected: $G^*$ can be written as cosets $gA$ for $g \in G$, whence
lifting the (projective) matrix representation $M(g)$ of $g$ simply
gives $\lambda M(g)$ for $\lambda$ a {\em phase factor} correponding
to the representation (or character as $A$ is always Abelian) of $A$.

What is happening should be clear: all of this is merely Part (i) of Theorem
\ref{cover2} at work. The phases $\lambda$ are precisely as
described in the theorem. The trivial phase 1 gives rows $2-12$, or
simply the ordinary representation of $G$ while the remaining 8
non-trivial phases give, in groups of 3 rows from char$(G^*)$, the
projective representations of $G$. And to determine to which cocycle the
projective representation belongs, we need and only need to determine
the the 1-dimensional irreps of $A$.
We shall show in Section 5 how to read out the actual cocycle values
$\alpha(g,h)$ for $g,h\in G$
directly with the knowledge of $A$ and $G^*$ without char$(G^*)$.

Enough said on the character tables. Let us proceed to analyse the
quiver diagrams. Detailed discussions had already been presented in
the case of the dihedral group in \cite{0010023}. Let us recapitulate the
key points. It is the group action on the Chan-Paton bundle that we
choose to be projective, the space-time action inherited from ${\cal
N}=4$ R-symmetry remain ordinary. In other words, ${\cal R}$ from
\eref{proj-tor} must still be a linear representation.

Now we evoke an obvious though handy result: 
the tensor product of an $\alpha$-projective representation 
with that of a $\beta$-representation gives
an $\alpha\beta$-projective representation (cf. \cite{Karp} p119), i.e., 
\beq\label{ab}
\Gamma_\alpha(g) \otimes \Gamma_\beta(g) = \Gamma_{\alpha\beta}(g).
\eeq
We recall that from \eref{proj-tor} and in the language of
\cite{AspinPles,LNV}, the bi-fundamental matter
content $a_{ij}^{\cal R}$ is given in terms of the irreducible representations $R_i$
of $G$ as
\beq
\label{proj-tor2}
{\cal R} \otimes R_i = \bigoplus\limits_j a_{ij}^{\cal R} R_j,
\eeq
(with of course ${\cal R}$ linear and $R_i$ projective representations).
Because ${\cal R}$ is an $\alpha=1$ (linear) representation, \eref{ab}
dictates that if $R_i$ in \eref{proj-tor2} is a $\beta$-representation,
then the righthand thereof must be written entirely in terms of 
$\beta$-representations $R_j$. In
other words, the various projective representations corresponding to
the different cocycles should not mix under \eref{proj-tor2}.
What this signifies for the matter matrix is that $a_{ij}^{\cal R}$ is
block-diagonal and the quiver diagram $Q(G^*,{\cal R})$ for $G^*$ {\em splits} into
precisely $|A|$ pieces, one of which is the ordinary (linear) quiver
for $G$ and the rest, the various quivers each corresponding to a
different value of the cocycle.

Thus motivated, let us present the quiver diagram for
$\Delta(3\times3^2)^*$ in \fref{f:del3_3}. The splitting does indeed
occur as desired, into precisely $|\IZ_3 \times \IZ_3|=9$ pieces, with
(i) being the usual $\Delta(3\times3^2)$ quiver
(cf. \cite{9811183,Muto}) and the rest, the quivers corresponding to
the 8 non-trivial projective representations.
\EPSFIGURE[ht]{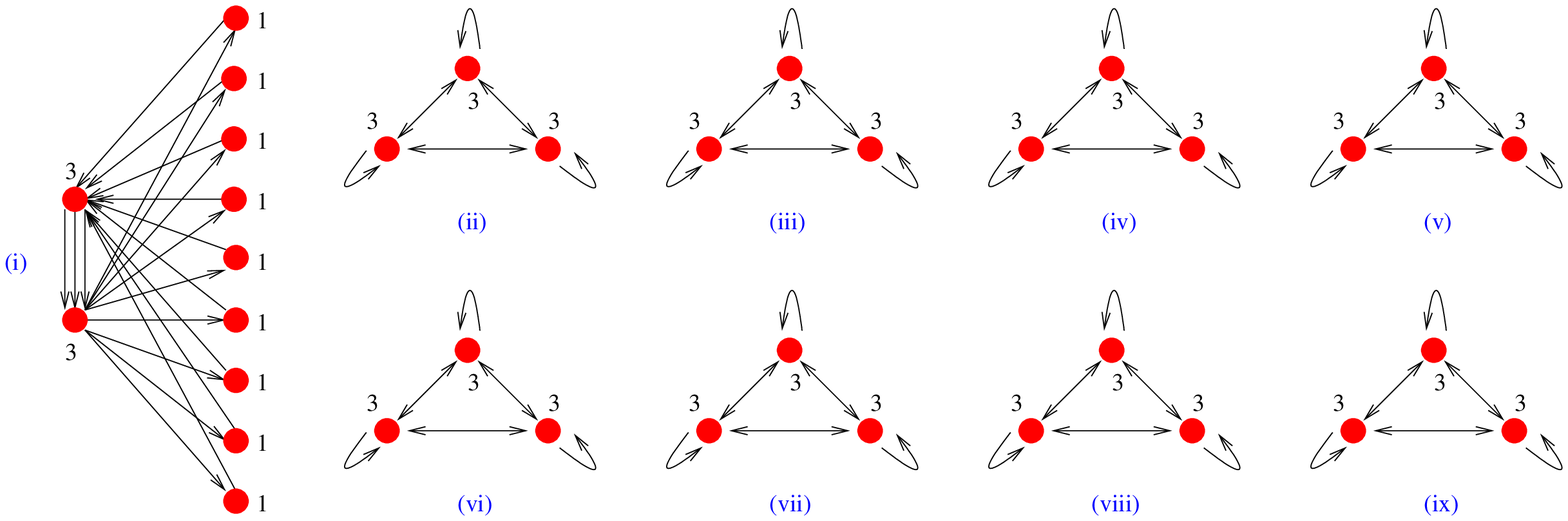,width=6in}
{
\label{f:del3_3}
The Quiver Diagram for $\Delta(3\times3^2)^*$ (the Space Invaders
Quiver): piece (i) corresponds to
the usual quiver for $\Delta(3\times3^2)$
while the remaining 8 pieces (ii) to (ix)
are for the cases of the 8 non-trivial discrete torsions (out of the
$\IZ_3 \times \IZ_3$) turned on.
}
\subsection{The General Method}
\index{Brane Probes!discrete torsion}
Having expounded upon the detailed example of $\Delta(3\times3^2)$ and
witnessed the subtleties, we now present, in an algorithmic manner,
the general method of computing the quiver diagram for an orbifold $G$
with discrete torsion turned on:
\begin{enumerate}
\item Compute the character table char$(G)$ of $G$;
\item Compute a covering group $G^*$ of $G$ and its character table 
	char$(G^*)$;
\item Judiciously re-order the rows and columns of char$(G^*)$:
	\begin{itemize}
	\item Columns must be arranged into cohorts of char$(G)$,
		i.e., group the columns which contain a corresponding
		column in char$(G)$ together;
	\item Rows must be arranged so that modulo the dimension of
		the irreps, the columns with conjugacy class number 1 
		must contain the character table of the Schur
		Multiplier $A = M(G)$ (recall that $G^*/A\cong G$);
	\item Thus char$(G)$ is a sub-matrix (up to repetition) of
		char$(G^*)$;
	\end{itemize}
\item Compute the (ordinary) matter matrix 
	$a_{ij}^{\cal R}$ and hence
	the quiver $Q(G^*,{\cal R})$ for a representation
	${\cal R}$ which corresponds to a linear representation of $G$.
\end{enumerate}
Now we have our final result:
\begin{theorem}
$Q(G^*,{\cal R})$
has $|M(G)|$ disconnected components (sub-quivers) in 1-1 correspondence with
the quivers $Q_{\alpha}(G,{\cal R})$ of $G$ for all possible cocycles
(discrete torsions) $\alpha \in A=M(G)$. Symbolically,
\[
Q(G^*,{\cal R}) = \bigsqcup_{\alpha \in A} Q_{\alpha}(G,{\cal R}).
\]
\end{theorem}
In particular, $Q(G^*,{\cal R})$ contains a piece for the trivial
$\alpha =1$ which is precisely the case without discrete torsion,
viz., $Q(G,{\cal R})$.

This algorithm facilitates enormously the investigation of the
matter spectrum of orbifold gauge theories with discrete torsion as
the associated quivers can be found without any recourse to explicit
evaluation of the cocycles and projective character tables.
Another fine feature of this new understanding is that, not only the 
matter content, but also the superpotential can be directly calculated 
by the explicit formulae in \cite{LNV} using the ordinary
Clebsch-Gordan coefficients of $G^*$. 

A remark is at hand. We have mentioned in Section 2 that the covering
group $G^*$ is not unique. How could we guarantee that the quivers
obtained at the end of the day will be independent of the choice of
the covering group? We appeal directly to the discussion in the
concluding paragraph of Subsection 4.1, where we remarked that using
the explicit form of \eref{proj-tor}, we see that the phase factor
$\lambda$ (being a $\IC$-number) always cancels out. In other
words, the linear representation of whichever $G^*$ we use,
when applied to orbifold projections \eref{proj-tor} shall result in the
same matrix form for the projective representations of $G$.
Whence we conclude that the
quiver $Q(G^*,{\cal R})$ obtained at the end will {\it ipso facto} be
independent of the choice of the covering group $G^*$.
\\ \\ \\ \\
\subsection{A Myriad of Examples}
With the method at hand, we move on to the host of other subgroups of
$SU(3)$ as tabulated in \cite{0010023}.
The character tables char$(G)$ and char$(G^*)$ will be left
to the appendix lest the reader be too distracted.
We present the cases of $\Sigma(60,168,1080)$, the exceptionals which
admit nontrivial discrete torsion and some first members of the Delta
series in \fref{f:sig60} to \fref{f:del3_5}.
\EPSFIGURE[h]{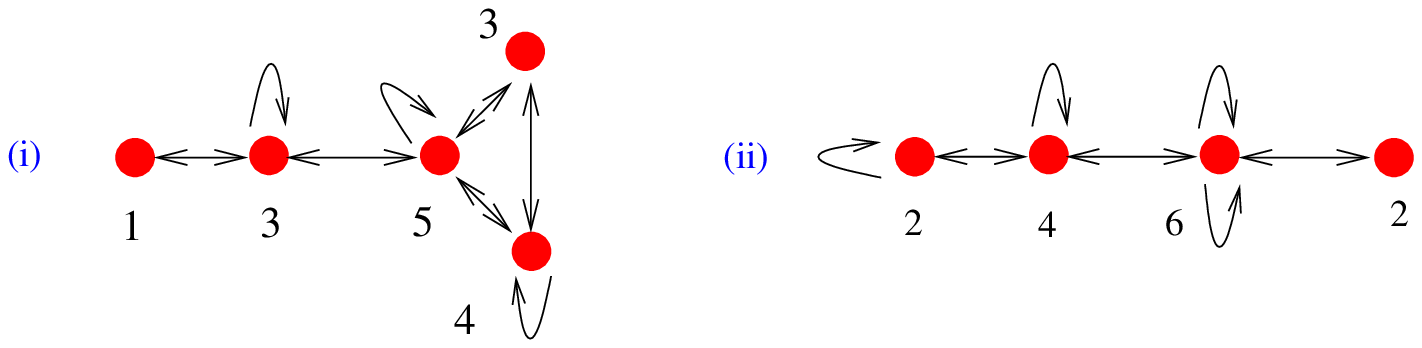,width=4.5in}
{\label{f:sig60}
The quiver diagram of $\Sigma(60)^*$: piece (i) is the ordinary quiver
of $\Sigma(60)$ and piece (ii) has discrete torsion turned on.
}
\EPSFIGURE[h]{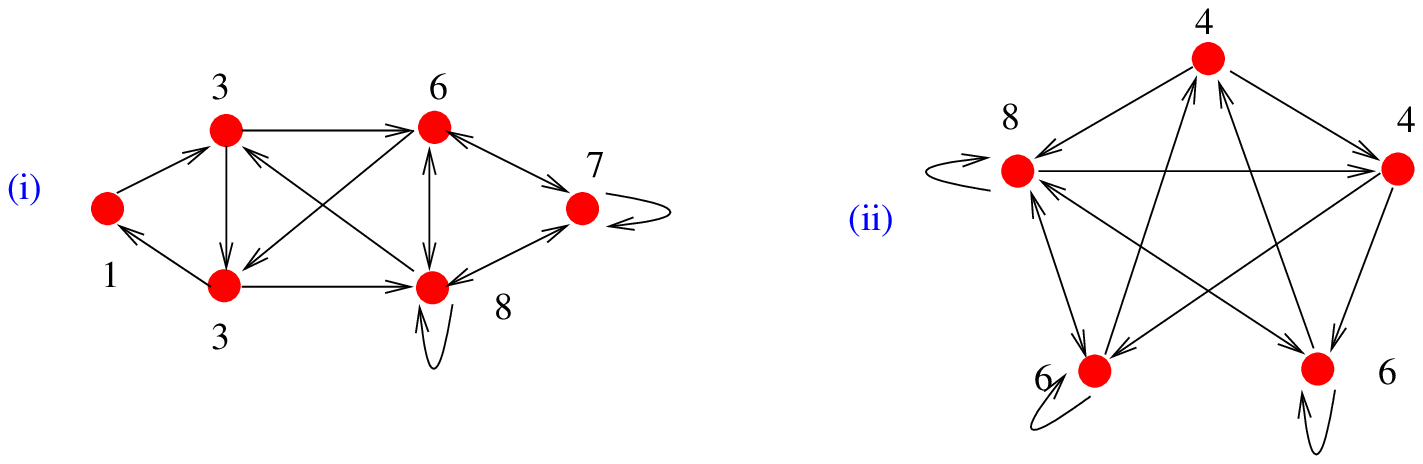,width=4.5in}
{\label{f:sig168}
The quiver diagram of $\Sigma(168)$: piece (i) is the ordinary quiver
of $\Sigma(168)$ and piece (ii) has discrete torsion turned on.
}
{\vspace{-1.0cm}}
\EPSFIGURE[ht]{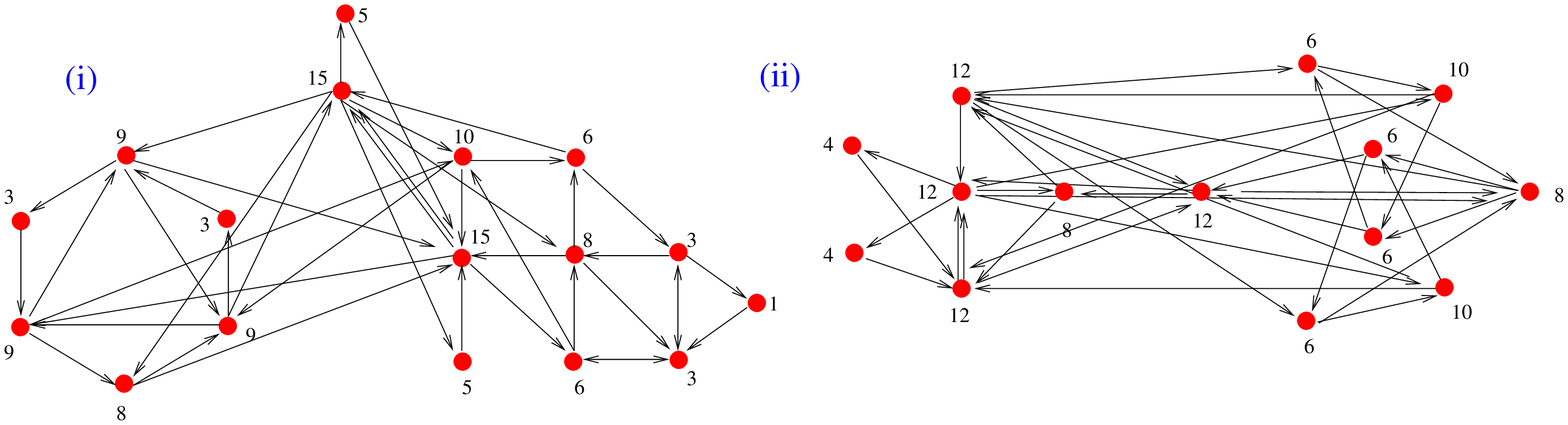,width=5.5in}
{\label{f:sig1080}
The quiver diagram of $\Sigma(1080)$: piece (i) is the ordinary quiver
of $\Sigma(1080)$ and piece (ii) has discrete torsion turned on.
}
\newpage
\EPSFIGURE[ht]{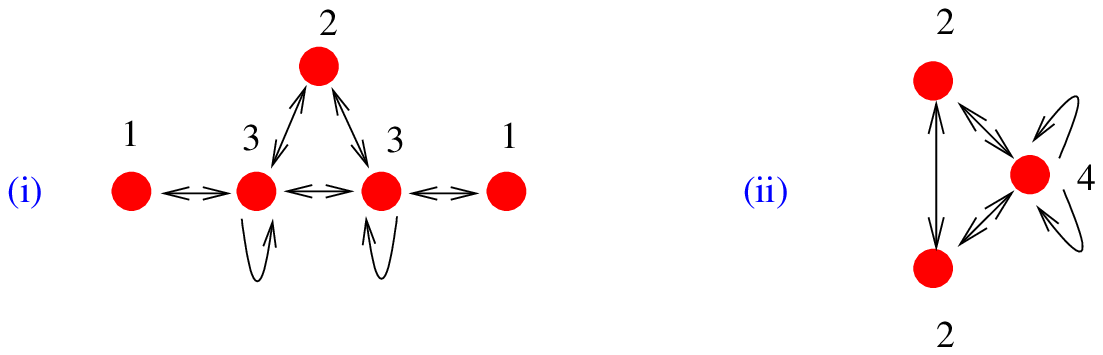,width=4.2in}
{\label{f:del6_2}
The quiver diagram of $\Delta(6\times 2^2)$: piece (i) is the ordinary quiver
of $\Delta(6\times 2^2)$ and piece (ii) has discrete torsion turned on.
}
\EPSFIGURE[ht]{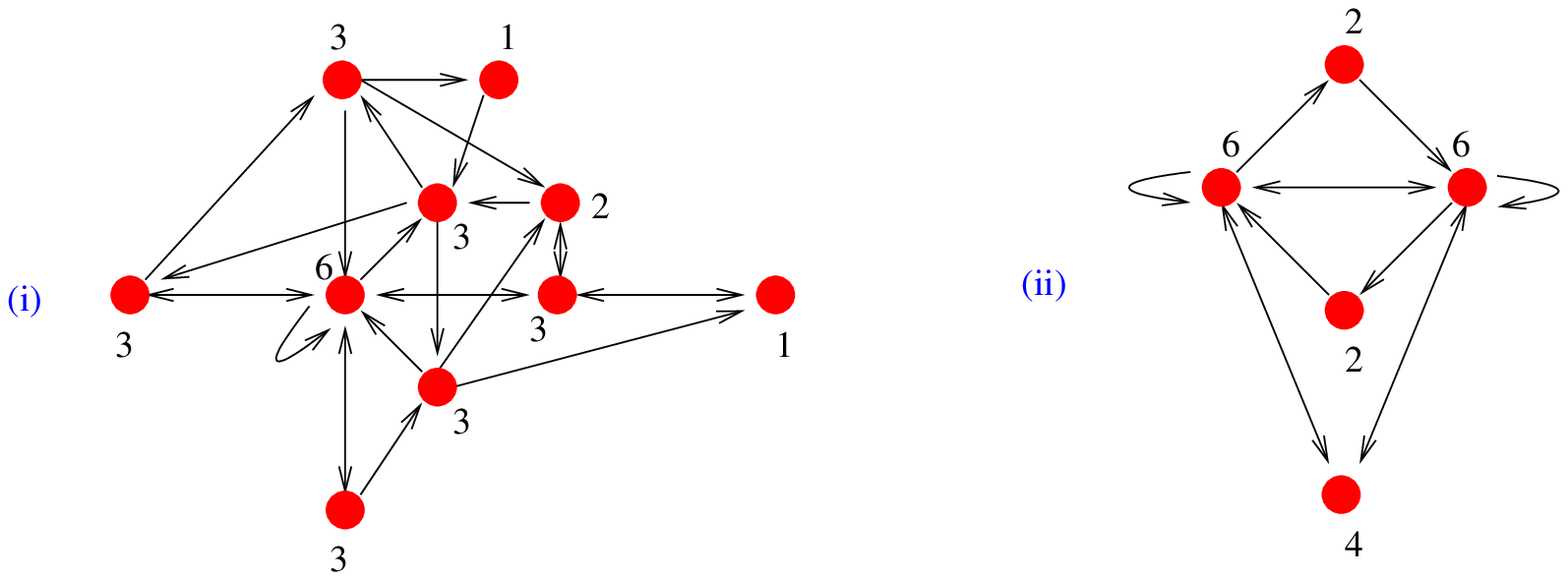,width=4.5in}
{\label{f:del6_4}
The quiver diagram of $\Delta(6\times 4^2)$: piece (i) is the ordinary quiver
of $\Delta(6\times 4^2)$ and piece (ii) has discrete torsion turned on.
}
\EPSFIGURE[ht]{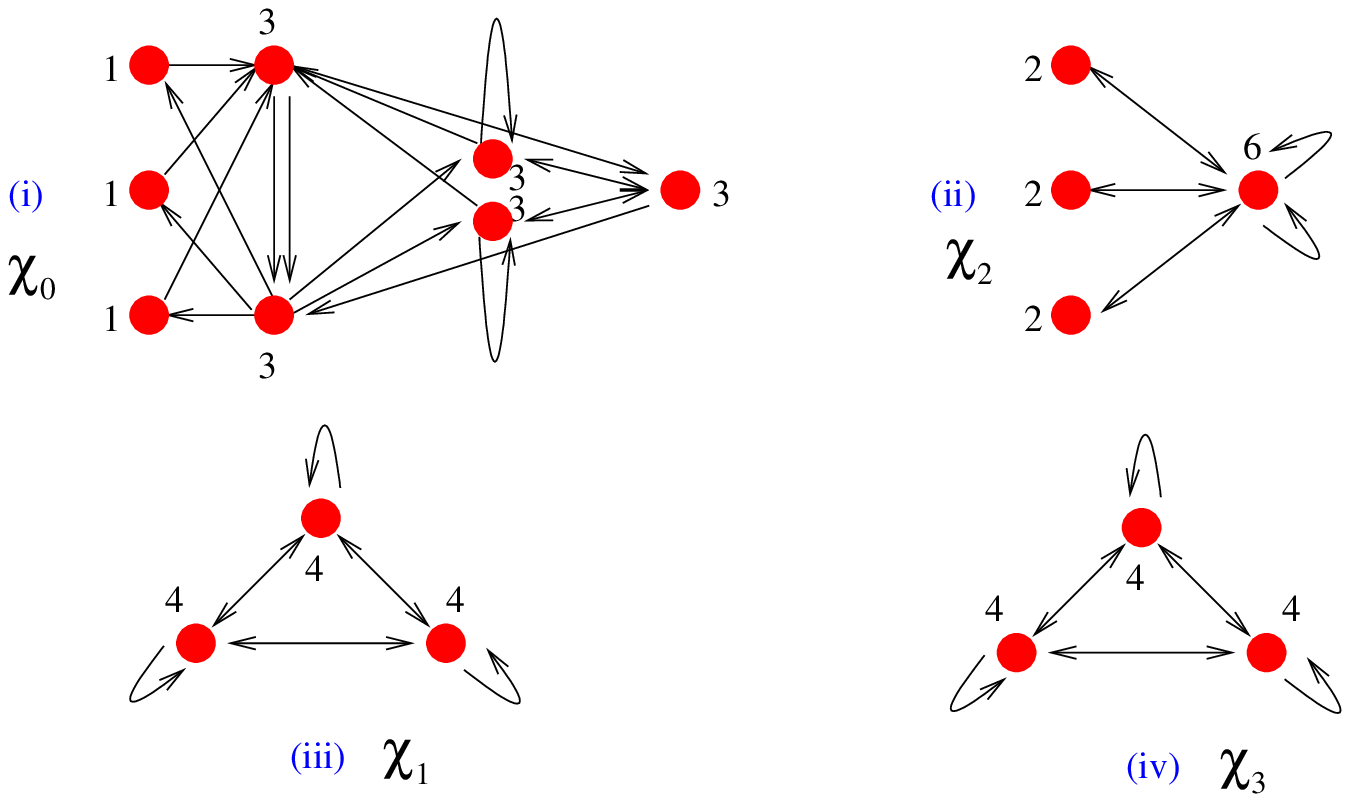,width=4.5in}
{\label{f:del3_4}
The quiver diagram of $\Delta(3\times 4^2)$: piece (i) is the ordinary quiver
of $\Delta(3\times 4^2)$ and pieces (ii-iv) have discrete torsion
turned on. We recall that the Schur Multiplier is $\IZ_4$.
}
\EPSFIGURE[ht]{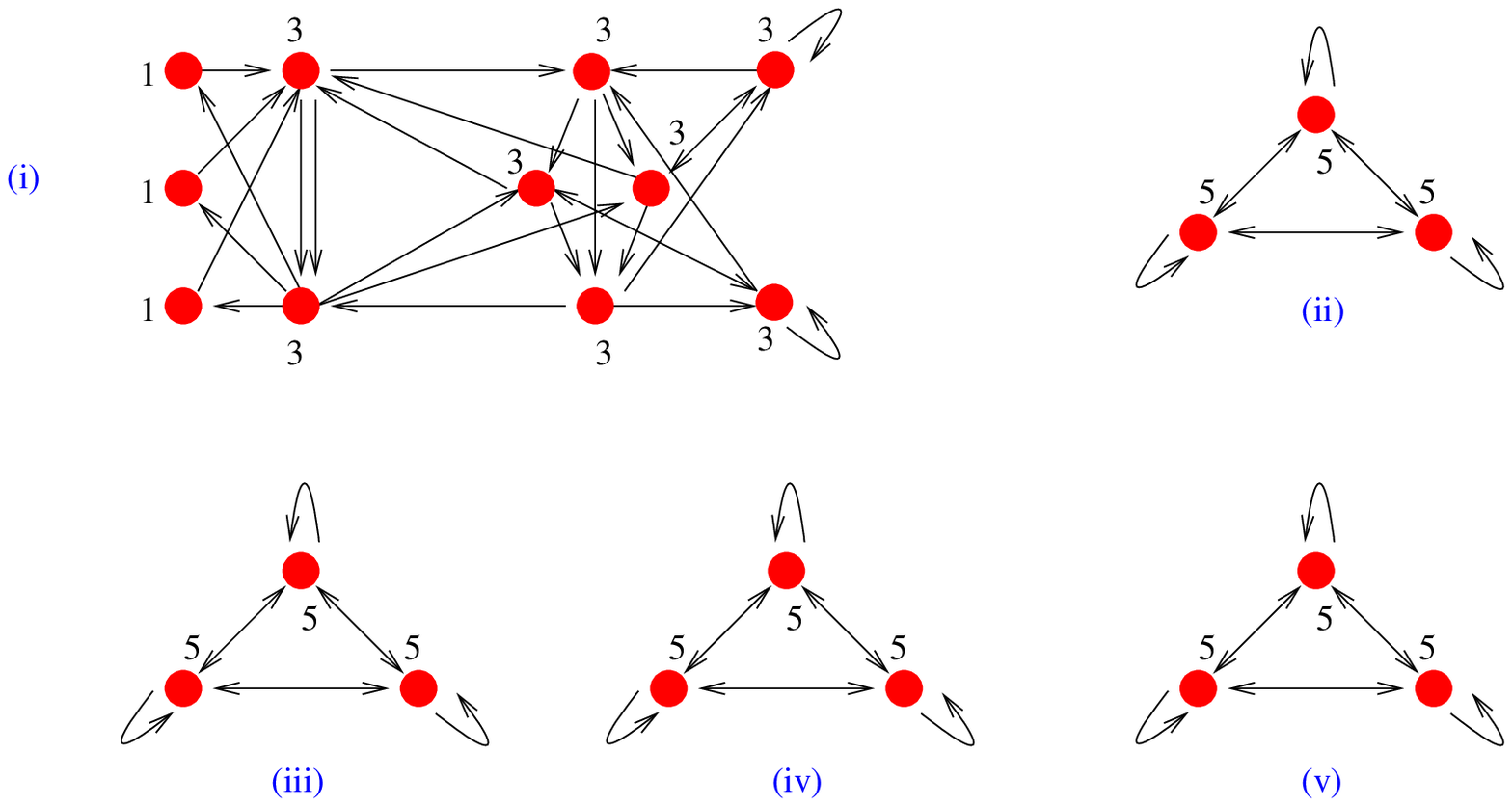,width=4.5in}
{\label{f:del3_5}
The quiver diagram of $\Delta(3\times 5^2)$: piece (i) is the ordinary quiver
of $\Delta(3\times 4^2)$ and pieces (ii-v) have discrete torsion
turned on. We recall that the Schur Multiplier is $\IZ_5$.
}
\section{Finding the Cocycle Values}
As advertised earlier, a useful by-product of the method is that we
can actually find the values of the 2-cocycles from the covering group.
Here we require even less information: only $G^*$ and not even
char$(G^*)$ is needed.

Let us recall some facts from Subsection 4.2.
The Schur multiplier is $A \subset Z(G^*)$, so every element therein has
its own conjugacy class in $G^*$. 
Hence for all linear representations of $G^*$, the character of $a_k
\in A$ will have the form $d \chi_i(a_k)$ where $d$ is the dimension
of that particular irrep of $G^*$ and $\chi_i(a_k)$ is the character
of ${a_k}$ in 
$A$ in its $i-th$ 1-dimensional irrep ($A$ is always Abelian and thus has
only 1-dimensional irreps).
This property has a very important consequence: merely reading out
the factor $\chi_i(a_k)$ from char$(G^*)$, we can
determine which linear representations will give which {\em projective}
representations of $G$. 
Indeed, two projective representations of $G$ belong to the same cocycle
{\em when and only when} the factor 
$\chi_i(a_k)$ is the same for every $a_k \in A$. 

Next we recall how to construct the matrix forms of projective
representations of $G$. 
$G^{\star}/A\equiv G$ implies that $G^*$ can be decomposed into cosets
$\bigcup\limits_{g \in G} g A$. Let $g a_i \in G^*$ correspond
canonically to $\tilde{g} \in G$ for some fixed $a_i \in A$; then
the matrix form of $\tilde{g}$ can be set to that of $g a_i$ and
furnishes the projective representation of $\tilde{g}$.
Different choices of $a_i$ will give different but projectively
equivalent projective representations of $G$.

Note that if we have
$\tilde{g_i}\tilde{g_j}=\tilde{g_k}$ in $G$, then in $G^*$, $g_i g_j= g_k
a_{ij}^{k}$, or
$(g_i a_i)(g_j a_j) = g_k a_k (a_{ij}^k a_i a_j a_k^{-1})$, but since
$(g_i a_i)$ is the projective matrix form for $\tilde{g_i} \in G$,
this is exactly the definition of the cocyle from which we read:
\beq
\label{cocycle}
\alpha(\tilde{g_i},\tilde{g_j})= \chi_p (a_{ij}^k a_i a_j a_k^{-1}),
\eeq
where $\chi_p(a)$ is the $p$-th character of the linear representation of
$a \in A$ defined above.

We can prove that \eref{cocycle} satisfies the 2-cocycle axioms (i)
and (ii).
Firstly notice that if $\tilde{g_i}= \II \in G$, 
we have $g_i = \II \in G^{\star}$; whence $a_{ij}^{k}= \delta_j^k~\forall~i$ and 
\[
(i)\quad \alpha(\II,\tilde{g_j})=\chi_p(\delta_j^k a_j a_k^{-1}) = \chi_p(\II)=1.
\]
Secondly if we assume that $\tilde{g_i}\tilde{g_j}=\tilde{g_q}$,
$\tilde{g_q}\tilde{g_k}=\tilde{g_h}$ and $\tilde{g_j}\tilde{g_k}=\tilde{g_l}$,
we have
$\alpha(\tilde{g_i},\tilde{g_j})\alpha(\tilde{g_i}\tilde{g_j},\tilde{g_k})
= \chi_p (a_{ij}^q a_i a_j a_q^{-1})
\chi_p (a_{qk}^h a_q a_k a_h^{-1}) = \chi_p(a_{ij}^k a_{qk}^h a_i a_j
a_k a_h^{-1})$

and
$\alpha(\tilde{g_i},\tilde{g_j}\tilde{g_k})\alpha(\tilde{g_j},\tilde{g_k})$
= $\chi_p (a_{jk}^l a_j a_k a_l^{-1}) \chi_p (a_{il}^h a_i a_l a_h^{-1})$
= $\chi_p (a_{il}^h a_{jk}^l a_i a_j a_k a_h^{-1})$.

However, because 
$(g_i g_j) g_k =g_q a_{ij}^q g_k= g_h a_{ij}^q a_{qk}^h
=g_i (g_j g_k)= g_i g_l a_{jk}^l = g_h a_{il}^h a_{jk}^l$ 
we have $a_{ij}^k a_{qk}^h=a_{il}^h a_{jk}^l$, and so
\[
(ii)\quad\alpha(\tilde{g_i},\tilde{g_j})\alpha(\tilde{g_i}\tilde{g_j},\tilde{g_k})
=\alpha(\tilde{g_i},\tilde{g_j}\tilde{g_k})\alpha(\tilde{g_j},\tilde{g_k}).
\]
Let us summarize the result. To read out the cocycle according to
\eref{cocycle} we need only two pieces of information: 
the choices of the
representative element in $G^*$ (i.e., $a_i\in A$), and
the definitions of $G^*$ which allows us to calculate the
$a_{ij}^k \in A$. We do not even need to calculate the character
table of $G^*$ to obtain the cocycle. Moreover, in a recent paper \cite{Craps} 
the values of cocycles are being used to construct 
boundary states. We hope our method shall make this above construction
easier.
\section{Conclusions and Prospects}
With the advent of discrete torsion in string theory, the hitherto
novel subject of projective representations has breathed out its
fragrance from mathematics into physics. However a short-coming has been
immediate: the necessary tools for physical computations have so far
been limited in the community due to the unavoidable fact that they,
if present in the mathematical literature, are obfuscated under often
too-technical theorems.

It has been the purpose of this writing, a companion to \cite{0010023},
to diminish the mystique of projective reprsentations in the context
of constructing gauge theories on D-branes probing orbifolds with
discrete torsion (non-trivial NS-NS B-fields) turned on.
In particular we have deviced an algorithm (Subsection 4.3),
culminating into Theorem 4.4, which computes the gauge theory data 
of the orbifold theory. The advantage of the method is its
directness: without recourse to the sophistry of twisted group algebras
and projective characters as had been suggested by some recent works
\cite{Doug-tor,DougFiol,AspinPles}, all methods so-far known in the
treatment of 
orbifolds (e.g. \cite{LNV,9811183}) are immediately generalisable.

We have shown that in computing the matter spectrum for an orbifold
$G$ with discrete torsion turned on, all that is
required is the ordinary charater table char$(G^*)$ of the covering
group $G^*$ of $G$. This table, together with the available character
table of $G$, immediately gives a quiver diagram which splits into
$|M(G)|$ disjoint pieces ($M(G)$ is the Schur Multiplier of
$G$), one of which is the ordinary quiver for $G$ and the rest, are
precisely the quivers for the various non-trivial discrete torsions.

A host of examples are then presented, demonstrating the systematic
power of the algorithm. In particular we have tabulated the results
for all the exceptional subgroups of $SU(3)$ as well as some first
members of the $\Delta$-series.

Directions for future research are self-evident. Brane setups for
orbifolds with discrete torsion have yet to be established. We
therefore need to investigate the groups satisfying BBM condition as
defined in \cite{9906031,9909125}, such as the intransitives of the form $\IZ
\times \IZ$ and $\IZ \times D$. Furthermore, we have given the
presentation of the covering groups of series such as $\IZ \times
\IZ$, $\IZ\times D,
\IZ\times E$ and $\Delta(3n^2), \Delta(6n^2)$. It will be interesting
to find the analytic results of the possible quivers.

More importantly, as we have reduced the problem of orbifolds with discrete
torsion to that of {\em linear} representations, we can instantly
extend the methods of \cite{LNV} to compute superpotentials and thence
further to an extensive and systematic study of non-commutative moduli
spaces in the spirit of \cite{BJL}.
So too do the families of toric varieties await us, methods utilised
in \cite{Tatar,0003085} eagerly anticipate their extension.
Indeed we have set a vessel adrift, it shall take the course in a vast
and unknown sea.

\chapter{Toric I: Toric Singularities and Toric Duality}
\section*{\center{{ Synopsis}}}
\label{chap:0003085}
\index{Singularity!Toric}
\index{Brane Probes!Toric}
{ T}he next three chapters shall constitute the last part of
Liber III; they shall be chiefly concerned with toric singularities
and D-brane probes thereupon. 

In this chapter, 
via partial resolution of Abelian orbifolds we present an algorithm
for extracting a consistent set of gauge theory data for an arbitrary
toric variety whose singularity a D-brane probes.
As illustrative examples, we
tabulate the matter content and superpotential for a D-brane living on
the toric del Pezzo surfaces as well as the zeroth Hirzebruch surface.
Moreover, we discuss the non-uniqueness of the general problem and present
examples of vastly different theories whose moduli spaces are
described by the same toric data. Our methods provide new tools
for calculating gauge theories which flow to the same universality
class in the IR. We shall call it ``Toric Duality'' \cite{0003085,DPF}.
\section{Introduction}
The study of D-branes as probes of geometry and topology of
space-time has by now been of wide practice (cf. e.g. \cite{Greene-Lec}).
In particular, the
analysis of the moduli space of gauge theories, their matter content,
superpotential and $\beta$-function, as world-volume theories of
D-branes sitting at geometrical singularities is still a widely
pursued topic. Since the pioneering work in \cite{DM}, where the
moduli and
matter content of D-branes probing ALE spaces had been extensively
investigated, much work ensued. The primary focus on (Abelian)
orbifold singularities of the type $\C^2/\Z_n$ was quickly
generalised using McKay's Correspondence, to arbitrary (non-Abelian)
orbifold singularities $\C^2/(\Gamma \subset SU(2))$, i.e., to
arbitrary ALE spaces, in \cite{Orb2}.

Several directions followed. With the realisation \cite{KS,Horizon} 
that these singularities provide various horizons,
\cite{DM,Orb2} was
quickly generalised to a treatment for arbitrary finite subgroups
$\Gamma \subset SU(N)$, i.e., to generic Gorenstein singularities, by
\cite{LNV}. The case of $SU(3)$ was then promptly studied in
\cite{9811183,Muto,Greene2} using this technique and a generalised McKay-type
Correspondence was proposed in \cite{9811183,9903056}.
Meanwhile, via T-duality transformations, certain orbifold
singularities can be
mapped to type II brane-setups in the fashion of \cite{HW}. The
relevant gauge theory data on the world volume can thereby be
conveniently read from configurations of NS-branes, D-brane stacks as
well as orientifold planes. For $\C^2$ orbifolds, the $A$ and $D$
series have been thus treated \cite{HW,Kapustin}, whereas for 
$\C^3$ orbifolds, the Abelian case of $\Z_k \times \Z_{k'}$ has been 
solved by the brane box models \cite{HZ,HU}. First examples
of non-Abelian $\C^3$ orbifolds have been addressed in
the previous chapters as well as \cite{ZD}.

Thus rests the status of orbifold theories. What we note in particular
is that once we specify the properties of the orbifold in terms of the
algebraic properties of the finite group, the gauge theory information
is easily extracted.
Of course, orbifolds are a small subclass of algebro-geometric
singularities. This is where we move on to {\bf toric varieties}.
Inspired by the linear $\sigma$-model approach of
\cite{GLSM}, which provides a rich structure of the moduli space,
especially in connexion with various geometrical phases of the
theory, the programme of utilising toric methods to study the
behaviour of the
gauge theory on D-branes which live on and hence resolve certain 
singularities was initiated in \cite{DGM}. In this light, toric
methods provide 
a powerful tool for studying the moduli space of the
gauge theory. In treating the F-flatness and D-flatness conditions for
the SUSY vacuum in conjunction, these methods show how branches of the
moduli space and hence phases of the theory may be parametrised by the
algebraic equations of the toric variety. Recent developments in
``brane diamonds,'' as an extension of the brane box rules,
have been providing great insight to such a wider class of toric
singularities, especially the generalised conifold, via blown-up
versions of the standard brane setups \cite{Aganagic}.
Indeed, with toric techniques much information could be
extracted as we can actually analytically describe patches of the
moduli space.

Now Abelian orbifolds have toric descriptions and the above
methodolgy is thus immediately applicable thereto. While bearing in
mind that though non-Abelian orbifolds have no toric descriptions, a single
physical D-brane has been placed on various general toric singularities.
Partial resolutions of $\C^3 / (\Z_2 \times \Z_2)$, such as the
\index{Conifold}
conifold and the suspended pinched point have been investigated in
\cite{Greene2,Muto3} and brane setups giving the field theory contents
are constructed by \cite{Unge,Park-con,Oh}. Groundwork for the next
family, coming from the toric orbifold $\C^3 / (\Z_3 \times \Z_3)$,
such as the del Pezzo surfaces and the zeroth Hirzebruch,
has been laid in \cite{Chris}. Essentially, given the gauge
theory data on the D-brane world volume, the procedure of transforming
this information (F and D terms) into toric data which parametrises
the classical moduli space is by now well-established.

One task is therefore immediately apparent to us: how do we proceed in
the reverse direction, i.e., {\em when we probe a toric singularity with a
D-brane, how do we know the gauge theory on its world-volume?}
We recall that in the case of orbifold theories, 
\cite{LNV} devised a general method to extract the
gauge theory data (matter content, superpotential etc.) from the
geometry data (the characters of the finite group $\Gamma$),
and {\it vice versa} given the geometry, brane-setups for example, conveniently
allow us to read out the gauge theory data. 
The same is not true for toric singularities, and the second half of
the above bi-directional convenience, namely, a general method which
allows us to treat the inverse problem of extracting gauge theory data
from toric data is yet pending, or at least not in circulation.

The reason for this shortcoming is, as we shall see later, that the
problem is highly non-unique. It is thus the purpose of this writing
to address this inverse problem: given the geometry data in terms of a
toric diagram, how does one read out (at least one) gauge theory data
in terms of the matter content and superpotential? We here present
precisely this algorithm which takes the matrices encoding the
singularity to the matrices encoding a good gauge theory of the
D-brane which probes the said singularity.

The structure of the chapter is as follows. In Section 2 we review the
procedure of proceeding from the gauge theory data to the toric data,
while establishing nomenclature. In Subsection 3.1, we demonstrate how
to extract the matter content and F-terms from the charge matrix of
the toric singularity. In Subsection 3.2, we exemplify our algorithm
with the well-known suspended pinched point before presenting in
detail in Subsection 3.3, the general algorithm of how to obtain the
gauge theory information from the toric data by the method of partial
resolutions. In Subsection 3.4, we show how to integrate back to
obtain the actual superpotential once the F-flatness equations are
extracted from the toric data.
Section 4 is then devoted to the illustration of our algorithm by
tabulating the D-terms and F-terms of D-brane world volume theory on
the toric del Pezzo surfaces and Hirzebruch zero.
We finally discuss in Section 5, the non-uniqueness of the inverse
problem and provide, through the studying of two types of ambiguities, ample
examples of rather different gauge theories flowing to the same toric
data. Discussions and future prospects are dealt with in Section 6.
\section{The Forward Procedure: Extracting Toric Data From Gauge Theories}
\index{Toric Variety!Forward Algorithm}
We shall here give a brief review of the procedures involved in
going from gauge theory data on the D-brane to toric data of the
singularity, using primarily the notation and concepts from \cite{DGM}.
In the course thereof special attention will be paid on 
how toric diagrams, SUSY fields and linear
$\sigma$-models weave together.

A stack of $n$ D-brane probes on algebraic singularities gives rise to
SUSY gauge
theories with product gauge groups resulting from the projection of
the $U(n)$ theory on the original stack by the geometrical structure
of the singularity.
For orbifolds $\C^k/\Gamma$, we can use the structure of the finite
group $\Gamma$ to
fabricate product $U(n_i)$ gauge groups \cite{DM,Orb2,LNV}.
For toric singularities,
since we have only (Abelian) $U(1)$ toroidal actions, we are so far
restricted to product $U(1)$ gauge groups\footnote{
	Proposals toward generalisations to D-brane stacks have
	been made \cite{Chris}.}.
In physical terms, we have
{\it a single D-brane probe}. Extensive work has been done in
\cite{Chris,DGM} to see how the geometrical structure of the variety
can be thus probed and how the gauge theory moduli may be
encoded. 
The subclass of toric singularities, namely Abelian orbifolds, has been
investigated to great detail \cite{DM,BL,DGM,Muto3,Chris}
and we shall make liberal usage of their properties throughout.

\begin{figure}
\centerline{\psfig{figure=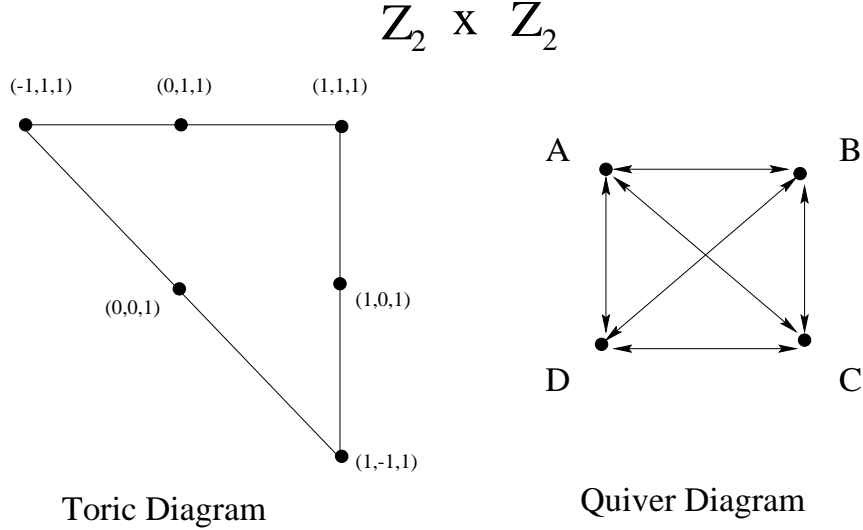,width=4.5in}}
\caption{The toric diagram for the singularity $\C^3/(\Z_2 \times
	\Z_2)$ and the quiver diagram for the gauge theory living on a
	D-brane probing it. We have labelled the nodes of the toric
	diagram by columns of $G_t$ and those of the quiver, with the
	gauge groups $U(1)_{\{A,B,C,D\}}$.}
\label{f:z2z2}
\end{figure}

Now let us consider the world-volume theory on the D-brane probe on
a toric singularity. Such a theory, as it is a SUSY gauge theory,
is characterised by its matter
content and interactions. The former is specified by quiver diagrams
which in turn give rise to {\bf D-term} equations; the latter is given
by a superpotential, whose partial derivatives with respect to the
various fields are the so-called {\bf F-term} equations. F and
D-flatness subsequently describe the (classical) moduli space of the theory.
The basic idea is that the D-term equations together with the FI-parametres,
in conjunction with the F-term equations, can be 
concatenated together into a matrix which gives the vectors forming
the dual cone of the toric variety which the D-branes probe. 
We summarise the algorithm of obtaining the toric data from the gauge
theory in the following, and to illuminate our abstraction and notation we
will use the simple example of the Abelian orbifold
$\C^3/(\Z_2 \times \Z_2)$ as given in \fref{f:z2z2}.
\begin{enumerate}
\index{Quivers!adjacency matrix}
\index{D-terms}
\item \underline{Quivers and D-Terms:}
	\begin{enumerate}
	\item The bi-fundamental matter content of the gauge theory
		can be conveniently
		encoded into a {\bf quiver diagram} ${\cal Q}$, which
		is simply the (possibly
		directed) graph whose {\bf adjacency matrix} $a_{ij}$
		is precisely the matrix of the
		bi-fundamentals. In the case of an Abelian
		orbifold\footnote{This is true
			for all orbifolds but of course only Abelian ones have
			known toric description.}
		prescribed by the group
		$\Gamma$, this diagram is the McKay Quiver (i.e., for
		the irreps $R_i$ 
		of $\Gamma$, $a_{ij}$ is such that $R \otimes R_i =
		\oplus_j a_{ij}R_j$
		for some fundamental representation $R$).
		We denote the set of nodes as ${\cal Q}_0 := \{v\}$
		and the set of the edges, ${\cal Q}_1 := \{a\}$.
		We let the number of nodes be $r$; for Abelian
		orbifolds, $r=|\Gamma|$ (and
		for generic orbifolds $r$ is the number of conjugacy
		classes of $\Gamma$).
		Also, we let the number of edges be $m$; this number
		depends on the number
		of supersymmetries which we have.
		The adjacency matrix (bi-fundamentals) is thus $r
		\times r$
		and the gauge group is $\prod\limits_{j=1}^r
		SU(w_j)$. For our example of $\Z_2 \times \Z_2$,
		$r=4$, 
		indexed as 4 gauge groups $U(1)_A \times U(1)_B 
		\times U(1)_C \times U(1)_D$ corresponding to the 4
		nodes, while $m=4 \times 3 = 12$,
		corresponding to the 12 arrows in \fref{f:z2z2}.
		The adjacency matrix for the quiver is
		${\tiny \left( \matrix{ 0 & 1 & 1 & 1 \cr
		1 & 0 & 1 & 1 \cr 1 & 1 & 0 & 1 \cr 1 & 1 & 1 & 0
    		\cr} \right)}$.
		Though for such simple
		\index{Conifold}
		examples as Abelian orbifolds and conifolds,
		brane setups and \cite{LNV} specify
		the values of $w_j$ as well as $a_{ij}$
		completely\footnote{For arbitrary
			orbifolds, $\sum\limits_j w_i n_i = |\Gamma|$
			where $n_i$ are 
			the dimensions of the irreps of $\Gamma$; for
			Abelian case, $n_i =1$.
			}, there is yet no
		discussion in the literature of obtaining the matter
		content and gauge group
		for generic toric varieties in a direct and systematic
		manner and 
		a partial purpose of this note is to present a
		solution thereof.
\index{Quivers!incidence matrix}
	\item From the $r \times r$ adjacency matrix, we construct a
		so-called $r \times m$ 
		{\bf incidence matrix} $d$ for ${\cal Q}$; this matrix is
		defined as 
		$d_{v,a} := \delta_{v,head(a)} - \delta_{v,tail(a)}$
		for $v \in {\cal Q}_0$
		and $a \in {\cal Q}_1$. Because each column of 
		$d$ must contain a 1, a $-1$ and the rest 0's by
		definition, one row
		of $d$ is always redundant; this physically signifies
		the elimination of
		an overall trivial $U(1)$ corresponding to the COM
		motion of the branes.
		Therefore we delete a row of $d$ to define the matrix
		$\Delta$ of dimensions $(r-1) \times m$; and we could
		always extract $d$ from $\Delta$ by adding a row so
		as to force each column to sum to zero. This matrix
		$\Delta$ thus contains almost as much information as
		$a_{ij}$ and once it is specified,
		the gauge group and matter content are also, with the
		exception that precise adjoints (those charged under
		the same gauge group factor and hence correspond to
		arrows that join a node to itself) are not manifest.
		For our example the $ 4 \times 12$ matrix $d$ is as
		follows and $\Delta$ is the top 3 rows:
		\[
		{\scriptsize
		d = \left(
		\begin{array}{c|cccccccccccc}
		&X_{AD}&X_{BC}&X_{CB}&X_{DA}&X_{AB}&X_{BA}&X_{CD}&
		X_{DC}&X_{AC}&X_{BD}&X_{CA}&X_{DB} \\
		A&-1&0&0&1&-1&1&0&0&-1&0&1&0 \\
		B&0&-1&1&0&1&-1&0&0&0&-1&0&1\\
		C&0&1&-1&0&0&0&-1&1&1&0&-1&0 \\ \hline 
		D&1&0&0&-1&0&0&1&-1&0&1&0&-1
		\end{array}
		\right)
		}
		\]
\index{Symplectic!moment map}
	\item The moment maps, arising in the sympletic-quotient
		language of the toric
		variety, are simply $\mu := d \cdot |x(a)|^2$ where
		$x(a)$ are the affine coordinates of the $\C^r$ for
		the torus $(\C^*)^r$ action.
		Physically, $x(a)$ are of course the bi-fundamentals
		in chiral multiplets (in our example they are 
		$X_{ij \in \{A,B,C,D\}}$ as labelled above)
		and the D-term equations for each
		$U(1)$ group is \cite{GLSM}
		\[
		D_i=-e^2(\sum_a d_{i a} |x(a)|^2- \zeta_i)
		\]
		with $\zeta_i$ the FI-parametres. In matrix form we
		have $\Delta \cdot |x(a)|^2 = \vec{\zeta}$ and see
		that D-flatness gives precisely the moment map.
		These $\zeta$-parametres will encode the resolution of
		the toric singularity as we shall shortly see.
	\end{enumerate}
\index{F-terms}
\item \underline{Monomials and F-Terms:}
	\begin{enumerate}
	\item From the super-potential $W$ of the SUSY gauge
		theory, one can write the F-Term equation
		as the system ${\partial \over {\partial X_j}} W = 0$.
		The remarkable fact is that we could solve the said
		system of equations and
		express the $m$ fields $X_i$ in terms of $r+2$ parametres
		$v_j$ which can be summarised by a matrix $K$.
		\beq
		\label{kmatrix}
		X_i =\prod_j v_j^{K_{ij}}, \qquad i=1,2,..,m;~~~j=1,2,..,r+2
		\eeq
		This matrix $K$ of dimensions $m \times (r+2)$ is the
		analogue of
		$\Delta$ in the sense that it encodes the F-terms and
		superpotential
		as $\Delta$ encodes the D-terms and the matter
		content.
		In the language of toric geometry $K$ defines a
		cone\footnote
		{
		We should be careful in this
		definition. Strictly speaking we have a lattice ${\bf
		M}=\Z^{r+2}$ with its dual lattice ${\bf N} \cong
		\Z^{r+2}$. Now let there be a set of $\Z_+$-independent
		vectors $\{\vec{k}_i\} \in {\bf M}$ and a cone is defined to
		be generated by these vectors as
		$\sigma := \{\sum_i a_i \vec{k}_i~|~a_i \in
		\R_{\ge 0}\}$; Our ${\bf M_+}$ should be ${\bf M}
		\cap \sigma$. In
		much of the literature ${\bf M_+}$ is taken to be
		simply ${\bf M'_+} := \{\sum_i a_i \vec{k}_i~|~a_i \in
		\Z_{\ge 0}\}$ in which case we must make sure that any
		lattice point contained in ${\bf M_+}$ but not in
		${\bf M'_+}$ must be counted as an independent
		generator and be added to the set of generators
		$\{\vec{k}_i\}$. After including all such points we
		would have ${\bf M'_+} = {\bf M_+}$.
		Throughout our analyses, our cone defined by $K$ as
		well the dual cone $T$ will be constituted by such a
		complete set of generators.
		}
		${\bf M_+}$ : a non-negative linear combination of $m$
		vectors $\vec{K_i}$ in an integral lattice $\Z^{r+2}$.\\
		For our example, the superpotential is
		\[
		\begin{array}{c}
		W = X_{AC}X_{CD}X_{DA} - X_{AC}X_{CB}X_{BA} + X_{CA}X_{AB}X_{BC}
		- X_{CA}X_{AD}X_{DC} \\
		+ X_{BD}X_{DC}X_{CB} - X_{BD}X_{DA}X_{AB}
		- X_{DB}X_{BC}X_{CD},
		\end{array}
		\]
		giving us 12 F-term equations and with the manifold of
		solutions parametrisable by $4+2$ new fields, whereby
		giving us the $12 \times 6$ matrix (we here show the
		transpose thereof, thus the horizontal direction
		corresponds to the original fields $X_i$ and the
		vertical, $v_j$):
		\[
		{\scriptsize
		K^t = \left(
		\begin{array}{c|cccccccccccc}
		& X_{AC} & X_{BD} & X_{CA} & X_{DB} & X_{AB} & X_{BA}
		& X_{CD} & X_{DC} & X_{AD} & X_{BC} & X_{CB} & X_{DA}
		\\ \hline
		v_1 =X_{AC} & 1 & 0 & 0 & 1 & 1 & 0 & 0 & 1 & 0 & 0 & 0 & 0 \\
		v_2 =X_{BD} & 0 & 1 & 1 & 0 & -1& 0 & 0 & -1& 0 & 0 & 0 & 0 \\
		v_3 =X_{BA} & 0 & 0 & 0 & 0 & 0 & 1 & 0 & 1 & 0 & 1 & 0 & 1 \\
		v_4 =X_{CD} & 0 & 0 & 0 & 0 & 1 & 0 & 1 & 0 & 0 & -1& 0 & -1\\
		v_5 =X_{AD} & 0 & 0 & -1& -1& 0 & 0 & 0 & 0 & 1 & 1 & 0 & 0 \\
		v_6 =X_{CB} & 0 & 0 & 1 & 1 & 0 & 0 & 0 & 0 & 0 & 0 & 1 & 1
		\end{array}
		\right).
		}
		\]
		For example, the third column reads $X_{CA} = v_2
		v_5^{-1} v_6$, i.e., $X_{AD} X_{CA} = X_{BD} X_{CB}$,
		which the the F-flatness condition ${\partial W \over
		{\partial X_{DC}}=0}$. The details of obtaining $W$
		and $K$ from each other are discussed in
		\cite{DGM,Chris} and Subsection 3.4.
	\item We let $T$ be the space of (integral) vectors dual to
		$K$, i.e., $K \cdot T \ge 0$ for all entries; this gives an
		$(r+2) \times c$ matrix for some positive integer
		$c$. Geometrically, this is the definition of a dual
		cone ${\bf N_+}$ composed of vectors $\vec{T_i}$
		such that $\vec{K} \cdot \vec{T} \ge 0$. The physical
		meaning for doing so is that $K$ may have negative
		entries which may give rise to unwanted singularities
		and hence we define a new set of $c$ fields
		$p_i$ ({\it a priori} we do not know the number $c$
		and we present the standard algorithm of finding dual
		cones in Appendix \ref{append:0003085}). 
		Thus we reduce (\ref{kmatrix})
		further into
		\beq
		\label{p_i}
			v_j=\prod_{\alpha}p_{\alpha}^{T_{j\alpha}}
		\eeq
		whereby giving $X_i = 
		\prod_j v_j^{K_{ij}}= \prod_{\alpha} 
		p_{\alpha}^{\sum_j K_{ij} T_{j\alpha}}$ with 
		$\sum_j K_{ij} T_{j\alpha}\geq 0$.
		For our $\Z_2 \times Z_2$ example, $c=9$ and
		\[
		{\scriptsize
		T_{j\alpha} = \left(
		\begin{array}{c|ccccccccc}
		& p_1 & p_2 & p_3 & p_4 & p_5 & p_6 & p_7
			& p_8 & p_9 \\ \hline
		X_{AC} & 1 & 1 & 0 & 0 & 0 & 0 & 0 & 0 & 1 \cr
		X_{BD} & 0 & 1 & 1 & 0 & 0 & 0 & 0 & 0 & 1 \cr
		X_{BA} & 0 & 0 & 1 & 1 & 1 & 0 & 0 & 0 & 0 \cr
		X_{CD} & 0 & 0 & 1 & 0 & 1 & 1 & 0 & 0 & 0 \cr
		X_{AD} & 0 & 0 & 0 & 0 & 0 & 1 & 1 & 0 & 1 \cr
		X_{CB} & 0 & 0 & 0 & 0 & 0 & 1 & 1 & 1 & 0
		\end{array}
		\right)
		}
		\]
	\item These new variables $p_\alpha$ are the
		matter fields in Witten's linear $\sigma$-model.
		How are these fields charged? We have
		written $r+2$ fields $v_j$ in terms of $c$ fields
		$p_\alpha$, and hence need $c-(r+2)$ relations to reduce
		the independent variables. Such a reduction can be
		done via the introduction of the new gauge group
		$U(1)^{c-(r+2)}$ acting on the $p_i$'s so as to give a
		new set of D-terms. The charges of these fields can be
		written as $Q_{k \alpha}$. The gauge invariance condition of
		$v_i$ under $U(1)^{c-(r+2)}$, by (\ref{p_i}), demands
		that the $(c - r - 2) \times c$ matrix
		$Q$ is such that $\sum_{\alpha}
		T_{j\alpha}Q_{k\alpha}=0$. This then defines for us our
		charge matrix $Q$ which is the cokernel of $T$:
		\[
		TQ^t=(T_{j \alpha})(Q_{k\alpha})^t = 0, \qquad
			j=1,..,r+2;~~\alpha=1,..,c;~~k=1,..,(c-r-2)
		\]
		For our example, the charge matrix is $(9-4-2) \times
		9$ and one choice is
		${\scriptsize Q_{k \alpha} = \left( 
		\matrix{
		0 & 0 & 0 & 1 & -1& 1 & -1& 0 & 0 \cr
		0 & 1 & 0 & 0 & 0 & 0 & 1 & -1& -1 \cr
		1 &-1 & 1 & 0 & -1& 0 & 0 & 0 & 0 \cr }
		\right)}$.
	\item In the linear $\sigma$-model language, the F-terms and
		D-terms can be treated in the same footing, i.e., as
		the D-terms (moment map) of the new fields $p_\alpha$; with the
		crucial difference being that the former must be set
		exactly to zero\footnote{Strictly speaking, we could
			have an F-term set to a non-zero constant. An example
			of this situation could be when there is a term $a
			\phi + \phi \tilde{Q} Q$ in the superpotential for
			some chargeless field $\phi$ and charged fields
			$\tilde{Q}$ and $Q$. The F-term for $\phi$
			reads $\tilde{Q} Q = -
a$ and not 0. However, in
			our context $\phi$ behaves like an integration
			constant and
			for our purposes, F-terms are set exactly to zero.}
		while the latter are to be resolved by
		arbitrary FI-parameters.\\
		Therefore in addition to finding the charge matrix $Q$
		for the new fields $p_\alpha$ coming from the original
		F-terms as done above, we must also find the
		corresponding charge matrix $Q_D$ for the $p_i$ coming from
		the original D-terms.
		We can find $Q_D$ in two steps. Firstly, we know the
		charge matrix for $X_i$ under $U(1)^{r-1}$, which is
		$\Delta$. By (\ref{kmatrix}), we transform the charges
		to that of the $v_j$'s, by introducing an $(r-1) \times
		(r+2)$ matrix $V$ so that $V \cdot K^t =
		\Delta$. To see this, let the charges of $v_j$ be
		$V_{lj}$ then by (\ref{kmatrix}) we have $\Delta_{li}
		= \sum\limits_j V_{lj} K_{ij} = V \cdot K^t$.
		A convenient $V$ which does so for our $\Z_2 \times
                \Z_2$ example is
                ${\scriptsize \left( 
                \matrix{
                1 & 0 & -1& 0 & 1 & 0 \cr
                0 & 1 & 1 & 0 & 0 & -1\cr
                -1& 0 & 0 & 1 & 0 & 1 \cr}
                \right)_{(4-1)\times (4+2)}}$.
		Secondly, we use (\ref{p_i}) to transform the
		charges from $v_j$'s to our final variables $p_\alpha$'s,
		which is done by introducing an
		$(r + 2) \times c$ matrix $U_{j \alpha}$ so that
                $U \cdot T^t = {\rm Id}_{(r+2) \times (r+2)}$. In our
		example, one choice for $U$ is
                ${\scriptsize U_{j \alpha} = \left( 
                \matrix{
                1 & 0 & 0 & 0 & 0 & 0 & 0 & 0 & 0 \cr
                -1& 1 & 0 & 0 & 0 & 0 & 0 & 0 & 0 \cr
                0 & 0 & 0 & 1 & 0 & 0 & 0 & 0 & 0 \cr
                0 & 0 & 0 & 0 & 0 & 1 & -1& 0 & 0 \cr
                0 & -1& 0 & 0 & 0 & 0 & 0 & 0 & 1 \cr
                0 & 0 & 0 & 0 & 0 & 0 & 0 & 1 & 0 }
                \right)_{(4+2) \times 9}}$.\\
		Threfore, combining the two steps, we obtain $Q_D = V \cdot
		U$ and in our example, ${\scriptsize (V \cdot U)_{l
		\alpha} = \left(
		\matrix{ 1 & -1& 0 & -1& 0 & 0 & 0 & 0 & 1 \cr
			-1& 1 & 0 & 1 & 0 & 0 & 0 & -1& 0 \cr
			-1& 0 & 0 & 0 & 0 & 1 & -1& 1 & 0} \right)}$.
	\end{enumerate}
\item Thus equipped with the information from the two sides: the
	F-terms and D-terms, and with the two required charge matrices
	$Q$ and  $V \cdot U$ obtained, finally we concatenate them to
	give a $(c-3) \times c$ matrix $Q_t$.
	The transpose of the kernel of $Q_t$, with (possible
	repeated columns) gives rise to a matrix $G_t$.
	The columns of this resulting $G_t$
	then define the vertices of the toric diagram
	describing the polynomial corresponding to the singularity on which
	we initially placed our D-branes.
	Once again for our example,
	$Q_t = {\scriptsize \left(
	\begin{array}{ccccccccc|c}
	0 & 0 & 0 & 1 & -1& 1 & -1& 0 & 0 &  0\cr
	0 & 1 & 0 & 0 & 0 & 0 & 1 & -1& -1&  0 \cr
	1 & -1& 1 & 0 & -1& 0 & 0 & 0 & 0 &  0 \cr \hline
	1 & -1& 0 & -1& 0 & 0 & 0 & 0 & 1 & \zeta_1\cr
	-1& 1 & 0 & 1 & 0 & 0 & 0 & -1& 0 & \zeta_2\cr
	-1& 0 & 0 & 0 & 0 & 1 & -1& 1 & 0 & \zeta_3
	\end{array}
	\right)}$ and
	$G_t = {\scriptsize \left(
	\matrix{
	0 & 1 & 0 & 0 & -1& 0 & 1 & 1 & 1 \cr
	1 & 1 & 1 & 0 & 1 & 0 & -1& 0 & 0 \cr
	1 & 1 & 1 & 1 & 1 & 1 & 1 & 1 & 1}
	\right)}$. The columns of $G_t$, up to repetition, are
	precisely marked in the toric diagram for $\Z_2 \times \Z_2$
	in \fref{f:z2z2}.
\end{enumerate}

Thus we have gone from the F-terms and the D-terms of the gauge theory
to the nodes of the toric diagram. In accordance with
\cite{Fulton}, $G_t$ gives the algebraic variety whose equation is
given by the maximal ideal in the polynomial ring \\
$\C[YZ,XYZ,Z,X^{-1}YZ,XY^{-1}Z,XZ]$ (the exponents $(i,j,k)$
in $X^iY^jZ^k$ are exactly the columns), which is $uvw=s^2$, upon
defining $u=(YZ)(XYZ)^2(Z)(XZ)^2; v=(YZ)^2(Z)^2(X^{-1}YZ)^2;
w=(Z)^2(XY^{-1}Z)(XZ)^2$ and \\$s = (YZ)^2 (XYZ) (Z)^2 (X^{-1}YZ)
(XY^{-1}Z) (XZ)^2$; this is precisely $\C^3/(\Z_2 \times \Z_2)$.
In physical terms this equation
parametrises the moduli space obtained from the F and D flatness of
the gauge theory.

We remark two issues here. In the case of there being no
superpotential we could still define $K$-matrix. In this case, with
there being no F-terms, we simply take $K$ to be the identity. This
gives $T=$Id and $Q=0$. Furthermore $U$ becomes Id and $V=\Delta$,
whereby making $Q_t = \Delta$ as expected because all information
should now be contained in the D-terms.
Moreover, we note that the very reason we can construct a $K$-matrix
is that all of the equations in the F-terms we deal with are in the
form $\prod\limits_i X_i^{a_i} = \prod\limits_j X_j^{b_j}$; this
holds in general if every field $X_i$ appears twice and precisely twice in the
superpotential. More generic situations would so far transcend the
limitations of toric techniques.

Schematically, our procedure presented above at length, what it
means is as follows: we begin with
two pieces of physical data: (1) matrix $d$ from the quiver encoding
the gauge groups and D-terms and (2) matrix $K$ encoding the F-term
equations. From these we extract the matrix $G_t$ containing the toric
data by the flow-chart:
\index{Toric Variety!Forward Algorithm}
\[
\begin{array}{ccccccc}
\mbox{Quiver} \rightarrow d	& \rightarrow	&\Delta	& & & & \\
	&	&\downarrow	&	&	&	&	\\
\mbox{F-Terms} \rightarrow K	& \stackrel{V \cdot K^t =
	\Delta}{\rightarrow}
		& V	 & & & & \\
\downarrow	&	& \downarrow	& & & & \\
T = {\rm Dual}(K)	& \stackrel{U \cdot T^t = {\rm
	Id}}{\rightarrow} & U & \rightarrow & VU & &\\
\downarrow	&	&	&	& \downarrow	& & \\
Q = [{\rm Ker}(T)]^t	&	& \longrightarrow	& & Q_t =
	\left( \begin{array}{c}
					Q \\ VU \end{array} \right) &
						\rightarrow & G_t =
	[{\rm Ker}(Q_t)]^t \\

\end{array}
\]
\index{Toric Variety!Inverse Algorithm}
\section{The Inverse Procedure: Extracting Gauge Theory Information
	from Toric Data}
As outlined above we see that wherever possible, the gauge theory of a D-brane
probe on certain singularities such as Abelian orbifolds, conifolds, etc., can
be conveniently encoded into the matrix $Q_t$ which essentially concatenates
the information contained in the D-terms and F-terms of the original gauge theory.
The cokernel of this matrix is then a list of vectors which prescribes the toric
diagram corresponding to the singularity.
It is natural to question ourselves whether the converse could be done, i.e.,
whether given an arbitrary singularity which affords a toric description, we could
obtain the gauge theory living on the D-brane which probes the said singularity.
This is the inverse problem we projected to solve in the introduction.

\index{D-terms}
\index{F-terms}
\index{Brane Probes!Toric}
\subsection{Quiver Diagrams and F-terms from Toric Diagrams}
Our result must be two-fold: first, we must be able to extract the
D-terms, or
in other words the quiver diagram which then gives the gauge group and
matter content; second, we must extract the F-terms, which we can
subsequently integrate back to give the superpotential. These two
pieces of data then suffice to specify the gauge theory.
Essentially we wish to trace the arrows in the above flow-chart from
$G_t$ back to $\Delta$ and $K$. The general methodology seems
straightforward:
\begin{enumerate}
\item Read the column-vectors describing the nodes of the given toric
	diagram, repeat the
	appropriate columns to obtain $G_t$ and then set $Q_t={\rm
	Coker}(G_t)$; 
\item Separate the D-term ($V \cdot U$) and F-term ($Q_t$) portions
	from $Q_t$;
\item From the definition of $Q$, we obtain\footnote{As
		mentioned before we must ensure that such a $T$ be chosen
		with a complete set of $\Z_+$-independent generators;}
	$T = \ker (Q)$.
\item Farka's Theorem \cite{Fulton} guarantees that the dual of a
	convex polytope remains convex whence we could invert and have
	$K = {\rm Dual}(T^t)$; Moreover the duality theorem gives that
	Dual(Dual($K)) = K$, thereby facilitating the inverse procedure. 
\item Definitions $U \cdot T^t = \rm{Id}$ and 
	$V \cdot K^t = \Delta$ $\Rightarrow$
	$(V \cdot U) \cdot (T^t \cdot K^t) = \Delta$.
\end{enumerate}

We see therefore that once the appropriate $Q_t$ has been found, the relations
\beq
\label{Delta}
K= {\rm Dual}(T^t) \qquad \Delta = (V \cdot U) \cdot (T^t \cdot K^t)
\eeq
retrieve our desired $K$ and $\Delta$.
The only setback of course is that the appropriate $Q_t$
is NOT usually found.
Two ambiguities are immediately apparent to us:
(A) In step 1 above, there is really no way to know a priori which of
	the vectors we should repeat when writing into the $G_t$ matrix;
(B) In step 2, to separate the D-terms and the F-terms, i.e.,
	which rows constitute $Q$ and which constitute $V \cdot U$
	within $Q_t$, seems arbitrary.
We shall in the last section discuss these ambiguities in more detail and
actually perceive it to be a matter of interest. Meanwhile, in light thereof,
we must find an alternative, to find a canonical method which avoids
such ambiguities and gives us a consistent gauge theory which has such
well-behaved properties as having only bi-fundamentals etc.; this is where
we appeal to partial resolutions.

Another reason for this canonical method is compelling. The astute
reader may question as to how could we guarantee, in our mathematical
excursion of performing the inverse procedure, that the gauge theory
we obtain at the end of the day is one that still lives on the
world-volume of a D-brane probe? Indeed, if we na\"{\i}vely traced back the
arrows in the flow-chart, bearing in mind the said ambiguities, we
have no {\it a fortiori} guarantee that we have a brane theory at
all. However, the method via partial resolution of Abelian orbifolds
(which are themselves toric) does give us assurance. When we are
careful in tuning the FI-parametres so as to stay inside
cone-partitions of the space of these parametres (and avoid flop
transitions) we do still have the resulting theory being physical
\cite{Chris}. Essentially this means that
with prudence we tune the FI-parametres in the allowed domains from
a parent orbifold theory, thereby giving a subsector theory which
still lives on the D-brane probe and is well-behaved.
Such tuning we shall practice in the following.

The virtues of this appeal to resolutions
are thus twofold: not only do we avoid ambiguities, we are further
endowed with physical theories. Let us thereby present this canonical
mathod.

\index{Resolution!blow up}
\index{Resolution!partial}
%
\subsection{A Canonical Method: Partial Resolutions of Abelian Orbifolds}
%
Our programme is standard \cite{Chris}: theories on the Abelian orbifold
singularity of the form $\C^k / \Gamma$ for 
$\Gamma(k,n) = \Z_n \times \Z_n \times ... \Z_n$ ($k-1$ times) are well
studied. The complete information (and in
particular the full $Q_t$ matrix) for $\Gamma(k,n)$ is well known: $k=2$ is
the elliptic model, $k=3$, the Brane Box, etc.
In the toric context, $k=2$ has been analysed in great detail by
\cite{DM}, $k=3,n=2$ in e.g. \cite{Unge,Park-con,Oh}, $k=3,n=3$ in
\cite{Chris}. Now we know that given any toric diagram of
dimension $k$, we can embed it into such a $\Gamma(k,n)$-orbifold for
some sufficiently large $n$; and we choose the smallest such $n$ which
suffices. This embedding is always possible because the toric diagram
for the latter is the $k$-simplex of length $n$ enclosing lattice
points and any toric diagram, being a collection of lattice points,
can be obtained therefrom via deletions of a subset of points.
This procedure is known torically as {\bf partial resolutions} of $\Gamma(k,n)$.
The crux of our algorithm is that the deletions in the toric diagram
corresponds to the turning-on of the FI-parametres, and which in turn
induces a method to determine a $Q_t$ matrix for our original
singularity from that of $\Gamma(n,k)$.

\begin{figure}
\centerline{\psfig{figure=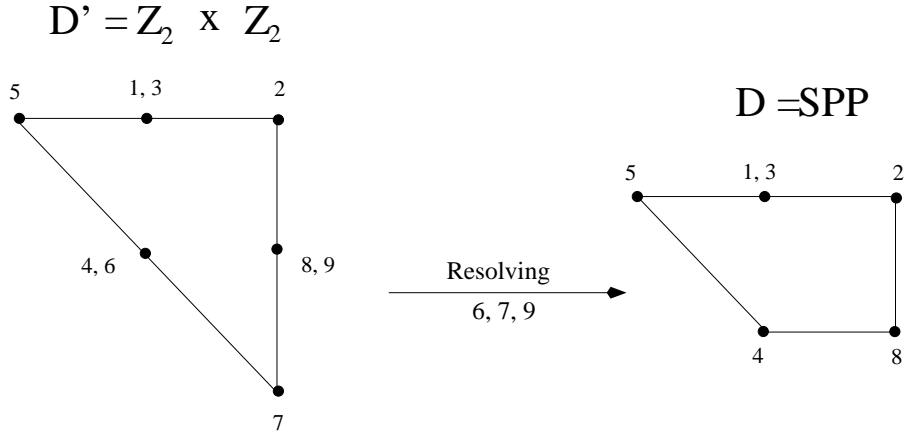,width=4.7in}}
\caption{The toric diagram showing the resolution of the 
	$\C^3 / (\Z_2 \times \Z_2)$ singularity to the suspended pinch
	point (SPP). The numbers $i$ at the nodes refer to the $i$-th
	column of the matrix $G_t$ and physically correspond 
	to the fields $p_i$ in the linear $\sigma$-model.}
\label{f:SPP}
\end{figure}

We shall first turn to an illustrative example of the suspended
pinched point singularity (SPP) and then move on to discuss generalities.
\index{Conifold}
\index{Suspended Pinched Point}
The SPP and conifold as resolutions of $\Gamma(3,2)=\Z_2 \times \Z_2$
have been extensively studied
in \cite{Park-con}. The SPP, given by $xy = zw^2$, can be obtained from
the $\Gamma(3,2)$
orbifold, $xyz = w^2$, by a single $\IP^1$ blow-up. This is shown torically in
\fref{f:SPP}. Without further ado let us demonstrate our procedure.
\begin{enumerate}
\item \underline{Embedding into $\Z_2 \times \Z_2$:}
	Given the toric diagram $D$ of SPP, we recognise that it can be
	embedded minimally into the diagram $D'$ of $\Z_2 \times
	\Z_2$. Now information on $D'$ is readily at hand
	\cite{Park-con}, as presented
	in the previous section. Let us re-capitulate:
	\[
	{\scriptsize
	Q'_t := \left(
	\matrix{p_1 & p_2 & p_3 & p_4 & p_5 & p_6 & p_7 & p_8 & p_9 & \cr
		0 & 0 & 0 & 1 & -1 & 1 & -1 & 0 & 0 & 0 \cr
		0 & 1 & 0 & 0 & 0 & 0 & 1 & -1 & -1 & 0 \cr
		1 & -1 & 1 & 0 & -1 & 0 & 0 & 0 & 0 & 0 \cr
		1 & -1 & 0 & -1 & 0 & 0 & 0 & 0 & 1 & \zeta_1 \cr
		-1 & 1 & 0 & 1 & 0 & 0 & 0 & -1 & 0 & \zeta_2 \cr
		-1 & 0 & 0 & 0 & 0 & 1 & -1 & 1 & 0 & \zeta_3 \cr}
	\right),
	}
	\]
	and
	\[
	{\scriptsize
	G'_t := {\rm coker}(Q'_t) = \left(
	\matrix{p_1 & p_2 & p_3 & p_4 & p_5 & p_6 & p_7 & p_8 & p_9 \cr
		0 & 1 & 0 & 0 & -1 & 0 & 1 & 1 & 1 \cr
		1 & 1 & 1 & 0 & 1 & 0 & -1 & 0 & 0 \cr
		1 & 1 & 1 & 1 & 1 & 1 & 1 & 1 & 1 \cr}
	\right),
	}
	\]
	which is drawn in \fref{f:z2z2}. The fact that the last row of
	$G_t$ has the same number (i.e., these three-vectors are all
	co-planar) ensures that $D'$ is Calabi-Yau \cite{Greene-Lec}.
	Incidentally, it would be very helpful for one to catalogue the
	list of $Q_t$ matrices of $\Gamma(3,n)$ for $n = 2,3...$ which
	would suffice for all local toric singularities of Calabi-Yau
	threefolds.\\
	In the above definition of $Q'_t$ we have included an extra column
	$(0,0,0,\zeta_1,\zeta_2,\zeta_3)$ so as to specify that the
	first three rows of $Q'_t$ are F-terms (and hence exactly
	zero) while the last three rows are D-terms
	(and hence resolved by FI-parametres $\zeta_{1,2,3}$). We
	adhere to the notation in \cite{Park-con} and label the columns
	(linear $\sigma$-model fields) as $p_1 ... p_9$;
	this is shown in \fref{f:SPP}.
\item \underline{Determining the Fields to Resolve by Tuning $\zeta$:}
	We note that if we turn on a single FI-parametre we
	would arrive at the SPP; this is the resolution of $D'$ to
	$D$. The subtlety is that one may need to eliminate more than
	merely the 7th column as there is more than one field
	attributed to each node in the toric diagram and eliminating
	column 7 some other columns corresponding to the adjacent
	nodes (namely out of 4,6,8 and 9) may also be eliminated.
	We need a judicious choice of $\zeta$ for a consistent blowup.
	To do so we must solve for fields $p_{1,..,9}$ and tune the
	$\zeta$-parametres such that 
	at least $p_7$ acquires non-zero VEV (and whereby resolved).
	Recalling that the D-term equations are actually linear
	equations in the modulus-squared of the fields, we shall
	henceforth define $x_i := |p_i|^2$ and consider linear-systems
	therein. 
	Therefore we perform Gaussian row-reduction on $Q'$ and solve
	all fields in terms of $x_7$ to give:
	$\vec{x} = \{x_1,x_2,x_1 + \zeta_2 + \zeta_3,
	{{2\,x_1 - x_2 + x_7 - \zeta_1 + \zeta_2}\over 2},
	2\,x_1 - x_2 + \zeta_2 + \zeta_3,
	{{2\,x_1 - x_2 + x_7 + \zeta_1 + \zeta_2 + 2\,\zeta_3}\over 2},x_7,
	{{x_2 + x_7 - \zeta_1 - \zeta_2}\over 2},
	{{x_2 + x_7 + \zeta_1 + \zeta_2}\over 2} \}$.\\
	The nodes far away from $p_7$ are clearly unaffected by the
	resolution, thus the fields corresponding thereto continue to have
	zero VEV. This means we solve the above set of solutions
	$\vec{x}$ once again, setting $x_{5,1,3,2} = 0$, with
	$\zeta_{1,2,3}$ being the variables, giving upon back
	substitution,
	$\vec{x} = \{0, 0, 0, {{x_7 - \zeta_1 - \zeta_3}\over 2}, 0, {{x_7 + 
	\zeta_1 + \zeta_3}\over 2}, x_7, {{x_7 - \zeta_1 + \zeta_3}\over 2},\\
	{{x_7 + \zeta_1 - \zeta_3}\over 2} \}$. Now we have an arbitrary
	choice and we set $\zeta_3 = 0$ and $x_7 = \zeta_1$ to make
	$p_4$ and $p_8$ have zero VEV. This makes $p_{6,7,9}$ our candidate
	for fields to be resolved and seems perfectly reasonable
	observing \fref{f:SPP}. The constraint on our choice is that
	all solutions must be $\ge 0$ (since the $x_i$'s are VEV-squared).
\item \underline{Solving for $G_t$:}
	We are now clear what the resolution requires of us: in order to
	remove node $p_7$ from $D'$ to give the SPP, 
	we must also resolve 6, 7 and 9.
	Therefore we immediately obtain $G_t$ by
	directly removing the said columns from $G'_t$:
	\[
	{\scriptsize
	G_t := {\rm coker}(Q_t) = \left(
	\matrix{p_1 & p_2 & p_3 & p_4 & p_5 & p_8 \cr 
		0 & 1 & 0 & 0 & -1 & 1 \cr
		1 & 1 & 1 & 0 & 1 & 0 \cr
		1 & 1 & 1 & 1 & 1 & 1  \cr}
	\right),
	}
	\]
	the columns of which give the toric diagram $D$ of the SPP, as
	shown in \fref{f:SPP}.
\item \underline{Solving for $Q_t$:}
	Now we must perform linear combination on the rows of $Q'_t$ to
	obtain $Q_t$ so as to force columns 6, 7 and 9 zero.
	The following constraints must be born in mind. 
	Because $G_t$ has 6 columns and
	3 rows and is in the null space of $Q_t$, which itself must
	have $9-3$ columns (having eliminated $p_{6,7,9}$), we must
	have $6-3=3$ rows for $Q_t$. Also, the row
	containing $\zeta_1$ must be eliminated as this is precisely 
	our resolution chosen above (we recall that the FI-parametres
	are such that $\zeta_{2,3} = 0$ and are hence unresolved,
	while $\zeta_1 > 0$ and must be removed from the D-terms for
	SPP).\\
	We systematically proceed. Let there be variables
	$\{a_{i=1,..,6}\}$ so that $y := \sum_i a_i {\rm
	row}_i(Q'_t)$
	is a row of $Q_t$. Then (a) the 6th, 7th and 9th
	columns of $y$ must be set to 0 and moreover (b) with these
	columns removed $y$ must be in the nullspace spanned by the
	rows of $G_t$. We note of course that since $Q'_t$ was in the
	nullspace of $G'_t$ initially, that the operation of
	row-combinations is closed
	within a nullspace, and that the columns to be set to 0 in
	$Q'_t$ to give $Q_t$ are precisely those removed in $G'_t$ to
	give $G_t$, condition (a) automatically implies (b).
	This condition (a) translates to the equations
	$\{a_1+a_6=0,-a_1+a_2-a_6=0,-a_2+a_4=0 \}$ which
	afford the solution $a_1 = -a_6; a_2=a_4=0$. The fact that
	$a_4=0$ is comforting, because it eliminates the row containing
	$\zeta_1$. We choose $a_1 = 1$. Furthermore
	we must keep row 5 as $\zeta_2$ is yet unresolved
	(thereby setting $a_5 = 1$).
	This already gives two of the 3 anticipated rows of $Q_t$: row$_5$ and
	row$_1$ - row$_6$. The remaining row must corresponds to an
	F-term since we have exhausted the D-terms, this we choose
	to be the only remaining variable: $a_3 = 1$.
	Consequently, we arrive at the matrix
	\[
	{\scriptsize
	Q_t = \left(
	\matrix{p_1 & p_2 & p_3 & p_4 & p_5 & p_8 & \cr
		1 & -1 & 1 & 0 & -1 & 0 & 0 \cr
		-1 & 1 & 0 & 1 & 0 &  -1 & \zeta_2 \cr
		-1 & 0 & 0 & -1 & 1 & 1 & \zeta_3 \cr }		
	\right).
	}
	\]
\item \underline{Obtaining $K$ and $\Delta$ Matrices:}
	The hard work is now done. We now recognise from $Q_t$ that
	$Q = (1,-1,1,0,-1,0)$, giving
	\[{\scriptsize
	T_{j \alpha} := {\rm ker}(Q) = \left(
	\matrix{0 & 0 & 0 & 0 & 0 & 1 \cr
		1 & 0 & 0 & 0 & 1 & 0 \cr
		0 & 0 & 0 & 1 & 0 & 0 \cr
		-1 & 0 & 1 & 0 & 0 & 0 \cr
		1 & 1 & 0 & 0 & 0 & 0 \cr}
	\right);
	\qquad
	K^t := {\rm Dual}(T^t) = \left(
	\matrix{1 & 0 & 0 & 0 & 0 & 0 \cr
		0 & 0 & 1 & 0 & 1 & 0 \cr
		0 & 1 & 0 & 0 & 0 & 0 \cr
		0 & 0 & 1 & 1 & 0 & 0 \cr
		0 & 0 & 0 & 1 & 0 & 1 \cr}
	\right).
	}\]	
	Subsequently we obtain
	${\tiny
	T^t \cdot K^t = \left(
	\matrix{0 & 0 & 0 & 0 & 1 & 1 \cr
		0 & 0 & 0 & 1 & 0 & 1 \cr
		0 & 0 & 1 & 1 & 0 & 0 \cr
		0 & 1 & 0 & 0 & 0 & 0 \cr
		0 & 0 & 1 & 0 & 1 & 0 \cr
		1 & 0 & 0 & 0 & 0 & 0 \cr}
	\right),
	}$
	which we do observe indeed to have every entry positive semi-definite.
	Furthermore we recognise from $Q_t$ that
	${\tiny
	V \cdot U = \left(
	\matrix{-1 & 1 & 0 & 1 & 0 & -1 \cr
		-1 & 0 & 0 & -1 & 1 & 1 \cr}
	\right),
	}$
	whence we obtain at last, using (\ref{Delta}),
	\[
	{\scriptsize
	\Delta = \left(
	\matrix{-1 & 1 & 0 & 1 & -1 & 0 \cr
		1 & -1 & 1 & 0 & 0 & -1 \cr}
	\right)
	\qquad \Rightarrow \qquad
	d = \left(
	\begin{array}{c|cccccc}
			& X_1 & X_2 & X_3 & X_4 &  X_5 & X_6 \\
		U(1)_A & -1 & 1 & 0 & 1 & -1 & 0 \cr
		U(1)_B & 1 & -1 & 1 & 0 & 0 & -1 \cr \hline
		U(1)_C & 0 & 0 & -1 & -1 & 1 & 1 \end{array}
	\right),
	}
	\]
	giving us the quiver diagram (included in \fref{f:SPPquiver}
	for reference), matter content and gauge group
	of a D-brane probe on SPP in agreement with \cite{Park-con}. We
	shall show in the ensuing sections that the superpotential we
	extract has similar accordance.
\end{enumerate}
\begin{figure}
\centerline{\psfig{figure=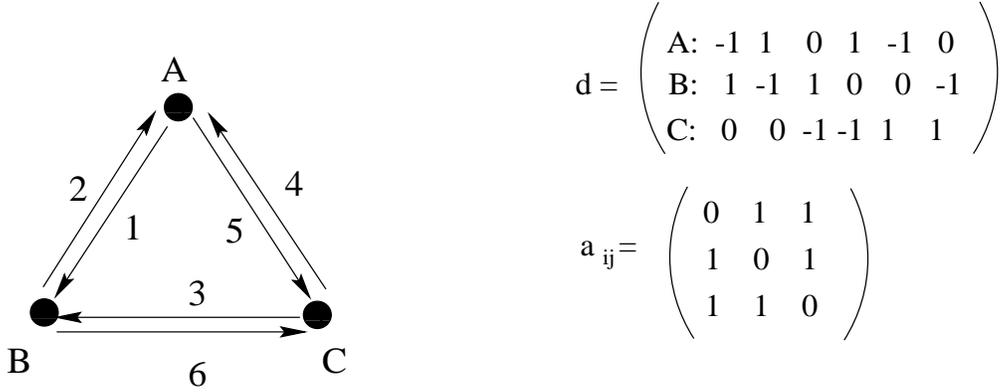,width=5.5in}}
\caption{The quiver diagram showing the matter content of a D-brane
	probing the SPP singularity. We have not marked in the
	chargeless field $\phi$ (what in a non-Abelian theory would
	become an adjoint) because thus far the toric techniques do not
	yet know how to handle such adjoints.}
\label{f:SPPquiver}
\end{figure}
\index{Toric Variety!Inverse Algorithm}
\subsection{The General Algorithm for the Inverse Problem}
Having indulged ourselves in this illustrative example of the SPP, we
proceed to outline the general methodology of obtaining the gauge
theory data from the toric diagram.
\begin{enumerate}
\item \underline{Embedding into $\C^k / (\Z_n)^{k-1}$:}
	We are given a toric diagram $D$ describing an algebraic variety of
	complex dimension $k$ (usually we are concerned with local
	Calabi-Yau singularities of $k=2,3$ so that branes living
	thereon give ${\cal N} = 2,1$ gauge theories). We immediately 
	observe that $D$ could always be embedded into $D'$,
	the toric diagram of the orbifold 
	$\C^k / (\Z_n)^{k-1}$ for some sufficiently large integer $n$.
	The matrices $Q'_t$ and $G'_t$ for $D'$ are standard.
	Moreover we know that the matrix $G_t$ for our original variety
	$D$ must be a submatrix of $G'_t$. Equipped with $Q'_t$ and  $G'_t$
	our task is to obtain $Q_t$; and as an additional check we
	could verify that $Q_t$ is indeed in the nullspace of $G_t$.
\item \underline{Determining the Fields to Resolve by Tuning $\zeta$:}
	$Q'_t$ is a $k \times a$ matrix\footnote{We henceforth
		understand that there is
		an extra column of zeroes and $\zeta$'s.}
	(because $D'$ and $D$ are dimension $k$) for some
	$a$; $G'_t$, being its nullspace, is thus $(a-k) \times a$.
	$D$ is a partial resolution of $D'$. 
	In the SPP example above, we performed a single resolution
	by turning on one FI-parametre, generically however, we could
	turn on as many $\zeta$'s as the embedding permits.
	Therefore we let $G_t$ be $(a-k) \times (a-b)$
	for some $b$ which depends on the number of resolutions.
	Subsequently the $Q_t$ we need is $(k-b) \times (a-b)$.\\
	Now $b$ is determined directly by examining $D'$ and $D$;
	it is precisely the number of fields $p$ associated to those
	nodes in $D'$ we wish to eliminate to arrive at $D$.
	Exactly which $b$ columns are to be eliminated is determined
	thus: we perform Gaussian row-reduction on $Q'_t$ so as to
	solve the $k$ linear-equations in $a$ variables $x_i :=
	|p_i|^2$, with F-terms set to 0 and D-terms to FI-parametres.
	The $a$ variables are then expressed in terms of the
	$\zeta_i$'s and the set $B$ of $x_i$'s corresponding to the nodes
	which we definitely know will disappear as we resolve $D'
	\rightarrow D$. The subtlety is that in eliminating $B$, some
	other fields may also acquire non zero VEV and be eliminated;
	mathematically this means that Order$(B) < b$.\\
	Now we make a judicious choice of which fields will remain and
	set them to zero and impose this further on the solution
	$x_{i=1,..,a} = x_i(\zeta_j;B)$ from above until Order$(B) =
	b$, i.e., until we have found all the fields we need to
	eliminate. We know this occurs and that our choice was correct
	when all $x_i \ge 0$ with those equaling 0 corresponding to
	fields we do not wish to eliminate as can be observed from the
	toric diagram. If not, we modify our initial choice and repeat
	until satisfaction. This procedure then determines the $b$
	columns which we wish to eliminate from $Q'_t$.
\item \underline{Solving for $G_t$ and $Q_t$:}
	Knowing the fields to eliminate, we must thus perform linear
	combinations on the $k$ rows of $Q'_t$
	to obtain the $k-b$ rows of $Q_t$ based upon the two constraints
	that (1) the $b$ columns must be all reduced to zero 
	(and thus the nodes can be removed) and that
	(2) the $k-b$ rows (with $b$ columns removed) are in the
	nullspace of $G_t$. As mentioned in our SPP example, condition
	(1) guarantees (2) automatically.\\
	In other words, we need to solve for $k$ variables 
	$\{ x_{i = 1,..,k} \}$ such that
	\beq
	\label{elimcol}
	\sum\limits_{i=1}^k x_i~(Q'_t)_{ij} = 0 
		\qquad {\rm for}~~ j=p_1, p_2,  ... p_b \in B.
	\eeq
	Moreover, we immediately obtain $G_t$ by eliminating the $b$
	columns from $G'_t$. Indeed, as discussed earlier,
	(\ref{elimcol}) implies that
	$\sum\limits_{i=1}^k \sum\limits_{j \ne p_{1...b}} 
	x_i~(Q'_t)_{ij}~(G_t)_{mj} = 0$ for $m=1,...,a-k$ and hence
	guarantees that the $Q_t$ we obtain is in the nullspace of $G_t$.\\
	We could phrase equation (\ref{elimcol}) for $x_i$ in matrix
	notation and directly evaluate
	\beq
	\label{qt}
	Q_t = NullSpace(W)^t \cdot \tilde{Q'_t}
	\eeq
	where $\tilde{Q'_t}$ is $Q'_t$ with the appropriate columns
	($p_{1...b}$) removed and $W$ is the matrix constructed from
	the deleted columns.
\item \underline{Obtaining the $K$ Matrix (F-term):}
	Having obtained the $(k-b) \times (a-b)$ matrix $Q_t$ for the
	original variety $D$, we proceed with ease. Reading from the
	extraneous column of FI-parametres, we recognise matrices $Q$ 
	(corresponding to the rows that have zero in the extraneous
	column) and $V \cdot U$ (corresponding to those with
	combinations of the unresolved $\zeta$'s in the last column).
	We let $V \cdot U$ be $c \times (a-b)$ whereby
	making $Q$ of dimension $(k-b-c) \times (a-b)$. The number $c$
 	is easily read from the embedding of $D$ into $D'$ as the
	number of unresolved FI-parametres.\\
	From $Q$, we compute the kernel $T$, a matrix of dimensions 
	$(a-b)-(k-b-c) \times (a-b) = (a-k+c) \times (a-b)$ as well as
	the matrix $K^t$ of dimensions $(a-k+c) \times d$ describing the
	dual cone to that spanned by the columns of $T$. The integer
	$d$ is uniquely determined from the dimensions of $T$ in
	accordance with the algorithm of finding dual cones presented
	in Appendix \ref{append:0003085}. 
	From these two matrices we compute $T^t \cdot
	K^t$, of dimension $(a-b) \times d$.
\item \underline{Obtaining the $\Delta$ Matrix (D-term):}
	Finally, we use (\ref{Delta}) to compute
	$(V \cdot U) \cdot (T^t \cdot K^t)$, arriving
	at our desired matrix $\Delta$ of dimensions $c \times d$,
	the incidence matrix of our quiver diagram. The number of
	gauge groups we have is therefore $c+1$ and the
	number of bi-fundamentals, $d$.\\
	Of course one may dispute that finding the kernel $T$ of $Q$
	is highly non-unique as
	any basis change in the null-space would give an equally valid $T$.
	This is indeed so. However we note that it is really the
 	combination $T^t \cdot K^t$ that we need. This is a dot-product
	in disguise, and by the very definition of
	the dual cone, this combination remains invariant under basis changes.
	Therefore this step of obtaining the quiver $\Delta$ from the
	charge matrix $Q_t$ is a unique	procedure.
\end{enumerate}
\subsection{Obtaining the Superpotential}
Having noticed that the matter content can be conveniently obtained,
we proceed to address the interactions, i.e., the
F-terms, which require a little more care. The matrix $K$ which our
algorithm extracts encodes the F-term equations and must at least be such
that they could be integrated back to a single function: the
{\it superpotential.}

Reading the possible F-flatness equations from $K$ is {\it ipso facto}
straight-forward. The subtlety exists in how to find the right
candidate among many different linear relations.
As mentioned earlier, $K$ has dimensions $m \times (r-2)$ with $m$
corresponding to the fields that will finally manifest in the
superpotential, $r-2$, the fields that solve them according to
(\ref{kmatrix}) and (\ref{p_i}); of course, $m \ge r-2$.
Therefore we have
$r-2$ vectors in $\Z^m$, giving generically $m-r+2$ linear relations
among them. Say we have ${\rm row}_1 + {\rm row}_3 -  {\rm row}_7 =
0$, then we simply write down $X_1 X_3 = X_7$ as one of the candidate
F-terms.
In general, a relation $\sum\limits_i a_i K_{ij} = 0$ with $a_i \in
\Z$ implies an F-term
$\prod\limits_i X_i^{a_i} = 1$ in accordance with (\ref{kmatrix}).
Of course, to find all the linear relations, we simply find the
$\Z$-nullspace of $K^t$ of dimension $m-r+2$.

Here a great ambiguity exists, as in our previous calculations of
nullspaces: any linear combinations therewithin may suffice to give a
new relation as a candidate F-term\footnote{Indeed each linear
	relation gives a possible
	candidate and we seek the correct ones. For the sake of
	clarity we shall call candidates ``relations'' and reserve the
	term ``F-term'' for a successful candidate.}.
Thus educated guesses are called for in order to find
the set of linear relations which may be most conveniently integrated
back into the superpotential. Ideally, we wish this back-integration
procedure to involve no extraneous fields (i.e., integration
constants\footnote{By constants we really mean functions since we are
	dealing with systems of partial differential equations.})
other than the $m$ fields which appear in the K-matrix. Indeed, as we shall
see, this wish may not always be granted and sometimes 
we must include new fields. In this case,
the whole moduli space of the gauge theory will be larger than
the one encoded by our toric data and the new fields parametrise new
branches of the moduli in the theory.

Let us return to the SPP example to enlighten ourselves before generalising.
We recall from subsection 3.2, that ${\scriptsize K=\left(
	\begin{array}{c|cccccc}
		& X_1 & X_2 & X_3 & X_4 & X_5 & X_6 \\ \hline
		v_1 & 1 & 0 & 0 & 0 & 0 & 0 \cr
		v_2 & 0 & 0 & 1 & 0 & 1 & 0 \cr
		v_3 & 0 & 1 & 0 & 0 & 0 & 0 \cr
		v_4 & 0 & 0 & 1 & 1 & 0 & 0 \cr
		v_5 & 0 & 0 & 0 & 1 & 0 & 1  \end{array}
\right)}$ from which we can read out only one relation
$X_3 X_6 - X_4 X_5=0$ using the rule described in the paragraph
above. Of course there can be only one relation because the nullspace
of $K^t$ is of dimension $6-5=1$.

Next we must calculate the charge under the gauge groups which this
term carries. We must ensure that the superpotential, being a term in a
Lagrangian, be a gauge invariant, i.e., carries no overall charge
under $\Delta$.
From ${\scriptsize d = \left(
	\begin{array}{c|cccccc}
	& X_1 &  X_2 & X_3 & X_4 & X_5 & X_6 \\ \hline
U(1)_A  & -1 &   1 &  0 & 1 &  -1 &  0 \\
U(1)_B  & 1  &  -1 &  1 & 0 &   0 & -1 \\
U(1)_C  & 0  &   0 &  -1& -1 &  1 &  1 \end{array} \right)}$
we find the charge of $X_3 X_6$ to be $(q_A,q_B,q_C) =
(0+0,1+(-1),(-1)+1) = (0,0,0)$; of course by our very construction,
$X_4 X_5$ has the same charge. Now we have two choices:
(a) to try to write the superpotential using only the six fields; or 
(b) to include some new field $\phi$ which also has charge $(0,0,0)$.
For (a) we can try the ansatz  $W=X_1 X_2 (X_3 X_6 -X_4  X_5)$ which
does give our F-term upon partial derivative with respect to $X_1$ or
$X_2$. However, we would also have a new F-term $X_1 X_2 X_3=0$ by
${\partial \over {\partial X_6}}$, which is inconsistent with our $K$
since columns 1, 2 and 3 certainly do not add to 0.

This leaves us with option (b), i.e., $W= \phi (X_3 X_6 -X_4  X_5)$ say.
In this case, when $\phi=0$ we not only obtain our F-term,
we need not even correct the matter content $\Delta$. This branch of
the moduli space is that of our original theory.
However, when $\phi \neq 0$, we must have $X_3=X_4=X_5=X_6=0$. 
Now the D-terms read $|X_1|^2 - |X_2|^2 = -\zeta_1 = \zeta_2$,
so the moduli space is: $\{\phi \in \C, X_1 \in \C \}$ such that
$\zeta_1+\zeta_2=0$ for otherwise there would be no moduli at all.
We see that we obtain another branch of moduli space.
As remarked before, this is a general phenomenon when we include
new fields: the whole moduli space will be larger than the one encoded
by the toric data. As a check, we see that our example is exactly that
given in \cite{Park-con}, after the identification with their notation,
$Y_{12} \rightarrow X_6, X_{24}\rightarrow X_3, Z_{23}\rightarrow X_1,
Z_{32}\rightarrow X_2, Y_{34}\rightarrow X_4, X_{13}\rightarrow X_5,
Z_{41}\rightarrow\phi$ and $(X_1 X_2 - \phi) \rightarrow \phi$.
We note that if we were studying a non-Abelian extension to the toric
theory, as by brane setups (e.g. \cite{Park-con}) or by stacks
of probes (in progress from \cite{Chris}),
the chargeless field $\phi$ would manifest
as an adjoint field thereby modifying our quiver diagram. Of course since
the study of toric methods in physics is so far restricted to product
$U(1)$ gauge groups, such complexities do not arise. To avoid
confusion we shall henceforth mark only the bi-fundamentals in our quiver
diagrams but will write the chargeless fields explicit in the
superpotential.

Our agreement with the results of \cite{Park-con} is very reassuring. It
gives an excellent example demonstrating that our canonical resolution
technique and the inverse algorithm do indeed, in response to what was
posited earlier, give a theory living on
a D-brane probing the SPP (T-dual to the setup in \cite{Park-con}).
However, there is a subtle point we would like to mention. There
exists an ambiguity in writing the superpotential when the chargeless
field $\phi$ is involved. Our algorithm gives $W = \phi (X_3 X_6 - X_4
X_6)$ while \cite{Park-con} gives $W = (X_1 X_2 - \phi)(X_3 X_6 - X_4
X_6)$. Even though they have identical moduli, it is the latter which
is used for the brane setup. Indeed, the toric methods by definition 
(in defining $\Delta$ from $a_{ij}$) do not handle chargeless fields
and hence we have ambiguities. Fortunately our later
examples will not involve such fields.

The above example of the SPP was a na\"{\i}ve one as we need only to
accommodate a single F-term. We move on to a more complicated example.
Suppose we are now given
${\tiny
d = \left(
\matrix{ & X_1&X_2&X_3&X_4&X_5&X_6&X_7&X_8&X_9&X_{10} \cr
 A&-1&0&0&-1&0&0&0&1&0&1 \cr B&1&-1&0&0&0&-1&0&0&1&0 \cr C&0&0&1&
  0&1&0&1&-1&-1&-1 \cr D&0&1&-1&1&-1&1&-1&0&0&0 \cr  }
\right)
}$ and
${\tiny
K = \left(
\matrix{
X_1&X_2&X_3&X_4&X_5&X_6&X_7&X_8&X_9&X_{10} \cr
 1&0&1&0&0&0&1&0&0&0 \cr 0&1&1&0&0&0&0&1&0&0 \cr 
1&0&0&1&0&0&0&0&1&0 \cr 0&1&0&1&0&1&0&0
  &0&0 \cr 0&0&1&0&1&0&1&0&0&0 \cr 0&0&0&0&0&1&1&0&0&1 \cr  }
\right)
}$. We shall see in the next section, that these arise for
the del Pezzo 1 surface. 
Now the nullspace of $K$ has dimension $10-6=4$, we could obtain a host of
relations from various linear combinations in this space.
One relation is obvious: $X_2 X_7 - X_3 X_6 = 0$. The charge it
carries is $(q_A,q_B,q_C,q_D) = (0+0,-1+0,0+1,1+(-1)) = (0,-1,1,0)$
which cancels that of $X_9$. Hence $X_9 (X_2 X_7 - X_3 X_6)$ could be
a term in $W$. Now ${\partial \over{\partial X_2}}$ thereof gives $X_7
X_9$ and from $K$ we see that $X_7 X_9 - X_1 X_5 X_{10} = 0$,
therefore, $- X_1 X_2 X_5 X_{10}$ could be another term in $W$. We repeat
this procedure, generating new terms as we proceed and introducing new
fields where necessary. We are fortunate that in this case we can
actually reproduce all F-terms without recourse to artificial
insertions of new fields: $W = X_{2} X_{7} X_{9} - X_{3} X_{6} X_{9}
- X_{4} X_{8} X_{7} - X_{1} X_{2} X_{5} X_{10} + X_{3} X_{4} X_{10} 
+ X_{1} X_{5} X_{6} X_{8}$.

Enlightened by these examples, let us return to some remarks upon
generalities.
Making all the exponents of the fields positive, the F-terms can then
be written as 
\beq
\label{F-term}
\prod\limits_i X_i^{a_i} = \prod\limits_j X_j^{b_j},
\eeq
with $a_i, b_j \in \Z^+$. Indeed if we were to have another field
$X_k$ such that $k \not\in \{i\}, \{j\}$ then the term
$X_k \left(\prod\limits_i X_i^{a_i} -  \prod\limits_j X_j^{b_j}
\right)$, on the condition that $X_k$ appears only this once, must be an
additive term in the superpotential $W$. This is because the
F-flatness condition ${\partial W \over {\partial {X_k}}} = 0$ implies
(\ref{F-term}) immediately. Of course judicious observations are called
for to (A) find appropriate relations (\ref{F-term}) and (B) find 
$X_k$ among our $m$ fields. Indeed (B) may not even be possible and
new fields may be forced to be introduced, whereby making the moduli
space of the gauge theory larger than that encodable by the toric data.

In addition, we must ensure that each term in $W$ be chargeless under
the product gauge groups.
What this means for us is that 
for each of the terms $X_k \left(\prod\limits_i X_i^{a_i} -  
\prod\limits_j X_j^{b_j}\right)$ we must have
${\rm Charge}_s(X_k) + \sum\limits_i a_i {\rm Charge}_s(X_i) = 0$
for $s=1,..,r$ indexing through our $r$ gauge group factors (we note
that by our very construction, for each gauge group, the charges 
for $\prod\limits_i X_i^{a_i}$ and for $\prod\limits_j X_j^{b_j}$ 
are equal).
If $X_k$ in fact cannot be found among our $m$ fields, it must be
introduced as a new field $\phi$ with appropriate charge.
Therefore with each such relation (\ref{F-term}) read from $K$, we
iteratively perform this said procedure, checking
$\Delta_{sk} + \sum\limits_i a_i \Delta_{si} = 0$ at each step, until
a satisfactory superpotential is reached. The right choices throughout
demands constant vigilance and astuteness.

\index{del Pezzo Surfaces}
\section{An Illustrative Example: the Toric del Pezzo Surfaces}
As the $\C^3 / (\Z_2 \times \Z_2)$ resolutions were studied in great
detail in \cite{Park-con}, we shall use the data from \cite{Chris} to
demonstrate the algorithm of finding the gauge theory from toric
diagrams extensively presented in the previous section.

\begin{figure}
\centerline{\psfig{figure=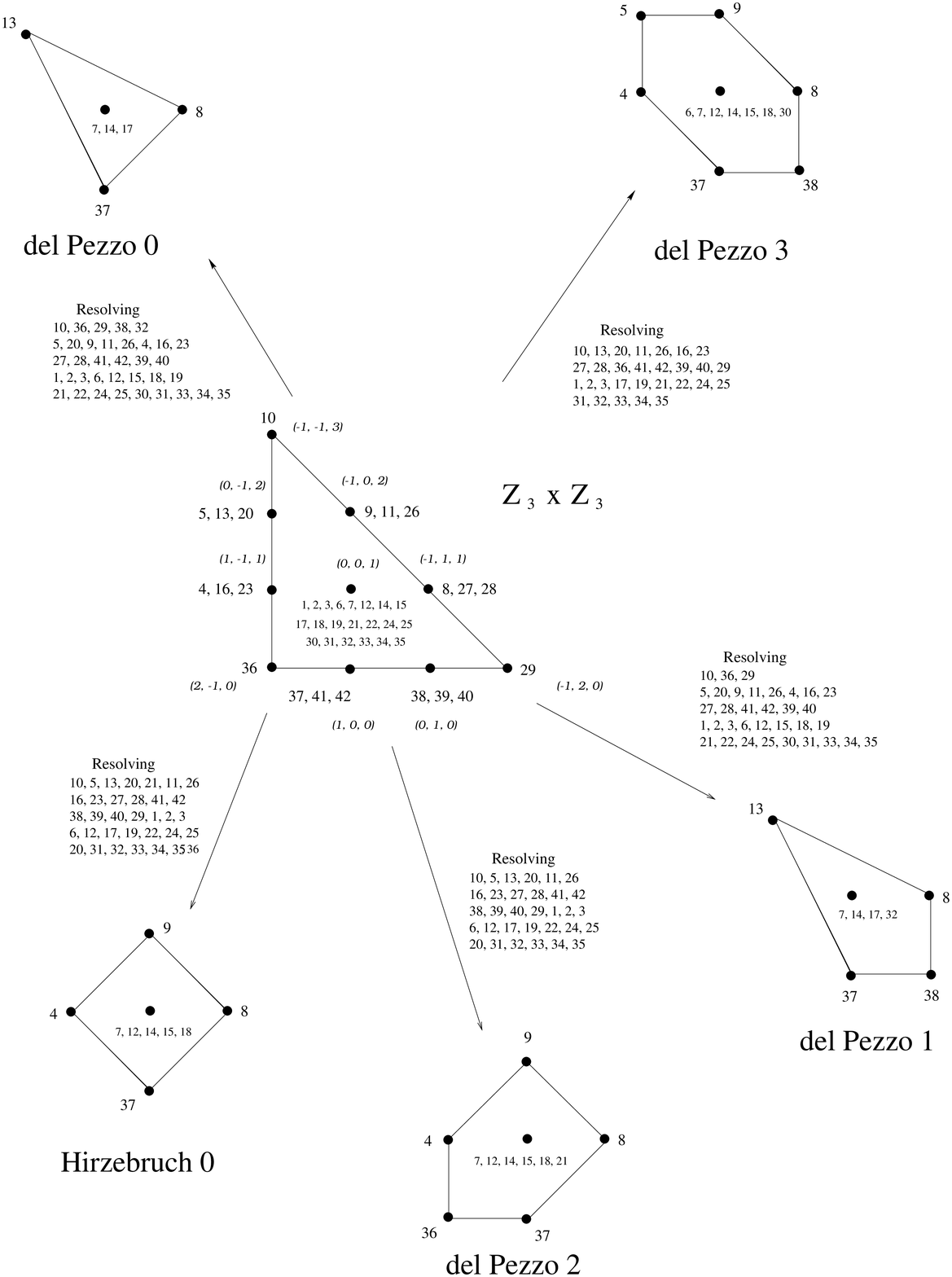,width=5.0in}}
\caption{The resolution of the Gorenstein singularity $\C^3/(\Z_3
	\times \Z_3)$ to the three toric del Pezzo surfaces as well as
	the zeroth Hirzebruch surface. We have
	labelled explicitly which columns (linear $\sigma$-model
	fields) are to be associated to each node in the toric
	diagrams and especially which columns are to be eliminated
	(fields acquiring non-zero VEV) in the various
	resolutions. Also, we have labelled the nodes of the parent
	toric diagram with the coordinates as given in the matrix
	$G_t$ for $\C^3/(\Z_3 \times \Z_3)$.}
\label{f:dP}
\end{figure}

The toric diagram of the dual cone of the (parent) quotient
singularity $\C^3 / (\Z_3 \times \Z_3)$ as well as those of its
resolution to the three toric del Pezzo surface are presented in
\fref{f:dP}.

\begin{figure}
\centerline{\psfig{figure=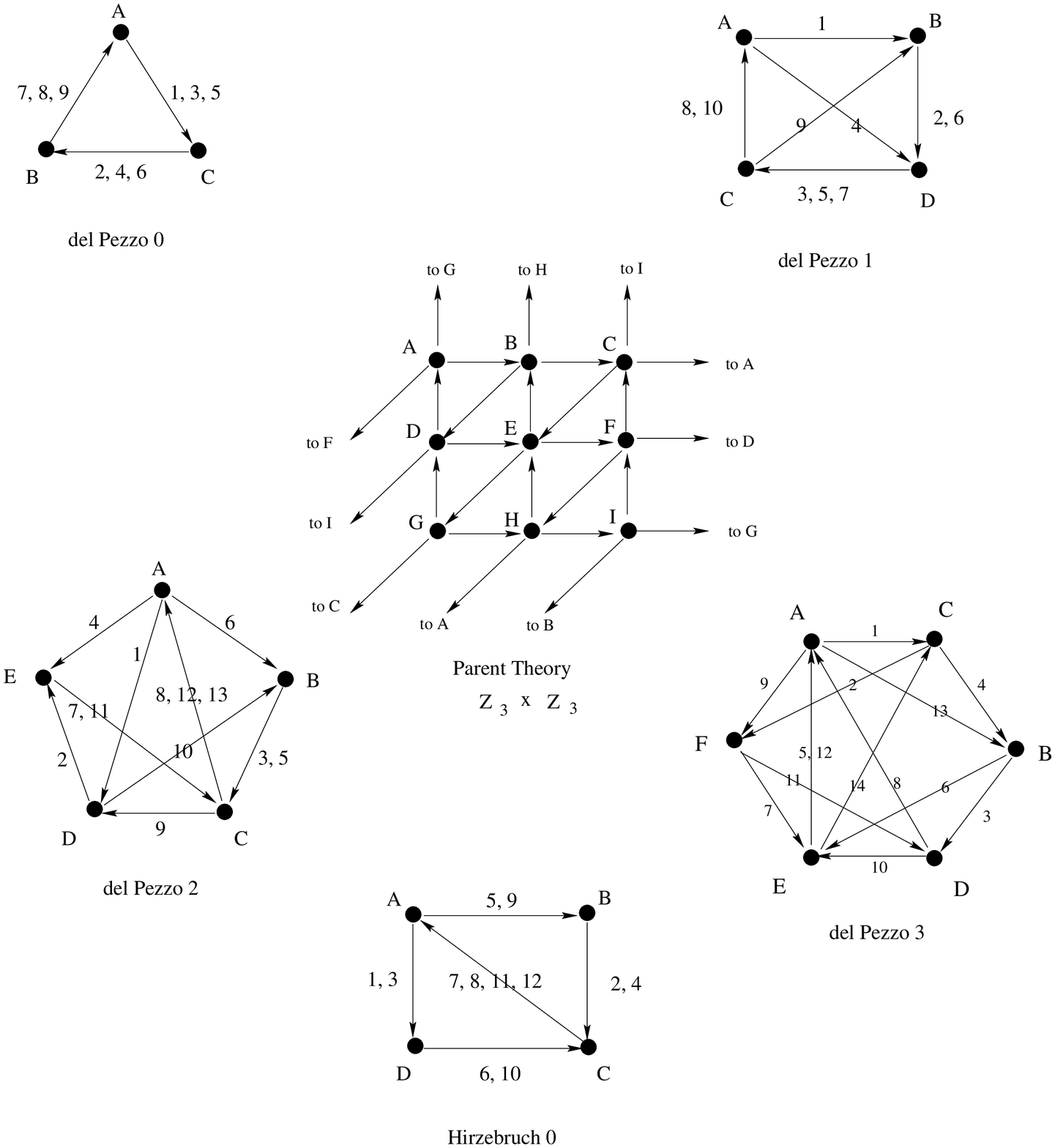,width=6.0in}}
\caption{The quiver diagrams for the matter content of the
	brane world-volume gauge theory on the 4 toric del Pezzo
	singularities as well as the zeroth Hirzebruch surface.
	We have specifically labelled the $U(1)$ 
	gauge groups (A, B, ..) and the bi-fundamentals (1, 2, ..)
	in accordance with our conventions in presenting the various
	matrices $Q_t$, $\Delta$ and $K$. As a reference we have also
	included the quiver for the parent $\Z_3 \times \Z_3$ theory.}
\label{f:dPquiver}
\end{figure}

\underline{del Pezzo 1:}
Let us commence our analysis with the first toric del Pezzo
surface\footnote{Now some may identify the toric diagram of del Pezzo
	1 as given by nodes (using the notation in \fref{f:dP}) $(1,-1,1)$, 
	$(2,-1,0)$, $(-1,1,1)$, $(0,0,1)$ and $(-1,0,2)$ instead of
	the one we have chosen in the convention of \cite{Chris},
	with nodes  $(0,-1,2)$, $(0,0,1)$, $(-1,1,1)$, $(1,0,0)$ and
	$(0,1,0)$. But of course these two $G_t$ matrices describe the
	same algebraic variety. The former corresponds to
	${\rm Spec}\left(\C[XY^{-1}Z,X^2Y^{-1},X^{-1}YZ,Z,X^{-1}Z^2]\right)$ 
	while the latter corresponds to
	${\rm Spec}\left(\C[Y^{-1}Z^2,Z,X^{-1}YZ,X,Y]\right)$. The
	observation that $(X^2Y^{-1})=(X)(X^{-1}YZ)^{-1}(Z)$,
	$(XY^{-1}Z) = (X)(Y)^{-1}(Z)$ and $(X^{-1}Z^2) =
	(Y^{-1}Z^2)(Y)(X^{-1})$ for the generators of the polynomial
	ring gives the equivalence. In other words, there is an
	$SL(5,\Z)$ transformation between the 5 nodes of the two toric
	diagrams.}. From its toric diagram, we see that the minimal
$\Z_n \times\Z_n$ toric diagram into which it embeds is $n=3$. As a
reference, the toric diagram for $\C^3 / (\Z_3 \times \Z_3)$ is given
in \fref{f:dP} and the quiver diagram, given later in the convenient
brane-box form, in \fref{f:dPquiver}. Luckily, the
matrices $Q'_t$ and $G'_t$ for this Abelian quotient is given in
\cite{Chris}. Adding the extra column of FI-parametres we present
these matrices below\footnote{In \cite{Chris}, a canonical
	ordering was used; for our purposes we need not belabour this point
	and use their $Q'_{total}$ as $Q'_t$. This is perfectly legitimate as
	long as we label the columns carefully, which we have done.}:
\[
{\tiny
\begin{array}{l}
G'_t = \left(
\matrix{p_{1}& p_{2}& p_{3}& p_{4}& p_{5}& p_{6}& p_{7}&
	p_{8}& p_{9}& p_{10}& p_{11}& p_{12}& p_{13}& 
  	p_{14}& p_{15}& p_{16}& p_{17}& p_{18}& p_{19}& p_{20} 
	& p_{21}& p_{22}& p_{23}& p_{24} \cr 
0&0&0&1&0&0&0&-1&-1&-1&-1&0&0&0&0&1&0&0&0&0&0&0&1&0
   \cr 0&0&0&-1&-1&0&0&1&0&-1&0&0&-1&0&0&-1&0&0&0&-1&0&0&-1&0
   \cr 1&1&1&1&2&1&1&1&2&3&2&1&2&1&1&1&1&1&1&2&1&1&1&1 \cr  }
\right. \cdots \cdots
\\ \\ \\
\qquad \qquad \qquad \qquad \cdots \cdots
\qquad \qquad
\left.
\matrix{
 p_{25}& p_{26}& p_{27}& p_{28}& p_{29}& p_{30}& p_{31}& p_{32}& 
p_{33}& p_{34}& p_{35}& p_{36}&  p_{37}& p_{38}& p_{39}& p_{40}
& p_{41}& p_{42} & \cr
 0&-1&-1&-1&-1&0&0&0&0&0&0&2&1&0&0&0&1&1 \cr 0&0&1&1&2&0
  &0&0&0&0&0&-1&0&1&1&1&0&0 \cr 1&2&1&1&0&1&1&1&1&1&1&0&0&
  0&0&0&0&0 \cr  }
\right)
\end{array}
}
\]
and
\[
{\tiny
\begin{array}{l}
Q'_t =
\left(
\matrix{  p_{1}& p_{2}& p_{3}& p_{4}& p_{5}& p_{6}& p_{7}&
	p_{8}& p_{9}& p_{10}& p_{11}& p_{12}& p_{13}& 
  	p_{14}& p_{15}& p_{16}& p_{17}& p_{18}& p_{19}& p_{20} 
	& p_{21}& p_{22}& p_{23}& p_{24}& \cr 
	1&0&0&0&0&0&0&0&0&0&0&0&0&-1&0&0&0&0&0&0&1&0&0&0&  0  \cr
  0&1&0&0&0&0&0&0&0&0&0&0&0&-1&0&0&0&0&0&0&1&0&0&0&  0  \cr0&0
  &1&0&0&0&0&0&0&0&0&0&0&-1&0&0&0&0&0&0&1&0&0&0&  0  \cr0&0&0&
  1&0&0&0&0&0&0&0&0&0&-1&0&0&0&0&0&0&1&0&0&0&  0  \cr0&0&0&0&1
  &0&0&0&0&0&0&0&0&-1&0&0&0&0&0&0&1&0&0&0&  0  \cr0&0&0&0&0&1&
  0&0&0&0&0&0&0&-1&0&0&0&0&0&0&1&0&0&0&  0  \cr0&0&0&0&0&0&1&0
  &0&0&0&0&0&-1&0&0&0&0&0&0&1&0&0&0&  0  \cr0&0&0&0&0&0&0&1&0&
  0&0&0&0&-1&0&0&0&0&0&0&0&0&0&0&  0  \cr0&0&0&0&0&0&0&0&1&0&0
  &0&0&-1&0&0&0&0&0&0&0&0&0&0&  0  \cr0&0&0&0&0&0&0&0&0&1&0&0&
  0&-1&0&0&0&0&0&0&0&0&0&0&  0  \cr0&0&0&0&0&0&0&0&0&0&1&0&0&-1
  &0&0&0&0&0&0&0&0&0&0&  0  \cr0&0&0&0&0&0&0&0&0&0&0&1&0&-1&0&
  0&0&0&0&0&0&0&0&0&  0  \cr0&0&0&0&0&0&0&0&0&0&0&0&1&-1&0&0&0
  &0&0&0&0&0&0&0&  0  \cr0&0&0&0&0&0&0&0&0&0&0&0&0&0&1&0&0&0&0
  &0&-1&0&0&0&  0  \cr0&0&0&0&0&0&0&0&0&0&0&0&0&0&0&1&0&0&0&0&
  -1&0&0&0&  0  \cr0&0&0&0&0&0&0&0&0&0&0&0&0&0&0&0&1&0&0&0&-1&0
  &0&0&  0  \cr0&0&0&0&0&0&0&0&0&0&0&0&0&0&0&0&0&1&0&0&-1&0&0&
  0&  0  \cr0&0&0&0&0&0&0&0&0&0&0&0&0&0&0&0&0&0&1&0&-1&0&0&0&  0  \cr0
  &0&0&0&0&0&0&0&0&0&0&0&0&0&0&0&0&0&0&1&-1&0&0&0&  0  \cr0&0&
  0&0&0&0&0&0&0&0&0&0&0&0&0&0&0&0&0&0&0&1&0&0&  0  \cr0&0&0&0&
  0&0&0&0&0&0&0&0&0&0&0&0&0&0&0&0&0&0&1&0&  0  \cr0&0&0&0&0&0&
  0&0&0&0&0&0&0&0&0&0&0&0&0&0&0&0&0&1&  0  \cr0&0&0&0&0&0&0&0&
  0&0&0&0&0&0&0&0&0&0&0&0&0&0&0&0&  0  \cr0&0&0&0&0&0&0&0&0&0&
  0&0&0&0&0&0&0&0&0&0&0&0&0&0&  0  \cr0&0&0&0&0&0&0&0&0&0&0&0&
  0&0&0&0&0&0&0&0&0&0&0&0&  0  \cr0&0&0&0&0&0&0&0&0&0&0&0&0&0&
  0&0&0&0&0&0&0&0&0&0&  0  \cr0&0&0&0&0&0&0&0&0&0&0&0&0&0&0&0&
  0&0&0&0&0&0&0&0&  0  \cr0&0&0&0&0&0&0&0&0&0&0&0&0&0&0&0&0&0&
  0&0&0&0&0&0&  0  \cr0&0&0&0&0&0&0&0&0&0&0&0&0&0&0&0&0&0&0&0&
  0&0&0&0&  0  \cr0&0&0&0&0&0&0&0&0&0&0&0&0&0&0&0&0&0&0&0&0&0&
  0&0&  0  \cr0&0&0&0&0&0&0&0&0&0&0&0&0&0&0&0&0&0&0&0&0&0&0&0
  &  0  \cr-1&1&0&0&0&0&0&0&0&0&0&0&0&0&0&0&0&0&0&0&0&0&0&0& \zeta_{1} \cr0
  &1&-1&-1&2&-1&0&1&0&-1&0&0&0&0&-1&1&0&0&0&0&0&0&0&0& \zeta_{2} \cr0&
  1&0&0&0&-1&0&0&0&0&0&0&0&0&0&0&0&0&0&0&0&0&0&0& \zeta_{3} \cr1&-2&1
  &0&-2&2&0&-1&0&1&0&0&0&0&0&0&0&0&0&0&0&0&0&0& \zeta_{4} \cr-1&1&-1&1
  &0&-1&0&1&0&0&0&0&0&0&0&0&0&0&0&0&0&0&0&0& \zeta_{5} \cr0&0&0&1&-2
  &1&0&-1&0&1&0&0&0&0&0&-1&0&0&0&0&0&1&0&0& \zeta_{6} \cr0&-1&1&0&0&0
  &0&0&0&0&0&0&0&0&0&0&0&0&0&0&0&0&0&0& \zeta_{7} \cr0&1&-1&0&2&-1&0
  &1&0&-1&0&0&0&0&0&0&0&0&0&0&0&-1&0&0& \zeta_{8} \cr }
\cdots \cdots \right. 
\end{array}
}
\]

\[
{\tiny
\begin{array}{l}
\qquad \qquad \qquad \qquad \cdots \cdots
\qquad \qquad
\left.
\matrix{
 p_{25}& p_{26}& p_{27}& p_{28}& p_{29}& p_{30}& p_{31}& p_{32}& 
p_{33}& p_{34}& p_{35}& p_{36}&  p_{37}& p_{38}& p_{39}& p_{40}
& p_{41}& p_{42} & \cr 
 -1&0&0&0&0&0&0&0&0&1&-1&0&0&0&1&-1&-1&1&  0  \cr-1&0&0&0&0&
  0&0&0&1&0&-1&0&0&0&1&-1&-1&1&  0  \cr-1&0&0&0&0&0&0&-1&2&0&-1&0
  &0&0&2&-2&-2&2&  0  \cr-1&0&0&0&0&0&0&-1&1&1&-1&0&0&0&2&-1&-2&1
  &  0  \cr-2&0&0&0&0&0&0&-1&2&1&-2&0&0&0&3&-2&-2&2&  0  \cr-2&0&0&0&0&0
  &0&0&2&0&-1&0&0&0&1&-1&-1&1&  0  \cr-2&0&0&0&0&0&0&0&1&1&-1&0&
  0&0&1&-1&-1&1&  0  \cr-1&0&0&0&0&0&0&0&1&0&0&0&0&0&0&-1&0&1&  0  \cr-1
  &0&0&0&0&0&0&0&0&1&-1&0&0&0&1&-1&0&1&  0  \cr-1&0&0&0&0&0&0&-1
  &1&0&-1&0&0&0&2&-1&-1&2&  0  \cr-1&0&0&0&0&0&0&-1&2&-1&0&0&0&0
  &1&-1&-1&2&  0  \cr0&0&0&0&0&0&0&-1&1&0&0&0&0&0&1&-1&-1&1&  0  \cr0&0
  &0&0&0&0&0&-1&0&0&0&0&0&0&1&0&-1&1&  0  \cr1&0&0&0&0&0&0&0&-1
  &0&0&0&0&0&-1&1&1&-1&  0  \cr1&0&0&0&0&0&0&0&-1&-1&1&0&0&0&-1&
  2&0&-1&  0  \cr1&0&0&0&0&0&0&-1&0&-1&1&0&0&0&-1&1&0&0&  0  \cr1&0&0&0
  &0&0&0&-1&0&0&0&0&0&0&0&0&0&0&  0  \cr0&0&0&0&0&0&0&0&0&-1&1
  &0&0&0&-1&1&0&0&  0  \cr0&0&0&0&0&0&0&0&0&-1&0&0&0&0&0&1&0&0
  &  0  \cr-1&0&0&0&0&0&0&0&0&1&-1&0&0&0&1&-1&0&0&  0  \cr-1&0&0&0&0&0
  &0&0&1&0&-1&0&0&0&1&0&-1&0&  0  \cr-1&0&0&0&0&0&0&0&1&-1&0&0&0
  &0&0&0&0&0&  0  \cr0&1&0&0&0&0&0&-1&0&0&-1&0&0&0&1&-1&0&1&  0  \cr0&
  0&1&0&0&0&0&-1&1&-1&0&0&0&0&0&-1&0&1&  0  \cr0&0&0&1&0&0&0&0&
  -1&1&-1&0&0&0&0&-1&1&0&  0  \cr0&0&0&0&1&0&0&0&0&0&0&0&0&0&-1&
  -1&1&0&  0  \cr0&0&0&0&0&1&0&-1&0&1&-1&0&0&0&1&-1&0&0&  0  \cr0&0&0&0
  &0&0&1&-1&1&0&-1&0&0&0&1&-1&-1&1&  0  \cr0&0&0&0&0&0&0&0&0&0&0
  &1&0&0&0&1&-1&-1&  0  \cr0&0&0&0&0&0&0&0&0&0&0&0&1&0&-1&1&0&-1
  &  0  \cr0&0&0&0&0&0&0&0&0&0&0&0&0&1&-1&0&1&-1&  0  \cr0&0&0&0&0&0&0
  &0&0&0&0&0&0&0&0&0&0&0& \zeta_{1} \cr0&0&0&0&0&0&0&0&0&0&0&0&0&0&0
  &0&0&0& \zeta_{2} \cr0&0&0&0&0&0&0&0&0&0&0&0&0&0&0&0&0&0& \zeta_{3} \cr0&0&0&0
  &0&0&0&0&0&0&0&0&0&0&0&0&0&0& \zeta_{4} \cr0&0&0&0&0&0&0&0&0&0&0&0
  &0&0&0&0&0&0& \zeta_{5} \cr0&0&0&0&0&0&0&0&0&0&0&0&0&0&0&0&0&0& \zeta_{6} \cr0
  &0&0&0&0&0&0&0&0&0&0&0&0&0&0&0&0&0& \zeta_{7} \cr0&0&0&0&0&0&0&0&0
  &0&0&0&0&0&0&0&0&0&\zeta_{8} \cr }
\right)
\end{array}
}
\]

According to our algorithm, we must perform Gaussian row-reduction on
$Q'_t$ to solve for 42 variables $x_i$. When this is done we find that we
can in fact express all variables in terms of 3 $x_i$'s together with
the 8 FI-parametres $\zeta_i$. We choose these three $x_i$'s to be
$x_{10,29,36}$ corresponding to the 3 outer vertices which we know must be
resolved in going from $\C^3 / (\Z_3 \times \Z_3)$ to del Pezzo 1.

Next we select the fields which must be kept and set them to zero in
order to determine the range for $\zeta_i$. Bearing in mind the toric
diagrams from \fref{f:dP}, these fields we
judiciously select to be: $p_{13, 8, 37, 38}$.
Setting $x_{13,8,37,38} = 0$  gives us the solution
$\{ \zeta_6=0; x_{29} = \zeta_7 = \zeta_3 = \zeta_1 - \zeta_5;
x_{10} = \zeta_4 + \zeta_5 + \zeta_3; x_{36} = \zeta_7 - \zeta_8
\}$, which upon back-substitution to the solutions $x_i$ we obtained
from $Q'_t$, gives zero for $x_{13, 8, 37, 38}$ (which we have
chosen by construction) as well as $x_{7,14,17,32}$; 
for all others we obtain positive
values. This means precisely that all the other fields
are to be eliminated and these 8 columns 
\{ 13, 8, 37, 38, 7,14,17,32 \}
are to be kept while the remaining 42-8=34 are to
be eliminated from $Q'_t$ upon row-reduction to give $Q_t$.
In other words, we have found our set $B$ to be
\{1,2,3,4,5,6,9,10,11,12,15,16,18,19,20,21,22,23,24,25,
26,27,28,29,30,31,33,34,35,36,39,40,41,42\}
and thus according to (\ref{qt}) we immediately obtain
\[
{\scriptsize
Q_t = \left(
\matrix{ 
p_{7}& p_{8}& p_{13}& p_{14}& p_{17}& p_{32}& p_{37}& p_{38} & \cr
1&0&0&0&0&-1&0&0&\zeta_2 + \zeta_8 \cr 0&0&0&0&-1&1&0&0&\zeta_6 \cr -1&0&
  0&1&0&0&0&0&\zeta_1 + \zeta_3 + \zeta_5 \cr 0&0&1&-1&0&-1&0&1&0 \cr -1&1&1&-1&-1
  &0&1&0&0 \cr  }
\right).
}
\]
We note of course that 5 out of the 8 FI-parametres have been
eliminated automatically; this is to be expected since in resolving 
$\C^3 / (\Z_3 \times \Z_3)$ to del Pezzo 1, we remove precisely 5
nodes. Obtaining the D-terms and F-terms is now straight-forward.
Using (\ref{Delta}) and re-inserting the last row we obtain the
D-term equations (incidence matrix) to be
\[
{\scriptsize
d = \left(
\matrix{ X_1&X_2&X_3&X_4&X_5&X_6&X_7&X_8&X_9&X_{10} \cr
 -1&0&0&-1&0&0&0&1&0&1 \cr 1&-1&0&0&0&-1&0&0&1&0 \cr 0&0&1&
  0&1&0&1&-1&-1&-1 \cr 0&1&-1&1&-1&1&-1&0&0&0 \cr  }
\right)
}
\]
From this matrix we immediately observe that there are 4 gauge groups,
i.e., $U(1)^4$ with 10 matter fields $X_i$ which we have labelled in
the matrix above. In an equivalent notation we rewrite $d$ as the
adjacency matrix of the quiver diagram (see \fref{f:dPquiver})
for the gauge theory:
\[
{\scriptsize
a_{ij} = \left(
\matrix{0&0&2&0\cr 1&0&1&0\cr 0&0&0&3\cr 1&2&0&0}
\right).
}
\]
The K-matrix we obtain to be:
\[
{\scriptsize
K^t = \left(
\matrix{
X_1&X_2&X_3&X_4&X_5&X_6&X_7&X_8&X_9&X_{10} \cr
 1&0&1&0&0&0&1&0&0&0 \cr 0&1&1&0&0&0&0&1&0&0 \cr 
1&0&0&1&0&0&0&0&1&0 \cr 0&1&0&1&0&1&0&0
  &0&0 \cr 0&0&1&0&1&0&1&0&0&0 \cr 0&0&0&0&0&1&1&0&0&1 \cr  }
\right)
}
\]
which indicates that the original 10 fields $X_i$ can be expressed in
terms of 6. This was actually addressed in the previous section and we
rewrite that pleasant superpotential here:
\[
W = X_{2} X_{7} X_{9} - X_{3} X_{6} X_{9} - X_{4} X_{8} 
	X_{7} - X_{1} X_{2} X_{5} X_{10} + X_{3} X_{4} X_{10} 
	+ X_{1} X_{5} X_{6} X_{8}.
\]

\underline{del Pezzo 2:}
Having obtained the gauge theory for del Pezzo 1, we now repeat the
above analysis for del Pezzo 2. Now we have the
FI-parametres restricted as
$\{ p_{36} = \zeta_2 = 0; \zeta_3 = \zeta_4; x_{29} = \zeta_4
+\zeta_6; x_{10} = \zeta_1 + \zeta_4 \}$, making the set to be
eliminated as $B = \{$ 1, 2, 3, 5, 6, 10, 11, 13, 16, 17, 19, 20,
22, 23, 24, 25, 26, 27, 28, 29, 30, 31, 32, 33, 34, 35, 
38, 39, 40, 41, 42 $\}$. Whence, we obtain
\[
{\scriptsize
Q_t = \left(
\matrix{
p_{4} & p_{7} & p_{8} & p_{9} & p_{12} & p_{14} & p_{15}
 & p_{18} & p_{21} & p_{36} & p_{37} & \cr
 0&1&0&0&0&0&0&0&-1&0&0&\zeta_{4} + \zeta_{6} + \zeta_{8} \cr 
1&-1&1&0&0&-1&0&0& 0&0&0&\zeta_{7} \cr
 0&-1&0&0&0&1&0&0&0&0&0&\zeta_{1} + \zeta_{3} + \zeta_{5} \cr -1&1&-1&0&0
  &1&-1&0&1&0&0&\zeta_{2} \cr 0&-1&0&1&0&0&-1&0&0&0&1&0 \cr 0&-1&1&1&0
  &-1&0&0&-1&1&0&0 \cr -1&1&-1&0&0&1&-1&1&0&0&0&0 \cr -1&1&-1&0&1
  &0&0&0&0&0&0&0 \cr  }
\right),
}
\]
and observe that 4 D-terms have been resolved, as 4 nodes have been
eliminated from $\C^3 / (\Z_3 \times \Z_3)$.
From this we easily extract (see \fref{f:dPquiver})
\[
{\scriptsize
d = 
\left(
\matrix{
X_{1} & X_{2} & X_{3} & X_{4} & X_{5} & X_{6} & X_{7} & X_{8} & X_{9} & 
   X_{10} & X_{11} & X_{12} & X_{13} \cr
 -1 & 0 & 0 & -1 & 0 & -1 & 0 & 1 & 0 & 0 & 0 & 1 & 1 \cr 0 & 0 & -1
    & 0 & -1 & 1 & 0 & 0 & 0 & 1 & 0 & 0 & 0 \cr 0 & 0 & 1 & 0 & 1 & 0 & 1 & 
   -1 & -1 & 0 & 1 & -1 & -1 \cr 1 & -1 & 0 & 0 & 0 & 0 & 0 & 0 & 1 & -1 & 0
    & 0 & 0 \cr 0 & 1 & 0 & 1 & 0 & 0 & -1 & 0 & 0 & 0 & -1 & 0 & 0 \cr  } 
\right);
}
\]
moreover, we integrate the F-term matrices
\[
{\scriptsize
K^t =
\left(
\matrix{
X_{1} & X_{2} & X_{3} & X_{4} & X_{5} & X_{6} & X_{7} & X_{8} & X_{9} & 
   X_{10} & X_{11} & X_{12} & X_{13} \cr
 0&1&1&0&0&0&0&1&0&0&0&1&0 \cr 1&0&1&0&0&0&0&0&0&0&1&1
  &0 \cr 1&0&0&1&0&1&0&0&1&0&0&0&0 \cr 0&1&0&1&0&1&0&0&0&1&0&0
  &0 \cr 0&1&1&1&1&0&0&0&0&0&0&0&0 \cr 0&0&1&0&1&0&1&0&0&0&1&0
  &0 \cr 0&0&0&1&1&0&0&0&1&0&0&0&1 \cr  }
\right)
}
\]
to obtain the superpotential
\[
\begin{array}{c}
W = X_{2} X_{9} X_{11} - X_{9} X_{3} X_{10} - X_{4} X_{8} X_{11} -
X_{1} X_{2} X_{7} X_{13} + X_{13} X_{3} X_{6} \\
- X_{5} X_{12} X_{6}+
X_{1} X_{5} X_{8} X_{10} + X_{4} X_{7} X_{12}.
\end{array}
\]

\underline{del Pezzo 3:}
Finally, we shall proceed to treat del Pezzo 3.
Here we have the range of the FI-parametres to be
$\{ \zeta_1 = \zeta_6 = \zeta_6 = 0; x_{29} = \zeta_3 = -\zeta_5;
x_{10} = \zeta_4; \zeta_2 = x_{36}; \zeta_8 = -\zeta_2 - \zeta_{10}
\}$, which gives the set $B$ as
\{1, 2, 3, 10, 11, 13, 16, 17, 19, 20, 21, 22, 23, 24, 25, 
26, 27, 28, 29, 31, 32, 33, 34, 35, 36, 39, 40, 41, 42\}, 
and thus according to (\ref{qt}) we immediately obtain
\[
{\scriptsize
Q_t = \left(
\matrix{
p_{4}&p_{5}&p_{6}&p_{7}&p_{8}&p_{9}&p_{12}&p_{14}&p_{15}&p_{18}
&p_{30}&p_{37}&p_{38} &\cr
 0&0&0&1&0&0&0&0&0&0&-1&0&0&\zeta_2  + \zeta_4  + \zeta_8  \cr 1&0&0&-1&1&0
  &0&-1&0&0&0&0&0&\zeta_7  \cr -1&0&0&1&-1&0&0&1&-1&0&1&0&0&\zeta_6  \cr 0
  &0&-1&0&0&0&0&1&0&0&0&0&0&\zeta_3  + \zeta_5  \cr 0&0&1&-1&0&0&0&0&0&0
  &0&0&0&\zeta_1  \cr 0&1&-1&0&0&0&0&0&0&0&-1&0&1&0 \cr -1&1&-1&0&0&0
  &0&1&-1&0&0&1&0&0 \cr -1&0&0&1&-1&0&0&1&-1&1&0&0&0&0 \cr -1&0
  &0&1&-1&0&1&0&0&0&0&0&0&0 \cr 1&-1&1&-1&0&1&0&-1&0&0&0&0&0&0 \cr  }
\right)
}
\]
We note indeed that 3 out of the 8 FI-parametres have been
automatically resolved, as we have removed 3 nodes from the toric
diagram for $\C^3 / (\Z_3 \times \Z_3)$.
The matter content (see \fref{f:dPquiver}) is encoded in
\[
{\scriptsize
d = \left(
\matrix{
 X_{1} & X_{2} & X_{3} & X_{4} & X_{5} & X_{6} & X_{7} & X_{8} & X_{9} & 
   X_{10} & X_{11} & X_{12} & X_{13} & X_{14} \cr 
 -1&0&0&0&1&0&0&1&-1&0&0&1&-1&0 \cr 0&0
   &-1&1&0&-1&0&0&0&0&0&0&1&0 \cr 1&-1&0&-1&0
   &0&0&0&0&0&0&0&0&1 \cr 0&0&1&0&0&0&0&-1&0
   &-1&1&0&0&0 \cr 0&0&0&0&-1&1&1&0&0&1&0&-1
   &0&-1 \cr 0&1&0&0&0&0&-1&0&1&0&-1&0&0&0 \cr 
    }
\right),
}
\]
and from the F-terms
\[
{\scriptsize
K^t = \left(
\matrix{
 X_{1} & X_{2} & X_{3} & X_{4} & X_{5} & X_{6} & X_{7} & X_{8} & X_{9} & 
   X_{10} & X_{11} & X_{12} & X_{13} & X_{14} \cr 
 1&0&0&0&0&1&1&1&0&0&0&0&0&0 \cr 0&1&0&0&0&1&0&1&0&0&0&
  1&0&0 \cr 1&0&0&0&0&0&0&0&1&0&0&0&1&1 \cr 0&1&0&1&0&0&0&0&1&0
  &0&0&1&0 \cr 0&1&1&0&0&1&0&0&1&0&0&0&0&0 \cr 0&0&1&0&0&1&1&0
  &0&0&1&0&0&0 \cr 0&0&1&0&1&0&0&0&1&0&0&0&0&1 \cr 0&0&0&0&0&1
  &1&1&0&1&0&0&0&0 \cr  }
\right)
}
\]
we integrate to obtain the superpotential
\[
\begin{array}{c}
W = X_{3} X_{8} X_{13} - X_{8} X_{9} X_{11} - X_{5} X_{6} X_{13} - 
X_{1} X_{3} X_{4} X_{10} X_{12} \\
+ X_{7} X_{9} X_{12} + X_{1} X_{2} X_{5} X_{10} X_{11} + 
X_{4} X_{6} X_{14} - X_{2} X_{7} X_{14}.
\end{array}
\]
Note that we have a quintic term in $W$; this is an interesting
interaction indeed.

\underline{del Pezzo 0:}
Before proceeding further, let us attempt one more example, viz., the
degenerate case of the del Pezzo 0 as shown in \fref{f:dP}. This time
we note that the ranges for the FI-parametres are $\{
\zeta_{5}=-x_{29}+\zeta_{6}-A; \zeta_{6}=x_{29}-B; x_{29}=B+C; 
\zeta_{8}=-x_{36}+B; x_{36}=B+C+D; x_{10}=A+E \}$ for some positive $A,
B, C, D$ and $E$, that $B =$ \{
1, 2, 3, 4, 5, 6, 9, 10, 11, 12, 15, 16, 18, 19, 20, 21, 22, 
23, 24, 25, 26, 27, 28, 29, 30, 31, 32, 33, 34, 35, 36, 38, 
39, 40, 41, 42 \} and whence the charge matrix is
\[
{\scriptsize
Q_t = \left(
\matrix{
p_{7} & p_{8} & p_{13} & p_{14} & p_{17} & p_{37} & \cr
 1&0&0&0&-1&0&\zeta_{2} + \zeta_{6} + \zeta_{8} \cr
-1&0&0&1&0&0&\zeta_{1} + \zeta_{3} + \zeta_{5} \cr -1&1&1&-1&-1&1&0
\cr  }
\right).
}
\]
We extract the matter content (see \fref{f:dPquiver}) as
$
{\scriptsize
d = \left(
\matrix{
X_{1}& X_{2}& X_{3}& X_{4}& X_{5}& X_{6}& X_{7}& X_{8}& X_{9} \cr
-1 & 0 & -1 & 0 & -1 & 0 & 1 & 1 & 1 \cr
0 & 1 & 0 & 1 & 0 & 1 & -1 & -1 & -1 \cr
1 & -1 & 1 & -1 & 1 & -1 & 0 & 0 & 0 \cr}
\right),
}
$
and the F-terms as
$
{\scriptsize
K^t = \left(
\matrix{ 
X_{1}& X_{2}& X_{3}& X_{4}& X_{5}& X_{6}& X_{7}& X_{8}& X_{9} \cr
1 & 1 & 0 & 0 & 0 & 0 & 1 & 0 & 0 \cr
1 & 0 & 1 & 0 & 1 & 0 & 0 & 0 & 0 \cr
 0 & 1 & 0 & 1 & 0 & 1 & 0 & 0 & 0 \cr 
0 & 0 & 1 & 1 & 0 & 0 & 0 & 1 & 0 \cr 0 & 
  0 & 0 & 0 & 1 & 1 & 0 & 0 & 1 \cr  }
\right),
}
$
and from the latter we integrate to obtain the superpotential
\[
W = X_{1} X_{4} X_{9} - X_{4} X_{5} X_{7} - X_{2} X_{3} X_{9} - 
	X_{1} X_{6} X_{8} + X_{2} X_{5} X_{8} + X_{3} X_{6} X_{7}.
\]
Of course we immediately recognise the matter content (which gives a
triangular quiver which we shall summarise below in \fref{f:dPquiver})
as well as the
superpotential from equations (4.7-4.14) of \cite{DGM}; it is simply
the theory on the Abelian orbifold $\C^3/\Z_3$ with action 
$(\alpha \in \Z_3) : (z_1, z_2, z_3) \rightarrow (e^{{2 \pi i} \over
3} z_1, e^{{2 \pi i} \over 3} z_2, e^{{2 \pi i} \over 3} z_3)$. 
Is our del Pezzo 0
then $\C^3/\Z_3$? We could easily check from the $G_t$ matrix (which
we recall is obtained from $G'_t$ of $\C^3 / (\Z_3 \times \Z_3)$
by eliminating the columns corresponding to the set $B$):
\[
{\scriptsize
G_t = \left(
\matrix{ 0 & -1 & 0 & 0 & 0 & 1 \cr 0 & 1 & -1 & 0 & 0 & 0 \cr
1 & 1 & 2 & 1 & 1 & 0 \cr  }
\right).
}
\]
These columns (up to repeat) correspond to monomials $Z, X^{-1}YZ,
Y^{-1}Z^2,X$ in the polynomial ring $\C[X,Y,Z]$. Therefore we need to find the
spectrum (set of maximal ideals) of the ring $\C[Z,
X^{-1}YZ,Y^{-1}Z^2,X]$, which is given by the minimal polynomial
relation: $(X^{-1}YZ) \cdot (Y^{-1}Z^2)\cdot X = (Z)^3$. This means,
upon defining $p = X^{-1}YZ; q = Y^{-1}Z^2; r=X$ and $s=Z$, our del
Pezzo 0 is described by $p q r = s^3$ as an algebraic variety 
in $\C^4({p,q,r,s})$, which is precisely $\C^3/\Z_3$. Therefore we
have actually come through a full circle in resolving $\C^3 / (\Z_3
\times \Z_3)$ to $\C^3/\Z_3$ and the validity of our algorithm
survives this consistency check beautifully. Moreover, since we know
that our gauge theory is exactly the one which lives on a D-brane probe on
$\C^3/\Z_3$, this gives a good check for physicality: that our careful
tuning of FI-parametres via canonical partial resolutions does give a
physical D-brane theory at the end.
We tabulate the matter content $a_{ij}$ and the superpotential $W$ for
the del Pezzo surfaces below, and the quiver diagrams, in \fref{f:dPquiver}.
\[
{\scriptsize
\begin{array}{|c|c|c|c|}
\hline
& \mbox{{\bf del Pezzo 1}} & \mbox{{\bf del Pezzo 2}} & \mbox{{\bf del Pezzo 3}} \\
\hline
\mbox{Matter } a_{ij}= &
	\left(\matrix{0&0&2&0\cr 1&0&1&0\cr 0&0&0&3\cr
		1&2&0&0}\right)&
	\left(\matrix{ 0 & 1 & 0 & 1 & 1 \cr 0 & 0 & 2 & 0 & 0 \cr
		 3 & 0 & 0 & 1 & 0 \cr 0
    		& 1 & 0 & 0 & 1 \cr 0 & 0 & 2 & 0 & 0 \cr} \right) &
	\left(\matrix{ 0&0&1&1&0&1 \cr 0&0&0&1&1&0\cr
	 	0 & 1 & 0 & 0& 0 & 1 \cr 1 & 0 & 0 & 0 & 1 & 0 \cr
		2 & 0 & 1 & 0 & 0 & 0 \cr 0 & 0 & 0
    		& 1 & 1 & 0 \cr  } \right)
\\ \hline
\mbox{Superpotential } W=  & \begin{array}{c}
	X_{2} X_{7} X_{9} - X_{3} X_{6} X_{9} \\
	- X_{4} X_{8} X_{7} - X_{1} X_{2} X_{5} X_{10} \\
	+ X_{3} X_{4} X_{10} + X_{1} X_{5} X_{6} X_{8} \end{array} &
	\begin{array}{c}
	X_{2} X_{9} X_{11} - X_{9} X_{3} X_{10} \\
	- X_{4} X_{8} X_{11} - X_{1} X_{2} X_{7} X_{13} \\
	+ X_{13} X_{3} X_{6} - X_{5} X_{12} X_{6} \\
	+ X_{1} X_{5} X_{8} X_{10} + X_{4} X_{7} X_{12}
	\end{array} &
	\begin{array}{c}
	X_{3} X_{8} X_{13} - X_{8} X_{9} X_{11} \\
	- X_{5} X_{6} X_{13} - X_{1} X_{3} X_{4} X_{10} X_{12} \\
	+ X_{7} X_{9} X_{12} + X_{1} X_{2} X_{5} X_{10} X_{11} \\
	+ X_{4} X_{6} X_{14} - X_{2} X_{7} X_{14}
	\end{array}
\\
\hline
\end{array}
}
\]
\[
{\scriptsize
\begin{array}{|c|c|c|}
\hline
& \mbox{{\bf del Pezzo 0}} \cong \C^3/\Z_3 & 
  \mbox{{\bf Hirzebruch 0}} \cong \IP^1 \times \IP^1 := F_0 = E_1 \\
\hline
\mbox{Matter }a_{ij} & \left(\matrix{0&3&0 \cr 0&0&3 \cr 3&0&0\cr}\right) &
	\left(\matrix{ 0 & 2 & 0 & 2 \cr 0 & 0 & 2 & 0 \cr 
	4 & 0 & 0 & 0 \cr 0 & 0 & 2 & 0 \cr  }\right)
\\ \hline
\mbox{Superpotential }W & \begin{array}{c}
	X_{1} X_{4} X_{9} - X_{4} X_{5} X_{7} \\
	- X_{2} X_{3} X_{9} - X_{1} X_{6} X_{8} \\
	+ X_{2} X_{5} X_{8} + X_{3} X_{6} X_{7}
	\end{array} &
	\begin{array}{c}
	 X_{1}X_{8}X_{10}- X_{3}X_{7}X_{10} \\
	- X_{2}X_{8}X_{9}- X_{1}X_{6}X_{12} \\
	+ X_{3}X_{6}X_{11}+ X_{4}X_{7}X_{9} \\
	+ X_{2}X_{5}X_{12}- X_{4}X_{5}X_{11}
	\end{array}
\\ \hline
\end{array}
}
\]
Upon comparing \fref{f:dP} and \fref{f:dPquiver}, we notice that as we go
from del Pezzo 0 to 3, the number of points in the toric diagram
increases from 4 to 7, and the number of gauge groups (nodes in the
quiver) increases from 3 to 6. This is consistent with the observation
for ${\cal N} = 1$ theories that the number of gauge groups equals the
number of perimetre points (e.g., for del Pezzo 1, the four nodes 13,
8, 37 and 38) in the toric diagram. Moreover, as discussed in
\cite{AHK}, the rank of the global symmetry group ($E_i$ for del Pezzo
$i$) which must exist for
these theories equals the number of perimetre point minus 3; it would
be an intereting check indeed to see how such a symmetry manifests
itself in the quivers and superpotentials.

\underline{Hirzebruch 0:}
Let us indulge ourselves with one more example, namely the 0th
Hirzebruch surface, or simply $\IP^1 \times \IP^1 := F_0 := E_1$.
The toric diagram is drawn in \fref{f:dP}. Now the FI-parametres are
$\{ \zeta_4 = -x_{29}-x_{36}-\zeta_5-\zeta_8-A; \zeta_5 =
-A-B; \zeta_7 = x_{10}+x_{29}+x_{36}+\zeta_8-C; \zeta_8 =
-x_{10}-x_{29}-x_{36}+D; D = A+B; C = A+B; A = x_{10}-E; x_{10} = E+F;
x_{29} = B+G \}$ for positive $A,B,C,D,E,F$ and $G$. Moreover, $B=$ \{
1, 2, 3, 5, 6, 10, 11, 13, 16, 17, 19, 20, 21, 22, 23, 24, 25, 26, 27,
28, 29, 30, 31, 32, 33, 34, 35, 36, 38, 39, 40, 41, 42 \}. We note
that this can be obtained directly by partial resolution of fields 
21 and 36 from
del Pezzo 2 as is consistent with \fref{f:dP}. Therefrom we obtain the
charge matrix
\[
{\scriptsize
Q_t = \left(
\matrix{
p_{4}& p_{7}& p_{8}& p_{9}& p_{12}& p_{14}& p_{15}& p_{18}& p_{37} & \cr
 -1&2&-1&0&0&1&-1&0&0&\zeta_2 + \zeta_4 + \zeta_6 + \zeta_8 \cr
1&-1&1&0&0&-1&0&0&0&\zeta_7 \cr
 0&-1&0&0&0&1 &0&0&0&\zeta_1 + \zeta_3 + \zeta_5 \cr 
0&-1&0&1&0&0&-1&0&1&0 \cr 
-1&1&-1&0&0&1&-1&1&0&0 \cr
-1&1&-1&0&1&0&0&0&0&0 \cr  }
\right),
}
\]
from which we have the matter content
${\scriptsize
d = \left(
\matrix{
X_{1}& X_{2}& X_{3}& X_{4}& X_{5}& X_{6}& X_{7}& X_{8}& X_{9}& 
X_{10}& X_{11}& X_{12} \cr
-1 & 0 & -1 & 0 & -1 & 0 & 1 & 1 & -1 & 0 & 1 & 1 \cr 
0 & -1 & 0 & -1 & 1 & 0 & 0 & 0 & 1 & 0 & 0 & 0 \cr 
0 & 1 & 0 & 1 & 0 & 1 & -1 & -1 & 0 & 1 & -1 & -1 \cr  
1 & 0 & 1 & 0 & 0 & -1 & 0 & 0 & 0 & -1 & 0 & 0 \cr}
\right)
}$ the quiver for which is presented in \fref{f:dPquiver}.
The F-terms are
\[{\scriptsize
K^t = \left(
\matrix{
X_{1}& X_{2}& X_{3}& X_{4}& X_{5}& X_{6}& X_{7}& X_{8}& X_{9}& 
X_{10}& X_{11}& X_{12} \cr
1 & 1 & 0 & 0 & 0 & 0 & 1 & 0 & 0 & 0 & 1 & 0 \cr 
1 & 0 & 1 & 0 & 1 & 0 & 0 & 0 & 1 & 0 & 0 & 0 \cr 
1 & 1 & 1 & 1 & 0 & 0 & 0 & 0 & 0 & 0 & 0 & 0 \cr
0 & 1 & 0 & 1 & 0 & 1 & 0 & 0 & 0 & 1 & 0 & 0 \cr
0 & 0 & 1 & 1 & 0 & 0 & 0 & 1 & 0 & 0 & 0 & 1 \cr
0 & 0 & 0 & 0 & 1 & 1 & 1 & 1 & 0 & 0 & 0 & 0 \cr  } \right)
},\]
from which we obtain
\[
W = X_{1}X_{8}X_{10}- X_{3}X_{7}X_{10}- X_{2}X_{8}X_{9}- X_{1}X_{6}X_{12}+ 
X_{3}X_{6}X_{11}+ X_{4}X_{7}X_{9}+ X_{2}X_{5}X_{12}- X_{4}X_{5}X_{11},
\]
a perfectly acceptable superpotential with only cubic interactions. We include
these results with our table above.

\index{Toric Duality}
\section{Uniqueness?}
In our foregoing discussion we have constructed in detail an algorithm
which calculates the matter content encoded by $\Delta$ and superpotential
encoded in $K$, given the toric diagram of the singularity which the
D-branes probe. As abovementioned, though this algorithm gives
one solution for the quiver and the $K$-matrix once the matrix $Q_t$
is determined, the general inverse
process of going from toric data to gauge theory data, 
is highly {\bf non-unique} and a classification of all possible
theories having the same toric description would be
interesting\footnote{We thank R. Plesser for pointing this issue out
	to us.}.
Indeed, by the very structure of our
algorithm, in immediately appealing to the partial resolution of gauge
theories on $\Z_n \times \Z_n$ orbifolds which are well-studied, we
have granted ourselves enough extraneous information to determine a
unique $Q_t$ and hence the ability to proceed with ease (this was the
very reason for our devising the algorithm).

However, generically we do not have any such luxury.
At the end of subsection 3.1, we have already mentioned two types of
ambiguities in the inverse problem. Let us refresh our minds. They
were (A) the {\bf F-D ambiguity} which is the inability to decide, simply by
observing the toric diagram, which rows of the charge matrix $Q_t$ are
D-terms and which are F-terms and (B) the {\bf repetition ambiguity} which
is the inability to decide which columns of $G_t$ to repeat once
having read the vectors from the toric diagram. Other ambiguities
exist, such as in each time when we compute nullspaces, but we shall here
discuss to how ambiguities (A) and (B) manifest themselves and
provide examples of vastly different gauge theories having the same
toric description. There is another point which we wish to emphasise: 
as mentioned at the end of subsection
3.1, the resolution method guarantees, upon careful tuning of the
FI-parametres, that the resulting gauge theory does originate from the
world-volume of a D-brane probe. Now of course, by taking liberties
with experimentation of these ambiguities we are no longer protected by
physicality and in general the theories no longer live on the D-brane.
It would be a truly interesting exercise to check which
of these different theories do.

\underline{F-D Ambiguity:}
First, we demonstrate type (A) by returning to our old
friend the SPP whose charge matrix we had earlier presented.
Now we write the
same matrix without specifying the FI-parametres:
\[
{\scriptsize
Q_t = \left(
\matrix{ 1 & -1 & 1 & 0 & -1 & 0 \cr -1 & 1 & 0 & 1 & 0 & -1 \cr
	 -1 & 0 & 0 & -1 & 1 & 1 \cr  } 
\right)
}
\]
We could apply the last steps of our algorithm to this matrix as
follows.
\begin{enumerate}
\def\theenumi{\alph{enumi}}\def\labelenumi{(\theenumi)}
\item If we treat the first row as $Q$ (the F-terms) and the second and
	third as $V \cdot U$ (the D-terms) we obtain the gauge theory
	as discussed in subsection 3.3 and in \cite{Park-con}.
\item If we treat the second row as $Q$ and first with the third as
	$V \cdot U$, we obtain ${\scriptsize d = \left(
	\matrix{ -1 & 0 & 1 & -1 & 1 & 0 \cr 
		1 & 0 & 0 & 1 & -2 & -1 \cr
		0 & 0 & -1 & 0 & 1 & 1}
	\right)}$ which is an exotic theory indeed with a field ($p_5$)
	charged under three gauge groups.\\
	Let us digress a moment to address the stringency of the
	requirements upon matter contents. By the
	very nature of finite group representations, any orbifold theory
	must give rise to only adjoints and bi-fundamentals because its matter
	content is encodable by an adjacency matrix due to tensors of
	representations of finite groups.
	The corresponding incidence matrix $d$, has (a) only 0 and $\pm 1$
	entries specifying the particular bi-fundamentals and (b) has
	each column containing precisely one 1, one $-1$ and
	with the remaining entries 0. However more exotic
	matter contents could arise from more generic toric
	singularities, such as fields charged under 3
	or more gauge group factors; these would then have $d$ matrices with
	conditions (a) and (b) relaxed\footnote{Note that we still
		require that each column sums to 0 so as to be able to	
		factor out an overall $U(1)$.}.
	Such exotic quivers (if we could even
	call them quivers still) would give
	interesting enrichment to those well-classified families as discussed
	in \cite{9911114}.\\
	Moreover we must check the anomaly cancellation
	conditions. These could be rather involved; even though for
	$U(1)$ theories they are a little simpler, we still need to
	check {\em trace anomalies} and {\em cubic anomalies}.
	In a trace-anomaly-free theory, for each node in the
	quiver, the number of incoming arrows must equal the number of
	outgoing (this is true for a $U(1)$ theory which is what
	toric varieties provide; for a discussion on this see
	e.g. \cite{9811183}). In matrix language this means that each
	row of $d$ must sum to 0.\\
	Now for a theory with only bi-fundamental matter with $\pm 1$
	charges, since $(\pm 1)^3 = \pm 1$, the cubic is equal
	to the trace anamaly; therefore for these theories we need
	only check the above row-condition for $d$. For more exotic
	matter content, which we shall meet later, we do need to
	perform an independent cubic-anomaly check.\\
	Now for the above $d$, the second row does not sum to zero
	and whence we do unfortunately have a
	problematic anomalous theory. Let us push on to see whether
	we have better luck in the following.
\item Treating row 3 as the F-terms and the other two as the D-terms
	gives\\
	${\scriptsize d = \left(
	\matrix{ 0 & -1 & 1 & -1 & 1 & 0 \cr
		0 & 1 & 0 & 1 & -2 & -1 \cr 
		0 & 0 & -1 & 0 & 1 & 1}
	\right)}$ which has the same anomaly problem as the one above.
\item Now let rows 1 and 2 as the F-terms and the 3rd, as the D-terms,
	we obtain ${\scriptsize d = \left(
	\matrix{X_1 & X_2 & X_3 & X_4 & X_5 \cr
		0 & 1 & 1 & -1 & -1 \cr
		0 & -1 & -1 & 1 & 1}
	\right)}$, which is a perfectly reasonable matter content.
	Integrating
	${\scriptsize K = \left(
	\matrix{ 1 & 0 & 1 & 0 & 0 \cr 0 & 1 & 0 & 1 & 0 \cr
	1 & 0 & 0 & 1 & 0 \cr 0 & 0 & 1 & 0 & 1 \cr  }
	\right)}$ gives the
	superpotential $W = \phi(X_1 X_2 X_5 - X_3 X_4)$ for some
	field $\phi$ of charge $(0,0)$ (which could be an adjoint for
	example; note however that we can not use $X_1$ even though it
	has charge $(0,0)$ for otherwise the F-terms would be
	altered). This theory is perfectly legitimate. We compare the
	quiver diagrams of theories (a) (which we recall from
	\fref{f:SPPquiver}) and this present example in
	\fref{f:compare}. As a check, let us define the gauge
	invariant quantities: $a=X_2 X_4$, $b=X_2 X_5$, $c=X_3 X_4$,
	$d=X_3 X_5$ and $e=X_1$. Then we have the algebraic relations 
	$ad=bc$ and $eb = c$, from which we immediately obtain $ad =
	eb^2$, precisely the equation for the SPP.
\begin{figure}
\centerline{\psfig{figure=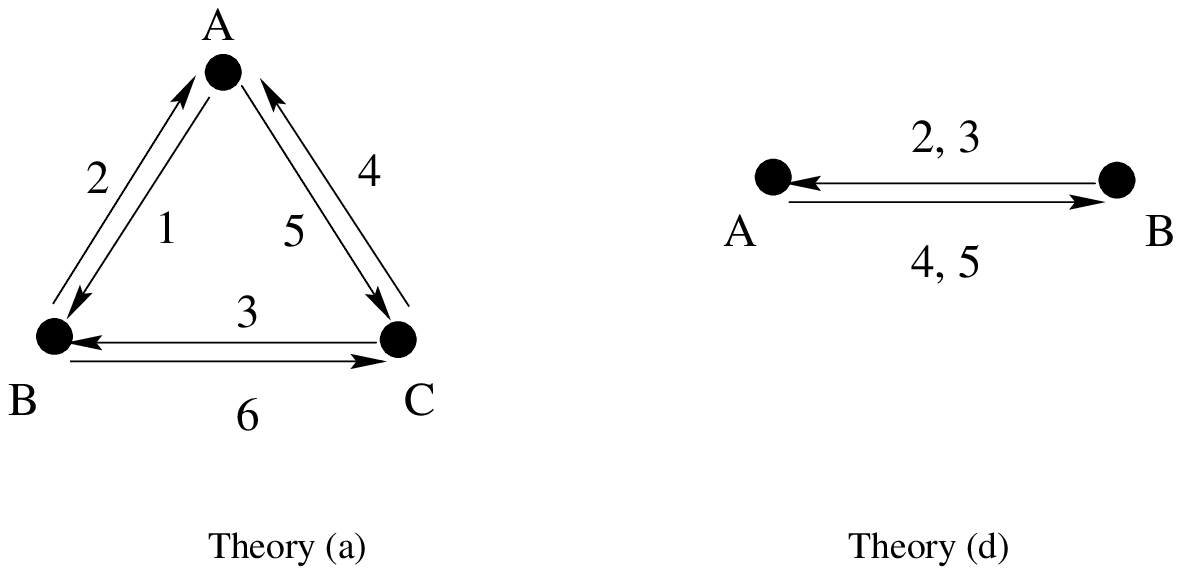,width=5.0in}}
\caption{The vastly different matter contents of theories (a) and (d),
	both anomaly free and flow to the toric diagram of the
	suspended pinched point in the IR.}
\label{f:compare}
\end{figure}
\item As a permutation on the above, treating rows 1 and 3 as the
	F-terms gives a theory equivalent thereto.
\item Furthermore, we could let rows 2 and 3 be $Q$ giving us
	${\scriptsize d = \left(
	\matrix{0 & 1 & -1 & -1 & -1 \cr
		0 & -1 & 1 & 1 & 1}
	\right)}$, but this again gives an anomalous matter content.
\item Finally, though we cannot treat all rows as F-terms, we can
	however treat all of them as D-terms in which $Q_t$ is simply
	$\Delta$ as remarked at the end of Section 2 before the flow
	chart. In this case we have the matter content
	${\scriptsize d = \left(
	\matrix{ 1 & -1 & 1 & 0 & -1 & 0 \cr -1 & 1 & 0 & 1 & 0 & -1 \cr
	 -1 & 0 & 0 & -1 & 1 & 1 \cr  1 & 0 & -1 & 0 & 0 & 0} 
	\right)}$ which clearly is both trace-anomaly free (each row
	adds to zero) and cubic-anomaly-free (the cube-sum of the each
	row is also zero). The superpotential, by our very choice, is
	of course zero. Thus we have a perfectly legitimate theory
	without superpotential but with an exotic field (the first column)
	charged under 4 gauge groups.			
\end{enumerate}
We see therefore, from our list of examples above, that for the simple
case of the SPP we have 3 rather different theories (a,d,g) with
contrasting matter content and superpotential which
share the same toric description.

\underline{Repetition Ambiguity:}
As a further illustration, let us give one example of type (B)
ambiguity.
First let us eliminate all repetitive columns from the $G_t$ of SPP,
giving us:
\[
{\scriptsize G_t = \left(
\matrix{ 1 & 0 & 0 & -1 & 1 \cr 1 & 1 & 0 & 1 & 0 \cr 
1 & 1 & 1 & 1 & 1 \cr} \right), }
\]
which is perfectly allowed and consistent with \fref{f:SPP}.
Of course many more possibilities for
repeats are allowed and we could redo the following analyses for each
of them.
As the nullspace of our present choice of $G_t$, 
we find $Q_t$, and we choose, in light
of the foregoing discussion, the first row to represent the D-term:
\[
{\scriptsize Q_t = \left(
\matrix{ -1 & 1 & -1 & 0 & 1 & \zeta \cr 1 & -2 & 0 & 1 & 0 & 0 \cr}
\right).}
\]
Thus equipped, we immediately retrieve, using our algorithm,
\[
{\scriptsize
d = \left(
	\matrix{X_1 & X_2 & X_3 & X_4 & X_5 \cr
		1 & -1 & 1 & -1 & 0 \cr -1 & 1 & -1 & 1 & 0}
\right)
\qquad
K^t = \left(
	\matrix{ 1 & 0 & 0 & 0 & 0\cr 0 & 0 & 2 & 0 & 1\cr
	0 & 1 & 0 & 0 & 0\cr 0 & 0 & 1 & 1 & 1\cr  }
\right)
\qquad
T = \left(
	\matrix{ 0 & 0 & 0 & 0 & 1 \cr -1 & 0 & 0 & 1 & 0 \cr
	0 & 0 & 1 & 0 & 0 \cr 2 & 1 & 0 & 0 & 0 \cr  } 
\right).
}
\]
We see that $d$ passes our anomaly test, with
the same bi-fundamental matter content as theory (d). The
superpotential can be read easily from $K$ (since there is only one
relation) as $W = \phi (X_5^2 - X_3 X_4)$.
As a check, let us define the gauge invariant quantities: $a=X_1 X_2$,
$b=X_1 X_4$, $c=X_3 X_2$, $d=X_3 X_4$ and $e=X_5$. These have among
themselves the
algebraic relations $ad=bc$ and $e^2 = d$, from which we immediately
obtain $bc = ae^2$, the equation for the SPP.
Hence we have yet another interesting anomaly free theory, 
which together with our theories (a), (d) and (g)
above, shares the toric description of the SPP.

Finally, let us indulge in one more demonstration. Now let us treat
both rows of our $Q_t$ as D-terms, whereby giving a theory with no
superpotential and the exotic matter content
${\scriptsize d = \left(
\matrix{ -1 & 1 & -1 & 0 & 1 \cr 1 & -2 & 0 & 1 & 0 \cr
	0 & 1 & 1 & -1 & -1}
\right)}$ with a field (column 2) charged under 3 gauge groups. Indeed
though the rows sum to 0 and trace-anomaly is avoided, the cube-sum of
the second row gives $1^3 + 1^3 + (-2)^3=-6$ and we do have a cubic
anomaly.

In summary, we have an interesting phenomenon indeed! Taking so immediate an
advantage of the ambiguities in the above has already produced quite a
few examples of vastly different gauge theories flowing in the IR to
the same universality class by having their moduli spaces identical.
The vigilant reader may raise two issues. First, as mentioned earlier,
one may take the pains to check whether these theories do indeed live on
a D-brane. Necessary conditions such as that the theories may be obtained from
an ${\cal N}=4$ theory must be satisfied. Second, the matching of
moduli spaces may not seem so strong since they are on a classical
level. However, since we are dealing with product $U(1)$ gauge groups
(which is what toric geometry is capable to dealing with so far), the
classical moduli receive no quantum corrections\footnote{We thank
	K. Intriligator for pointing this out.}. Therefore the
matching of the moduli for
these various theories do persist to the quantum regime, which hints
at some kind of ``duality'' in the field theory. We shall
call such a duality {\bf toric duality}. It would be interesting to
investigate how, with non-Abelian versions of the theory (either by
brane setups or stacks of D-brane probes), this toric duality may be
extended.

\section{Conclusions and Prospects}
The study of resolution of toric singularities by D-branes is by now
standard. In the concatenation of the F-terms and D-terms from the
world volume gauge theory of a single D-brane at the singularity, the
moduli space could be captured by the algebraic data of the toric
variety. However, unlike the orbifold theories, the inverse problem
where specifying the structure of the singularity specifies the
physical theory has not yet been addressed in detail.

We recognise that in contrast with D-brane probing orbifolds, where
knowing the group structure and its space-time action uniquely
dictates the matter content and superpotential, such flexibility is
not shared by generic toric varieties due to the highly non-unique
nature of the inverse problem. It has been the purpose and main
content of the current writing to device an {\bf algorithm} which
constructs the matter content (the incidence matrix $d$) and the
interaction (the F-term matrix $K$) of a well-behaved gauge theory 
given the toric diagram $D$ of the singularity at hand.

By embedding $D$ into the Abelian orbifold $\C^k/(\Z_n)^{k-1}$ and
performing the standard partial resolution techniques, we have
investigated how the induced action upon the charge matrices
corresponding to the toric data of the latter gives us a convenient
charge matrix for $D$ and have constructed a programmatic methodology
to extract the matter content and superpotential of {\it one} D-brane world
volume gauge theory probing $D$.
The theory we construct, having its origin from an orbifold, is nicely
behaved in that it is anomaly free, with bi-fundamentals only and
well-defined superpotentials.
As illustrations we have tabulated the results for
all the toric del Pezzo surfaces and the zeroth Hirzebruch surface.

Directions of further work are immediately clear to us. From the
patterns emerging from del Pezzo surfaces 0 to 3, we could speculate
the physics of higher (non-toric) del Pezzo cases. For example, we
expect del Pezzo $n$ to have $n+3$ gauge groups.
Moreover, we could attempt to fathom how our resolution techniques
translate as Higgsing in brane setups, perhaps with recourse to
diamonds, and realise the various theories
on toric varieties as brane configurations.

Indeed, as mentioned, the inverse problem is highly non-unique; we
could presumably attempt to classify all the different theories
sharing the same toric singularity as their moduli space. In light of
this, we have addressed two types of ambiguity: that in having
multiple fields assigned to the same node in the toric diagram and
that of distinguishing the F-terms and D-terms in the charge
matrix. In particular we have turned this ambiguity to a matter of
interest and have shown, using our algorithm, how vastly different theories,
some with quite exotic matter content, may have the same toric
description. This commonality would correspond to a duality wherein
different gauge theories flow to the same universality class in the
IR. We call this phenomenon {\bf toric duality}.
It would be interesting indeed how this duality may
manifest itself as motions of branes in the corresponding setups.
Without further ado however, let us pause here awhile and leave such
investigations to forthcoming work.
\chapter{Toric II: Phase Structure of Toric Duality}
\section*{\center{{ Synopsis}}}
\label{chap:0104259}
The previous chapter mentioned the concept of ``Toric Duality.'' Here,
we systematically study possible causes arising from our ``Inverse
Algorithm.''

{ H}arnessing the unimodular degree of freedom in the definition
of any toric diagram, we present a method of constructing inequivalent
gauge theories which are world-volume theories of D-branes probing
the same toric singularity. These theories are various {\em phases} in
partial resolution of Abelian orbifolds. As examples, two phases are
constructed for both the zeroth Hirzebruch and the second del Pezzo
surfaces.
Furthermore, we investigate the general conditions that
distinguish these different gauge theories with the same 
(toric) moduli space \cite{0104259}.
\section{Introduction}
The methods of toric geometry have been a crucial tool to the
understanding of many fundamental aspects of string theory on
Calabi-Yau manifolds (cf. e.g. \cite{Greene-Lec}). 
In particular, the connexions between toric
singularities and
the manufacturing of various gauge theories as D-brane world-volume
theories have been intimate.

Such connexions have been motivated by a myriad of sources. 
As far back as
1993, Witten \cite{GLSM} had shown, via the so-called gauged linear
sigma model, that the Fayet-Illiopoulos parametre $r$ in the D-term of
an ${\cal N}=2$ supersymmetric field theory with $U(1)$ gauge groups
can be tuned as an order-parametre which extrapolates between
the Landau-Ginzburg and Calabi-Yau phases of the theory, whereby
giving a precise viewpoint to the LG/CY-correspondence. What this
means in the context of Abelian gauge theories is that whereas for $r
\ll 0$, we have a Landau-Ginzberg description of the theory,
by taking $r \gg 0$, the space of classical vacua obtained from D- 
and F-flatness is described by a Calabi-Yau manifold, and in particular
a toric variety.

With the advent of D-brane technologies, vast amount of work has been
done to study the dynamics of world-volume theories on D-branes
probing various geometries. Notably, in \cite{DM}, D-branes
have been used to probe Abelian singularities of the form
$\IC^2/\IZ_n$. Methods of studying the moduli space of the SUSY
theories describable by quiver diagrams have been developed by the
recognition of the Kronheimer-Nakajima ALE instanton construction,
especially the moment maps used therein \cite{Orb2}.

Much work followed \cite{KS,Horizon,LNV}. 
A key advance was made in \cite{DGM}, where,
exemplifying with Abelian $\IC^3$ orbifolds, a detailed method was
developed for capturing the various phases of the moduli space of the
quiver gauge theories as toric varieties.
In another vein, the huge factory built after the brane-setup approach
to gauge theories \cite{HW} has been continuing to elucidate the
T-dual picture of branes probing singularities (e.g. \cite{HZ,
HU,9811183}). Brane setups for toric resolutions of $\IZ_2 \times
\IZ_2$, including the famous conifold, were addressed in
\cite{Park-con,Greene2}. The general question of how to construct the quiver
gauge theory for an arbitrary toric singularity was still pertinent.
With the AdS/CFT correspondence emerging \cite{KS,Horizon},
the pressing need for the question arises again: 
given a toric singularity, how
does one determine the quiver gauge theory having the former as its
moduli space? 

The answer lies in ``Partial Resolution of Abelian Orbifolds'' and was
introduced and exemplified for the toric resolutions of the $\IZ_3 \times
\IZ_3$ orbifold \cite{DGM,Chris}. The method was subsequently presented in an
algorithmic and computationally feasible fashion in \cite{0003085} and
was applied to a host of examples in \cite{Sarkar}.

One short-coming about the inverse procedure of going from the toric
data to the gauge theory data is that it is highly non-unique and in
general, unless one starts by partially resolving an orbifold
singularity, one would not be guaranteed with a physical world-volume
theory at all! Though the non-uniqueness was harnessed in \cite{0003085}
to construct families of quiver gauge theories with the same toric
moduli space, a phenomenon which was dubbed ``toric duality,'' the
physicality issue remains to be fully tackled.

The purpose of this writing is to analyse toric duality within the
confinement of the canonical method of partial resolutions. Now we are
always guaranteed with a world-volume theory at the end and this
physicality is of great assurance to us. We find indeed that with the
restriction of physical theories, toric duality is still very much at
work and one can construct D-brane quiver theories that flow to the
same moduli space.

We begin in \S 2 with a seeming paradox which initially motivated our
work and which {\it ab initio} appeared to present a challenge to
the canonical method. In \S 3 we resolve the paradox by introducing the
well-known mathematical fact of toric isomorphisms. Then in \S 4, we
present a detailed analysis, painstakingly tracing through each step
of the inverse procedure to see how much degree of freedom one is
allowed as one proceeds with the algorithm. We consequently arrive at
a method of extracting torically dual theories which are all physical;
to these we refer as ``phases.''
As applications of these ideas in \S 5 we re-analyse the examples in
\cite{0003085}, viz., the toric del Pezzo surfaces as well as the zeroth
Hirzebruch surface and find the various phases of the quiver gauge
theories with them as moduli spaces. Finally in \S 6 we end with
conclusions and future prospects.
\section{A Seeming Paradox}
In \cite{0003085} we noticed the emergence of the phenomenon of ``Toric
Duality'' wherein the moduli space of vast numbers of gauge theories could
be parametrised by the {\em same} toric variety. Of course, as we
mentioned there, one needs to check extensively whether these theories
are all physical in the sense that they are world-volume theories of
some D-brane probing the toric singularity.

Here we shall discuss an issue of more immediate concern to the
physical probe theory. We recall that using the method of {\em partial
resolutions of Abelian orbifolds} \cite{0003085,DGM,Chris,Park-con}, we could
always extract a canonical theory on the D-brane probing the
singularity of interest.

However, a discrepancy of results seems to have risen between
\cite{0003085} and \cite{Horizon} on the precise world-volume theory of
a D-brane probe sitting on the zeroth Hirzebruch surface; let us
compare and contrast the two results here.
\begin{itemize}
\item Results from \cite{0003085}: The matter contents of the theory are
given by (on the left we present the quiver diagram and on the right,
the incidence matrix that encodes the quiver):
$$
\begin{array}{lcr}
\mbox{
\epsfxsize=4cm
\epsfysize=4cm
\epsfbox{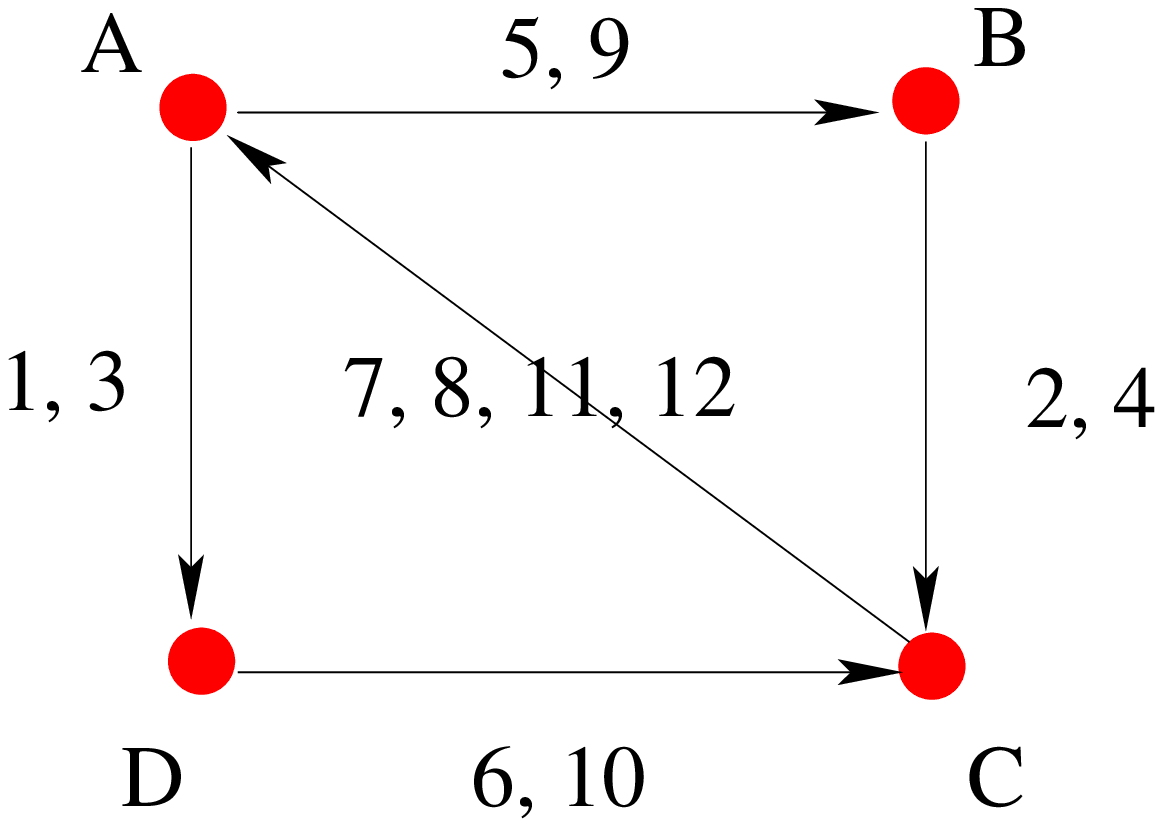}
}
&&
d =
\tmat{ 
\ba{c|cccccccccccc} 
	& X_1 & X_2 & X_3 & X_4 & X_5 & X_6 & X_7 & X_8 & X_9 & X_{10}
	& X_{11} & X_{12}\\ \hline
	A & -1 & 0 & -1 & 0 & -1 & 0 & 1 & 1 & -1 & 0 & 1 & 1 \\ 
	B & 0 & -1 & 0 & -1 & 1 & 0 & 0 & 0 & 1 & 0 & 0 & 0 \\
	C & 0 & 1 & 0 & 1 & 0 & 1 & -1 & -1 & 0 & 1 & -1 & -1 \\ 
	D & 1 & 0 & 1 & 0 & 0 & -1 & 0 & 0 & 0 & -1 & 0 & 0 \ea}
\end{array}
$$
and the superpotential is given by
\beq
W = X_{1}X_{8}X_{10}- X_{3}X_{7}X_{10}- X_{2}X_{8}X_{9}- X_{1}X_{6}X_{12}+\\ 
X_{3}X_{6}X_{11}+ X_{4}X_{7}X_{9}+ X_{2}X_{5}X_{12}- X_{4}X_{5}X_{11}.
\label{HirzeUs}
\eeq
\item Results from \cite{Horizon}: The matter contents of the theory are
given by (for $i=1,2$):
$$
\ba{lcr}
\epsfxsize=4cm
\epsfysize=4cm
\epsfbox{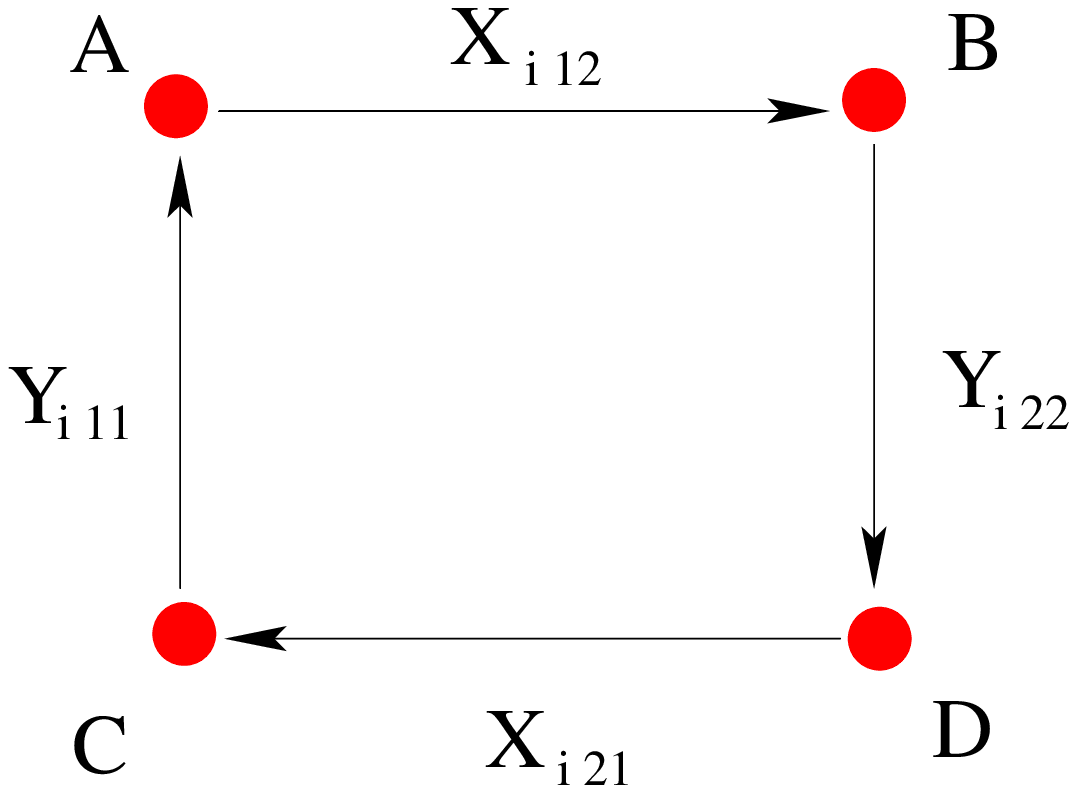}
\qquad \qquad
&&
d = 
\mat{\begin{array}{c|cccc}
         & X_{i~12}   & X_{i~21}    & Y_{i~11}  & Y_{i~22}   \\
\hline
A  &    -1     &     0    &   1     &    0  \\
B  &    1     &    0    &    0    &    -1   \\
C  &    0    &     1    &    -1    &    0   \\
D  &    0     &    -1    &   0    &    1   \\
\end{array}
}
\ea
$$
and the superpotential is given by 
\beq
W = \epsilon^{ij}
\epsilon^{kl}X_{i~12}Y_{k~22}X_{j~21}Y_{l~11}.
\label{HirzeMor}
\eeq
\end{itemize}

Indeed, even though both these theories have arisen from the canonical
partial resolutions technique and hence are world volume theories of a
brane probing a Hirzebruch singularity, we see clearly that they
differ vastly in both matter content and superpotential! Which is the
``correct'' physical theory?

In response to this seeming paradox, let us refer to \fref{f:Hirze}.
\EPSFIGURE[!ht]{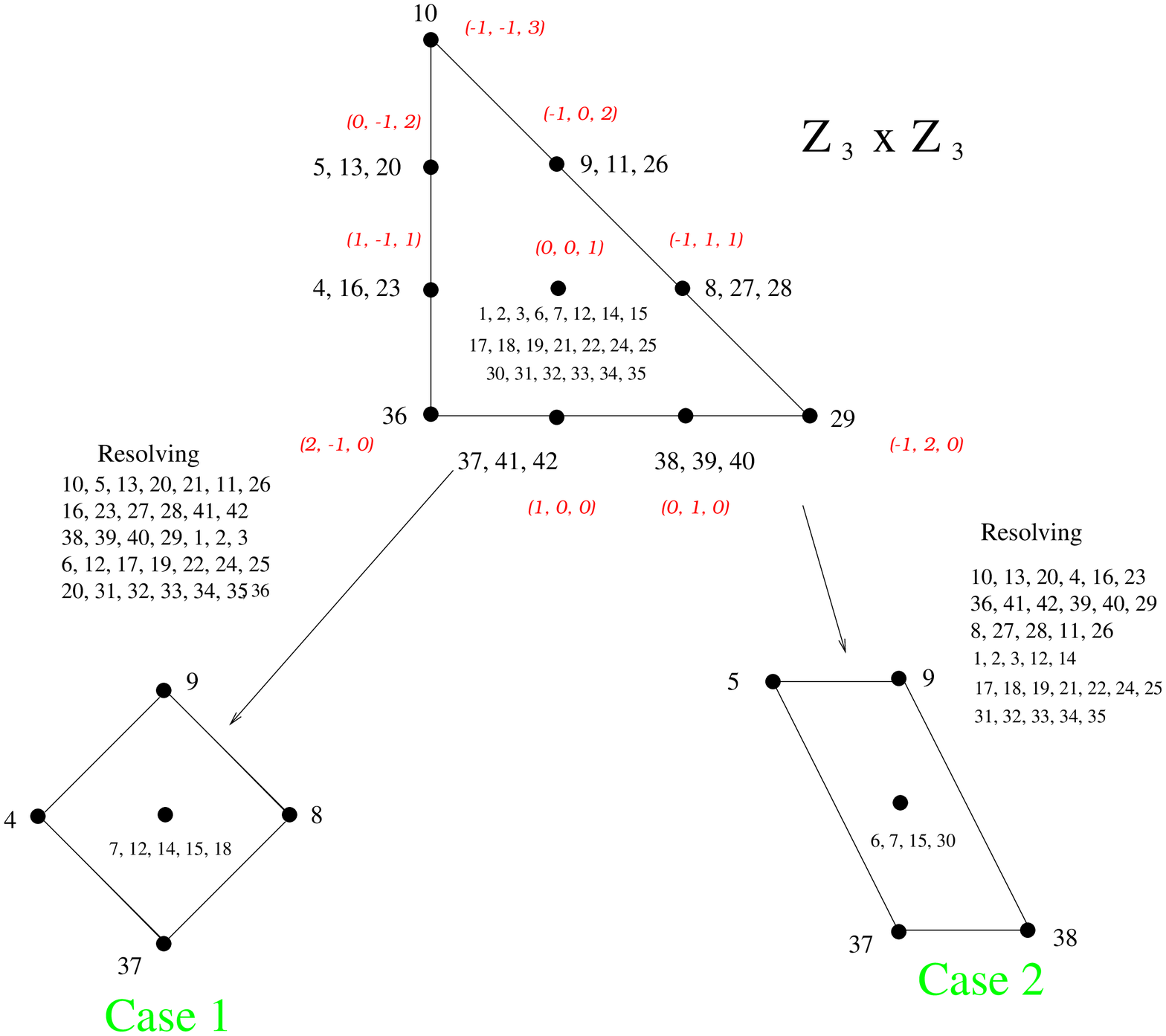,width=4.0in}
{
Two alternative resolutions of $\IC^2/\IZ_3 \times
\IZ_3$ to the Hirzebruch surface $F_0$: Case 1 from \cite{0003085} and
Case 2 from \cite{Horizon}.
\label{f:Hirze}
}
Case 1 of course was what had been analysed in \cite{0003085}
(q.v. ibid.) and presented in \eref{HirzeUs}; 
let us now consider case 2. Using the canonical
algorithm of \cite{Chris,0003085}, we obtain the matter content (we have
labelled the fields and gauge groups with some foresight)
$$
d_{ia} = \left(
\ba{c|cccccccc}
	& X_1 & X'_1 & X'_2 & Y_1 & Y_2 & Y'_1 & Y_2 & Y'_2 \\ \hline
D & 0 & 1 & 1 & 0 & 0 & -1 & 0 & -1 \\
A & -1 & 0 & 0 & 1 & 1 & 0 & -1 & 0 \\
B & 1 & -1 & -1 & 0 & 0 & 0 & 1 & 0 \\
C & 0 & 0 & 0 & -1 & -1 & 1 & 0 & 1
\ea
\right)
$$
and the dual cone matrix
$$
K_{ij}^T =
\left(
\ba{c|cccccccc}
	& X_1 & X'_1 & X'_2 & Y_1 & Y_2 & Y'_1 & X_2 & Y'_2 \\ \hline
p_1 & 1 & 0 & 0 & 0 & 0 & 1 & 0 & 0 \\
p_2 & 0 & 1 & 0 & 1 & 0 & 0 & 0 & 0 \\ 
p_3 & 1 & 0 & 0 & 0 & 0 & 0 & 1 & 0 \\
p_4 & 0 & 1 & 1 & 0 & 0 & 0 & 0 & 0 \\
p_5 & 0 & 0 & 1 & 0 & 1 & 0 & 0 & 0 \\
p_6 & 0 & 0 & 0 & 0 & 0 & 1 & 0 & 1
\ea
\right)
$$
which translates to the F-term equations
$$
X_1 Y'_2 = p_1 p_3 p_6 = Y'_1 X_2; \quad
X'_1 Y_2 = p_2 p_4 p_5 = Y_1 X'_2.
$$
What we see of course, is that with the field redefinition
$X_i \leftrightarrow X_{i~12}, X'_i \leftrightarrow Y_{i~22},
Y_i \leftrightarrow Y_{i~11}$ and $Y'_i \leftrightarrow X_{i~21}$
for $i=1,2$, the above results are in exact agreement with the results
from \cite{Horizon} as presented in \eref{HirzeMor}.

This is actually of no surprise to us because upon closer inspection
of \fref{f:Hirze}, we see that the toric diagram for Cases 1 and 2
respectively has the coordinate points
$$
G1_t=
\left(
  \matrix{ -1 & 1 & 1 & 0 & -1 \cr 0 & 
     -1 & 0 & 0 & 1 \cr 2 & 1 & 0 & 1 & 1 \cr  } 
\right)
\qquad
G2_t=
\left(
  \matrix{ 0 & -1 & 1 & 0 & 0 \cr 
     -1 & 0 & 0 & 1 & 0 \cr 2 & 2 & 0 & 0 & 1 \cr  } 
\right).
$$
Now since the algebraic equation of the toric variety is given by
\cite{Fulton}
$$
V(G_t) = Spec_{Max}\left( \C[X_i^{G_t^\vee \cap \IZ^3}]\right),
$$
we have checked that, using a reduced Gr\"obner polynomial basis algorithm to
compute the variety \cite{Sturmfels}, the equations are
identical up to redefinition of variables.

Therefore we see that the two toric diagrams in Cases 1 and 2 of
\fref{f:Hirze} both describe the zeroth Hirzebruch surface as they
have the same equations (embedding into $\IC^9$). Yet due to the
particular choice of the diagram, we end up with strikingly different
gauge theories on the D-brane probe despite the identification of the
moduli space in the IR. This is indeed a curiously strong
version of ``toric duality.''

Bearing the above in mind, in this chapter, 
we will analyse the degrees of freedom in the Inverse
Algorithm expounded upon in \cite{0003085}, 
i.e., for a given toric singularity, how many different physical gauge
theories (phase structures), resulting from various partial
resolutions can one have for a D-brane probing such a singularity? 
To answer this question, first in
\S 2 we present the concept of toric isomorphism and give the
conditions for different toric data to correspond to the same toric 
variety. Then in \S 3 we follow the Forward Algorithm and give the 
freedom at each step from a given set of gauge theory data all the way
to the output of the toric data. 
Knowing these freedoms, we can identify the sources that may give rise
to different gauge theories in the Inverse Algorithm starting
from a prescribed toric data.
In section 4, we apply the above results and analyse the different
phases for the partial resolutions of the
$\IZ_3\times \IZ_3$ orbifold singularity, in particular,
we found that there are
two inequivalent phases of gauge theories respectively for the zeroth
Hirzebruch surface and the second del Pezzo surface.
Finally, in section 5, we give discussions for further investigation.
\index{Toric Duality}
\index{Toric Variety!isomorphism}
\section{Toric Isomorphisms}
Extending this observation to generic toric singularities, we expect
classes of inequivalent toric diagrams corresponding to the same
variety to give rise to inequivalent gauge theories on the D-brane
probing the said singularity. An immediate question is naturally
posed: ``is there a classification of these different theories and is
there a transformation among them?''

To answer this question we resort to the following result. 
Given $M$-lattice cones $\sigma$ and $\sigma'$, let the linear span of
$\sigma$ be lin$\sigma = \IR^n$ and that of $\sigma'$ be $\IR^m$. 
Now each cone gives rise to a semigroup which is the intersection of
the dual cone $\sigma^\vee$ with the dual lattice $M$, i.e., $S_\sigma
:= \sigma^\vee \cap M$ (likewise for $\sigma'$). Finally the toric
variety is given as the maximal spectrum of the polynomial ring of
$\IC$ adjoint the semigroup, i.e., $X_\sigma := Spec_{Max}
\left( \IC[S_\sigma]\right)$.
\begin{definition}We have these types of isomorphisms:
\begin{enumerate}
\item We call $\sigma$ and $\sigma'$ {\em cone isomorphic}, denoted
	$\sigma \cong_{cone} \sigma'$, if $n=m$ and there is
	a unimodular transformation $L:\IR^n \rightarrow \IR^n$ with
	$L(\sigma) = \sigma'$;
\item we call $S_\sigma$ and $S_{\sigma'}$ {\em monomial isomorphic},
	denoted $S_\sigma \cong_{mon} S_{\sigma'}$, if there
	exists mutually inverse monomial homomorphisms between the two
	semigroups.
\end{enumerate} 
\end{definition}
Thus equipped, we are endowed with the following
\begin{theorem}\label{iso}
(\cite{Ewald}, VI.2.11) The following conditions are equivalent:
$$
(a)~~\sigma \cong_{cone} \sigma' \Leftrightarrow
(b)~~S_\sigma \cong_{mon} S_{\sigma'} \Leftrightarrow
(c)~~X_\sigma \cong X_{\sigma'}
$$
\end{theorem}
What this theorem means for us is simply that, for the $n$-dimensional
toric variety, an $SL(n;\IZ)$ transformation\footnote{Strictly
	speaking, by unimodular we mean $GL(n;\IZ)$ matrices with
	determinant $\pm 1$; we shall denote these loosely by $SL(n;\IZ)$.
	}
on the original lattice
cone amounts to merely co\"ordinate transformations on the polynomial
ring and results in the same toric variety. This, is precisely what we
want: different toric diagrams giving the same variety.

The necessity and sufficiency of the condition in Theorem \ref{iso}
is important. Let us think of one example to
illustrate. Let a cone be defined by $(e_1,e_2)$, we know this
corresponds to $\IC^2$. Now if we apply the transformation
$$
(e_1,e_2) \left[ \begin{array}{cc} 2  & 0 \\ -1 & 1 \end{array}
\right]=(2e_1-e_2,e_2),
$$
which corresponds to the variety $xy=z^2$, i.e., $\IC^2/\IZ_2$, which of
course is not isomorphic to $\IC^2$. The reason for this is obvious:
the matrix we have chosen is certainly not unimodular.
\index{Toric Duality}
\section{Freedom and Ambiguity in the Algorithm}
In this section, we wish to step back and address the issue in fuller
generality. Recall that the procedure of obtaining the moduli space
encoded as toric data once given the gauge
theory data in terms of product $U(1)$ gauge groups, D-terms from matter
contents and F-terms from the superpotential, has been well developed
\cite{Horizon,DGM}. Such was called the {\bf forward algorithm} in
\cite{0003085}. On the other hand the {\bf reverse algorithm} of
obtaining the gauge theory data from the toric data has been discussed
extensively in \cite{Chris,0003085}.

It was pointed in \cite{0003085} that both the forward and reverse
algorithm are highly non-unique, a property which could actually be
harnessed to provide large classes of gauge theories having the same
IR moduli space.
In light of this so-named ``toric duality''
it would be instructive for us to investigate 
how much freedom do we have at each step in the algorithm.
We will call two data related by such a freedom {\em equivalent} to
each other. Thence further we could see how freedoms at every step
accumulate and appear in the final toric data. Modulo such
equivalences we believe that the data should be uniquely determinable.
\index{Toric Variety!Forward Algorithm}
\index{Vacuum Moduli Space}
\subsection{The Forward Algorithm}
We begin with the forward algorithm of extracting toric data from
gauge data. A brief review is at hand.
To specify the gauge theory, we require three pieces of information:
the number of $U(1)$ gauge fields, the charges of matter fields and the
superpotential.
The first two are summarised by the so-called charge matrix
$d_{li}$ where $l=1,2,...,L$ with $L$ the number of $U(1)$ gauge fields and
$i=1,2,...,I$ with $I$ the number of matter fields.
When using the forward algorithm to find the vacuum manifold (as a
toric variety), we need to solve the D-term and F-term flatness equations.
The D-terms are given by $d_{li}$ matrix while the 
F-terms are encoded in a matrix $K_{ij}$ with 
$i,1,2,...,I$ and $j=1,2,...,J$ where $J$ is the number of 
independent parameters needed to solve the F-terms. By gauge data then
we mean the matrices $d$ (also called the incidence matrix) and the
$K$ (essentially the dual cone); the forward algorithm takes these as input.
Subsequently we trace a flow-chart:
$$
\begin{array}{ccccccc}
\mbox{D-Terms} \rightarrow d     & \rightarrow   &\Delta & & & & \\
        &       &\downarrow     &       &       &       &       \\
\mbox{F-Terms} \rightarrow K    & \stackrel{V \cdot K^T =
        \Delta}{\rightarrow}
                & V      & & & & \\
\downarrow      &       & \downarrow    & & & & \\
T = {\rm Dual}(K)       & \stackrel{U \cdot T^T = {\rm
        Id}}{\rightarrow} & U & \rightarrow & VU & &\\
\downarrow      &       &       &       & \downarrow    & & \\
Q = [{\rm Ker}(T)]^T    &       & \longrightarrow       & & Q_t =
        \left( \begin{array}{c}
                                        Q \\ VU \end{array} \right) &
                                                \rightarrow & G_t =
        [{\rm Ker}(Q_t)]^T \\

\end{array}
$$
arriving at a final matrix $G_t$ whose columns are the vectors which
prescribe the nodes of the toric diagram.

What we wish to investigate below is how much procedural freedom we have at
each arrow so as to ascertain the non-trivial toric dual
theories. Hence, if $A_1$ is the matrix whither one arrives from a
certain arrow, then we would like to find the most general
transformation taking $A_1$ to another solution $A_2$ which would give
rise to an identical theory. It is to this transformation that we shall 
refer as ``freedom'' at the particular step.
\subsection*{Superpotential: the matrices $K$ and $T$}
The solution of F-term equations gives rise to a dual cone $K_1 = K_{ij}$
defined by $I$ vectors in $\IZ^J$.
Of course, we can choose different
parametres to solve the F-terms and arrive at another dual cone $K_2$. 
Then, $K_1$ and $K_2$, being integral cones, are equivalent if
they are unimodularly related, i.e., $ K_2^T= A \cdot K_1^T$ for
$A\in GL(J,\IZ)$ such that $\det(A)=\pm 1$.
Furthermore, the order of the $I$ vectors in
$\IZ^J$ clearly does not matter, so we can permute them by a matrix
$S_I$ in the symmetric group ${\cal S}_I$.
Thus far we have two freedoms, multiplication by $A$ and $S$:
\beq\label{K_tran}
K_2^T= A \cdot K_1^T \cdot S_I,
\eeq
and $K_{1,2}$ should give equivalent theories.

Now, from $K_{ij}$ we can find its dual matrix
$T_{j \alpha}$ (defining the cone $T$) where $\alpha=1,2,...,c$ and $c$ is the
number of vectors of the cone $T$ in $\IZ^J$, as constrained by
\beq \label{T_def}
K \cdot T \geq 0
\eeq
and such that $T$ also spans an integral cone.
Notice that finding dual cones, as given in a algorithm in
\cite{Fulton}, is actually unique up to
permutation of the defining vectors. 
Now considering the freedom of $K_{ij}$ as in
(\ref{K_tran}), let $T_2$ be the dual of $K_2$ and $T_1$ that of
$K_1$,  we have
$K_2 \cdot T_2= S_I^T \cdot  K_1 \cdot A^T \cdot T_2 \geq 0$,
which means that 
\beq
\label{T_tran}
T_1 = A^T \cdot T_2 \cdot S_c.
\eeq
Note that here $S_c$ is the permutation of the $c$ vectors of the cone $T$ in
and not that of the dual cone in \eref{K_tran}.
\subsection*{The Charge Matrix $Q$}
The next step is to find the charge matrix $Q_{k \alpha}$ where
$\alpha=1,2,...,c$ and $k=1,2,...,c-J$. This matrix is defined by
\beq
\label{Q_def}
T \cdot Q^T=0.
\eeq
In the same spirit as the above discussion, from \eref{T_tran} we have
$T_1 \cdot Q_1^T=A^T \cdot T_2 \cdot S_c \cdot Q_1^T=0$.
Because $A^T$ is a invertible matrix, this has a solution when and
only when $T_2 \cdot S_c \cdot Q_1^T=0$. Of course this is equivalent
to $T_2 \cdot S_c \cdot Q_1^T \cdot B_{kk'}=0$
for some invertible $(c-J)\times (c-J)$ matrix $B_{kk'}$. So the freedom for
matrix $Q$ is
\beq
\label{Q_tran}
Q_2^T=  S_c \cdot Q_1^T \cdot  B.
\eeq
We emphasize a difference from \eref{T_def}; there we required
both matrices $K$ and $T$ to be integer where here \eref{Q_def} does not
possess such a constraint. Thus the only condition for the matrix $B$ is
its invertibility.
\subsection*{Matter Content: the Matrices $d$, $\widetilde{V}$ and $U$}
Now we move onto the D-term and the integral $d_{li}$ matrix.
The D-term equations are $d \cdot |X|^2 = 0$ for matter fields $X$.
Obviously, any transformation on $d$ by an invertible matrix
$C_{L\times L}$ does not change the D-terms. Furthermore, any
permutation $S_I$ of the order the fields $X$, so long as it is
consistent with the $S_I$ in \eref{K_tran}, is also game.
In other words, we have the freedom:
\beq
\label{d_tran}
d_2 =C \cdot d_1 \cdot S_I.
\eeq
We recall that a matrix $V$ is then determined from $\Delta$, which is
$d$ with a row deleted due to the centre of mass degree of freedom.
However, to not to spoil the above freedom enjoyed by matrix $d$ in
\eref{d_tran}, we will make a slight
amendment and define the matrix $\widetilde{V}_{lj}$ by
\beq
\label{til_V_def}
\widetilde{V}\cdot K^T= d.
\eeq 
Therefore, whereas in \cite{DGM,0003085} where $V\cdot K^T= \Delta$ 
was defined, we generalise $V$ to $\widetilde{V}$ by \eref{til_V_def}. 
One obvious way to obtain $\widetilde{V}$ from $V$ is to add one row such
that the sum of every column is zero. However, there is a caveat:
when there exists a vector $h$ such that
$$
h \cdot K^T=0,
$$
we have the freedom to add $h$ to any row of $\widetilde{V}$.
Thus finding the freedom of $\widetilde{V}_{lj}$ is a little more
involved. From \eref{K_tran} we have
$d_2= \widetilde{V}_2\cdot K_2^T=\widetilde{V}_2\cdot 
A \cdot K_1^T \cdot S_I$
and 
$d_2 =C \cdot d_1\cdot S_I =C \cdot \widetilde{V}_1\cdot K_1^T\cdot S_I$.
Because $S_I$ is an invertible square matrix, we have
$(\widetilde{V}_2\cdot A-C \cdot \widetilde{V}_1)\cdot K_1^T=0$,
which means 
$
\widetilde{V}_2\cdot A-C \cdot \widetilde{V}_1= CH_{K_1}
$
for a matrix $H$ constructed by having the aforementioned vectors $h$
as its columns.
When $K^T$ has maximal rank, $H$ is zero and this is in fact the more
frequently encountered situation.
However, when $K^T$ is not maximal rank, so as to give
non-trivial solutions of $h$,
we have that $\widetilde{V}_1 $ and $\widetilde{V}_2$ are equivalent if
\beq
\label{til_V_tran}
\widetilde{V}_2=C \cdot (\widetilde{V}_1+H_{K_1} ) \cdot A^{-1}.
\eeq

Moving on to the matrix $U_{j\alpha}$ defined by
\beq
\label{U_def}
U\cdot T^T= \II_{jj'},
\eeq
we have from \eref{T_tran}
$\II_{jj'}=U_1\cdot T_1^T=U_1\cdot S_c^T \cdot T_2^T \cdot A$,
whence $A^{-1}=U_1\cdot S_c^T \cdot T_2^T$ and 
$\II = A \cdot U_1\cdot S_c^T \cdot T_2^T$. This gives
$(A \cdot U_1\cdot S_c^T-U_2)\cdot T_2^T=0$
which has a solution
$A \cdot U_1\cdot S_c^T-U_2=H_{T_2}$
where $H_{T_2} \cdot T_2^T=0$ is precisely as defined in analogy of
the $H$ above. Therefore the freedom on $U$ is subsequently 
\beq
\label{U_tran}
U_2= A \cdot (U_1-H_{T_1})\cdot S_c^T,
\eeq
where $H_{T_1}=A^{-1} H_{T_2} (S_c^T)^{-1}$ and
$H_{T_1} \cdot T^T_1= (A^{-1} H_{T_2} (S_c^T)^{-1})(S_c^T \cdot T_2^T \cdot A)
=0$.
Finally using \eref{til_V_tran} and \eref{U_tran}, we have
\beq
\label{VU_tran}
(\widetilde{V}_2 \cdot U_2) =C \cdot (\widetilde{V}_1+H_{K_1})
 \cdot A^{-1} 
\cdot A \cdot (U_1-H_{T_1})\cdot S_c^T=
C \cdot (\widetilde{V}_1+H_{K_1})(U_1-H_{T_1}) \cdot S_c^T,
\eeq
determining the freedom of the relevant combination
$(\widetilde{V} \cdot U)$.

Let us pause for an important observation that in most cases 
$H_{K_1} = 0$, as we shall see in the examples later.
From \eref{Q_def}, which propounds the existence of a non-trivial
nullspace  for $T$, we see that one can indeed obtain
a non-trivial $H_{T_1}$ in terms of the combinations of the rows of
the charge matrix $Q$, whereby simplifying \eref{VU_tran} to
\beq
\label{VU_tran_1}
(\widetilde{V}_2 \cdot U_2)=C \cdot (\widetilde{V}_1 \cdot U_1+
H_{VU_1}) \cdot S_c^T,
\eeq
where every row of $H_{VU_1}$ is linear combination of rows of $Q_1$ and
the sum of its columns is zero.
\subsection*{Toric Data: the Matrices $Q_t$ and $G_t$}
At last we come to $\widetilde{Q}_t$, which is given by adjoining
$Q$ and $\widetilde{V}\cdot U$.  The freedom is of course, by
combining all of our results above,
\beq
\label{Qt_tran}
(\widetilde{Q}_t)_2 =\left( \begin{array}{c}  Q_2 \\ 
\widetilde{V}_2 \cdot U_2 \end{array} \right)
=\left( \begin{array}{c} B^T \cdot Q_1 \cdot S_c^T\\ 
C \cdot (\widetilde{V}_1 \cdot U_1+H_{VU_1}) \cdot S_c^T
\end{array} \right)
=\left( \begin{array}{c} B^T \cdot Q_1 \\ 
C \cdot (\widetilde{V}_1 \cdot U_1+H_{VU_1})
\end{array} \right) \cdot S_c^T
\eeq
Now $\widetilde{Q}_t$ determines the nodes of the toric diagram
$(G_t)_{p \alpha}$ ($p=1,2,..,(c-(L-1)-J)$ and $\alpha=1,2,...,c$) by 
\beq
\label{Gt_def}
Q_t \cdot G_t^T =0;
\eeq
\index{Vacuum Moduli Space}
The columns of $G_t$ then describes the toric diagram of the algebraic
variety for the vacuum moduli space and is the output of the algorithm.
From \eref{Gt_def} and  \eref{Qt_tran} we find that if 
$(\widetilde{Q}_t)_1 \cdot (G_t)_1^T=0$, i.e., 
$Q_1 \cdot (G_t)_1^T=0$ and $\widetilde{V}_1 \cdot U_1 \cdot
(G_t)_1^T=0,$ we automatically have the freedom $(\widetilde{Q}_t)_2 
\cdot (S_c^T)^{-1}\cdot (\widetilde{G}_t)_1^T=0$. This means that
at most we can have
\beq
\label{Gt_tran}
(G_t)_2^T=(S_c^T)^{-1}\cdot (G_t)_1^T \cdot D,
\eeq
where $D$ is a $GL(c-(L-1)-J,\IZ)$ matrix with $\det(D)=\pm 1$ which
is exactly the unimodular freedom for toric data as given by Theorem
\ref{iso}.

One immediate remark follows. From \eref{Gt_def} we obtain
the nullspace of $Q_t$ in $\IZ^c$. It seems that we can choose an
arbitrary basis so that $D$ is a $GL(c-(L-1)-J,\IZ)$ matrix with
the only condition that $\det(D)\neq 0$.
However, this is not stringent enough: in fact, when we find
cokernel $G_t$, we need to find the {\em integer basis} for the null
space, i.e., we need to find the basis such that any integer null vector can be 
decomposed into a linear combination of the columns of $G_t$. If we
insist upon such a choice, the only remaining freedom\footnote{
We would like to express our gratitude to M. Douglas for clarifying this point
to us.} is that $\det(D)=\pm 1$, viz, unimodularity.
\index{Toric Variety!Inverse Algorithm}
\subsection{Freedom and Ambiguity in the Reverse Algorithm}
Having analysed the equivalence conditions in last subsection,
culminating in \eref{Qt_tran} and \eref{Gt_tran}, we now proceed in
the opposite direction and address the ambiguities in the reverse
algorithm.
\subsection*{The Toric Data: $G_t$}
We note that the $G_t$ matrix produced by the forward algorithm is
not minimal in the sense that certain columns are repeated, which
after deletion, constitute the toric diagram. 
Therefore, in our reverse algorithm, 
we shall first encounter such an ambiguity in deciding which columns
to repeat when constructing $G_t$ from the nodes of the toric
diagram. This so-called {\em repetition ambiguity} was discussed in
\cite{0003085} and different choices of repetition may indeed
give rise to different gauge theories. It was pointed out ({\it
loc.~cit.}) that arbitrary repetition of the columns certainly does not
guarantee physicality. By physicality we mean that
the gauge theory arrived at the end of the day should be
{\em physical} in the sense of still being a D-brane world-volume theory.
What we shall focus here however, is the
inherent symmetry in the toric diagram, given by \eref{Gt_tran}, that
gives rise to the same theory. This is so that we could find truly
inequivalent {\em physical} gauge theories not related by
such a transformation as \eref{Gt_tran}.
\subsection*{The Charge Matrix: from $G_t$ to $Q_t$}
From \eref{Gt_def} we can solve for $Q_t$. However, for
a given $G_t$, in principle we can have two solutions $(Q_t)_1$ and
$(Q_t)_2$ related by
\beq \label{Qt_tran_rev}
(Q_t)_2 = P (Q_t)_1,
\eeq
where $P$ is a $p\times p$ matrix with $p$ the number of rows of
$Q_t$. Notice that the set of such transformations $P$ is much larger
than the counterpart in the forward algorithm given in
\eref{Qt_tran}. This is a second source of ambiguity in the reverse
algorithm. More explicitly,
we have the freedom to arbitrarily divide the $Q_t$ into two parts, viz.,
the D-term part $\widetilde{V}U$ and the F-term part $Q$.
Indeed one may find a matrix $P$ such that $(Q_t)_1$ and $(Q_t)_2$ satisfy
\eref{Qt_tran_rev} but not matrices $B$ and $C$ in order to satisfy 
\eref{Qt_tran}.
Hence different choices of $Q_t$ and different division therefrom into
D and F-term parts give rise to different gauge theories.
This is what we called {\em FD Ambiguity} in \cite{0003085}. Again,
arbitrary division of the rows of $Q_t$ was pointed out to not to
ensure physicality. As with the discussion on the repetition ambiguity
above, what we shall pin down is the freedom due to the linear algebra
and not the choice of division.
\subsection*{The Dual Cone and Superpotential: from $Q$ to $K$}
The nullspace of $Q$ is the matrix $T$. The issue is the same as
discussed at the paragraph following \eref{Gt_tran} and one can
uniquely determine $T$ by imposing that its columns give an integral
span of the nullspace.
Going further from $T$ to its dual $K$, this is again a unique
procedure (while integrating back from $K$ to obtain the superpotential
is certainly not).
In summary then, these two steps give no sources for ambiguity.
\subsection*{The Matter Content: from $\widetilde{V}U$ to $d$ matrix}
The $d$ matrix can be directly calculated as \cite{0003085}
\beq \label{d_def}
d=(\widetilde{V}U)\cdot T^T \cdot K^T.
\eeq
Substituting the freedoms in \eref{K_tran}, \eref{T_tran} and
\eref{VU_tran} we obtain
$$
\begin{array}{rcl}
d_2 & = & (\widetilde{V}_2 \cdot U_2) \cdot T_2^T \cdot K_2^T
=C \cdot [(\widetilde{V}_1 \cdot U_1)+H_{VU_1}] \cdot S_c^T
\cdot (S_c^T)^{-1} \cdot T_1^T \cdot A^{-1} \cdot A \cdot K_1^T \cdot S_I
\\ & & \\
& = & C \cdot (\widetilde{V}_1 \cdot U_1)\cdot T_1^T\cdot K_1^T\cdot S_I
+ C \cdot H_{VU_1} \cdot T_1^T\cdot K_1^T\cdot S_I
=C \cdot d_1 \cdot S_I,
\end{array}
$$
which is exactly formula (\ref{d_tran}). 
This  means that the matter matrices are equivalent up to a
transformation and there is no source for extra ambiguity.
\index{Resolution!partial}
\index{Toric Duality}
\index{del Pezzo Surfaces}
\section{Application: Phases of $\IZ_3 \times \IZ_3$ Resolutions}
In \cite{0003085} we developed an algorithmic outlook to the Inverse
Procedure and applied it to the construction of gauge theories on the
toric singularities which are partial resolutions of $\IZ_3 \times
\IZ_3$. The non-uniqueness of the method allowed one to obtain many
different gauge theories starting from the same toric variety, theories
to which we referred as being toric duals.
The non-uniqueness mainly comes from three
sources: (i) the repetition of the vectors in the toric data $G_t$
(Repetition Ambiguity), (ii) the different
choice of the null space basis of $Q_t$ and (iii) the different
divisions of the rows of $Q_t$ (F-D Ambiguity).
Many of the possible choices in the above will generate unphysical
gauge theories, i.e., not world-volume theories of D-brane probes. 
We have yet to catalogue the exact conditions which guarantee
physicality.

However, {\em Partial Resolution} of Abelian orbifolds, which stays
within subsectors of the latter theory, does indeed constrain the
theory to be physical. To these physical theories we shall refer as
{\bf phases} of the partial resolution. As discussed in \cite{0003085}
any $k$-dimensional toric diagram can be embedded into $\IZ_n^{k-1}$
for sufficiently large $n$, one obvious starting point to
obtain different phases of a D-brane gauge theory is to try various
values of $n$. We leave some relevances of general $n$ to the Appendix.
However, because the algorithm of finding dual cones becomes
prohibitively computationally intensive even for $n \ge 4$, this
approach may not be immediately fruitful.

Yet armed with Theorem \ref{iso} we have an alternative. We can
certainly find all possible unimodular transformations of the given
toric diagram which still embeds into the same $\IZ_n^{k-1}$ and then
perform the inverse algorithm on these various {\it a fortiori}
equivalent toric data and observe what physical theories we obtain at
the end of the day. In our two examples in \S 1, we have essentially
done so; in those cases we found that two inequivalent gauge theory
data corresponded to two unimodularly equivalent toric data for the
examples of $\IZ_5$-orbifold and the zeroth Hirzebruch surface $F_0$.

The strategy lays itself before us. Let us illustrate with the same
examples as was analysed in \cite{0003085}, namely the partial
resolutions of $\IC^3/(\IZ_3\times \IZ_3)$, i.e., $F_0$ and the toric del
Pezzo surfaces $dP_{0,1,2,3}$. We need to (i) find all $SL(3;\IZ)$
transformations of the toric diagram $G_t$ of these five singularities
that still remain as sub-diagrams of that of $\IZ_3\times \IZ_3$ and
then perform the inverse algorithm; therefrom, we must (ii) select
theories not related by any of the freedoms we have discussed above
and summarised in \eref{Qt_tran}.
\subsection{Unimodular Transformations within $\IZ_3\times \IZ_3$}
We first remind the
reader of the $G_t$ matrix of $\IZ_3\times \IZ_3$ given in
\fref{f:Hirze}, its columns are given by vectors:
$(0, 0, 1)$, $(1, -1, 1)$, $(0, -1, 2)$, $(-1, 1, 1)$, $(-1, 0, 2)$, 
$(-1, -1, 3)$, $(1, -1, 1)$, $(-1, 2, 0)$, $(1, 0, 0)$, $(0, 1, 0)$.
Step (i) of our above strategy can be immediately performed. Given the
toric data of one of the resolutions $G_t'$ with $x$ columns, 
we select $x$ from the above 10 columns of $G_t$ and check whether any
$SL(3;\IZ)$ transformation relates any permutation thereof
unimodularly to $G_t'$. We shall at the end find that there are
three different cases for $F_0$, five for $dP^0$, twelve for $dP_1$,
nine for $dP_2$ and only one for $dP_3$. The (unrepeated) $G_t$
matrices are as follows:
$$
\ba{|c|l|}
\hline
(F_0)_1 &  (0,0,1), (1,-1,1), (-1,1,1), (-1,0,2), (1,0,0) \\
(F_0)_2 & (0,0,1), (0,-1,2), (0,1,0), (-1,0,2), (1,0,0)\\
(F_0)_3 & (0,0,1), (1,-1,1), (-1,1,1), (0,-1,2), (0,1,0)\\
\hline
(dP_0)_1 & (0,0,1), (1,0,0),  (0,-1,2), (-1,1,1)\\
(dP_0)_2 & (0,0,1), (1,0,0), (-1,-1,3), (0,1,0)\\
(dP_0)_3 & (0,0,1), (-1,2,0), (1,-1,1), (0,-1,2)\\
(dP_0)_4 & (0,0,1), (0,1,0), (1,-1,1), (-1,0,2)\\
(dP_0)_5 & (0,0,1), (2,-1,0), (-1,1,1), (-1,0,2)\\
\hline
(dP_1)_1 & (1, 0, 0), (0, 1, 0), (-1, 1, 1), (0, -1, 2), (0, 0, 1)\\
(dP_1)_2 & (-1, -1, 3), (0, -1, 2), (1, 0, 0), (0, 1, 0), (0, 0, 1)\\
(dP_1)_3 & (0, -1, 2), (1, -1, 1), (1, 0, 0), (-1, 1, 1), (0, 0, 1)\\
(dP_1)_4 & (0, -1, 2), (1, -1, 1), (0, 1, 0), (-1, 2, 0), (0, 0, 1)\\
(dP_1)_5 & (0, -1, 2), (1, -1, 1), (0, 1, 0), (-1, 0, 2), (0, 0, 1)\\
(dP_1)_6 & (0, -1, 2), (1, -1, 1), (-1, 2, 0), (-1, 1, 1), (0, 0, 1)\\
(dP_1)_7 & (0, -1, 2), (1, 0, 0), (-1, 1, 1), (-1, 0, 2), (0, 0, 1)\\
(dP_1)_8 & (1, -1, 1), (2, -1, 0), (-1, 1, 1), (-1, 0, 2), (0, 0, 1)\\
(dP_1)_9 & (1, -1, 1), (1, 0, 0), (0, 1, 0), (-1, 0, 2), (0, 0, 1)\\
(dP_1)_{10} & (1, -1, 1), (0, 1, 0), (-1, 1, 1), (-1, 0, 2), (0, 0, 1)\\
(dP_1)_{11} & (2, -1, 0), (1, 0, 0), (-1, 1, 1), (-1, 0, 2), (0, 0, 1)\\
(dP_1)_{12} & (-1, -1, 3), (1, 0, 0), (0, 1, 0), (-1, 0, 2), (0, 0,1)\\
\hline
(dP_2)_1 & (2, -1, 0), (1, -1, 1), (-1, 0, 2), (-1, 1, 1), (1, 0, 0), (0, 0, 1)\\
(dP_2)_2 & (-1, -1, 3), (0, -1, 2), (1, 0, 0), (0, 1, 0), (-1, 0, 2), (0, 0, 1)\\
(dP_2)_3 & (0, -1, 2), (1, -1, 1), (1, 0, 0), (0, 1, 0), (-1, 1, 1), (0, 0, 1)\\
(dP_2)_4 & (0, -1, 2), (1, -1, 1), (1, 0, 0), (0, 1, 0), (-1, 0, 2), (0, 0, 1)\\
(dP_2)_5 & (0, -1, 2), (1, -1, 1), (1, 0, 0), (-1, 1, 1), (-1, 0, 2), (0, 0, 1)\\
(dP_2)_6 & (0, -1, 2), (1, -1, 1), (0, 1, 0), (-1, 2, 0), (-1, 1, 1), (0, 0, 1)\\
(dP_2)_7 & (0, -1, 2), (1, -1, 1), (0, 1, 0), (-1, 1, 1), (-1, 0, 2), (0, 0, 1)\\
(dP_2)_8 & (0, -1, 2), (1, 0, 0), (0, 1, 0), (-1, 1, 1), (-1, 0, 2), (0, 0, 1)\\
(dP_2)_9 & (1, -1, 1), (1, 0, 0), (0, 1, 0), (-1, 1, 1), (-1, 0, 2), (0, 0, 1)\\
\hline
dP_3 & (0, -1, 2), (1, -1, 1), (1, 0, 0), (0, 1, 0), 
	(-1, 1, 1), (-1, 0, 2), (0, 0, 1)\\
\hline
\ea
$$
The reader is referred to \fref{f:F0} to \fref{f:dP3} 
for the toric diagrams of the data
above. The vigilant would of course recognise $(F_0)_1$ to be Case 1
and $(F_0)_2$ as Case 2 of \fref{f:Hirze} as discussed in \S 2 and
furthermore $(dP_{0,1,2,3})_1$ to be the cases addressed in \cite{0003085}.
\vspace{1in}
\subsection{Phases of Theories}
The Inverse Algorithm can then be readily applied to the above toric
data; of the various unimodularly equivalent toric diagrams of the del Pezzo
surfaces and the zeroth Hirzebruch, the details of which fields remain
massless at each node (in the
notation of \cite{0003085}) are also presented in those figures
immediately referred to above.
\EPSFIGURE[!ht]{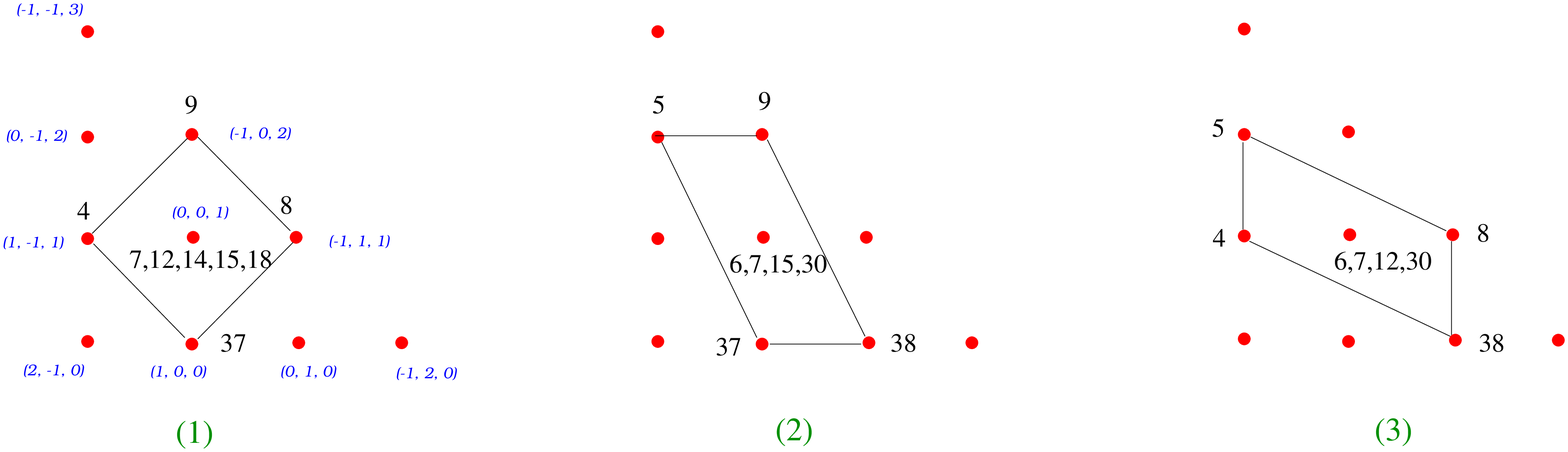,width=6in}
{The 3 equivalent representations of the toric diagram of the zeroth
Hirzebruch surface as a resolution of $\IZ_3\times \IZ_3$. We see that
(2) and (3) are related by a reflection about the $45^o$ line (a symmetry
inherent in the parent $\IZ_3 \times \IZ_3$ theory) and we
have the two giving equivalent gauge theories as expected.
\label{f:F0}
}
\EPSFIGURE[!ht]{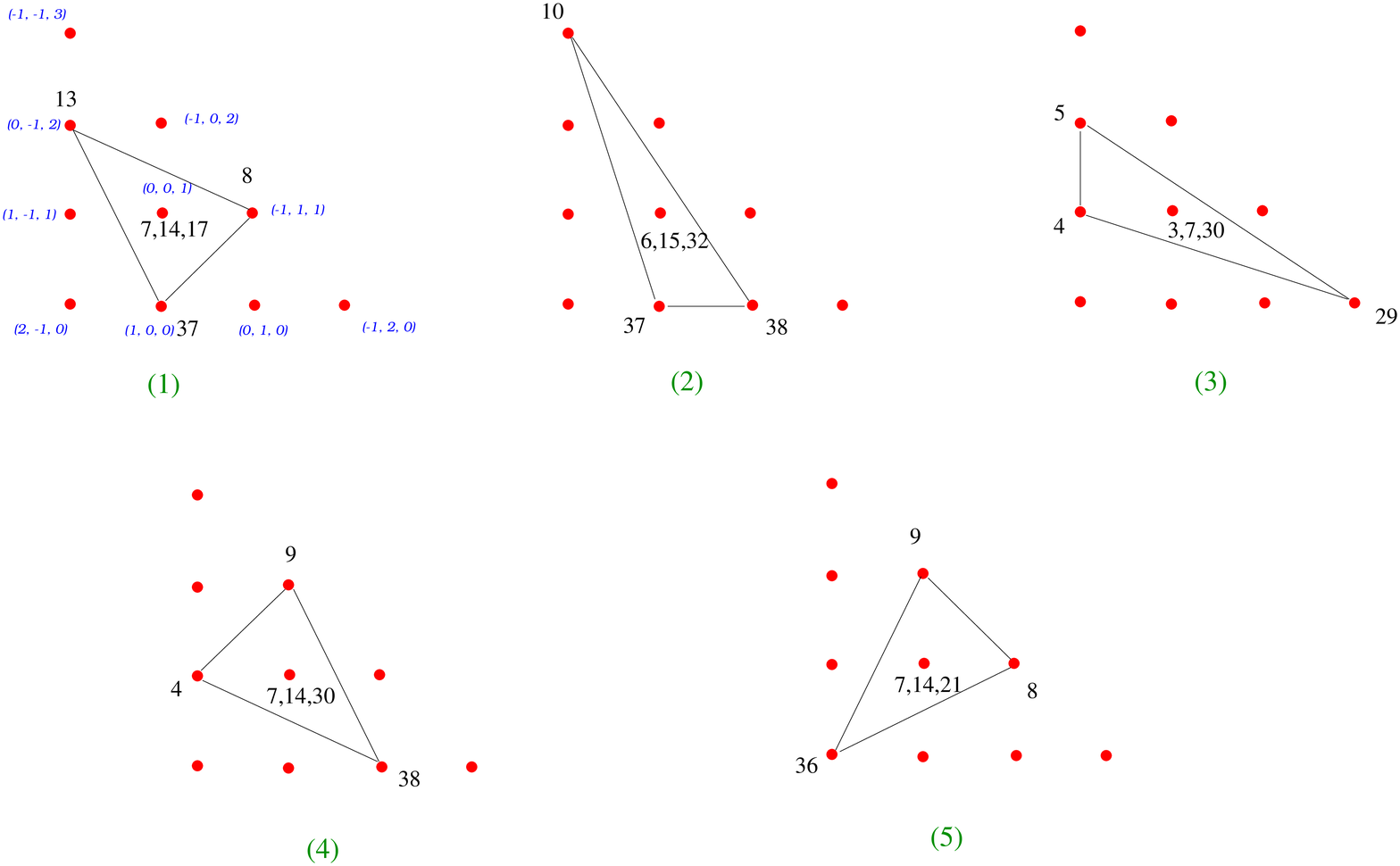,width=5.8in}
{The 5 equivalent representations of the toric diagram of the zeroth
del Pezzo surface as a resolution of $\IZ_3\times \IZ_3$. Again (1)
and (4) (respectively (2) and (3)) are related by the $45^o$
reflection, and hence give equivalent theories. In fact further
analysis shows that all 5 are equivalent.
\label{f:dP0}
}
\EPSFIGURE[!ht]{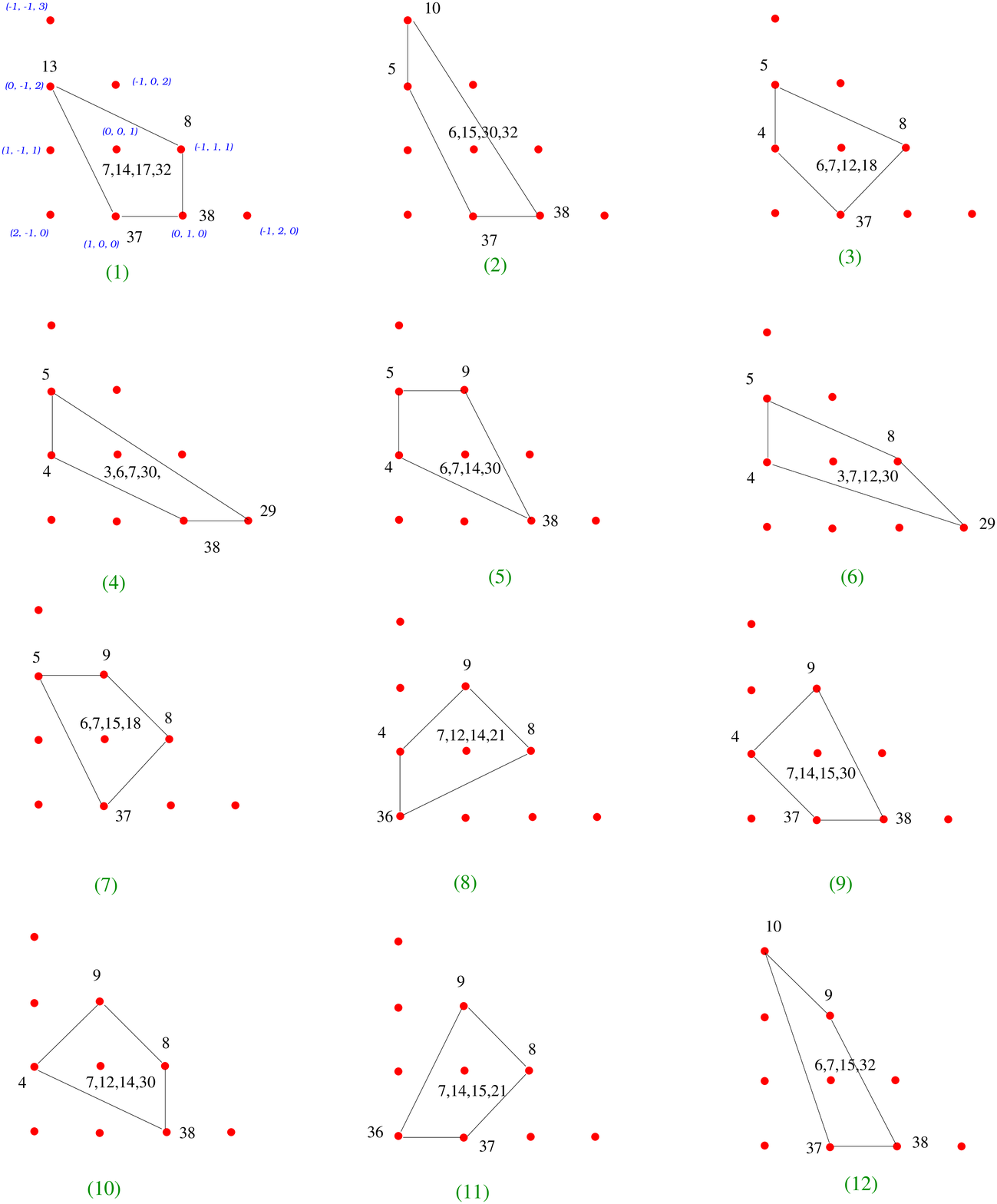,width=6in}
{The 12 equivalent representations of the toric diagram of the first
del Pezzo surface as a resolution of $\IZ_3\times \IZ_3$. The pairs
(1,5); (2,4); (3,9); (6,12); (7,10) and (8,11) are each reflected by
the $45^o$ line and give mutually equivalent gauge theories
indeed. Further analysis shows that all 12 are equivalent.
\label{f:dP1}
}
\EPSFIGURE[!ht]{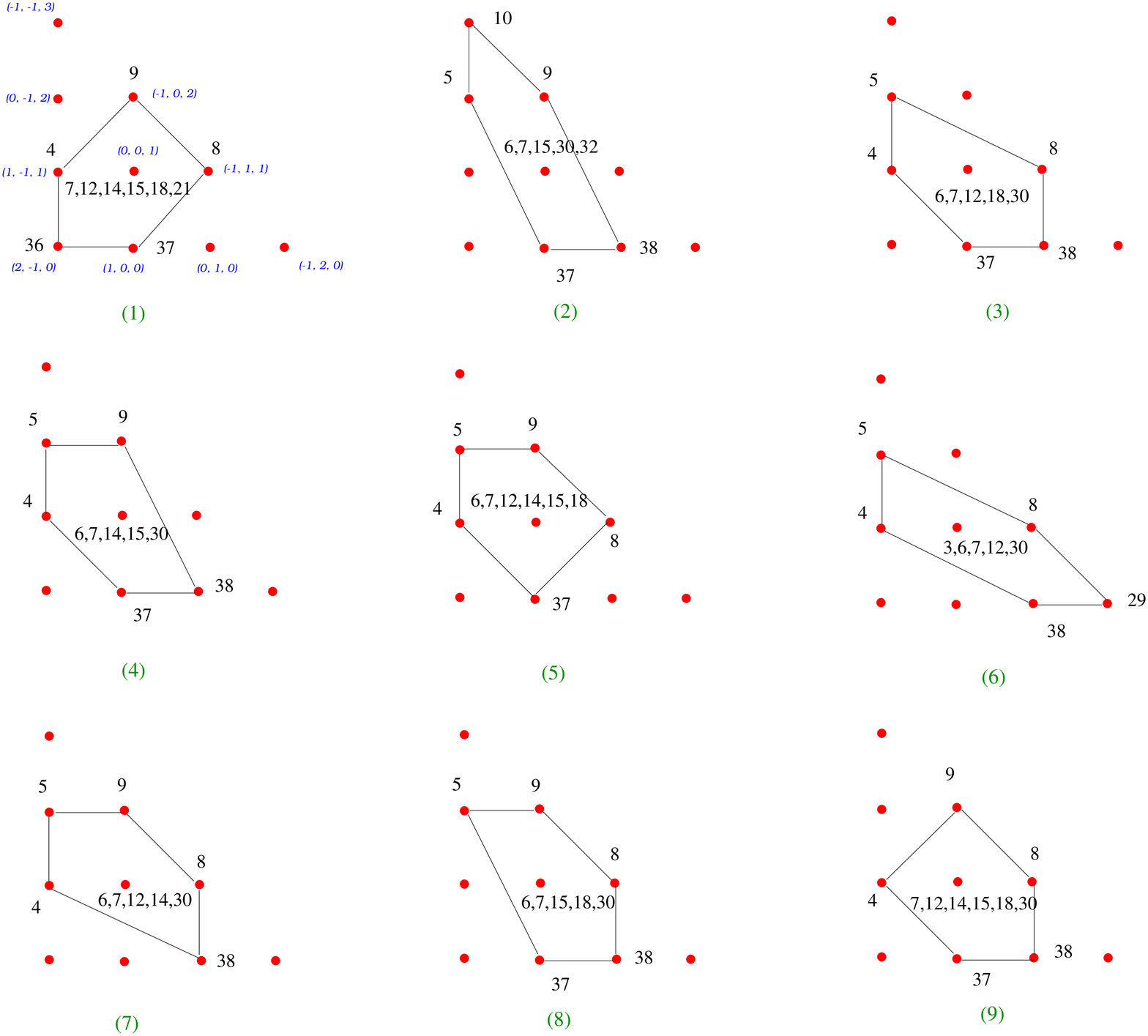,width=6in}
{The 9 equivalent representations of the toric diagram of the second
del Pezzo surface as a resolution of $\IZ_3\times \IZ_3$. The pairs
(2,6); (3,4); (5,9) and (7,8) are related by $45^o$ reflection while
(1) is self-reflexive and are hence give pairwise equivalent
theories. Further analysis shows that there are two phases given
respectively by (1,5,9) and (2,3,4,6,7,8).
\label{f:dP2}
}
\EPSFIGURE[!ht]{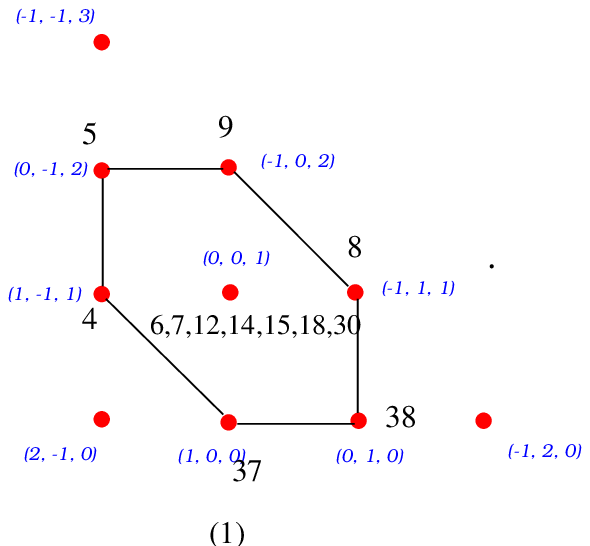,width=5in}
{The unique representations of the toric diagram of the third
del Pezzo surface as a resolution of $\IZ_3\times \IZ_3$.
\label{f:dP3}
}

Subsequently, we arrive at a number of D-brane gauge theories;
among them, all five cases for $dP^0$ are equivalent (which is in
complete consistency with the fact that $dP^0$ is simply
$\IC^3/\IZ_3$ and there is only one
nontrivial theory for this orbifold, corresponding to the decomposition 
${\bf 3}\rightarrow 1 + 1 + 1$). 
For $dP_1$, all twelve cases give back
to same gauge theory (q.v. Figure 5 of \cite{0003085}). 
For $F_0$, the three cases give two inequivalent
gauge theories as given in \S 2. Finally
for $dP_2$, the
nine cases again give two different theories. For reference we
tabulate the D-term matrix $d$ and F-term matrix $K^T$ below. If more
than 1 theory are equivalent, then we select one representative from
the list, the matrices for the rest are given by transformations
\eref{K_tran} and \eref{d_tran}.
\newpage
\[
\hspace{-0.3in}
\setlength{\arraycolsep}{0.1mm}
\ba{|c|c|c|}
\hline
\mbox{Singularity} & \mbox{Matter Content }d & \mbox{Superpotential}\\
\hline
(F_0)_1 & {\tiny \ba{c|cccccccccccc} 
	& X_1 & X_2 & X_3 & X_4 & X_5 & X_6 & X_7 & X_8 & X_9 & X_{10}
	& X_{11} & X_{12}\\ \hline
	A & -1 & 0 & -1 & 0 & -1 & 0 & 1 & 1 & -1 & 0 & 1 & 1 \\ 
	B & 0 & -1 & 0 & -1 & 1 & 0 & 0 & 0 & 1 & 0 & 0 & 0 \\
	C & 0 & 1 & 0 & 1 & 0 & 1 & -1 & -1 & 0 & 1 & -1 & -1 \\ 
	D & 1 & 0 & 1 & 0 & 0 & -1 & 0 & 0 & 0 & -1 & 0 & 0 \ea } 
	&
{\footnotesize
\ba{r}
X_{1}X_{8}X_{10}- X_{3}X_{7}X_{10}- X_{2}X_{8}X_{9}- X_{1}X_{6}X_{12}+\\ 
X_{3}X_{6}X_{11}+ X_{4}X_{7}X_{9}+ X_{2}X_{5}X_{12}- X_{4}X_{5}X_{11}
\ea
}
\\ \hline
(F_0)_{2,3} & {\tiny \ba{c|cccccccc}
	& X_{112} & Y_{122} & Y_{222} & Y_{111} & Y_{211} & X_{121} &
X_{212} & X_{221} \\ \hline
	A & -1 & 0 & 0 & 1 & 1 & 0 & -1 & 0 \\ 
	B & 1 & -1 & -1 & 0 & 0 & 0 & 1 & 0 \\ 
	C & 0 & 0 & 0 & -1 & -1 & 1 & 0 & 1 \\
	D & 0 & 1 & 1 & 0 & 0 & -1 & 0 & -1 \ea }
& 
{\tiny
\epsilon^{ij} \epsilon^{kl}X_{i~12}Y_{k~22}X_{j~21}Y_{l~11}
}
\\ \hline \hline
(dP_0)_{1,2,3,4,5} & {\tiny \ba{c|ccccccccc}
	& X_1 & X_2 & X_3 & X_4 & X_5 & X_6 & X_7 & X_8 & X_9 \\ \hline
	A & -1 & 0 & -1 & 0 & -1 & 0 & 1 & 1 & 1 \cr 
	B & 0 & 1 & 0 & 1 & 0 & 1 & -1 & -1 & -1 \cr 
	C & 1 & -1 & 1 & -1 & 1 & -1 & 0 & 0 & 0 \ea } 
&
{\footnotesize
\ba{r}
X_{1} X_{4} X_{9} - X_{4} X_{5} X_{7} - X_{2} X_{3} X_{9} - \\
X_{1} X_{6} X_{8} + X_{2} X_{5} X_{8} + X_{3} X_{6} X_{7}
\ea
}
\\ \hline \hline
(dP_1)_{1,2,...,12} &
	{\tiny	\ba{c|cccccccccc}
	& X_1 & X_2 & X_3 & X_4 & X_5 & X_6 & X_7 & X_8 & X_9 & X_{10}
	\\ \hline 
	A & -1 & 0 & 0 & -1 & 0 & 0 & 0 & 1 & 0 & 1 \cr
	B & 1 & -1 & 0 & 0 & 0 & -1 & 0 & 0 & 1 & 0 \cr
	C & 0 & 0 & 1 & 0 & 1 & 0 & 1 & -1 & -1 & -1 \cr 
	D & 0 & 1 & -1 & 1 & -1 & 1 & -1 & 0 & 0 & 0 \ea }
& 
{\footnotesize
\ba{r}
X_{2} X_{7} X_{9} - X_{3} X_{6} X_{9} - X_{4} X_{8} 
        X_{7} - X_{1} X_{2} X_{5} X_{10} \\
        + X_{3} X_{4} X_{10} + X_{1} X_{5} X_{6} X_{8}
\ea}
\\ \hline \hline
(dP_2)_{1,5,9} &
	{\tiny \ba{c|ccccccccccccc}
	& X_1 & X_2 & X_3 & X_4 & X_5 & X_6 & X_7 & X_8 & X_9 & X_{10}
	& X_{11} & X_{12} & X_{13} \\ \hline
	A & -1 & 0 & 0 & -1 & 0 & -1 & 0 & 1 & 0 & 0 & 0 & 1 & 1 \cr
	B & 0 & 0 & -1 & 0 & -1 & 1 & 0 & 0 & 0 & 1 & 0 & 0 & 0 \cr
	C & 0 & 0 & 1 & 0 & 1 & 0 & 1 & -1 & -1 & 0 & 1 & -1 & -1 \cr 
	D & 1 & -1 & 0 & 0 & 0 & 0 & 0 & 0 & 1 & -1 & 0 & 0 & 0 \cr 
	E & 0 & 1 & 0 & 1 & 0 & 0 & -1 & 0 & 0 & 0 & -1 & 0 & 0 \ea}
&
{\footnotesize
\ba{r}
X_{2} X_{9} X_{11} - X_{9} X_{3} X_{10} - X_{4} X_{8} X_{11} -
X_{1} X_{2} X_{7} X_{13} + X_{13} X_{3} X_{6} \\
- X_{5} X_{12} X_{6}+
X_{1} X_{5} X_{8} X_{10} + X_{4} X_{7} X_{12}
\ea
}
\\ \hline
(dP_2)_{2,3,4,6,7,8} &
	{\tiny \ba{c|cccccccccccc}
	& X_1 & X_2 & X_3 & X_4 & X_5 & X_6 & X_7 & X_8 & X_9 & X_{10}
	& X_{11} \\ \hline
	A & -1 & 0 & -1 & 0 & 0 & 0 & 1 & 0 & 0 & 0 & 1 \cr
	B & 1 & -1 & 0 & 0 & -1 & 0 & 0 & 0 & 1 & 0 & 0 \cr
	C & 0 & 0 & 1 & -1 & 0 & 1 & 0 & 0 & -1 & 0 & 0 \cr
	D & 0 & 0 & 0 & 0 & 0 & -1 & -1 & 1 & 0 & 1 & 0 \cr
	E & 0 & 1 & 0 & 1 & 1 & 0 & 0 & -1 & 0 & -1 & -1 \ea}
&
{\footnotesize
\ba{r}
X_5 X_8 X_6 X_9 + X_1 X_2 X_{10} X_7 + X_{11} X_3 X_4 \\
- X_4 X_{10} X_6 - X_2 X_8 X_7 X_3 X_9 - X_{11} X_1 X_5
\ea
}
\\ \hline \hline
(dP_3)_1 & {\tiny \ba{c|cccccccccccccc}
	& X_1 & X_2 & X_3 & X_4 & X_5 & X_6 & X_7 & X_8 & X_9 & X_{10}
	& X_{11}& X_{12}& X_{13}& X_{14} \\ \hline
	A & -1 & 0 & 0 & 0 & 1 & 0 & 0 & 1 & -1 & 0 & 0 & 1 & -1 & 0 \cr 
	B & 0 & 0 & -1 & 1 & 0 & -1 & 0 & 0 & 0 & 0 & 0 & 0 & 1 & 0\cr 
	C & 1 & -1 & 0 & -1 & 0 & 0 & 0 & 0 & 0 & 0 & 0 & 0 & 0 & 1\cr 
	D & 0 & 0 & 1 & 0 & 0 & 0 & 0 & -1 & 0 & -1 & 1 & 0 & 0 & 0\cr 
	E & 0 & 0 & 0 & 0 & -1 & 1 & 1 & 0 & 0 & 1 & 0 & -1 & 0 & -1\cr 
	F & 0 & 1 & 0 & 0 & 0 & 0 & -1 & 0 & 1 & 0 & -1 & 0 & 0 & 0 \ea}
&
{\footnotesize
\ba{r}
X_{3} X_{8} X_{13} - X_{8} X_{9} X_{11} - X_{5} X_{6} X_{13} - 
X_{1} X_{3} X_{4} X_{10} X_{12} \\
+ X_{7} X_{9} X_{12} + X_{1} X_{2} X_{5} X_{10} X_{11} + 
X_{4} X_{6} X_{14} - X_{2} X_{7} X_{14}
\ea
}
\\ \hline
\ea
\]
The matter content for these above theories are represented as quiver
diagrams in \fref{f:quiver} (multi-valence arrows are labelled with a
number) and the superpotentials, in the table below.
\EPSFIGURE[h]{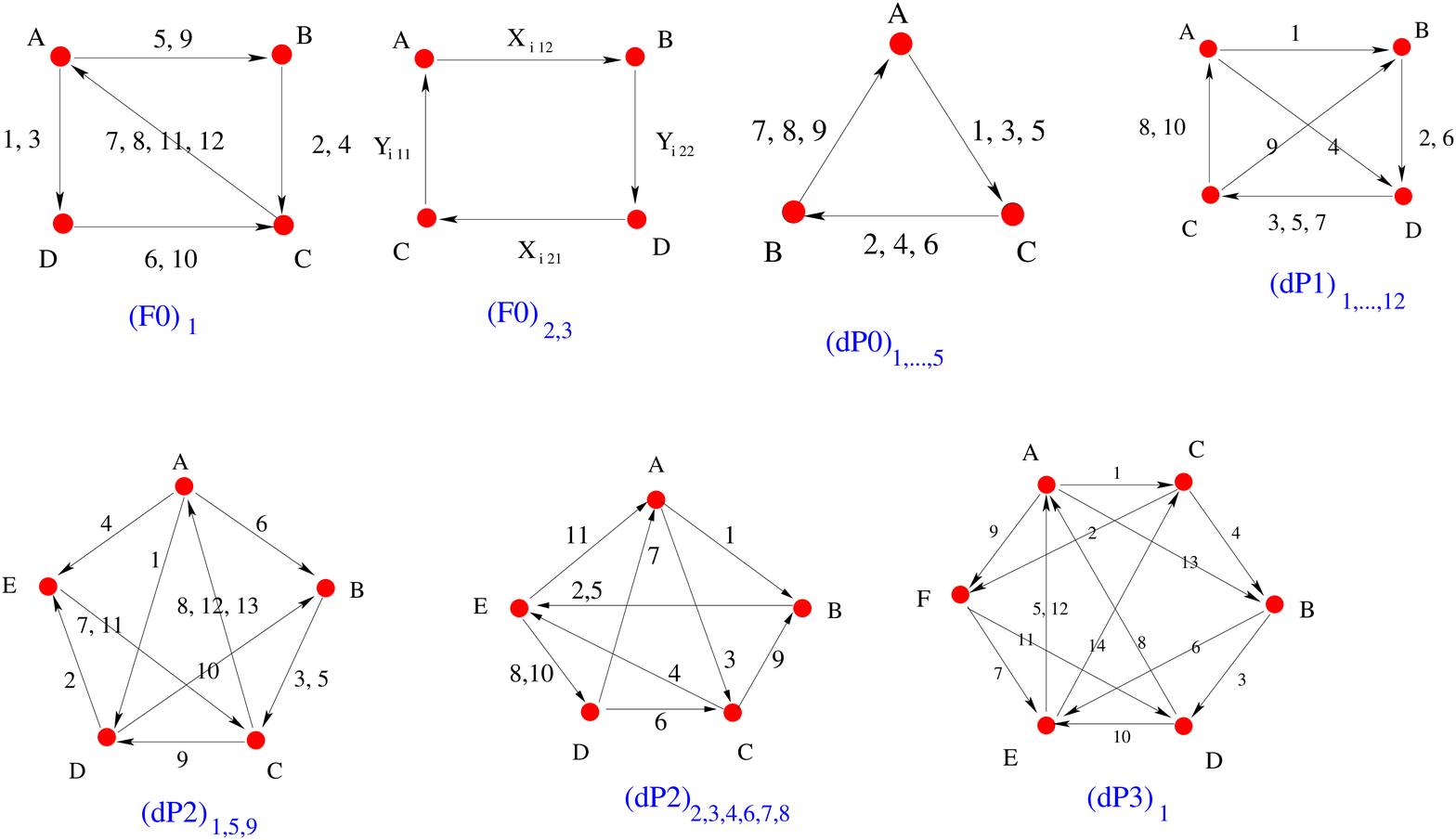,width=6.5in}
{The quiver diagrams for the various phases of the gauge theory for
the del Pezzo surfaces and the zeroth Hirzebruch surface.
\label{f:quiver}
}

In all of the above discussions, we have restricted ourselves to the
cases of $U(1)$ gauge groups, i.e., with only a single brane probe;
this is because such is the only case to which the toric technique can
be applied.
However, after we obtain the matter contents and superpotential for
$U(1)$ gauge groups, we should have some idea for multi-brane probes. 
One obvious generalization is to replace the $U(1)$ with 
$SU(N)$ gauge groups directly. For the matter content, the
generalization is not so easy. 
A field with charge $(1,-1)$ under gauge groups 
$U(1)_A\times U(1)_B$ and zero for others generalised to a
bifundamental $(N,\bar{N})$ of $SU(N)_A\times SU(N)_B$. However, for
higher charges, e.g., charge 2, we simply do not know 
what should be the generalization in the multi-brane case (for a
discussion on generalised quivers cf. e.g. \cite{9911114}).
Furthermore, a field with zero charge under all $U(1)$ groups, 
generalises to an adjoint of one $SU(N)$ gauge group in the
multi-brane case, though we do not know which one.

The generalization of the superpotential is also not so
straight-forward. For
example, there is a quartic term in the conifold with nonabelian 
gauge group \cite{Park-con,Greene2}, 
but it disappears when we go to the $U(1)$
case. The same phenomenon can happen when treating the generic toric
singularity.

For the examples we give in this chapter however, we do not
see any obvious obstruction in the matter contents and superpotential;
they seem to be special enough to be trivially generalized to the
multi-brane case; they are all charge $\pm 1$ under no more than 2
groups. We simply
replace $U(1)$ with $SU(N)$ and $(1,-1)$ fields with bifundamentals
while keeping the superpotential invariant. Generalisations to
multi-brane stack have also been discussed in \cite{Chris}.
\section{Discussions and Prospects}
It is well-known that in the study of the world-volume gauge
theory living on a D-brane probing an orbifold singularity
$\IC^3/\Gamma$,
different choices of decomposition into irreducibles of the space-time
action of $\Gamma$ lead to different matter content and interaction in
the gauge theory and henceforth different moduli spaces (as different
algebraic varieties). 
This strong relation between the decomposition 
and algebraic variety has been shown explicitly for Abelian orbifolds
in \cite{Aspin-Reso}. It seems that there is only one gauge
theory for each given singularity. 

A chief motivation and purpose of this chapter is the realisation that
the above strong statement can not be
generalised to arbitrary (non-orbifold) singularities and in
particular toric singularities.
It is possible that there are several gauge
theories on the D-brane probing the same singularity. 
The moduli space of these inequivalent theories are indeed
by construction the same, as dictated by the geometry of the singularity.

In analogy to the freedom of decomposition into irreps of the group
action in the orbifold case, there too exists a freedom in toric
singularities: any toric diagram is defined only up to a unimodular
transformation (Theorem \ref{iso}). We harness this toric isomorphism
as a tool to create inequivalent gauge theories which live on the
D-brane probe and which, by construction, flow to the same (toric) moduli
space in the IR.

Indeed, these theories constitute another sub-class of examples of
{\em toric duality} as proposed in \cite{0003085}. A key point to note
is that unlike the general case of the duality (such as F-D
ambiguities and repetition ambiguities as discussed therein) 
of which we have hitherto little control, these
particular theories are all physical (i.e., guaranteed to be
world-volume theories) by virtue of their being obtainable from the
canonical method of partial resolution of Abelian orbifolds.
We therefore refer to them as {\em phases} of partial resolution.

As a further tool, we have re-examined the Forward and Inverse
Algorithms developed in \cite{Chris,0003085,DGM} of extracting the
gauge theory data and toric moduli space data from each other. In
particular we have taken the pains to show what {\em degree of freedom} can
one have at each step of the Algorithm. This will serve to
discriminate whether or not two theories are physically equivalent
given their respective matrices at each step.

Thus equipped, we have re-studied the partial resolutions of the Abelian
orbifold $\IC^3 / (\IZ_3 \times \IZ_3)$, namely the 4 toric del Pezzo
surfaces $dP_{0,1,2,3}$ and the zeroth Hirzebruch surface $F_0$.
We performed all possible $SL(3;\IZ)$ transformation of these toric
diagrams which are up to permutation still embeddable in $\IZ_3 \times
\IZ_3$ and subsequently initiated the Inverse Algorithm therewith.
We found at the end of the day, in addition to the physical theories
for these examples presented in \cite{0003085}, an additional one for
both $F_0$ and $dP_2$. Further embedding can of course be done, viz.,
into $\IZ_n \times \IZ_n$ for $n > 3$; it is expected that more phases
would arise for these computationally prohibitive cases, for example
for $dP_3$.

A clear goal awaits us: because for the generic (non-orbifold) 
toric singularity
there is no concrete concept corresponding to the different decomposition of
group action, we do not know at this moment how to classify the phases
of toric duality.
We certainly wish, given a toric singularity, to know (a)
how many inequivalent gauge theory are there and (b) what are 
the corresponding matter contents and superpotential.
It will be a very interesting direction for further
investigation.

Many related questions also arise. For example, by
the AdS/CFT correspondence, we need to understand how to
describe these different gauge theories on the 
supergravity side while the underline geometry is same.
 Furthermore the $dP^2$ theory can be
described in the brane setup by $(p,q)$-5 brane webs
\cite{AHK}, so we want to ask how to
understand these different phases in such brane setups.
Understanding these will help us to get the gauge theory in
higher del Pezzo surface singularities. 

Another very pertinent issue is to clarify the meaning of ``toric
duality.'' So far it is merely an equivalence of moduli spaces of
gauge theories in the IR. It would be very nice if we could make this
statement stronger. For example, could we find the explicit mappings
between gauge invariant operators of various toric-dual theories? 
Indeed, we believe that the study of toric duality and its phase 
structure is worth further pursuit.
\index{Seiberg Duality}
\index{Hanany-Witten!brane diamonds}
\chapter{Toric III: Toric Duality and Seiberg Duality}
\section*{\center{{ Synopsis}}}
\label{chap:0109063}
What then is Toric Duality, as proposed in our previous two chapters?

We use field theory and brane diamond techniques to
demonstrate that Toric Duality is Seiberg duality for ${\cal N}=1$
theories with toric moduli spaces.
This resolves the puzzle concerning the 
physical meaning of
Toric Duality.

Furthermore, using this strong connection we arrive at
three new phases which can not be thus far obtained by the so-called
``Inverse Algorithm'' applied to partial resolution of
$\C^3/(\Z_3 \times \Z_3)$. The standing proposals of Seiberg duality as 
diamond duality in the work by Aganagic-Karch-L\"ust-Miemiec
are strongly supported and
new diamond configurations for these
singularities are obtained as a byproduct.
We also make some remarks about the relationships between
Seiberg duality and Picard-Lefschetz monodromy \cite{0109063}.
\section{Introduction}
Witten's gauge linear sigma approach \cite{GLSM} to ${\cal N}=2$
super-conformal theories has provided deep insight not only to the
study of the phases of the field theory but also to the understanding
of the mathematics of Geometric Invariant Theory 
quotients in toric geometry. Thereafter, the
method was readily applied to the study of the ${\cal N}=1$ supersymmetric 
gauge theories on D-branes at singularities 
\cite{DGM,Horizon,Park-con,Chris}.
Indeed the classical moduli space of the gauge theory corresponds
precisely to the spacetime which the D-brane probes transversely. In 
light of this therefore, toric geometry has been widely used in the study 
of the moduli space of vacua of the gauge theory living on D-brane probes.

The method of encoding the gauge theory data into the moduli
data, or more specifically, the F-term and D-term information into the
toric diagram of the algebraic variety describing the moduli space,
has been well-established \cite{DGM,Horizon}. The reverse, of
determining the SUSY gauge theory data in terms of a given toric
singularity upon which the D-brane probes, has also been addressed
using the method partial resolutions of abelian quotient
singularities. Namely, a general non-orbifold singularity is regarded as 
a partial resolution of a worse, but orbifold, singularity. This ``Inverse 
Procedure'' was formalised into a linear optimisation algorithm, easily 
implementable on computer, by \cite{0003085}, and was subsequently checked 
extensively in \cite{Sarkar}.

One feature of the Inverse Algorithm is its non-uniqueness, viz., that
for a given toric singularity, one could in theory construct countless
gauge theories. This means that there are classes of gauge theories which 
have identical toric moduli space in the IR. Such a salient feature was 
dubbed in \cite{0003085} as {\bf toric duality}. Indeed in a follow-up work, 
\cite{0104259} attempted to analyse this duality in detail, concentrating 
in particular on a method of fabricating dual theories which are physical, 
in the sense that they can be realised as world-volume theories on 
D-branes. Henceforth, we shall adhere to this more restricted meaning of 
toric duality.

Because the details of this method will be clear in later examples we 
shall not delve into the specifics here, nor shall we devote too much 
space reviewing the algorithm. Let us highlight the key points. The gauge 
theory data of D-branes probing Abelian orbifolds is well-known (see e.g. 
the appendix of \cite{0104259}); also any toric diagram can be embedded
into that of such an orbifold (in particular any toric local Calabi-Yau 
threefold $D$ can be embedded into $\IC^3/(\IZ_n \times \IZ_n)$ for 
sufficiently large $n$. We can then obtain the subsector of orbifold 
theory that corresponds the gauge theory constructed for $D$. This is the 
method of ``Partial Resolution.''

A key point of \cite{0104259} was the application of the well-known 
mathematical fact that the toric diagram $D$ of any toric variety has an 
inherent ambiguity in its definition: namely any unimodular transformation 
on the lattice on which $D$ is defined must leave $D$ invariant. In other 
words, for threefolds defined in the standard lattice $\IZ^3$, any 
$SL(3;\IC)$ transformation on the vector endpoints of the defining toric 
diagram gives the same toric variety. Their embedding into the diagram of 
a fixed Abelian orbifold on the other hand, certainly is different. Ergo, 
the gauge theory data one obtains in general are vastly different, even 
though per constructio, they have the same toric moduli space.

What then is this ``toric duality''? How clearly it is defined
mathematically and yet how illusive it is as a physical phenomenon.
The purpose of the present writing is to make the first leap toward
answering this question. In particular, we shall show, using brane
setups, and especially brane diamonds, that known cases for toric
duality are actually interesting realisations of Seiberg Duality.
Therefore the mathematical equivalence of moduli spaces for different 
quiver gauge theories is related to a real physical equivalence of the 
gauge theories in the far infrared.

The chapter is organised as follows. In Section 2, we begin with an
illustrative example of two torically dual cases of a generalised
conifold. These are well-known to be Seiberg dual theories as seen
from brane setups. Thereby we are motivated to conjecture in Section 3
that toric duality is Seiberg duality. We proceed to check this
proposal in Section 4 with all the known cases of torically dual
theories and have successfully shown that the phases of the partial
resolutions of $\IC^3/(\IZ_3 \times \IZ_3)$ constructed in
\cite{0003085} are indeed Seiberg dual from a field theory analysis.
Then in Section 6 we re-analyse these examples from the perspective of
brane diamond configurations and once again obtain strong support of
the statement. From rules used in the diamond dualisation, 
we extracted a so-called ``quiver duality'' which explicits Seiberg
duality as a transformation on the matter adjacency matrices. Using
these rules we are able to extract more phases of theories not yet
obtained from the Inverse Algorithm. In a more geometrical vein, in
Section 7, we remark the
connection between Seiberg duality and Picard-Lefschetz and point out
cases where the two phenomena may differ. Finally we finish with
conclusions and prospects in Section 8.

While this manuscript is about to be released, we became aware of the
nice work \cite{BP}, which discusses similar issues.
\section{An Illustrative Example}
We begin with an illustrative example that will demonstrate how Seiberg 
Duality is realised as toric duality.
\subsection{The Brane Setup}
\index{Conifold!generalised}
\index{Hanany-Witten}
The example is the well-known generalized conifold described 
as the hypersurface $xy=z^2w^2$ in $\IC^4$, and which can be obtained 
as a $\IZ_2$ quotient of the famous conifold $xy=zw$ by the action 
$z\to -z, w\to -w$. The gauge theory on the D-brane sitting at such a 
singularity can be established by orbifolding the conifold gauge theory in 
\cite{Kle-Wit}, 
as in \cite{uranga}. Also, it can be derived by another method 
alternative to the Inverse Algorithm, namely performing a T-duality to a 
brane setup with NS-branes and D4-branes \cite{uranga,dandm}. Therefore 
this theory serves as an excellent check on our methods.

The setup involves stretching D4 branes (spanning 01236) between 2 pairs 
of NS and NS$'$ branes (spanning 012345 and 012389, respectively), with 
$x^6$ parameterizing a circle. These configurations are analogous to 
those in \cite{Methods}.
There are in fact two inequivalent brane setups (a) and (b) (see 
\fref{z2w2brane}), differing in the way the NS- and NS$'$-branes are 
ordered in the circle coordinate.
%
\EPSFIGURE[h]{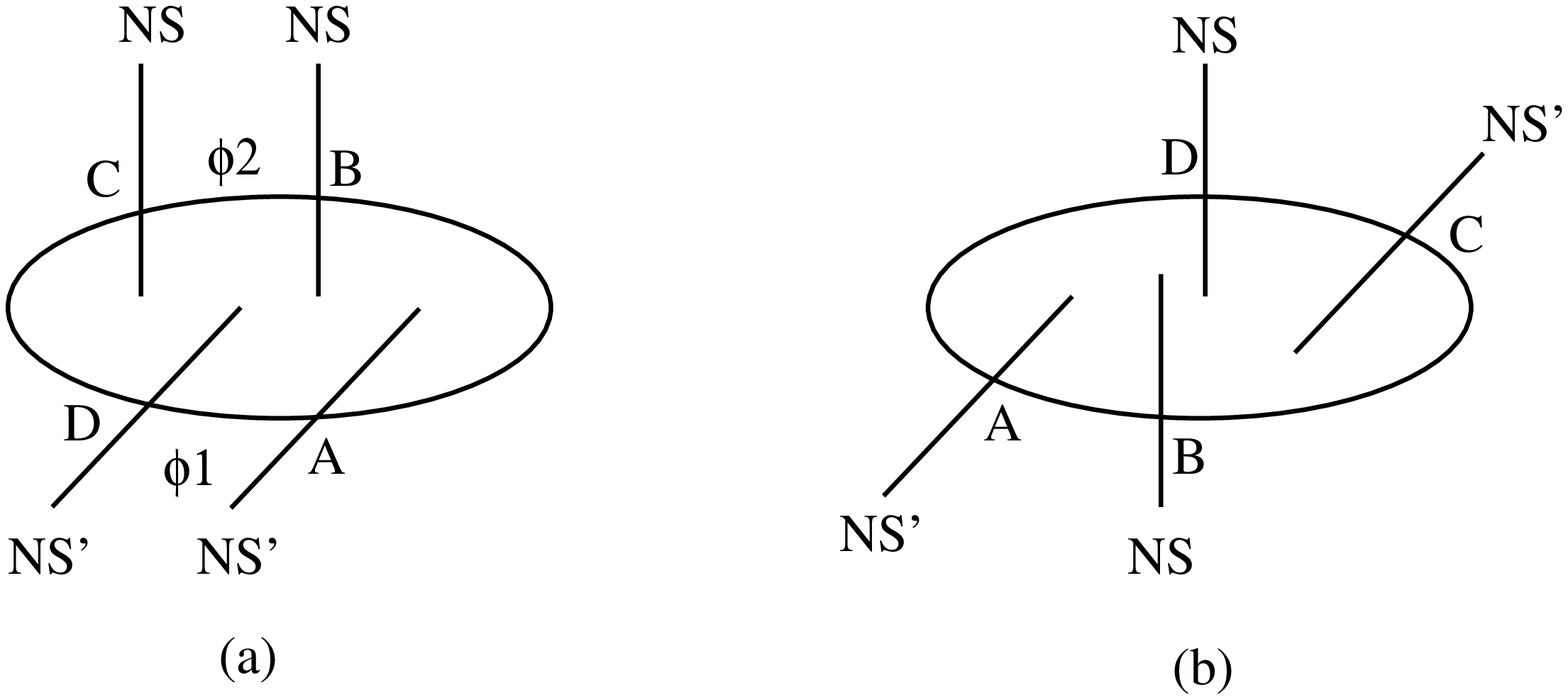,width=14cm}
{The two possible brane setups for the generalized conifold $xy=z^2
w^2$. They are related to each other passing one NS-brane through an
NS'-brane. $A_i,B_i,C_i,D_i$ $i=1,2$ are
bifundamentals while  $\phi_1,\phi_2$ are
two adjoint fields.
\label{z2w2brane}
}
Using standard rules \cite{HW,Methods}, we see from the figure that there are 
4 product gauge groups (in the Abelian case, it is simply $U(1)^4$. As for 
the matter content, theory (a) has 8 bi-fundamental chiral multiplets 
$A_i$, $B_i$, $C_i$, $D_i$ $i=1,2$ (with charge $(+1,-1)$ and $(-1,+1)$ 
with respect to adjacent $U(1)$ factors) and 2 adjoint chiral multiplets
$\phi_{1,2}$ as indicated. On the other hand (b) has only 8 
bi-fundamentals, with charges as above. The superpotentials are 
respectively \cite{amiandofer,uranga}
\barrayn
(a)~~~ & W_a= & -A_1 A_2 B_1 B_2 +B_1 B_2 \phi_2-C_1 C_2 \phi_2 
	+C_1 C_2 D_1 D_2 - D_1 D_2  \phi_1 + A_1 A_2 \phi_1,
\\
(b)~~~ & W_b= & A_1 A_2 B_1 B_2-B_1 B_2  C_1 C_2 +  C_1 C_2 D_1 D_2 
	-D_1 D_2 A_1 A_2  
\earrayn
With some foresight, for comparison with the results later, we rewrite 
them as
\be
\label{orbi-coni-1-1} 
W_a=(B_1 B_2 - C_1 C_2)(\phi_2- A_1 A_2)+ (A_1 A_2 - D_1 D_2)
(\phi_1 -C_1 C_2)
\ee
\be
\label{orbi-coni-2-1}
W_b = (A_1 A_2 -C_1 C_2) (B_1 B_2 -D_1 D_2)
\ee
\index{Resolution!partial}
\subsection{Partial Resolution} 
Let us see whether we can reproduce these field theories with the Inverse 
Algorithm. The toric diagram for $xy=z^2w^2$ is given in the very left of 
\fref{z2w2toric}. Of course, the hypersurface is three complex-dimensional 
so there is actually an undrawn apex for the toric diagram, and each of 
the nodes is in fact a three-vector in $\IZ^3$. Indeed the fact that it is 
locally Calabi-Yau that guarantees all the nodes to be coplanar. The 
next step is the realisation that it can be embedded into the well-known 
toric diagram for the Abelian orbifold $\IC^3/(\IZ_3 \times \IZ_3)$ 
consisting of 10 lattice points. The reader is referred to
\cite{0003085,0104259} 
for the actual co\"ordinates of the points, a detail which, though crucial, 
we shall not belabour here.

The important point is that there are six ways to embed our toric diagram 
into the orbifold one, all related by $SL(3;\IC)$ transformations. This is 
indicated in parts (a)-(f) of \fref{z2w2toric}. We emphasise that these 
six diagrams, drawn in red, are {\em equivalent} descriptions of $xy=z^2w^2$ 
by virtue of their being unimodularly related; therefore they are all 
candidates for toric duality.
\EPSFIGURE[h]{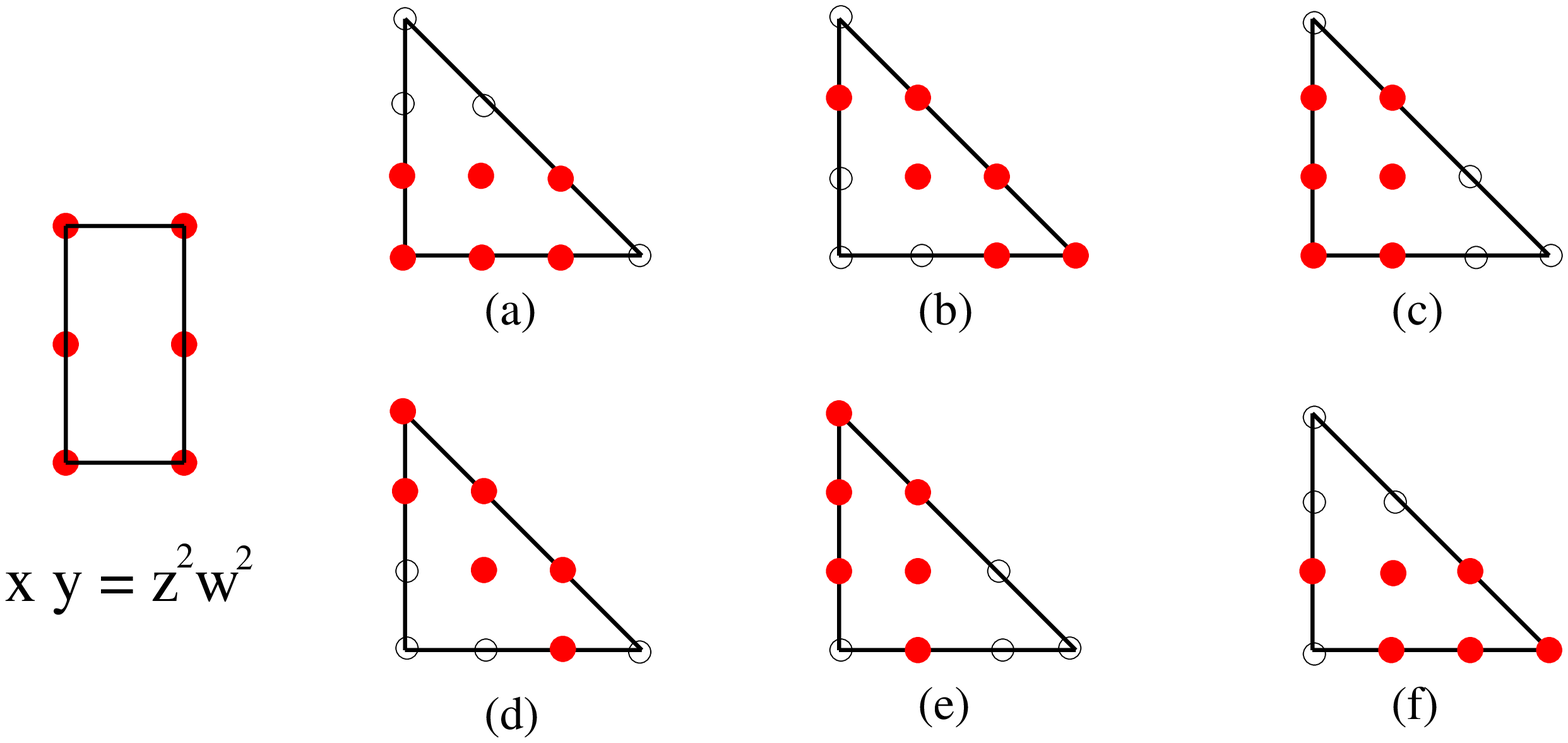,width=14cm}
{The standard toric diagram for the generalized conifold $xy=uv=z^2$
(far left). To the right are six $SL(3;\IC)$ transformations (a)-(f)
thereof (drawn in red) and hence are equivalent toric diagrams for the
variety. We embed these six diagrams into the Abelian orbifold
$\IC^3/(\IZ_3 \times \IZ_3)$ in order to perform partial resolution and
thus the gauge theory data.
\label{z2w2toric}
}

Now we use our 
Inverse Algorithm, by partially resolving $\IC^3/(\IZ_3 \times 
\IZ_3)$, to obtain the gauge theory data for the D-brane probing 
$xy=z^2w^2$. In summary, after exploring the six possible partial 
resolutions, we find that cases (a) and (b) give identical results, while 
(c,d,e,f) give the same result which is inequivalent from (a,b). Therefore 
we conclude that cases (a) and (c) are inequivalent torically dual 
theories for $xy=z^2 w^2$. In the following we detail the data for these 
two contrasting cases. We refer the reader to \cite{0003085,0104259}
for details and notation.
\EPSFIGURE[h]{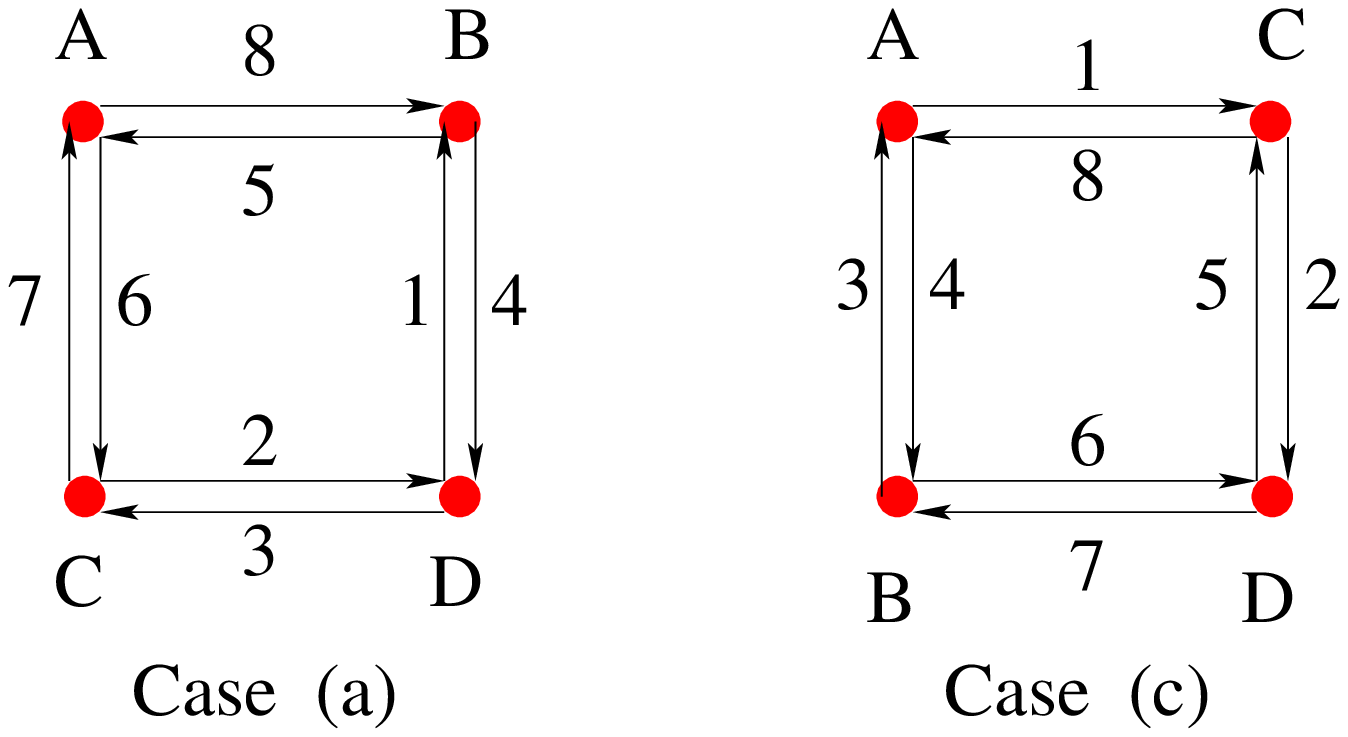,width=10cm}
{The quiver diagram encoding the matter content of Cases (a) and (c)
of \fref{z2w2toric}.
\label{z2w2quiver}
}
\subsection{Case (a) from Partial Resolution}
For case (a), the matter content is encoded the $d$-matrix  
which indicates the charges of the 8 bi-fundamentals under the 4 gauge
groups. This is the incidence matrix for the quiver diagram drawn in part 
(a) of \fref{z2w2quiver}. 
$$
\tmat{
         &  X_1 & X_2 & X_3 & X_4 & X_5 & X_6 & X_7 & X_8  \cr
U(1)_A   &   0  &  0  &  0  &  0  &  1  &  -1 &  1  &  -1  \cr
U(1)_B   &   1  &  0  &  0  &  -1 & -1  &  0  &  0  &   1  \cr
U(1)_C   &   0  & -1  &  1  &  0  &  0  &   1 & -1  &  0   \cr
U(1)_D   &   -1 & 1   &  -1 &  1  &  0  &  0  &  0  &  0   \cr
}
$$
On the other hand, the F-terms are encoded in the $K$-matrix
$$
\tmat{
  X_1 & X_2 & X_3 & X_4 & X_5 & X_6 & X_7 & X_8  \cr
1 &0 &1 &0 &0 &0 &0 &0  \cr
1 &1 &0 &0 &0 &0 &0 &0  \cr
0 &1 &0 &1 &0 &0 &0 &0  \cr
0 &0 &0 &0 &1 &0 &1 &0  \cr
0 &0 &0 &0 &1 &1 &0 &0  \cr
0 &0 &0 &0 &0 &1 &0 &1  \cr
}
$$
From $K$ we get two relations $X_5 X_8 =X_6 X_7 $ and $X_1 X_4=X_2 X_3$ 
(these are the relations one must impose on the quiver to obtain the final 
variety; equivalently, they correspond to the F-term constraints arising 
from the superpotential). Notice that here each term is chargeless under 
all 4 gauge groups, so when we integrate back to get the superpotential,
we should multiply by chargeless quantities also\footnote{In more general
situations the left- and right-hand sides may not be singlets, but transform 
in the same gauge representation.}.

The relations must come from the F-flatness $\frac{\partial}{\partial X_i} 
W = 0$ and thus we can use these relations to integrate back to the 
superpotential $W$. However we meet some ambiguities
\footnote{The ambiguities arise because in the abelian case (toric 
language) the adjoints are chargeless. In fact, no ambiguity arises if 
one performs the Higgsing associated to the partial resolution in the 
non-abelian case. We have performed this exercise in cases (a) and (c), 
and verified the result obtained by the different argument offered in the 
text.}. In principle we can have two 
different choices:
\barrayn
(i)~~~~ & W_1= & (X_5 X_8 -X_6 X_7)(X_1 X_4-X_2 X_3)  \\
(ii)~~~~~& W_2= & \psi_1 (X_5 X_8 -X_6 X_7) +\psi_2 (X_1 X_4-X_2 X_3)
\earrayn
where for now $\psi_i$ are simply chargeless fields.

We shall evoke physical arguments to determine which is correct. Expanding 
(i) gives $W_1=X_5 X_8 X_1 X_4-X_6 X_7 X_1 X_4-X_5 X_8 X_2X_3+ 
X_6 X_7 X_2 X_3$. Notice the term $X_6 X_7 X_1 X_4$: there is no common 
gauge group under which there four fields are charged, i.e. these 4 
arrows (q.~v. \fref{z2w2quiver}) do not intersect at a single node. This 
makes (i) very unnatural and exclude it.

Case (ii) does not have the above problem and indeed all four fields 
$X_5 ,X_8 ,X_6, X_7$ are charged under the $U(1)_A$ gauge group, so 
considering $\psi_1$ to be an adjoint of $U(1)_A$, we do obtain a physically 
meaningful interaction. Similarly $\psi_2$ will be the adjoint of $U(1)_D$, 
interacting with $X_1 ,X_4 ,X_2, X_3$.

However, we are not finish yet. From \fref{z2w2quiver} we see that 
$X_5, X_8,X_1, X_4$ are all charged under $U(1)_B$, while $X_6, X_7,X_2 ,X_3$ 
are all charged under $U(1)_C$. From a physical point of view, there 
should be some interaction terms between these fields. Possibilities are
$X_5 X_8 X_1 X_4$ and $X_6 X_7 X_2 X_3$. To add these terms into $W_2$ is 
very easy, we simply perform the following replacement:\footnote{Here we  
choose the sign purposefully for later convenience. However, we do need, 
for the cancellation of the unnatural interaction term $X_1 X_4 X_6 X_7$, 
that they both have the same sign.} $\psi_1\longrightarrow \psi_1-X_1 
X_4,~~~~~~\psi_2\longrightarrow \psi_2-X_6 X_7.$ Putting everything 
together, we finally obtain that Case (a) has matter content as described 
in \fref{z2w2quiver} and the superpotential
\beq
\label{z2w2casea}
W=(\psi_1-X_1 X_4)(X_5 X_8 -X_6 X_7) +(\psi_2-X_6 X_7)(X_1 X_4-X_2 X_3)
\eeq

This is precisely the theory (a) from the brane setup in the last
section! Comparing \eref{z2w2casea} with (\ref{orbi-coni-1-1}), we see
that they are exact same under the following redefinition of variables:
$$
\ba{ccccc}
B_1,B_2  \Longleftrightarrow   X_5, X_8  &\qquad &
C_1,C_2  \Longleftrightarrow   X_6,  X_7  &\qquad &
D_1,D_2  \Longleftrightarrow   X_2, X_3  \cr
A_1,A_2  \Longleftrightarrow   X_1,  X_4 &\qquad &
\phi_2 \Longleftrightarrow   \psi_1  & \qquad&
\phi_1 \Longleftrightarrow   \psi_2  \cr 
\ea
$$

In conclusion, case (a) of our Inverse Algorithm reproduces the
results of case (a) of the brane setup.
\subsection{Case (c) from Partial Resolution}
For case (c), the matter content is given by the quiver in
\fref{z2w2quiver}, which has the charge matrix $d$ equal to
$$
\tmat{
         &  X_1 & X_2 & X_3 & X_4 & X_5 & X_6 & X_7 & X_8  \cr
U(1)_A   &   -1 &  0  & -1  &  1  &  0  &  0  &  0  &  1   \cr
U(1)_B   &   0  &  0  &  1  &  -1 &  0  & -1  &  1  &  0   \cr
U(1)_C   &   1  & -1  &  0  &  0  &  1  &  0  &  0  & -1   \cr
U(1)_D   &   0  & 1   &  0  &  0  & -1  &  1  & -1  &  0   \cr
}
$$
This is precisely the matter content of case (b) of the brane setup.
The F-terms are given by
$$
K=
\tmat{
  X_1 & X_2 & X_3 & X_4 & X_5 & X_6 & X_7 & X_8  \cr
0 &1  &0  &1  &0  &0  &0  &0  \cr
1 &0  &0  &0  &0  &0  &1  &0  \cr
1 &0  &0  &0  &0  &1  &0  &0  \cr
0 &1  &1  &0  &0  &0  &0  &0  \cr
0 &0  &1  &0  &1  &0  &0  &0  \cr
0 &0  &0  &0  &0  &1  &0  &1  \cr
}
$$
From it we can read out the relations $X_1 X_8= X_6 X_7$ and $X_2 X_5 = 
X_3 X_4$. Again there are two ways to write down the superpotential
\barrayn
(i)~~~~ & W_1= & (X_1 X_8 -X_6 X_7)(X_3 X_4-X_2 X_5)  \\
(ii)~~~~~& W_2= & \psi_1 (X_1 X_8 -X_6 X_7) +\psi_2 (X_3 X_4-X_2 X_5)
\earrayn
In this case, because $X_1$, $X_8$, $X_6$, $X_7$ are not charged under
any common gauge group, it is impossible to include any adjoint field 
$\psi$ to give a physically meaningful interaction and so (ii) is unnatural. We are left the
superpotential $W_1$. Indeed, comparing with (\ref{orbi-coni-2-1}),
we see they are identical under the redefinitions
$$
\ba{ccc}
A_1,A_2 \Longleftrightarrow X_1, X_8 &\qquad &
B_1,B_2 \Longleftrightarrow X_3, X_4 \\
C_1,C_2 \Longleftrightarrow X_6, X_7 &\qquad &
D_1,D_2 \Longleftrightarrow X_2, X_5 \\
\ea
$$
Therefore we have reproduced case (b) of the brane setup.

\medskip

What have we achieved? We have shown that toric duality due to inequivalent 
embeddings of unimodularly related toric diagrams for the generalized 
conifold $xy = z^2w^2$ gives two inequivalent physical world-volume 
theories on the D-brane probe, exemplified by cases (a) and (c). On the 
other hand, there are two T-dual brane setups for this singularity, also 
giving two inequivalent field theories (a) and (b). Upon comparison, case (a)
(resp. (c)) from the Inverse Algorithm beautifully corresponds to case (a) 
(resp. (b)) from the brane setup. Somehow, a seemingly harmless trick in 
mathematics relates inequivalent brane setups. In fact we can say much 
more.
\index{Seiberg Duality}
\index{Toric Duality}
\section{Seiberg Duality versus Toric Duality}
As follows from \cite{Methods}, the two theories from the brane setups are 
actually related by Seiberg Duality \cite{SeibergDual}, as pointed out in 
\cite{uranga} (see also \cite{Unge,Uranga2}. Let us first review the 
main features of this famous duality, for unitary gauge groups.

Seiberg duality is a non-trivial infrared equivalence of ${\cal 
N}=1$ supersymmetric field theories, which are different in the 
ultraviolet, but flow the the same interacting fixed point in the 
infrared. In particular, the very low energy features of the different 
theories, like their moduli space, chiral ring, global symmetries, agree 
for Seiberg dual theories. Given that toric dual theories, by definition, 
have identical moduli spaces, etc , it is natural to propose a connection 
between both phenomena.

The prototypical example of Seiberg duality is ${\cal N}=1$ $SU(N_c)$ 
gauge theory with $N_f$ vector-like fundamental flavours, and no 
superpotential. The global chiral symmetry is $SU(N_f)_L\times SU(N_f)_R$, 
so the matter content quantum numbers are 
\[
\ba{c|ccc}
	& SU(N_c)  &  SU(N_f)_L  &  SU(N_f)_R \\ \hline
Q   	& \fund  &   \fund   &      1 \\
Q'	& \antifund &    1     &     \antifund \\
\ea
\]
In the conformal window, $3N_c/2\leq N_f\leq 3N_c$, the theory flows to an 
interacting infrared fixed point. The dual theory, flowing to the same 
fixed point is given ${\cal} N=1$ $SU(N_f-N_c)$ gauge theory with $N_f$ 
fundamental flavours, namely
\[
\ba{c|ccc}
     	&SU(N_f-N_c)   	&SU(N_f)_L   	&SU(N_f)_R \\ \hline
q      	& \fund       	&\antifund	&1 \\
q'     	& \antifund	&1        	&\fund \\
M       &1         	&\fund     	&\antifund \\
\ea
\]
and superpotential $W = M q q'$. From the matching of chiral rings, the 
`mesons' $M$ can be thought of as composites $QQ'$ of the original quarks.

\medskip

It is well established \cite{Methods}, that in an ${\cal N}=1$ (IIA) brane
setup for the four dimensional theory such as \fref{z2w2brane}, Seiberg 
duality is realised as the crossing of 2 non-parallel NS-NS$'$ branes. In 
other words, as pointed out in \cite{uranga}, cases (a) and (b) are in 
fact a Seiberg dual pair. Therefore it seems that the results from the 
previous section suggest that toric duality is a guise of Seiberg duality, 
for theories with moduli space admitting a toric descriptions. It is 
therefore the intent of the remainder of this chapter to examine and support
\begin{conjecture}
Toric duality is Seiberg duality for ${\cal N}=1$ theories with toric
moduli spaces. 
\end{conjecture}
\section{Partial Resolutions of $\IC^3/(\IZ_3 \times \IZ_3)$ and Seiberg 
duality}
Let us proceed to check more examples. So far the other known examples
of torically dual theories 
are from various partial resolutions of $\IC^3/(\IZ_3 \times \IZ_3)$. In 
particular it was found in \cite{0104259} that the (complex) cones over the 
zeroth Hirzebruch surface as well as the second del Pezzo surface each 
has two toric dual pairs. We remind the reader of these theories.
\subsection{Hirzebruch Zero}
There are two torically dual theories for the cone over the zeroth
Hirzebruch surface $F_0$. The toric and quiver diagrams are given in
\fref{f:F0seiberg}, the matter content and interactions are
\beq
\label{F0}
\hspace{-1.0cm}
\ba{c|c|c}
& \mbox{Matter Content }d & \mbox{Superpotential}\\
\hline
$I$ & {\tiny \ba{c|cccccccccccc} 
	& X_1 & X_2 & X_3 & X_4 & X_5 & X_6 & X_7 & X_8 & X_9 & X_{10}
	& X_{11} & X_{12}\\ \hline
	A & -1 & 0 & -1 & 0 & -1 & 0 & 1 & 1 & -1 & 0 & 1 & 1 \\ 
	B & 0 & -1 & 0 & -1 & 1 & 0 & 0 & 0 & 1 & 0 & 0 & 0 \\
	C & 0 & 1 & 0 & 1 & 0 & 1 & -1 & -1 & 0 & 1 & -1 & -1 \\ 
	D & 1 & 0 & 1 & 0 & 0 & -1 & 0 & 0 & 0 & -1 & 0 & 0 \ea } 
	&
{\footnotesize
\ba{r}
X_{1}X_{8}X_{10}- X_{3}X_{7}X_{10}- X_{2}X_{8}X_{9}- X_{1}X_{6}X_{12}+\\ 
X_{3}X_{6}X_{11}+ X_{4}X_{7}X_{9}+ X_{2}X_{5}X_{12}- X_{4}X_{5}X_{11}
\ea
}
\\ \hline
$II$ & {\tiny \ba{c|cccccccc}
	& X_{112} & Y_{122} & Y_{222} & Y_{111} & Y_{211} & X_{121} &
X_{212} & X_{221} \\ \hline
	A & -1 & 0 & 0 & 1 & 1 & 0 & -1 & 0 \\ 
	B & 1 & -1 & -1 & 0 & 0 & 0 & 1 & 0 \\ 
	C & 0 & 0 & 0 & -1 & -1 & 1 & 0 & 1 \\
	D & 0 & 1 & 1 & 0 & 0 & -1 & 0 & -1 \ea }
& 
{\tiny
\epsilon^{ij} \epsilon^{kl}X_{i~12}Y_{k~22}X_{j~21}Y_{l~11}
}
\ea
\eeq
\EPSFIGURE[h]{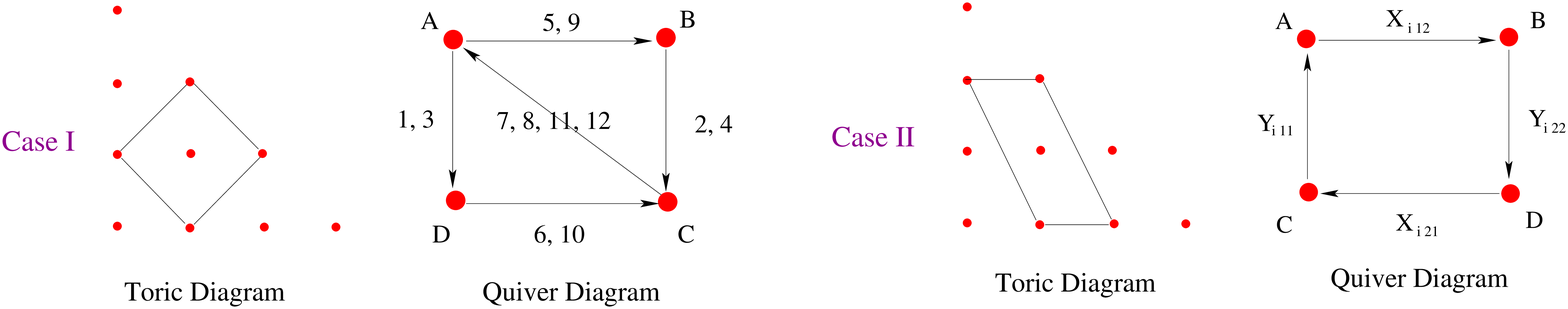,width=7.0in}
{The quiver and toric diagrams of the 2 torically dual theories
corresponding to the cone over the zeroth Hirzebruch surface $F_0$.
\label{f:F0seiberg}
}

\medskip

Let us use the field theory rules from Section 3 on Seiberg Duality to examine
these two cases in detail. The charges of the matter content for case II, 
upon promotion from $U(1)$ to $SU(N)$ 
\footnote{Concerning the $U(1)$ factors, these are in fact generically 
absent, since they are anomalous in the original $\IZ_3\times \IZ_3$ 
singularity, and the Green-Schwarz mechanism canceling their anomaly 
makes them massive \cite{LRA} (see \cite{sagnan,DM,intri} for an analogous 
6d phenomenon). However, there is a well-defined sense in which one can 
use the abelian case to study the toric moduli space \cite{Horizon}.}
(for instance, following the partial resolution in the non-abelian case, 
as in \cite{Horizon,Park-con}), can be re-written as (redefining fields 
$(X_i,Y_i,Z_i,W_i) := (X_{i~12},Y_{i~22},X_{i~21},Y_{i~11})$ with $i=1,2$ 
and gauge groups $(a,b,c,d) := (A,C,B,D)$ for convenience):
$$
\ba{c|cccc}
	&	SU(N)_a & SU(N)_b  & SU(N)_c  & SU(N)_d \\
\hline
X_i	& 	\fund   & \antifund & & \\
Y_i     & 		& \fund   & \antifund & \\
Z_i     &		&	& \fund & \antifund \\
W_i	&	\antifund &	&	& \fund
\ea
$$
The superpotential is then
$$
W_{II} = X_1 Y_1 Z_2 W_2 - X_1 Y_2 Z_2 W_1 - X_2 Y_1 Z_1 W_2 + X_2 Y_2
Z_1 W_1.
$$
Let us dualise with respect to the $a$ gauge group. This is a $SU(N)$ 
theory with $N_c = N$ and $N_f = 2 N$ (as there are two $X_i$'s).
The chiral symmetry is however broken from $SU(2N)_L\times  SU(2N)_R$ to 
$SU(N)_L \times SU(N)_R$, which moreover is gauged as $SU(N)_b \times
SU(N)_d$. Ignoring the superpotential $W_{II}$, the dual theory would be:
\beq
\label{F0dualmatter}
\ba{c|cccc}
	&	SU(N)_{a'} & SU(N)_b  & SU(N)_c  & SU(N)_d \\
\hline
q_i	&	\antifund  & \fund & &\\
Y_i     &		& \fund  & \antifund & \\
Z_i	&		&	& \fund & \antifund \\
q_i'	&	\fund	& & & \antifund \\
M_{ij}	&		& \antifund & & \fund
\ea
\eeq
We note that there are $M_{ij}$ giving 4 bi-fundamentals for $bd$. They
arise from the Seiberg mesons in the bi-fundamental of the enhanced chiral 
symmetry $SU(2N) \times SU(2N)$, once decomposed with respect to the 
unbroken chiral symmetry group.
The superpotential is
$$
W' = M_{11} q_1 q_1' - M_{12} q_2 q_1' - M_{21} q_1 q_2' + 
M_{22} q_2 q_2'.
$$
The choice of signs in $W'$ will be explained shortly. 

Of course, $W_{II}$ is not zero and so give rise to a deformation in
the original theory, analogous to those studied in e.g. \cite{amiandofer}. In 
the dual theory, this deformation simply corresponds to $W_{II}$ rewritten in 
terms of mesons, which can be thought of as composites of the original 
quarks, i.e., 
$M_{ij} = W_i X_j$. Therefore we have 
$$
W_{II} = M_{21} Y_1 Z_2 - M_{11} 
Y_2 Z_2  - M_{22} Y_1 Z_1  + M_{12} Y_2 Z_1
$$
which is written in the new variables. The rule for the signs is that e.g. 
the field $M_{21}$ appears with positive sign in $W_{II}$, hence it 
should appear with negative sign in $W'$, and analogously for others.
Putting them together we 
get the superpotential of the dual theory
\beq
\label{F0dualsup}
\ba{ll}
W_{II}^{dual} & = W_{II} + W' = \\
& M_{11} q_1 q_1' - M_{12} q_2 q_1' - M_{21}
q_1 q_2' + M_{22} q_2 q_2' + M_{21} Y_1 Z_2 - M_{11} Y_2 Z_2  - M_{22}
Y_1 Z_1  + M_{12} Y_2 Z_1
\ea
\eeq
Upon the field redefinitions
$$
\ba{cccc}
M_{11} \rightarrow X_7 \qquad &
M_{12} \rightarrow X_8 \qquad &
M_{21} \rightarrow X_{11} \qquad & 
M_{22} \rightarrow X_{12}\\
q_1 \rightarrow X_4 \qquad &
q_2 \rightarrow X_2 \qquad &
q_{1'} \rightarrow X_9 \qquad &
q_{2'} \rightarrow X_5 \\
Y_1 \rightarrow X_6 \qquad &
Y_2 \rightarrow X_{10}\qquad &
Z_1 \rightarrow X_1\qquad &
Z_2 \rightarrow X_3
\ea
$$
we have the field content \eref{F0dualmatter} and superpotential
\eref{F0dualsup} matching precisely with case I in \eref{F0}. We
conclude therefore that the two torically dual cases I and II obtained
from partial resolutions are indeed Seiberg duals!
\subsection{del Pezzo 2}
Encouraged by the results above, let us proceed with the cone over the
second del Pezzo surface, which also have 2 torically dual theories.
The toric and quiver diagrams are given in \fref{f:dP2seiberg}.
\beq
\label{dP2}
\hspace{-1.0cm}
\setlength{\arraycolsep}{0.1mm}
\ba{c|c|c}
& \mbox{Matter Content }d & \mbox{Superpotential}\\
\hline
$I$ &
	{\tiny \ba{c|ccccccccccccc}
	& Y_1 & Y_2 & Y_3 & Y_4 & Y_5 & Y_6 & Y_7 & Y_8 & Y_9 & Y_{10}
	& Y_{11} & Y_{12} & Y_{13} \\ \hline
	A & -1 & 0 & 0 & -1 & 0 & -1 & 0 & 1 & 0 & 0 & 0 & 1 & 1 \cr
	B & 0 & 0 & -1 & 0 & -1 & 1 & 0 & 0 & 0 & 1 & 0 & 0 & 0 \cr
	C & 0 & 0 & 1 & 0 & 1 & 0 & 1 & -1 & -1 & 0 & 1 & -1 & -1 \cr 
	D & 1 & -1 & 0 & 0 & 0 & 0 & 0 & 0 & 1 & -1 & 0 & 0 & 0 \cr 
	E & 0 & 1 & 0 & 1 & 0 & 0 & -1 & 0 & 0 & 0 & -1 & 0 & 0 \ea}
&
{\footnotesize
\ba{r}
Y_{2} Y_{9} Y_{11} - Y_{9} Y_{3} Y_{10} - Y_{4} Y_{8} Y_{11} -
Y_{1} Y_{2} Y_{7} Y_{13} + Y_{13} Y_{3} Y_{6} \\
- Y_{5} Y_{12} Y_{6}+
Y_{1} Y_{5} Y_{8} Y_{10} + Y_{4} Y_{7} Y_{12}
\ea
}
\\ \hline
$II$ &
	{\tiny \ba{c|cccccccccccc}
	& X_1 & X_2 & X_3 & X_4 & X_5 & X_6 & X_7 & X_8 & X_9 & X_{10}
	& X_{11} \\ \hline
	A & -1 & 0 & -1 & 0 & 0 & 0 & 1 & 0 & 0 & 0 & 1 \cr
	B & 1 & -1 & 0 & 0 & -1 & 0 & 0 & 0 & 1 & 0 & 0 \cr
	C & 0 & 0 & 1 & -1 & 0 & 1 & 0 & 0 & -1 & 0 & 0 \cr
	D & 0 & 0 & 0 & 0 & 0 & -1 & -1 & 1 & 0 & 1 & 0 \cr
	E & 0 & 1 & 0 & 1 & 1 & 0 & 0 & -1 & 0 & -1 & -1 \ea}
&
{\footnotesize
\ba{r}
X_5 X_8 X_6 X_9 + X_1 X_2 X_{10} X_7 + X_{11} X_3 X_4 \\
- X_4 X_{10} X_6 - X_2 X_8 X_7 X_3 X_9 - X_{11} X_1 X_5
\ea
}
\ea
\eeq
\EPSFIGURE[h]{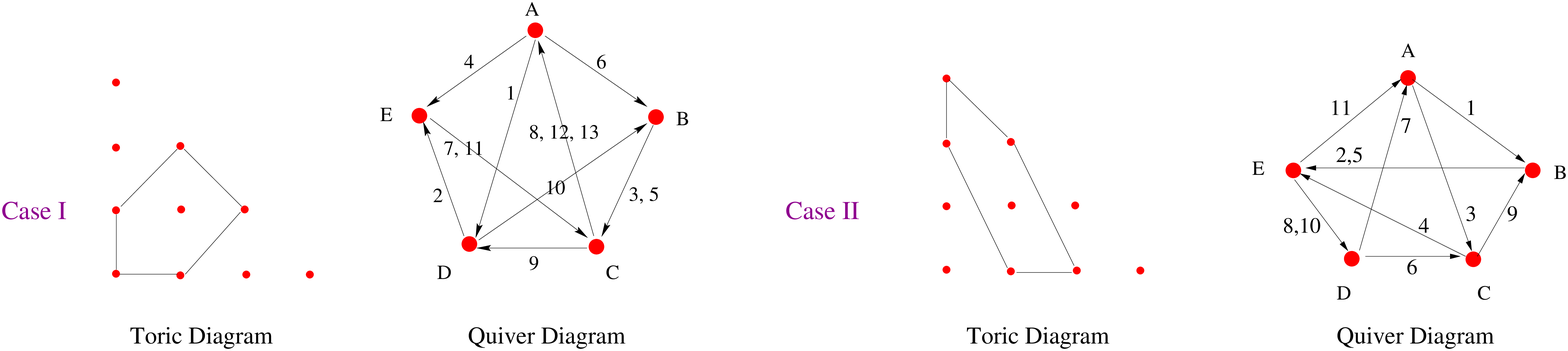,width=7in}
{The quiver and toric diagrams of the 2 torically dual theories
corresponding to the cone over the second del Pezzo surface.
\label{f:dP2seiberg}
}
Again we start with Case II. Working analogously, upon dualisation on node 
$D$ neglecting the superpotential, the matter content of II undergoes the 
following change:
\beq
\hspace{-1cm}
{\small
\setlength{\arraycolsep}{0.1mm}
\label{dp2matter}
\ba{c|ccccc}
	&  SU(N)_A  & SU(N)_B & SU(N)_C & SU(N)_D & SU(N)_E \\
\hline
X_1   &  \antifund  & \fund  & & &  \\
X_2 & & \antifund & & & \fund \\
X_5 & & \antifund & & & \fund \\
X_3 & \antifund & & \fund & & \\
X_4 & & & \antifund & & \fund \\
X_9 & & \fund & \antifund & & \\
X_{11} & \fund & & & & \antifund \\
X_6 & & & \fund & \antifund &    \\
X_7 & \fund & & & \antifund &   \\
X_8 & & & & \fund & \antifund  \\
X_{10} & & & & \fund & \antifund \\
\ea
\stackrel{\mbox{dual on D}}{\Longrightarrow}
\ba{c|ccccc}
	&  SU(N)_A  & SU(N)_B & SU(N)_C & SU(N)_D & SU(N)_E \\
\hline
X_6   &  \antifund  & \fund  & & &  \\
X_5 & & \antifund & & & \fund \\
X_3 & & \antifund & & & \fund \\
X_1 & \antifund & & \fund & & \\
X_4 & & & \antifund & & \fund \\
X_{10} & & \fund & \antifund & & \\
X_{13} & \fund & & & & \antifund \\
\widetilde{X}_6 & & & \antifund & \fund &\\
\widetilde{X}_7 & \antifund & & & \fund &\\
\widetilde{X}_8 & & &&  \antifund & \fund  \\
\widetilde{X}_{10} & & &&  \antifund & \fund\\
M_{EA,1}& \fund & & & & \antifund \\
M_{EA,2}& \fund & & & & \antifund \\
M_{EC,1}&  & &\fund & & \antifund \\
M_{EC,2}&  & & \fund& & \antifund \\
\ea
}
\eeq

Let us explain the notations in \eref{dp2matter}. Before Seiberg duality we 
have 11 fields $X_{1,\ldots,11}$. After the dualisation on gauge group  
$D$, the we obtain dual quarks (corresponding to bi-fundamentals conjugate 
to the original quark $X_6,X_7, X_8,X_{10}$) which we denote 
$\widetilde{X}_6, \widetilde{X}_7, \widetilde{X}_8, \widetilde{X}_{10}$. 
Furthermore we have added meson fields $M_{EA,1},M_{EA,2},M_{EC,1},M_{EC,2}$, 
which are Seiberg mesons decomposed with respect to the unbroken chiral 
symmetry group. 

As before, one should incorporate the interactions as a deformation of 
this duality. Na\"{\i}vely we have $15$ fields in the dual theory, but
as we will show 
below, the resulting superpotential provides a mass term for the fields 
$X_4$ and $M_{EC,2}$, which transform in conjugate representations. 
Integrating them out, we will be left with 13 fields, the number of 
fields in Case I. In fact, with the mapping
$$
\ba{c|c|c|c|c|c|c|c|c|c|c|c}
\mbox{dual of II}& X_1 & X_2 & X_5 & X_3 & X_4 &X_9 & X_{11} &
\widetilde{X}_6 & \widetilde{X}_7 & \widetilde{X}_8 &
\widetilde{X}_{10} \\ \hline
\mbox{Case I} & Y_6 & Y_5 & Y_3 & Y_1 & $massive$ & Y_{10} & Y_{13} &
Y_2 & Y_4 & Y_{11} & Y_7
\ea
$$
and
%
$$
\ba{c|c|c|c|c}
\mbox{dual of II}& M_{EA,1} & M_{EA,2} & M_{EC,1} & M_{EC,2} \\
\hline
\mbox{Case I} & Y_8 & Y_{12} & Y_9 & $massive$
\ea
$$
we conclude that the matter content of the Case II dualised on gauge
group $D$ is identical to Case I!

Let us finally check the superpotentials, and also verify the claim that 
$X_4$ and $M_{EC,2}$ become massive. Rewriting the 
superpotential of II from \eref{dP2} in terms of the dual variables 
(matching the mesons as composites $M_{EA,1}=X_8 X_7$, $M_{EA,2}= X_{10} 
X_7$, $M_{EC,1}=X_8 X_6$, $M_{EC,2}= X_{10} X_6$), we have 
\begin{eqnarray*}
W_{II} & = & X_5 M_{EC,1} X_9 + X_1 X_2 M_{EA,2} + X_{11} X_3 X_4 \\
& & -X_4 M_{EC,2}- X_2 M_{EA,1} X_3 X_9 - X_{11} X_1 X_5.
\end{eqnarray*} 

As is with the previous subsection, to the above we must add the meson
interaction terms coming from Seiberg duality, namely
\begin{eqnarray*}
W_{meson} & = & M_{EA,1} \widetilde{X}_7 \widetilde{X}_8 -
M_{EA,2} \widetilde{X}_7 \widetilde{X}_{10} -M_{EC,1} \widetilde{X}_6
\widetilde{X}_8 + M_{EC,2} \widetilde{X}_6 \widetilde{X}_{10},
\end{eqnarray*}
(notice again the choice of sign in $W_{meson}$). Adding this two 
together we have
\begin{eqnarray*}
W^{dual}_{II} & = & X_5 M_{EC,1} X_9 + X_1 X_2 M_{EA,2} + X_{11} X_3 X_4 \\
& & -X_4 M_{EC,2}- X_2 M_{EA,1} X_3 X_9 - X_{11} X_1 X_5 \\
& & + M_{EA,1} \widetilde{X}_7 \widetilde{X}_8 -
M_{EA,2} \widetilde{X}_7 \widetilde{X}_{10} -M_{EC,1} \widetilde{X}_6
\widetilde{X}_8 + M_{EC,2} \widetilde{X}_6 \widetilde{X}_{10}.
\end{eqnarray*}
Now it is very clear that both $X_4$ and $M_{EC,2}$ are massive
and should be integrated out:
$$
X_4=\widetilde{X}_6 \widetilde{X}_{10},~~~~
M_{EC,2}=X_{11} X_3.
$$
Upon substitution we finally have
\begin{eqnarray*}
W^{dual}_{II} & = & X_5 M_{EC,1} X_9 + X_1 X_2 M_{EA,2} + X_{11} X_3 
\widetilde{X}_6 \widetilde{X}_{10}  - X_2 M_{EA,1} X_3 X_9 \\ 
& & - X_{11} X_1 X_5  + M_{EA,1} \widetilde{X}_7 \widetilde{X}_8 -
M_{EA,2} \widetilde{X}_7 \widetilde{X}_{10} -M_{EC,1} \widetilde{X}_6
\widetilde{X}_8,
\end{eqnarray*}
which with the replacement rules given above we obtain
\begin{eqnarray*}
W_{II}^{dual} & = & Y_3 Y_9 Y_{10} + Y_6 Y_5 Y_{12} + Y_{13} Y_1 Y_2 Y_7 
 - Y_5 Y_1 Y_{10} Y_8 \\ & & - Y_{13} Y_6 Y_3  +Y_8 Y_4 Y_{11} 
- Y_{12} Y_4 Y_7 - Y_9 Y_2 Y_{11}.
\end{eqnarray*}
This we instantly recognise, by referring to \eref{dP2}, as the
superpotential of Case I.

In conclusion therefore, with the matching of matter content and 
superpotential, the two torically dual cases I and II of the cone over
the second del Pezzo surface are also Seiberg duals.
\index{Seiberg Duality}
\index{Hanany-Witten!brane diamonds}
\section{Brane Diamonds and Seiberg Duality}
Having seen the above arguments from field theory, let us support that
toric duality is Seiberg duality from yet another perspective, namely,
through brane setups. The use of this T-dual picture for D3-branes at 
singularities will turn out to be quite helpful in showing that toric 
duality reproduces Seiberg duality.

What we have learnt from the examples where a brane interval picture is
available (i.e. NS- and D4-branes in the manner of \cite{HW}) is that the 
standard Seiberg duality by brane crossing reproduces the different gauge 
theories obtained from toric arguments (different partial resolutions of a 
given singularity). Notice that the brane crossing corresponds, under 
T-duality, to a change of the $B$ field in the singularity picture, rather 
than a change in the singularity geometry \cite{uranga,Unge}. Hence, the 
two theories arise on the world-volume of D-branes probing the same 
singularity.

Unfortunately, brane intervals are rather limited, in that they can be
used to study Seiberg duality for generalized conifold singularities, 
$xy=w^kw^l$. Although this is a large class of models, not many examples 
arise in the partial resolutions of $\IC^3/(\IZ_3\times\IZ_3)$. Hence
the relation to toric duality from partial resolutions cannot be checked 
for most examples.

Therefore it would be useful to find other singularities for which a nice 
T-dual brane picture is available. Nice in the sense that there is a 
motivated proposal to realize Seiberg duality in the corresponding brane 
setup. A good candidate for such a brane setup is {\bf brane diamonds}, 
studied in \cite{Aganagic}. 

Reference \cite{HZ} (see also \cite{HSU,HU}) introduced brane box  
configurations of intersecting NS- and NS'-branes (spanning 012345 and 
012367, respectively), with D5-branes (spanning 012346) suspended among 
them. Brane diamonds \cite{Aganagic} generalized (and refined) this setup by 
considering situations where the NS- and the NS'-branes recombine and span 
a smooth holomorphic curve in the $4567$ directions, in whose holes 
D5-branes can be suspended as soap bubbles. Typical brane diamond pictures 
are as in figures in the remainder of the chapter. 

Brane diamonds are related by T-duality along $46$ to a large set of 
D-branes at singularities. With the set of rules to read off the 
matter content and interactions in \cite{Aganagic}, they provide a useful 
pictorial representation of these D-brane gauge field theories. In 
particular, they correspond to singularities obtained as the abelian 
orbifolds of the conifold studied in Section 5 of \cite{uranga}, and 
partial resolutions 
thereof. Concerning this last point, brane diamond configurations admit 
two kinds of deformations: motions of diamond walls in the directions 57, 
and motions of diamond walls in the directions 46. The former T-dualize to 
geometric sizes of the collapse cycles, hence trigger partial resolutions 
of the singularity (notice that when a diamond wall moves in 57, the 
suspended D5-branes snap back and two gauge factors recombine, leading to 
a Higgs mechanism, triggered by FI terms). The later do not  modify the 
T-dual singularity geometry, and correspond to changes in the B-fields in 
the collapsed cycles.

The last statement motivates the proposal made in \cite{Aganagic} for Seiberg 
duality in this setup. It corresponds to closing a diamond, while keeping 
it in the 46 plane, and reopening it with the opposite orientation. The 
orientation of a diamond determines the chiral multiplets and interactions
arising from the picture. The effect of this is shown in fig 7 of 
\cite{Aganagic}: The rules are 
\begin{enumerate} 
\item When the orientation of a diamond is flipped, the arrows going in 
or out of it change orientation; 
\item one has to include/remove additional arrows to ensure a good `arrow 
flow' (ultimately connected to anomalies, and to Seiberg mesons)
\item Interactions correspond to closed loops of arrows in the brane 
diamond picture.
\item In addition to these rules, and based in our experience with Seiberg 
duality, we propose that when in the final picture some mesons appear in 
gauge representations conjugate to some of the original field, the 
conjugate pair gets massive.
\end{enumerate} 
\EPSFIGURE[h]{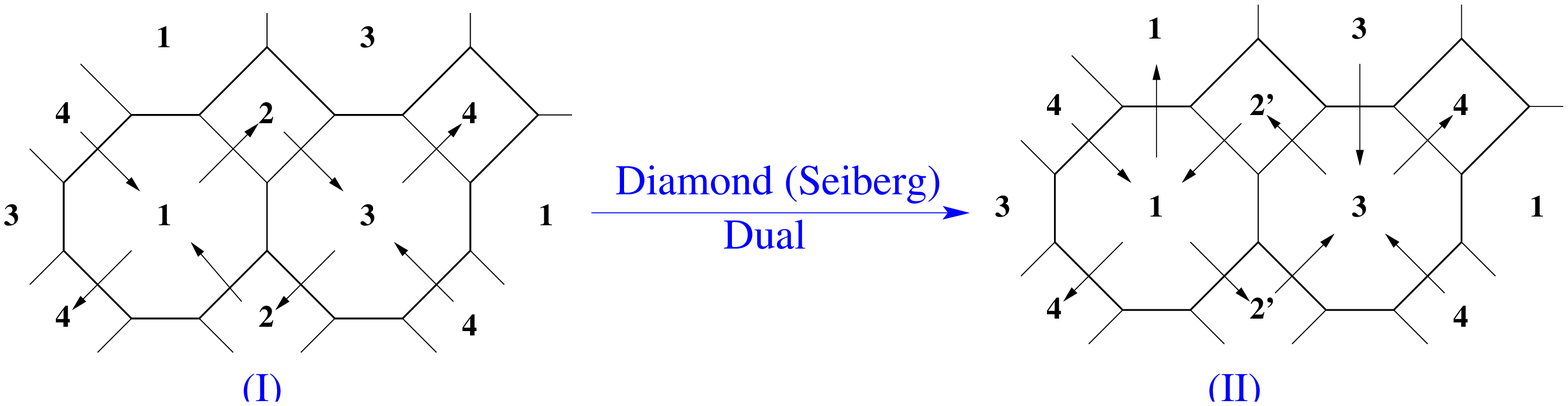,width=4in}
{Seiberg duality from the brane diamond construction for the generalized 
conifold $xy=z^2w^2$. Part (I) corresponds to the brane interval picture 
with alternating ordering of NS- and NS$'$-branes, whereas part (II) 
matches the other ordering.
\label{figz2w2}
}

These rules reproduce Seiberg duality by brane crossing in cases where a 
brane interval picture exists. In fact, one can reproduce our previous 
discussion of the $xy=z^2w^2$ in this language, as shown in figure 
\fref{figz2w2}. 
Notice that in analogy with the brane interval case the diamond transition 
proposed to reproduce Seiberg duality does not involve changes in the 
T-dual singularity geometry, hence ensuring that the two gauge theories 
will have the same moduli space. 

Let us re-examine our aforementioned examples.
\subsection{Brane diamonds for D3-branes at the cone over $F_0$}
Now let us show that diamond Seiberg duality indeed relates the two gauge 
theories arising on D3-branes at the singularity which is a complex cone 
over $F_0$.
The toric diagram of $F_0$ is similar to that of the conifold, only that 
it has an additional point (ray) in the middle of the square. Hence, it can 
be obtained from the conifold diagram by simply refining the lattice (by a 
vector $(1/2,1/2)$ if the conifold lattice is generated by $(1,0)$, 
$(0,1)$). This implies \cite{Aspin-Reso}) that the space can be obtained as 
a $\IZ_2$ quotient of the conifold, specifically modding $xy=zw$ by the 
action that flips all coordinates.

Performing two T-dualities in the conifold one reaches the brane diamond
picture described in \cite{Aganagic} 
(fig. 5), which is composed by two-diamond 
cell with sides identified, see Part (I) of \fref{diamcon}.
\EPSFIGURE[h]{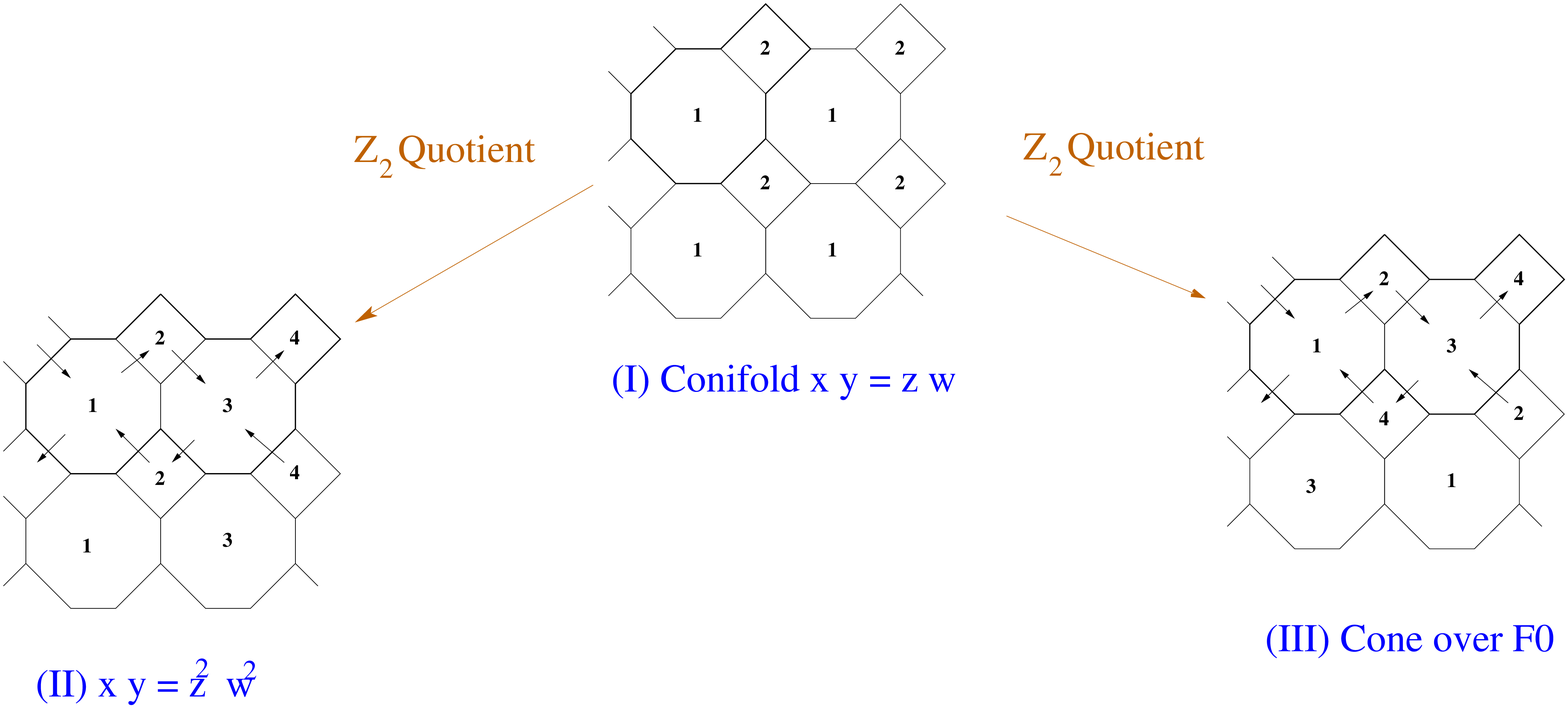,width=5in}
{
(I) Brane diamond for the conifold. Identifications in the
infinite periodic array of boxes leads to a two-diamond unit cell, whose
sides are identified in the obvious manner.
From (I) we have 2 types of $\IZ_2$ quotients: 
(II) Brane diamond for the $\IZ_2$ quotient of the conifold
$xy=z^2 w^2$, which is a case of the so-called generalised conifold. 
The identifications of sides are trivial, not tilting. The
final spectrum is the familiar non-chiral spectrum for a brane interval
with two NS and two NS' branes (in the alternate configuration);
(III) Brane diamond for the $\IZ_2$ quotient of the conifold
yielding the complex cone over $F_0$. The identifications of sides are
shifted, a fact related to the specific `tilted' refinement of the toric 
lattice.
\label{diamcon}
}
However, we are interested not in the conifold but on a $\IZ_2$ quotient 
thereof. Quotienting a singularity amounts to including more diamonds in 
the unit cell, i.e. picking a larger unit cell in the periodic array. 
There are two possible ways to do so, corresponding to two different 
$\IZ_2$ quotients of the conifold. One corresponds to the generalized 
conifold $xy=z^2w^2$ encountered above, and whose diamond picture is given 
in Part (II) of \fref{diamcon} for completeness. The second possibility
is shown in Part (III) of \fref{diamcon} and does correspond to the 
T-dual of the complex cone over $F_0$, so we shall henceforth concentrate 
on this case. Notice that the identifications of sides of the unit cell 
are shifted. The final spectrum agrees with the quiver before eq (2.2)
in \cite{0003085}. Moreover, following \cite{Aganagic}, these fields have 
quartic interactions, associated to squares in the diamond picture, with 
signs given by the orientation of the arrow flow. They match the ones in 
case II in (\ref{F0}).

Now let us perform the diamond duality in the box labeled 2. Following the 
diamond duality rules above, we obtain the result shown in  
\fref{diamF0dual}. Careful comparison with the spectrum and interactions 
of case I in (\ref{F0}), and also with the Seiberg dual computed in 
Section 4.1 shows that the new diamond picture reproduces the toric dual / 
Seiberg dual of the initial one. Hence, brane diamond configurations 
provide a new geometric picture for this duality.

\EPSFIGURE[h]{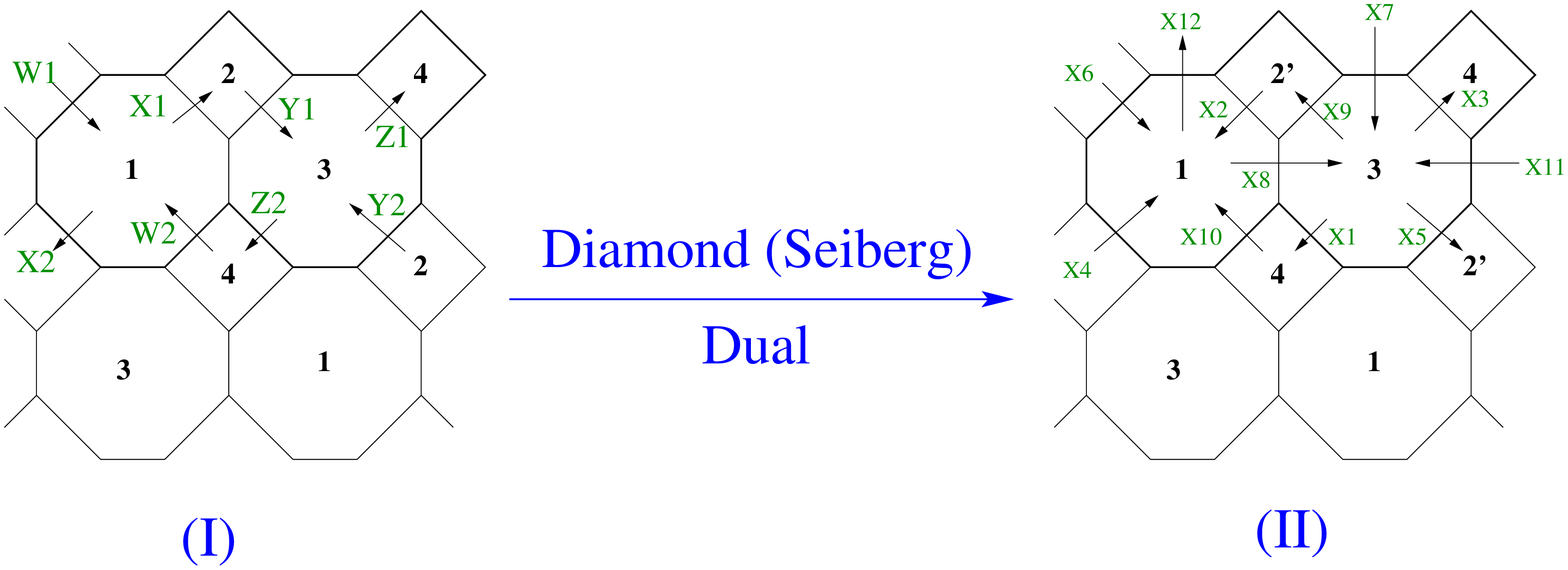,width=4in}
{Brane diamond for the two cases of the cone over $F_0$. (I) is as in 
\fref{diamcon} and (II) is the result after the diamond duality. The 
resulting spectrum and interactions are those of the toric dual (and 
also Seiberg dual) of the initial theory (I).
\label{diamF0dual}
}
\subsection{Brane diamonds for D3-branes at the cone over $dP_2$}
The toric diagram for $dP_2$ shows it cannot be constructed as a quotient 
of the conifold. However, it is a partial resolution of the orbifolded 
conifold described as $xy=v^2$, $uv=z^2$ in $\IC^5$ (we refer the
reader to \fref{condP2}.
This is a $\IZ_2\times \IZ_2$ quotient of the conifold whose brane 
diamond, shown in Part (I) of \fref{conz2z2}, contains 8 diamonds in its 
unit cell.
\EPSFIGURE[h]{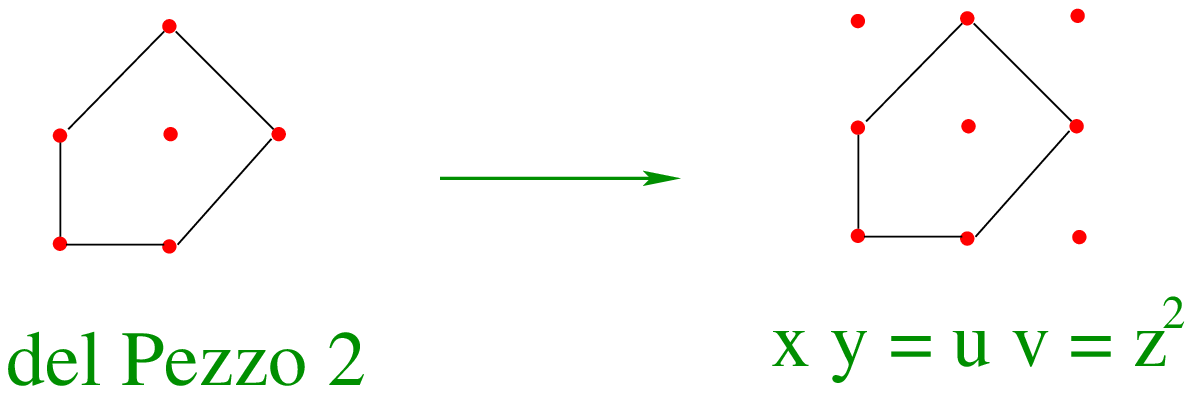,width=4in}
{Embedding the toric diagram of dP2 into the orbifolded 
conifold described as $xy=v^2$, $uv=z^2$.
\label{condP2}
}
\EPSFIGURE[h]{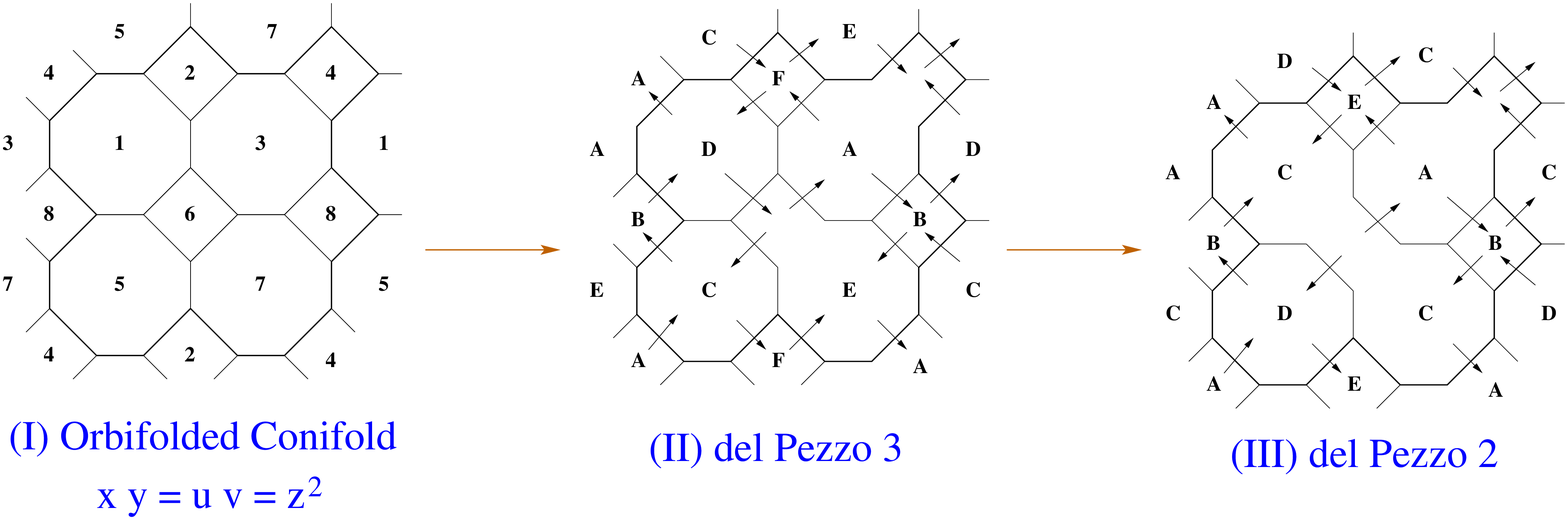,width=5.5in}
{(I) Brane diamond for a $\IZ_2\times \IZ_2$ orbifold of the
conifold, namely $xy=z^2; uv=z^2$. From this we can partial resolve to
(II) the cone over $dP3$ and thenceforth again to (III) the cone over
$dP_2$, which we shall discuss in the context of Seiberg duality.
\label{conz2z2}
}
Partial resolutions in the brane diamond language correspond to partial
Higgsing, namely recombination of certain diamonds. As usual, the 
difficult part is to identify which diamond recombination corresponds to 
which partial resolution. A systematic way proceed would 
be\footnote{As an aside, let us remark that the use of brane diamonds to 
follow partial resolutions of singularities may provide an alternative to 
the standard method of partial resolutions of orbifold singularities  
\cite{Horizon,0003085}. The existence of a brane picture for partial 
resolutions of orbifolded conifolds may turn out to be a useful advantage 
in this respect.}:
\begin{enumerate}
\item Pick a diamond recombination;
\item Compute the final gauge theory;
\item Compute its moduli space, which should be the partially resolved
	singularity. 
\end{enumerate}
However, instead of being systematic, we prefer a shortcut and simply 
match the spectrum of recombined diamond pictures with known results of 
partial resolutions. In order to check we pick the right resolutions, it 
is useful to discuss the brane diamond picture for some intermediate step 
in the resolution to $dP_2$. A good intermediate point, for which the 
field theory spectrum is known is the complex cone over $dP_3$.

By trial and error matching, the diamond recombination which reproduces the 
world-volume spectrum for D3-branes at the cone over $dP_3$ (see 
\cite{0003085,0104259}), is shown in Part (II) of \fref{conz2z2}.
Performing a further resolution, chosen so as to match known results, one 
reaches the brane diamond picture for D3-branes on the cone over $dP_2$, 
shown in Part (III) of \fref{conz2z2}. More specifically, the spectrum and 
interactions in the brane diamond configuration agrees with those of case 
I in (\ref{dP2}).

This brane box diamond, obtained in a somewhat roundabout way, is our
starting point to discuss possible dual realizations. In fact, recall that
there is a toric dual field theory for $dP_2$, given as case II in
(\ref{dP2}). After some inspection, the desired effect is obtained by
applying diamond Seiberg duality to the diamond labeled B. The
corresponding process and the resulting diamond picture are shown in
\fref{pezzo2dual}. Two comments are in order: notice that in
applying diamond duality using the rules above, some vector-like pairs of 
fields have to be removed from the final picture; in fact one can check by 
field theory Seiberg duality that the superpotential makes them massive. 
Second, notice that in this case we are applying duality in the direction 
opposite to that followed in the field theory analysis in Section 4.2; it 
is not difficult to check that the field theory analysis works in this 
direction as well, namely the dual of the dual is the original theory. 
\EPSFIGURE{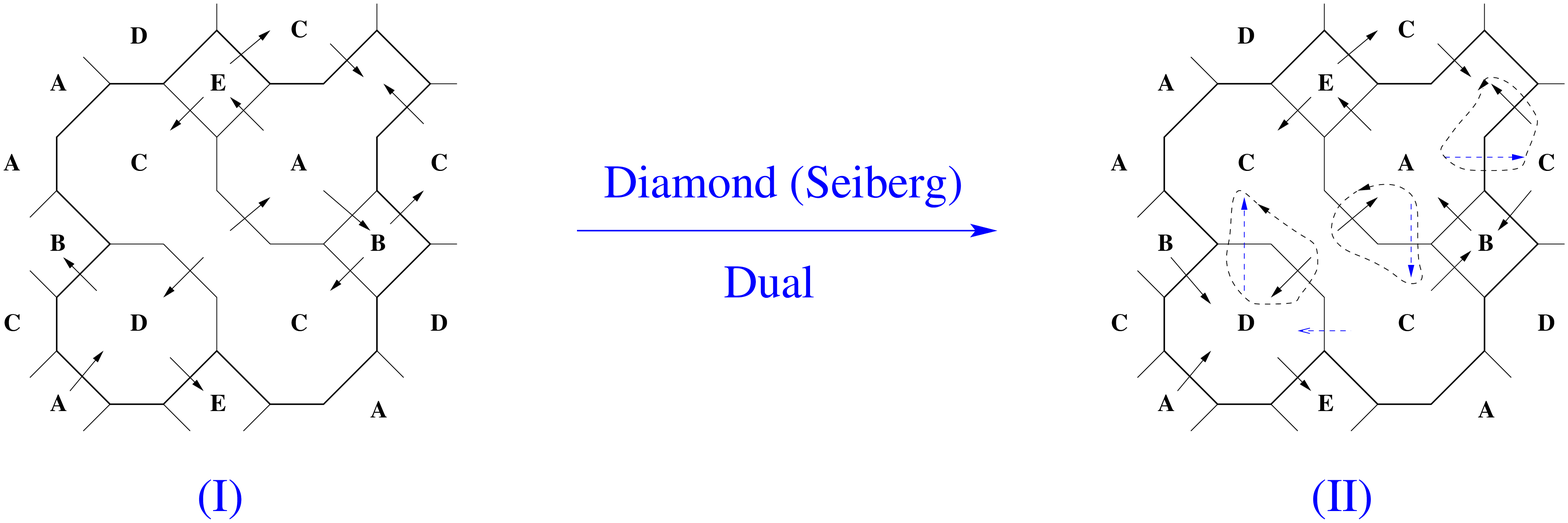,width=5in}
{The brane diamond setup for the Seiberg dual configurations of the
cone over $dP_2$. (I) is as in \fref{conz2z2} and (II) is the results
after Seiberg (diamond) duality and
gives the spectrum for the toric dual theory.
The added meson fields are drawn in dashed blue lines.
Notice that applying the diamond dual rules
carelessly one gets some additional vectorlike pairs, shown in the
picture within dotted lines. Such multiplets presumably get massive in the
Seiberg dualization, hence we do not consider them in the quiver.
\label{pezzo2dual}
}
Therefore this new example provides again a geometrical realization of 
Seiberg duality, and allows to connect it with Toric Duality.

\medskip

We conclude this Section with some remarks. The brane diamond picture 
presumably provides other Seiberg dual pairs by picking different gauge 
factors. All such models should have the same singularities as moduli 
space, and should be toric duals in a broad sense, even though all such 
toric duals may not be obtainable by partial resolutions of $\IC^3/(\IZ_3
\times \IZ_3)$. From this viewpoint we learn that Seiberg duality can 
provide us with new field theories and toric duals beyond the reach of 
present computational tools. This is further explored in Section 7.

A second comment along the same lines is that Seiberg duality on nodes for 
which $N_f\neq 2N_c$ will lead to dual theories where some gauge factors 
have different rank. Taking the theory back to the `abelian' case, some 
gauge factors turn out to be non-abelian. Hence, in these cases, even 
though Seiberg duality ensures the final theory has the same singularity 
as moduli space, the computation of the corresponding symplectic 
quotient is beyond the standard tools of toric geometry. Therefore, 
Seiberg duality can provide (`non-toric') gauge theories with toric moduli 
space.
\index{Seiberg Duality}
\section{A Quiver Duality from Seiberg Duality}
If we are not too concerned with the superpotential,
when we make the 
Seiberg duality transformation, we can obtain the matter content 
very easily at the level of the quiver diagram. What we obtain are
rules for a so-called ``quiver duality'' which is a rephrasing of the
Seiberg duality transformations in field (brane diamond) theory in the 
language of quivers. Denote $(N_c)_i$ the number of colors at the $i^{th}$ 
node, and $a_{ij}$ the number of arrows from the node $i$ to the $j$ (the 
adjacency matrix)
The rules on the quiver to obtain Seiberg dual theories are
\begin{enumerate}
\item	Pick the dualisation node $i_0$. Define the following sets of 
        nodes: $I_{in} :=$ nodes having arrows going into $i_0$; 
	$I_{out} := $ those having arrow coming from $i_0$ and
	$I_{no} :=$ those unconnected with $i_0$. The node $i_0$
	should not be included in this classification. 
\item   Change the rank of the node $i_0$ from $N_c$ to $N_f-N_c$ where 
        $N_f$ is the number of vector-like flavours, $N_f=\sum\limits_{i\in 
        I_{in}} a_{i,i_0}= \sum\limits_{i\in I_{out}} a_{i_0,i}$ 
%
\item	Reverse all arrows going in or out of $i_0$, therefore
	\[
	a^{dual}_{ij} = a_{ji} \qquad \mbox{ if either }i,j = i_0
	\]
\item	Only arrows linking $I_{in}$ to $I_{out}$ will be changed and
	all others remain unaffected.
\item	For every pair of nodes $A$, $B$, $A \in I_{out}$ and  $B \in 
        I_{in}$, change the number of arrows $a_{AB}$ to
	\[
	a^{dual}_{AB} = a_{AB} - a_{i_0 A} a_{B i_0} \qquad \mbox{ for
	} A \in I_{out},~~B \in I_{in}.
	\]
	If this quantity is negative, we simply take it to mean 
        $-a^{dual}$ arrow go from $B$ to $A$. 
\end{enumerate}
These rules follow from applying Seiberg duality at the field theory 
level, and therefore are consistent with anomaly cancellation. In 
particular, notice the for any node $i\in I_{in}$, we have replaced 
$a_{i,i_0} N_c$ fundamental chiral multiplets by $-a_{i,i_0}(N_f-N_c) + 
\sum_{j\in I_{out}} a_{i,i0} a_{i_0,j}$ which equals $-a_{i,i_0}(N_f-N_c) 
+ a_{i,i0} N_f=a_{i,i_0} N_c$, and ensures anomaly cancellation in the 
final theory. Similarly for nodes $j\in I_{out}$.

It is straightforward to apply these rules to the quivers in the by now 
familiar examples in previous sections.

\medskip

In general, we can choose an arbitrary node to perform the above Seiberg 
duality rules. However, not every node is suitable for a toric description.
The reason is that, if we start from a quiver whose every node has the 
same rank $N$, after the transformation it is possible that this no longer 
holds. We of course wish so because due to the very definition of the 
$\IC^*$ action for toric varieties, toric descriptions are possible iff 
all nodes are $U(1)$, or in the non-Abelian version, $SU(N)$. If for 
instance we choose to Seiberg dualize a node with $3N$ flavours, the dual 
node will have rank $3N-N=2N$ while the others will remain with rank $N$, 
and our description would no longer be toric. For this reason we must 
choose nodes with only $2N_f$ flavors, if we are to remain within toric 
descriptions.

One natural question arises: if we Seiberg-dualise every possible allowed 
node, how many different theories will we get? Moreover how many of these
are torically dual? Let we re-analyse the examples we have thus far 
encountered.
\index{del Pezzo Surfaces}
\index{Toric Duality}
\subsection{Hirzebruch Zero}
Starting from case $(II)$ of $F_0$ (recall \fref{F0}) all of four nodes 
are qualified to yield toric Seiberg duals (they each have 2 incoming and 2 
outgoing arrows and hence $N_f=2N$). Dualising any one will give to case 
$(I)$ of $F_0$. On the other hand, from $(I)$ of $F_0$, we see that only 
nodes $B,D$ are qualified to be dualized. Choosing either, we get back to 
the case $(II)$ of $F_0$. In another word, cases $(I)$ and $(II)$ are 
closed under the Seiberg-duality transformation. In fact, this is a very 
strong evidence that there are only two toric phases for $F_0$ no matter 
how we embed the diagram into higher $\IZ_k \times \IZ_k$ singularities. 
This also solves the old question \cite{0003085,0104259} that the Inverse 
Algorithm does not in principle tell us how many phases we could have. Now 
by the closeness of Seiberg-duality  transformations, we do have a way to 
calculate the number of possible phases. Notice, on the other hand, the 
existence of non-toric phases.
\subsection{del Pezzo 0,1,2}
Continuing our above calculation to del Pezzo singularities, we see that
for $dP_0$ no node is qualified, so there is only one toric phase which is 
consistent with the standard result \cite{0104259} as a resolution ${\cal 
O}_{\IP^2}(-1) \rightarrow \IC^3/\IZ_3$. 
For $dP_1$, nodes $A,B$ are qualified (all notations coming from 
\cite{0104259}), but the dualization gives back to same theory, so it too 
has only one phase.

For our example $dP_2$ studied earlier (recall \fref{dP2}), there are 
four points $A,B,C,D$ which are qualified in case (II). Nodes $A,C$ give 
back to case (II) while nodes $B,D$ give rise to case (I) of $dP_2$. On 
the other hand, for case (I), three nodes $B,D,E$ are qualified. Here nodes 
$B,E$ give case (II) while node $D$ give case (I). In other words, cases 
(I) and (II) are also closed under the Seiberg-duality transformation, so 
we conclude that there too are only two phases for $dP_2$, as presented 
earlier.
\subsection{The Four Phases of $dP_3$}
Things become more complex when we discuss the phases of $dP_3$. As we 
remarked before, due to the running-time limitations of the Inverse 
Algorithm, only one phase was obtained in \cite{0104259}. However, one may
expect this case to have more than just one phase, and in fact a recent 
paper has given another phase \cite{HI}. Here, using the closeness 
argument we give evidence that there are four (toric) phases for $dP_3$. 
We will give only one phase in detail. Others are similarly obtained.
\EPSFIGURE[h]{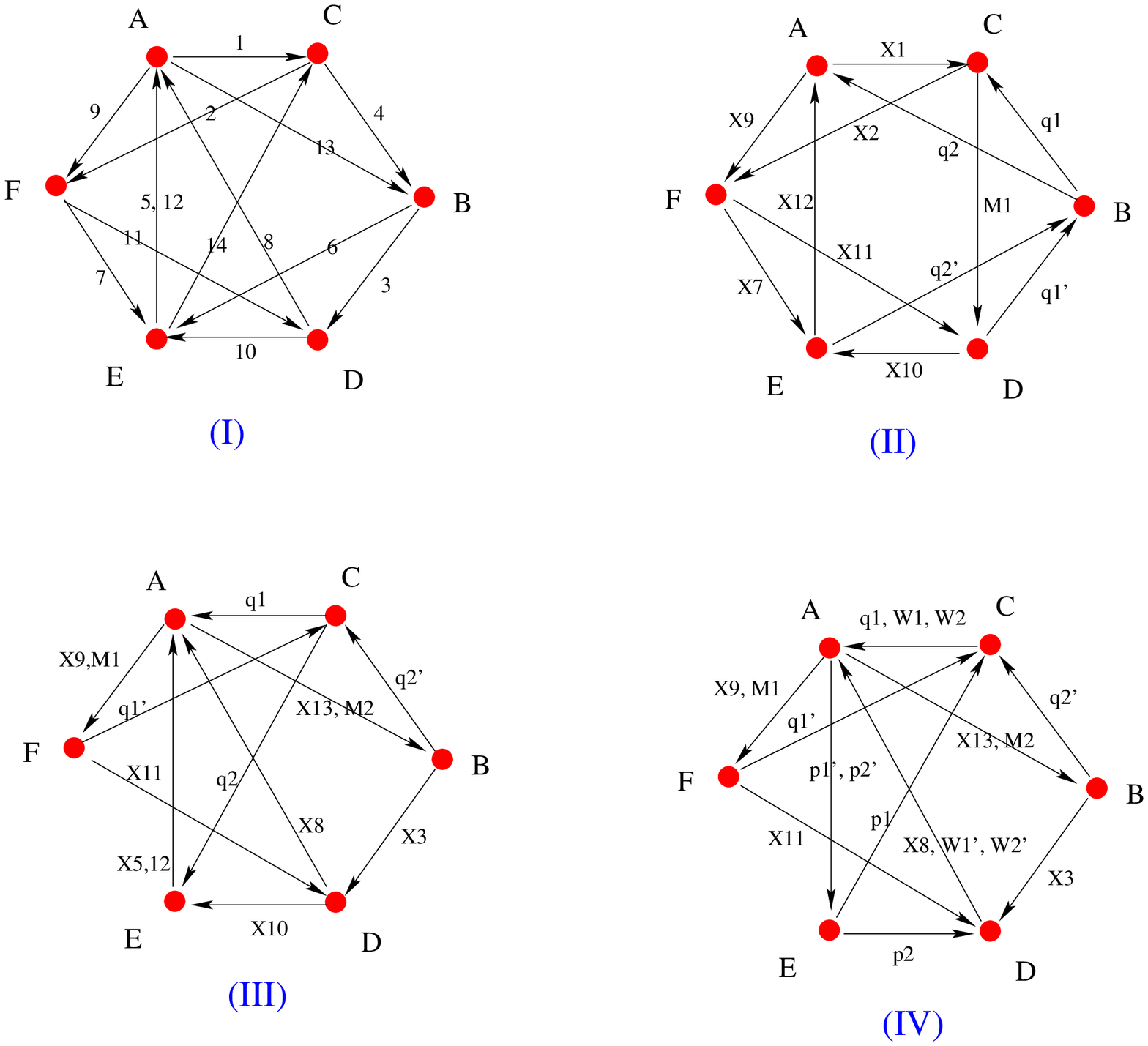,width=5in}
{The four Seiberg dual phases of the cone over $dP_3$.
\label{f:dP3seiberg}
}
Starting from case (I) given in \cite{0104259} and dualizing node $B$,
(we refer the reader to \fref{f:dP3seiberg}) 
we get the charge (incidence) matrix $d$ as 
\[
\tmat{
 & q_1 & q_2 & q'_1 & q'_2 & X_1 & X_2 & X_7 & X_9 & X_{10} & X_{11} & 
M_1 &  X_{14}& M_2 & X_8 & M'_1 & X_5 & X_{12} &
M'_2 \cr
A & 0 & 1   & 0 & 0  & -1 & 0 & 0 & -1 & 0 & 0 & 0 & 0 & 0 & 1 & -1 &1&1
&-1  \cr
B & -1 & -1 & 1  & 1 & 0 & 0  & 0 & 0 & 0 & 0 & 0 & 0 & 0 & 0 & 0 &0&0&0
\cr
C & 1 & 0 &   0 & 0  & 1 & -1 & 0 & 0 & 0 & 0 & -1 & 1 & -1 & 0 & 0 &0 & 0
&0 \cr
D & 0 & 0   & -1 & 0 & 0 & 0 & 0 & 0 & -1 & 1 &  1 & 0  & 0 & -1 & 1 &0& 0
&0 \cr
E & 0 & 0   & 0 & -1 & 0 & 0 & 1 & 0 & 1 & 0 & 0 & -1 & 1 & 0 & 0 &-1 & -1
& 1 \cr
F & 0 & 0   & 0 & 0  & 0 & 1 & -1 & 1 &0  & -1 & 0 &  0 & 0 & 0 & 0 &0 & 0
& 0\cr
}
\]
where
$$
M_1= X_4 X_3,~~~M_2=X_4 X_6,~~~ M'_1= X_{13} X_3,~~~ M'_2= X_{13} X_6
$$
are the added mesons. Notice that $X_{14}$ and $M_2$ have opposite charge.
In fact, both are massive and will be integrate out. Same for
pairs $(X_8,M'_1)$ and $(X_5,M'_2)$.

Let us derive the superpotential. Before dual transformation, the 
superpotential is \cite{0003085}
\begin{eqnarray*}
W_{I} & = & X_3 X_8 X_{13} - X_8 X_9 X_{11} - X_5 X_6 X_{13} - X_1 X_3 X_4 
X_{10} X_{12} \\
&  & X_7 X_9 X_{12} + X_4 X_6 X_{14} + X_1 X_2 X_5 X_{10} X_{11} -
X_2 X_7 X_{14}
\end{eqnarray*}
After dualization, superpotential is rewritten as
\begin{eqnarray*}
W' & = & M'_1 X_8  - X_8 X_9 X_{11} - X_5 M'_2 - X_1 M_1 
X_{10} X_{12} \\
&  & X_7 X_9 X_{12} + M_2 X_{14} + X_1 X_2 X_5 X_{10} X_{11} -
X_2 X_7 X_{14}.
\end{eqnarray*}
It is very clear that fields $X_8,M'_1,X_5,M'_2,X_{14},M_2$ are
all massive. Furthermore, we need to add the meson part
\begin{eqnarray*}
W_{meson} & = & M_1 q'_1 q_1 - M_2 q_1 q'_2 -M'_1 q'_1 q_2 + M'_2 q'_2 q_2
\end{eqnarray*}
where we determine the sign as follows: since the term $M'_1 X_8$ in 
$W'$ in positive, we need term $M'_1 q'_1 q_2$ to be negative. 
After integration all massive fields, we get the superpotential as
\begin{eqnarray*}
W_{II} & = &   -q'_1 q_2  X_9 X_{11} - X_1 M_1 X_{10} X_{12}+
X_7 X_9 X_{12}   + X_1 X_2 q'_2 q_2 X_{10} X_{11} -
X_2 X_7 q_1 q'_2 +M_1 q'_1 q_1.
\end{eqnarray*}
The charge matrix now becomes
\[
\tmat{
 & q_1 & q_2 & q'_1 & q'_2 & X_1 & X_2 & X_7 & X_9 & X_{10} & X_{11} & 
M_1 &   X_{12}  \cr
A & 0 & 1   & 0 & 0  & -1 & 0 & 0 & -1 & 0 & 0 & 0 & 1 \cr
B & -1 & -1 & 1  & 1 & 0 & 0  & 0 & 0 & 0 & 0 & 0 &  0\cr
C & 1 & 0 &   0 & 0  & 1 & -1 & 0 & 0 & 0 & 0 & -1 & 0 \cr
D & 0 & 0   & -1 & 0 & 0 & 0 & 0 & 0 & -1 & 1 &  1 & 0 \cr
E & 0 & 0   & 0 & -1 & 0 & 0 & 1 & 0 & 1 & 0 & 0 &  -1\cr
F & 0 & 0   & 0 & 0  & 0 & 1 & -1 & 1 &0  & -1 & 0 & 0 \cr
}
\]
This is in precise agreement with \cite{HI}; very re-assuring indeed!

Without further ado let us present the remaining cases.
The charge matrix for the third one (dualising node $C$ of (I)) is
\[
\tmat{
 & q_1 & q'_1 & q'_2 & q_2 & X_5 & X_{12} & X_3 & X_8 & X_9 & M_1 & X_{10}
& 
X_{11} & X_{13} & M_2 \cr
A & 1 & 0 & 0 & 0 & 1 & 1 & 0 & 1 & -1 & -1 & 0 & 0 & -1 & -1 \cr
B & 0 & 0 & -1 & 0 & 0 & 0 & -1 & 0 & 0 & 0 & 0 & 0 & 1 & 1 \cr
C & -1 & 1 & 1 & -1 & 0 & 0 & 0 & 0 & 0 & 0 & 0 & 0 & 0 & 0 \cr
D & 0 & 0 & 0 & 0 & 0 & 0 & 1 & -1 & 0 & 0 & -1 & 1 & 0 & 0 \cr
E & 0 & 0 & 0 & 1 & -1 & -1 & 0 & 0 & 0 & 0 & 1 & 0 & 0 & 0 \cr
F & 0 & -1 & 0 & 0 & 0 & 0 & 0 & 0 & 1 & 1 & 0 & -1 & 0 & 0 \cr
}
\]
with superpotential
\begin{eqnarray*}
W_{III} & = & X_3 X_8 X_{13} - X_8 X_9 X_{11} - X_5 q_2 q'_2  X_{13} 
-M_2 X_3 X_{10} X_{12} \\
& & + q_2 q'_1 X_9 x_{12} + M_1 X_5 X_{10} X_{11} -M_1 q_1 q'_1 
+M_2 q_1 q'_2.
\end{eqnarray*}

Finally the fourth case (dualising node $E$ of (III)) has the charge
matrix
\[
\tmat{
 & q_1 & W_1 & W_2 & q'_1 & q'_2 & X_3 & X_8 & W'_1 & W'_2 & X_9 & M_1 &
X_{11}& X_{13} & M_2 & p_1 & p'_1 & p'_2 & p_2 \cr
A & 1 & 1 & 1 & 0 & 0 & 0 & 1 & 1 & 1 & -1 & -1 & 0 &-1 & -1 & 0 & -1 & -1
& 0\cr
B & 0 & 0 & 0 & 0 & -1 & -1 & 0 & 0 & 0 & 0 & 0 & 0 & 1 & 1 & 0 & 0 & 0 &
0\cr
C & -1 & -1 & -1 & 1 & 1 & 0 & 0 & 0 & 0 & 0 & 0 & 0 & 0 & 0 & 1 & 0 & 0 &
0\cr
D & 0 & 0 & 0 & 0 & 0 & 1 & -1 & -1 & -1 & 0 & 0 & 1 & 0 & 0 & 0 & 0 & 0 &
1\cr
E & 0 & 0 & 0 & 0 & 0 & 0 & 0 & 0 & 0 & 0 & 0 & 0 & 0 & 0 & -1 & 1 & 1 &
-1 \cr
F & 0 & 0 & 0 & -1 & 0 & 0 & 0 & 0 & 0 & 1 & 1 & -1 & 0 & 0 & 0 & 0 & 0 &
0  \cr
}
\]
with superpotential
\begin{eqnarray*}
W_{IV} & = & X_3 X_8 X_{13} - X_8 X_9 X_{11} - W_1 q'_2 X_{13} -M_2 X_3
W'_2
+q'_1 X_9 W_2 + M_1 W'_1 X_{11} \\
& & -M_1 q_1 q'_1+M_2 q_1 q'_2+W_1 p_1 p'_1 - W_2 p_1 p'_2 -W'_1 p_2 p'_1
+W'_2 p_2 p'_2
\end{eqnarray*}
\index{Picard-Lefschetz Theory!and Seiberg Duality}
\section{Picard-Lefschetz Monodromy and Seiberg Duality}
In this section let us make some brief comments about Picard-Lefschetz
theory and Seiberg duality, a relation between which has been within
the literature \cite{VO}.
It was argued in \cite{ItoDual} that at least in the case of 
D3-branes placed on ADE conifolds \cite{gubser,lopez}
Seiberg duality for ${\cal N}=1$ SUSY gauge theories 
can be geometrised into Picard-Lefschetz monodromy.
Moreover in \cite{HI}
Toric Duality is interpreted as 
Picard-Lefschetz monodromy action on the 3-cycles.

On the level of brane setups, this interpretation seems to be reasonable.
Indeed, consider a brane crossing process in a brane 
interval picture. Two branes separated in $x^6$ approach, are exchanged, 
and move back. The T-dual operation on the singularity corresponds to 
choosing a collapsed cycle, decreasing its B-field to zero, and continuing 
to negative values. This last operation is basically the one generating 
Picard-Lefschetz monodromy at the level of homology classes. Similarly, 
the closing and reopening of diamonds corresponds to continuations past 
infinite coupling of the gauge theories, namely to changes in the T-dual 
B-fields in the collapsed cycles.

It is the purpose of this section to point out the observation that
while for restricted classes of theories the two phenomena are the
same, in general Seiberg duality and a na\"{\i}ve application of
Picard-Lefschetz (PL) monodromy do not
seem to coincide. We leave this issue here as a puzzle, which we shall
resolve in an upcoming work.

The organisation is as follows.
First we briefly introduce the
concept of Picard-Lefschetz monodromy for the convenience of the
reader and to establish some notation.
Then we give two examples: the first is one with two Seiberg dual
theories not related by PL and the second, PL dual theories not
related by Seiberg duality.
\subsection{Picard-Lefschetz Monodromy}
We first briefly remind
the reader of the key points of the PL theory \cite{Arnold}.
Given a singularity on a manifold $M$ and a basis $\{ \Delta_i \} \subset 
H_{n-1}(M)$ for its vanishing $(n-1)$-cycles,
going around these vanishing cycles induces 
a monodromy, acting on arbitrary cycles
$a \in H_\bullet(M)$; moreover this action is computable in terms of
intersection $a \circ \Delta_i$ of the cycle $a$ with the basis:
\begin{theorem}
The monodromy group of a singularity is generated by the
Picard-Lefschetz operators $h_i$, corresponding to a 
basis $\{ \Delta_i \} \subset H_{n-1}$ of vanishing cycles.
In particular for
any cycle $a \in H_{n-1}$ (no summation in $i$)
\[
h_i(a) = a + (-1)^{\frac{n(n+1)}{2}} (a \circ \Delta_i) \Delta_i.
\]
\end{theorem}
More concretely, the PL monodromy operator
$h_i$ acts as a matrix $(h_i)_{jk}$ on the basis $\Delta_j$:
$$
h_i(\Delta_j)=(h_i)_{jk} \Delta_k.
$$

Next we establish the relationship between this geometric concept and 
a physical interpretation. 
According geometric engineering, when a
D-brane wraps a vanishing cycle in the basis, 
it give rise to a simple factor in the product gauge group. Therefore
the total number of vanishing
cycles gives the number of gauge group factors. Moreover, 
the rank of each particular factor
is determined by how many times it wraps that cycle.

For example, an original theory with gauge group $\prod\limits_j
SU(M_j)$ is represented by the brane wrapping the cycle $\sum\limits_j
M_j\Delta_j$.
Under PL monodromy, the cycle undergoes the
transformation
$$
\sum\limits_j M_j\Delta_j
\Longrightarrow \sum\limits_j M_j (h_i)_{jk} \Delta_k.
$$
Physically, the final gauge theory is 
$\prod\limits_k SU(\sum_j M_j(h_i)_{jk})$.

The above shows how the rank of the gauge theory changes under 
PL. To determine the theory completely, we also need to
see how the matter content transforms. In geometric engineering,
the matter content is given by intersection of these cycles
$\Delta_j$. Incidentally, our Inverse Algorithm gives a nice way and
alternative method of computing such intersection matrices of cycles.

Let us take $a = \Delta_j$, then
\[
h_i(\Delta_j) = \Delta_j + (\Delta_j \circ \Delta_i) \Delta_i.
\]
This is particularly useful to us because $(\Delta_j \circ \Delta_i)$,
as is well-known, is the anti-symmetrised adjacency matrix of the
quiver (for a recent discussion on this, see \cite{HI}). Indeed this
intersection matrix of (the blowup of) the vanishing homological 
cycles specifies the matter content as prescribed by D-branes wrapping
these cycles in the mirror picture. Therefore we have $(\Delta_j \circ
\Delta_i) = [a_{ji}] := a_{ji} - a_{ij}$ for $j \ne i$ and for $i=j$,
we have the self-intersection numbers $(\Delta_i \circ \Delta_i)$.
Hence we can safely write (no summation in $i$)
\beq
\label{PLmatter}
\Delta_j^{dual} = h_i(\Delta_j) = \Delta_j + [a_{ji}] \Delta_i
\eeq
for $a_{ji}$ the quiver (matter) matrix when Seiberg dualising on the
node $i$; we have also used the notation $[M]$ to mean the
antisymmetrisation $M - M^t$ of matrix $M$.
Incidentally in the basis prescribed by $\{ \Delta_i 
\}$, we have the explicit form of the Picard-Lefschetz operators in terms 
of the quiver matrix (no summation over indices):
$(h_i)_{jk} = \delta_{jk} + [a_{ji}] \delta_{ik}$.

From \eref{PLmatter} we have
\beq
\label{quiverPL}
\ba{rcl}
[a^{dual}_{jk}] & := & \Delta^{dual}_j \circ \Delta^{dual}_k =
	(\Delta_j + [a_{ji}] \Delta_i) \circ (\Delta_k + [a_{ki}]
	\Delta_i)\\
	& = & [a_{jk}] + [a_{ki}][a_{ji}] + [a_{ji}][a_{ik}] + 
		[a_{ji}][a_{ki}]\Delta_i \circ \Delta_i\\
	& = & [a_{jk}] + c_i [a_{ij}] [a_{ki}]
\ea
\eeq
where $c_i := \Delta_i \circ \Delta_i$, are constants depending only on
self-intersection.

We observe that our quiver duality rules obtained from field
theory (see beginning of Section 6) seem to resemble
\eref{quiverPL}, i.~e.~when $c_i = 1$ and $j,k \ne i$. 
However the precise relation of trying to reproduce Seiberg duality
with PL theory still remains elusive.
\subsection{Two Interesting Examples}
However the situation is not as simple. In the following we shall argue
that while Seiberg duality and a straightforward 
Picard-Lefschetz transformation
certainly do have common features and that in restricted classes of
theories such as those in \cite{ItoDual}, for general singularities the
two phenomena may bifurcate.

We first present two theories related by Seiberg duality that cannot
be so by Picard-Lefschetz. Consider the standard $\IC^3/\IZ_3$ theory
with $a_{ij} = \tmat{0 & 0 & 3 \cr 3 & 0 & 0 \cr 0 & 3 & 0 \cr}$ and
gauge group $U(1)^3$, given in (a) of \fref{z3}.
Let us Seiberg-dualise on node $A$ to obtain a theory (b), with
matter content $a_{ij}^{dual} = \tmat{ 0
& 3 & 0 \cr 0 & 0 & 6 \cr 3 & 0 & 0 \cr}$ and gauge group $SU(2)
\times U(1)^2$. Notice especially that the rank of the gauge 
group factors in part (b) are $(2,1,1)$ while those in  part (a) are
$(1,1,1)$. Therefore theory (b) has total rank 4 while (a) has only 3.
Since geometrically PL only shuffles the vanishing cycles and
certainly preserves their number, we see that (a) and (b) cannot be
related by PL even though they are Seiberg duals.
\EPSFIGURE{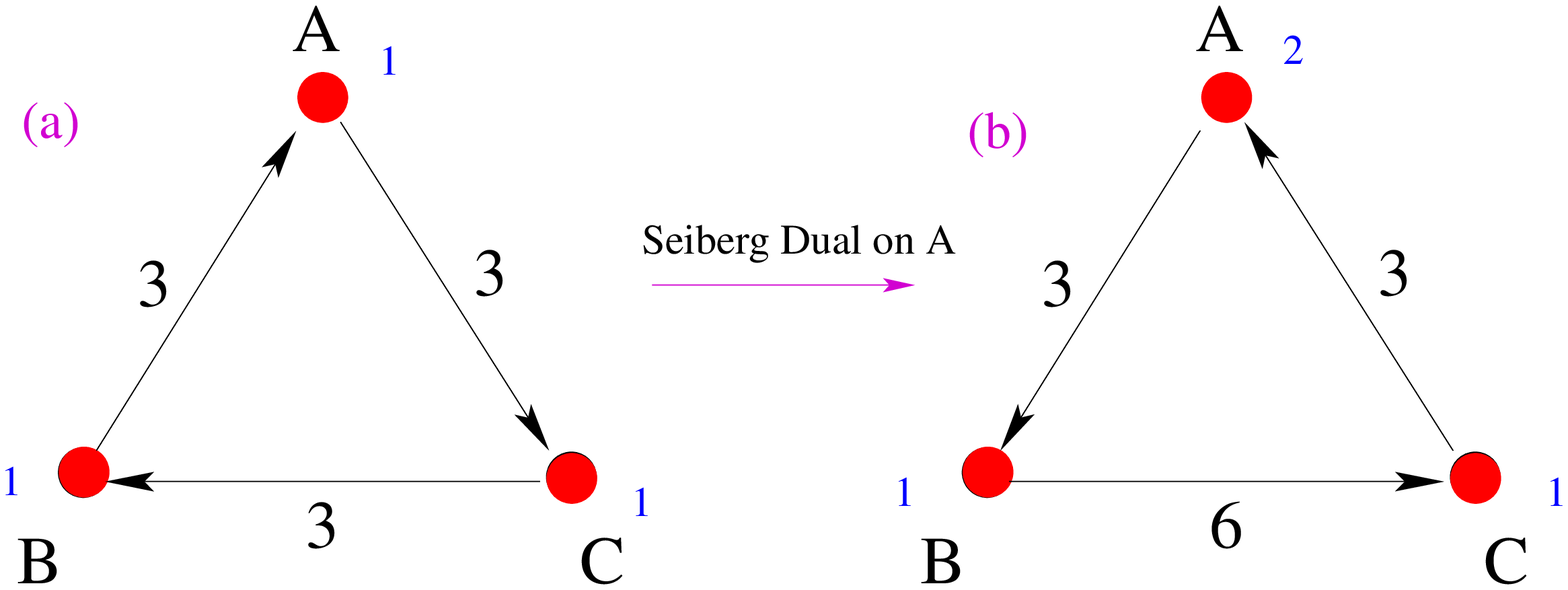,width=5in}
{Seiberg Dualisation on node $A$ of the $\IC^3/\IZ_3$ orbifold
theory. The subsequent theory cannot be obtained by a Picard-Lefschetz
monodromy transformation.
\label{z3}
}

On the other hand we give an example in the other direction, namely
two Picard-Lefschetz dual theories which are not Seiberg
duals. Consider the case given in \fref{f:NAdP3}, this is a phase of
the theory for the complex cone over dP3 as given in \cite{HI2}. This
is PL dual to any of the 4 four phases in \fref{f:dP3seiberg} 
in the previous
section by construction with $(p,q)$-webs. Note that the total rank
remains 6 under PL even though the number of nodes changed. However
Seiberg duality on any of the allowed node on any of the 4 phases
cannot change the number of nodes. Therefore, this example in
\fref{f:NAdP3} is not Seiberg dual to the other 4.

What we have learnt in this short section is that Seiberg duality and
a na\"{\i}ve application of Picard-Lefschetz monodromy seem to have
discrepancies for general singularities. The resolution of this
puzzle will be delt with in a forthcoming work.
\EPSFIGURE[h]{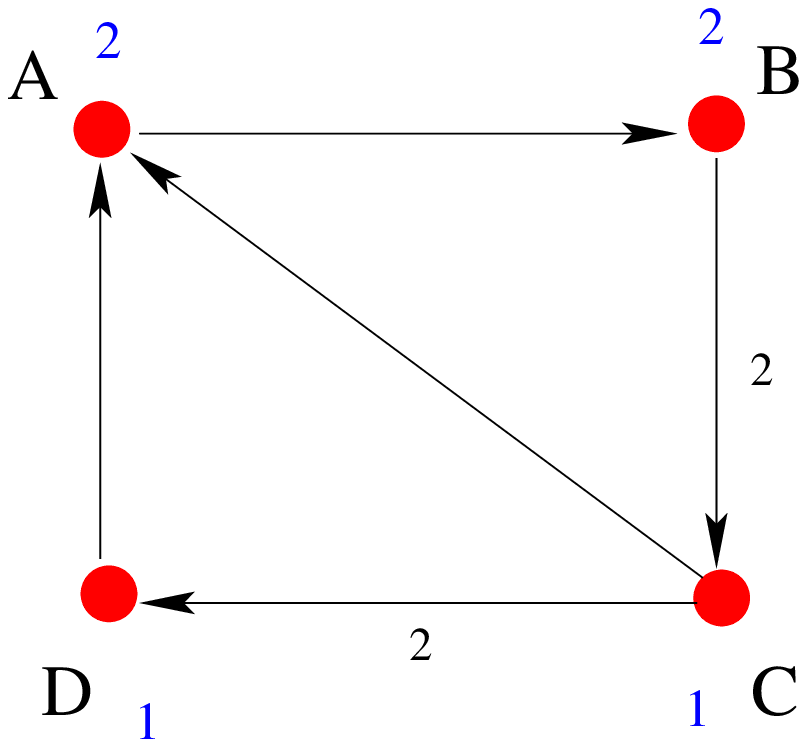,width=4.0in}
{A non-Abelian phase of the complex cone over $dP_3$. This example is
Picard-Lefschetz dual to the other 4 examples in \fref{f:dP3seiberg} 
but not Seiberg dual thereto.
\label{f:NAdP3}
}
\section{Conclusions}
In \cite{0003085,0104259} a mysterious duality between classes of
gauge theories on D-branes probing toric singularities was
observed. Such a Toric Duality identifies the infrared moduli space of
very different theories which are candidates 
for the world-volume theory on D3-branes at threefold singularities. 
On the other hand, \cite{uranga,Unge} have recognised certain
brane-moves for brane configurations of certain toric singularities as
Seiberg duality.

In this chapter we take a unified view to the above. Indeed we
have provided a physical interpretation for toric duality.
The fact that the gauge theories share by definition the same moduli space 
motivates the proposal that they are indeed physically equivalent in the 
infrared. In fact, we have shown in detail that toric dual gauge theories 
are connected by Seiberg duality.

This task has been facilitated by the use of T-dual configurations of NS 
and D-branes, in particular brane intervals and brane diamonds
\cite{Aganagic}. These 
constructions show that the Seiberg duality corresponds in the singularity 
picture to a change of B-fields in the collapsed cycles. Hence, the 
specific gauge theory arising on D3-branes at a given singularity, depends 
not only on the geometry of the singularity, but also on the B-field data. 
Seiberg duality and brane diamonds provide us with the tools to move 
around this more difficult piece of the singular moduli space, and 
probe different phases.

This viewpoint is nicely connected with that in \cite{0003085,0104259}, 
where toric duals were obtained as different partial resolutions of a 
given orbifold singularity, $\IC^3/(\IZ_3\times \IZ_3)$, leading to 
equivalent geometries (with toric diagrams equivalent up to unimodular 
transformations). Specifically, the original orbifold singularity has a 
specific assignments of B-fields on its collapsed cycles. Different 
partial resolutions amount to choosing a subset of such cycles, and 
blowing up the rest. Hence, in general different partial resolutions 
leading to the same geometric singularity end up with different 
assignments of B-fields. This explains why different gauge theories, 
related by Seiberg duality, arise by different partial resolutions.

In particular we have examined in detail the toric dual theories for
the generalised conifold $xy = z^2w^2$, the partial resolutions of
$\IC^3/(\IZ_3\times \IZ_3)$ exemplified by the complex cones over the
zeroth Hirzebruch surface as well as the second del Pezzo surface.
We have shown how these theories are equivalent under the above scheme
by explicitly having 
\begin{enumerate}
\item unimodularly equivalent toric data;
\item the matter content and superpotential related by Seiberg duality;
\item the T-dual brane setups related by brane-crossing and diamond
	duality.
\end{enumerate}
The point d'appui of this work is to show that the above three phenomena
are the same.

As a nice bonus, the physical understanding of toric duality 
has allowed us to construct 
new toric duals in cases where the partial resolution technique provided 
only one phase. Indeed the exponential running-time of the Inverse
Algorithm currently prohibits larger embeddings and partial
resolutions. Our new perspective greatly facilitates the calculation
of new phases. As an example we have constructed three new phases for
the cone over del Pezzo three one of which is in reassuring agreement
with a recent work \cite{HI} obtained from completely different methods.

Another important direction is to understand the physical meaning of 
Picard-Lefschetz transformations. 
As we have pointed out in Section 7, PL transformation and
Seiberg duality are really two different concepts even though they
coincide for certain restricted classes of theories. 
We have provided examples of two theories which are related by one but
not the other. Indeed we must pause to question ourselves. For those
which are Seiberg dual but not PL related, what geometrical action
does correspond to the field theory transformation. On the other 
hand, perhaps more importantly, for those
related to each other by PL transformation but not by
Seiberg duality, what kind of duality is realized in the 
dynamics of field theory? Does there exists a new kind of dynamical
duality not yet uncovered??

\chapter{Appendices}
%
%
\index{Finite Groups!$SU(2)$ subgroups!characters}
\section{Character Tables for the Discrete Subgroups of $SU(2)$}
\label{append:9811183.I}
{ H}enceforth we shall use $\Gamma_i$ to index the
representations and the numbers 
in the first row of the character tables shall refer to the order of each conjugacy 
class, or what we called $r_\gamma$.

\noindent
{\tiny
{\large $\widehat{A}_{n}=$ Cyclic ${\IZ}_{n+1}$ \\}
\[

\]

\noindent
{\normalsize
\\
For reference, next to each of the binary groups,
we shall also include the character table of the corresponding ordinary cases,
which are in $SU(2)/\IZ_2$.\\
}

\noindent
{\large $\widehat{D}_{n}$ = Binary Dihedral}
\[
\begin{array}{|c|c|c|c|c|c|c|c|c|}
\hline
	   & 1 & 1  & 2 & \cdots  & 2 & n & n \\ \hline
\Gamma_{1} & 1 & 1  & 1 & \cdots  & 1 & 1 & 1 \\ \hline
\Gamma_{2} & 1 & 1  & 1 & \cdots  & 1 & -1 & -1 \\ \hline
\Gamma_{3} & 1 & (-1)^n & (-1)^1  & \cdots  & (-1)^{n-1} & i^n & -i^n \\ \hline
\Gamma_{4} & 1 & (-1)^n & (-1)^1  & \cdots  & (-1)^{n-1} & -i^n  & i^n \\ \hline
\Gamma_{5} & 2 & (-2)^1 & 2\cos \frac{\pi}{n}  & \cdots  & 
	2\cos\frac{\pi(n-1)}{n}  & 0 & 0 \\ \hline
	\vdots &\vdots  &\vdots  &\cdots  &\vdots  &\vdots  &\vdots  &\vdots \\ \hline
\Gamma_{n+1} & 2 & (-2)^{n-1} & 2\cos\frac{\pi(n-1)}{n}  &
	\cdots  & 2\cos\frac{\pi (n-1)^{2}}{n}  & 0 & 0 \\ \hline
\end{array}%
\]
}
\section{Matter Content for ${\cal N}=2$ SUSY Gauge
\label{append:9811183.II}
Theory ($\Gamma \subset SU(2)$)}
Only the fermionic matrices are presented here; as can be seen from the
decomposition, twice the fermion $a_{ij}$ subtracted by $2\delta_{ij}$ 
should give the bosonic counterparts, which follows from supersymmetry.
In the ensuing, {\bf 1} shall denote the (trivial) principal representation,
${\bf 1^{'}}$ and ${\bf 1^{''}}$, dual (conjugate) pairs of 1 dimensional
representations.

{\tiny

\[
%
\]
}

\setlength{\textwidth}{8.0in}           
\oddsidemargin -0.5in
\evensidemargin -0.5in
\index{Finite Groups!$SU(3)$ subgroups!characters}
\section{Classification of Discrete Subgroups of $SU(3)$}
\label{append:9811183.III}
\subsection* {Type I: The $\Sigma$ Series}
These are the analogues of the $SU(2)$ crystallographic groups and their 
double covers, i.e., the E series. We have:
\[
\begin{array}{ccc}
\Sigma_{36}, \Sigma_{72}, \Sigma_{216}, \Sigma_{60}, \Sigma_{168},
\Sigma_{360} & \subset & SU(3) / (\IZ_3 $ center$) \\
\Sigma_{36 \times 3}, \Sigma_{72 \times 3}, \Sigma_{216 \times 3}, 
\Sigma_{60 \times 3}, \Sigma_{168 \times 3}, \Sigma_{360 \times 3} & \subset & SU(3)
\end{array}
\]

\subsubsection*{Type Ia: $\Sigma \subset SU(3) / \IZ_3 $}
The character tables for the center-removed case have been given by 
\cite{Fairbairn}.
{\tiny

\[
%
\]
}

\subsubsection*{Type Ib: $\Sigma \subset $ full $SU(3)$}
The character tables are computed, using \cite {Prog}, from the generators
presented in \cite {Yau}. In what follows, we define 
$e_n = \exp{\frac{2 \pi i}{n}}$.

{\tiny
\[
%
\]
}

\subsubsection*{Type Ic: $\Sigma \subset$ both $SU(3)$ and $SU(3)/\IZ_3$}

{\tiny
\[
%
\]
}

\subsection*{Type II: The $\Delta$ series}
These are the analogues of the dihedral subgroups of $SU(2)$ (i.e., the
D series).\\
{\large $\Delta_{3n^2}$}
\[
\begin{array}{cccc}
		  	& $Number of classes$	& $Subgroup of$ & $Some Irreps$ \\ \hline
n = 0 $ mod $ 3 	& 8 + \frac{1}{3}n^2   	& $Full $ SU(3)  
			& 9  $ {\bf 1}'s$ \frac13n^2-1 $ {\bf 3}'s$ \\
n \ne 0 $ mod $ 3 	& \frac{1}{3}(8 + n^2)   & $Full $ SU(3) $ and $ SU(3)/\IZ_3  
			& 3 $ {\bf 1}'s$, \frac{1}{3}(n^2 -1) $ {\bf 3}'s$	\\
\end{array}%
\]
\section{Matter content for $\Gamma \subset SU(3)$}
\label{append:9811183.IV}
Note here that since the ${\cal N} = 1$ theory is chiral, the fermion
matter matrix need not be symmetric. A graphic representation for some of these
theories appear in figures 3, 4 and 5.

{\tiny

\[
%
\]
}
%
%
\section{Steinberg's Proof of Semi-Definity}
\label{append:9911114}
We here transcribe Steinberg's proof of the semi-definity of the scalar product
with respect to the generalised Cartan matrix, in the vector space
$V = \{x_i \in \Z_+ \}$ of labels \cite{Steinberg}.
Our starting point is (\ref{aij}), which we re-write here as
\[
	r_d \otimes r_i = \bigoplus\limits_{j} a_{ij} r_j
\]
First we note that, if $\bar{i}$ is the dual representation to $i$,
then $a_{ij} = a_{\bar{j}\bar{i}}$ by taking the conjugates (dual)
of both sides of (\ref{aij}). Whence we have
\begin{lemma}
\label{app:dimaij} For $d_i = \dim{r_i}$,
	$d d_i = \sum\limits_j a_{ij} d_j = \sum\limits_j a_{ji} d_j$.
\end{lemma}
The first equality is obtained directly by taking the dimension of
both sides of (\ref{aij}) as in (\ref{dimaij}). To see the second
we have $d d_i = d d_{\bar{i}}$ (as dual representations have the
same dimension) which is thus equal to $\sum\limits_j a_{\bar{i}j} d_j$,
and then by the dual property $a_{ij} = a_{\bar{j}\bar{i}}$ above
becomes $\sum\limits_{\bar{j}} a_{\bar{j}i} d_{\bar{j}} =
\sum\limits_j a_{ji} d_j$. QED.

Now consider the following for the scalar product:
\begin{eqnarray*}
2 \sum_{ij} c_{ij} x_i x_j & = & 2 \sum_{ij} (d \delta_{ij} - a_{ij}) x_i x_j
		= 2 (d \sum_i x_i^2 - \sum_{ij} a_{ij} x_i x_j)	\\
	& = & 2 (\sum_i (d - a_{ii}) x_i^2 - \sum_{i \ne j} a_{ij} x_i x_j) \\
	& = & 2 \sum_i \frac{1}{2} (\frac{1}{d_i} \sum_j a_{ij} d_j +
		\frac{1}{d_i} \sum_j a_{ji} d_j - a_{ii}) x_i^2 -
		\sum_{i \ne j} a_{ij} x_i x_j)~~\mbox{(by Lemma \ref{app:dimaij})} \\
	& = & \sum_{i \ne j} (a_{ij} + a_{ji}) \frac{d_j}{d_i} x_i^2 - 2 a_{ij} x_i x_j
		= \sum_{i < j} (a_{ij} + a_{ji}) (\frac{d_j}{d_i} x_i^2 
		+ \frac{d_i}{d_j} x_j^2- 2 x_i x_j) \\
	& = & \sum_{i < j} (a_{ij} + a_{ji}) \frac{(d_j x_i - d_i x_j)^2}{d_i d_j} \ge 0
\end{eqnarray*}
From which we conclude
\begin{proposition} {\rm (Steinberg)}
	In the vector space of positive labels, the scalar product is positive 
	semi-definite, i.e., $\sum\limits_{ij} c_{ij} x_i x_j \ge 0$.
\end{proposition}
%
%
\index{Finite Groups!$SU(3)$ subgroups!$Z$-$D$ type}
\section{Conjugacy Classes for $Z_k \times D_{k'}$}
\label{append:9906031}
Using the notation introduced in \sref{sec:group}, we see that the 
conjugation within $G$ gives
\begin{equation}
(q,\tilde{q},\tilde{n},k)^{-1}(m,\tilde{m},n,p) (q,\tilde{q},\tilde{n},k)=
\left\{  
\begin{array}{l}
  (\tilde{m}+q-\tilde{q},m-q+\tilde{q},n,2k-p) $ for $ n=0,\tilde{n}=0 \\
  (m-q+\tilde{q}, \tilde{m}+q-\tilde{q},n,2k+p) $ for $ n=0,\tilde{n}=1 \\
 (\tilde{m},m,n,-p) $ for $ n=1,\tilde{n}=0  \\
  (m,\tilde{m},n,p) $ for $ n=1,\tilde{n}=1.
        \end{array}
\right.
\label{conj}
\end{equation}
Also, we present the multiplication rules in $G$ for reference:
\begin{eqnarray}
(m,\tilde{m},0,p_1)(n,\tilde{n},0,p_2) & = & (m+\tilde{n},\tilde{m}+n,1,p_2-p_1)
\nonumber \\
(m,\tilde{m},0,p_1)(n,\tilde{n},1,p_2) & = & (m+\tilde{n},\tilde{m}+n,0,p_2+p_1-k')
\nonumber \\
(m,\tilde{m},1,p_1)(n,\tilde{n},0,p_2) & = & (m+n,\tilde{m}+\tilde{n},0,p_2-p_1-k')
\nonumber \\
(m,\tilde{m},1,p_1)(n,\tilde{n},1,p_2) & = & (m+n,\tilde{m}+\tilde{n},1,p_2+p_1-k')
\label{multi}
\end{eqnarray}

First we focus on the conjugacy class of elements such that $n=0$.
From (\ref{orbit}) and (\ref{conj}), we see that if two 
elements are within the same conjugacy class, then they must have the same 
$m+\tilde{m} \bmod k$.
Now we need to distinguish between two cases:
\begin{itemize}
\item (I) if $\frac{2k'}{(k,2k')}=\even$, the orbit conditions conserve 
the parity of $p$, making even and odd $p$ belong to different conjugacy classes;
\item (II) if $\frac{2k'}{(k,2k')}=\odd$, the orbit conditions change $p$ and 
we find that all $p$ belong to the same conjugacy class 
they have the same value for $m+\tilde{m}$.
\end{itemize}
In summary then, for $\frac{2k'}{(k,2k')}=\even$, we have $2k$ 
conjugacy classes each of which has $\frac{k'k}{(k,2k')}$ elements;
for $\frac{2k'}{(k,2k')}=\odd$, we have $k$ 
conjugacy classes each of which has $\frac{2k'k}{(k,2k')}$ elements. 

Next we analyse the conjugacy class corresponding to $n=1$.
For simplicity, we divide the interval $[0,k)$ by factor $(k,2k')$ and define 
\[
V_{i} = \left[\frac{ik}{(k,2k')},\frac{(i+1)k}{(k,2k')}\right)
\]
with $i=0,...,(k,2k')-1$. 
Now from (\ref{orbit}), we can always fix $m$ to belong $V_{0}$. 
Thereafter, $\tilde{m}$ and $p$ can change freely within $[0,k)/[0,2k')$. 
Again, we have two different cases.
(I) If $\frac{2k'}{(k,2k')}=\even$, for every subinterval $V_{i}$ 
we have $2k_{0}$ (we define $k_0:=2\frac{k}{(k,2k')}$) conjugacy classes each containing
only one element, namely, 
\[
(m,\tilde{m}=m+\frac{ik}{(k,2k')},n=1,p=k'-\frac{ik'}{(k,2k')} {\rm ~or~}
2k'-\frac{ik'}{(k,2k')}).
\]
Also we have a total of 
$k_{0}\frac{2k'-2}{2}+\frac{k_{0}(k_{0}-1)}{2}
2k'=k_{0}(k'-1)+k'k_{0}^2+k_{0}k'=k'k_{0}^2-k_{0}$
conjugacy classes of 2 elements, namely
$(m,\tilde{m},n=1,p)$ and
$(\tilde{m}-\frac{ik}{(k,2k')},m+\frac{ik}{(k,2k')},n=1,-p-i\frac{2k'}{(k,2k')})$.
Indeed, the total number of conjugacy classes is 
$2k+(k,2k')(2k_{0})+(k,2k')(k'k_{0}^2-k_{0})=4k+k(\frac{k'k}{(k,2k')}-1)$, giving
the order of $G$ as expected.
Furthermore, there are $4k$ 1-dimensional irreducible representations 
and $k(\frac{k'k}{(k,2k')}-1)$ 2-dimensional irreducible representations. This is consistent since
$\sum_i\dim{\bf r}_i = 1^2\cdot 4k+2^2\cdot k(\frac{k'k}{(k,2k')}-1)=
\frac{4k'k^2}{(k,2k')} = |G|$.

We summarize case (I) into the following table:
\[
\begin{array}{c|c|c|c}
      	& C_{n=0}^{m+\tilde{m}(\bmod k),p=\odd / \even}  &
        C_{n=1,V_{i}}^{\tilde{m}=m+\frac{ik}{(k,2k')},p=
			(k'-\frac{ik'}{(k,2k')}) /
        		(2k'-\frac{ik'}{(k,2k')})} 
	& C_{n=1,V_{i}}^{(m,\tilde{m},p)=
		(\tilde{m}-\frac{ik}{(k,2k')},
		m+\frac{ik}{(k,2k')},-p-i\frac{2k'}{(k,2k')})} \\ \hline
|C| & \frac{k'k}{(k,2k')}  & 1 & 2 \\ \hline
\#C &  2k  & 2k & k(\frac{k'k}{(k,2k')}-1)
\end{array}
\]

Now let us treat case (II), where $\frac{2k'}{(k,2k')}$ is odd (note that
in this case we must have $k$ even). Here, for $V_{i}$ and $i$ even, the situation
is as (I) but for $i$ odd there are no one-element conjugacy classes. We tabulate the
conjugacy classes in the following:
\[
\begin{array}{c|c|c|c}
	& C_{n=0}^{m+\tilde{m}(\bmod k),\rm{any~}p}  &
        C_{n=1,V_{i},i=\even}^{\tilde{m}=m+\frac{ik}{(k,2k')},
		p=(k'-\frac{ik'}{(k,2k')})
        	/(2k'-\frac{ik'}{(k,2k')})}
	& C_{n=1,V_{i}}^{(m,\tilde{m},p)=
		(\tilde{m}-\frac{ik}{(k,2k')},
		m+\frac{ik}{(k,2k')},-p-i\frac{2k'}{(k,2k')})} \\ \hline
|C| & \frac{2k'k}{(k,2k')}  & 1 & 2 \\ \hline
\#C &  k  & 2\frac{k}{2}=k & \frac{(k,2k')}{2}[(k'k_{0}^2-k_{0})+k'k_{0}^2]=
k(\frac{k'k}{(k,2k')}-\frac{1}{2})
\end{array}
\]
\index{Finite Groups!Schur multiplier}
\section{Some Explicit Computations for $M(G)$}
\label{append:0010023.A}
\subsection{Preliminary Definitions}
We begin with a few rudimentary definitions \cite{Karp}.
Let $H$ be a subgroup of $G$ and let $g \in G$. For any cocycle
$\alpha \in  Z^2(G,\IC^*)$ we define an induced action
$g \cdot \alpha \in Z^2(g H g^{-1}, \IC^*)$ thereon as
$g \cdot \alpha(x,y) = \alpha(g^{-1} x g, g^{-1} y g),~~\forall~~x,y 
\in gHg^{-1}$.
Now, it can be proved that the mapping
\[
c_{g} : M(H) \rightarrow M(gHg^{-1}), \;\;\; c_{g}(\alpha) := g \cdot \alpha
\]
is a homomorphism, which we call {\bf cocycle conjugation} by $g$.

On the other hand we have an obvious concept of restriction:
for $S \subseteq L$ subgroups of $G$, we denote by  Res$_{L,S}$ the
restriction map $M(L) \rightarrow M(S)$. Thereafter we define
stability as:
\begin{definition}
Let $H$ and $K$ be arbitrary subgroups of $G$. An element
$\alpha \in M(H)$ is said to be {\bf K-stable} if
\[
{\rm Res}_{H,gHg^{-1}\cap H}(\alpha) = 
{\rm Res}_{gHg^{-1},gHg^{-1}\cap H}(c_{g}(\alpha)) ~~\forall~~ g \in K.
\] 
\end{definition}
The set of all K-stable elements of $M(H)$ will be denoted
by $M(H)^{K}$ and it forms a subgroup of $M(H)$ known
as the K-stable subgroup of $M(H)$.

When $K \subseteq N_{G}(H)$ all the above concepts\footnote{$N_{G}(H)$
	is the normalizer of $H$ in $G$, i.e., the set of all elements
	$g \in G$ such that $g H g^{-1} = H$. When $H$ is a normal
	subgroup of $G$ we obviously have $N_{G}(H)=G$.}
	coalesce and we have the following important lemma:
\begin{lemma}
\label{lemma_stab}{\rm (\cite{Karp} p299)}
If $H$ and $K$ are subgroups of $G$ such that $K \subseteq N_{G}(H)$,
then $M(H)^{K}$ is the K-stable subgroup of $M(H)$ with respect to
the action of $K$ on $M(H)$ induced by the action of $K$ on $H$
by conjugation. In other words,
\[
M(H)^{K}=\{\alpha \in M(H), \;\;   \alpha(x,y)=c_{g}(\alpha)(x,y)
~~\forall~~ g \in K, ~~\forall~~ x,y \in H\}.
\]
\end{lemma}

Finally let us present a useful class of subgroups:
\begin{definition}
A subgroup $H$ of a group $G$ is called a {\bf Hall subgroup} of $G$
if the order of $H$ is coprime with its index in $G$,
i.e. $\gcd(|H|,|G/H|)=1$.
\end{definition}
For these subgroups we have:
\begin{theorem}{\rm (\cite{Karp} p334)}
\label{normalhall}
If $N$ is a normal Hall subgroup of $G$. Then
\[
M(G) \cong M(N)^{G/N} \times M(G/N).
\]
\end{theorem}

The above theorem is really a corollary of a more general case of
semi-direct products:
\begin{theorem}\label{sequence}{\rm (\cite{Karp3} p33)}
Let $G = N \rtimes T$ with $N \triangleleft G$, then \\
$(i) \quad M(G) \cong M(T) \times \tilde{M}(G)$; \\
$(ii) \quad \mbox{The sequence }
	1 \rightarrow H^1(T,N^*) \rightarrow \tilde{M}(G)
	{\stackrel{{\rm Res}}{\rightarrow}} M(N)^T \rightarrow H^2(T,N^*)
	\mbox{ is exact,}$\\
where $\tilde{M}(G) := \ker{\rm Res}_{G,N}$, $N^* := \Hom(N,\IC^*)$ and
$H^{i=1,2}(T,N^*)$ is the cohomology defined with respect to the
conjugation action by $T$ on $N^*$.
\end{theorem}
Part (ii) of this theorem actually follows from the
Lyndon-Hochschild-Serre spectral sequence into which we shall not delve.

One clarification is needed at hand. Let us define the first
$A$-valued cohomology group for $G$, which we shall utilise later in our
calculations. Here the 1-cocycles are the set of functions $Z^1(G,A) :=
\{f:G\rightarrow A|f(xy) = (x\cdot f(y))f(x)\quad\forall x,y \in G\}$,
where $A$ is being acted upon ($x \cdot A \rightarrow A$ for $x \in G$) 
by $G$ as a $\IZ G$-module. These are known as {\em crossed
homomorphisms}. On the other hand, the 1-coboundaries are what is
known as the principal crossed homomorphisms, $B^1(G,A) := \{f_{a \in
A}(x) = (x \cdot a) a^{-1} \}$ from which we define $H^1(G,A) :=
Z^1(G,A) / B^1(G,A)$.

Alas, {\it caveat emptor}, we have defined in subsection 2.2,
$H^2(G,A)$. There, the action of $G$ on $A$ (as in the case of the Schur
Multiplier) is taken to be trivial, we must be careful, in
the ensuing, to compute with respect to non-trivial actions such as
conjugation. In our case the conjugation action of $t \in T$ on
$\chi \in \Hom(N,\IC^*)$ is given by $\chi(t n t^{-1})$ for $n\in N$.
\subsection{The Schur Multiplier for $\Delta_{3n^2}$}
\subsubsection{Case I: $\gcd(n,3)=1$}
Thus equipped, we can now use theorem \ref{normalhall} at our ease to
compute the Schur multipliers the first case of
the finite groups $\Delta_{3 n^2}$. Recall that
$\IZ_{n} \times \IZ_{n} \triangleleft \Delta(3n^2)$ or explicitly
\[
\Delta_{3 n^2} \cong (\IZ_{n} \times \IZ_{n}) \rtimes \IZ_{3}.
\]
Our crucial observation is that when $\gcd(n,3)=1$, $\IZ_{n} \times
\IZ_{n}$ is in fact a normal Hall subgroup of $\Delta_{3 n^2}$ with
quotient group $\IZ_{3}$. Whence Theorem \ref{normalhall} can be
immediately applied to this case when  $n$ is coprime to 3:
\[
M(\Delta_{3 n^2})=(M(\IZ_n\times\IZ_n))^{\IZ_3} \times M(\IZ_3) 
		= (M(\IZ_n\times\IZ_n))^{\IZ_3},
\]
by recalling that the Schur Multiplier of all cyclic groups is
trivial and that of $\IZ_n\times\IZ_n$ is $\IZ_n$ \cite{Karp}.
But, $\IZ_3 \subseteq N_{\Delta_{3 n^2}}(\IZ_{n} \times
\IZ_{n})=\Delta_{3 n^2}$, and hence by Lemma \ref{lemma_stab} it suffices
to compute the $\IZ_3$-stable subgroup of $\IZ_n$ by cocycle conjugation.

Let the quotient group $\IZ_3$ be $\gen{z|z^3=\II}$ and similarly, if
$x, y, x^n=y^n=\II$ are the generators of $\IZ_{n} \times \IZ_{n}$,
then a generic element thereof becomes $x^a y^b, a,b = 0,\ldots,n-1$.
The group conjugation by $z$ on such an element gives
\beq\label{conj-tor}
z^{-1} x^a y^b z = x^b y^{-a-b} \qquad z x^a y^b z^{-1} = x^{-a-b} y^a.
\eeq
It is easy now to check that if $\alpha$ is
a generator of the Schur multiplier $\IZ_{n}$, we have an induced action
\[
c_{z}(\alpha)(x^a y^b,x^{a'} y^{b'}):=
\alpha(z^{-1} x^a y^b z,z^{-1} x^{a'} y^{b'} z)=
\alpha(x^b y^{-(a+b)}, x^{b'} y^{-(a'+b')})
\]
by Lemma \ref{lemma_stab}.

However, we have a well-known result \cite{Klein}:
\begin{proposition}\label{zeta}
For the group $\IZ_{n} \times \IZ_{n}$, the explicit generator of the
Schur Multiplier is given by
\[
\alpha(x^a y^b, x^{a'} y^{b'})=\omega_{n}^{a b' - a' b}.
\]
\end{proposition}
Consequently, $\alpha(x^b y^{-(a+b)}, x^{b'} y^{-(a'+b')})=
\alpha(x^a y^b,x^{a'} y^{b'})$ whereby making the $c_z$-action trivial
and causing $(M(\IZ_n \times \IZ_n)^{\IZ_3} \cong M(\IZ_n \times
\IZ_n) = \IZ_n$. From this we conclude part
I of our result: $M(\Delta_{3 n^2}) = \IZ_{n}$ for
$n$ coprime to 3.
\subsubsection{Case II: $\gcd(n,3) \neq 1$}
Here the situation is much more involved. Let us appeal to Part (ii)
of Theorem \ref{sequence}. We let $N = \IZ_n \times \IZ_n$ and $T =
\IZ_3$ as above and define $U := \Hom(\IZ_n \times \IZ_n,\IC^*))$;
the exact sequence then takes the form
\beq\label{seq2}
1 \rightarrow H^1(\IZ_3,U) \rightarrow \tilde{M}(\Delta_{3n^2}) \rightarrow \IZ_n 
 \rightarrow H^2(\IZ_3,U)
\eeq
using the fact that the stable subgroup $M(\IZ_n \times
\IZ_n)^{\IZ_3} \cong \IZ_n$ as shown above.
Some explicit calculations are now called for.

As for $U$, it is of course isomorphic to $\IZ_n \times \IZ_n$ since
for an Abelian group $A$, $\Hom(A,\IC^*) \cong A$ (\cite{Karp3}
p17). We label the elements thereof as $(p,q)(x^a y^b) := \omega_n^{a
p + b q}$, taking $x^a y^b \in \IZ_n \times \IZ_n$ to $\IC^*$.

We recall that the conjugation by $z \in \IZ_3$ on $\IZ_n \times
\IZ_n$ is \eref{conj-tor}. Therefore, by the remark at the end of the
previous subsection, $z$ acts on $U$ as: $(z \cdot (p,q))(x^a y^b) 
:= (p,q)(z(x^a
y^b)z^{-1}) = \omega_n^{a' p + b' q}$ with $a' = -a-b$ and $b' = a$
due\footnote{Note that we must be careful to let the order of conjugation
	be the opposite of that in the cocycle conjugation.} 
to \eref{conj-tor}, whence 
\beq\label{conjZ3}
z \cdot (p,q) = (q-p,-p),\quad\mbox{ for }(p,q) \in U. 
\eeq

Some explicit calculations are called for.
First we compute $H^1(\IZ_3,U)$. $Z^1$ is
generically composed of functions such that $f(z) = (p,q)$
(and also $f(\II) = \II$ and $f(z^2) = (z \cdot f(z)) f(z)$ by the
crossed homomorphism condition, and is subsequently equal to
$(q,p+q)$ by \eref{conjZ3}. Since no
further conditions can be imposed, $Z^1 \cong \IZ_n \times \IZ_n$.
Now $B^1$ consists of all functions of the form  $(z \cdot
(p,q))(p,q)^{-1} = (q-2p,-p-q)$, these are to be identified
with the trivial map in $Z^1$. We can re-write these elements as 
$(p':=q-2p,-p'-3p) = (\omega_n^a\omega_n^{-b})^{p'}(\omega_n^b)^{-3p}$,
and those in $Z^1$ we re-write as
$(\omega_n^a\omega_n^{-b})^{p'}(\omega_n^b)^{q'}$ as we are free to do.
Therefore if $\gcd(3,n)=1$, then 
$H^1 := Z^1/B^1$ is actually trivial because in mod $n$, $3p$ also
ranges the full $0,\cdots, n-1$, whereas if $\gcd(3,n) \ne 1$ then 
$H^1 := Z^1/B^1 \cong \IZ_3$.

The computation for $H^2(\IZ_3,U)$ is a little more
involved, but the idea is the same. First we determine $Z^2$ as
composed of $\alpha(z_1,z_2)$ constrained by the cocycle condition (with
respect to conjugation which differs from \eref{alpha} where the trivial
action was taken)
\[
\alpha(z_1,z_2) \alpha(z_1 z_2,z_3) = (z_1 \cdot
\alpha(z_2,z_3)) \alpha(z_1,z_2 z_3) \qquad z_1,z_2,z_3 \in \IZ_3.
\] 
Again we
only need to determine the following cases: $\alpha(z,z) :=
(p_1,q_1); \alpha(z^2,z^2) := (p_2,q_2);$ $\alpha(z^2,z) :=
(p_3,q_3); \alpha(z,z^2) := (p_4,q_4)$. The cocycle
constraint gives $(p_1,q_1) = (q_4,-q_3);$ $(p_2,q_2) = (-q_3-q_4,-q_4);
(p_3,q_3) = (-q_4,q_3); (p_4,q_4) = (p_4,q_4)$, giving
$Z^2 \cong \IZ_n \times \IZ_n$.
The coboundaries are given by $(\delta t)(z_1,z_2) = (z_1 \cdot t(z_2))
t(z_1) t(z_1 z_2)^{-1}$ (for any mapping $t : \IZ_3 \rightarrow \IZ_n
\times \IZ_n$ which we define to take values $t(z) = (r_1,s_1)$ and
$t(z^2) = (r_2,s_2)$)), making $(\delta t)(z,z) = (s_1-r_2,-r_1+s_1-s_2);
(\delta t)(z^2,z^2) (-s_2+r_2-r_1,r_2-s_1);
(\delta t)(z^2,z) = (-s_1+r_2,r_1-s_1+s_2);
(\delta t)(z,z^2) = (s_2-r_2+r_1,s_1-r_2)$. Now, the transformation
$r_2 = s_1 + q_4; r_1 = s_1 - s_2 - p_4 + q_4$ makes this set of
values for $B^2$ completely identical to those in $Z^2$, whence we
conclude that $B^2 \cong \IZ_n \times \IZ_n$.
In conclusion then $H^2 := Z^2 / B^2 \cong \II$.

The exact sequence \eref{seq2} then assumes the simple form of
\[
1\rightarrow \left\{\ba{cc}\IZ_3, &\gcd(n,3)\neq 1 \\
	\II, & \gcd(n,3)=1\ea \right\}\rightarrow \tilde{M}(G)
 \rightarrow \IZ_n \rightarrow 1,
\]
which means that if $n$ does not
divide 3, $\tilde{M}(G) \cong \IZ_n$, and otherwise $\tilde{M}(G) /
\IZ_3 \cong \IZ_n$. Of course, in conjunction with Part (i) of Theorem
\ref{sequence}, we immediately see that the first case makes Part I of
our discussion (when $\gcd(n,3)= 1$) a special case of our present
situation.

On the other hand, for the remaining case of $\gcd(n,3)\neq 1$, we
have $M(\Delta_{3n^2})/\IZ_3 \cong \IZ_n$, which means that
$M(\Delta_{3n^2})$, being an Abelian group, can only be $\IZ_{3n}$ or
$\IZ_n \times \IZ_3$. The exponent of the former is $3n$, while the
later (since 3 divides $n$), is $n$, but by Theorem \ref{exponent}, the
exponent squared must divide the order, which is $3n^2$, whereby
forcing the second choice.

Therefore in conclusion we have our {\it theorema egregium}:
\[
M(\Delta_{3n^2}) = \left\{\ba{cc}\IZ_n \times \IZ_3, &\gcd(n,3)\neq 1 \\
	\IZ_n, & \gcd(n,3)=1 \ea \right.
\]
as reported in Table \eref{SU3}.
\subsection{The Schur Multiplier for $\Delta_{6n^2}$}
Recalling that $n$ is even, we have
$\Delta_{6n^2} \cong (\IZ_n \times \IZ_n) \rtimes
S_3$ with $\IZ_n \times \IZ_n$ normal and thus we are once more aided
by Theorem \ref{sequence}.

We let $N := \IZ_n \times \IZ_n$ and $T := S_3$ and the exact sequence
assumes the form
\[
1 \rightarrow H^1(S_3,U) \rightarrow \tilde{M}(\Delta_{6n^2})
\rightarrow (\IZ_n)^{S_3} \rightarrow H^2(S_3,U)
\]
where $U := \Hom(\IZ_n \times \IZ_n,\IC^*)$ as defined in the previous
subsection.

By calculations entirely analogous to the case for $\Delta_{3n^2}$, we
have $(\IZ_n)^{S_3} \cong \IZ_2$. This is straight-forward to show.
Let $S_3 := \gen{z,w | z^3 = w^2 = \II, zw = wz^2}$. We see that it
contains $\IZ^3 = \gen{z|z^3=\II}$ as a subgroup, which we have
treated in the previous section. In addition to \eref{conj}, we have
\[
w^{-1}x^a y^b w = x^{-1-b}y^b = w x^a y^b w^{-1}.
\]
Using the form of the cocycle in Proposition \ref{zeta}, we see that 
$c_w(\alpha) = \alpha^{-1}$. Remembering that $c_z(\alpha) = \alpha$
from before, we see that the $S_3$-stable part of consists of
$\alpha^m$ with $m=0$ and $n/2$ (recall that in our case of
$\Delta(6n^2)$, $n$ is even), giving us a $\IZ_2$.

Moreover we have $H^1(S_3,U) \cong \II$. This is again easy to show.
In analogy to \eref{conjZ3}, we have
\[
w \cdot (p,q) = (-q, q-p), \quad\mbox{ for }(p,q) \in U,
\]
using which we find that $Z^1$ consists of $f : S_3 \rightarrow U$
given by $f(z) = (l_1,3k_2-l_1)$ and $f(w) = (2k_2,k_2)$. In addition
$B^1$ consists of $f(z) = (k-2l,-l-k)$ and $f(w) = (-2l,-l)$. Whence
we see instantly that $H^1$ is trivial.
 
Now in fact $H^2(S_3,U) \cong \II$ as well (the involved details of these
computations are too pathological to be even included in an appendix
and we have resisted the urge to write an appendix for the appendix). 

The exact sequence then forces immediately that
$\tilde{M}(\Delta_{6n^2}) \cong \IZ_2$. Moreover, since $M(S_3) \cong
\II$ (q.v. e.g. \cite{Karp}), by Part (i) of Theorem
\ref{sequence}, we conclude that
\[
M(\Delta_{6n^2}) \cong \IZ_2
\]
as reported in Table \eref{SU3}.
\section{Intransitive subgroups of $SU(3)$}
\label{append:0010023.B}
The computation of the Schur Multipliers for the non-Abelian 
intransitive subgroups of $SU(3)$ involves some subtleties related to the
precise definition and construction of the groups.

Let us consider the case of combining the generators of $\IZ_n$ with
these of $\widehat{D_{2m}}$ to construct the intransitive subgroup
$<\IZ_n, \widehat{D_{2m}}>$. We can take the generators of
$\widehat{D_{2m}}$ to be 
\[
\label{matrix_1}
\alpha=\left( \begin{array}{ccc} \omega_{2m}  & 0 & 0 \\
0 & \omega_{2m}^{-1} & 0 \\ 0 & 0  & 1  \end{array} \right) ,~~~~
\beta=\left( \begin{array}{ccc} 0 & i & 0 \\ i & 0 & 0 \\ 
 0 & 0  & 1 \end{array} \right)
\]
and that of $\IZ_n$ to be
\[
\gamma=\left(  \begin{array}{ccc} \omega_n & 0 & 0 \\
0 & \omega_n & 0 \\ 0 & 0 & \omega_n^{-2}  \end{array} \right).
\]
 
The group $<\IZ_n, \widehat{D_{2m}}>$ is not in general the direct 
product of $\IZ_n$ and $\widehat{D_{2m}}$.
More specifically, when $n$ is odd $<\IZ_n, \widehat{D_{2m}}> = 
\IZ_n \times \widehat{D_{2m}}$. For $n$ even however, we notice that
$\alpha^m=\beta^2=\gamma^{n/2}$. Accordingly, we conclude that 
 $<\IZ_n, \widehat{D_{2m}}> = (\IZ_n \times \widehat{D_{2m}})/ \IZ_2$
for $n$ even where the central $\IZ_2$ is generated by $\gamma^{n/2}$.
Actually the conditions are more refined: 
when $n = 2(2k+1)$ we have $\IZ_n=\IZ_2 \times
\IZ_{2k+1}$ and so $(\IZ_2 \times \widehat{D_{2m}})/ \IZ_2 = 
\IZ_{2k+1} \times \widehat{D_{2m}}$.
Thus the only non-trivial case is when $n=4k$.

This subtlety in the group structure holds for all the cases where
$\IZ_n$ is combined with binary groups $\widehat{G}$. When $n~{\rm
mod}~4 \neq 0$, $<\IZ_n, \widehat{G}>$ is the direct product 
of $\widehat{G}$ with either $\IZ_n$ or $\IZ_{n/2}$. For $n~{\rm mod}~4 = 0$ 
it is the quotient group $(\IZ_n \times \widehat{G})/\IZ_2$.
In summary
\[
<\IZ_n, \widehat{G}> = \left\{\ba{lc} 
\IZ_n  \times \widehat{G} & n~{\rm mod}~2 = 1 \\
\IZ_{n/2}  \times \widehat{G} & n~{\rm mod}~4 = 2 \\
(\IZ_n \times \widehat{G})/\IZ_2  & n~{\rm mod}~4 = 0 
\ea \right. .
\]

The case of $\IZ_n$ combined with the ordinary dihedral group
$D_{2m}$ is a bit different however.
The matrix forms of the generators are 
\[
\label{matrix_2}
\alpha=\left( \begin{array}{ccc} \omega_{m}  & 0 & 0 \\
0 & \omega_{m}^{-1} & 0 \\ 0 & 0  & 1  \end{array} \right) ,~~~~
\beta=\left( \begin{array}{ccc} 0 & 1 & 0 \\ 1 & 0 & 0 \\ 
 0 & 0  & -1 \end{array} \right) ,~~~~
\gamma=\left( \begin{array}{ccc} \omega_n & 0 & 0 \\
0 & \omega_n & 0 \\ 0 & 0 & \omega_n^{-2}  \end{array} \right)
\]
where $\alpha$ and $\beta$ generate $D_{2m}$ and $\gamma$ generates
$\IZ_n$.

From these we notice that when both $n$ and $m$ are even,
$\alpha^{m/2}=\gamma^{n/2}$ and $<\IZ_n,D_{2m}>$ is not a direct
product. After inspection, we find that 
\[
<\IZ_n, D_{2m}> = \left\{\ba{lc} 
\IZ_n  \times D_{2m} & m~{\rm mod}~2 = 1 \\
\IZ_n  \times D_{2m} & m~{\rm mod}~2 = 0, n~{\rm mod}~2 = 1  \\
\IZ_{n/2} \times D_{2m} &  m~{\rm mod}~2 = 0, n~{\rm mod}~4 = 2 \\
(\IZ_n \times D_{2m})/\IZ_2  &  m~{\rm mod}~2 = 0, n~{\rm mod}~4 = 0
\ea \right. .
\]
The Schur Multipliers of the 
direct product cases are immediately computable by consulting Theorem
\ref{directprod}. For example,
$M(\IZ_n \times \widehat{D_{2m}}) \cong M(\IZ_n) \times M(\widehat{D_{2m}})
\times (\IZ_n \otimes \widehat{D_{2m}})$ by Theorem
\ref{directprod}, the last term of
which in turn equates to
$\Hom(\IZ_n,\widehat{D_{2m}}/\widehat{D_{2m}}')$. This is 
$\Hom(\IZ_n,\IZ_2 \times \IZ_2) \cong \IZ_{\gcd(n,2)} \times
\IZ_{\gcd(n,2)}$ for $m$ even and 
$\Hom(\IZ_n,\IZ_4) \cong \IZ_{\gcd(n,4)}$ for $m$ odd.
By similar token, we have that $M(\IZ_n \times D_{2m})$ for
even $m$ is $\IZ_2 \times \Hom(\IZ_n,\IZ_2 \times \IZ_2) \cong
\IZ_2 \times \IZ_{\gcd(n,2)} \times \IZ_{\gcd(n,2)}$
and $\Hom(\IZ_n,\IZ_2) \cong \IZ_{\gcd(n,2)}$ 
for odd $m$. Likewise $M(\IZ_n \times \widehat{E_{6,7,8}}) =
\Hom(\IZ_n,\IZ_{3,2,1})$.
\index{Finite Groups!$SU(3)$ subgroups!character}
\index{Finite Groups!$SU(3)$ subgroups!Schur multiplier}
\section{Ordinary and Projective Representations of Some Discrete
Subgroups of $SU(3)$}
\label{append:0011192}
We here present, for the reference of the reader, the (ordinary)
character tables of the groups as well as the covering groups thereof,
of the examples which we studied in Section 4 of Chapter \ref{chap:dis2}.
\vspace{2.0cm}
\[
{\tiny
\Sigma(60) \quad
\ba{|c|c|c|c|c|}
\hline
1 & 12 & 12 & 15 & 20 \\ \hline 1 & 1 & 1 & 1 & 1 \\ \hline 3 & -
\omega_5^2 - \omega_5^{-2} & -\omega_5 - \omega_5^{-1} & -1 & 0 \\ \hline 3 & -\omega_5 - 
    \omega_5^{-1} & -\omega_5^2 - \omega_5^{-2} & -1 & 0 \\ \hline 4 & 
     -1 & -1 & 0 & 1 \\ \hline 5 & 0 & 0 & 1 & -1 \\ \hline
\ea
\qquad
\Sigma(60)^* \quad
\ba{|c|c|c|c|c|c|c|c|c|}
\hline
    1 & 1 & 12 & 12 & 12 & 12 & 30 & 20 & 20 \\ \hline 1 & 1 & 1 & 1 & 1 & 1 & 1 & 
    1 & 1 \\ \hline 3 & 3 & -\omega_5^2 - \omega_5^{-2} & -\omega_5^2 - 
    \omega_5^{-2} & -\omega_5 - \omega_5^{-1} & -\omega_5 - \omega_5^{-1} & 
     -1 & 0 & 0 \\ \hline 3 & 3 & -\omega_5 - \omega_5^{-1} & -\omega_5 - \omega_5^{-1} & -
      \omega_5^2 - \omega_5^{-2} & -\omega_5^2 - 
    \omega_5^{-2} & -1 & 0 & 0 \\ \hline 4 & 4 & -1 & -1 & -1 & 
     -1 & 0 & 1 & 1 \\ \hline 5 & 5 & 0 & 0 & 0 & 0 & 1 & -1 & -1 \\ \hline 2 & -2 & 
      -\omega_5^2 - \omega_5^{-2} & \omega_5^2 + 
    \omega_5^{-2} & -\omega_5 - \omega_5^{-1} & \omega_5 + \omega_5^{-1} & 0 & 
     -1 & 1 \\ \hline 2 & -2 & -\omega_5 - \omega_5^{-1} & \omega_5 + \omega_5^{-1} & -
      \omega_5^2 - \omega_5^{-2} & \omega_5^2 + 
    \omega_5^{-2} & 0 & -1 & 1 \\ \hline 4 & -4 & 1 & -1 & 1 & -1 & 0 & 1 & 
     -1 \\ \hline 6 & -6 & -1 & 1 & -1 & 1 & 0 & 0 & 0 \\ \hline
\ea
}
\]

\[
{\tiny
\Sigma(168) \quad
\ba{|c|c|c|c|c|c|}
\hline
1 & 21 & 42 & 56 & 24 & 24 \\ \hline 1 & 1 & 1 & 1 & 1 & 1 \\ \hline 3 & 
     -1 & 1 & 0 & a & \bar{a} \\ \hline 3 & -1 & 1 & 0 & \bar{a}
     & a \\ \hline 6 & 2 & 0 & 0 & -1 & -1 \\ \hline 7 & -1 & 
     -1 & 1 & 0 & 0 \\ \hline 8 & 0 & 0 & -1 & 1 & 1 \\ \hline
\ea
\qquad
\Sigma(168)^* \quad
\ba{|c|c|c|c|c|c|c|c|c|c|c|}
\hline
    1 & 1 & 42 & 42 & 42 & 56 & 56 & 24 & 24 & 24 & 24 \\ \hline 1 & 
    1 & 1 & 1 & 1 & 1 & 1 & 1 & 1 & 1 & 1 \\ \hline 3 & 3 & 
     -1 & 1 & 1 & 0 & 0 & a & a & \bar{a} & \bar{a}
     \\ \hline 3 & 3 & -1 & 1 & 1 & 0 & 0 & \bar{a} & \bar{a}
     & a & a \\ \hline 6 & 6 & 2 & 0 & 0 & 0 & 0 & -1 & -1 & -1 & 
     -1 \\ \hline 7 & 7 & -1 & -1 & 
     -1 & 1 & 1 & 0 & 0 & 0 & 0 \\ \hline 8 & 8 & 0 & 0 & 0 & -1 & 
     -1 & 1 & 1 & 1 & 1 \\ \hline 4 & -4 & 0 & 0 & 0 & 1 & -1 & 
     -a & a & -\bar{a} & \bar{a} \\ \hline 4 & 
     -4 & 0 & 0 & 0 & 1 & -1 & -\bar{a} & \bar{a} & 
     -a & a \\ \hline 6 & -6 & 0 & -{\sqrt{2}} & {\sqrt{2}} & 0 & 0 & 
     -1 & 1 & -1 & 1 \\ \hline 6 & -6 & 0 & {\sqrt{2}} & -{
       \sqrt{2}} & 0 & 0 & -1 & 1 & -1 & 1 \\ \hline 8 & -8 & 0 & 0 & 0 & 
     -1 & 1 & 1 & -1 & 1 & -1 \\ \hline
\ea
\qquad
a := \frac{-1 + \sqrt{7} i}{2}
}
\]

\[
{\tiny
\Sigma(1080) \quad
\ba{|c|c|c|c|c|c|c|c|c|c|c|c|c|c|c|c|c|}
\hline
1 & 1 & 1 & 45 & 45 & 45 & 72 & 72 & 72 & 72 & 72 & 72 & 90 & 90 & 90 & 120 & 120 \\ \hline 1 & 1 & 1 & 1 & 1 & 1 & 1 & 1 & 1 & 1 & 1 & 1 & 1 & 1 & 1 & 1 & 1 \\ \hline 3 &
   3\bar{A} & 3A & -A & -\bar{A} & -1 & X & Y & Z & W & \bar{Z} & \bar{W} & \bar{A} & A & 1 & 0 & 0 \\ \hline 3 & 3
   \bar{A} & 3A & -A & -\bar{A} & -1 & Y & X & W & Z & \bar{W} & \bar{Z} & \bar{A} & A & 1 & 0 & 0 \\ \hline 3 & 3
   A & 3\bar{A} & -\bar{A} & -A & -1 & X & Y & \bar{Z} & \bar{W} & Z & W & A & \bar{A} & 1 & 0 & 0 \\ \hline 3 & 3
   A & 3\bar{A} & -\bar{A} & -A & -1 & Y & X & \bar{W} & \bar{Z} & W & Z & A & \bar{A}
    & 1 & 0 & 0 \\ \hline 5 & 5 & 5 & 1 & 1 & 1 & 0 & 0 & 0 & 0 & 0 & 0 & -1 & -1 & -1 & 2 & -1 \\ \hline 5 & 5 & 5 & 1 & 1 & 1 & 0 & 0 & 0 & 0 & 0 & 0 & -1 & -1 & -1 & 
    -1 & 2 \\ \hline 6 & 6\bar{A} & 6A & 2A & 2\bar{A} & 2 & 1 & 1 & \bar{A} & \bar{A}
    & A & A & 0 & 0 & 0 & 0 & 0 \\ \hline 6 & 6A & 6\bar{A} & 2\bar{A} & 2A & 2 & 1 & 1 & A & A & \bar{A} & \bar{A} & 0 & 0 & 0 & 0 & 0 \\ \hline 8 & 8 & 8 & 0 & 0 & 0 & X & Y & Y & X & Y & X & 0 & 0 & 0 & -1 & -1 \\ \hline 8 & 8 & 8 & 0 & 0 & 0 & Y & X & X & Y & X & Y & 0 & 0 & 0 & 
    -1 & -1 \\ \hline 9 & 9 & 9 & 1 & 1 & 1 & -1 & -1 & -1 & -1 & -1 & -1 & 1 & 1 & 1 & 0 & 0 \\ \hline 9 & 9\bar{A} & 9A & A & \bar{A} & 1 & -1 & 
    -1 & -\bar{A} & -\bar{A} & -A & -A & \bar{A} & A & 1 & 0 & 0 \\ \hline 9 & 9A & 9\bar{A} & \bar{A} & A & 1 & -1 & 
    -1 & -A & -A & -\bar{A} & -\bar{A} & A & \bar{A} & 1 & 0 & 0 \\ \hline 10 & 10 & 10 & -2 & -2 & 
    -2 & 0 & 0 & 0 & 0 & 0 & 0 & 0 & 0 & 0 & 1 & 1 \\ \hline 15 & 15\bar{A} & 15A & -A & -\bar{A} & -1 & 0 & 0 & 0 & 0 & 0 & 0 & -\bar{A} & -A & -1 & 0 & 0 \\ \hline 15 & 15A & 15\bar{A} & -\bar{A} & -A & -1 & 0 & 0 & 0 & 0 & 0 & 0 & -A & -\bar{A} & -1 & 0 & 0 \\ \hline
\ea
\qquad
\ba{l}
A := \omega_3;\\
B := \omega_5;\\
C := \omega_{15};\\
X := -B-\bar{B};\\
Y := -B^2 - \bar{B}^2;\\
Z := -C - C^4;\\
W := -\bar{C}^2 - C^7;
\ea
}
\]

$\Sigma(1080)^*$
{\tiny 
\[
\vspace{-0.2in}
D := B+\bar{B}, E := B^2 + \bar{B}^2, F
:= \bar{C} + \bar{C}^4, G := C^2 + \bar{C}^7, H := \omega_{24}, J :=
\bar{H}^7 - H^{11}, K := \bar{H}^5 - H
\]
}
\[
\setlength{\arraycolsep}{0.2mm}
{\tiny
\ba{|c|c|c|c|c|c|c|c|c|c|c|c|c|c|c|c|c|c|c|c|c|c|c|c|c|c|c|c|c|c|c|}
\hline
   1 & 1 & 1 & 1 & 1 & 1 & 90 & 90 & 90 & 72 & 72 & 72 & 72 & 72 & 72 & 72 & 72 & 72 & 72 & 72 & 72 & 90 & 90 & 90 & 90 & 90 & 90 & 120 & 120 & 120 & 120 \\ \hline 1 & 
   1 & 1 & 1 & 1 & 1 & 1 & 1 & 1 & 1 & 1 & 1 & 1 & 1 & 1 & 1 & 1 & 1 & 1 & 1 & 1 & 1 & 1 & 1 & 1 & 1 & 1 & 1 & 1 & 1 & 1 \\ \hline 3 & 3 & 3\bar{A} & 3
   \bar{A} & 3A & 3A & -A & -\bar{A} & -1 & X & X & Y & Y & Z & Z & W & W & \bar{Z} & \bar{Z} & \bar{W} & 
   \bar{W} & \bar{A} & \bar{A} & A & A & 1 & 1 & 0 & 0 & 0 & 0 \\ \hline 3 & 3 & 3\bar{A} & 3\bar{A} & 3A & 3A & 
    -A & -\bar{A} & -1 & Y & Y & X & X & W & W & Z & Z & \bar{W} & \bar{W} & \bar{Z} & \bar{Z} & \bar{A} & \bar{A} & A & A & 1 & 1 & 0 & 0 & 0 & 0 \\ \hline 3 & 3 & 3A & 3A & 3\bar{A} & 3\bar{A} & -\bar{A} & -A & 
    -1 & X & X & Y & Y & \bar{Z} & \bar{Z} & \bar{W} & \bar{W} & Z & Z & W & W & A & A & \bar{A} & \bar{A} & 1 & 1 & 0 & 0 & 0 & 0 \\ \hline 3 & 3 & 3A & 3A & 3\bar{A} & 3\bar{A} & -\bar{A} & -A & -1 & Y & Y & X & X & \bar{W} & \bar{W} & \bar{Z} & \bar{Z} & W & W & Z & Z & A & A & \bar{A} & \bar{A} & 1 & 1 & 0 & 0 & 0 & 0 \\ \hline 5 & 5 & 5 & 5 & 5 & 5 & 1 & 1 & 1 & 0 & 0 & 0 & 0 & 0 & 0 & 0 & 0 & 0 & 0 & 0 & 0 & -1 & -1 & -1 & -1 & -1 & -1 & 2 & 2 & -1 & 
    -1 \\ \hline 5 & 5 & 5 & 5 & 5 & 5 & 1 & 1 & 1 & 0 & 0 & 0 & 0 & 0 & 0 & 0 & 0 & 0 & 0 & 0 & 0 & -1 & -1 & -1 & -1 & -1 & -1 & -1 & -1 & 2 & 2 \\ \hline 6 & 6 & 6
   \bar{A} & 6\bar{A} & 6A & 6A & 2A & 2\bar{A} & 2 & 1 & 1 & 1 & 1 & \bar{A} & \bar{A} & \bar{A} & \bar{A} & A & A & A & A & 0 & 0 & 0 & 0 & 0 & 0 & 0 & 0 & 0 & 0 \\ \hline 6 & 6 & 6A & 6A & 6\bar{A} & 6\bar{A} & 2
   \bar{A} & 2A & 2 & 1 & 1 & 1 & 1 & A & A & A & A & \bar{A} & \bar{A} & \bar{A} & \bar{A} & 0 & 0 & 0 & 0 & 0 & 0 & 0 & 0 & 0 & 0 \\ \hline 8 & 8 & 8 & 8 & 8 & 8 & 0 & 0 & 0 & X & X & Y & Y & Y & Y & X & X & Y & Y & X & X & 0 & 0 & 0 & 0 & 0 & 0 & -1 & 
    -1 & -1 & -1 \\ \hline 8 & 8 & 8 & 8 & 8 & 8 & 0 & 0 & 0 & Y & Y & X & X & X & X & Y & Y & X & X & Y & Y & 0 & 0 & 0 & 0 & 0 & 0 & -1 & -1 & -1 & 
    -1 \\ \hline 9 & 9 & 9 & 9 & 9 & 9 & 1 & 1 & 1 & -1 & -1 & -1 & -1 & -1 & -1 & -1 & -1 & -1 & -1 & -1 & -1 & 1 & 1 & 1 & 1 & 1 & 1 & 0 & 0 & 0 & 0 \\ \hline 9 & 9 & 9
   \bar{A} & 9\bar{A} & 9A & 9A & A & \bar{A} & 1 & -1 & -1 & -1 & -1 & -\bar{A} & -\bar{A} & -\bar{A} & -\bar{A} & -A & -A & -A & -A & \bar{A} & \bar{A} & A & A & 1 & 1 & 0 & 0 & 0 & 0 \\ \hline 9 & 9 & 9A & 9A & 9
   \bar{A} & 9\bar{A} & \bar{A} & A & 1 & -1 & -1 & -1 & -1 & -A & -A & -A & -A & -\bar{A} & -\bar{A} & -\bar{A} & -\bar{A} & A & A & \bar{A} & \bar{A} & 1 & 1 & 0 & 0 & 0 & 0 \\ \hline 10 & 10 & 10 & 10 & 10 & 10 & -2 & -2 & 
    -2 & 0 & 0 & 0 & 0 & 0 & 0 & 0 & 0 & 0 & 0 & 0 & 0 & 0 & 0 & 0 & 0 & 0 & 0 & 1 & 1 & 1 & 1 \\ \hline 15 & 15 & 15\bar{A} & 15\bar{A} & 15
   A & 15A & -A & -\bar{A} & -1 & 0 & 0 & 0 & 0 & 0 & 0 & 0 & 0 & 0 & 0 & 0 & 0 & -\bar{A} & -\bar{A} & -A & -A & -1 & 
    -1 & 0 & 0 & 0 & 0 \\ \hline 15 & 15 & 15A & 15A & 15\bar{A} & 15\bar{A} & -\bar{A} & -A & 
    -1 & 0 & 0 & 0 & 0 & 0 & 0 & 0 & 0 & 0 & 0 & 0 & 0 & -A & -A & -\bar{A} & -\bar{A} & -1 & -1 & 0 & 0 & 0 & 0 \\ \hline 4 & -4 & -4 & 4 & 
    -4 & 4 & 0 & 0 & 0 & 1 & -1 & 1 & -1 & 1 & -1 & 1 & -1 & 1 & -1 & 1 & -1 & 0 & 0 & 0 & 0 & 0 & 0 & 1 & -1 & -2 & 2 \\ \hline 4 & -4 & -4 & 4 & 
    -4 & 4 & 0 & 0 & 0 & 1 & -1 & 1 & -1 & 1 & -1 & 1 & -1 & 1 & -1 & 1 & -1 & 0 & 0 & 0 & 0 & 0 & 0 & -2 & 2 & 1 & -1 \\ \hline 6 & -6 & -6A & 6A & -6
   \bar{A} & 6\bar{A} & 0 & 0 & 0 & -1 & 1 & -1 & 1 & -A & A & -A & A & -\bar{A} & \bar{A} & -\bar{A} & \bar{A} & J & -J & K & -K & -{\sqrt{2}} & {\sqrt{2}} & 0 & 0 & 0 & 0 \\ \hline 6 & -6 & -6A & 6A & -6\bar{A} & 6\bar{A} & 0 & 0 & 0 & -1 & 1 & 
    -1 & 1 & -A & A & -A & A & -\bar{A} & \bar{A} & -\bar{A} & \bar{A} & -J & J & -K & K & {\sqrt{2}} & -{
      \sqrt{2}} & 0 & 0 & 0 & 0 \\ \hline 6 & -6 & -6\bar{A} & 6\bar{A} & -6A & 6A & 0 & 0 & 0 & -1 & 1 & -1 & 1 & -\bar{A} & 
   \bar{A} & -\bar{A} & \bar{A} & -A & A & -A & A & -K & K & -J & J & -{\sqrt{2}} & {\sqrt{2}} & 0 & 0 & 0 & 0 \\ \hline 6 & -6 & -6
   \bar{A} & 6\bar{A} & -6A & 6A & 0 & 0 & 0 & -1 & 1 & -1 & 1 & -\bar{A} & \bar{A} & -\bar{A} & \bar{A} & -A & A & -A & A & K & -K & J & -J & {\sqrt{2}} & -{\sqrt{2}} & 0 & 0 & 0 & 0 \\ \hline 8 & -8 & -8 & 8 & 
    -8 & 8 & 0 & 0 & 0 & D & X & E & Y & E & Y & D & X & E & Y & D & X & 0 & 0 & 0 & 0 & 0 & 0 & -1 & 1 & -1 & 1 \\ \hline 8 & -8 & -8 & 8 & 
    -8 & 8 & 0 & 0 & 0 & E & Y & D & X & D & X & E & Y & D & X & E & Y & 0 & 0 & 0 & 0 & 0 & 0 & -1 & 1 & -1 & 1 \\ \hline 10 & -10 & -10 & 10 & 
    -10 & 10 & 0 & 0 & 0 & 0 & 0 & 0 & 0 & 0 & 0 & 0 & 0 & 0 & 0 & 0 & 0 & -{\sqrt{2}} & {\sqrt{2}} & {\sqrt{2}} & -{\sqrt{2}} & -{\sqrt{2}} & {\sqrt{2}} & 1 & 
    -1 & 1 & -1 \\ \hline 10 & -10 & -10 & 10 & -10 & 10 & 0 & 0 & 0 & 0 & 0 & 0 & 0 & 0 & 0 & 0 & 0 & 0 & 0 & 0 & 0 & {\sqrt{2}} & -{\sqrt{2}} & -{\sqrt{2}} & {
     \sqrt{2}} & {\sqrt{2}} & -{\sqrt{2}} & 1 & -1 & 1 & -1 \\ \hline 12 & -12 & -12A & 12A & -12\bar{A} & 12
   \bar{A} & 0 & 0 & 0 & X & D & Y & E & \bar{Z} & \bar{F} & \bar{W} & G & Z & F & W & \bar{G} & 0 & 0 & 0 & 0 & 0 & 0 & 0 & 0 & 0 & 0 \\ \hline 12 & -12 & -12A & 12A & -12\bar{A} & 12\bar{A} & 0 & 0 & 0 & Y & E & X & D & \bar{W} & G & \bar{Z} & \bar{F} & W & \bar{G} & Z & F & 0 & 0 & 0 & 0 & 0 & 0 & 0 & 0 & 0 & 0 \\ \hline 12 & -12 & -12
   \bar{A} & 12\bar{A} & -12A & 12A & 0 & 0 & 0 & X & D & Y & E & Z & F & W & \bar{G} & \bar{Z} & \bar{F} & 
   \bar{W} & G & 0 & 0 & 0 & 0 & 0 & 0 & 0 & 0 & 0 & 0 \\ \hline 12 & -12 & -12\bar{A} & 12\bar{A} & -12A & 12
   A & 0 & 0 & 0 & Y & E & X & D & W & \bar{G} & Z & F & \bar{W} & G & \bar{Z} & \bar{F} & 0 & 0 & 0 & 0 & 0 & 0 & 0 & 0 & 0 & 0 \\ \hline
\ea
}
\]

\[
{\tiny
\Delta(6 \times 2^2) = \quad
\ba{|c|c|c|c|c|}
\hline
1 & 3 & 6 & 6 & 8 \\ \hline 1 & 1 & 1 & 1 & 1 \\ \hline 1 & 1 & 
    -1 & -1 & 1 \\ \hline 2 & 2 & 0 & 0 & -1 \\ \hline 3 & -1 & 
    -1 & 1 & 0 \\ \hline 3 & -1 & 1 & -1 & 0 \\ \hline 
\ea
\qquad
\Delta(6\times 2^2)^* = \quad
\ba{|c|c|c|c|c|c|c|c|}
\hline
1 & 1 & 6 & 6 & 6 & 12 & 8 & 8 \\ \hline 1 & 1 & 1 & 1 & 1 & 
   1 & 1 & 1 \\ \hline 1 & 1 & 1 & -1 & -1 & 
    -1 & 1 & 1 \\ \hline 2 & 2 & 2 & 0 & 0 & 0 & -1 & 
    -1 \\ \hline 3 & 3 & -1 & -1 & -1 & 1 & 0 & 0 \\ \hline 3 & 3 & 
    -1 & 1 & 1 & -1 & 0 & 0 \\ \hline 2 & -2 & 0 & -e^
     {\frac{i }{4}\pi } - e^{\frac{3i }{4}\pi }
   & e^{\frac{i }{4}\pi } + 
   e^{\frac{3i }{4}\pi } & 0 & -1 & 1 \\ \hline 2 & 
    -2 & 0 & e^{\frac{i }{4}\pi } + 
   e^{\frac{3i }{4}\pi } & -e^{\frac{i }{4}\pi } - e^{\frac{3i }{4}\pi }
   & 0 & -1 & 1 \\ \hline 4 & -4 & 0 & 0 & 0 & 0 & 1 & -1 \\ \hline
\ea
}
\]

\[
\setlength{\arraycolsep}{0.4mm}
\ba{lcl}
\Delta(6 \times 4^2) = & & \Delta(6\times 4^2)^* = \\
{\tiny
\ba{|c|c|c|c|c|c|c|c|c|c|}
\hline
   1 & 3 & 3 & 3 & 6 & 12 & 12 & 12 & 12 & 32 \\ \hline 1 & 1 & 1 & 1 & 1 & 1 & 
    1 & 1 & 1 & 1 \\ \hline 1 & 1 & 1 & 1 & 1 & -1 & -1 & -1 & 
     -1 & 1 \\ \hline 2 & 2 & 2 & 2 & 2 & 0 & 0 & 0 & 0 & -1 \\ \hline 3 & 3 & -1 & 
     -1 & -1 & -1 & 1 & 1 & -1 & 0 \\ \hline 3 & 3 & -1 & -1 & -1 & 1 & -1 & 
     -1 & 1 & 0 \\ \hline 3 & -1 & -1 - 2i  & -1 + 2i  & 1 & -1 & i
     & -i  & 1 & 0 \\ \hline 3 & -1 & -1 + 2i  & -1 - 2i  & 1 & 
     -1 & -i  & i  & 1 & 0 \\ \hline 3 & -1 & -1 - 2i  & -1 + 
    2i  & 1 & 1 & -i  & i  & -1 & 0 \\ \hline 3 & -1 & -1 + 
    2i  & -1 - 2i  & 1 & 1 & i  & -i  & -1 & 0 \\ \hline 6 & 
     -2 & 2 & 2 & -2 & 0 & 0 & 0 & 0 & 0 \\ \hline
\ea
}
&~~~&
{\tiny
\ba{|c|c|c|c|c|c|c|c|c|c|c|c|c|c|c|}
\hline
    1 & 1 & 3 & 3 & 6 & 6 & 12 & 24 & 12 & 12 & 12 & 12 & 24 & 32 & 32 \\ \hline 
    1 & 1 & 1 & 1 & 1 & 1 & 1 & 1 & 1 & 1 & 1 & 1 & 1 & 1 & 1 \\ \hline 1 & 1 & 1 &
    1 & 1 & 1 & 1 & -1 & -1 & -1 & -1 & -1 & 
     -1 & 1 & 1 \\ \hline 2 & 2 & 2 & 2 & 2 & 2 & 2 & 0 & 0 & 0 & 0 & 0 & 0 & 
     -1 & -1 \\ \hline 3 & 3 & 3 & 3 & -1 & -1 & -1 & -1 & 1 & 1 & 1 & 1 & 
     -1 & 0 & 0 \\ \hline 3 & 3 & 3 & 3 & -1 & -1 & -1 & 1 & -1 & -1 & -1 & 
     -1 & 1 & 0 & 0 \\ \hline 3 & 3 & -1 & -1 & -1 - 2i  & -1 + 
    2i  & 1 & -1 & i  & i  & -i  & -i
      & 1 & 0 & 0 \\ \hline 3 & 3 & -1 & -1 & -1 + 2i  & -1 - 
    2i  & 1 & -1 & -i  & -i  & i  & i
     & 1 & 0 & 0 \\ \hline 3 & 3 & -1 & -1 & -1 - 2i  & -1 + 
    2i  & 1 & 1 & -i  & -i  & i  & i  & 
     -1 & 0 & 0 \\ \hline 3 & 3 & -1 & -1 & -1 + 2i  & -1 - 
    2i  & 1 & 1 & i  & i  & -i  & -i  & 
     -1 & 0 & 0 \\ \hline 6 & 6 & -2 & -2 & 2 & 2 & 
     -2 & 0 & 0 & 0 & 0 & 0 & 0 & 0 & 0 \\ \hline 2 & -2 & 
     -2 & 2 & 0 & 0 & 0 & 0 & i {\sqrt{2}} & -i {\sqrt{2}} & 
    -i {\sqrt{2}} & i {\sqrt{2}} & 0 & -1 & 1 \\ \hline 2 & -2 & 
     -2 & 2 & 0 & 0 & 0 & 0 & -i {\sqrt{2}} & i {\sqrt{2}} & 
    i {\sqrt{2}} & -i {\sqrt{2}} & 0 & -1 & 1 \\ \hline 4 & -4 & 
     -4 & 4 & 0 & 0 & 0 & 0 & 0 & 0 & 0 & 0 & 0 & 1 & -1 \\ \hline 6 & -6 & 2 & 
     -2 & 0 & 0 & 0 & 0 & -{\sqrt{2}} & {\sqrt{2}} & -{\sqrt{2}} & {
      \sqrt{2}} & 0 & 0 & 0 \\ \hline 6 & -6 & 2 & -2 & 0 & 0 & 0 & 0 & {
      \sqrt{2}} & -{\sqrt{2}} & {\sqrt{2}} & -{\sqrt{2}} & 0 & 0 & 0 \\ \hline 
\ea
}
\ea
\]

\[
\setlength{\arraycolsep}{0.2mm}
\ba{lcl}
\Delta(3 \times 4^2) = & & \Delta(3 \times 4^2)^* = \\
{\tiny
\ba{|c|c|c|c|c|c|c|c|}
\hline
1 & 3 & 3 & 3 & 3 & 3 & 16 & 16 \\ \hline 1 & 1 & 1 & 1 & 1 & 1 & 1 & 1 \\ \hline 1 &
    1 & 1 & 1 & 1 & 1 & \omega_3 & \bar{\omega_3} \\ \hline 1 & 1 & 1 & 1 & 1 & 1 & 
    \bar{\omega_3} & \omega_3 \\ \hline 3 & -1 & -1 & 3 & -1 & 
     -1 & 0 & 0 \\ \hline 3 & 1 & 1 & -1 & -1 - 2i  & -1 + 
    2i  & 0 & 0 \\ \hline 3 & 1 & 1 & -1 & -1 + 2i  & -1 - 
    2i  & 0 & 0 \\ \hline 3 & -1 - 2i  & -1 + 2i  & 
     -1 & 1 & 1 & 0 & 0 \\ \hline 3 & -1 + 2i  & -1 - 2i  & 
     -1 & 1 & 1 & 0 & 0 \\ \hline
\ea}
&~~~&
{\tiny
\ba{|c|c|c|c|c|c|c|c|c|c|c|c|c|c|c|c|c|c|}
\hline
    1 & 1 & 1 & 1 & 12 & 12 & 6 & 6 & 12 & 12 & 16 & 16 & 16 & 16 & 16 & 16 &
    16 & 16 \\ \hline 1 & 1 & 1 & 1 & 1 & 1 & 1 & 1 & 1 & 1 & 1 & 1 & 1 & 1 & 1 & 
    1 & 1 & 1 \\ \hline 1 & 1 & 1 & 1 & 1 & 1 & 1 & 1 & 1 & 1 & \omega_3 & \omega_3 & \omega_3 & \omega_3 &
    \bar{\omega}_3 & \bar{\omega}_3 & \bar{\omega}_3 & \bar{\omega}_3 \\ \hline 
1 & 1 & 1 & 1 & 1 & 1 & 1 & 1 & 1 & 1 & \bar{\omega}_3 & 
    \bar{\omega}_3 & \bar{\omega}_3 & \bar{\omega}_3 & \omega_3 &
    \omega_3 & \omega_3 & \omega_3 
\\ \hline 3 & 3 & 3 & 3 & -1 & -1 & 3 & 3 & -1 & 
     -1 & 0 & 0 & 0 & 0 & 0 & 0 & 0 & 0 \\ \hline 3 & 3 & 3 & 3 & 1 & 1 & -1 & 
     -1 & -1 - 2i  & -1 + 
    2i  & 0 & 0 & 0 & 0 & 0 & 0 & 0 & 0 \\ \hline 3 & 3 & 3 & 3 & 1 & 1 & 
     -1 & -1 & -1 + 2i  & -1 - 
    2i  & 0 & 0 & 0 & 0 & 0 & 0 & 0 & 0 \\ \hline 3 & 3 & 3 & 3 & -1 - 
    2i  & -1 + 2i  & -1 & 
     -1 & 1 & 1 & 0 & 0 & 0 & 0 & 0 & 0 & 0 & 0 \\ \hline 3 & 3 & 3 & 3 & -1 + 
    2i  & -1 - 2i  & -1 & 
     -1 & 1 & 1 & 0 & 0 & 0 & 0 & 0 & 0 & 0 & 0 \\ \hline 2 & -2 & 2 & 
     -2 & 0 & 0 & 2 & -2 & 0 & 0 & -\omega_3 & \omega_3 & \omega_3 & -\omega_3 & -\bar{\omega}_3 & 
     -\bar{\omega}_3 & \bar{\omega}_3 & \bar{\omega}_3 \\ \hline 2 & 
     -2 & 2 & -2 & 0 & 0 & 2 & -2 & 0 & 0 & -\bar{\omega}_3 & 
\bar{\omega}_3 & \bar{\omega}_3 & -\bar{\omega}_3 & -\omega_3 & -\omega_3 & \omega_3 & \omega_3 \\ \hline 2 & 
     -2 & 2 & -2 & 0 & 0 & 2 & -2 & 0 & 0 & -1 & 1 & 1 & -1 & -1 & 
     -1 & 1 & 1 \\ \hline 6 & -6 & 6 & -6 & 0 & 0 & 
     -2 & 2 & 0 & 0 & 0 & 0 & 0 & 0 & 0 & 0 & 0 & 0 \\ \hline 4 & 4i  & 
     -4 & -4i  & 0 & 0 & 0 & 0 & 0 & 0 & \bar{\omega}_3 & 
-\bar{\omega}_{12} & \bar{\omega}_{12} & -\bar{\omega}_3 & \omega_3 &
-\omega_3 & \bar{\omega}_{12}^5 & 
-\bar{\omega}_{12}^5 \\ \hline 4 & 4i  & -4 & -4
    i  & 0 & 0 & 0 & 0 & 0 & 0 & \omega_3 & -\bar{\omega}_{12}^5 & 
\bar{\omega}_{12}^5 & -\omega_3 & \bar{\omega}_3 & -\bar{\omega}_3 & 
\bar{\omega}_{12} & -\bar{\omega}_{12} \\ \hline 4 & 4i  & -4 & -4
    i  & 0 & 0 & 0 & 0 & 0 & 0 & 1 & -i  & i  & -1 & 1 & 
     -1 & i  & -i  \\ \hline 4 & -4i  & -4 & 4
    i  & 0 & 0 & 0 & 0 & 0 & 0 & \bar{\omega}_3 & \bar{\omega}_{12} 
& -\bar{\omega}_{12} & -\bar{\omega}_3 & \omega_3 & -\omega_3 & -\bar{\omega}_{12}^5 
& \bar{\omega}_{12}^5 \\ \hline 4 & -4i  & -4 & 4
    i  & 0 & 0 & 0 & 0 & 0 & 0 & \omega_3 & \bar{\omega}_{12}^5 &
-\bar{\omega}_{12}^5 & 
-\omega_3 & \bar{\omega}_3 & -\bar{\omega}_3 & -\bar{\omega}_{12} & 
    \bar{\omega}_{12} \\ \hline 4 & -4i  & -4 & 4
    i  & 0 & 0 & 0 & 0 & 0 & 0 & 1 & i  & -i  & -1 & 1 & 
     -1 & -i  & i  \\ \hline 
\ea
}
\ea
\]

\[
\setlength{\arraycolsep}{0.2mm}
\ba{lcl}
\Delta(3 \times 5^2) = && \Delta(3 \times 5^2)^* = \\
{\tiny
\ba{|c|c|c|c|c|c|c|c|c|c|c|}
\hline
    1 & 3 & 3 & 3 & 3 & 3 & 3 & 3 & 3 & 25 & 25 \\ \hline 1 & 1 & 1 & 1 & 1 & 1 & 
    1 & 1 & 1 & 1 & 1 \\ \hline 1 & 1 & 1 & 1 & 1 & 1 & 1 & 1 & 1 & E & \bar{E} \\ \hline 
1 & 1 & 1 & 1 & 1 & 1 & 1 & 1 & 1 & \bar{E} 
& E \\ \hline 3 & A & A & B & B & C & D & \bar{C} & \bar{D} & 0 & 0 \\ \hline 
3 & A & A & B & B & \bar{C} & \bar{D} & C & D & 0 & 0 \\ \hline 
3 & B & B & A & A & \bar{D} & C & D & 
    \bar{C} & 0 & 0 \\ \hline 3 & B & B & A & A & D & \bar{C} & \bar{D} 
& C & 0 & 0 \\ \hline 3 & C & \bar{C} & D & 
    \bar{D} & B & A & B & A & 0 & 0 \\ \hline 3 & \bar{D} & D & C & 
\bar{C} & A & B & A & B & 0 & 0 \\ \hline 3 & \bar{C} & C & \bar{D} 
& D & B & A & B & A & 0 & 0 \\ \hline 3 & D & 
\bar{D} & \bar{C} & C & A & B & A & B & 0 & 0 \\ \hline
\ea}
&~~~&
{\tiny
\ba{|c|c|c|c|c|c|c|c|c|c|c|c|c|c|c|c|c|c|c|c|c|c|c|}
\hline
   1 & 1 & 1 & 1 & 1 & 15 & 15 & 15 & 15 & 15 & 15 & 15 & 15 & 25 & 25
& 25 & 25 & 25 & 25 & 25 & 25 & 25 & 25 \\
 \hline 1 & 1 & 1 & 1 & 1 & 
   1 & 1 & 1 & 1 & 1 & 1 & 1 & 1 & 1 & 1 & 1 & 1 & 1 & 1 & 1 & 1 & 1 &
1 \\ \hline
 1 & 1 & 1 & 1 & 1 & 1 & 1 & 1 & 1 & 1 & 1 & 1 & 1 & E & 
   E & E & E & E & {\bar{E}} & {\bar{E}} & {\bar{E}} & {\bar{E}} & {\bar{E}
   } \\ \hline 1 & 1 & 1 & 1 & 1 & 1 & 1 & 1 & 1 & 1 & 1 & 1 & 1 & {\bar{E}} & {\bar{E}} & {\bar{E}} & {\bar{E}
   } & {\bar{E}} & E & E & E & E & E \\ \hline 3 & 3 & 3 & 3 & 3 & A & A & B & B & C & D & {\bar{C}} & {\bar{D}
   } & 0 & 0 & 0 & 0 & 0 & 0 & 0 & 0 & 0 & 0 \\ \hline 3 & 3 & 3 & 3 & 3 & A & A & B & B & {\bar{C}} & {\bar{D}
   } & C & D & 0 & 0 & 0 & 0 & 0 & 0 & 0 & 0 & 0 & 0 \\ \hline 3 & 3 &
3 & 3 & 3 & B & B & A & A & {\bar{D}} & C & D & {\bar{C}
   } & 0 & 0 & 0 & 0 & 0 & 0 & 0 & 0 & 0 & 0 \\ \hline 3 & 3 & 3 & 3 & 3 & B & B & A & A & D & {\bar{C}} & {\bar{D}
   } & C & 0 & 0 & 0 & 0 & 0 & 0 & 0 & 0 & 0 & 0 \\ \hline 3 & 3 & 3 & 3 & 3 & C & {\bar{C}} & D & {\bar{D}
   } & B & A & B & A & 0 & 0 & 0 & 0 & 0 & 0 & 0 & 0 & 0 & 0 \\ \hline 3 & 3 & 3 & 3 & 3 & {\bar{D}} & D & C & {\bar{C}
   } & A & B & A & B & 0 & 0 & 0 & 0 & 0 & 0 & 0 & 0 & 0 & 0 \\ \hline 3 & 3 & 3 & 3 & 3 & {\bar{C}} & C & {\bar{D}
   } & D & B & A & B & A & 0 & 0 & 0 & 0 & 0 & 0 & 0 & 0 & 0 & 0 \\ \hline 3 & 3 & 3 & 3 & 3 & D & {\bar{D}} & {\bar{C}
   } & C & A & B & A & B & 0 & 0 & 0 & 0 & 0 & 0 & 0 & 0 & 0 & 0 \\ \hline 5 & 5{\bar{F}^2} & 5F & 5{\bar{F}} & 5
   {\bar{F}^2} & 0 & 0 & 0 & 0 & 0 & 0 & 0 & 0 & -1 & -{\bar{F}} & -{\bar{F}^2} & -{\bar{F}^2} & -F & -1 & -F & 
    -{\bar{F}^2} & -{\bar{F}^2} & -{\bar{F}} \\ \hline 5 & 5{\bar{F}^2} & 5F & 5{\bar{F}} & 5
   {\bar{F}^2} & 0 & 0 & 0 & 0 & 0 & 0 & 0 & 0 & -E & -{\bar{G}^7} & -{\bar{G}} & -{\bar{G}^4} & -{G^2
    } & -{\bar{E}} & -{\bar{G}^7} & -G & -{\bar{G}^4} & -{\bar{G}^2} \\ \hline 5 & 5{\bar{F}^2} & 5F & 5
   {\bar{F}} & 5{\bar{F}^2} & 0 & 0 & 0 & 0 & 0 & 0 & 0 & 0 & -{\bar{E}} & -{\bar{G}^2} & -{\bar{G}^4} & 
    -G & -{\bar{G}^7} & -E & -{G^2} & -{\bar{G}^4} & -{\bar{G}} & -{\bar{G}^7} \\ \hline 5 & 5
   {\bar{F}^2} & 5{\bar{F}} & 5F & 5{\bar{F}^2} & 0 & 0 & 0 & 0 & 0 & 0 & 0 & 0 & -1 & -F & -{\bar{F}^2
    } & -{\bar{F}^2} & -{\bar{F}} & -1 & -{\bar{F}} & -{\bar{F}^2} & -{\bar{F}^2} & -F \\ \hline 5 & 5
   {\bar{F}^2} & 5{\bar{F}} & 5F & 5{\bar{F}^2} & 0 & 0 & 0 & 0 & 0 & 0 & 0 & 0 & -E & -{G^2} & -
    {\bar{G}^4} & -{\bar{G}} & -{\bar{G}^7} & -{\bar{E}} & -{\bar{G}^2} & -{\bar{G}^4} & -G & -
    {\bar{G}^7} \\ \hline 5 & 5{\bar{F}^2} & 5{\bar{F}} & 5F & 5{\bar{F}^2} & 0 & 0 & 0 & 0 & 0 & 0 & 0 & 0 & 
    -{\bar{E}} & -{\bar{G}^7} & -G & -{\bar{G}^4} & -{\bar{G}^2} & -E & -{\bar{G}^7} & -{\bar{G}
    } & -{\bar{G}^4} & -{G^2} \\ \hline 5 & 5{\bar{F}} & 5{\bar{F}^2} & 5{\bar{F}^2} & 5
   F & 0 & 0 & 0 & 0 & 0 & 0 & 0 & 0 & -1 & -{\bar{F}^2} & -F & -{\bar{F}} & -{\bar{F}^2} & -1 & -{\bar{F}^2
    } & -{\bar{F}} & -F & -{\bar{F}^2} \\ \hline 5 & 5{\bar{F}} & 5{\bar{F}^2} & 5{\bar{F}^2} & 5
   F & 0 & 0 & 0 & 0 & 0 & 0 & 0 & 0 & -E & -{\bar{G}} & -{G^2} & -{\bar{G}^7} & -{\bar{G}^4} & -{
    \bar{E}} & -G & -{\bar{G}^2} & -{\bar{G}^7} & -{\bar{G}^4} \\ \hline 5 & 5{\bar{F}} & 5{\bar{F}^2} & 5
   {\bar{F}^2} & 5F & 0 & 0 & 0 & 0 & 0 & 0 & 0 & 0 & -{\bar{E}} & -{\bar{G}^4} & -{\bar{G}^7} & -{\bar{G}^2
    } & -G & -E & -{\bar{G}^4} & -{\bar{G}^7} & -{G^2} & -{\bar{G}} \\ \hline 5 & 5F & 5{\bar{F}^2} & 5
   {\bar{F}^2} & 5{\bar{F}} & 0 & 0 & 0 & 0 & 0 & 0 & 0 & 0 & -1 & -{\bar{F}^2} & -{\bar{F}} & -F & -
    {\bar{F}^2} & -1 & -{\bar{F}^2} & -F & -{\bar{F}} & -{\bar{F}^2} \\ \hline 5 & 5F & 5{\bar{F}^2} & 5
   {\bar{F}^2} & 5{\bar{F}} & 0 & 0 & 0 & 0 & 0 & 0 & 0 & 0 & -E & -{\bar{G}^4} & -{\bar{G}^7} & -{
    G^2} & -{\bar{G}} & -{\bar{E}} & -{\bar{G}^4} & -{\bar{G}^7} & -{\bar{G}^2} & -G \\ \hline 5 & 5F & 5
   {\bar{F}^2} & 5{\bar{F}^2} & 5{\bar{F}} & 0 & 0 & 0 & 0 & 0 & 0 & 0
	& 0 & -{\bar{E}} & -G & -{\bar{G}^2}
	 & -{\bar{G}^7} & -{\bar{G}^4} & -E & -{\bar{G}} & -{G^2} & -{\bar{G}^7} & -{\bar{G}^4
    } \\ \hline
\ea
}
\ea
\]

$
A := -\omega_5 - \bar{\omega}_5, B := -\omega_5^2 - \bar{\omega}_5^2,
C := \bar{\omega}_5 - 2\bar{\omega}_5^2, D := 2\omega_5 +
\bar{\omega}_5^2;
E := \omega_3, F := \bar{\omega}_5, G := \omega_{15}.
$
\index{Toric Variety!dual cone}
\section{Finding the Dual Cone}
\label{append:0003085}
Let us be given a convex polytope $C$,  with the edges specifying
the faces of which given by the matrix $M$ whose columns are the 
vectors corresponding to these edges.
Our task is to find the dual cone $\tilde{C}$ of $C$, or more
precisely the matrix $N$ such that
\[
N^t \cdot M \ge 0 \qquad \mbox{for all entries.}
\]
There is a standard algorithm, given in \cite{Fulton}.
Let $M$ be $n \times p$, i.e., there are $p$ $n$-dimensional vectors
spanning $C$. We note of course that $p \ge n$ for convexity.
Out of the $p$ vectors, we choose $n-1$. This gives us an $n \times
(n-1)$ matrix of co-rank 1, whence we can extract a 1-dimensional
null-space (as indeed the initial $p$ vectors are all linearly
independent) described by a single vector $u$.

Next we check the dot product of $u$ with the remaining $p-(n-1)$
vectors. If all the dot products are positive we keep $u$, and if
all are negative, we keep $-u$, otherwise we discard it.

We then select another $n-1$ vectors and repeat the above until all
combinations are exhausted. The set of vectors we have kept, $u$'s or
$-u$'s then form the columns of $N$ and span the dual cone
$\tilde{C}$.

We note that this is a very computationally intensive algorithm, the
number of steps of which depends on 
$\left( \begin{array}{c} p \\ n -1 \\ \end{array} \right)$ which
grows exponentially.

A subtle point to remark. In light of what we discussed in a footnote
in the paper on the difference between ${\bf M_+} = {\bf M} \cap
\sigma$ and ${\bf M'_+}$, here we have computed the dual of
$\sigma$. We must ensure that $\Z_+$-independent lattice points inside
the cones be not missed.
\index{Resolution!partial}
\index{Brane Probes!Orbifolds}
\section{Gauge Theory Data for $\IZ_n\times \IZ_n$}
\label{append:0104259}
For future reference we include here the gauge theory data for the 
$\IZ_n\times \IZ_n$ orbifold, so that, as mentioned in \cite{0003085},
any 3-dimensional toric singularity may exist as a partial resolution
thereof.

We have $3n^2$ fields denoted as $X_{ij},Y_{ij},Z_{ij}$ and
choose the decomposition ${\bf 3}\rightarrow (1,0)+(0,1)+(-1,-1)$.
The matter content (and thus the $d$ matrix) 
is well-known from standard brane box
constructions, hence we here focus on
the superpotential \cite{HSU} (and thus the $K$ matrix):
$$
X_{ij}Y_{i(j+1)}Z_{(i+1)(j+1)}-Y_{ij}X_{(i+1)j}Z_{(i+1)(j+1)},
$$
from which the F-terms are
\beq
\label{F-term2}
\ba{crcl}
\frac{\partial W}{\partial X_{ij}}: & 
 Y_{i(j+1)}Z_{(i+1)(j+1)} & = & Z_{i(j+1)}Y_{(i-1)j}  \\
&  & & \\
\frac{\partial W}{\partial Y_{ij}}: & 
Z_{(i+1)j} X_{i(j-1)}& =& X_{(i+1)j}Z_{(i+1)(j+1)} \\ & & &  \\
\frac{\partial W}{\partial Z_{(i+1)(j+1)}}: & 
X_{ij}Y_{i(j+1)}& =& Y_{ij}X_{(i+1)j}.
\ea
\eeq

Now let us solve \eref{F-term2}. First we have
$Y_{i(j+1)}=Y_{ij}X_{(i+1)j}/X_{ij}$.
Thus if we take $Y_{i0}$ and $X_{ij}$ as the independent 
variables, we have
\beq
\label{solve_Y}
Y_{i(j+1)}=\frac{\prod_{l=0}^{j} X_{(i+1)l}}{\prod_{l=0}^{j} X_{il}}
Y_{i0}.
\eeq
There is of course the periodicity which gives
\beq
\label{constraint_1}
Y_{in}=Y_{i0} \Longrightarrow
 \prod_{l=0}^{n-1} X_{(i+1)l}=\prod_{l=0}^{n-1} X_{il}.
\eeq
Next we use $X_{ij}$ to solve the $Z_{ij}$ as
$Z_{i(j+1)}=Z_{ij} X_{(i-1)(j-1)}/X_{ij},$ whence
\beq
\label{solve_Z}
Z_{i(j+1)}=\frac{\prod_{l=0}^{j} X_{(i-1)(l-1)}}{\prod_{l=0}^{j} X_{il}}
Z_{i0}.
\eeq
As above,
\beq
\label{constraint_2}
Z_{in}=Z_{i0} \Longrightarrow
 \prod_{l=0}^{n-1} X_{(i-1)(l-1)}=\prod_{l=0}^{n-1} X_{il}.
\eeq
Putting the solution of $Y,Z$ into the first equation of \eref{F-term2}
we get
$$
\frac{\prod_{l=0}^{j} X_{(i+1)l}}{\prod_{l=0}^{j} X_{il}}
Y_{i0}
\frac{\prod_{l=0}^{j} X_{(i)(l-1)}}{\prod_{l=0}^{j} X_{(i+1)l}}
Z_{(i+1)0}
=
\frac{\prod_{l=0}^{j} X_{(i-1)(l-1)}}{\prod_{l=0}^{j} X_{il}}
Z_{i0}
\frac{\prod_{l=0}^{j-1} X_{il}}{\prod_{l=0}^{j-1} X_{(i-1)l}}
Y_{(i-1)0},
$$
which can be simplified as $Y_{i0} Z_{(i+1)0} X_{i(n-1)}=
Z_{i0} Y_{(i-1)0} X_{(i-1)(n-1)}$,
or $X_{i(n-1)}=X_{(i-1)(n-1)}
\frac{Y_{(i-1)0}}{Y_{i0}}\frac{Z_{i0}}{Z_{(i+1)0}}$.
From this we solve
\beq
\label{X_n-1}
X_{i(n-1)}=X_{0(n-1)} \prod_{l=0}^{i-1}
\frac{Y_{l0}}{Y_{(l+1)0}}\frac{Z_{(l+1)0}}{Z_{(l+2)0}}.
\eeq
The periodicity gives
\beq
\label{constraint_3}
\prod_{l=0}^{n-1}
\frac{Y_{l0}}{Y_{(l+1)0}}\frac{Z_{(l+1)0}}{Z_{(l+2)0}}=1.
\eeq
Now we have the independent variables $Y_{i0}$
$Z_{i0}$ and  $X_{ij}$ for $j\neq n-1$ and $X_{0(n-1)}$,
plus three constraints (\ref{constraint_1})
(\ref{constraint_2}) (\ref{constraint_3}). In fact, considering the
periodic condition for $X$, (\ref{constraint_1}) is equivalent to
(\ref{constraint_2}). Furthermore considering the periodic conditions
for $Z_{i0}$ and $Y_{i0}$, (\ref{constraint_3}) is trivial.
So we have only one constraint. Putting
the expression (\ref{X_n-1}) into (\ref{constraint_1}) we get
$
\prod_{l=0}^{n-2} X_{(i+1)l} 
\frac{Y_{i0}}{Y_{(i+1)0}}\frac{Z_{(i+1)0}}{Z_{(i+2)0}}
=\prod_{l=0}^{n-2} X_{il}
\Rightarrow
\prod_{l=0}^{n-2} X_{(i+1)l}\frac{1}{Y_{(i+1)0}Z_{(i+2)0}}
=\prod_{l=0}^{n-2} X_{il} \frac{1}{Y_{i0}Z_{(i+1)0}}.
$

From this we can solve the $X_{i(n-1)}$ for $i\neq 0$ as
\beq
\label{X_n_2}
X_{i(n-2)}=(\prod_{l=0}^{n-2} X_{0l}) \frac{Y_{i0}Z_{(i+1)0}}{Y_{00}Z_{10}}
(\prod_{l=0}^{n-2} X_{il})^{-1}.
\eeq
The periodic condition does not give new constraints. 

Now we have finished solving the F-term and can summarise the results
into the $K$-matrix. We use the following independent variables:
$Z_{i0}$, $Y_{i0}$ for $i=0,1,...,n-1$; $X_{ij}$ for $i=0,1,...,n-1$
 $j=0,1,...,n-3$ and $X_{0(n-2)}$ $X_{0(n-1)}$, so the total number of
variables is $2n+n(n-2)+2=n^2+2$. This is usually too large to
calculate. For example, even when $n=4$, the $K$ matrix is $48\times
18$. The standard method to
find the dual cone $T$ from $K$ needs to analyse some $48!/(17!31!)$
vectors, which
is computationally prohibitive.

\addcontentsline{toc}{chapter}{Index}
\printindex
\end{document}